\documentclass{memoir}

\usepackage{amsmath,amsthm,amsfonts,amssymb,stmaryrd,mathrsfs}
\usepackage{enumerate,rotating,multirow,caption}
\usepackage{framed}
\usepackage{bbm}
\usepackage{pifont}
\usepackage{makeidx}

\usepackage{bussproofs}
\usepackage[all]{xy} 

\newcommand{\cat}[1][C]{\mathbf{#1}}
\newcommand{\hcat}[1][C]{\widehat{\mathbf{#1}}}
\newcommand{\Coalg}{\mathbf{CoAlg}}
\newcommand{\Alg}{\mathbf{Alg}}
\newcommand{\Set}{\mathbf{Set}}
\newcommand{\BA}{\mathbf{BA}}
\newcommand{\A}{\mathbf{A}}
\newcommand{\DL}{\mathbf{DL}}
\newcommand{\BDL}{\mathbf{BDL}}
\newcommand{\Pos}{\mathbf{Pos}}

\newcommand{\BAE}{\mathbf{BAE}}
\newcommand{\DLE}{\mathbf{DLE}}
\newcommand{\AEs}{\mathbf{AE}}
\newcommand{\HA}{\mathbf{HA}}
\newcommand{\Grp}{\mathbf{Grp}}
\newcommand{\Vect}{\mathbf{Vect}}
\newcommand{\el}[1][T]{\mathbf{El}(#1)}
\newcommand{\atob}[1][T]{\mathbf{Atom}(#1)}
\newcommand{\ob}{^{\mathbf{ob}}}
\newcommand{\mor}{^{\mathbf{mor}}}
\newcommand{\rel}[1][\mathbf{C}]{\mathbf{Rel}(#1)}

\newcommand{\pow}{\mathcal{P}}

\newcommand{\powf}{\mathcal{P}_{\omega}}
\newcommand{\cpow}{\mathsf{P}}
\newcommand{\cpowf}{\mathsf{P}_\omega}
\newcommand{\uf}{\mathsf{Uf}}
\newcommand{\pf}{\mathsf{Pf}}
\newcommand{\ups}{\mathcal{U}}

\newcommand{\Base}{\mathsf{B}}
\newcommand{\powc}{\mathcal{Q}}
\newcommand{\Bag}{\mathcal{B}}

\newcommand{\Free}{\mathsf{F}}
\newcommand{\Forg}{\mathsf{U}}
\newcommand{\Yon}{\mathsf{Y}}
\newcommand{\Filt}{\mathcal{F}}
\newcommand{\Inc}{\mathsf{I}}
\newcommand{\Hom}{\mathrm{Hom}}
\newcommand{\Sub}{\mathrm{Sub}}
\newcommand{\Nat}{\mathrm{Nat}}
\newcommand{\Diag}{\mathscr{D}}
\newcommand{\DiagF}{\mathscr{F}}

\newcommand{\rex}{\mathsf{Rex}}
\newcommand{\filtC}{\mathsf{Acc}}
\newcommand{\CC}{\mathsf{CoCo}}
\newcommand{\Id}{\mathsf{Id}}

\newcommand{\polyFunc}[1][\Sigma]{\mathsf{S}_{#1}}
\newcommand{\Gr}{\mathsf{Gr}}
\newcommand{\Cofree}{\mathsf{C}}
\newcommand{\LHey}{L_{\mathrm{Hey}}}
\newcommand{\LRL}{L_{\mathrm{RL}}}
\newcommand{\TRL}{T_{\mathrm{RL}}}
\newcommand{\THey}{T_{\mathrm{Hey}}}
\newcommand{\semTHey}{\delta^{\mathrm{Hey}}}
\newcommand{\semTRL}{\delta^{\mathrm{RL}}}

\newcommand{\lngS}{\mathcal{L}_S}
\newcommand{\plLang}[1][\Sigma]{\mathsf{L}_{#1}}

\newcommand{\lngT}{\mathcal{L}_T}

\newcommand{\KKV}{\mathsf{KKV}}
\newcommand{\logic}[1][L]{\mathbf{#1}}

\newcommand{\two}{\mathbbm{2}}
\newcommand{\inv}{^{-1}}
\newcommand{\ce}{^{\sigma}}
\newcommand{\ced}{^{\pi}}
\newcommand{\lang}{\mathcal{L}}
\newcommand{\op}{^\mathrm{op}}

\newcommand{\fto}{\rightarrow}
\newcommand{\lsem}{\llbracket}
\newcommand{\rsem}{\rrbracket}

\newcommand{\dia}{\diamondsuit}
\newcommand{\epi}{\twoheadrightarrow}
\newcommand{\mono}{\rightarrowtail}
\newcommand{\inc}{\hookrightarrow}
\newcommand{\ari}{\mathrm{ar}}
\newcommand{\restrict}{\upharpoonright}
\newcommand{\Lor}{\bigvee}
\newcommand{\Land}{\bigwedge}

\newcommand{\rng}{\mathsf{rng}}
\newcommand{\colim}{\operatornamewithlimits{colim}}
\newcommand{\Lan}{\mathrm{Lan}}
\newcommand{\Ran}{\mathrm{Ran}}
\newcommand{\im}{\mathrm{im}}

\newcommand{\up}{\hspace{2pt}\uparrow\hspace{-2pt}}
\newcommand{\down}{\hspace{2pt}\downarrow\hspace{-2pt}}
\newcommand{\eup}{^{\uparrow}}
\newcommand{\edown}{^{\downarrow}}
\newcommand{\vect}[1][x]{\overrightarrow{#1}}
\newcommand{\supp}{\mathrm{supp}}
\newcommand{\dom}{\mathrm{dom}}
\newcommand{\cod}{\mathrm{cod}}

\newcommand{\init}[1][T]{\mu{#1}}
\newcommand{\term}[1][T]{\nu{#1}}
\newcommand{\deriv}[1][EL]{\vdash_{\mathrm{#1}}}
\newcommand{\Ax}{\mathsf{Ax}}
\newcommand{\ba}[1][A]{\mathbb{#1}}
\newcommand{\id}{\mathrm{Id}}
\newcommand{\pb}{\mathrm{pb}}
\newcommand{\xto}{\xrightarrow}
\newcommand{\lRes}{\char `\\}
\newcommand{\rRes}{/}
\newcommand{\kop}[1][k]{\langle #1 \rangle}

\newcommand{\eA}{^\mathcal{A}}
\newcommand{\eAs}{^{\mathcal{A}^\sigma}}

\newcommand{\cho}{\mathsf{Choice}}

\newcommand{\Slift}[1][X]{\hspace{2pt}\bar{S}\hspace{-1pt}#1\hspace{-1pt}}

\newcommand{\Tlift}[1][X]{\hspace{1pt}\bar{T\hspace{-2pt}}#1\hspace{-1pt}}
\newcommand{\lift}[2]{\hspace{1pt}\overline{#1}\hspace{-1pt}#2\hspace{-1pt}}
\newcommand{\Smem}[1][X]{\hspace{2pt}\bar{S}\hspace{-4pt}\in_{#1}\hspace{-2pt}}
\newcommand{\Tmem}[1][X]{\hspace{2pt}\bar{T}\hspace{-4pt}\in_{#1}\hspace{-2pt}}
\newcommand{\cmark}{\ding{52}}
\newcommand{\xmark}{\ding{54}}

\theoremstyle{plain}
\newtheorem{theorem}{Theorem}[chapter]
\newtheorem{lemma}[theorem]{Lemma}
\newtheorem{proposition}[theorem]{Proposition}
\newtheorem{corollary}[theorem]{Corollary}

\newtheorem{propdef}[theorem]{Proposition and Definition}

\theoremstyle{definition}
\newtheorem{definition}[theorem]{Definition}
\newtheorem{example}[theorem]{Example}
\newtheorem{remark}[theorem]{Remark}

\makeindex

\begin{document}
\begin{titlingpage}
\pretitle{\begin{center}\LARGE University of London \\ Imperial College of Science, Technology and Medicine \\ Department of Computing \\ \vskip 12em  }
\posttitle{ \par\end{center}}

\title{\HUGE \textbf{Completeness-via-canonicity for coalgebraic logics}}

\author{\LARGE Fredrik Paul Herbert Dahlqvist}

\predate{\begin{center}\vskip 12em Submitted in part fulfilment of the requirements for the degree of \\
Doctor of Philosophy in Computing of the University of London and \\
the Diploma of Imperial College, }
\postdate{\par\end{center}}
\date{April 2014}

\maketitle
\end{titlingpage}

\pagenumbering{roman}
\pagestyle{plain}
\begin{abstract}
This thesis aims to provide a suite of techniques to generate completeness results for coalgebraic logics with axioms of arbitrary rank. We have chosen to investigate the possibility to generalize what is arguably one of the most successful methods to prove completeness results in `classical' modal logic, namely \emph{completeness-via-canonicity}. This technique is particularly well-suited to a coalgebraic generalization because of its clean and abstract algebraic formalism. In the case of classical modal logic, it can be summarized in two steps, first it isolates the purely algebraic problem of canonicity, i.e. of determining when a variety of boolean Algebras with Operators (BAOs) is closed under canonical extension (i.e. canonical). Secondly, it connects the notion of canonical varieties to that of canonical models to explicitly build models, thereby proving completeness.

The classical algebraic theory of canonicity is geared towards \emph{normal} logics, or, in algebraic terms, BAOs (or generalizations thereof). Most coalgebraic logics are not normal, and we thus develop the algebraic theory of canonicity for Boolean Algebra with Expansions (BAEs), or more generally for Distributive Lattice Expansions (DLEs). We present new results about a class of expansions defined by weaker preservation properties than meet or join preservation, namely (anti)-$k$-additive and (anti-)$k$-multiplicative expansions. We show how canonical and Sahlqvist equations can be built from such operations.

In order to connect the theory of canonicity in DLEs and BAEs to coalgebraic logic, we choose to work in the abstract formulation of coalgebraic logic. An abstract coalgebraic logic is defined by a functor $L:\BA\to\BA$, and we can heuristically separate these logics in two classes. In the first class the functor $L$ is relatively simple, and in particular can be interpreted as defining a BAE. This class includes the predicate lifting style of coalgebraic logics. In the second class the functor $L$ can be very complicated and the whole theory requires a different approach. This class includes the nabla style of coalgebraic logics.

For simple functors, we develop results on strong completeness and then prove strong completeness-via-canonicity in the presence of canonical frame conditions for strongly complete abstract coalgebraic logics. In particular we show coalgebraic completeness-via-canonicity for Graded Modal Logic, Intuitionistic Logic, the distributive full Lambek calculus, and the logic of trees of arbitrary branching degrees defined by the $\mathsf{List}$ functor. These results are to the best of our knowledge, new.

For a complex functor $L$ we use an indirect approach via the notion of \emph{functor presentation}. This allows us to represent $L$ as the quotient of a much simpler polynomial functor. Polynomial functors define BAEs and can thus be treated as objects in the first class of functors, in particular we can apply all the above mentioned techniques to the logics defined by such functors. We develop techniques that ensure that results obtained for the simple \emph{presenting} logic can be transferred back to the complicated \emph{presented} logic. We can then prove strong-completeness-via-canonicity in the presence of canonical frame conditions for coalgebraic logics which do not define a BAE, such as the nabla coalgebraic logics.
\end{abstract}

\pagestyle{empty}
\begin{center}\textbf{Acknowledgements}\end{center}
This thesis is the fruit of a long discussion (four years!) with my supervisor Dirk Pattinson, to whom I am immensely grateful for a number of things: for taking me on as his student, for passing on a great deal of technical knowledge and teaching me how to write a paper, for taking the time to see me every week (at least when we were not at opposite ends of the world), for his amazing ability to think, talk and solve problems for hours on end without need for food or drink, and for always being up for a pint! I would also like to thank Ian Hodkinson for stepping in after Dirk's move down under and his invaluable help when I was trying to understand canonicity. I am also very grateful to Alexander Kurz and Corina C\^{i}rstea for accepting to examine my thesis, and for doing it so conscientiously.

My thanks of course to Tanja, firstly for always encouraging me, despite the massive pay cut, to pursue what I really wanted to do; secondly for bringing Ingrid to the world, by far the most wonderful thing to have happened during my PhD years; and finally for her patience and sacrifices during the long writing up period. I also thank my family, and my parents in particular, for their constant support, understanding and encouragements, in the face of what might have appeared as a strange career move.

I must also thank all the people who made going to work such a fun and interesting experience, in no particular order Bj{\"o}rn, Ray, Florian, Victor, Tony, Rumi, Clemens, Nick, Faris and Rob.

\clearpage

\pagestyle{empty}
\begin{center}\textbf{Dedication}\end{center}

I declare that this thesis was composed by myself, that the work contained herein
is my own except where explicitly stated otherwise in the text, and that this work
has not been submitted for any other degree or professional qualification except
as specified.
\vspace{8ex}

\hfill (Fredrik Dahlqvist)

\vspace{70ex}
The copyright of this thesis rests with the author and is made available under a Creative Commons 
Attribution Non-Commercial No Derivatives licence. Researchers are free to copy, distribute or 
transmit the thesis on the condition that they attribute it, that they do not use it for commercial 
purposes and that they do not alter, transform or build upon it. For any reuse or redistribution, 
researchers must make clear to others the licence terms of this work
\clearpage

\tableofcontents

\mainmatter
\pagenumbering{arabic}
\chapter{Prolegomenon}

\section{What this works sets out to achieve}

\subsection{Introductory remarks}

Coalgebras have gained popularity as an elegant and general framework to study and represent a wide variety of dynamical systems in computer science (see \cite{Rutten00}) and even in physics (see \cite{2009:abramsky:coalgChu}). The fact that they encompass so many of the most frequently used models of computation suggests that they provide the correct level of abstraction to formalize the dynamics of state based systems.

In parallel to the research on coalgebras \emph{per se}, the field of coalgebraic logic has emerged as a unifying framework for the many types of (modal) logics used in knowledge representation and to reason about dynamical systems (see \cite{DirkOverview} for an overview). One of the great insights in the relationship between coalgebras and coalgebraic logics, is that the class of all $T$-coalgebras for a functor $T$ can always be characterised logically by its one-step behaviour, i.e. axioms and rules with nesting depth of modal operators uniformly equal to 1 (see \cite{Schroder06}). However, once the transition type (i.e. the functor $T$) has been described logically in such a way, one may be interested in subclasses of $T$-coalgebras which are characterised by more complex axioms (such as transitivity for example) which we will refer to as \emph{frame conditions}. The problem of logically characterising subclasses of the class of all $T$-coalgebras for an arbitrary functor $T$ is by and large still open. To our knowledge, Pattinson and Schr{\"{o}}der's \cite{BeyondRank1} offer the only attempt at making progress in this direction and offer a solution for some of the standard frame conditions of classical modal logic.

This thesis aims to provide a suite of techniques to generate completeness results for coalgebraic logics with axioms of arbitrary rank. As practitioners of modal logic know all too well, there is no panacea for proving completeness in the presence of frame conditions, and the generality of coalgebraic logic only makes this observation more true. Thus what we ambition to provide is not a general theory of completeness in the presence of frame conditions, but rather \emph{some} techniques to deal with \emph{some} frame conditions in \emph{some} coalgebraic logics. In this spirit, we have chosen to investigate the possibility to generalize what is arguably one of the most successful methods to prove completeness results in `classical' modal logic (i.e. the normal modal logic of e.g. \cite{2001:ModalLogic}), namely \emph{completeness-via-canonicity}. This technique is particularly well-suited to a coalgebraic generalization because of its clean and abstract algebraic formalism. In the case of classical modal logic, it can be summarized in two steps, first it isolates the purely algebraic problem of determining when a variety of boolean Algebras with Operators (BAOs) is canonical, i.e. closed under canonical extension. Secondly, it connects the notion of canonical varieties to that of canonical models, i.e. the algebraic and syntactic world to the coalgebraic and semantic world, and explicitly builds models proving completeness.

There are two challenges in trying to generalize classical completeness-via-canonicity to the coalgebraic framework. Firstly, the classical algebraic theory of canonicity (see for example in chronological order \cite{Jonsson51, Ribeiro52, 1975:Sahlqvist,1989:Goldblatt,Jonsson94,1995:Goldblatt,deRijkeVenema95, 2001:GehrkeHarding,2004:GehrkeJonsson,2005:GehrkeVenema,2006:VenemaAC}) is geared towards \emph{normal} logics, or, in algebraic terms, BAOs (or generalizations thereof). Most coalgebraic logics are not normal, and we thus need to develop the algebraic theory of canonicity for boolean Algebra with Expansions (BAEs), i.e. boolean algebras with additional operators that have \emph{a priori} no preservation properties. Our key findings on this topic are presented in Chapter 2 and can be summarized as follows:
\begin{itemize}
\item Some results for non-normal expansions can already be found in \cite{2001:GehrkeHarding,2004:GehrkeJonsson,2005:GehrkeVenema,2006:VenemaAC}, and we generalize them further to isotone and antitone expansions of arbitrary arities

\item We study the canonical extension of expansions which are (anti-)$k$-additive (see \cite{1970:Henkin}) or (anti-)$k$-multiplicative. To the best of our knowledge, our results on this class of expansions are new. A very interesting application is Graded Modal Logic (GML) whose modalities are $k$-additive (and whose dual modalities are $k$-multiplicative).
\item We show how we can define canonical identities via a generalization of the notion of Sahlqvist identity to non-normal modal logics whose expansions are (anti-)$k$-additive or (anti-)$k$-multiplicative, or (anti)-preserve up-directed joins, or (anti-)preserve down-directed meets.
\end{itemize}

The second challenge is to relate this purely algebraic theory to the coalgebraic semantic of coalgebraic logic. As we shall detail later in this Chapter, there are three flavours of coalgebraic logic: the predicate lifting, the nabla, and the abstract flavour. We choose to place ourselves in the abstract framework for most of this thesis because it is at this level of abstraction that results are easiest to get and cleanest to present. Moreover, the abstract flavour subsumes the other two. An abstract coalgebraic logic is essentially defined by a functor $L:\BA\to\BA$, and we can heuristically separate these logics in two classes. In the first class the functor $L$ is relatively simple, and in particular can be interpreted as defining a BAE; this case can be thought of as the abstraction of the predicate lifting flavour of coalgebraic logic. In the second class the functor $L$ can be very complicated and the whole theory requires a different approach; this case can be thought of as the abstraction of the nabla flavour of coalgebraic logic.

The main results concerning the first class of abstract coalgebraic logics are
\begin{itemize}
\item In Theorem \ref{ch5:thm:strongcomplRelational} we show strong completeness for a very large class of positive coalgebraic logic, which we call \emph{relational logics} and which includes positive modal logic, intuitionistic logic and all distributive substructural logics. This results generalises results published in \cite{2015:self}.
\item In the case of boolean coalgebraic logics, we prove a suite of Theorems \ref{ch5:thm:deltahatsurj}-\ref{ch5:thm:strgcomp3} which guarantee the existence of (quasi)-canonical models, and thus strong completeness. The key requirement is that the functor defining the semantics must weakly preserve cofiltered limits.
\item For functors which do not weakly preserve cofiltered limits, we present a new technique which we call \emph{semantic completion}, which extends the semantics in a generic way in order for Theorems \ref{ch5:thm:deltahatsurj}-\ref{ch5:thm:strgcomp3} to be used.
\item We clarify the connection between the syntactic notion of \emph{canonical extension} and the semantic notion of \emph{canonical model} in Theorems \ref{ch5:thm:CanExtJonTarksiExt} and \ref{ch5:thm:CanExtJonTarksiExtBAE} in a coalgebraic way. To our knowledge this is the first time this question is addressed, let alone answered.
\item With the connection between canonical extensions and canonical models clarified, we present coalgebraic completeness-via-canonicity results for functors defining DLEs and BAEs in Theorems \ref{ch5:thm:strongcompcan} and \ref{ch5:thm:strongcompcan2}.
\end{itemize}

The second class of abstract coalgebraic logic requires an indirect approach via the notion of \emph{functor presentation}. If the functor $L$ presenting the logic is very complex, as is typically the case in the nabla flavour of coalgebraic logic, the main technical tool for dealing with this complexity is to represent $L$ as the quotient of a much simpler polynomial functor. Polynomial functors define BAEs and can thus be treated as objects in first class of functors, in particular we can apply all the above mentioned techniques to the logics defined by such functors. We develop techniques that ensure that results obtained for the simple \emph{presenting} logic can be transferred back to the complicated \emph{presented} logic. The main results of this line of research are:
\begin{itemize}
\item We study functor presentation in great detail and show how it is related to the classical category theoretic construction presenting a $\Set$-functor as a colimit of representable functors. We show a very close relationship between the notion of functor presentation and the notion of presentation by generators and relations for an algebraic variety. In particular, we define notions of $\lambda$-presented and $\lambda$-generated functors, and show that they are equivalent to being $\lambda$-presentable or $\lambda$-generatable in a category of functor. The latter concepts being precisely those which abstract the usual notion of being $\lambda$-presented or $\lambda$-generated for an algebraic variety.
\item We show in Chapter 4 how natural transformations (such as functor presentations) can be used to (1) translate between abstract coalgebraic logics (Theorem \ref{ch4:thm:syntaxthm}), (2) translate between their semantic domains (Theorem \ref{ch4:thm:semthm}), (3) translate syntax and semantics simultaneously whilst preserving satisfiability (Theorem \ref{ch4:thm:compSem}), and finally (4) translate between proof systems in a way that preserve derivability in a precise way (Theorem \ref{ch4:thm:prooftrans}).
\item  By using presentations by polynomial functors, all our translation results, and our completeness-via-canonicity results on the BAEs defined by polynomial functors, we show a step-by-step technique for proving strong-completeness-via-canonicity in the presence of canonical frame conditions for coalgebraic logics which do not define BAEs, such as the nabla coalgebraic logic for the finitary powerset functor (Section \ref{sec:transmeth}).
\end{itemize}


\subsection{Structure, summary and main contributions of the thesis.}

\begin{itemize}
\item \textbf{Chapter 1: Prolegomenon}

\begin{itemize} 
\item \textbf{Section \ref{ch1:sec:boolstruct}: boolean structures}, in which all the basic constructs on boolean algebras are introduced concretely, as well as in the appropriate categorical setting. The question of when the category $\Alg_{\BA}(L)$ (for a varietor $L:\BA\to\BA$) is monadic over $\Set$ is raised and the answer, which depends on results from the next Section, is presented.
\item \textbf{Section \ref{ch1:sec:rel}: Relations, categorically}: in which the notions of relation, relation lifting, equivalence relation and fully invariant relation are defined and developed in the context of regular categories. Although the results of Propositions \ref{ch1:prop:liftingwpb} and \ref{ch1:prop:liftingProp} are known, the direct proofs we present are, to our knowledge, new and provide a simple and yet purely categorical handle on the key properties of relation liftings which are used extensively in the nabla and abstract flavours of coalgebraic logic. The study of the (Barr-) exactness of categories of algebras is a relatively straightforward exercise on regular categories, but provides a nice new framework for the algebraic semantics of abstract coalgebraic logics developed in Section \ref{ch1:sec:coalglog}. Many ideas originate from the thesis of Hughes \cite{HughesPhD} on categories of algebras and coalgebras. The notion of fully invariant relation was defined in \cite{HughesPhD}, but our construction of the fully invariant closure of a relation (Proposition \ref{ch1:prop:fullInvClosure}) is, to the best of our knowledge, new.
\item \textbf{Section \ref{ch1:sec:coalglog}: Coalgebraic logics}, in which coalgebraic languages, their semantics and their axiomatizations are defined. We pay particular attention to working out the details of the abstract flavour of coalgebraic logic and its connections with the other two flavours. We present an algebraic semantic for abstract coalgebraic logics in great detail and at what we believe to be the correct level of abstraction, i.e. working over the category $\BA$. Most of this material is new, and generalizes/categorifies earlier work on the topic such as \cite{AlgSem2004}), relying heavily on the notions developed in the previous section, notably the notion of fully invariant closure.
\end{itemize}

\item \textbf{Chapter 2: Algebraic Canonicity}
\begin{itemize}
\item \textbf{Section \ref{ch2:sec:canext}: Canonical Extensions}, in which basic concepts such as canonical extensions, compactness, closed and open elements, etc are introduced. Whilst all the results shown in this section are known, we have opted for a purely algebraic presentation, i.e. we make no reference to duality (Priestley or Stone spaces) as is usually the case in the literature. The basic properties of the canonical extensions of maps on distributive lattices or boolean algebras, developed in e.g. \cite{1994:GehrkeJonsson,2001:GehrkeHarding,2004:GehrkeJonsson,2006:VenemaAC}, are generalised to $n$-ary monotone maps, i.e. maps $f$ such that for each argument, $f$ is either order preserving (isotone) or order reversing (antitone) in this argument. 
\item \textbf{Section \ref{ch2:sec:algprop}: Algebraic properties of the canonical extension}, in which known results results from \cite{1994:GehrkeJonsson,2001:GehrkeHarding,2004:GehrkeJonsson,2006:VenemaAC} are extended to the case of maps which are (anti-)$k$-additive or (anti-) $k$-multiplicative in Theorem \ref{ch2:thm:complkadd}. This Theorem generalises known results about join and meet-preserving maps. Lemma \ref{ch2:lem:ineq} which shows how canonical extension and function composition interact is of paramount importance for the rest of the Chapter. 
\item \textbf{Section \ref{ch2:sec:topol}: Topology to the rescue}, in which topological techniques developed in \cite{2004:GehrkeJonsson} and \cite{2006:VenemaAC} are presented and extended to the case of $n$-ary maps which are isotone or antitone in each argument (Theorem \ref{ch2:thm:topolChar}). We also show that (anti-)$k$-additive and (anti-)$k$-multiplicative maps are smooth (Theorem \ref{ch2:thm:kaddSmooth}); to the best of our knowledge this result is new. These topological results, and in particular the \emph{principle of matching topologies} of \cite{2006:VenemaAC}, will form the backbone of our study of canonicity. They provide us with conditions under which the converse of the inequality of Lemma \ref{ch2:lem:ineq} holds, i.e. conditions under which canonical extension and function composition commute. This property is key  to building canonical terms.
\item \textbf{Section \ref{ch2:sec:can}: Canonicity}, in which the notions of canonical equation, stable and expanding terms are defined. Our presentation broadly follows that of \cite{Jonsson94}, defining terms as maps and studying their stability under canonical extension. The key result is the Principle of Matching Topology, an idea of \cite{2006:VenemaAC}, which we use to generate a large number of conditions under which terms are stable under canonical extension. The results are presented in Table \ref{ch2:table}.
\item \textbf{Section \ref{ch2:sec:Sahl}: Sahlqvist identities}, in which we attempt to clarify and formalise the notion of quasi-equation which is the key technical tool used in \cite{Jonsson94} to define Sahlqvist identities. We suggest an abstract definition of Sahlqvist identities for a general BAE, based on algebraic properties of terms. We then present concrete - i.e. syntactically defined - Sahlqvist identities; first the classical Sahlqvist identities of modal logic, then a more general definition of Sahlqvist identities for BAEs whose operations satisfy one of the preservation properties listed in Table \ref{ch2:table}. We use Tables \ref{ch2:table} and \ref{ch2:table2} as the basis of a method for building these general Sahlqvist identities. We illustrate the method by examples, including an example of a general Sahlqvist formula for the classical modal logic, which is not a classical Sahlqvist formula.
\item \textbf{Section \ref{ch2:sec:Ex}: Applications.} In this section we present interesting examples of DLEs and BAEs and define notions of Sahlqvist identities for them. We start with the case of Graded Modal Logic (GML) which is characterized by BAEs whose expansions are $k$-additive. Canonical equations and Sahlqvist identities can therefore be defined by using the results on $k$-additive maps developed in the Chapter. Next, we present Intuitionistic Logic (IL) and the distributive Lambek calculus as examples of positive modal logics with binary expansions whose arguments are all either isotone or antitone. The logics provide examples of expansions which preserve joins or preserve meets or anti-preserve joins in their arguments. We then use results developed in the Chapter to define Sahlqvist identities in these logics. 
\end{itemize}
\item \textbf{Chapter 3: Functor Presentations}
\begin{itemize}
\item \textbf{Section \ref{ch3:sec:base}: The base transformation}, where the base transformation, which is key to the nabla flavour of coalgebraic logic, is defined and discussed in a very general categorical setting, viz. for any $\Set$-valued functor on a locally small, well-powered category with pullbacks of monos (i.e. intersections). Theorem \ref{ch3:thm:baseNat} is a generalisation of results from \cite{Gumm05}, and characterises functors for which the base transformation is natural.
\item \textbf{Section \ref{ch3:sec:funcPresIntro}: The Concept of Functor Presentation}, in which the most basic concepts of the theory of functor presentation are introduced: the category of elements of a functor and the associated expression of a functor as a colimit of representables (Theorem \ref{ch3:thm:fstpresthm}). The analogy between functor presentations and algebraic varieties, which will be the unifying thread of the Chapter, is first hinted at.
\item \textbf{Section \ref{ch3:sec:tools}: Some Essential Tools and Concepts}, in which the concepts of locally presentable and accessible categories, left Kan extensions, canonical and cofinal diagrams are introduced. Proposition \ref{ch3:prop:LanF0} shows how how a $\lambda$-accessible functor is determined uniquely, via the left Kan extension, by its restriction to $\lambda$-presentable objects.
\item \textbf{Section \ref{ch3:sec:sizeBoundPres}: Size-Bound Functor Presentations.} In this large section we develop the analogy between $\Set$-valued functors on a small category and elements of an algebraic variety. We define two notions of a functor being $\lambda$-presented and show that they are equivalent in Proposition \ref{ch3:prop:StrongPresIffPres}. The key result of this section is Proposition \ref{ch3:prop:presfunc} which shows that a functor is finitely presented iff it is a finitely presentable object in $\Set^{\cat}$. We then define two notions of a functor being $\lambda$-generated and determine a condition on $\cat$ that ensures that they agree. The notion of a `generatable' object is our terminology for the notion of generated of \cite{LPAC}. Proposition \ref{ch3:prop:geniffgen} is the equivalent for generated functors of Proposition  \ref{ch3:prop:presfunc}, i.e. it shows that a functor is finitely generated iff it is a finitely generatable object in $\Set^{\cat}$. A summary of the analogy between functors in $\Set^{\cat}$ and elements of an algebraic variety is presented in Table \ref{ch3:table1}. The section ends with a re-interpretation of some of its results in terms of various notions of cocompletion.
\item \textbf{Section \ref{ch3:sec:presaccessfunc}: Presentations of accessible functors }, where the theory developed for $\Set$-valued functors over a small category is extended to the case of accessible $\Set$-valued functors on accessible categories. We show how the left Kan extension construction and the accessibility assumption allows us to lift all result from the previous section to accessible categories. In particular we define and characterise presented/presentable and generated/generatable accessible functor and show that the category of $\lambda$-accessible functors is locally finitely presentable (Proposition \ref{ch3:prop:accfuncLocFinPres}). We also present a class of accessible categories over which finitely presentable and finitely generatable functors coincide (Proposition \ref{ch3:prop:locfinvar}). We then introduce the so-called canonical presentation, and show a generic method for removing the redundancy of this presentation via the notion of atomic elements in the category of elements, itself related to the concept of base. We call the presentation obtained in this way the minimal presentation (already present in some form in \cite{AAC})).
\item \textbf{Section \ref{ch3:sec:liftpres}: Lifting presentations using adjunctions}, in which we show how we can extend the result of the previous section from the case of functors $\cat\to\Set$ to the case of functors $\cat\to\cat$ when there exists an adjunction $\Free\dashv\Forg: \Set\to\cat$. We show how any finitary functor $T:\cat\to\cat$ can be given such a presentation if (1) the adjunction is finitary, (2) the adjunction is of descent type, (3) $T$ preserves regular epis, and (4) the composition of regular epis is a regular epi in $\cat$. All these conditions are met by our main application, namely endofunctors on $\BA$. 
\end{itemize}
\item \textbf{Chapter 4: Translations}
\begin{itemize}
\item \textbf{Section \ref{ch4:sec:syntrans} Syntax translation}. In this section we examine how a natural transformations between functors presenting two coalgebraic languages defines a translation map between these languages. We show how this translation, which is in fact a natural transformation, is defined, that it exists, and that it is a regular epi-transformation whenever the natural transformation between the functors is in Theorem \ref{ch4:thm:syntaxthm}
\item \textbf{Section \ref{ch4:sec:semtrans} Models and semantic translations}, in which we show how a natural transformation between two functors defining different coalgebraic semantics defines translation between the two types of models. We show in Theorem \ref{ch4:thm:semthm} how the cofree coalgebra for the two functors are related by this translation map, and in particular that if the natural transformation is epi, so is the translation map. Next, we show how the syntax and semantic translations can be combined in a compatible way and prove that the truth predicate is preserved by such compatibles translations. We then show how to implement the notion of compatible semantics in the case of a the nabla logics associated with functors $S,T:\Set\to\Set$ for which there exists an epi-transformation $q: S\epi T$. This involves presenting nabla logic in the abstract style and some detailed examination of the connection between derivability under the proof systems $\KKV(S)$ and $\KKV(T)$ associated with $S$ and $T$.
\item \textbf{Section \ref{ch4:sec:prooftrans}: Proof-theoretic translation} in which proof systems are translated via the syntax translation defined in Section \ref{ch4:sec:syntrans}. Given two functors $K,L$ presenting two abstract coalgebraic logics, a epi-transformation $q: K\epi L$ and a set of equations for the $L$-logic, we show how this set of equations can be lifted to the $K$ language. In this way we show how the abstract Hilbert systems defined in Section \ref{ch1:subsec:algsem} can be lifted along an epi-transformation. The key result is Theorem \ref{ch4:thm:prooftrans} which shows that there is a very tight relationship between the variety defined by the $L$- equations and the variety defined by the lifted $K$-equations.
\end{itemize}
\item \textbf{Chapter 5}
\begin{itemize}
\item \textbf{Section \ref{ch5:sec:weakComp}: Weak completeness.} We start this Chapter by introducing classical results about weak completeness in the abstract flavour of coalgebraic logic, and in particular the connection with the injective nature of the semantic natural transformation. For an abstract coalgebraic logic given by a triple $(L:\BA\to\BA, T:\Set\to\Set, \delta: L\pow\to\pow T)$, the precise connection between the free $L$-algebra over the boolean algebra $\Free V$ and the cofree coalgebra over the set (of colours) $\mathcal{Q}V$ is established in Theorem \ref{ch5:thm:strgcomp1}. This result is probably folklore, but left us perplexed for a long time until the connection between the adjunctions $\uf\dashv\pow$ and $\Free\dashv\Forg$ became clear. This greatly clarifies the role of propositional variables and valuations. We also review the notion of filtration and of expressivity in the abstract setting.
\item \textbf{Section \ref{ch5:sec:strongComp}: Strong completeness.} This Section starts by reformulating and developing some recent results from \cite{2009:DirkStrongComp} and \cite{2012:KurzStrongComp} about strong completeness in coalgebraic logic, most notably the coalgebraic J\'{o}nsson-Tarski Theorem (Theorem \ref{ch5:thm:jontarski}). We then give a list of Theorems under which strong completeness is guaranteed and illustrate them with important Examples which are developed in some detail. The case of positive logics (i.e. logics defined over $\DL$) is treated first and strong completeness is shown in Theorem \ref{ch5:thm:strongcomplRelational} for a very wide class of logics which generalise positive modal logics and which we call \emph{relational logics}. This Theorem generalises a result from \cite{2015:self}. The case of boolean logics is treated next, and a particularity of $\Set$, namely that all epis are split, can be exploited to provide conditions for strong completeness in Theorem \ref{ch5:thm:deltahatsurj}-\ref{ch5:thm:strgcomp3}. We study Examples of boolean logics in detail, with a particular emphasis on establishing the injectivity and surjectivity of the semantic transformation and of its transpose. One of our main tools for this is the construction of retractions and sections, which is a technique we have not seen being used elsewhere in this context, and our direct proofs of weak completeness for modal logic and graded modal logics are to our knowledge new. In the case of boolean logics, the preservation of cofiltered limits is a key requirement on the semantic functor in order to get strong completeness and we present a generic technique based on the right Kan extension construction to modify a functor in a way that ensures this preservation property. This new technique, which we call \emph{semantic completion} is applied to the case of graded modal logic, where it is show to yield strong completeness. 
\item \textbf{Section \ref{ch5:sec:compCanon}: Completeness-via-canonicity}. This section gathers all the results of the thesis, into proofs of completeness-via-canonicity in the presence of canonical frame conditions. It starts by showing when the extension obtained by the coalgebraic J\'{o}nsson-Tarski Theorem coincides with the canonical extension defined in Chapter 2. To our knowledge, this question was never addressed, let alone answered, previously and Theorems \ref{ch5:thm:CanExtJonTarksiExt} and  \ref{ch5:thm:CanExtJonTarksiExtBAE} provide some clarity about the exact relation between the \emph{syntactic} notion of canonical extension, and the \emph{semantic} notion of canonical model at an abstract, coalgebraic, level. We then give coalgebraic completeness-via-canonicity results in the case of logics defined by endofunctors in $\DL$ or $\BA$ which define DLEs or BAEs (Theorems \ref{ch5:thm:strongcompcan}-\ref{ch5:thm:strongcompcan2}). We illustrate the results with Examples which include classical and graded modal logics, as well as intuitionistic logic and the distributive Lambek calculus (these examples have been published in \cite{2015:self}). We finish the Chapter with the more complicated case of logics whose defining endofunctor must be presented as the quotient of a polynomial functor. We detail a procedure for proving completeness-via-canonicity in this case too by exploiting the results of Chapter 4. The method is illustrated in the case of the nabla logic for the finitary covariant powerset functor $\cpowf$.
\end{itemize}
\end{itemize}

\section{Notation and Conventions}
All important terms will be written in bold font where they are first introduced, the corresponding entry in the index will usually link to this place in the text. In terms of notational convention, we have tried to adhere to the following guiding principles:
\begin{enumerate}
\item Categories are denoted in bold font, e.g. $\cat,\Set,\BA,...$ and we reserve the letters $\cat[I]$ and $\cat[J]$ for filtered or directed (index) categories
\item Functors come in a variety of shapes depending on custom and requirements. However, functors defining diagram will be denoted by a calligraphic font, typically $\mathscr{D}$. Many functors will be denoted using a sans-serif font, in particular free and forgetful functors will be written as $\Free,\mathsf{G}$ and $\Forg,\mathsf{V}$ respectively, and the identity functor on a category $\cat$ as $\Id_{\cat}$. We will use three types of powerset functors: $\pow: \Set\op\to\BA$, $\mathcal{Q}:\Set\op\to\Set$ and $\mathsf{P}:\Set\to\Set$.
\item Natural transformations will be denoted by lower-case Greek letters, typically $ \delta,\epsilon,\zeta,\eta,\theta$.
\item Sets will be denoted by $X,Y,Z$ and subsets by $U,V,W$, whilst elements will be denoted by $x,y,z$
\item Objects in a general category will be denoted by $A,B,C$ and morphisms will be written using lower case Roman letters, typically, $f,g,h,k$. The letters $i,m$ will often be used for monomorphisms, the letters $e,p,q$ for epimorphisms.
\item We follow the notation of \cite{KKV:2012:Journal} for anything related to the nabla-style of coalgebraic logic. In particular, formulas will be denoted by $a,b,c,...$ throughout. Propositional variables will be written as $p,q,r,...$. 
\item When the need arises to differentiate between a boolean algebra an its carrier, the former will typically be written as $\mathbb{A}$ and the latter by $A$.
\item The category of algebras for an endofunctor $L:\cat\to\cat$ will be denoted $\Alg_{\cat}(L)$. Dually the category of coalgebras for an endofunctor $T:\cat\to\cat$ will be denoted $\Coalg_{\cat}(T)$. We will drop the subscript category if $\cat=\Set$ and there is no ambiguity. In the case of algebra for a monad $T:\cat\to\cat$, we stick to conventional notation $\cat^T$. Generic structure maps and transition maps will be written using lower case Greek letters.
\item We have opted for the lower case spelling of `boolean', both because it will occur very frequently in the text, and because we feel that it is a greater honour to George Boole to have his name turned into a regular adjective rather than a unusual, capitalized, one.
\end{enumerate}

\section{Reasoning structures}\label{ch1:sec:boolstruct}

One of the attractive features of coalgebraic logic is its relative parametricity in the choice of algebraic structure for the syntax of the logics. As long as a dual adjunction exists with the category which the coalgebraic models will inhabit, we are free to choose the `reasoning kernel' of our logic. In other words, the framework of coalgebraic logic is parametric in the choice of a fundamental reasoning structure, as long as it comes with a dual adjunction to an appropriate `semantic' category. Several such adjunctions are presented in \cite{JacobsExpressivity}. In this work we will focus on two reasoning structures: the category of boolean algebras ($\BA$) and the category of distributive lattices ($\DL$). They will play a key role both as the main algebraic environment in which to study canonicity, and as the base category in which the `abstract' version of coalgebraic logic is formulated. This latter use will require a fair amount of categorical information about $\BA$ and $\DL$, and algebras defined over them. Throughout this work we use $\A$ to denote either $\BA$ or $\DL$, thus a statement about $\A$ will be equally applicable to either categories.

\subsection{The category $\BA$}

Let us briefly summarize some of the fundamental properties of the category $\BA$ whose objects are boolean algebras and whose morphisms are boolean homomorphisms. We assume all the basic facts about boolean algebras (BAs), for details we refer the reader to the excellent introduction by Givant and Halmos \cite{GivantHalmos}. Perhaps the most important feature of $\BA$ is that, together with the familiar (faithful) forgetful functor $\Forg:\BA\to\Set$, it forms a concrete category $(\BA,\Forg)$ with all the attributes of a finitary algebraic category. Since these attributes are common to all such categories, we will simply list the following useful facts and refer the reader to e.g. \cite{cats,LPAC,AlgTheor,Borceux,Borceux2,MacLane} for all the details
\begin{enumerate}[{Fact} 1]
\item  $\Forg$ has a left adjoint $\Free:\Set\to\BA$ which builds the free boolean algebra over a given set (of propositional variable).
\item $\Forg$ is therefore continuous\index{Continuous functor}, i.e. it preserves all limits. In fact, $\Forg$ \emph{creates} all limits.
\item As a consequence of the above, $\BA$ is complete, since $\Set$ is.
\item All of the above facts are a consequence of the fact that $\Forg$ is \textbf{monadic}\index{Monadic} (see e.g. Proposition 20.12 of \cite{cats}), i.e. that $\Forg\Free$ is a monad and that $\BA$ is isomorphic to the category of $\Set^{\Forg\Free}$ of $\Forg\Free$-algebras (in the sense of algebra for a monad).
\item Dually, $\Free$ is co-continuous\index{Co-continuous}, i.e. it preserves all the colimits.
\item \label{ch1:fact:descent}Since $\Free\dashv\Forg$ is monadic, it is in particular of descent type (see \cite{1993:KellyPower,2011:KurzEqPresMon}) and thus its counit $\epsilon:\Free\Forg\to\Id_{\BA}$ is a regular epi-transformation.
\item \label{ch1:fact:filtcol} $\Forg$ \emph{creates} filtered colimits (see e.g. \cite{MacLane} p.213).
\item $\BA$ is a \textbf{regular category}\index{Regular category}, i.e. it is finitely complete, it has coequalizers of kernel pairs and the pullback of a regular epimorphism along any morphism is a regular epimorphism.
\item In particular, $\BA$ has regular-epi-mono factorisation.
\item Regular epimorphisms are those whose underlying set function is surjective
\item Like all finitary varieties (see Corollary 3.7 of \cite{LPAC}), $\BA$ is a a locally finitely presentable category (see Chapter 3 for details), and in particular is cocomplete.
\item The free boolean algebras over finite sets of variables are regular projectives\index{Regular projective}, i.e. projective with respect to regular epis, and they form a dense collection, i.e. every boolean algebra is a canonical colimit of free boolean algebras
\item Moreover, $\BA$ is a locally finite variety (see \cite{2001:NicksBro}), i.e. finitely generated  boolean algebras are finite (i.e. have a finite underlying set) (see \cite{GivantHalmos} for a proof of this). \label{ch1:fact:locallyfinite}
\end{enumerate}

Apart from the last fact, we would like to emphasise that theses facts apply to any finitary variety, in particular to the boolean algebras expansions (BAE) which we will soon consider. 

Recall that, as is the case in any finitary variety, boolean algebras have a \textbf{presentation by generators and relations}\index{Presentation by generators and relations}. Formally, this means that for any $A$ in $\BA$, we can find a \emph{set} $X$, from which freely generated terms are built, and a pair of jointly monic arrows $e_1,e_2: E\to \Forg\Free X$ which selects freely generated terms to be identified, such that $A$ is the coequalizer of the adjoint transpose maps $\hat{e}_1, \hat{e}_2: \Free E\to \Free X$ (i.e. the maps freely generated from $e_1,e_2$). 
\[
\xymatrix
{
FE\ar@<1ex>[r]^{\hat{e}_1}\ar@<-1ex>[r]_{\hat{e}_2} & \Free X\ar@{->>}[r]^{q} & A
}
\]
If $X$ is finite, then $A$ is said to be \textbf{finitely generated}\index{Finitely generated!algebraic object}. If both $X$ and $E$ are finite, then $A$ is said to be \textbf{finitely presented}\index{Finitely presented!algebraic object}. Note that $q$ is a regular epi, and indeed we could abstract away from $e_1,e_2$ and write $A$ as a regular quotient of a free algebra, however the information on the cardinality of $E$ is then a lot less transparent. Typically, an algebra will have numerous presentations by generators and relations, but to see that such a presentation  always exists, note that from Fact \ref{ch1:fact:descent}, every boolean algebra $A$ is a regular quotient of a free boolean algebra:
\[
\xymatrix
{
\ker \epsilon_A\ar@<1ex>[r]^{p_1}\ar@<-1ex>[r]_{p_2} & \Free\Forg A\ar@{->>}[r]^{\epsilon_A} & A
}
\]
where $p_1,p_2$ are the projections of the kernel pair $\ker\epsilon_A$ (all regular epis are the coequalizers of their kernel pairs in a regular category). This presentation is trivial in the sense that if we consider the adjoint transpose $\hat{\epsilon}_A:\Forg A\to\Forg A$ of $\epsilon_A$, it is easy to check that $\hat{\epsilon}_A=\id_{\Forg A}$, which makes this presentation pretty poor in terms of information compression.

We conclude with a comment on colimits in $\BA$. As was have just stated, $\BA$ is cocomplete, but unlike limits which are essentially built in $\Set$ (by Fact 2), colimits are less straightforward (unless, by Fact 7, they are filtered).  Coproducts in $\BA$ are usually built as products in the dual category of Stone spaces, however, they can be built directly in $\BA$ as an example of `free product' and this can be done particularly naturally if we use presentations by generators and relations. Consider $A,B$ in $\BA$ and assume that they have presentations as coequalizers of $e_1,e_2: FE\to FX$ and $e_1',e_2': FE'\to FX'$ respectively. Now consider the following commutative diagram:

\[
\xymatrix@C=12ex
{
\Free E\ar@<-1ex>[d]_{e_1} \ar@<1ex>[d]^{e_2} \ar[r]^{i_E}& \Free E+\Free E'\ar@<-1ex>@{-->}[d]_{i_E\circ e_1+i_{E'}\circ e_1'} \ar@<1ex>@{-->}[d]^{i_{E}\circ e_2+i_{E'}\circ e_2'} & \Free E'\ar@<-1ex>[d]_{e_1'} \ar@<1ex>[d]^{e_2'}\ar[l]_{i_{E'}}\\
\Free X\ar@{->>}[d]_{q}\ar[r]_(.3){i_{\Free X}} & \Free X+\Free X'\simeq\Free(X+X')\ar@{-->>}[d]^{i_A\circ q+i_B\circ q'} & \Free X'\ar@{->>}[d]^{q'}\ar[l]^(.3){i_{\Free X'}} \\
A\ar[r]_{i_A} & A+B & A \ar[l]^{i_B}
}
\]
It is straightforward to check that $i_A\circ q+i_B\circ q'$ is indeed the coequalizer of the two arrows above it: note first that by the universal property of coproducts it is a cocone of the diagram defining the coequalizer, since it is a rival cocone for $\Free X+\Free X'$ and the diagram commutes. To see that it is a limiting cocone, assume that some $C$ coequalizes $i_E\circ e_1+i_{E'}\circ e_1'$ and $i_{E}\circ e_2+i_{E'}\circ e_2'$, then it is easy to check that by commutativity of the diagram if would also coequalize $e_1,e_2$ and $e_1',e_2'$. There would therefore exist unique arrows $A\to C$ and $B\to C$, and this would in turn trigger the existence of a unique arrow $A+B\to C$ proving that $A+B$ is indeed the coequalizer. What this means, is that the coproduct $A+B$ can be described as the boolean algebra given by the presentation whose set of generators is $X+X'$, i.e. the disjoint union of the generators of $A$ and $B$, and whose relations are given by the coproduct of the relations defining $A$ and $B$. 

Coequalizers are more straightforward: let $f,g:A\to B$ be two boolean homomorphisms. We define the equivalence relation $\equiv$ (more on equivalence relations in $\BA$ in the next section) as the smallest equivalence relation on the set $\Forg B$ generated by 
\[b\equiv b'\iff \exists a\in A \text{ s.th. }b=f(a), b'=g(a)\]
for $b,b'\in \Forg B$. The relation $\equiv$ is, by construction, an equivalence relation on the carrier of $B$, but it is in fact also an equivalence relation on the boolean algebra $B$, i.e. a sub-boolean algebra of $B\times B$\footnote{We use the term \emph{equivalence relation} where many authors use the term \emph{congruence}. The reason for this choice of nomenclature is twofold. Firstly, we consider relations in a categorical setting, i.e. as subobjects in a given category, thus relations naturally inherit the structure of the ambient category. Secondly, the term \emph{congruence} will be used heavily in the context of the congruence rule of coalgebraic logics. This rule is ultimately related to the notion of a subobject in a category of algebras, but we like to keep the two notions separate.}. This is an easy consequence of the fact that $f,g$ are boolean homomorphism. Indeed if $b_1\equiv b_1'$ and $b_2\equiv b_2'$, then there exist $a_1, a_2$ such that $f(a_1)=b_1, g(a_1)=b_1', f(a_2)=b_2, g(a_2)=b_2'$ and thus $f(a_1\wedge a_2)=f(a_1)\wedge f(a_2)=b_1\wedge b_2$ and $g(a_1\wedge a_2)=g(a_1)\wedge g(a_2)=b_1'\wedge b_2'$, i.e. $b_1\wedge b_2\equiv b_1'\wedge b_2'$. A similar argument shows that $\equiv$ is closed under $\vee$ and $\neg$. The quotient $B/\equiv$ is defined as the set of equivalence classes $[b], b\in B$ of $\equiv$ together with the obvious operations, for example $[b_1]\wedge [b_2]=[b_1\wedge b_2]$ etc. These operations are independent of the choice of representative precisely because $\equiv$ is a boolean sub-algebra of $B\times B$.  The coequalizer of $f,g$ is then the morphism $q: B\epi B/\equiv$ mapping an element $b\in B$ to its equivalence class $[b]$. It is a boolean homomorphism by construction of $B/\equiv$. 

We conclude this examination of $\BA$ by introducing the dual adjunction which lies at the heart of its use in coalgebraic logic, namely the adjunction $\uf\dashv\pow:\BA\to\Set$, where $\uf$ is the functor associating to each boolean algebra its set of ultrafilters and to each $\BA$-morphism its inverse image, and $\pow$ is the functor associating to each set the boolean algebra of its subsets (under set operations) and to each function its inverse image.

Note that for each $A\in\BA$, $\uf A=\Hom(A,\mathbbm{2})$, where $\mathbbm{2}=\{\top,\bot\}$ is the two element boolean algebra, and for each set $\Forg\pow(X)=\Hom(X,2)$, where $2$ is the two elements set.

\subsection{The category $\DL$}\label{ch1:sub:DL}

The category of distributive lattices is also a finitary algebraic category and all the facts listed above for $\BA$ also apply to $\DL$ with the obvious free-forgetful adjunction $\Free\dashv\Forg:\DL\to\Set$. The construction of colimits works just as in $\BA$. It is interesting and important to note that Fact \ref{ch1:fact:locallyfinite} also holds in $\DL$, i.e. finitely generated distributive lattices are finite (see e.g. \cite{1975:balbesdistributive}). Local finiteness will prove to be extremely important in Chapters 3 and 5, and this justifies to a great extent our focus on $\BA$ and $\DL$, but not for example on $\HA$, the category of Heyting algebras.

The fundamental dual adjunction corresponding to $\uf\dashv\pow$ in the case of $\BA$ is the adjunction $\pf\dashv\ups:\DL\to\Pos$ where $\Pos$ is the category of posets and monotone maps, $\pf$ is the functor sending a distributive lattice to its poset of prime filters and a $\DL$-morphism to its inverse image (which is easily seen to be monotone), and $\ups$ is the functor sending a poset to the distributive lattice of its downsets and a monotone map to its inverse image.

Note that in the case of boolean algebra, the two adjunctions, i.e. $\Free\dashv\Forg$ and $\uf\dashv\pow$ involve the same categories (although one is covariant and the other contravariant). However, for $\DL$ the situation is different since the prime filter functor $\pf$ takes values in $\Pos$. However, since each distributive lattice is a poset, there exists a forgetful functor $\Forg:\DL\to\Pos$. Moreover, this functor has a left adjoint $\Free:\Pos\to\DL$ which builds the free distributive lattice over a poset (see \cite{1967:Monteiro}). As it turns out this functor yields a `finitary version' of the canonical extension functor which we will describe in Chapter 2.


\subsection{Boolean algebra and distributive lattice expansions}

We now turn to the algebraic structures that underpin most of modal logic: BAOs and BAEs, as well as their distributive lattice counterparts. In this subsection, as in Chapter 2, we will use the convention of writing an object of $\A$ (i.e. $\BA$ or $\DL)$ as $\ba$ and their underlying set as $A$ if there is any danger of confusion between the two entities. Recall that, given a finitary signature $(\Sigma, \ari: \Sigma\to \mathbb{N})$, we have:

\begin{definition} A $\Sigma$-\textbf{Boolean Algebra Expansion}\index{Boolean algebra expansion}\index{BAE|see{boolean algebra expansion}} (BAE), or expanded boolean algebra, is a boolean algebra $\mathbb{A}$ together with maps $f_s: A^{\ari(s)}\to A$ for each $s\in\Sigma$, where $A$ is the underlying set of $\ba$. In other words, a BAE is a pair $\mathcal{A}=(\mathbb{A}, (f_s)_{s\in\Sigma})$. A function $f_s: A^{\ari(s)}\to A$ which preserves the bottom element $\bot\in A$ and joins in each of its arguments, is called an \textbf{operator}\index{Operator}. A BAE whose set of extending functions are all operators is called a \textbf{Boolean Algebra with Operators} (BAO)\index{Boolean algebra with operators}\index{BAE|see{Boolean algebra with operators}}. We similarly define a $\Sigma$-\textbf{Distributive Lattice Expansion}\index{Distributive lattice expansion}\index{DLE|see{Distributive lattice expansion}} (DLE) as a distributive lattice $\ba$ together with maps $f_s: A^{\ari(s)}\to A$ for each $s\in\Sigma$.
\end{definition}

The theory of BAOs is well understood (see the seminal \cite{Jonsson51} and \cite{Jonsson94,deRijkeVenema95, 2006:VenemaAC}), but is tailored to the study of classical (relational) modal logics.  In the context of coalgebraic logics we are interested in a larger class of interpretations and this is reflected in the fact that - in its predicate lifting flavour (see Section \ref{ch1:sec:coalglog}) - the `operators' of coalgebraic modal logics are not, well, operators; they do not in general preserve $\bot$ or joins in each of their argument. We will therefore consider the weaker structure of BAEs where the functions $f_s: A^{\ari(s)}\to A$ have, in general, no preservation properties. One purpose of Chapter 2 will be to isolate some properties of these functions, other than `operator-ness', which are relevant and useful to the study of canonicity.

For now, let us develop some lightweight categorical infrastructure. For a given signature $\Sigma$, the collection of all $\Sigma$-BAEs (resp. DLEs) forms a category which we will denote by $\BAE(\Sigma)$ (resp. $\DLE(\Sigma)$), or simply $\BAE$ (resp. $\DLE$) when there is no ambiguity about the signature. This category is defined as follows: the objects are of course the BAEs (resp. DLEs) and a morphism $\phi$ between two BAEs (resp. DLEs) $\mathcal{A}=(\ba, (f_s)_{s\in\Sigma})$ and $\mathcal{B}=(\mathbb{B}, (g_s)_{s\in\Sigma})$ is a $\BA$- (resp. $\DL$-) morphism $\phi:\mathbb{A}\to\mathbb{B}$ such that for all $s\in \Sigma$, $$\phi\circ f_s=g_s\circ \langle\phi,\ldots\phi\rangle$$
where $\langle\phi,\ldots\phi\rangle$ is the $n$-fold product of $\phi$ and $n=\ari(f_s)=\ari(g_s)$. 

An important fact about the categories $\BAE(\Sigma)$ and $\DLE(\Sigma)$ is that they can be characterised as categories of algebras over $\BA$ and $\DL$ respectively. To describe this relation, let $\A$ denote either $\BA$ or $\DL$ and $\AEs(\Sigma)$ (or $\AEs$ if there is no ambiguity over the signature) denote either $\BAE(\Sigma)$ or $\DLE(\Sigma)$. We will talk about `AEs' (expansions of objects of $\A$) if we want to refer to either DLEs or BAEs. We now build a polynomial endofunctor $\mathsf{S}_\Sigma: \A\to \A$:
\begin{equation}\label{ch1:eq:free}
\mathsf{S}_\Sigma(\ba)=\Free\left(\coprod_{s\in\Sigma} (A)^{\ari(s)}\right)
\end{equation}
(where $\Free\dashv\Forg$ are the obvious free/forgetful functors) and we have:
\begin{proposition}\label{ch1:prop:catequ}
The category $\Alg_{\A}(\mathsf{S}_\Sigma)$ is equivalent to $\AEs(\Sigma)$.
\end{proposition}
\begin{proof}
Given an object $\ba$ of $\A$, the freeness of the construction of $\mathsf{S}_\Sigma(\ba)$ means that the structure map $\alpha: \mathsf{S}_\Sigma(\ba)\to \ba$ of any $\mathsf{S}_\Sigma$-algebra in $\A$ is in unique correspondence with its adjoint transpose $\tilde{\alpha}: \coprod_{s\in\Sigma} (A)^{\ari(s)}\to A$, where $A=\Forg \ba$, i.e. by a collection of maps $f_s: A^{\ari(s)}\to  A, s\in\Sigma$. This collection of maps together with $\ba$ clearly define an object of $\AEs(\Sigma)$. Conversely, given such an object $(\ba, (f_s)_{s\in\Sigma})$, the coproduct morphism $ \coprod_{s\in\Sigma}f_s: \coprod_{s\in\Sigma}(A)^{\ari(s)}\to A$ can be freely extended to a $\A$-morphism $\alpha:\mathsf{S}_\Sigma(\ba)\to \ba$. These two operations define functors from $\Alg_{\A}(\mathsf{S}_\Sigma)$ to $\AEs(\Sigma)$ and back, and the compositions of these two functors are clearly isomorphic to the identities on $\Alg_{\A}(\mathsf{S}_\Sigma)$ and $\AEs(\Sigma)$ respectively.
\end{proof}
Following the characterisation of the coproduct in $\A$ as a free product, the fact that free boolean algebras are presented without any relations, and that products are constructed in $\Set$, it is easy to see that Eq. (\ref{ch1:eq:free}) can be re-written as $\polyFunc \ba=\coprod_{s\in \Sigma} (\ba)^{\ari(s)}$, where the product and coproduct are now taken in $\A$. In other words we can view $\polyFunc$ as the formal equivalent in $\BA$ or $\DL$ of its $\Set$ counterpart.

We will now examine how the categories $\BAE(\Sigma)$ and $\DLE(\Sigma)$ are related to the categories $\BA$ and $\DL$. Using the same notation as above, we define $\mathsf{V}:\AEs\to\A$ as the obvious forgetful functor, i.e. we make $\AEs$ concrete over $\A$. Initial algebras for endofunctors on $\A$ will be important in the following Proposition and we cite without proof the following  classic result, for details see for example \cite{Awodey} 10.12.
\begin{theorem}\label{ch1:thm:initAlg}
Let $\cat$ be a category with an initial element $0$ and $\omega$-colimits (i.e. colimits of diagrams whose index category is the totally ordered set of natural numbers), and let $T:\cat\to\cat$, then if $T$ preserves $\omega$-colimits it has an initial algebra.
\end{theorem}

\begin{corollary}\label{ch1:cor:initAlgFin}
Let $T:\cat\to\cat$ be an finitary endofunctor on a locally finitely presentable category, then $T$ has an initial algebra.
\end{corollary}
\begin{proof}
Every locally presentable is cocomplete, and thus has an initial object. Moreover, a diagram indexed by $\omega$ is trivially directed, thus preserved by finitary functors. The conclusion then follows from Theorem \ref{ch1:thm:initAlg}
\end{proof}

\begin{proposition}\label{ch1:prop:BAERightAdj}
The forgetful functor $\mathsf{V}:\AEs\to\A$  has a left adjoint $\mathsf{G}: \A\to\AEs$.
\end{proposition}
\begin{proof}
It is enough to show that for any boolean algebra $A$ (we will use the forgetful functor explicitly to denote the carrier in this proof), the functor $\polyFunc(-)+A$ has an initial algebra, since this amounts to the existence of a free $\polyFunc$-algebra over $A$, and by Proposition \ref{ch1:prop:catequ} this will mean that $\AEs$ has free objects. We will use Corollary \ref{ch1:cor:initAlgFin} above and show that $\polyFunc$ is finitary, i.e. preserves directed colimits. We are here using some notions which will be fully developed in Chapter 3, but we will only need one very basic concept, namely that finite sets are finitely presentable, i.e. their $\hom$ functors preserves directed colimits. Let $(I,\leq)$ be a directed set and let $\Diag:I\to\A$ be a directed diagram in $\A$, we then have
\begin{align*}
\polyFunc (\colim_i A_i)&=\Free\left(\coprod_{s\in\Sigma} (\Forg \colim_i A_i)^{\ari(s)}\right)\\
&\stackrel{1}{=}\Free\left(\coprod_{s\in\Sigma} (\colim_i \Forg A_i)^{\ari(s)}\right)\\
&\stackrel{2}{=}\Free\left(\coprod_{s\in\Sigma} \hom(\ari(\sigma),\colim_i \Forg A_i)\right)\\
&\stackrel{3}{=}\Free\left(\coprod_{s\in\Sigma} \colim_i \hom(\ari(\sigma),\Forg A_i)\right)\\
&\stackrel{4}{=}\Free\left(\colim_i \coprod_{s\in\Sigma}  \hom(\ari(\sigma),\Forg A_i)\right)\\
&\stackrel{5}{=}\colim_i \Free\left(\coprod_{s\in\Sigma}  (\Forg A_i)^{\ari(s)}\right)\\
&=\colim_i\polyFunc A_i
\end{align*}
where (1) follows from Fact \ref{ch1:fact:filtcol},(2) is a rewriting of the $\hom$ set in $\Set$, (3) follows from the fact that since the signature is finitary $\ari(s)$ is finite and finite sets are finitely presentable, (4) is a special case of the general rule that colimits commute, and finally (5) follows from the fact that $\Free$ is cocontinuous by virtue of being left-adjoint. Thus $\polyFunc$ is finitary, as is $\polyFunc(-)+A$ for any boolean algebra $A$. By corollary \ref{ch1:cor:initAlgFin}, $\polyFunc(-)+A$ has an initial algebra $\init[(\polyFunc(-)+A)]$ for every $A$.

So let us define $\mathsf{G}:\A\to\AEs$ by $\mathsf{G}(A)=\init[(\polyFunc(-)+A)]$ and let us check that we do indeed have $\mathsf{G}\dashv \mathsf{V}$, i.e. that for any $\A$-morphism $f: A \to \mathsf{V}(B, \alpha)$ where $\alpha: \polyFunc(B)\to B$ (i.e. any morphism $A\to B)$, there exists a unique $\Alg_{\A}(\polyFunc)$-morphism $\mathsf{G}(A)\to(B,\alpha)$. This is immediate by initiality of $\mathsf{G}(A)$ as can be seen from the following diagram
\[
\xymatrix@C=12ex
{
\polyFunc(\mathsf{G}A)+A=\polyFunc(\init[(\polyFunc(-)+A)]\ar@{-->}[r]^(0.7){[\polyFunc\hat{f}+\Id_A]}\ar[d] & \polyFunc B+A\ar[d]^{[\alpha+f]} \\
\mathsf{G}A=\init[(\polyFunc(-)+A)]\ar@{-->}[r]_(0.6){\hat{f}} & B
}
\]
and $\hat{f}$ defines an $\Alg_{\A}(\polyFunc)$-morphism.
\end{proof}

\begin{proposition}\label{ch1:prop:adj}
$\mathsf{G}\circ \Free\dashv \Forg\circ \mathsf{V}$.
\end{proposition}
\begin{proof}
Let us consider $A$ in $\Set$, $(B,\alpha)$ in $\Alg_{\A}(\mathsf{S}_\Sigma)$ and $f: A\to \Forg\circ \mathsf{V}((B,\alpha))$. We need to show that $f$ determines a unique morphism $g: \mathsf{G}\circ \Free(A)\to (B,\alpha)$. But since $\Free\dashv\Forg$, $f$ determines a unique map $h: \Free(A) \to \mathsf{V}((B,\alpha))$ and since $\mathsf{G}\dashv\mathsf{V}$, this $h$ in turns determines a unique $g:\mathsf{G}\circ \Free(A)\to (B,\alpha)$.
\end{proof}

This is clearly a much more general result, i.e. adjunctions compose.

\subsection{Algebras in $\BA$ and $\DL$}\label{ch1:subsec:LAlg}

We have just encountered a particular class of algebras in $\BA$ and $\DL$ which are defined by certain finitary polynomial endofunctors. In fact, algebras in $\A$ (where again $\A$ is either $\BA$ of $\DL$) for general finitary functors  are a very useful tool to study coalgebraic logic. However, they introduce a nesting of algebraic constructions (i.e. algebras defined over algebras) which is not completely straightforward. We have just seen in Proposition \ref{ch1:prop:BAERightAdj} that for finitary polynomial functors $\mathsf{S}_{\Sigma}: \A\to\A$, free $\mathsf{S}_{\Sigma}$-algebras exist. This makes finitary polynomial functors \textbf{varietors}\index{Varietor}. A varietor is an endofunctor $L:\cat\to\cat$ such that the forgetful functor $\Forg:\Alg_{\cat}(L)\to\cat$, has a left adjoint $\Free:\cat\to\Alg_{\cat}(L)$, i.e. such that $\Alg_{\cat}(L)$ has free objects. As we have seen from Corollary \ref{ch1:cor:initAlgFin}, every finitary functor on a locally presentable category is a varietor. The following property of varietors is crucial.

\begin{theorem}[\cite{cats} Theorem 20.56]
If $T:\cat\to\cat$ is a varietor, then $\Alg_{\cat}(T)$ is monadic over $\cat$
\end{theorem}

For any varietor $L:\A\to\A$, we thus have two monadic functors: $\Forg: \A\to\Set$ and $\mathsf{V}: \Alg_{\A}(L)\to\A$. However, as shown in Example 20.45 of \cite{cats}, monadic functors do \emph{not} in general compose, and in particular, $\Forg\mathsf{V}:\Alg_{\A}(L)\to\Set$ is not, in general, monadic. Since monadicity is a desirable feature, it is important to identify which additional feature on $\Forg$ and $\mathsf{V}$ will make their composition monadic. This feature is detailed in Exercise 23.F of \cite{cats}, and is the notion of \textbf{varietal functor}\index{Varietal}. A regularly monadic functor $\Forg:\cat\to\cat[D]$ (i.e. $\Forg$ is monadic, $\cat[D]$ has regular epi-mono factorisations, and $\Forg$ preserves regular epis)\index{Regularly monadic} is varietal if $\cat[D]$ has pullbacks and $\Forg$ creates coequalizers of kernel pairs, also known as exact sequences. We will go through all these terms in more details in the next section; for now, let us just state the following three propositions. The first is a re-wording of Proposition \ref{ch1:prop:MonadicExactOverSet}, the second is a re-wording of Proposition \ref{ch1:prop:AlgExact}.

\begin{proposition}\label{ch1:prop:Setvarietal}
Every monadic functor over $\Set$ is varietal.
\end{proposition}

\begin{proposition}\label{ch1:prop:Lvarietal}
If $L:\cat\to\cat$ is a varietor over a regular category, and $L$ preserves regular epis, then $\Forg:\Alg_{\cat}(L)\to\cat$ is varietal.
\end{proposition}

\begin{proposition}[\cite{cats} 23.F (f)]\label{ch1:prop:varietalCompose}
Varietal functors are monadic and compose.
\end{proposition}

These Propositions provide an answer the question of knowing when $\Forg\mathsf{V}:\Alg_{\A}(L)\to\Set$ is monadic: it is monadic if $L$ preserves regular epimorphisms. Note that no assumption is needed for $\Forg$ to be varietal, it is essentially a feature of exact sequences in $\Set$ which gives this nice behaviour for free. Note also that preservation of regular epis has also been identified as a desirable feature of endofunctors on algebraic varieties in \cite{Thesis:Rob}, where this condition guarantees the existence of a notion of functor presentation. 

\section{Relations, categorically}\label{ch1:sec:rel}

This section will be fairly technical and aims to introduce and develop the familiar $\Set$ notions of relations, relation liftings, equivalence relations and fully invariant relations, in a setting which is abstract enough to capture the categories encountered in the previous section, namely $\BA$, $\DL$, and $\Alg_{\A}(L)$ for a varietor $L:\A\to \A$ (where $\A$ is $\BA$ or $\DL$). The most natural and convenient environment to achieve this is to work in the framework of \emph{regular categories}. The special case of the category of $\Set$ will of course be very useful in its own right, particularly in the context of the nabla and abstract flavours of coalgebraic logic.

\subsection{Relations}

There is a large literature on the topic of relations in categories other than $\Set$, see e.g. \cite{1991:KellyRel}, which considers various generalization of the concept and various classes of suitable categories, but the classical theory of relations in a category other than $\Set$ takes place in regular categories. Which suits us fine since $\BA$ and $\DL$ are regular category. The key features of regular categories which allow them to have a good notion of relations is that (1) they have regular epi/mono factorisation, (2) the class of regular epimorphisms is closed under pullback, i.e. the factorisation system is \emph{stable}.

A relation $R\subseteq X\times Y$ between $X,Y$ in $\Set$ can be seen as a subobject of $X\times Y$, and this notion can clearly be generalized to other categories (for more details on subobjects, see Section \ref{ch3:sec:base}). For a well-powered regular category $\cat$, let use write $\rel$ for the category whose objects are the same objects as $\cat$ but whose morphism $R: A\to B$ are subobjects of $A\times B$. We need to show how morphisms  in $\rel$ are composed. Let $R:A\to B$ and $S: B\to C$ be $\rel$-morphisms, we compose $S\circ R$ in the same way as spans (since relations are spans), i.e. we build the pullback of $R\stackrel{p_2^R}{\to} B\stackrel{p_1^S}{\leftarrow}C$ where $p_i^R, p_i^S$, $i=1,2$ are the $i^{th}$ projections of $R$ and $S$ respectively. We then take the regular epi-mono factorisation of the map from the limiting cone of this pullback to $A\times C$ (which exists by universality of products) and define $S\circ R$ to be the subobject of $A\times C$ thus defined.

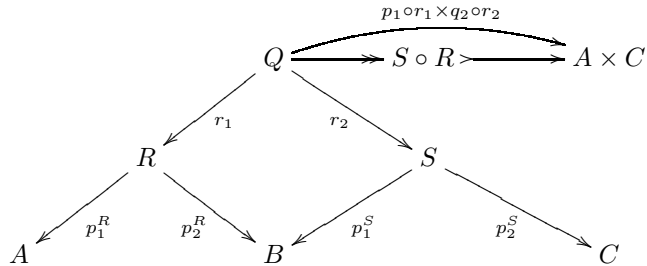
\[
\xymatrix@C=8ex
{
& & Q\ar@/^1pc/[rr]^{p_1\circ r_1\times q_2\circ r_2} \ar[dr]_{r_2}\ar[dl]^{r_1}\ar@{->>}[r]& S\circ R\hspace{1ex}\ar@{>->}[r] & A\times C\\
& R\ar[dr]_{p_2^R}\ar[dl]^{p_1^R} & & S\ar[dr]_{p_2^S}\ar[dl]^{p_1^S}\\
A & & B & & C
}
\]

It can be shown that composition defined in this way is associative iff the class of regular epis is stable under pullback, which is the case by definition in a regular category. This explains why regular categories are a particularly natural setting in which to generalize relations.

There exists a very useful alternative description of a relation. In a finitely complete category, relations $R\mono A\times A$ are in one-to-one correspondence with pairs of maps $e_1,e_2:R\to A$ which are jointly monic, i.e. such that for any $f,g: B\to R$, if $e_1\circ f=e_1\circ g$ and $e_2\circ f=e_2\circ g$, then $f=g$. This correspondence is clear from the following diagrams
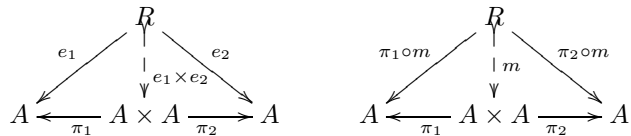
\[
\xymatrix
{
& R\ar@{>-->}[d]^(0.6){e_1\times e_2}\ar[dr]^{e_2}\ar[dl]_{e_1} & & & R\ar@{>-->}[d]^{m}\ar[dr]^{\pi_2\circ m}\ar[dl]_{\pi_1\circ m}\\
A & A\times A\ar[l]^{\pi_1}\ar[r]_{\pi_2} & A & A & A\times A\ar[l]^{\pi_1}\ar[r]_{\pi_2} & A
}
\]
where it is not difficult to check that $e_1\times e_2$ is monic if $e_1,e_2$ are jointly monic, and conversely, that $\pi_1\circ m, \pi_2\circ m$ are jointly monic if $m$ is monic. In what follows, having these equivalent descriptions of what a relation is will prove very useful.

\subsection{Relation liftings}

With $\rel$ defined, it is natural to consider relation liftings for endofunctor $L:\cat\to\cat$. In fact, being in a regular category ensures that what is done in $\Set$, i.e. essentially taking a direct image, can be transferred to $\cat$. Formally, given a relation $R: A\to B$ and a $\cat$-endofunctor $L$, we define its relation lifting $\lift{L}{R}$ as the subobject of $LA\times LB$ defined as the image of $LR$ in $LA\times LB$, i.e. we build the following regular epi-mono factorisation:
\[
\xymatrix@C=12ex
{
LA & LR\ar[l]_{Lp_1}\ar[r]^{Lp_2}\ar@{->>}[d]\ar@/^1pc/@{-->}[dd]^<<<<{Lp_1\times Lp_2} & LB\\
& \lift{L}{R}\ar@{>->}[d] & \\
& LA\times LB\ar[uul]^{\pi_1}\ar[uur]_{\pi_2}
}
\]

A proof of the following Proposition can be found in \cite{1991:CWK}, but involves advanced concepts of category theory. We can in fact prove it using the simple machinery of regular categories.

\begin{proposition}\label{ch1:prop:liftingwpb}
Let $\cat$ be a regular category, and let $R\mono A\times B$, $S\mono B\times C$ be relations in $\cat$. Assume that $L:\cat\to\cat$ preserves regular epis and weak pullbacks, then $\lift{L}{S}\circ \lift{L}{R}=\lift{L}{(S\circ R)}$.
\end{proposition}
\begin{proof}
We first show that $\lift{L}{(S\circ R)}\mono \lift{L}{S}\circ \lift{L}{R}$.
By definition, the relation $S\circ R$ is defined by building the pullback of $R\stackrel{p_2^R}{\to} B\stackrel{p_1^S}{\leftarrow} S$ where $p_i^R, i=1,2$ are the projection from  $R$ to $A,B$, and similarly for $S$. By the regular epi-mono factorisation of $Lp_i^R, Lp_i^S, i=1,2$ we first construct $\lift{L}{S}$ and $\lift{L}{R}$. The relation $\lift{L}{S}\circ \lift{L}{R}$ is then defined by building the pullback  of $\lift{L}{R}\stackrel{r_2}{\to} LB\stackrel{s_1}{\leftarrow} \lift{L}{S}$ where $r_1,r_2$ are the projections from $\lift{L}{R}$ to $LA$ and $LB$ respectively, and $s_1,s_2$ are the projections from $\lift{L}{S}$ to $LB$ and $LC$ respectively.

It is easy to see that $L$ applied to the first pullback forms a cone for the diagram defining the latter:
\[
\xymatrix
{
L \pb(p^R_2,p^S_1)\ar[dd]\ar[rr]\ar@{-->}[dr]^{u} & & LS\ar@{->>}[d]^{q^S}\ar@/^2pc/[dd]^{Lp_1^S}\\
& \pb(r_2,s_1)\ar[d]_{\pi_2} \ar[r]^(0.6){\pi_1} & \lift{L}{S}\ar[d]^{s_1}\\
LR\ar@{->>}[r]_{q^R}\ar@/_2pc/[rr]_{Lp_2^R} & \lift{L}{R}\ar[r]_{r_2} & LB
}
\]
and there must therefore exist a unique morphism $u: L \pb(p^R_2,p^S_1)\to \pb(r_2,s_1)$. Since $L$ preserves regular epi, we have a regular epi
\[
L \pb(p^R_2,p^S_1)\epi L(S\circ R)\epi\lift{L}{(S\circ R)}
\]
where the first regular epi is just $L$ applied to the regular epi factor defining $S\circ R$, and the second regular epi is by definition of relation liftings. Combining our observations so far we have a square:
\[
\xymatrix
{
L \pb(p^R_2,p^S_1)\ar@{->>}[r]\ar[d]_{u}& \lift{L}{(S\circ R)}\ar@{>->}[dd]\\
\pb(r_2,s_1)\ar@{->>}[d]\\
\lift{L}{S}\circ \lift{L}{R}\hspace{1ex}\ar@{>->}[r] & LA\times LC
}
\]
and since regular epis are strong, there must exist a unique diagonal fill-in, which, moreover, must be monic, i.e. $\lift{L}{(S\circ R)}\mono \lift{L}{S}\circ \lift{L}{R}$ as claimed.

For the converse direction, it is enough to show that $u$ is a regular epi, and we can then use the same square as above to find a diagonal fill-in $\lift{L}{S}\circ\lift{L}{R}\mono \lift{L}{(S\circ R)}$. To see that $u$ is indeed a regular epi, we use the fact that $L$ weakly preserves pullbacks. We build the following two pullbacks

\[
\xymatrix
{
L \pb(p^R_2,p^S_1)\ar@/^1pc/[drrr]\ar@/_1pc/[dddr]\ar@<2pt>[dr]^{h}\\
& \pb(L p^R_2,L p^S_1)\ar[dd]\ar[rr]\ar@{-->}[dr]^{v}\ar@<1ex>@{-->}[ul]^{w} & & LS\ar@{->>}[d]^{q^S}\ar@/^2pc/[dd]^{Lp_1^S}\\
& & \pb(r_2,s_1)\ar[d]_{\pi_2} \ar[r]^(0.6){\pi_1} & \lift{L}{S}\ar[d]^{s_1}\\
& LR\ar@{->>}[r]_{q^R}\ar@/_2pc/[rr]_{Lp_2^R} & \lift{L}{R}\ar[r]_{r_2} & LB
}
\]
it is not hard to verify by                                                           taking intermediary pullbacks and using the pullback lemma that $v$ is a regular epi (see \cite{Borceux2} Lemma 2.1.2). Moreover, since $L$ weakly preserves pullbacks there must exist a (not necessarily unique) morphism $h:L \pb(p^R_2,p^S_1)\to \pb(L p^R_2,L p^S_1)$. By the fact that $\pb(L p^R_2,L p^S_1)$ is the actual pullback of the same diagram, there must exist a unique arrow $w: \pb(L p^R_2,L p^S_1)\to L \pb(p^R_2,p^S_1)$, and by the universal property of the pullback it is clear that $h\circ w=\id$. It is easy to check that morphisms with a right inverse such as $h$ are coequalizers (in this case of $\id$ and $h\circ w$), and thus $h$ is a regular epi. We can therefore conclude that $u=v\circ h$ is a regular epi, since regular epis compose in regular categories (\cite{Borceux2} Corollary 2.1.5), and this concludes the proof.

\end{proof}

The following proposition is an application of the previous one, and is particularly important for the nabla and abstract versions of coalgebraic logics, since it allows well-defined inductive definitions.

\begin{proposition}\label{ch1:prop:liftingProp}
Let $L:\cat\to\cat$ where $\cat$ is regular, then
\begin{enumerate}[(i)]
\item Every $\cat$-morphism $f:A\to B$ induces a relation $\Gr(f)\mono A\times B$ and $\lift{L}{\Gr(f)}=\Gr(Lf)$
\item Every relation $R\mono A\times B$ has a converse relation $R\op\mono B\times A$ and $\lift{L}{(R\op)}=(\lift{L}{R})\op$
\item If $L$ weakly preserves pullbacks and preserves regular epis, then for any $A_1\mono B_1$, $A_2\mono B_2$ and $R\mono B_1\times B_2$ \[\lift{L}{(R\cap (A_1\times A_2))}=\lift{L}{R}\cap (LA_1\times LA_2)\]
\end{enumerate}
\end{proposition}
\begin{proof}
(i) The relation $R_f$ defined by a morphism is obtained by the pullback 
\[
\xymatrix
{
R_f \ar[r]^{p_1}\ar[d]_{p_2} & B\ar[d]^{\id_B}\\
A\ar[r]_{f} & B
}
\]
or, alternatively and equivalently, as the equalizer of $\pi_2, f\circ \pi_1: A\times B\to B$. Note that since the identity map is an isomorphism, and the pullback of an iso is an iso, $p_2$ must be an iso. If we apply $L$ to the previous diagram we get:
\[
\xymatrix
{
LR_f\ar@/^1pc/[drr]^{Lp_1} \ar@/_1pc/[ddr]_{Lp_2}\ar@{-->}[dr]^{u}\\
& R_{Lf} \ar[r]^{q_1}\ar[d]_{q_2} & LB\ar[d]^{\id_{LB}=L\id_B}\\
& LA\ar[r]_{Lf} & LB
}
\]
The morphism $q_2$ must be an iso, and since any functor preserves isos, $Lp_2$ is also an iso. Therefore $u$ must be an iso, and its image is isomorphic to its domain, i.e. $\lift{L}{R_f}=R_{Lf}$.

(ii) The converse relation $R\op$ of a relation $R\mono A\times B$ is also obtained by a pullback
\[
\xymatrix
{
R\op \ar[r]\ar@{>->}[d] & R\ar@{>->}[d]\\
B\times A\hspace{1ex}\ar[r]_{\pi_2\times \pi_1} & A\times B
}
\]
And since $\pi_1\times \pi_2$ is an iso, the proof follows the same lines as (i).

For (iii), let $m: R\mono B_1\times B_2$, $i_1: A_1\mono B_1$ and $i_2: A_2\mono B_2$. We will first show that 
\[R\cap (A_1\times A_2)=\Gr(i_2)\op\circ R\circ \Gr(i_1)\] 
To achieve this, we first show that $R\cap (A_1\times B_2)=R\circ \Gr(i_1)$. To see this let us first build the appropriate pullback:
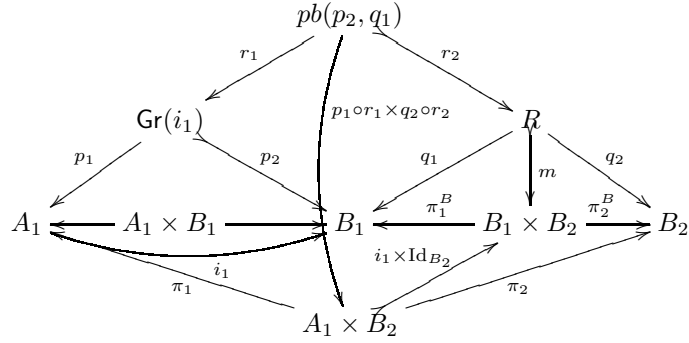
\[
\xymatrix
{
& & pb(p_2, q_1)\ar@{>->}[dr]^{r_2}\ar[dl]_{r_1}\ar@/_1pc/[ddd]^(0.3){p_1\circ r_1\times q_2\circ r_2}\\
& \Gr(i_1)\ar[dl]_{p_1}\ar@{>->}[dr]^{p_2} & & R\ar@{>->}[d]^{m}\ar[dl]_{q_1} \ar[dr]^{q_2}\\
A_1\ar@/_1pc/[rr]_(0.6){i_1} & A_1\times B_1 \ar[l]\ar[r] & B_1 & B_1\times B_2\ar[l]_{\pi_1^B}\ar[r]^{\pi_2^B} & B_2\\
& & A_1\times B_2\ar[ull]^{\pi_1}\ar[urr]_{\pi_2}\ar@{>->}[ur]^{i_1\times \id_{B_2}}
}
\]
It is not too hard to check that by definition of $\Gr$, $p_1$ must be an iso, and that $p_2\circ p_1\inv=i_1$. Let us now show that $(pb(p_2,q_1), r_1,r_2)$ is the pullback of $A_1\times B_2\mono B_1\times B_2\leftarrowtail R$. It is clearly a cone by commutativity of the above diagram. Now let $g: A\to R, h_1\times h_2: A\to A_1\times B_2$ such that $m\circ g=(i_1\times \id_{B_2})\circ (h_1\times h_2)$ be another cone for the same diagram. We then have
\begin{align*}
q_1\circ g & = \pi_1^B\circ m \circ g\\
& = \pi_1^B \circ (i_1\times \id_{B_1})\circ (h_1\times h_2)\\
& = i_1\circ \pi_1\circ (h_1\times h_2) \\
& = p_2\circ p_1\inv\circ \pi_1\circ (h_1\times h_2)
\end{align*}
i.e. $(A,g, p_1\inv\circ \pi_1\circ (h_1\times h_2))$ forms a cone for the diagram defining $pb(p_2,q_1)$ and so there must exist a unique morphism $u: A\to \pb(p_2,q_1)$ such that $r_2\circ u=g$ and $r_1\circ u= p_1\inv\circ \pi_1\circ (h_1\times h_2)$. To see that this cone is also a cone for the intersection $R\cap A_1\times B_2$, we now just need to check that $(p_1\circ r_1\times q_2\circ r_2)\circ u= (h_1\times h_2)$ (since $r_2\circ u=g$ is already satisfied). We just calculate:
\begin{align*}
(p_1\circ r_1\times q_2\circ r_2)\circ u & =p_1\circ r_1\circ  u\times q_2\circ r_2\circ u\\
&= p_1\circ p_1\inv\circ \pi_1\circ (h_1\times h_2)\times q_2\circ g\\
& =\pi_1\circ (h_1\times h_2)\times \pi_2^B\circ m\circ g\\
&= h_1\times \pi_2^B\circ (i_1\circ \id_{B_2})\circ (h_1\circ h_2)\\
&= h_1\times h_2
\end{align*} 
as required. Thus $pb(p_2,q_1)=R\cap (A_1\times B_2)$, and in particular $p_1\circ r_1\times q_2\circ r_2$ must be a mono, and therefore there is no regular epi part to its factorization and we can conclude 
\[R\circ \Gr(i_1)=R\cap (A_1\times B_2)\]

A completely analogous proof shows that $\Gr(i_2)\op\circ R=R\cap B_1\times A_2$. And thus the composition $\Gr(i_2)\op\circ R\circ \Gr(i_1)$ is defined by first building the pullback of $R\cap A_1\times B_2$ with $R\cap B_1\times A_2$, and then taking its image. But since both the morphisms defining this pullback are monos, we get 
\[\Gr(i_2)\op\circ R\circ \Gr(i_1)=R\cap( A_1\times A_2)\]

To complete the proof, we use Proposition \ref{ch1:prop:liftingwpb} on the above composition, as well as (i) and (ii) to get 
\begin{align*}
\lift{L}{(R\cap( A_1\times A_2))}&=\lift{L}{(\Gr(i_2)\op\circ R\circ \Gr(i_1))}\\
&= \lift{L}{\Gr(i_2)\op}\circ \lift{L}{R}\circ \lift{L}{\Gr(i_1)}\\
&=\Gr(Li_2)\op\circ \lift{L}{R}\circ \Gr(Li_1)
\end{align*}
and the same argument as above shows that the latter expression is in fact $\lift{L}{R}\cap (LA_1\times LA_2)$.
\end{proof}

\subsection{Equivalence relations and exactness}

A particular type of relation is the notion of \textbf{equivalence relation}\index{Equivalence relation}, sometimes called `congruence', but we prefer equivalence relation as it avoids confusion with the congruence rule, which, as we shall see a bit later, is related to the idea of a relation in $\BA$, but not specifically to that of an \emph{equivalence} relation in $\BA$. 
Categorically, an equivalence relation $R$ on an object $A$ in a regular category $\cat$ is a subobject $E\stackrel{e_1\times e_2}{\mono} A\times A$ which is an internal equivalence relation on $A$, where the axioms defining an equivalence are categorified as follows:
\begin{itemize}
\item reflexivity: the diagonal $\Delta_A$ is a subobject of $E$
\item symmetry: $e_1\times e_2=e_2\times e_1$ as subobjects of $E$
\item transitivity: if $E\stackrel{e'_1}{\leftarrow}E\times_A E\stackrel{e'_2}{\to} E$ is the pullback of $E\stackrel{e_1}{\to} A\stackrel{e_2}{\leftarrow} E$, then $(e_1\circ e_1')\times (e_2\circ e_2')$ is a subobject of $e_1\times e_2$.
\end{itemize}
Kernel pairs provide the typical example of equivalence relation.

\begin{lemma}\label{ch1:lem:kpRequiv}
Let $\cat$ be a regular category, and let $f:A\to B$ be a $\cat$-morphism, then $\ker f$ defined by the pullback
\[
\xymatrix
{
\ker f \ar[r]^{p_1}\ar[d]_{p_2} & A\ar[d]^{f}\\
A\ar[r]_{f} & B
}
\]
is an equivalence relation on $A$.
\end{lemma}
\begin{proof}
This is essentially an exercise on pullbacks. We first need to show that $\ker f$ is a relation on $A$. To see this note first that by universality of the product, there exists a unique arrow $p_1\times p_2: \ker f\to A\times A$. To see that this arrow is monic, consider two morphisms $h,g:C\to\ker f$ with the property that 
\[(p_1\times p_2)\circ g=(p_1\times p_2)\circ h\]
Since $f\circ p_1=f\circ p_2$, it is clear that
\[f\circ p_1\circ g=f\circ p_2\circ g\]
and similarly for $h$, i.e. the arrows $p_1\circ g,p_2\circ g$ and $p_1\circ h,p_2\circ h$ make $C$ a cone for the limit defining the pullback, thus there must exist a single map $C\to \ker f$ making the obvious diagram commute. But $g$ and $h$ are also two such maps, thus we must have $g=h$.

We now show $\ker f$ is an equivalence relation. For the reflexivity note that $A$ with two copies of the identity map $\id_A$ forms a cone of the diagram defining the pullback $\ker f$, and there must therefore exist a unique map $h: A\to \ker f$. Moreover, by the same argument as above, it is easy to see that the unique map $\Delta_A: A\to A\times A$ is monic, and since $p_1\times p_2$ is also monic, we must have $h$ monic, and by unicity $\Delta_A=(p_1\times p_2)\circ h$, which is to say that $\Delta_A\subset (p_1\times p_2)$ as subobjects of $A\times A$. Symmetry is immediate, since $\ker f$ together with the projection maps $p_1,p_2$ in the reverse order forms a cone which must be isomorphic to $\ker f$. Finally for the transitivity, if we build the pullback $\ker f\stackrel{p_1'}{\leftarrow}\ker_f\times_A\ker f\stackrel{p_2'}{\to}\ker f$ of $\ker f\stackrel{p_1}{\to} A\stackrel{p_2}{\leftarrow}\ker f$, then it is clear that $p_1\circ p_1'$ and $p_2\circ p_2'$ define a cone for the diagram defining the pullback $\ker f$, and thus there must exist a unique mediating morphism $\ker f\times_A\ker f\to\ker f$ and by the same argument as above this morphism must be a mono, and in consequence $(p_1\circ p_1'\times p_2\circ p_2')$ is a subobject of $p_1\times p_2$ are required.
\end{proof}

In fact, in many categories of interest, and all categories of interest to us, the concepts of kernel pairs and equivalence relation coincide.

\begin{definition}
Let $\cat$ be a regular category, an equivalence relation $R\mono A\times A$ in $\cat$ is called \textbf{effective}\index{Effective equivalence relation} if it is the kernel pair of some morphism. A regular category where all equivalence relations are effective is called \textbf{exact}\index{Exact category}. Finally, a diagram of the shape
\[
\xymatrix
{
\ker q \ar@<1ex>[r]^{e_1} \ar@<-1ex>[r]_{e_2} & B\ar@{->>}[r]^q & Q
}
\]
where $q$ is the coequalizer of its kernel pair, is called an \textbf{exact sequence}\index{Exact sequence}.
\end{definition}

\begin{lemma}\label{ch1:lem:equivForg}
Let $\Free\dashv\Forg: \cat\to\cat[D]$ be an adjunction between regular categories,  if $R\mono A\times A$ is an equivalence relation in $\cat[D]$, then $\Forg R\mono \Forg (A\times A)=\Forg A\times\Forg A$ is an equivalence relation in $\cat$. In other words, right adjoints preserve equivalence relations.
\end{lemma}
\begin{proof}
This is a simple consequence of the fact that $\Forg$ preserves monos and limits.
\end{proof}

\begin{proposition}\label{ch1:prop:MonadicExactOverSet}
If $(\cat, \Forg)$ is a concrete category over $\Set$ and $\Forg$ is monadic, then $\Forg$ creates exact sequences and $\cat$ is exact.
\end{proposition}
\begin{proof}
Notice first that $\Set$ is exact: if $R\subset X\times X$ is an equivalence relation in $\Set$, consider the map $q: X\epi X/R$, where $X/R$ is the quotient of $X$ under $R$. It is easy to see that $q$ is the coequalizer of the component maps $e_1,e_2: R\to A$ of $R$ and that $R=\ker q$. Note that the diagram defined by $e_1,e_2, q$ is an exact sequence.

Next, we will show that the coequalizer of an equivalence relation in $\Set$ is absolute, i.e. preserved by any functor. To see this we need to find a section $s$ of $q$ and a section $t$ of $e_1$ such $e_2\circ t=s\circ q$. Since every epi in $\Set$ has a section, we can always find a section $s$ of $q$. Given such a section $s$, we build $t:X\to R$ as follows:
\[
t=(\id_X\times (s\circ q))\circ\Delta_X
\] 
This map is well defined since by definition of $q$, $(x,s\circ q(x))\in R$ for any choice of $s$, and it is clearly a section of $e_1$ such that $e_2\circ t=s\circ q$. It follows that $q$ is an absolute coequalizer.

Finally, since $\Forg$ is monadic, by Beck's theorem, $\Forg$ \emph{creates} absolute coequalizers. In particular this means that if we take any equivalence relation $e_1\times e_2: R\mono A\times A$ in $\cat$, then by Lemma \ref{ch1:lem:equivForg}, $\Forg e_1\times \Forg e_2:\Forg R\subset\Forg A\times\Forg A$ is an equivalence class in $\Set$, and is thus equal to the kernel pair of its coequalizer. Moreover, since this coequalizer is absolute and $\Forg$ is monadic, there must exist $Q$ in $\cat$ and $q: A\to Q$ such that $q$ is the coequalizer of  $e_1,e_2$ and $\Forg q: \Forg A\to \Forg Q$ is the coequalizer of $\Forg e_1,\Forg e_2$. Finally, since $\Forg$ is monadic, it also \emph{creates} all limits, and in particular since $\Forg R$ is the kernel pair of $\Forg q$, $R$ is the kernel pair of $q$.
\end{proof}

This proposition tells us that $\BA$ and $\DL$ are exact categories, as is and $\Alg_{\Set}(T)$ whenever $T$ is a varietor. In what follows, we will be confronted with relations in $\Alg_{\A}(L)$ for an endofunctor $L:\A\to\A$ where $\A$ is $\BA$ or $\DL$. In this situation it will no longer be possible to use the very nice, but exceptional, property of coequalizers of equivalence relation in $\Set$ to conclude that $\Alg_{\A}(L)$ is exact. We can however guarantee that $\Alg_{\A}(L)$ is exact for a large class of varietors $L:\A\to\A$, as we will now show.

\begin{lemma}\label{ch1:lem:coeqkp}
Let $\cat$ be a regular category. Then 
\begin{enumerate}[(i)]
\item Every regular epi is the coequalizer of its kernel pair.
\item Let $f: A\to B$ and let $f^*$ be the regular epi part of its regular epi-mono factorization, then $\ker f\simeq \ker f^*$.
\item If $\cat$ is exact, then any equivalence relation is the kernel pair of its coequalizer.
\end{enumerate}
\end{lemma}
\begin{proof}
To show (i), let $p: A\epi B$ be a regular epi, i.e. there exists $g,h: C\to A$ such that $p$ is the coequalizer of $g,h$. Now consider its kernel pair $p_1,p_2:\ker p\to A$. We then have 
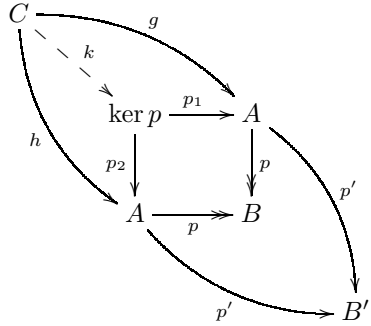
\[
\xymatrix
{
C\ar@/^1pc/[drr]^{g}\ar@/_1pc/[ddr]_{h}\ar@{-->}[dr]^{k}\\
& \ker p\ar[r]^{p_1}\ar[d]_{p_2} & A\ar@{->>}[d]^{p}\ar@/^1pc/[ddr]^{p'} \\
& A\ar@{->>}[r]_{p}\ar@/_1pc/[drr]_{p'}& B\\
& & & B'
}
\]
Since $C$ together with $g,h$ constitutes a cone for the pullback diagram defining $\ker p$, there exists a unique mediating morphism $k:C\to\ker p$. Now, let $p': A\to B'$ be the coequalizer of $p_1,p_2$. Clearly since $p'\circ p_1=p'\circ p_2$, then $p'\circ p_1\circ k=p'\circ g=p'\circ p_1\circ h=p'\circ h$ and thus there must exist a unique morphism $u: B\to B'$ since $p$ is the coequalizer of $g,h$. Conversely, since $p\circ p_1=p\circ p_2$ by construction, we must have a unique arrow $u': B'\to B$ since $p'$ is the coequalizer of $p_1,p_2$. Since $u,u'$ must be inverse of each other we have $B\simeq B'$.

For (ii), it is easy to see than an arrow is mono iff its kernel pair is the identity. With this in mind, consider the following diagram:
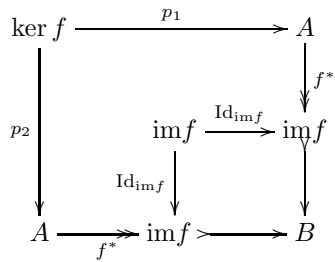
\[
\xymatrix
{
\ker f\ar[rr]^{p_1}\ar[dd]_{p_2} & & A\ar@{->>}[d]^{f^*}\\
& \im f\ar[d]_{\id_{\im f}}\ar[r]^{\id_{\im f}} & \im f\ar@{>->}[d]\\
A\ar@{->>}[r]_{f^*} & \im f\hspace{1ex}\ar@{>->}[r] & B
}
\]
and it is easy to see that $p_1,p_2:\ker f\rightrightarrows A$ is the kernel pair of $f^*$.

For (iii), take any equivalence relation $e_1\times e_2:R\mono A\times A$. Since $\cat$ is exact, we are guaranteed that $R=\ker f$ for some $f: A\to B$. Now, take the regular epi-mono factorisation of $f$, and it is easy to see from (ii) that $R=\ker f^*$ where $f^*$ is the regular epi part of the factorisation. 
\[
\xymatrix
{
R=\ker f\ar[r]^(0.6){e_1}\ar[d]_{e_2} & A\ar@{->>}[d]^{f^*}\ar@/^1pc/[ddr]^{f} \\
A\ar@{->>}[r]_{f^*}\ar@/_1pc/[drr]_{f} & B^* \ar@{>->}[dr]\\
& & B
}
\]
Since $f^*$ is regular epi, by (i) it is the coequalizer of its kernel pair, i.e. $f^*$ is the coequalizer of $e_1,e_2$ and $R$ is its kernel pair by construction.
\end{proof}

\begin{proposition}[\cite{HughesPhD}]\label{ch1:prop:regfact}
Let $\cat$ be a regular category and let $L:\cat\to\cat$ preserve regular epis, then $\Alg_{\cat}(L)$ has regular epi-mono factorisation preserved and reflected by $\Forg: \Alg_{\cat}(L)\to \cat$.
\end{proposition}
\begin{proof}
Let $f:(A,\alpha)\to (B,\beta)$. Since $\cat$ is regular we have a regular epi-mono factorization for $\Forg f: A\to B$ as $f=i\circ f^*$. Since $L$ preserves regular epis, $Lf^*$ is a regular epi, and is in particular a strong epi. If we now consider the following diagram
\[
\xymatrix
{
L A\ar@{->>}[r]^{Lf^*}\ar@{->>}[d]_{f^*\circ\alpha} & L\im f\ar[d]^{\beta\circ L i}\ar@{-->}[dl]\\
\im f\hspace{1ex}\ar@{>->}[r]_{i} & B
}
\]
which provides us with a structure map for the mediating $L$-algebra $L\im f\to \im f$. This shows that $\Alg_{\cat}(L)$ has regular-epi factorisation reflected by $\Forg$. That it is preserved by $\Forg$ is immediate.
\end{proof}

\begin{proposition}\label{ch1:prop:AlgRegCat}
Let $\cat$ be a regular category and $L:\cat\to\cat$ be a varietor preserving regular epis, then $\Alg_{\cat}(L)$ is regular.
\end{proposition}
\begin{proof}
We need to check that $\Alg_{\cat}(L)$ is (i) finitely complete, (ii) has coequalizers of kernel pairs, and (iii) that regular epis in $\Alg_{\cat}(L)$ are stable under pullback.

\begin{enumerate}[(i)]
\item Since $L$ is a varietor, $\Forg: \Alg_{\cat}(L)\to\cat$ is monadic, and in particular creates all limits which exists in $\cat$. Since $\cat$ is regular, it is finitely complete and thus so is $\Alg_{\cat}(L)$.
\item By Proposition \ref{ch1:prop:regfact} we know that $\Alg_{\cat}(L)$ has regular epi-mono factorisation, and by Lemma \ref{ch1:lem:coeqkp} (ii), we know that the kernel pair of a morphism is isomorphic to the kernel pair of its regular epi factor. We can therefore assume without loss of generality that the morphism $q: (A,\alpha)\to (B,\beta)$ whose kernel pair we will consider is a regular epi. By (i) we know that we can build its kernel pair $p_1,p_2:(\ker q,\kappa)\rightrightarrows (A,\beta)$ in $\Alg_{\cat}(L)$. Let us now check that $q$ is the coequalizer of $p_1,p_2$. For this consider another map $c: (A,\alpha)\to (C,\gamma)$ coequalizing $p_1,p_2$ (in $\Alg_{\cat}(L)$). Since $\cat$ is regular, $q$ coequalizes its kernel pair $p_1,p_2$, and thus there must exist a unique arrow $u: B\to C$ with $u\circ q=c$. We then have the following diagram:
\[
\xymatrix
{
L\ker q \ar@<1ex>[r]^{Lp_1}\ar@<-1ex>[r]_{Lp_2}\ar@{-->}[dd]_{\kappa} & LA\ar@{->>}[rr]^{Lq} \ar[dd]^{\alpha}\ar[dr]^{Lc}& & LB\ar[dd]^{\beta}\ar@{-->}[dl]^{Lu}\\ 
& & LC\ar[dd]^(0.3){\gamma} &\\
\ker q \ar@<1ex>[r]^{p_1}\ar@<-1ex>[r]_{p_2} & A\ar@{->>}[rr]_(0.4){q}\ar[dr]_{c} & & B\ar@{-->}[dl]^{u}\\
& & C &
}
\]
All we now need to check is that $u$ defines a morphism in $\Alg_{\cat}(L)$. This follows from the fact that $L$ preserves regular epis, and that $Lq$ is therefore an epi. Indeed, we necessarily have 
\begin{align*}
\gamma\circ Lc&=\gamma\circ Lu\circ Lq\\
&=c\circ\alpha\\
&=u\circ q\circ \alpha\\
&=u\circ \beta\circ Lq
\end{align*}
and thus $\gamma\circ Lu=u\circ \beta$ as required, since $Lq$ is epi.

\item Finally, we need to show that regular epis are stable under pullback, so let $p: (A,\alpha)\epi(B,\beta)$ be a regular epi and let $f: (C,\gamma)\to (B,\beta)$ be an arbitrary morphism. By (i), we know that the pullback of $p$ along $f$ in $\Alg_{\cat}(L)$ is created by the pullback of $p$ along $f$ in $\cat$, and since $\cat$ is regular, we know that it must be a regular epi in $\cat$. Let us call it $p': (A\times_B C, \delta)\epi (A,\alpha)$. By assumption on $L$, $Lp'$ is also a regular epi, and it remains to show that $p'$ and $Lp'$ combine to form a regular epi in $\Alg_{\cat}(L)$, i.e. a coequalizer of some morphisms. To this effect, let us consider the kernel pair $k_1,k_2: (\ker p',\kappa)\rightrightarrows (A\times_B C, \delta)$ of $p'$. We now proceed exactly as in (ii) to show that $p'$ is the coequalizer of $k_1,k_2$ in $\Alg_{\cat}(L)$, and thus $p'$ is a regular epi in $\Alg_{\cat}(L)$.

\end{enumerate}
\end{proof}

\begin{proposition}\label{ch1:prop:AlgExact}
Let $\cat$ be an exact category and let $L:\cat\to\cat$ be a varietor which preserves regular epimorphisms, then $\Forg: \Alg_{\cat}(L)\to \cat$ creates exact sequences and $\Alg_{\cat}(L)$ is exact.
\end{proposition}
\begin{proof}
Showing that $\Forg$ creates exact sequences in $\Alg_{\cat}(L)$ uses the same proof as (ii) of Proposition \ref{ch1:prop:AlgRegCat}, i.e. we lift an exact sequence
\[
\xymatrix
{
\ker q \ar@<1ex>[r]^{p_1}\ar@<-1ex>[r]_{p_2}& A\ar@{->>}[r]^{q} & B
}
\]
 in $\cat$ by first using the fact that $\Forg$ creates the kernel pair in $\Alg_{\cat}(L)$ and then using the fact $L$ preserves regular epi to show that $q:(A,\alpha)\to(B,\beta)$ is the coequalizer of $p_1,p_2$.
\[
\xymatrix
{
L\ker q \ar@<1ex>[r]^{Lp_1}\ar@<-1ex>[r]_{Lp_2}\ar[dd]_{\kappa} & LA\ar@{->>}[rr]^{Lq} \ar[dd]^{\alpha}\ar[dr]^{Lc}& & LB\ar[dd]^{\beta}\ar@{-->}[dl]^{Lu}\\ 
& & LC\ar[dd]^(0.3){\gamma} &\\
\ker q \ar@<1ex>[r]^{p_1}\ar@<-1ex>[r]_{p_2} & A\ar@{->>}[rr]_(0.4){q}\ar[dr]_{c} & & B\ar@{-->}[dl]^{u}\\
& & C &
}
\]
which shows that any exact sequence in $\cat$ can be lifted to an exact sequence in $\Alg_{\cat}(L)$. 

To see that $\Alg_{\cat}(L)$ is exact, notice first that by Proposition \ref{ch1:prop:AlgRegCat} it is regular, so we need only show that all equivalence relations are effective. Let $e_1\times e_2:(R,\rho)\mono (A\times A, (\alpha\circ L\pi_1)\times (\alpha\circ L\pi_2))$ be an equivalence relation in $\Alg_{\cat}(L)$.
By Lemma \ref{ch1:lem:equivForg}, $\Forg(R,\rho)= R$ is an equivalence relation in $\cat$, and since $\cat$ is exact, $R$ is the kernel pair of a morphism. Since $L$ is a varietor, $\Forg$ is monadic and thus creates limits in $\Alg_{\cat}(L)$, and in particular the kernel pair $(\ker q, \kappa)$ where $\kappa: L\ker q\to \ker q$ defines the unique structure map of $\ker q$ as an $L$-algebra. Finally, since $\Forg$ is monadic it reflects isomorphism (see \cite{cats} Proposition 20.12), and since $\Forg (R, \rho)=\Forg (\ker q,\kappa)$, we get $(R, \rho)=(\ker q,\kappa)$ as desired.
\end{proof}

\subsection{Fully invariant equivalence relations}\label{ch1:subsec:fullyinvariant}

Whilst, equivalence relations are ubiquitous in the whole of mathematics, the stronger notion of \emph{fully invariant} equivalence relation is key in universal algebra, and will be very important in the development of an algebraic semantics for the abstract flavour of coalgebraic logic in the next section. The notion of a fully invariant relation is touched upon in \cite{HughesPhD} (we essentially use his definition 3.5.2.) but our construction of the fully invariant closure of a relation in a categorical setting is, to the best of our knowledge, new. We start by the basic definitions.
\begin{definition}
Let $A$ be an object of some well-powered category $\cat$. We say that a subobject $m: B\mono A$ in $\Sub(A)$ is \textbf{fully invariant}\index{Fully invariant!subobject} if for any endomorphism $f: A\to A$, there exist a unique endomorphism $\tilde{f}:B\to B$ such that
\[
\xymatrix
{
B\hspace{1ex}\ar@{>->}[r]^{m}\ar@{-->}[d]_{\tilde{f}} & A\ar[d]^{f}\\
B\hspace{1ex}\ar@{>->}[r]_{m} & A
}
\]
commutes. In particular, if $e_1\times e_2: R\mono A\times A$ is a relation on $A$, we say that $R$ is a \textbf{fully invariant relation}\index{Fully invariant!relation}  if for any endomorphism $f:A\to A$, there exists a unique morphism $\tilde{f}: R\to R$ making the following diagram commute:
\[
\xymatrix
{
R\hspace{1ex}\ar@{>->}[r]^{e_1\times e_2}\ar@{-->}[d]_{\tilde{f}} & A\times A\ar[d]^{f\times f}\\
R\hspace{1ex}\ar@{>->}[r]_{e_1\times e_2} & A\times A
}
\]
\end{definition}
Let us give some examples of this interesting concept.

\begin{example}
\begin{enumerate}
\item Let $\cat=\Grp$ be the category of group. In this case, the concept of a fully invariant subgroup $H$ of a group $G$ is well known. In a cyclic group $\mathbb{Z}_p$, every subgroup $H$ is fully invariant. Indeed, since $H$ must itself be cyclic, $H=\mathbb{Z}_m$ with $m$ a divisor of $p$, and $H$ can be viewed as the powers of $m$. An endomorphism $f$ of $\mathbb{Z}_p$ is uniquely determined by $f(1)=n$, and thus $f[H]$ are the powers of $mn$ which is included or equal to $H$ itself. Another example of fully invariant subgroup is the commutator subgroup $G'$ of any group $G$ as is easily seen: if $a,b\in G$ then 
\[
f([a,b])=f(aba\inv b\inv)=f(a)f(b)f(a)\inv f(b)\inv=[f(a),f(b)]
\] 
As an example of fully invariant relation, consider $R\subset G\times G$ defined by $(a,b)\in R$ if $b=a\inv$ for $a,b\in G$. If $G$ is abelian, this $\Set$ relation is a subgroup of $G\times G$, and thus a relation in $\Grp$. It is easy to see that it is fully invariant.
\item Let $\cat=\Vect$. It is clear that no proper subvector space $U$ of a vector space $V$ can be fully invariant. Indeed, consider the endomorphism $f$ projecting everything onto a one dimensional subspace which is orthogonal to $U$; clearly $f[U]\nsubseteq U$.
\item Let $A$ be an object of $\DL$ and let us define a relation $R$ on $A$ by $(a,b)\in R$ iff $a\wedge b=a$, for $a,b\in A$ (i.e. $R=\hspace{3pt}\leq_A$). Then it is easy to check by applying the distributivity rule that $R$ defines a relation in $\DL$, and that this relation is fully invariant. This does not hold for boolean algebras, since the relation is not closed under negation.
\item The kind of example which will interest us most is as follows. Let $\cat=\Alg_{\Set}(\polyFunc[\BA])$ where $\polyFunc[\BA]:\Set\to\Set$ is a polynomial functor whose signature can be used to define boolean algebras, for example $\{\bot, \neg,\wedge,\vee\}$.  Consider $\Free_{\BA} V$, the free $\polyFunc[\BA]$-algebra over a set $V$, and define a relation $R$ on $\Free_{\BA} V$ as $a R b$ for $a,b\in \Free_{\BA} V$ if $\BA\deriv a=b$, where $\BA\deriv$ means `is derivable using equational logic and the axioms of boolean algebras'. It is easy to see that $R$ is an equivalence relation in $\Alg_{\Set}(\polyFunc[\BA])$. Moreover, by adjunction, any endomorphism $f:\Free_{\BA}V\to\Free_{\BA} V$ is equivalent to a map $\hat{f}: V\to \Forg_{\BA}\Free_{\BA} V$ (where $\Forg_{\BA}$ is the obvious forgetful functor), i.e. $f$ is just a substitution instance. It is thus clear from the rules of equational logic that $R$ is a fully invariant.
\end{enumerate}
\end{example}

Let us now show how to build the fully invariant closure of a relation. Strictly speaking we will need two constraints on the ambient category (apart from being well-powered): it must be cocomplete and locally small. However, since we are dealing with relations, and relations are well behaved in regular category, it is not unreasonable to place ourselves in such a category. In particular, well-poweredness and local smallness then coincide. 

\begin{proposition}\label{ch1:prop:fullInvClosure}
Let $\cat$ be a well-powered cocomplete regular category, $A$ be an object of $\cat$ and $e_1\times e_2: R\mono A\times A$ be a relation on $R$. Then the kernel pair $E=\ker q$ of the following coequalizer is the fully invariant closure of $R$:
\[
\xymatrix@C=18ex
{
\coprod\limits_{f\in\hom(A,A)} R\ar@<1ex>[r]^<<<<<<<<<<<<<{\coprod\limits_{f\in\hom(A,A)}f\circ e_1}\ar@<-1ex>[r]_<<<<<<<<<<<<<{\coprod\limits_{f\in\hom(A,A)}f\circ e_2} & A\ar@{->>}[r]^{q} & Q
}
\]
\end{proposition}
\begin{proof}
Clearly, this coequalizer and the coproducts defining it exist by the cocompleteness assumption, and $E=\ker q$ is a relation on $A$ (in fact an equivalence relation). To see that it is a fully invariant relation, consider any endomorphism $g: A\to A$, and the diagram
\[
\xymatrix@C=18ex@R=8ex
{
\coprod\limits_{f\in\hom(A,A)} R\ar@<1ex>[r]^<<<<<<<<<<<<<{\coprod\limits_{f\in\hom(A,A)}f\circ e_1}\ar@<-1ex>[r]_<<<<<<<<<<<<<{\coprod\limits_{f\in\hom(A,A)}f\circ e_2} & A\ar@{->>}[r]^{q}\ar[d]^{g} & Q\ar@{-->}[d]^{u} \\
\coprod\limits_{f\in\hom(A,A)}R\ar@<1ex>[r]^<<<<<<<<<<<<<{\coprod\limits_{f\in\hom(A,A)}f\circ e_1}\ar@<-1ex>[r]_<<<<<<<<<<<<<{\coprod\limits_{f\in\hom(A,A)}f\circ e_2} & A\ar@{->>}[r]^{q} & Q
}
\]
We would like to show that there exists a unique arrow $u: Q\to Q$ making the diagram commute. We do this by showing that $q\circ g$ coequalizes the parallel pair of the top row. To see this, note that, by definition, $q$ coequalizes any pair $f\circ e_1, f\circ e_2$ for $f\in \hom(A,A)$. In particular it must coequalize all such pairs factoring through $g$, i.e. all the pairs of the type $g\circ f\circ e_1, g\circ f\circ e_2$, and thus $q$ coequalizes $g\circ \coprod\limits_{f\in\hom(A,A)}f\circ e_i, i=1,2$. So $u:Q\to Q$ exists and is unique, which in turn implies that there must exist a unique $v: E\to E$ making the following diagram commute (since the top row of the diagram can be seen as a cone for the diagram defining $E=\ker q$).
\[
\xymatrix
{
E\ar@{-->}[d]_{v}\ar@<1ex>[r]^{p_1}\ar@<-1ex>[r]_{p_2} & A\ar@{->>}[r]^{q}\ar[d]^{g} & Q\ar[d]^{u}\\
E\ar@<1ex>[r]^{p_1}\ar@<-1ex>[r]_{p_2} & A\ar@{->>}[r]^{q} & Q
}
\]
This shows that $E$ is fully invariant. 

We can easily see that $E$ contains $R$ (i.e. that there exists a mono from $R$ to $E$). Note first that since $q\circ e_1=q\circ e_2$ (since $\id_A\in\hom(A,A)$), $R$ is a cone for the diagram defining $E$, so there must exist a unique morphism $u: R\to E$, Moreover, this morphism must satisfy $p_1\times p_2\circ u=e_1\times e_2$, which means that $u$ must be a mono since $e_1\times e_2$ is a mono. To see that $E$ is the smallest such fully invariant relation, simply notice that the coequalizer $q'$ of any other fully invariant relation $E'$ containing $R$ would have to coequalize the pair $f\circ e_1, f\circ e_2$ for any $f\in\hom(A,A)$ as is clear from the following diagram
\[
\xymatrix@C=14ex
{
R\ar@{>->}[d]\ar@{>->}[dr]^{e_1\times e_2}\\
E'\hspace{3pt}\ar@{>->}[r]^{p_1'\times p_2'}\ar@{-->}[d] & A\times A\ar[d]^{f\times f} \\
E'\hspace{3pt}\ar@{>->}[r]^{p_1'\times p_2'} & A\times A\ar@{->>}[r]^{q'} & Q'
}
\]
where $p_1',p_2'$ are the projections of the from the relation $E'$ to $A$. In consequence, there must exist a unique morphism $Q\to Q'$, and thus a unique morphism $u: E=\ker q\mono E'=\ker q'$ such that $(p_1'\times p_2')\circ u=p_1\times p_2$. This morphism $u$ is a mono since $p_1\times p_2$ is a mono (Lemma \ref{ch1:lem:kpRequiv}).
\end{proof}

\section{Coalgebraic logics}\label{ch1:sec:coalglog}

We introduce coalgebraic logics by following the pedagogically helpful trichotomy of Kupke and Pattinson's review paper \cite{DirkOverview}, namely that coalgebraic logics come in three flavours: the \textbf{predicate lifting, nabla}, and \textbf{abstract}\index{Predicate Lifting!style} flavour. The latter unifying the former two. 
The existing literature on coalgebraic logic is overwhelmingly boolean (see \cite{JacobsExpressivity,2012:PositiveCoalgExpress,2013:PositiveCoalg} for some exceptions). The predicate lifting and nabla flavours of coalgebraic logic in particular can, for all intents and purposes, be considered to be boolean logics. The abstract flavour of coalgebraic logic however, explicitly parametrizes the choice of reasoning structure, and indeed \cite{JacobsExpressivity} gives several interesting example of non-boolean coalgebraic logics. We will therefore proceed as follows: the predicate lifting and nabla styles will be presented as strictly boolean, whereas the abstract style will be presented in its full generality, i.e. parametric in a choice of reasoning structure.

\subsection{Three styles of language}\label{ch1:sec:coalgLang}

We start with the \textbf{predicate lifting} style of coalgebraic logic. Given a finitary signature $(\Sigma,\ari), \ari:\Sigma\to\mathbb{N}$ and a set $V$ of propositional variables, the \textbf{predicate lifting language}\index{Predicate Lifting!language} $\plLang(V)$ (or simply $\plLang$ if there is no ambiguity) has a syntax given by
\[
a::=p\mid \bot\mid \neg a\mid a\wedge a\mid \sigma(a,\ldots,a)
\]
where $p\in V$, and $\sigma\in\Sigma$. Note that we are using the notational convention of \cite{KKV:2012:Journal} where the lower case Roman letters $a,b,c$ stand for formulae. Algebraically, if we denote by $\Sigma_{BA}$ the signature we've chosen for the propositional (i.e. boolean) part of the language (in this instance $\Sigma_{BA}=\{\bot,\neg,\wedge\}$ with the usual arities), then the language $\plLang$ is the free algebra over $V$ for the $\Set$-endofunctor $S_{\Sigma_{BA}+\Sigma}$ defined as 
\[
\mathsf{S}_{\Sigma_{BA}+\Sigma} X=\coprod_{\sigma\in \Sigma_{BA}+\Sigma} X^{\ari(\sigma)}
\]
We know that $\plLang$ exists by Corollary \ref{ch1:cor:initAlgFin} and the fact that $\mathsf{S}_{\Sigma_{BA}+\Sigma}$ is a finitary polynomial functor.

The \textbf{$\nabla$-style}\index{Nabla!style} of coalgebraic logic (also known as Moss style, or coalgebraic logic for the cover modality) has a very different flavour. We refer to \cite{KKV:2012:Journal} for a very good and very thorough overview of this quirky but very powerful logic. Since the language involves objects of many types, the notational conventions are crucial to avoid any confusion. We will scrupulously follow the conventions of \cite{KKV:2012:Journal}. We start by fixing a weak-pullback preserving and standard\footnote{\textbf{Standard}\index{Standard functor} means that the functor preserves monomorphisms. As shown in \cite{1969:Trnkova} and \cite{Gumm05}, for the purpose of studying coalgebras defined by a $\Set$-endofunctor $T$ we can always assume that $T$ preserves monomorphisms.} $\Set$-endofunctor $T$ which will be used both to build the syntax of the logic and to define its coalgebraic semantics. We define $T_\omega X= 
\bigcup \{TY \mid Y \subseteq X, Y \mbox{ finite}\}$, the \textbf{finitary version}\index{Finitary version} of $T$ (more on finitary functors in Chapter 3 and on the finitary version in Chapter 5). The \textbf{nabla language}\index{Nabla!language}\index{Nabla!operator} induced by $T$ is written as $\lngT$ and is given by:
\[ 
a::= p \mid \neg a \mid \Land \phi \mid \Lor \phi \mid \nabla \alpha
\]
where $p \in V$,  $\phi \in\powf\lngT$ and $\alpha \in T_\omega\lngT$. Elements of type $\powf\lngT$ will always be denoted by $\phi,\psi,\cdots$ whilst elements of type $T_\omega\lngT$ will always be denoted by $\alpha,\beta,\cdots$. Note that $\bigwedge$ and $\bigvee$ are operators of type $\powf\lngT\to\lngT$ and of course corresponds to the usual meet and join operations, the constant $\bot$ is defined as $\bigvee\emptyset$ and $\top$ as $\bigwedge\emptyset$. Once again, we can view the language as a free algebra over $V$ for a $\Set$-endofunctor, this time we use $L_T=\Id + \powf + \powf + T_\omega$, and $\lngT$ is the free $L_T$-algebra over $V$. It is clear that since $L_T$ is a coproduct of finitary functors, it is itself finitary, and thus $\lngT$ exists by Corollary \ref{ch1:cor:initAlgFin}. As is customary, the semantics of $\lngT$ will be defined inductively, which means that we need to define what subformulas are in this rather unusual language, which is not immediately obvious for $\nabla$-terms. For this we need the notion of \textbf{base} \index{Base} to which we will return at the beginning of Chapter 3. For any $\alpha \in T_\omega \lngT$ we define the base of $\alpha$ by $\Base_T(\alpha) = \bigcap \{U\subseteq \lang\mid \alpha \in T U\}$. The base of $\alpha$ is the set of immediate subformulas of $\nabla\alpha$. We can now define the set $Sfml(a)$ of subformulas\index{Subformula} of a formula $a$ recursively in the obvious way for formulas whose outermost operators are boolean connectives, and by
\[
Sfml(\nabla\alpha)=\{\nabla\alpha\}\cup\bigcup_{a\in\Base_T(\alpha)}Sfml(a)
\]
for $\nabla$-terms.

Finally, the \textbf{abstract style}\index{Abstract!style} of coalgebraic logic is both a generalization and a unification of the previous two styles. One of its key features is that it hides the fundamental structure of the formulas by defining formulas as elements of a free algebra for an endofunctor defined not on $\Set$, but on a choice of reasoning structure, typically $\BA$ or $\DL$. In other words all the terms are understood to be elements of an algebraic structure, and equivalence in this structure is hidden away. The basic setup, which can be traced back to \cite{AlgSem2004} and was subsequently developed in \cite{2005:UfExtCoalg} and \cite{JacobsExpressivity}, is as follows:
\begin{equation}\label{ch1:diag:fundamentalSituation}
\xymatrix
{
\cat\ar@/^1pc/[rr]^{F} \ar@(l,u)^{L} \ar@<5pt>[d]^{\Forg} & \perp&\cat[D]\op\ar@/^1pc/[ll]^{G}\ar@(r,u)_{T\op}\\
\Set\ar@<5pt>[u]^{\Free}_{\dashv}
}
\end{equation}
The left hand-side of the diagram is the syntactic side, and the right-hand side the semantic one. The category $\cat$ represents a choice of `reasoning kernel', i.e. of logical operations which we consider to be fundamental, whilst $L$ is a syntax constructing functor which builds terms over the reasoning kernel. The category $\cat$ is a finitary variety and there exist and free-forgetful adjunction $\Free\dashv\Forg:\cat\to\Set$ which is $\Forg$ is monadic and creates directed colimits. On the semantics side, objects in $\cat[D]$ are the carriers of models and $T$ specifies the coalgebras on these carriers in which the operations defined by $L$ will be interpreted. The functors $F$ and $G$ (where $F\dashv G$) relate the syntax and the semantics. Note that we only need a dual adjunction, not a full duality.

A typical example of this situation is the adjunction $\pf\dashv\ups: \DL\to\Pos\op$, where $\pf$\index{Prime filter!functor} is the functor sending a distributive lattice to its poset of prime filters, and $\DL$-morphisms to their inverse images, and $\ups$\index{Up-set!functor} is the functor sending a poset to the distributive lattice of its up-sets and monotone maps to their inverse images. In the case where a distributive lattice is a boolean algebra, it is well-known that prime filters are maximal, i.e. ultrafilters, and the partial order on the set of ultrafilter is thus discrete, i.e. ultrafilters \index{Ultrafilter!functor} are only related to themselves. Downsets are thus simply subsets and the adjunction $\pf\dashv\ups$ becomes the well-known adjunction $\uf\dashv\pow:\BA\to\Set\op$ where $\pow$ is the contravariant powerset functor \index{Powerset!contravariant functor} and powersets are endowed with the usual boolean structure. More situations of the type of Diagram (\ref{ch1:diag:fundamentalSituation}) can be found in \cite{JacobsExpressivity}.

Let us for the moment focus only on the syntax builder $L$. It defines an \textbf{abstract coalgebraic language}\index{Abstract!language} which is defined for a set of propositional variables $V$ as the free $L$-algebra over $\Free V$. We therefore require that $L$ be a varietor. If we denote by $\mathsf{G}\dashv\mathsf{V}$ the adjunction $\cat\to\Alg_{\cat}(L)$, then the abstract language is given by $(\mathsf{G}\Free V,\langle-\rangle)$, where $\langle-\rangle$ denotes the structure map of the free algebra. Note that we can equally well consider the language as the initial algebra of the functor $L'=L(-)+\Free V$. Indeed, we can easily put an $L'$-algebra structure on $\mathsf{G}\Free V$ via the map $\langle-\rangle+\eta^{\mathsf{G}}_{\Free V}$, where $\eta^{\mathsf{G}}$ is the unit of $\mathsf{G}\dashv\mathsf{V}$. So the two descriptions are equivalent, and being able to use either will prove very useful indeed. 

We can once again use the notion of \emph{base} to describe the subformulas of a formula $\langle\alpha\rangle\in\mathsf{G}\Free V, \alpha\in L\mathsf{G}\Free V$. We use the same notational convention for abstract coalgebraic languages as for nabla languages. Some care must be taken, since we are no longer working in $\Set$, however as we shall see in Chapter 3, the notion of base is well-behaved when $L$ preserves all intersections. We can then define subformulas inductively, just as in the case of the nabla logic, with the usual clauses for propositional connectives, and with 
\begin{align*}
Sfml(\langle\alpha\rangle)=\{\langle\alpha\rangle\}\cup \bigcup_{a\in\Base_L(\alpha)}Sfml(a)
\end{align*}

This abstract language covers the previous two styles of languages in the following sense. Let us first look at the predicate lifting language $\plLang$ for a signature $\Sigma$. We define $L:\BA\to\BA$ as 
\[L=\Free\left(\coprod_{\sigma\in\Sigma}(\Forg(-))^{\ari(\sigma)}\right)=\Free \mathsf{S}_{\Sigma}\Forg
\]
In other words we use the adjunction $\Free\dashv\Forg$ to lift $\mathsf{S}_\Sigma$ from $\Set\to\Set$ to $\BA\to\BA$. We can then see that
\[\mathsf{G}\Free V=\plLang/\equiv\]
where $a \equiv b$ if $a$ and $b$ are provably equivalent using equational logic and the axioms of boolean algebras. In other words, the abstract language for $L$ is just the predicate lifting language quotiented by boolean equivalence. Note that $\mathsf{G}\Free V$ exists by Proposition \ref{ch1:prop:BAERightAdj}.

Similarly, we can define an abstract language corresponding to the quotienting of the nabla language under boolean equivalence. For this we define $L=\Free T\Forg$ and we then have
\[\mathsf{G}\Free V=\lngT/\equiv\]
Again, we are in the presence of a functor in $\BA$ which is the `lifted' version of a functor in $\Set$. $\mathsf{G}\Free V$ exists when $L$ is a varietor, which is for example the case when $T$ is finitary (since $\Free$ is a left adjoint it preserves all colimits, and in particular filtered ones).

A functor $L$, defining an abstract coalgebraic language, does not have to be of the form $\Free T\Forg$ for some $\Set$ endofunctor $T$. It can be much more general, and encode much more than simply quotienting term building in $\Set$ under boolean equivalence. For example it can encode some extra axioms which we would like to take as standard for all the logics we consider. For example, we might want to encode the axiom $\mathbf{K}$ in any functor $L$ describing classical modal logics. As a rule of thumb, everything we want as standard, should be encoded in $L$, whilst formulas we want to experiment with should be kept out of $L$. Alternatively, we can use the rule of thumb that $L$ should contain all the axioms necessary to axiomatize the class of all coalgebras for a certain functor of interest, whilst extra axioms, kept out of the functor $L$, can be used to describe proper classes (covarieties) of such coalgebras. Time to move on to the semantic side.

\subsection{Three styles of semantics}\label{ch1:sec:semantics}

All three types of coalgebraic languages are, not surprisingly, interpreted in coalgebras.  Given a standard $\Set$-endofunctor $T$, a \textbf{coalgebra}\index{Coalgebra} is a pair $(W,\gamma)$ where $W$ is a set (of worlds) and $\gamma:W\to TW$ is a transition map ($T$ defines the `transition type').

We start with the predicate lifting style of coalgebraic logic which owes its name to the device used to interpret its operators. Let us fix once again a signature $(\Sigma,\ari)$ and let $\plLang$ be the associated language as defined above. Each modal operator $\sigma\in\Sigma$ is interpreted by a \textbf{predicate lifting}\index{Predicate lifting}, i.e. a natural transformation $\lsem\sigma\rsem:\powc^n\fto \powc T$ where $\powc:\Set\op\to\Set$ is the contravariant powerset functor. Intuitively, predicate liftings `lift' $n$-tuples of predicates (i.e. subsets, hence the powerset functor $\mathcal{Q}$) to a predicate on transitions (hence $\mathcal{Q} T$). A $T$-\textbf{coalgebraic model}\index{Model!coalgebraic} - or $T$-model (or just model if there is no ambiguity about $T$) - is a triple $\mathcal{M}=(W,\gamma,\pi)$ where $\pi: W\to\pow(V)$ is a valuation. The notion of \textbf{truth} of a formula $a$ at a point $w\in W$ is defined inductively in the usual manner for propositional variables and boolean operators, and by\index{Semantics!predicate lifting style}
$$\mathcal{M},w\models\sigma(a_1,\ldots,a_n)\mbox{ iff }\gamma(w)\in\lsem\sigma\rsem_W(\lsem a_1\rsem,\ldots,\lsem a_n\rsem)$$
for modal operators, where $\lsem a_i\rsem$ is the interpretation of $a_i$ in $W$. A formula $a$ is \textbf{satisfiable}\index{Satisfiability!in the predicate lifting style} in $\mathcal{M}$ if there exists $w\in W$ such that $\mathcal{M},w\models a$. A \textbf{coalgebraic frame}\index{Frame!coalgebraic} - or $T$-frame - is just a $T$-coalgebra $(W,\gamma)$ and a formula $a$ is \textbf{valid}\index{Validity} on the frame if for any valuation $\pi$, $a$ is true at every point in the model $(W,\gamma,\pi)$.

Let us now turn to the interpretation of the nabla style of coalgebraic logic. As we mentioned earlier, the same functor is used both for the syntax and the semantics. So let $T$ be a weak pullback preserving $\Set$-endofunctor, and let $\lngT$ be the language defined by $T$ as above. In order to define the semantics of nabla formulas we need the notion of relation lifting, developed in the previous section for general regular categories, in $\Set$. Note that the ubiquitous requirement that a functor should preserve regular epis in order for liftings to be well-behaved is automatically satisfied in $\Set$ since every epi has a section (assuming the axiom of choice). Let us briefly specialize the definition of relation lifting from the previous section to the category $\Set$. Let $X_1, X_2$ be sets and let $R\subseteq X_1\times X_2$ be a relation between $X_1$ and $X_2$. By construction, the lifting $\Tlift[R]$ of $R$ is 
$\Tlift[R]=(T\pi_1\times T\pi_2)[TR]$. Alternatively and equivalently, $\Tlift[R]$ is the relation between $TX_1$ and $TX_2$ defined by 
\[
\Tlift[R]=\{(\alpha,\beta)\in TX_1\times TX_2\mid \exists z\in TR \text{ s.th. }T\pi_1(z)=\alpha, T\pi_2(z)=\beta\}
\]
where $\pi_i: R\to X_i, i=1,2$ is the projection onto the $i^{th}$ component. If $T$ preserves weak-pullbacks (regular epis is automatic), then $T$-liftings satisfy Propositions \ref{ch1:prop:liftingwpb} and \ref{ch1:prop:liftingProp} and we can then define the semantics of $\lngT$. Given a $T$-model $\mathcal{M} = (W, \gamma, \pi)$, we define the validity relation $\models\subseteq W\times\lngT$ inductively for any $a\in \lngT$ and world $w \in W$ as usual for atomic propositions and propositional connectives, together with \index{Semantics!nabla style}
\[
\mathcal{M}, w \models \nabla \alpha\mbox{ iff } (\gamma(w),\alpha) \in \Tlift[\models]
\]
where $(\Tlift[\models])\subseteq TW \times T\lngT$ is the relation lifting of the truth-relation $\models \subseteq W \times \lngT$. Note that for this definition to indeed be inductive, it must be the case that we only need to know $\models$ on the subformulas of $\nabla\alpha$, i.e. on the base of $\alpha$. This is where Propositions \ref{ch1:prop:liftingwpb} and \ref{ch1:prop:liftingProp} come into play: since we're assuming that $T$ preserves weak pullbacks we have
\begin{align*}
(\gamma(w),\alpha)\in\Tlift[\models]&\text{ iff }(\gamma(w),\alpha)\in\Tlift[\models]\hspace{1pt}\restrict (TW\times T\Base_T(\alpha))\\
&\text{ iff }(\gamma(w),\alpha)\in\Tlift[(\models\hspace{1pt}\restrict(W\times\Base_T(\alpha)))]
\end{align*}
i.e. we only need to know how $\models$ relates states to subformulas of $\nabla\alpha$, which is what was required for an inductive definition.

Finally, let us describe the semantics of the abstract version of coalgebraic logic. Again, this semantics generalises the two types of coalgebraic semantics we have defined above. Recall that the fundamental situation of Diagram (\ref{ch1:diag:fundamentalSituation})
\[
\xymatrix
{
\cat\ar@/^1pc/[rr]^{F} \ar@(l,u)^{L} & \perp&\cat[D]\op\ar@/^1pc/[ll]^{G}\ar@(r,u)_{T\op}
}
\]
where $L$ describes the syntax, and $T$ describes the semantic domain, i.e. $T$-coalgebras. Intuitively, the semantic should be something which associates to an $L$-formula, a `predicate' on the carrier of a $T$-frame $(W,\gamma)$, namely the denotation of the formula. In the abstract setting `predicates' are given by the functor $G$. Indeed in practise the elements of $GW$ can often be understood as morphisms into a two-element structure $\mathbbm{2}$. Down-sets on a poset $W$ for example, are equivalent to monotone maps $W\to\mathbbm{2}$, where $\mathbbm{2}$ in the obvious two-element poset, and similarly subsets are equivalent to their characteristic function. 

Since the semantic is given inductively, we can assume that to interpret an $L$-formula we need to interpret an $L$-construct of subformulas which have already been interpreted, i.e. an element of the type $LG W$. In this perspective it therefore natural to define an abstract semantic as a device which translates $L$-constructs of interpreted sub-formulas to their denotation, i.e. a natural transformation of the type\index{Semantics!abstract style}
\[\delta: L G\to G T\]
We will call such a natural transformation a \textbf{semantic transformation} \index{Semantic transformation}. Recall that the language is the free $L$-algebra over $\Free V$. Its carrier is equivalently given by the initial algebra $\init[L']$ where $L'=L(-)+\Free V$. The semantic natural transformation $\delta$ yields an interpretation maps $\lsem-\rsem_W: \init[L']\to G W$ as follows. Given a coalgebra $(W,\gamma)$, if we can endow $G W$ with an $L'$-algebra structure, then we get a unique interpretation map $\lsem-\rsem_W: \init[L']\to G W$ by initiality. To define an $L'$-algebra structure on $G W$, we need $\delta$ and a map $\Free V\to G W$, or equivalently, a function $v: V\to\Forg G W$. Such functions are abstract versions of the familiar valuations as is clear by considering the case where $F=\uf, G=\pow$. Given such a valuation, we get:
\[
\xymatrix@C=12ex
{
L\init[L']+\Free V\ar@{-->}[r]^{L\lsem-\rsem_W+\id_{\Free V}}\ar[dd] & L\pow W\ar[d]^{\delta_W+\id_{\Free V}} +\Free V\\
& \pow TW+\Free V\ar[d]^{\pow\gamma+\hat{v}}\\
\init[L']\ar@{-->}[r]_{\lsem-\rsem_W} & \pow W
}
\]
where $\hat{v}$ is the adjoint transpose of the valuation. The natural transformation $\delta$ inserted in this way does precisely what we had in mind, namely inductively define the semantics of formulas in $\mathsf{G}\Free V$.

Let us briefly show how this type of semantics covers the two cases above. In the case of the predicate lifting semantics interpreted over $T$-coalgebras, recall that we had defined $L$ as $\Free \polyFunc\Forg$ where $\polyFunc$ is the polynomial functor on $\Set$ induced by the signature $\Sigma$. The abstract definition of the semantics requires a natural transformation 
\[
\delta: L\pow=\Free \mathsf{S}_\Sigma\Forg\pow=\Free\mathsf{S}_\Sigma\mathcal{Q} \to \pow T
\]
which, by adjunction, is equivalent to a natural transformation 
\[
\tilde{\delta}: \mathsf{S}_\Sigma\mathcal{Q}\to \Forg\pow T=\mathcal{Q}T
\]
which, in turn, is equivalent to a collection of natural transformations
\[
\lsem\sigma\rsem: \mathcal{Q}^{\ari(\sigma)}\to \mathcal{Q}T
\]
for each $\sigma\in\Sigma$. In other words, we recover precisely the concept of predicate lifting. The predicate lifting style of semantics is thus simply the adjoint transpose of the abstract style of semantics for `polynomial syntax' functors.

Let us now show that the abstract semantics also subsumes the nabla-style of semantics. Recall first that in this case we had put $L: \Free T\Forg$. Again we write $L'=\Free T\Forg(-)+\Free V$, and recall that the interpretation maps should be given by:
\[
\xymatrix@C=12ex
{
\Free T\Forg \init[L']+\Free V\ar@{-->}[r]^{L\lsem-\rsem_W+\id_{\Free V}}\ar[dd] & \Free T\mathcal{Q} W\ar[d]^{\delta_W+\id_{\Free V}} +\Free V\\
& \pow TW+\Free V\ar[d]^{\pow\gamma+\hat{v}}\\
\init[L']\ar@{-->}[r]_{\lsem-\rsem_W} & \pow W
}
\]
for a valuation $v: V\to\mathcal{Q}W$ and a well-chosen semantic transformation $\delta$.
By using the adjoint transpose maps where required we can re-write the diagram above as
\[
\xymatrix@C=12ex
{
T\Forg \init[L']+V\ar[dd]\ar@{-->}[r]^{T\lsem-\rsem_W+\id_V} & T\mathcal{Q}W+V\ar[d]^{\tilde{\delta}_W+\id_V}\\
& \mathcal{Q}TW+V\ar[d]^{\mathcal{Q}\gamma+v}\\
\Forg\init[L]\ar@{-->}[r]_{\lsem - \rsem_W} & \mathcal{Q}W
}
\]
where we have used the same notation for the interpretation maps and its adjoint transpose to clarify the notation. Notice now that for a purely propositional formula $p\in\lngT$ we have
\[
w\models p\mbox{ iff }w\in v(p)\mbox{ iff }(w,p)\in (\in;\lsem-\rsem_W\op)
\]
For a formula $\nabla\alpha\in\lngT$ of modal depth equals to one, we can use this equivalence and the nabla-style of semantics to conclude that
\begin{align*}
w\models \nabla\alpha & \Leftrightarrow (\gamma(w),\alpha)\in\Tlift[(\models\hspace{1pt}\restrict(W\times\Base_T(\alpha)))]\\
& \Leftrightarrow (\gamma(w),\alpha)\in \Tlift[((\in_W;\lsem-\rsem_W\op)\hspace{1pt}\restrict(W\times\Base_T(\alpha)))] & \mbox{Equiv. above} \\
& \Leftrightarrow (\gamma(w),\alpha)\in \Tmem[W]; (T\lsem - \rsem_W)\op \restrict TW\times T\Base_T(\alpha) &  \mbox{Prop.  \ref{ch1:prop:liftingProp}}\\
&\Leftrightarrow \gamma(w) \Tmem[W] T\lsem \alpha \rsem_W
\end{align*}
The argument can then be repeated for formulas of any modal depth. It therefore follows that if we define $\tilde{\delta}_W$ by
\[\tilde{\delta}_W: T\mathcal{Q} W\to\mathcal{Q}T W, A\mapsto\{t\in TW\mid t\Tmem[W] A\}\]
the diagram above will commute, and we thus have shown that the abstract style of semantics also covers the nabla case.

\subsection{Two styles of axiomatization}

One purpose of coalgebraic logics is to logically describe classes of coalgebras. In order to achieve this some axiomatization is needed. There are essentially two types of axiomatization: using \emph{rules} or using \emph{axioms}. For the predicate lifting style, either style can be used, and the choice is largely a matter of taste and convenience. The nabla flavour of coalgebraic logic is again rather different, and is arguably the most elegant type of coalgebraic logic in that it possesses a  generic, rule-based, axiomatization which is fully parametric in the choice of functor defining the coalgebraic semantics. Finally, the abstract style of coalgebraic logic is rarely used in practise to axiomatize classes of coalgebras but could be given either an axiom based or a rule based axiomatization, although we will opt for axioms only. 

\subsubsection{Rule and axiom-based axiomatizations in the predicate lifting style}

Let us once again start by examining the case of predicate-lifting style coalgebraic logic. Given a choice of semantics - i.e. the choice of a $\Set$-endofunctor $T$ - it has been shown in \cite{2003:Dirk, Schroder06} that it is always possible to find an axiomatization of the class $\Coalg(T)$ of all $T$-coalgebra in a Hilbert system using either rules or axioms. Moreover, the rules or axioms can be assumed to have a particularly simple form. We cite the key definitions and results, for more details see \cite{2003:Dirk, Schroder06, DirkOverview}.

\begin{definition}\index{One step rule}\index{Rank 1 axiom}
Let $(\Sigma,\ari)$ be a finitary signature, let $V$ be a set of propositional variables and let $\plLang$ be the associated predicate-lifting style language. For any subset $U\subset\plLang$ we define $\mathsf{Up}(U)$ as 
\[\mathsf{Up}(U)=\{\sigma(a_1,\ldots,a_n)\mid \sigma\in\Sigma, \ari(\sigma)=n, a_i\in U, 1\leq i\leq n\}\]
For any subset $U$ of $\plLang$ we define $\mathsf{Cl}(U)$ to be the set of clauses over $U$, i.e. the set of all formulas of the form $\bigvee_{i=1}^n \varepsilon_i b_i$ where $\varepsilon_i$ is either nothing or $\neg$ and $b_i\in U$, $1\leq i\leq n$.

A \textbf{one step rule}\index{One step rule} $a/b$ is a pair $(a,b)\in \Free V\times \mathsf{Cl}(\mathsf{Up}(V))$. A \textbf{rank 1 clause} is a formula $a\in \mathsf{Cl}(\mathsf{Up}(\Free V))$.
\end{definition}

\index{Rule-based axiomatization}\index{Axiom-based axiomatization} Given a collection $\mathsf{R}$ of rules, we define a very simple Hilbert deduction system whose axioms are given by some choice of axioms of propositional logic (considered as rules without premise) together with the rules of $\mathsf{R}$, and whose rules of inference are modus ponens, uniform substitution, and the \textbf{congruence rule}\index{Congruence rule}
\begin{prooftree}
\AxiomC{$a_i\leftrightarrow b_i, 1\leq i\leq n$}
\LeftLabel{(Congruence)}
\RightLabel{($\sigma\in\Sigma$)}
\UnaryInfC{$\sigma(a_1,\ldots,a_n)\leftrightarrow \sigma(b_1,\ldots, b_n)$}
\end{prooftree}
This deduction system defines a provability predicate $\mathsf{R}\deriv[ML]$ on $\plLang$, and, as usual, we define the \textbf{logic}\index{Logic} $\logic_{\mathsf{R}}$ as the set of formulas $a\in\plLang$ such that $\mathsf{R}\deriv[ML] a$. We use the notation $\deriv[ML]$ because the Hilbert system is essentially that of modal logic, the only difference being that the necessitation rule is replaced by the congruence rule.

Similarly, given a set of axioms $\Ax\subset \plLang$, we define the Hilbert deduction system whose axioms are given by some choice of axioms of propositional logic together with the axioms of $\Ax$, and whose inference rules are modus ponens, uniform substitution, and congruence. We denote the associated provability predicate on $\plLang$ as $\Ax\deriv[ML]$ and define the logic $\logic_{\Ax}$ as all the formulas $a\in\plLang$ such that $\Ax\deriv[ML] a$.

The main technical tool in the study of completeness of such Hilbert systems with respect to a choice of coalgebraic semantics, are the notions of \emph{one step soundness} and \emph{one step completeness}.

\begin{definition}\index{One step soundness}\index{One step completeness}
Given a predicate lifting style language $\plLang$ and an interpretation of this language in the class of $T$-coalgebra, a one step rule $a/b$  where $b=\bigvee_{i=1}^n \varepsilon_i b_i$ and each $b_i=\sigma_i(c_1^i,\ldots,c_{k_i}^i)$ with $c_1^i,\ldots,c_{k_1}^i\in  V$, is \textbf{one step sound} if for any $T$-model $\mathcal{M}=(W,\gamma,\pi)$, whenever $\mathcal{M}\models a$ 
\[\
\bigcup_{i=1}^n \varepsilon_i\lsem\sigma_i\rsem (\lsem c^i_1\rsem,\ldots,\lsem c^i_{k_i}\rsem)=TW
\]
where $\varepsilon_i$ here stands for nothing or complementation. In other words, if $a$ is true in the entire model, then any transition must make $b$ true too. 

Conversely, a rule a one step rule $a/b$ is \textbf{one step complete} if for every $T$-model $(W,\gamma,\pi)$ and every formula $d\in \mathsf{Cl}(\mathsf{Up}(V)$, $d=\bigvee_{i=1}^n\epsilon_i\sigma_i(c_1^i,\ldots,c_{k_i}^i)$,
\[\
\bigcup_{i=1}^n \varepsilon_i\lsem\sigma_i\rsem (\lsem c^i_1\rsem,\ldots,\lsem c^i_{k_i}\rsem)=TW
\]
implies that
\[
\{\sigma b\mid \sigma:\Free V\to\Free V, (W,\gamma)\models \sigma a\}\deriv[PL] d
\]
where $\deriv[PL]$ means `is derivable using propositional calculus only'. In other words, if a formula in $\mathsf{Cl}(\mathsf{Up}(V)$ (the format of the conclusion of one step rules) is true for any transition of the frame $(W,\gamma)$, then it must be derivable, using propositional logic, from the substitution instances of the conclusion of the rule which make the premise valid on $(W,\gamma)$. A rank 1 clause is one-step complete if it is one step complete when viewed as a one-step rule with empty premise.
\end{definition}

\begin{theorem}[\cite{Schroder06}]
Given a choice $T$ of semantics, and a predicate lifting style language $\plLang$, the set of all one-step sound one-step rules is complete with respect to $\Coalg(T)$.
\end{theorem}

As was shown in Proposition 15 of \cite{Schroder06}, for any one-step rule $a/b$, there exists a rank 1 clause $c$ which is equivalent to $a/b$ in the sense that $\{a,c\}\deriv[ML] b$ iff $\{a/b\}\deriv[ML] c$. We therefore have the corollary:

\begin{corollary}[\cite{Schroder06}]
Given a choice $T$ of semantics, and a predicate lifting style language $\plLang$, the set of all one-step sound rank-1 clauses is complete with respect to $\Coalg(T)$.
\end{corollary}

Note that this Theorem and its Corollary, only shows that complete axiomatizations exists in principle. Of course in practise, there exist more compact axiomatizations of many well-known endofunctor, and we refer the reader to e.g. Example 3.17 of \cite{2006:DirkLutzPSpace} for a list of such Hilbert systems (typically given in terms of one-step rules).

\subsubsection{The rule-based axiomatization of nabla logic}

Let us now detail the generic axiomatization of the nabla style of coalgebraic logics. From its genericity and its simplicity it warrants to be studied separately, and this what we will do in Chapter 4 too. We once again refer the reader to \cite{KKV:2012:Journal} for all the details about this axiomatization, and the proof of its completeness. For our purpose it will be enough to state that for a weak pullback preserving standard functor $T$, the 2-dimensional Hilbert system, which we call $\KKV(T)$, given by the axioms and rules \index{Axiomatization of nabla logics} of Table \ref{ch1:table:KKV} is sound and weakly complete with respect to the nabla-style of semantics in the class $\Coalg(T)$.

\begin{table}
\begin{center}
\begin{tabular}{| c c |}
\hline
& \\
\AxiomC{}
\UnaryInfC{$a\leq a$}
\DisplayProof
& 
\AxiomC{$a\leq c$}
\AxiomC{$c\leq b$}
\LeftLabel{(Cut)}
\BinaryInfC{$a\leq b$}
\DisplayProof 
\\ & \\
\AxiomC{$\{a\leq b\mid a\in\phi\}$}
\LeftLabel{$(\bigvee L)$}
\UnaryInfC{$\bigvee\phi\leq b$}
\DisplayProof
&
\AxiomC{$a\leq b$}
\LeftLabel{$(\bigvee R)$}
\RightLabel{$b\in\phi$}
\UnaryInfC{$a\leq\bigvee\phi$}
\DisplayProof
\\ & \\
\AxiomC{$a\leq b$}
\LeftLabel{$(\bigwedge L)$}
\RightLabel{$a\in\phi$}
\UnaryInfC{$\bigwedge\phi\leq b$}
\DisplayProof
&
\AxiomC{$\{a\leq b\mid b\in\phi\}$}
\LeftLabel{$(\bigwedge R)$}
\UnaryInfC{$a\leq\bigwedge\phi$}
\DisplayProof
\\ & \\
\AxiomC{$\bigwedge\{\phi\cup\{\neg a\}\}\leq \bigvee\psi$}
\LeftLabel{$(\neg E)$}
\UnaryInfC{$\bigwedge\phi\leq\bigvee\{\psi\cup\{a\}\}$}
\DisplayProof
&
\AxiomC{$\bigwedge\{\phi\cup\{a\}\}\leq\bigvee\psi$}
\LeftLabel{$(\neg I)$}
\UnaryInfC{$\bigwedge\phi\leq\bigvee\{\psi\cup\{\neg a\}\}$}
\DisplayProof
\\ & \\
\multicolumn{2}{|c|}{
\AxiomC{}
\LeftLabel{(Distributivity)}
\UnaryInfC{$\bigwedge\{\bigvee\phi\mid\phi\in X\}\leq\bigvee\{\bigwedge\rng(\gamma)\mid\gamma\in\cho(X)\}$}
\DisplayProof
} 
\\ & \\
\AxiomC{$\{a\leq b\mid(a,b)\in R\}$}
\LeftLabel{$(\nabla 1)$}
\RightLabel{$(\alpha,\beta)\in \bar{T}R$}
\UnaryInfC{$\nabla\alpha\leq\nabla\beta$}
\DisplayProof 
& \\
& \\
\LeftLabel{($\nabla2$)}
\AxiomC{$\{\nabla (T\Land)(\Phi)\leq b \mid \Phi\in SRD(A)\}$}
\UnaryInfC{$\Land\{\nabla\alpha\mid \alpha\in A\}\leq b$}
\DisplayProof 
&
\hspace{20pt}\LeftLabel{($\nabla3$)}
\AxiomC{$\{\nabla\alpha\leq b\mid\alpha\Tmem[]\Phi\}$}
\UnaryInfC{$\nabla (T\Lor)(\Phi)\leq b$}
\DisplayProof  
\\
& \\
\hline
\end{tabular}
\end{center}
\caption{The $\KKV$ system}
\label{ch1:table:KKV}
\end{table}

Table \ref{ch1:table:KKV} requires some additional information, in particular the following typing information: $a,b\in\lngT$, $\phi,\psi\in\powf\lngT$, $X\in\powf\powf\lngT$, $\alpha,\beta\in T_{\omega}\lngT$, $\Phi\in T_\omega\powf \lngT$, $A\in \powf T_\omega \lngT$. The set $\cho(X)$ is the set of choice functions on $X$, i.e. the maps $\gamma:X\to\lngT$ such that $\gamma(\phi)\in\phi$, and $\rng$ denotes the range of the function. $R\subseteq\lngT\times\lngT$ is any relation and $\bar{T}R$ is its lifting. Finally $SRD(A)$ is the set of  so-called `slim redistributions' of $A$. This last concept is important, and we therefore define it in extenso. A \textbf{redistribution}\index{Redistribution} of $A\in \powf T_\omega \lngT$ is an element $\Phi$ of $T_\omega\powf \lngT$ which `contains' all the elements of $A$ as lifted members, i.e. $\alpha\Tmem[]\Phi$ for all $\alpha\in A$. It is called \textbf{slim}\index{Silm redistribution} if it is build from the direct subformulae of the elements of $A$, i.e. if $\Phi\in T_\omega\powf (\bigcup_{\alpha\in A}\Base(\alpha))$.

Before we move on to axiomatizing the abstract flavour of coalgebraic logic, we would like to highlight the fact that $(\nabla 1)$ is a congruence rule, and that the use of an auxiliary relation $R$ in its premise allows for the system to be well-defined in the sense that $\leq$ is then not defined in terms of itself. A particularly natural choice of relation $R$ is to consider the restriction of $\leq$ to subformulas. We will return to this point in examining the abstract case.

\subsubsection{Axiomatizations in the abstract setting}\label{ch1:subsec:Hilbert}
Let us now turn to the abstract style of coalgebraic logic. We would like to define what we mean by an axiomatization in this setting too, and show that it subsumes what was done in the predicate lifting and nabla case, just as we did for the definition of the language and of the semantics. We will however delay this last point to the end of the chapter because it will involve concepts which we first need to introduce.

Let us place ourselves once again in the fundamental situation of Diagram (\ref{ch1:diag:fundamentalSituation}) and assume that the syntax constructor $L$ is regular epi-preserving. Recall that the abstract coalgebraic language associated with $L$ is defined as $\mathsf{G}\Free V$, the free $L$-algebra over $\Free V$. We must now abandon some of the generality of the abstract setting if we want to define a style of Hilbert systems which generalises the systems for the predicate lifting and nabla style, and make some assumptions on the category $\cat$. For reasons which will become clear in the rest of this work, we will consider the category $\cat=\DL$ as a good `minimal reasoning structure'. In fact we will consider Hilbert systems for the following choices of $\cat$: $\DL,\BDL $ and $\BA$.

For any of the above choices of fundamental reasoning structure we will define a relation $\deriv[ML]\subseteq \Forg\mathsf{V}\mathsf{G}\Free V\times \Forg\mathsf{V}\mathsf{G}\Free V$ whose intuitive meaning will be $a\deriv[ML]b$ iff `$b$ can be derived from $a$' in the abstract Hilbert system. The reasons for choosing a 2-dimensional Hilbert system is that in $\DL$ inequalities are the only meaningful statements, since there might not be a top or a bottom element. We start with axioms. In the 2-dimensional setting, axioms are of the form $a\deriv[ML] b$ for some $a,b\in  \Forg\mathsf{V}\mathsf{G}\Free V$. The propositional part of the system  is given, by increasing strength of the base category, by the following axioms (we use the same notation as in the $\KKV(T)$-axiomatization):

\noindent For $\DL:$
\begin{align*}
& \AxiomC{$\{a\deriv[ML] b\mid a\in\phi\}$}
\LeftLabel{$(\bigvee L)$}
\UnaryInfC{$\bigvee\phi \deriv[ML] b$}
\DisplayProof   \\[1em]
& \AxiomC{$a\deriv[ML] b$}
\LeftLabel{($\bigwedge L$)}
\RightLabel{($a\in\phi$)}
\UnaryInfC{$\bigwedge\phi \deriv[ML] b$}
\DisplayProof   \\[1em]
& \AxiomC{$a\deriv[ML] b$}
\LeftLabel{($\bigvee R$)}
\RightLabel{($b\in\phi$)}
\UnaryInfC{$a\deriv[ML] \bigvee\phi $}
\DisplayProof   \\[1em]
& \AxiomC{$\{a\deriv[ML] b\mid \in\phi\}$}
\LeftLabel{($\bigwedge R$)}
\UnaryInfC{$a\deriv[ML] \bigwedge\phi$}
\DisplayProof   \\[1em]
& \AxiomC{}
\LeftLabel{(Distributivity)}
\UnaryInfC{$\bigwedge\{\bigvee\phi\mid\phi\in X\}\deriv[ML]\bigvee\{\bigwedge\rng(\gamma)\mid\gamma\in\cho(X)\}$}
\DisplayProof  
\end{align*}

\noindent For $\BDL$, add
\begin{align*}
& \AxiomC{}
\LeftLabel{$(\top)$}
\UnaryInfC{$a\deriv[ML]\top$}
\DisplayProof   \\[1em]
& \AxiomC{}
\LeftLabel{$(\bot)$}
\UnaryInfC{$\bot\deriv[ML]a$}
\DisplayProof
\end{align*}

\noindent For $\BA$, add
\begin{align*}
& \AxiomC{$\bigwedge\{\phi\cup\{\neg a\}\}\deriv[ML] \bigvee\psi$}
\LeftLabel{$(\neg E)\updownarrow$}
\UnaryInfC{$\bigwedge\phi\deriv[ML]\bigvee\{\psi\cup\{a\}\}$}
\DisplayProof \\[1em]
& \AxiomC{$\bigwedge\{\phi\cup\{a\}\}\deriv[ML]\bigvee\psi$}
\LeftLabel{$(\neg I)\updownarrow$}
\UnaryInfC{$\bigwedge\phi\deriv[ML]\bigvee\{\psi\cup\{\neg a\}\}$}
\DisplayProof
\end{align*}
where $\updownarrow$ means that the rule can be used in either direction. We also include the following structural rules:
\begin{align*}
& \AxiomC{}
\LeftLabel{(Reflexivity)}
\UnaryInfC{$a\deriv[ML] a$}
\DisplayProof
& \AxiomC{$a\deriv[ML] c$}
\AxiomC{$c\deriv[ML] b$}
\LeftLabel{(Cut)}
\BinaryInfC{$a\deriv[ML] b$}
\DisplayProof 
\end{align*}

Finally, we need to add \textbf{substitution} and \textbf{congruence}. The standard notion of substitution\index{Substitution} can readily be transported to the abstract setting unchanged: given a map $\sigma: V\to \Forg\mathsf{V}\mathsf{G}\Free V$ in $\Set$, by using the two adjunctions connecting $\Set$ to $\Alg_{\cat}(L)$, we get an adjoint transpose $\hat{\sigma}:\mathsf{G}\Free V\to\mathsf{G}\Free V$, such a map is called a substitution, and given a term $a\in \mathsf{G}\Free V$, the application of $\hat{\sigma}$ to $a$ is traditionally denoted as $\hat{\sigma}a$ or $a\hat{\sigma}$ (we will use the former notation). We therefore add the rule
\begin{prooftree}
\AxiomC{$a\deriv[ML]b$}
\LeftLabel{(Substitution)}
\RightLabel{$\sigma: V\to \Forg\mathsf{V}\mathsf{G}\Free V$}
\UnaryInfC{$\hat{\sigma}a\deriv[ML]\hat{\sigma}b$}
\end{prooftree}

%
%


The notion of congruence\index{Congruence rule} is a little trickier to adapt to the abstract framework since $L$ is not necessarily polynomial, i.e. the terms are not necessarily built from $n$-ary operation symbols. Recall that the congruence rule in the predicate lifting setting was given by:
\begin{prooftree}
\AxiomC{$a_i\leftrightarrow b_i, 1\leq i\leq n$}
\LeftLabel{(Congruence)}
\RightLabel{($\sigma\in\Sigma$)}
\UnaryInfC{$\sigma(a_1,\ldots,a_n)\leftrightarrow \sigma(b_1,\ldots, b_n)$}
\end{prooftree}
The key to generalizing this rule is first to notice that the premise $a_i\leftrightarrow b_i, 1\leq i\leq n$ can be re-written as $(a_i, b_i)\in R, 1\leq i\leq n$ where $R\mono \plLang\times \plLang$ is the relation (in $\Set$) defined by $(a,b)\in R$ iff $\Ax\vdash a\leftrightarrow b$. It is then easy to realize that the premise can be understood as a witness of the lifting of this relation $R$ by a polynomial functor. We can thus rewrite the congruence rule as
\begin{prooftree}
\AxiomC{$\left((a_1,\ldots, a_n),(b_1,\ldots,b_n)\right)\in \lift{\polyFunc}{R}$}
\LeftLabel{(Congruence)}
\UnaryInfC{$\sigma(a_1,\ldots,a_n),\sigma(b_1,\ldots, b_n)\in R$}
\end{prooftree}
Prefixing by $\sigma$ is essentially the structure map of the language $\plLang$, viewed as an initial $\polyFunc(-)+\Free V$-algebra. The congruence rule written in this way is clearly ready to be generalized to our Hilbert system for abstract coalgebraic logic. We re-define the relation $R$ as the relation on $\Forg\mathsf{V}\mathsf{G}\Free V$ defined by $aRb$ iff $a\deriv[ML] b$ and $b\deriv[ML] a$. Using the notion of relation lifting in a regular category developed in the previous section, we can then consider the lifting, $\lift{L}{R}$ of $R$ and put \index{Congruence} for any $\alpha,\beta\in L\mathsf{V}\mathsf{G}\Free V$:
\begin{prooftree}
\AxiomC{$(\alpha,\beta)\in \lift{L}{R}$}
\LeftLabel{(Congruence)}
\UnaryInfC{$(\langle\alpha\rangle,\langle\beta\rangle)\in R$}
\end{prooftree}
where $\langle-\rangle: L\mathsf{G}\Free V\to \mathsf{G}\Free V$ is the structure map of the language.  We will call this rule the \textbf{abstract congruence rule}, and note its similarity with the rule $(\nabla 1)$ of the $\KKV$ system. The reader will be justifiably concerned that the derivability relation $\deriv[ML]$ is now defined in terms of itself via the relation $R$. However, by using the notion of base, mentioned earlier and developed in detail in Chapter 3, we can tighten the abstract congruence rule and make it more similar to the congruence rule in the predicate lifting set-up which only uses in its premise the `ingredients' necessary to build the terms in its conclusion. Since limits in $\cat$ are created in $\Set$ (recall that in Diagram (\ref{ch1:diag:fundamentalSituation}) we assume $\Forg$ to be monadic), and since the base is essentially a limit, the base $\Base_{\mathsf{G}\Free V}(\alpha)$ is created in $\Set$. Assuming that $L$ preserves weak pullbacks, then it also preserves monos by Proposition \ref{ch3:prop:MonoPres}, and this property is clearly preserved by $\Forg$. Moreover, if we assume that $\Forg L$ is finitary, then by Proposition \ref{ch3:prop:weakPBinter} it preserves all intersections. In consequence under the relatively `normal' assumption of weak-pullback preservation and finitarity, we have for any $\alpha\in L\mathsf{G}\Free V$, that $\alpha\in Li[L\Base_L(\alpha)]$, where $i: \Base_L{\alpha}\mono\mathsf{G}\Free V$ is the obvious injection, and $Li[L\Base_L(\alpha)]$ is the image of $Li$ (which we can always take in a regular category). Since $L$ is assumed to preserve regular epis we get, by Proposition \ref{ch1:prop:liftingProp}, the following congruence rule:
\begin{prooftree}
\AxiomC{$(\alpha,\beta)\in \lift{L}{(R\restrict \Base_L(\alpha)\times \Base_L(\beta))}$}
\LeftLabel{(Congruence)}
\UnaryInfC{$(\langle\alpha\rangle,\langle\beta\rangle)\in R$}
\end{prooftree}
In other words, we need only find a witness in the $L$-lifting of the much smaller relation $(R \restrict \Base_L(\alpha)\times \Base_L(\beta))$, just as was the case in the predicate lifting style Hilbert system. The definition of $\Ax\deriv[ML]$ is thus an inductive definition. Formally, given a set of axioms $\Ax\subseteq(\Forg\mathsf{V}\mathsf{G}\Free V)^2$, we define the abstract Hilbert system associated with it as the system whose axioms are those of $\Ax$, and whose rules are (1) the propositional rules defined above and corresponding to the appropriate base category (1) the reflexivity and cut rules, (3) closure under substitution instances, and (4) the abstract congruence rule we have just detailed.

The `propositional rules' essentially enforce that the relation $\deriv[ML]$ is in fact a relation in $\cat$, i.e. a subobject of $(\mathsf{V}\mathsf{G}\Free V)^2$. Similarly, the congruence rule guarantees that $\deriv[ML]$ is in fact a relation in $\Alg_{\cat}(L)$, i.e. a subobject of $(\mathsf{G}\Free V)^2$ as we will now show. Note first that since $\cat$ is assumed to be monadic over $\Set$, it is regular. Moreover, since $L$ is a regular epi preserving varietor $\Alg_{\cat}(L)$ is regular (by Proposition \ref{ch1:prop:AlgRegCat}), and relations can thus be defined in $\Alg_{\cat}(L)$. It is also monadic over $\cat$ (by Proposition \ref{ch1:prop:Lvarietal}) and limits in $\Alg_{\cat}(L)$ are therefore created in $\cat$; and in particular the product of two $L$-algebras $\theta: LA\to A$ and $\xi: LB \to B$ is defined by the structure map $L(A\times B)\to A\times B$ via the unique map $\theta\circ L \pi_1\times \xi\circ L \pi_2$ where $\pi_1,\pi_2$ are the obvious projections. By definition of the product in $\Alg_{\cat}(L)$, and of relation lifting we have the following commutative diagram:
\begin{equation}\label{ch1:diag:cong}
\xymatrix@C=12ex
{
 & \lift{L}{R}\hspace{1ex}\ar@{>->}[dr] & \\
LR\hspace{1ex}\ar@{>->}[r]\ar[d]\ar@{->>}[ur] & L(A\times A)\ar[d]_{\theta\circ L \pi_1\times \theta\circ L \pi_2}\ar[r]^{L\pi_1\times L\pi_2} &  LA\times LA\ar[dl]^{\theta\times\theta} \\
R\hspace{1ex}\ar@{>->}[r] & A\times A
}
\end{equation}

Recall that the epi $LR\to\lift{L}{R}$ is regular i.e. the underlying $\Set$ map is surjective. The meaning of the congruence rule is now much clearer: it simply states that if $(\alpha,\beta)\in \lift{L}{R}\hspace{1pt}$ for some $\alpha,\beta\in LA$, then $(\theta(\alpha),\theta(\beta))\in R$, i.e. the congruence rule allows us to specify a relation on $L(A\times A)\to A\times A$ in $\Alg_{\cat}(L)$ by using a relation on $LA$ in $\cat$. We have thus shown:
\begin{proposition}\label{ch1:prop:congRule}
Let $L:\cat\to \cat$ be a regular epi-preserving varietor. A relation $R\mono A\times A$ in $\cat$ is a relation in $\Alg_{\cat}(L)$ on the $L$-algebra $\theta: LA\to A$ iff the congruence rule is satisfied, i.e. iff $(\theta(\alpha),\theta(\beta))\in R$ whenever $(\alpha,\beta)\in \lift{L}{R}$.
\end{proposition} 
Thus, the congruence rule gives us a criterion on pairs of modal formulas (i.e. elements of $L\mathsf{G}\Free V\times L\mathsf{G}\Free V$) which enforces that the (equivalence) relation $R$ in $\cat$ is also an (equivalence) relation in $\Alg_{\cat}(L)$. 


\subsection{One style of algebraic semantics}\label{ch1:subsec:algsem}

As in the case of (classical) modal logic (see Chapter 5 of \cite{2001:ModalLogic}), it is very useful to define an algebraic semantics for the languages defined above. Fundamentally, the usefulness of an algebraic semantics lies in its close proximity with the languages themselves and with derivability. As we will now show, the algebraic semantics gives us an alternative description of derivability in the language of universal algebra, which is in many ways easier to manipulate than individual (modal) axiomatizations. This will be crucial in our discussion of translations in Chapter 4.

We saw earlier that the abstract description of coalgebraic languages and their semantics provides a unifying framework in which we don't have to worry about boolean reasoning. We therefore choose this level of abstraction to define an algebraic semantics (it is also interesting to look at varieties over a category different than $\Set$).

Let once again $L:\cat\to\cat$ be a regular epi-preserving varietor and let us consider the language $\mathsf{G}\Free V$. We will use the fact that if we define $L'=L(-)+\Free V$, the language can also be seen as $\init[L']$, the initial $L'$-algebra, to give a very simple and elegant definition of the algebraic semantic of $\init[L']$. Let $\theta:L'A\to A$ be an arbitrary $L'$-algebra, we define the \textbf{algebraic interpretation map}\index{Algebraic interpretation map}\index{Semantics!algebraic} $\lsem-\rsem_A$ as the unique morphism:
\[
\xymatrix@C=12ex
{
L'\init[L']\ar[d]\ar@{-->}[r]^{L'\lsem-\rsem_A}& L'A\ar[d]_{\theta}\\
\init[L']\ar@{-->}[r]_{\lsem-\rsem_A} & A
}
\]
Observe that by definition of $L'$, structure maps of the type $\theta:L'A=LA+\Free V\to A$ are in one to one correspondence with pairs of maps $\xi: LA\to A$ and $\hat{v}_A: \Free V\to A$. By adjunction, the latter are in one-to-one correspondence with functions $v_A: V\to\Forg A$, i.e. valuations. In other words, by considering $L'$ algebras rather than $L$ algebras we automatically include the notion of valuation which is essential to the definition of an algebraic semantics. In particular, every choice of $v_A: V\to\Forg A$ for a fixed $\xi: LA\to A$ defines a new $L'$-algebra and a new algebraic interpretation map into $A$. We will say that a formula $a\in\init[L']$ is \textbf{algebraically satisfiable}\index{Satisfiable!algebraically} if there exists an $L'$-algebra $\xi: L'A\to A$ such that $\lsem a\rsem_A\neq \bot$. Similarly, we will say that a formula $a\in\init[L']$ is \textbf{algebraically valid over a class $\mathcal{C}$ of $L$-algebras}\index{Validity!algebraic} - denoted $\mathcal{C}\models a$ - if for every algebra $\xi: LA\to A$ in $\mathcal{C}$ and every valuation $v_A: V\to\Forg A$, $\lsem a\rsem_A=\top$, where $\lsem-\rsem_A$ is defined as above by the catamorphism into $\xi+\hat{v}_A: L'A\to A$. Similarly, we say that an \emph{equation} $a=b$ between two abstract terms is algebraically valid over $\mathcal{C}$ - denoted $\mathcal{C}\models a=b$ if $\lsem a\rsem_A=\lsem b\rsem_B$ for all algebras $A$ in $\mathcal{C}$ and every valuation. The purpose of this section is to prove that an axiomatization is always sound and complete with respect to the algebraic semantics over a class of algebra which is naturally associated with the axiomatization.

Axiomatizations are often given in terms of a set $\Ax\subset\Forg\mathsf{V}\mathsf{G}\Free V$ of \emph{axioms}, but we will prefer to work with a set of \emph{equations} i.e. a set $E\subset\Forg\mathsf{V}\mathsf{G}\Free V\times \Forg\mathsf{V}\mathsf{G}\Free V$. From $E$ we will build two entities. First, on the semantic side, we will use $E$ to build an equivalence relation (in $\Alg_{\cat}(L)$) on $\mathsf{G}\Free V$. This equivalence relation will define, via its coequalizer, a class of $L$-algebras (more precisely a variety of $L$-algebras) which will form the semantic domain of our algebraic semantics. Secondly, on the proof theoretic side, we will use the set of equations to define an equational calculus, i.e. a set of rules to build derivations of equations. Finally, we will prove the adequacy of the semantics and proof-theoretic constructions.

\subsubsection*{Defining a variety from equations}

We want to build an equivalence relation on $\mathsf{G}\Free V$ whose interpretation will be that two terms $a,b$ are be related if they can be proved to be equal in the equational calculus we will define in a moment. Having defined relations in regular categories, we now know that this means a subobject of $\mathsf{G}\Free V\times \mathsf{G}\Free V$ viewed as an $L$-algebra. But all we have so far is a \emph{set} $\Ax$ of equations. In order to turn this set into a subobject of $\mathsf{G}\Free V\times \mathsf{G}\Free V$, and in particular into an $L$-algebra, we proceed as follows. The set $\Ax$ of equations naturally defines two functions 
\[e_1,e_2:\Ax\to \Forg\mathsf{V}\mathsf{G}\Free V\]
where $e_1$ picks the left-hand-side term of the equations and $e_2$ the right-hand-side term. It is natural to assume that $e_1,e_2$ are jointly monic in $\Set$, i.e. that if $e_1(a)=e_1(b)$ and $e_2(a)=e_2(b)$ then $a=b$. Indeed the premise $e_1(a)=e_1(b)$ and $e_2(a)=e_2(b)$ means that we have the same equation appearing twice in $\Ax$, and we can clearly eliminate the redundant copy without loss of generality. So we will assume $e_1,e_2$ jointly monic, i.e. $(e_1\times e_2)$ monic as we saw earlier. We would like to turn these arrows in $\Set$ into an equivalence relations in $\Alg_{\cat}(L)$, which suggests the following construction. First we use the adjunction to get two arrows in $\Alg_{\cat}(L)$, i.e.
\[
\hat{e}_1,\hat{e}_2: \mathsf{G}\Free\Ax\to \mathsf{G}\Free V
\]
Intuitively, $\hat{e}_1,\hat{e}_2$, create algebraic combinations of the original equations. Second, we take the kernel pair of their coequalizer, i.e. we define
\[
\xymatrix
{
\mathsf{G}\Free \Ax \ar@<1ex>[r]^{\hat{e}_1} \ar@<-1ex>[r]_{\hat{e}_2} & \mathsf{G}\Free V\ar@{->>}[r]^{q} & Q
}
\]
This coequalizer quotients $\mathsf{G}\Free V$ under the equations. Finally, we consider the kernel pair $\ker q$ of $q$ 
\[
\xymatrix
{
\ker q\ar[d]_{p_2}\ar[r]^{p_1} & \mathsf{G}\Free V\ar@{->>}[d]^{q} \\
\mathsf{G}\Free V\ar@{->>}[r]_{q} & Q
}
\]
This kernel pair clearly defines an equivalence relation in $\Alg_{\cat}(L)$ (see Lemma \ref{ch1:lem:kpRequiv}), and it captures all the pairs of terms which are quotiented by $q$. The regular quotient $q$ defines a variety of $L$-algebras which we will denote $\mathbb{V}_q$. To define a variety we need the following concept:
\begin{definition}\label{ch1:def:variety} Let $C$ be an object and $f:A\to B$ be a morphism of a category $\cat$. We say that $C$ is \textbf{orthogonal} to $f$, or that $A$ is in the \textbf{orthogonality class} of $f$ if for every morphism $g: A\to C$ there exist a unique morphism $u:B\to C$ such that $g=u\circ f$. The orthogonality relation is denoted by $C\perp f$.
\end{definition}
We refer the reader to \cite{Borceux} 5.4 or \cite{LPAC} 1.C for more details on this concept. We follow \cite{2000:Thesis:Kurz, HughesPhD} and define a \textbf{variety}\index{Variety} in a category $\cat$ as the orthogonality class of a regular epimorphism whose domain is a projective object. In all our applications we will consider categories of the form $\Alg_{\cat}(L)$ for a regular epi-preserving varietor $L$ and a regular category $\cat$. In this case it is easy to check by using the obvious adjunction $\mathsf{G}\dashv\mathsf{V}:\cat\to \Alg_{\cat}(L)$, that any object in $\Alg_{\cat}(L)$ of the form $\Free A$ for some $A$ in $\cat$ is projective if $A$ is projective in $\cat$. Note that in the case of an adjunction $\Free\dashv\Forg: \Set\to\cat$ where $\Set$ is understood to satisfy the axiom of choice, $\Free A$ is always projective, since all sets are projective objects. This means in particular that if $L:\cat\to\cat$ is a regular epi-preserving varietor, then $\mathsf{G}\Free V$ is projective, so our very general definition of variety does apply to the case we are most interested in, i.e. if we define $\mathbb{V}_q$ as the orthogonality class of $q$, then $\mathbb{V}_q$ is a variety. We refer the reader to Chapter 3 of \cite{HughesPhD} for a categorical version of Birkhoff's HSP theorem which uses the definition of variety that we have presented here.

Before moving on to the equational logic defined by $\Ax$, we would like to make the following important comment which will simplify proofs further on. If $q:\mathsf{G}\Free V\epi Q$ is a regular epi, then $q$ defies a variety, but $Q$ is in general not a member of this variety. Equivalently (see Theorem 3.5.3. in \cite{HughesPhD}), $\ker q$ is in general not fully invariant. If either of these conditions were satisfied, $Q$ would represent the `prototypical' or `most general' object in the variety. In particular, it would be the initial object in the category defined by the variety. Obtaining such a prototypical $Q$ simplifies many proofs considerably, since it is then enough to reason only about $Q$. 

\begin{proposition}\label{ch1:prop:Q*}
Let $\cat$ be a cocomplete well-powered regular category, let $P$ be a projective object in $\cat$ and $q:P\epi Q$ be the coequalizer of two jointly monomorphic arrows $p_1,p_2: E\to P$. If we denote by $q^*:P\epi Q^*$ the regular epi defined in Proposition \ref{ch1:prop:fullInvClosure} and which constructs the fully invariant closure of $\ker q$, then 
\[
\mathbb{V}_q=\mathbb{V}_{q^*}
\]
\end{proposition} 
\begin{proof}
By using the construction of Proposition \ref{ch1:prop:fullInvClosure} it is clear $q^*$ also coequalizes $p_1,p_2$ (since it coequalizes $f\circ p_1, f\circ p_2$ for all $f\in \hom(P,P)$) and we thus have a unique morphism $u$:
\[
\xymatrix
{
E\ar@<1ex>[r]^{p_1}\ar@<-1ex>[r]_{p_2} & P\ar@{->>}[r]^{q}\ar@{->>}[dr]_{q^*} & Q\ar@{-->}[d]^{u}\\
& & Q^*
}
\]
Note that since $\cat$ is regular and $q^*$ is a regular epi, then $u$ must be a regular epi (\cite{Borceux2} 2.1.5.). It is now easy to check that if $A$ is orthogonal to $q^*$, then it must also be orthogonal to $q$, i.e. for every $f: P\to A$
\[
\xymatrix
{
& Q\ar@{->>}[dr]^{u}\\
P\ar[dr]_{f}\ar@{->>}[rr]^{q^*}\ar@{->>}[ur]^{q} & & Q^*\ar@{-->}[dl]\\
& A
}
\]
For the converse implication, consider any $f: P\to P$, any $h: P\to A$, and assume that $A$ is orthogonal to $q$, i.e. for any $g: P\to A$, there exists a unique $v: Q\to A$ such that $v\circ q=g$. In particular if we take $g=h\circ f$ for some $h: P\to A$. We then have that
\[h\circ f\circ e_1=g\circ e_1=v\circ q\circ e_1=v\circ q\circ e_2=g\circ e_2=h\circ f\circ e_2\]
and thus $h: P\to A$ coequalizes the pair $f\circ e_1,f\circ e_2$. Since the choice of $f$ is irrelevant, we actually get that $h$ coequalizes $\coprod_{f\in\hom(P,P)}f\circ e_i, i=1,2$, and there must therefore exist a unique morphism $u: Q^* \to A$ such that $u\circ q^*=h$, by definition of $q^*$ and $Q^*$. In other words $A$ is orthogonal to $q^*$ which concludes the proof. 
\end{proof}

\subsubsection*{Equational logic in $\cat$}

We now define the equational logic\index{Equational logic} associated with $\Ax$ by specifying the following derivation rules. For any $a,b,c,d\in\mathsf{G}\Free V, \alpha,\beta\in L\mathsf{G}\Free V$
\begin{enumerate}[EL(1)]
\item Axioms:\label{ch1:ax:ax} $\Ax\deriv a=b$ if $(a,b)\in\Ax$
\item Reflexivity: \label{ch1:ax:ref} $\Ax\deriv a=a$
\item Symmetry: \label{ch1:ax:sym} if $\Ax\deriv a=b$ then $\deriv b=a$
\item Transitivity: \label{ch1:ax:tra} if $\Ax \deriv a=b$ and $\deriv b=c$ then $\deriv a=c$
\item Substitutions: \label{ch1:ax:sub} if $\Ax \deriv a=b$ and $\sigma:\mathsf{G}\Free V\to\mathsf{G}\Free V$ is a substitution, then $\Ax\deriv \sigma a=\sigma b$
\item Modal congruence: \label{ch1:ax:con}if $(\alpha,\beta)\in \lift{L}{(\Ax\deriv)}$ then $\Ax\deriv\langle\alpha\rangle=\langle\beta\rangle$
\item $\cat$-congruence: $\deriv$ is a relation in $\cat$. In particular, if $\cat=\DL$ \label{ch1:ax:boolCon} we have that $\Ax\deriv a=b$ and $\Ax\deriv c=d$ imply $\Ax\deriv a\wedge c=b\wedge d$, $\Ax\deriv a\vee c=b\vee d$. If $\cat=\BA$ we also have $\Ax\deriv \neg a=\neg b$
\end{enumerate}

\begin{proposition}\label{ch1:prop:smallest}
Assuming that $L$ preserves regular epimorphisms and weak pullbacks, the binary derivability predicate $\Ax\deriv$ defines the smallest fully invariant equivalence relation on $\mathsf{G}\Free V$ which contains $\Ax$.
\end{proposition}
\begin{proof}
Let us first show that $\Ax\deriv$ does indeed define a fully invariant equivalence relation on $(\mathsf{G}\Free V,\langle-\rangle)$ in $\Alg_{\cat}(L)$. Let $K$ denote the relation defined by $(a,b)\in K$ iff $\Ax\deriv a=b$. It is immediate by virtue of EL\ref{ch1:ax:boolCon}, that $m:K\mono \mathsf{G}\Free V\times\mathsf{G}\Free V$ in $\cat$. The congruence rule EL\ref{ch1:ax:con} ensures that $K$ is an $L$-algebra. Indeed, if we define \[\xi=\langle-\rangle\times\langle-\rangle\circ L\pi_1\times L\pi_2\circ L m\] then the modal congruence rule guarantees that the image of that map lies in $K$ (see Diagram \ref{ch1:diag:cong}). Hence $(K,\xi)\mono (\mathsf{G}\Free V\times\mathsf{G}\Free V, \langle-\rangle\circ L\pi_1\times\langle-\rangle\circ L\pi_2)$, i.e. $K$ is a sub $L$-algebra of $\mathsf{G}\Free V\times\mathsf{G}\Free V$. Finally, the rules EL\ref{ch1:ax:ref}-\ref{ch1:ax:tra} ensures that $K$ is an equivalence relation, and rule \ref{ch1:ax:ax} ensures that it contains $\Ax$. Thus $K$ is an equivalence relation in $\Alg_{\cat}(L)$ which contains $\Ax$. The fact that it is fully invariant is a direct consequence of EL\ref{ch1:ax:sub}.

Now let us show that $K$ is the \emph{smallest} fully invariant equivalence relation on $\mathsf{G}\Free V$ containing $\Ax$. Assume $K'$ is another fully invariant equivalence relation on $(\mathsf{G}\Free V,\langle-\rangle)$ containing $\Ax$, and let's show that $(a,b)\in K\Rightarrow (a,b)\in K'$. We proceed by induction on the length of derivation of $(a,b)$.

\textbf{Base case:} Clearly, if $(a,b)\in \Ax$, then $(a,b)\in K'$ by assumption. Similarly, if $(a,a)\in K$, then $(a,a)\in K'$ since $K'$ is an equivalence relation.

\textbf{Inductive case:} We go through the possible last steps of the derivation $\Ax\deriv a=b$. 

EL\ref{ch1:ax:sym}: the last step of the derivation is an application of the symmetry rule, i.e. there is a shorter proof of $\Ax\deriv b=a$, and by induction hypothesis, this means that $(b,a)\in K'$ and thus $(a,b)\in K'$ since it's an equivalence relation. 

EL\ref{ch1:ax:tra}: the proof is the same as in the previous case. 

EL\ref{ch1:ax:sub}: the last step of the derivation is an application of the substitution rule, i.e. there exists a shorter proof of $\Ax\deriv c=d$ and a substitution $\sigma:\mathsf{G}\Free V\to\mathsf{G}\Free V$ such that $\sigma c=a, \sigma d=b$. By the induction hypothesis we have $(c,d)\in K'$, and since by assumption $K'$ is fully invariant, we must also have $(\sigma c,\sigma d)\in K'$.

EL\ref{ch1:ax:con}: The congruence rule is quite different, and it is not immediately clear how the induction hypothesis works in this case. Since $L$ is assumed to preserve regular epis and weak pullbacks we can assume that the last rule applied was in fact the congruence rule restricted to the bases. Thus, what we need to apply this rule, is to have established derivability between enough formulas in the bases of $\alpha$ and $\beta$ - but not necessarily to have established the full relation $K\cap \Base_{L}(\alpha)\times\Base_{L}(\beta)$ - to be able to apply ${L}$ to a subobject of $R\mono K\cap \Base_{L}(\alpha)\times\Base_{L}(\beta)$ and get a witness in ${L}R\mono {L}(K\cap \Base_{L}(\alpha)\times\Base_{L}(\beta))$ which projects to $\alpha,\beta$, i.e. such that modulo some mono, we have $(\alpha,\beta)\in\lift{L}{R}$. In other words, if we apply this rule in a derivation, it means that we have shorter proofs of equality between terms in the bases of $\alpha,\beta$, which we can collect in a sub-relation $R\mono K\cap \Base_{L}(\alpha)\times\Base_{L}(\beta)$ whose lifting contains the witness we need. By the induction hypothesis we have $R\mono K'\cap \Base_{L}(\alpha)\times\Base_{L}(\beta)$. By using the properties of relation liftings from Propositions \ref{ch1:prop:liftingwpb} and  \ref{ch1:prop:liftingProp} we then get $(\alpha,\beta)\in\lift{L}{K'}\cap(L\Base_{L}(\alpha)\times L\Base_{L}(\beta))$ and thus $(\langle\alpha\rangle,\langle\beta\rangle)\in K'$ by the fact that $K'$ is a relation in $\Alg_{\cat}(L)$.

EL\ref{ch1:ax:boolCon}: immediate since $K'$ is an $L$-algebra, i.e. its carrier is a  $\cat$-object.
\end{proof}

\subsubsection*{Soundness and completeness results}
We now prove that provability in the abstract Hilbert system is equivalent to provability in the equational logic derivation system. We first translate axioms between the two system. Assuming that $\cat=\DL,\BDL$ or $\BA$, it makes sense to define the following translation of axioms in the Hilbert system into equations:
\[
(a,b)^t=(a\wedge b,a)
\]
And conversely from equations into Hilbert axioms:
\[
(a,b)_t=\{(a,b),(b,a)\}
\]
More generally, we will denote by $(\Ax)^t$ the set of equations obtained from a set of Hilbert-style axioms by applying the map $(-)^t$, and conversely for $\Ax_t$. Note that every axiom of $\Ax$ can be recovered from $((\Ax)^t)_t: $ assume $(a,b)\in\Ax$ then $(a\wedge b,a), (a,a \wedge b)\in ((\Ax)^t)_t$ and we get
\begin{prooftree}
\AxiomC{$a\deriv[ML] a\wedge b$}
\AxiomC{$b\deriv[ML]b$}
\RightLabel{$b\in\{a,b\}$}
\LeftLabel{$(\bigwedge L)$}
\UnaryInfC{$a\wedge b\deriv[ML] b$}
\LeftLabel{(Cut)}
\BinaryInfC{$a\deriv[ML]b$}
\end{prooftree}
Conversely, every equation in $\Ax$ can be recovered from $((\Ax)_t)^t$: if $(a,b)\in \Ax$ then $(a\wedge b, a),(b\wedge a,b)\in((\Ax)_t)^t$, and the original equality can be recovered from the fact that $a\wedge b=b\wedge a$ modulo $\cat$-equivalence and EL\ref{ch1:ax:sym} and EL\ref{ch1:ax:tra}.

\begin{proposition}\label{ch1:prop:MLEQconnection}
Let $L:\cat\to\cat$ be a finitary, weak pullback and regular epi-preserving functor and let $V$ be a set of variables. Consider $a,b\in\Forg\mathsf{V}\mathsf{G}\Free V$ and a set of axioms $\Ax\subseteq(\Forg\mathsf{V}\mathsf{G}\Free V)^2$ in the Hilbert system of Section \ref{ch1:subsec:Hilbert}
\[
\text{ if }a\deriv[ML]b\text{ then } (\Ax)^t\deriv a\wedge b=a
\]
Conversely, assuming a set of equations $\Ax\subseteq(\Forg\mathsf{V}\mathsf{G}\Free V)^2$, 
\[
\text{ if }\Ax\deriv a=b \text{ then } a\deriv[ML] b\text{ and }b\deriv[ML] a\text{ can be derived from }(\Ax)_t
\]
\end{proposition}
\begin{proof}
By induction on the size of the derivation trees. 

Let us show the first implication. If $a\deriv[ML]b$ is derived in 1-step, then $(a,b)\in\Ax$, i.e. $(a\wedge b,a)\in(\Ax)^t$ by definition of $(\Ax)^t$. 

Now for the inductive step. If the last step of the proof is propositional it is not hard to check that since $\deriv$ is defined on equivalence classes modulo equivalence in $\cat$, we can translate the last step into an equational proof. We show the example of the rule $(\bigvee L)$ in the case where $\phi=\{a,b\}$ and $a\vee b\deriv[ML]$ is the conclusion. By the induction hypothesis we can then build an equational derivation ending in
\begin{prooftree}
\AxiomC{$a\wedge c=a$}
\AxiomC{$b\wedge c=b$}
\LeftLabel{EQ\ref{ch1:ax:boolCon}}
\BinaryInfC{$(a\vee b)\wedge c=a\vee b$}
\end{prooftree}
where we've used the fact that $(a\wedge c)\vee (b\wedge c)$ and $(a\vee b)\wedge c$ are two representative of the same equivalence class under the axioms of $\cat$. The same goes for all propositional rules. The case of the Reflexivity rule is trivial. Assume now that the last rule applied was the Cut rule, i.e. that we have shorter derivations of $a\deriv[ML] b'$ and $b'\deriv[ML] b$. By the induction hypothesis, this means that we have derivations of $\deriv a\wedge b'=a$ and $\deriv b'\wedge b=b'$. We can then proceed as follows: 
\begin{prooftree}
\AxiomC{$a\wedge b'=a$}
\LeftLabel{EL\ref{ch1:ax:sym}}
\UnaryInfC{$a=a\wedge b'$}
\AxiomC{$b=b$}
\LeftLabel{EL\ref{ch1:ax:boolCon}}
\BinaryInfC{$a\wedge b=a\wedge b'\wedge b$}
\AxiomC{$a=a$}
\AxiomC{$b'\wedge b=b'$}
\LeftLabel{EL\ref{ch1:ax:boolCon}}
\BinaryInfC{$a\wedge b\wedge b' =a\wedge b'$}
\LeftLabel{EL\ref{ch1:ax:tra}}
\BinaryInfC{$a\wedge b = a\wedge b'$}
\AxiomC{$a\wedge b'=a$}
\LeftLabel{EL\ref{ch1:ax:tra}}
\BinaryInfC{$a\wedge b=a$}
\end{prooftree}
The case of the substitution rule follows trivially from the fact that substitutions are $L$-algebra morphisms. Finally, for the congruence rule, assume that $\alpha,\beta\in \lift{L}(R\restrict \Base_L(\alpha)\times\Base_L(\beta))$. By the induction hypothesis we have for any $(a,b)\in R\restrict\Base_L(\alpha)\times\Base_L(\beta)$, $\deriv a\wedge b=b$ and $\deriv b\wedge a=a$, i.e. $\deriv a=b$, and the result then follows from an application of EL\ref{ch1:ax:con}.

The second implication poses no difficulty. The case of axioms follows immediately from the definition of $(-)_t$ and if the equational proof ends in EL\ref{ch1:ax:ax}-EL\ref{ch1:ax:sym} the I.H. immediately provides the corresponding $\deriv[ML]$ proof. In the case of EL\ref{ch1:ax:tra}, we easily get a $\deriv[ML]$ proof from the I.H. and the Cut rule. The case of the substitution rule is trivial and Congruence is treated exactly as above but in the opposite direction. Finally it is not difficult to check that any equational proof ending in an application of EL\ref{ch1:ax:boolCon} can be translated into a pair of $\deriv[ML]$ proof by using the propositional rules pertaining to $\cat$. For example, if the last step of an equation derivation is 
\begin{prooftree}
\AxiomC{$a=c$}
\AxiomC{$b=d$}
\LeftLabel{EL\ref{ch1:ax:boolCon}}
\BinaryInfC{$a\wedge b=c\wedge d$}
\end{prooftree}
then by the I.H., we have proofs of $a\deriv[ML]c, c\deriv[ML]a,b\deriv[ML]d$ and $d\deriv[ML]a$ and it follows that we have a proof ending in 
\begin{prooftree}
\AxiomC{$b\deriv[ML]d$}
\RightLabel{$b\in\{a,b\}$}
\LeftLabel{$(\bigwedge L)$}
\UnaryInfC{$a\wedge b\deriv[ML] d$}
\AxiomC{$a\deriv[ML]c$}
\RightLabel{$a\in\{a,b\}$}
\LeftLabel{$(\bigwedge L)$}
\UnaryInfC{$a\wedge b\deriv[ML] c$}
\LeftLabel{$(\bigwedge R)$}
\BinaryInfC{$a\wedge b\deriv[ML] c\wedge d$}
\end{prooftree}
We similarly get a proof of $c\wedge d\deriv[ML] a\wedge b$ and the result follows.
\end{proof}

Let us now move to the algebraic semantic. Recall that the \emph{set} of equations $\Ax$ can be used to define a regular quotient $q: \mathsf{G}\Free V\to Q_{\Ax}$ via the coequalizer (in $\Alg_{\cat}(L)$) of the maps $\hat{e}_1,\hat{e}_2$ associated with the set $\Ax$. Moreover, using the construction of Proposition \ref{ch1:prop:fullInvClosure} and the result of Proposition \ref{ch1:prop:Q*}, we can assume that $Q_{\Ax}$ defines the smallest fully invariant equivalence relation containing the equations of  $\Ax$, and is thus orthogonal to $q$.

\begin{proposition}\label{ch1:prop:BirkComp}
Using the above notation, we have
\[\Ax\deriv a=b \iff Q_{\Ax}\models a=b \]
\end{proposition}
\begin{proof}
We just need to show that the relation $K\subseteq \mathsf{G}\Free V\times \mathsf{G}\Free V$ defined by $(a,b)\in K$ if $Q_{\Ax}\models a=b$ is the smallest fully invariant equivalence relation containing $\Ax$, and the result will then follow from Proposition \ref{ch1:prop:smallest}. For notational clarity we drop the $\Ax$ subscript for the remainder of the proof. The definition of the algebraic semantic means that for any valuation $v_Q$ on $Q$, 
\[\lsem a\rsem_{Q}=\lsem b\rsem_Q\]
where $\lsem-\rsem_Q:\init[L']\to Q$ is the unique $L'$-algebra morphism defined by initiality of $\init[L']$ and the fact that $v_Q$ endows $Q$ with an $L'$-algebra structure. We show that the equality \[\lsem a\rsem_{Q}=\lsem b\rsem_Q\] implies $q(a)=q(b)$. Since the equality holds for any valuation, it must, in particular, hold for the valuation defined by
\[
v_Q: \Free V \to Q, p\mapsto q\circ \eta_{\Free V}^{\mathsf{G}}(p)
\]
where $\eta^{\mathsf{G}}: \Id\to \mathsf{V}\mathsf{G}$ is the unit of the adjunction $\mathsf{G}\dashv\mathsf{V}: \cat\to\Alg_{\cat}(L)$. This valuation turns the $L$-algebra morphism $q$ into an $L'$-algebra morphism, $ \lsem a\rsem_{Q}=\lsem b\rsem_Q$ then implies $q(a)=q(b)$ by definition. Conversely, it is immediate that if $q(a)=q(b)$, then the fact that $Q\perp q$ implies that $\lsem a\rsem_Q=\lsem b\rsem_Q$ for any valuation on $Q$.

Thus $(a,b)\in K$ iff $(a,b)\in\ker q$. But we know from the construction in Proposition \ref{ch1:prop:fullInvClosure}, that $\ker q$ is the smallest fully invariant equivalence relation which contains $\Ax$, which is what we needed to show.
\end{proof}

\begin{proposition}\label{ch1:prop:variety}
Let $L:\cat\to\cat$ be a varietor, and consider a regular quotient $q:\mathsf{G}\Free V\epi Q$ defining a variety $\mathbb{V}_q$ such that $Q\perp q$, then
\[
Q\models a=b \iff\text{ for all }A\text{ in }\mathbb{V}_{q}, A\models a=b
\]
\end{proposition}
\begin{proof}
By assumption, $Q$ itself belongs to the variety $\mathbb{V}_{q}$ which is defined by $q$, thus the implication from right to left is trivial.

For the implication from left to right, it is enough to unravel the definitions. Consider any $L$-algebra $A$ in the variety $\mathbb{V}_{q}$ defined by $q$, and assume that $Q\models a=b$, i.e. $\lsem a\rsem_{Q}=\lsem b\rsem_{Q}$. As we've just seen in the proof of Proposition \ref{ch1:prop:BirkComp}, this in turn implies that  $q(a)=q(b)$. By definition of the variety, the unique $L$-algebra morphism $\lsem-\rsem_A:\init[L']\to A$ factors through $q$, and the result follows immediately.
\end{proof}

Gathering our results, we have the following theorem summing up the completeness of the algebraic semantics of coalgebraic logics.

\begin{theorem}\label{ch1:thm:algsem}
Under the conditions of Proposition \ref{ch1:prop:variety} the following are equivalent:
\begin{enumerate}[(i)]
\item $a\deriv[ML] b$ and $b\deriv[ML] a$ can be derived from $(\Ax)_t$
\item $\Ax\deriv a=b$
\item $Q_{\Ax}\models a=b$
\item for all $A$ in $\mathbb{V}_{q_\Ax}, A\models a=b$
\end{enumerate}
\end{theorem}

Alternatively, given a set $\Ax\subseteq (\Forg\mathsf{V}\mathsf{G}\Free V)^2$, we could have written $a\deriv[ML] b$ can be derived from $\Ax$ iff $(\Ax)^t\deriv a\wedge b=a$ iff $Q_{(\Ax)^t}\models a\wedge b=a$ iff for all $A$ in $\mathbb{V}_{q_{(\Ax)^t}}, A\models a\wedge b=a$. Note however, that by working with equations, and using the universal constructions detailed above, the collection of valid/derivable equations forms a fully invariant equivalence relation in $\cat$. In contrast, the set of derivable pairs of formula in a modal-style Hilbert system does not live in $\Alg_{\cat}(L)$ when $\cat=\BA$, and in particular it is not a subobject of the language (viewed as an $L$-algebra). In fact it does not even live in $\cat$, since this would imply a system where $a\deriv[ML] b$ iff $\neg a\deriv[ML] \neg b$.

\subsubsection{Comparing the systems}

We finish this chapter by showing that the Hilbert systems which we have defined for the abstract version of coalgebraic logic are enough to reason about derivability in the predicate lifting and nabla flavour of coalgebraic logic. Combined with the results of Theorem \ref{ch1:thm:algsem}, this will be very useful in Chapter 4. Let us first show that axiomatizations of the abstract style of coalgebraic logic subsume axiomatizations of the predicate lifting and nabla style. In order to achieve this, we will use some of the tools developed so far in order to understand precisely how the concrete and abstract languages are related. The intuitive relation is clear, but formalizing this relation is not totally trivial.

Let $\Sigma_{\BA}$ denote a signature which can accommodate a description of boolean algebras, for example $\Sigma_{\BA}=\{\bot,\neg,\wedge\}$ (as in the predicate lifting syntax),  or $\Sigma_{\BA}=\{\neg,\bigwedge,\bigvee\}$ (as in the nabla logic syntax). Let $\polyFunc[\BA]$ denote the $\Set$-endofunctor associated with this signature, i.e. the obvious polynomial functor, or the functor $\id+\powf+\powf$ in the nabla case.

Let also $T$ be a finitary functor. We can consider the nabla (or predicate lifting if $T$ is polynomial) style language associate with $T$, which is just the free $T+\polyFunc[\BA]$-algebra over a chosen set $V$ of propositional variables, as was discussed earlier. We wish to show precisely how this language is related to the abstract language defined by $T$, i.e. the free $\Free T\Forg$-algebra (in $\BA$) over $\Free V$. Let $\Free_{T+\BA}:\Set\to\Alg_{\Set}(T+\polyFunc[\BA])$ be the free functor defined by $T+\polyFunc[\BA]$. Since both $T$ and $\polyFunc[\BA]$ are finitary, so is their coproduct, which is therefore a varietor. Next, consider any set of equation axiomatizing boolean algebras, which we call $\mathsf{BA}$. We can assume without loss of generality, that the (three) variables used in these equations form a subset of $V$. We then proceed as above by considering the maps $e_1,e_2: \mathsf{BA}\to \Forg\Free_{T+\BA} V$ defined by these equations, taking their adjoint transpose $\hat{e}_1,\hat{e}_2$, and then their coequalizer $q$. Finally, we form the fully invariant closure of $\ker q$, and get a regular epi $q_{\BA}: \Free_{T+\BA}V\epi Q_{\BA}$ such that $Q_{\BA}\perp q_{\BA}$. 

\begin{proposition}\label{ch1:prop:concreteabstract}
The $T+\polyFunc[\BA]$-algebra $Q_{\BA}$ defined above is isomorphic to the free $\Free T\Forg$-algebra over $\Free V$.
\end{proposition}
\begin{proof}
By `is isomorphic to' we mean that by modifying its structure map, $Q_{\BA}$ (viewed as the carrier of a $T+\polyFunc[\BA]$-algebra structure) is the free $\Free T\Forg$-algebra over $\Free V$. To see this, we first need to show that it can be equipped with a $\Free T\Forg(-)+\Free V$ structure. Next we'll show that it is initial. By adjunction and the fact that $\Free$ preserved coproducts, to be equipped with $\Free T\Forg(-)+\Free V$ structure in $\BA$ is equivalent to being equipped with a $T\Forg(-)+V$ structure in $\Set$. Since $Q_{\BA}$ is, by construction, a boolean algebra, we can consider its $T$-algebra structure map $\alpha$ (we forget the $\polyFunc[\BA]$-structure for now) as a map with signature $T\Forg Q_{\BA}\to\Forg Q_{\BA}$, and we can also endow it with a map $v: V\to \Forg Q_{\BA}$ defined as $v=q_{\BA}\circ \eta^{T+\BA}$ where $\eta^{T+\BA}$ is the unit of the adjunction $\Free_{T+\BA}\dashv \Forg_{T+\BA}: \Set\to\Alg_{\Set}(T+\BA)$. By using the adjunction $\Free\dashv \Forg$, we then get that the dual transpose of $\alpha+v$ is a map
\[
\widehat{\alpha+v}=\hat{\alpha}+\hat{v}: \Free T\Forg Q_{\BA}+ \Free V\to Q_{\BA} 
\]
where $Q_{\BA}$ is naturally viewed as a boolean algebra (since it satisfies all their equations). This shows that $Q_{\BA}$ is a $\Free T\Forg(-)+\Free V$-algebra. 

To show that it is the initial such algebra, let us consider any other such algebra $\beta: \Free T\Forg B+\Free V\to B$. Since $B$ is a boolean algebra, we can define a trivial $\polyFunc[\BA]$-structure on $\Forg B$ by simply evaluating the boolean connectives, let us call this map $ev: \polyFunc[\BA]\Forg B\to \Forg B$. By combining $ev$ with the adjoint transpose $\tilde{\beta}$ of $\beta$, we can put a $T(-)+\polyFunc[\BA](-)+V$-algebra structure on $\Forg B$. We then get the following commutative diagram:
\[
\xymatrix@C=3ex
{
T \Free_{T+BA} V+\polyFunc[\BA] V+ V\ar[dd]_{\langle-\rangle_V+\eta^{T+\BA}}\ar@{->>}[rr]^{Tq_{\BA}+\id_V}\ar[dr]^{T!} & & T \Forg Q_{\BA}+\polyFunc[\BA] \Forg Q_{\BA}+ V\ar[dd]^{\alpha+b+v}\ar@{-->}[dl]\\
& T\Forg B+\polyFunc[\BA] B + V\ar[dd]^<<<<<<<<{ev+\tilde{\beta}}\\
\Free_{T+\BA} V\ar@{->>}[rr]_<<<<<<<<<<<<<<<<{q_{\BA}}\ar[dr]^{!} & & \Forg Q_{\BA}\ar@{-->}[dl]\\
& \Forg B
}
\] 
Where $!$ denotes the unique $T(-)+\polyFunc[\BA]+V$-algebra morphism which exists by virtue of $\Free_{T+\BA}$ being the initial such algebra, and $b$ is the boolean part of the structure map of $Q_{\BA}$. Finally, since $B$ is a boolean algebra, it is clear that $\Forg B$ must belong to the variety defined by the equations axiomatizing boolean algebras, and in particular we must have $\Forg B\perp q$, which proves the existence and uniqueness of the morphism from $\Forg Q_{\BA}$ to $\Forg B$, which in turns provides us by adjoint transpose with the unique $\Free T\Forg(-)+\Free V$-algebra morphism $Q_{\BA}\to B$. Note that the unique morphism $\Forg Q_{\BA}\to \Forg A$ is a boolean algebra morphism (i.e. is of the form $\Forg f$ for some $f$) by assumption on $q_{\BA}$. The $\Forg$ functor only really appears because the domain and codomain are viewed as carriers of algebra structures.
\end{proof}

Note that the construction above also gives us an expression for the structure map
of the abstract language, namely as the adjoint transpose of the $T$-algebra structure map of $Q_{\BA}$.  

Before we can present the main result of this subsection, we need to briefly examine substitution instances in the concrete and abstract setting. In the concrete case, a substitution is defined uniquely by a map from the set $V$ of propositional languages to the language, i.e.
\[
\sigma: V\to \Forg_{T+\BA}\Free_{T+\BA}(V)
\]
or, equivalently, as an endomorphism of $\hat{\sigma}:\Free_{T+\BA}(V)\to\Free_{T+\BA}(V)$. By using the construction of Proposition \ref{ch1:prop:concreteabstract} we can immediately extend this valuation to the abstract case.

\begin{lemma}\label{ch1:lem:substinst}
For any endomorphism $\sigma:\Free_{T+\BA}V\to\Free_{T+\BA} V$, there exists an endomorphism $\tilde{\sigma}: Q_{\BA}=\mathsf{G}\Free V\to \mathsf{G}\Free V$ such that the following square commutes
\[
\xymatrix
{
\Free_{T+\BA}V\ar[r]^{\sigma}\ar[d]_{q_{\BA}} & \Free_{T+\BA} V\ar[d]^{q_{\BA}}\\
Q_{\BA}\ar[r]_{\tilde{\sigma}} & Q_{\BA}
}
\]
\end{lemma}
\begin{proof}
We use the fact that $Q_{\BA}\perp q_{\BA}$ or, equivalently, the construction of $q_{\BA}$, to get the unique arrow by orthogonality:
\[
\xymatrix
{
\Free_{T+\BA}V\ar@{->>}[rr]^{q}\ar[dr]_{q\circ\sigma} & & Q_{\BA}\ar@{-->}[dl]^{\tilde{\sigma}}\\
& Q_{\BA}
}
\]
\end{proof}

Let us denote by $\deriv[\mathrm{PL}]$ the derivability relation for predicate lifting style coalgebraic logics, and recall that $\deriv[ML]$ is the derivability relation for abstract coalgebraic logics, as defined above. If $\Ax$ is a set of formulas in a predicate lifting language, let $[\Ax]$ denote the set $\{\top\deriv[ML][a]_{\sim_{\BA}}\mid a\in\Ax\}$ where $[a]_{\sim_{\BA}}$ is the class of terms which are $\BA$-equivalent to $a$.
\begin{theorem}
Let $T=\polyFunc[\Sigma+\BA]$ define a predicate lifting style language $\plLang$. If $\Ax$ defines a Hilbert system for $\plLang$ with axioms only, then for any $b\in \plLang$
\[
\Ax\deriv[\mathrm{PL}] b \hspace{3ex}\Leftrightarrow\hspace{3ex} \top\deriv[ML][b]\text{ can be derived from }[\Ax]
\]
\end{theorem}
\begin{proof}
$\Rightarrow$: by induction on the size of the proof. For the base case, we must have that $b$ is a substitution instance of an axiom in $\Ax$ or a tautology of propositional logic. If $b=\sigma a$ for some endomorphism $\sigma: \plLang\to\plLang$ and some $a\in A$, then it immediately follows from Lemma \ref{ch1:lem:substinst} that there exist an endomorphism $\tilde{\sigma}$ on $\mathsf{G}\Free V$ such that $\tilde{\sigma}([a])=[\sigma(a)]=[b]$, i.e. $[b]$ is a substitution instance of an axiom in $[\Ax]$. If $a$ is a tautology of propositional logic, then $[a]=[\top]$, which is an axiom in the abstract system.

For the inductive step, we go through the possible last steps of the derivation of $a$. We can ignore the rules of propositional calculus since they can assumed to be equivalent in both systems. If it was modus ponens, i.e.
\begin{prooftree}
\AxiomC{$c$}
\AxiomC{$c\to b$}
\LeftLabel{(M.P.)}
\BinaryInfC{$b$}
\end{prooftree}
then by using the induction hypothesis, we have proof $\top\deriv[ML] [c]$ and $\top\deriv[ML][c\to b]$, i.e. $\top\deriv[ML] \neg [c]\vee [b]$, from which it follows by application of $(\neg I)$ that $[c]\deriv[ML][b]$, and thus $\top\deriv[ML][b]$ by an application of the Cut rule. If the last applied rule was the congruence rule, i.e. 
\begin{prooftree}
\AxiomC{$a_i\leftrightarrow b_i, 1\leq i\leq n$}
\LeftLabel{(Congruence)}
\RightLabel{($\sigma\in\Sigma$)}
\UnaryInfC{$\sigma(a_1,\ldots,a_n)\leftrightarrow \sigma(b_1,\ldots, b_n)$}
\end{prooftree}
then by using the induction hypothesis, we have a proof of $\top\deriv[ML] [a_i]\leftrightarrow[b_i], 1\leq i\leq n$. By application of $(\neg I), (\bigwedge L)$ and Cut it is easy to see that we have proofs of $[b_i]\deriv[ML] [a_i]$, and symmetrically, we can derive proofs of $[a_i]\deriv[ML][b_i], 1\leq i\leq n$. As we have described earlier, this can be viewed as a witness of the lifted relation 
\[
\lift{\Free\polyFunc[\Sigma]\Forg}{\left( R\restrict(\Base_{\polyFunc}([\sigma([a_1],\ldots,[a_n])])\times\Base_{\polyFunc[\Sigma]}([\sigma([b_1],\ldots[,b_n])])\right )}
\] 
where $R\subset \mathsf{G}\Free V\times\mathsf{G}\Free V$ is defined by $[a]R[b] \Leftrightarrow [a]\deriv[ML][b]$, and we can then apply the abstract congruence rule of the $\deriv[ML]$ system to $\alpha=([a_1],\ldots,[a_n]), \beta=([b_1],\ldots,[b_n]])$
\begin{prooftree}
\AxiomC{$(\alpha,\beta)\in \lift{\Free\polyFunc[\Sigma]\Forg}{(R\restrict \Base_L(\alpha)\times \Base_L(\beta))}$}
\LeftLabel{(Congruence)}
\UnaryInfC{$(\langle\alpha\rangle,\langle\beta\rangle)\in R$}
\end{prooftree}
to conclude $\langle\alpha\rangle\deriv[ML][\langle\beta\rangle]$ and $[\langle\beta\rangle]\deriv[ML][\langle\alpha\rangle]$, where $\langle\alpha\rangle=[\sigma([a_1],\ldots,[a_n]])]$ and similarly for $\langle\beta\rangle$. From this $\top\deriv[ML]\langle\alpha\rangle\leftrightarrow\langle\beta\rangle$ then follows easily from the propositional rules.

$\Rightarrow$: the opposite direction is shown in exactly the same way but in reverse and hinges on the fact that $\deriv[PL]$ axiomatizes all propositionally valid formulas. In particular we are free to choose any representative $a$ in an equivalence class $[a]$, since if we choose another representatice $a'$ then there must exist a proof $\deriv[PL]a\leftrightarrow a'$.
\end{proof}

We have thus shown that the two-dimensional $\deriv[ML]$- system for abstract coalgebraic system subsumes the more familiar proof system of coalgebrac logic in the predicate lifting format. It is also clear that it generalises the $\KKV(T)$ proof system of the nabla style of predicate logic.

\chapter{Algebraic canonicity}

This chapter will be almost purely algebraic, and will pursue a line of inquiry which dates back to Tarski and J\'{o}nsson's seminal work on boolean Algebras with Operators \cite{Jonsson51}, namely the study of modal logics through algebraic means (although \cite{Jonsson51} does not explicitly mention modal logics). The idea behind this approach is simple: since propositional logic can be given an algebraic semantic in terms of boolean algebras, it is reasonable to give an algebraic semantic to modal logics - broadly construed as in the prolegomenon - in terms of boolean algebras extended with some additional maps to interpret the modal operators. This approach is very natural, and just as the language of propositional logic can be seen as the free BA over a set $V$ of propositional variables, modal languages can also be seen as free extended boolean algebras over $V$. Thus from a universal algebra perspective, the algebraic study of modal logic is the study of varieties defined by adding equations to these free constructions. In this way, algebraic modal logic offers an algebraic handle on the main concepts and problems of modal logic, and the reader is referred to the classic \cite{2001:ModalLogic} for an introduction and overview to the subject and to \cite{2006:VenemaAC} for more details.

The particular problem which we wish to tackle algebraically in this chapter, and to some extent throughout this thesis, is that of canonicity. A precursor to a systematic algebraic study of canonicity can clearly be found in \cite{Jonsson51}.  This area of research then followed two complementary paths. Very crudely, the first path is concerned with the canonicity of \emph{varieties} (through  e.g. closure properties), whilst the second path investigates syntactic criteria for \emph{formulas} to define canonical varieties. The first line of research was developed by Goldblatt who studied canonicity in the wider context of relation, cylindric and modal algebras as a way to understand the connection between first-order logic and equational logic. In \cite{1989:Goldblatt, 1995:Goldblatt}, he investigated when an elementary relational structure (i.e. a relational structure defined by a first order formula) gives rise to a canonical variety. This wide-ranging and powerful line of research uses dualities and many ideas for universal algebra in the spirit of the famous Golblatt-Thomason theorem of modal logic. The other track, which we follow here, can be traced back to the near simultaneous papers of J\'{o}nsson \cite{Jonsson94} and de Rijke and Venema \cite{deRijkeVenema95}. The focus of both papers is to identify classes of formulas that define canonical varieties, and in particular to prove the classical Sahlqvist completeness result of \cite{1975:Sahlqvist} algebraically (see \cite{2001:ModalLogic} for the `traditional' treatment of the subject). 

Since then, the main focus of investigations into this algebraic theory of canonicity has been to weaken or modify the boolean core of the theory or to consider other types of extensions than the canonical extensions. The work of  Gehrke for example has extended much of the theory to bounded distributive lattice expansions \cite{1994:GehrkeJonsson, 2004:GehrkeJonsson} and bounded lattices expansions \cite{2001:GehrkeHarding}. Sahlqvist results in these settings can be found in \cite{2005:GehrkeVenema}. The thesis of Teheux \cite{2008:Teheux} examines canonicity in the context of multi-valued modal algebras, i.e. the boolean part of the theory is generalised to MV-algebras. Another strand of research can be found in Givant and Venema's \cite{1999:GivantVenema} where the Dedekind-Mac Neille completion is considered rather than the canonical extension, or in the work of Vosmaer who studies canonical extensions of various lattice structures in relation to profinite completions (see \cite{2010:Vosmaer} for an overview). Finally, canonicity for lattices has been used to prove completeness with respect to relational semantics  for substructural logics, see for example \cite{2005:GerhkeSubstruct,2006:GerhkeSubstruct,2011:Suzuki,2011:coumansrelational}.

In this chapter we will study canonical extensions of BAEs and DLEs whose expansions satisfy a large range of algebraic properties, with the ambition of getting some new canonicity results for non-normal modal logics. The approach will be very similar to that taken in \cite{2004:GehrkeJonsson}, but with a focus on covering a wide range of types of expansions and providing methods for building canonical (in)equations. In particular we aim to present a theory that includes $n$-ary maps which are either isotone or antitone in their arguments and which includes the notion of $k$-additivity of \cite{1970:Henkin}.

\section{Canonical Extensions}\label{ch2:sec:canext}
\subsection{Canonical extension of distributive lattices and boolean algebras}
We now briefly present the salient facts about canonical extensions. For more details the reader is referred to \cite{GivantHalmos} for BAs, \cite{Jonsson51} for BAOs, and \cite{1994:GehrkeJonsson,2004:GehrkeJonsson} for DLEs. The main rationale for studying canonical extensions is to embed a lattice-based structure, typically a language quotiented by some axioms, into a similar structure which is more `set-like', i.e. whose elements can be viewed as parts of a set, or of a set with some additional structure. In this way, we can hope to establish a connection between the syntax and the semantics, i.e. build models from formulas. But what does being `set-like' mean? Two criteria emerge as being fundamental: completeness and being generated from below (i.e. by joins) by something akin to `elements'. Canonical extensions satisfy these two conditions. Our approach here will be purely algebraic, i.e. we will not make use of the duality theory of distributive lattices or boolean algebras.

We start with distributive lattices. For a distributive lattice $\ba$, the idea behind the construction of its canonical extension $\ba\ce$ is to build a completion of $\ba$ which is not `too big' and not `too different' from $\ba$. Technically, we want $\ba$ to be \textbf{dense} and \textbf{compact} in $\ba\ce$.

\paragraph*{Density.} To build a completion of $\ba$ it is natural to formally add to $\ba$ all meets, all joins, all meets of all joins, all joins of all meets, etc. In the case of the canonical extension we require that this procedure stops after two iterations (i.e. we want a $\Delta_1$-completion , see \cite{2013:Delta1Completions}). Intuitively, this prevents the completion from becoming `too big'. Based on this intuition we introduce the following terminology. Given a sub-lattice $\ba$ of a complete distributive lattice $\ba[C]$, we define the meets in $\ba[C]$ of elements of $\ba$ as the \textbf{closed elements}\index{Closed element} of $\ba[C]$ and denote the set of closed elements as $K(\ba)$ (or simply $K$ when there is no ambiguity). The sub-lattice $\ba$ is said to be \textbf{meet-dense}\index{Dense!meet-dense} in $\ba[C]$ if $K(\ba)=C$, i.e. if every element is closed. Dually, we define the joins in $\ba[C]$ of elements of $\ba$ as the \textbf{open elements}\index{Open element} of $\ba[C]$ and denote the set of open elements as $O(\ba)$. The sub-lattice $\ba$ is said to be \textbf{join-dense}\index{Dense!join-dense} in $\ba[C]$ if $O(\ba)=C$, i.e. if every element is open. Finally, we say that $\ba$ is \textbf{dense}\index{Dense} in $\ba[C]$ if 
\[
\ba[C]=O(K(\ba))=K(O(\ba))
\]
It is natural to describe the elements of $\ba$ as \text{clopens} in $\ba[C]$ since they are both open and closed.

\paragraph*{Compactness.} The canonical extension $\ba\ce$ of $\ba$ is also required not to be too different from $\ba$ in the sense that facts about arbitrary meets and joins of elements of $\ba$ in $\ba\ce$ must already be `witnessed' by finite meets and joins in $\ba$. Formally, if $\ba$ is a sub-lattice of $\ba[C]$, we say that $\ba$ is \textbf{compact}\index{Compact} in $\ba[C]$ if for every $X,Y\subseteq \ba$ such that
\[
\bigwedge X\leq \bigvee Y
\] 
there exists \emph{finite} subsets $X_0\subseteq X, Y_0\subseteq Y$ such that
\[
\bigwedge X_0\leq \bigvee Y_0
\]
An equivalent definition is that $\ba$ is compact in $\ba[C]$ if for every closed element $p\in K(\ba)$ and open element $u\in O(\ba)$ such that $p\leq u$, there exists an element $a\in\ba$ such that $p\leq a\leq u$.

We are now ready to define the \textbf{canonical extension}\index{Canonical extension} $\ba\ce$ of a distributive lattice $\ba$ as the complete distributive lattice such that $\ba$ is dense and compact in $\ba\ce$. We will show that canonical extensions exist and are unique up to isomorphism. We start with the following lemmas concerning infinite distributivity in canonical extensions.

\begin{lemma}\label{ch2:lem:InfDistrib1}
Let $\ba\ce$ be the canonical extension of a distributive lattice $\ba$, $U\subseteq O$ be a set of open elements and $P\subseteq K$ be a set of closed elements, then for any $x\in \ba\ce$
\begin{enumerate}
\item $x\wedge \bigvee U=\bigvee\{x\wedge u\mid u\in U\}$
\item $x\vee \bigwedge P=\bigwedge\{x\vee p\mid p\in P\}$
\end{enumerate}
\end{lemma}
\begin{proof}
We show the first distribution law, the second follows dually. Clearly, since $x\wedge u\leq a\wedge \bigvee U$ for each $u\in U$, we have $ \bigvee\{x\wedge u\mid u\in U\}\leq x\wedge \bigvee U$. For the opposite inequality, we will show that if a closed element $p\leq x\wedge \bigvee U$, then $p\leq \bigvee\{x\wedge u\mid u\in U\}$, and the result will then follow by density. If $p\leq x\wedge \bigvee U$, then $p\leq x$ and $p\leq  \bigvee U$, and since $ \bigvee U$ is a set of open elements, compactness gives us a finite subset $U_0\subseteq U$ such that $p\leq \bigvee U_0$. In consequence, by using the fact that $\ba\ce$ is distributive
\[
p\leq x\wedge\bigvee U_0= \bigvee\{x\wedge u\mid u\in U_0\}\leq \bigvee\{x\wedge u\mid u\in U\}
\]
which is want we wanted to show.
\end{proof}

Using a very similar proof, we can also show complete distributivity for directed sets of closed and open elements.

\begin{lemma}\label{ch2:lem:GenDistribDirected}
Let $\ba\ce$ be the canonical extension of a distributive lattice $\ba$, $U_i, i\in I$ be collection of an upward-directed sets of open elements and $P_j, j\in J$ be a collection of downward-directed sets of closed elements, then
\begin{enumerate}
\item $\bigwedge_i \bigvee U_i=\bigvee\{\bigwedge \im\phi\mid \phi: I\to O, \phi(i)\in U_i\}$
\item $\bigvee_j \bigwedge P_j=\bigwedge\{\bigvee \im\phi\mid \phi: J\to K, \phi(j)\in P_j\}$
\end{enumerate}
\end{lemma}
\begin{proof}
See Lemma 3.2. of \cite{2001:GehrkeHarding}.
\end{proof}

We can now show the following important infinite distribution laws.
\begin{lemma}\label{ch2:lem:InfDistrib2}
Let $\ba\ce$ be the canonical extension of a distributive lattice $\ba$, $U$ be a set of open elements and $P$ be a set of closed elements, then for any $q\in K, v\in O$
\begin{enumerate}
\item $q\wedge \bigvee P=\bigvee\{q\wedge p\mid p\in P\}$
\item $v\vee \bigwedge U=\bigwedge\{v\vee u\mid u\in U\}$
\end{enumerate}
\end{lemma}
\begin{proof}
\begin{align*}
q\wedge \bigvee P&=q\wedge\bigvee\{\bigwedge\{a\in A\mid a\geq p\}\mid p\in P\}& \text{Density}\\
&=q\wedge\bigwedge\{\bigvee \im\phi\mid \phi: P\to A, \phi(p)\geq p\} & \text{Lemma }\ref{ch2:lem:GenDistribDirected}\\
&=\bigwedge\{q\wedge\bigvee \im\phi\mid \phi: P\to A, \phi(p)\geq p\}\\
&=\bigwedge\{\bigvee\{q\wedge a\mid a\in\im\phi\}\mid \phi: P\to A, \phi(p)\geq p\}& \text{Lemma }\ref{ch2:lem:InfDistrib1}
\end{align*}
Note that we have used Lemma \ref{ch2:lem:GenDistribDirected} because each $\{a\in A\mid a\geq p\}$ is a down-directed set of clopen - and thus closed - elements, and Lemma \ref{ch2:lem:InfDistrib1} because each $\im\phi$ is open. We want to show that the last term of the derivation is equal to 
\[
\bigwedge\{\bigvee\im\psi\mid \psi: P\to A, \psi(p)\geq q\wedge p\}
\]
For every $\phi$, we can define the following $\psi: P\to A, p\mapsto q\wedge \phi(p)$ and we get
\[
\bigvee \{q\wedge a\mid a\in\im\phi\}=\bigvee\im\psi
\]
Thus 
\begin{align*}
\bigwedge\{\bigvee\im\psi\mid \psi: P\to A,& \psi(p)\geq q\wedge p\} \leq\\
& \bigwedge\{\bigvee\{q\wedge a\mid a\in\im\phi\}\mid \phi: P\to A, \phi(p)\geq p\}
\end{align*}
For the opposite inequality, we start with a $\psi$ which picks a clopen above each element $q\wedge p$, and we want to define a $\phi$ which picks a clopen above each $p$ and satisfies $\bigvee \{q\wedge a\mid a\in\im\phi\}=\bigvee\im\psi$. To define such a $\phi$, we choose for $p\in P$ any clopen $a'$ above $p$ such that $q\wedge a'\leq a=\psi(p)$. To see that such an element exists, we proceed by contraposition. Assume that no such element exist, i.e. that for all $a'\geq p$ it is the case that $a'\wedge q\nleq a$, i.e. that for all $a'\geq p$, $a\vee(a'\wedge q)\gneq a$. But then
\begin{align*}
& \bigwedge\{a\vee(a'\wedge q)\mid a'\geq p\}\gneq a\\
\Leftrightarrow &\hspace{1ex} a \vee \bigwedge\{a'\wedge q \mid a'\geq p\}\gneq a & \text{Lemma \ref{ch2:lem:InfDistrib1} (each }a'\wedge q\text{ if closed)}\\
\Leftrightarrow & \hspace{1ex} a\vee (q\wedge \bigwedge\{a'\mid a'\geq p\}\gneq a\\
\Leftrightarrow & \hspace{1ex} a\vee (q\wedge p)\gneq a & p\text{ is closed}
\end{align*}
which contradicts the hypothesis that $a=\psi(p)\geq q\wedge p$. Thus for each $p$ we can choose a clopen $a'$ above it such that $q\wedge a'\leq a=\psi(p)$, and this defines a choice function $\phi: P\to A$ such that
\[
\bigvee\im\psi=\bigvee \{q\wedge a'\mid a'\in\im\phi\}
\]
and we thus the get inequality
\begin{align*}
\bigwedge\{\bigvee\{q\wedge a\mid a\in\im\phi\}\mid \phi: P\to A, &\phi(p)\geq p\}\leq\\
&\bigwedge\{\bigvee\im\psi\mid \psi: P\to A, \psi(p)\leq q\wedge p\}
\end{align*}
Thus we have
\begin{align*}
q\wedge \bigvee P&=\bigwedge\{\bigvee\im\psi\mid \psi: P\to A, \psi(p)\geq q\wedge p\}\\
& =\bigvee \{\bigwedge \{a\in A\mid a\geq q\wedge p\}\mid p\in P\} & \text{Lemma }\ref{ch2:lem:GenDistribDirected}\\
&=\bigvee\{q\wedge p\mid p\in P\} & p\wedge q\text{ is closed}
\end{align*}
The other distribution rule follows dually.
\end{proof}

There are two explicit representations of the canonical extension of a distributive lattice. The first one requires the Prime Filter Theorem which is a non-constructive principle, albeit strictly weaker than the axiom of choice (see \cite{1971:PrimeIdlThmAC}). The second is constructive and based on Galois connections (see \cite{2001:GehrkeHarding}) but is not compatible with the adjunction $\pf\dashv\ups:\DL\to\Pos$ and the process of model building which we develop in Chapter 5. We will therefore follow the more traditional non-constructive approach. The key to representing canonical extensions is to show that certain classes of elements of the abstractly defined canonical extension $\ba\ce$ can be realized as familiar constructions on the original lattice $\ba$.

\begin{lemma}\label{ch2:lem:closedFilters}
Let $\ba\ce$ be the canonical extension of a distributive lattice $\ba$, then the sub-lattice $K$ of closed elements of $\ba\ce$ is isomorphic to the lattice of filters of $\ba$. Dually, the sub-lattice $O$ of open elements of $\ba\ce$ is isomorphic to the lattice of ideals of $\ba$.
\end{lemma}
\begin{proof}
For a full proof, see Lemma 3.3. of \cite{2001:GehrkeHarding}. The isomorphism maps a closed element of $\ba\ce$ to the set of elements of $\ba$ (clopens) above it, and, in the opposite direction, a filter of $\ba$ to its meet in $\ba\ce$. Dually, open elements are mapped to the set of elements of $\ba$ below it, and ideals of $\ba$ to their joins.
\end{proof}

Since every element of the canonical extension $\ba\ce$ is a join of closed element, we could try to represent arbitrary elements as formal joins of filters of $\ba$. However, using the Prime Filter Theorem, it is possible to isolate a subset of the closed elements which is join-dense in $\ba\ce$ and has a much more natural logical interpretation.

\begin{theorem}[Prime Filter Theorem for distributive lattices]\label{ch2:thm:PrimeFilterThm}
Suppose $\ba$ is a distributive lattice, $I$ is an ideal of $\ba$, $F$ is a filter of $\ba$, and $I\cap F=\emptyset$. then there exists a prime filter $G$ such that $F\subseteq G$ and $I\cap G=\emptyset$
\end{theorem}
\begin{proof}
See Section III.4 Theorem 1 in \cite{1975:balbesdistributive}.
\end{proof}

Recall that an element $x$ of a lattice $L$ is called \textbf{join irreducible}\index{Join irreducible} if it cannot be written as a (non-trivial) join of elements of $L$, i.e. if $x=y_1\vee y_2, y_1,y_2\in L$ implies $x=y_1$ or $x=y_2$. Similarly, $x$ is said to be \textbf{join prime}\index{Join prime} if $x\leq y_1\vee y_2$ implies that $x\leq y_1$ of $x\leq y_2$. Every join prime element is join irreducible and the converse can easily be seen to hold if the lattice is distributive.  If a lattice $L$ is complete, we can generalise these concepts as follows: we will say that $x\in L$ is \textbf{completely join irreducible}\index{Join irreducible!Completely} if $x=\bigvee Y$ implies $x=y$ for some $y\in Y$, and that $x$ is \textbf{completely join prime}\index{Join prime!Completely} is $x\leq \bigvee Y$ implies $x\leq y$ for some $y\in Y$. It is conventional to denote the set of completely join irreducible elements of a complete lattice $L$ as $J(L)$. The following corollary of Lemma \ref{ch2:lem:InfDistrib2} shows that the two concepts coincide in canonical extensions of distributive lattices.

\begin{lemma}\label{ch2:lem:compJoinPrime}
Let $\ba\ce$ be the canonical extension of a distributive lattice $\ba$, the completely join irreducible elements of $\ba\ce$ are completely join prime.
\end{lemma}
\begin{proof}
Let $p$ be completely join irreducible. Clearly, since every element of $A\ce$ is a join of closed element, $p$ must be closed. Assume now that $p\leq \bigvee X$ for some $X\subseteq \ba\ce$. By density, we can assume w.l.o.g. that $X$ is a set of closed elements. We now use Lemma \ref{ch2:lem:InfDistrib2} to conclude that
\[
p=p\wedge \bigvee X=\bigvee\{p\wedge x\mid x\in X\}
\]
and thus $p\leq x$ for some $x\in X$, since it is completely join irreducible.
\end{proof}

\begin{lemma}[\cite{2001:GehrkeHarding}]\label{ch2:lem:joinIrredPrimeFilters}
Let $\ba\ce$ be the canonical extension of a distributive lattice $\ba$, then the poset $J(\ba\ce)$ of completely join irreducible elements of $\ba\ce$ is isomorphic to $(\pf\ba)\op$, the poset of prime filters under reverse set-theoretic ordering. Moreover, every element of $\ba\ce$ can be written as a join of elements of $J(\ba\ce)$.
\end{lemma}
\begin{proof}
The isomorphism in $\Pos$ works as follows: we define $\phi: J(\ba\ce)\to (\pf\ba)\op, p\mapsto\{a\in\ba\mid p\leq a\}$ and $\psi: (\pf\ba)\op\to J(\ba\ce), F\mapsto\bigwedge F$. Clearly if $F\subseteq F'$ then $\psi(F)\geq \psi(F')$, and conversely if $p\leq p'$ then $\phi(p)\supseteq \phi(p')$, hence the $(-)\op$ order. It is not difficult to see that if both maps are well-defined, then $\psi\circ \phi(p)=p$. For the opposite direction we clearly have $F\subseteq \{a\in\ba\mid a\geq \bigwedge F\}$, to show the equality we use compactness: since $a$ is clopen it is in particular open and thus $\bigwedge F\leq a$ is witnessed by a finite set $F_0$ of elements of $F$, i.e. $\bigwedge F_0\leq a$, but since $F$ is a filter it follows that $\bigwedge F_0\in F$ and thus $a\in F$. Thus $\psi,\phi$ are inverse of each other provided that they are well-defined. The fact that $\phi$ is well-defined is easy: $\phi(p)$ is clearly a filter, moreover it is prime because $p$ is join irreducible. To see that $\psi(F)=\bigwedge F$ is completely join irreducible when $F$ is prime, let $\bigwedge F=\bigvee M$, where we can assume without loss of generality that $M$ is a set of closed elements. Clearly $m\leq \bigwedge F$ each $m\in M$, thus if we want to proceed by contradiction we should assume that $m<\bigwedge F$ for each $m\in M$. But then $\{a\in\ba\mid a\geq m\}\supsetneq F$ since $m$ is closed, and for each $m\in M$ we can therefore select an $a_m\geq m$ not in $F$. It follows that 
\[
\bigwedge F\leq \bigvee M\leq \bigvee_{m\in M} a_m
\]
which by compactness means that there must exist a finite set $M_0$ of $a_m$s such that $\bigvee_{m\in M_0}a_m\in F$, a contradiction since $F$ is prime.

The second part of the lemma requires the Prime Filter Theorem for distributive lattices, and is central to the `representability' of the canonical extension. Since every element of $\ba\ce$ is a join of closed elements, it is enough to show the property for closed elements. To show the property we'll show that if $k\neq k', k,k'\in K$ then there exists a completely join irreducible element $p$ witnessing the non-equality from below, i.e. such that $p\leq k, p\nleq k'$. Note that by density, if $k\neq k'$ there must exist an open set witnessing the inequality from above, i.e. $u\in O$ s.th. $k\nleq u, k'\leq u$. Using Lemma \ref{ch2:lem:closedFilters}, we know that $k$ uniquely defines a filter $F(k)=\{a\in \ba\mid k\leq a\}$ and $u$ uniquely defines an ideal $I(u)=\{a\in \ba\mid a\leq u\}$. If $a\in F(k)\cap I(k)$ then $k\leq a$ and $a\leq u$, and thus $k\leq u$ which contradicts the assumption that $k\nleq u$. Thus $F(k)\cap I(u)=\emptyset$, and we can apply the Prime Filter Theorem to get a prime filter $P\supseteq F(k)$ such that $P\cap I(k)=\emptyset$. By the first part of the proof, and Lemma \ref{ch2:lem:closedFilters}, we can map this prime filter back to $\ba\ce$ to get a completely join irreducible element $p$ such that
\[
p=\bigwedge P\leq \bigwedge F(k)=k\nleq \bigvee I(u)=u
\]
as desired.
\end{proof}

\begin{theorem}\label{ch2:thm:CanExtRepresentation}
The canonical extension of a distributive lattice is given by
\[
\ba\ce\simeq \ups\pf\ba
\]
and is unique up to isomorphism.
\end{theorem}
\begin{proof}
Using Lemma \ref{ch2:lem:joinIrredPrimeFilters} it suffices to show that $\ba\ce\simeq \mathcal{D} J(\ba\ce)$, where $\mathcal{D}$ is the down-set functor. We define
\[
\phi: \ba\ce\to \mathcal{D} J(\ba\ce): x\mapsto \{p\in J(\ba\ce)\mid p\leq x\}
\]
Conversely, we define
\[
\psi: \mathcal{D} J(\ba\ce)\to \ba\ce, X\mapsto \bigvee X
\]
It is clear that $\phi$ preserves meets, and it preserves joins by the fact that each $p$ is join prime. The fact that $\psi$ preserve joins is trivial and the fact that it preserves meets follows from the fact that $\ba\ce$ is distributive. Thus $\phi,\psi$ are $\DL$-morphisms. From the fact that every element of $\ba\ce$ is a join of element of $J(\ba\ce)$ (by Lemma \ref{ch2:lem:joinIrredPrimeFilters}), it is clear that $\psi\circ\phi(x)=x$ for all $x\in\ba\ce$. Conversely, since every join irreducible is join prime (by Lemma \ref{ch2:lem:compJoinPrime}), we have 
\begin{align*}
\phi\circ\psi(X)&=\phi(\bigvee X)\\
&=\{p\in J(A\ce)\mid p\leq \bigvee X\}\\
&=\{p\in J(A\ce)\mid p\leq x\text{ for some }x\in X\}\\
&=X
\end{align*}
where the last step uses the fact that $X$ is a down-set.

The unicity follows from the fact that every element in the canonical extension can be written as a join of completely join irreducibles, and that these are in bijective correspondence with the prime filters of $\ba$. This allows us to build a bijection between the sets of completely join irreducibles of any two canonical extensions, and thus between any two canonical extensions.
\end{proof}

Both functors $\pf$ and $\ups$ arise from the fact that $\mathbbm{2}$ is both a distributive lattice and a poset, and $\pf \ba$ is just the set $\Hom_{\DL}(\ba,\mathbbm{2})$ with the pointwise partial ordering, whilst $\ups X$ is the natural lattice structure on $\Hom_{\Pos}(X,\mathbbm{2})$). An alternative representation of $\ba\ce$ is thus given by the set of monotone maps from $\pf\ba$ (or, equivalently $J(\ba\ce)$ by Lemma \ref{ch2:lem:joinIrredPrimeFilters}) into $\mathbbm{2}$ via the inverse image of $0$. In the case of the boolean algebras to which we will now turn our attention, the down-sets functor becomes the power set functor which can also be characterized as inverse images under characteristic functions.

Each BA $\ba$ is a distributive lattice, and can thus be embedded into a unique (up to isomorphism) complete boolean algebra $\ba\ce$ in which it is both dense and compact. \index{Canonical extension! of a boolean algebra} Since prime filters in a boolean algebra are always maximal, i.e. ultrafilters, it follows that the partial order on $\pf\ba=\uf\ba$ is discrete, i.e. each ultrafilter is only related to itself, and thus $\uf\ba$ is really just a set. For this discrete partial order any map into $\mathbbm{2}$ is monotone, which explains why the adjunction $\pf\dashv\ups:\DL\to\Pos$ becomes $\uf\dashv\pow: \BA\to\Set$.

The following  results specialise the canonical extension construction to the case of boolean algebras.

\begin{proposition}An element of a boolean algebra is an atom iff it is completely join irreducible.
\end{proposition}
\begin{proof}
See Theorem 8.10.2. of \cite{2001:DunnAlgebraic}.
\end{proof}

Since every element of the canonical extension can be written as a join of completely join irreducibles (Lemma \ref{ch2:lem:joinIrredPrimeFilters}), the previous Proposition  shows that the canonical extension of a boolean algebra is \emph{atomic}. Thus the canonical extension of a boolean algebra is a Complete Atomic Boolean Algebra (CABA).  Moreover, we have:

\begin{theorem}[\cite{GivantHalmos} Ch. 14]\label{ch2:thm:GD}
A complete boolean algebra is atomic iff it is completely distributive.
\end{theorem}
In particular, the canonical extension of a boolean algebra is always completely distributive. In fact, this holds for any distributive lattice since the canonical extension $\ba\ce$ of a distributive lattice $\ba$ is isomorphic to $\ups\pf\ba$ which is completely distributive. 

\subsection{Canonical extension of expansions}

We now turn our attention to maps defined on distributive lattices and boolean algebras, fix some terminology and define their canonical extensions. We will use the same convention as in Section \ref{ch1:sec:boolstruct} and use $\A$ to denote either of the categories $\DL, \BDL$ or $\BA$. For any object $\ba$  in $\A$ with underlying set $A$, will say that a \emph{function} $f: A^m \to A$ is \textbf{isotone in its $i^{th}$ argument}\index{Isotone map} if whenever $a_i\leq b_i$
\[
f(a_1,\ldots,a_{i-1},a_i,a_{i+1},\ldots,a_m)\leq f(a_1,\ldots,a_{i-1},b_i,a_{i+1},\ldots,a_m)
\]
Similarly, we will say that $f$ is \textbf{antitone in its $i^{th}$ argument}\index{Antitone map} if whenever $a_i\leq b_i$
\[
f(a_1,\ldots,a_{i-1},a_i,a_{i+1},\ldots,a_m)\geq f(a_1,\ldots,a_{i-1},b_i,a_{i+1},\ldots,a_m)
\]
A map $f$ is \textbf{monotone in its $i^{th}$ argument}\index{Monotone map} if it is either isotone or antitone in this argument. Finally, we will say that $f$ is isotone (resp. antitone, resp monotone) if it is isotone (resp. antitone, resp. monotone) in \emph{all} of its arguments, and we will say that $f$ is \textbf{mixed} if it has both isotone and antitone arguments, i.e. monotone but neither isotone nor antitone.

The existence and unicity of canonical extensions for BAs was extended in \cite{Jonsson51} to any BAO $\mathcal{A}$ by defining the \textbf{extension of an operator}\index{Extension of map} $f$ of arity $n$ as either of the following alternatives
\begin{align}
f^\sigma(x)=\bigvee_{x\geq y\in K^n} \bigwedge_{y\leq z\in A^n}f(z)\label{ch2:eq:mapext11} \\
f\ced(x)=\bigwedge_{x\leq u\in O^n}\bigvee_{u\geq z\in A^n} f(z)\label{ch2:eq:mapext21}
\end{align}
Note that the original definition in \cite{Jonsson51, Jonsson94} was Eq. (\ref{ch2:eq:mapext11}), the dual possibility of Eq. (\ref{ch2:eq:mapext21}) is defined for example in \cite{2001:GehrkeHarding,2006:VenemaAC}. A map $f$ is called \textbf{smooth}\index{Smooth map} if $f\ce=f^\pi$, we will return to this notion shortly. The canonical extension of a BAO $\mathcal{A}$ is traditionally defined as $\mathcal{A}\ce=(\mathbb{A}\ce, (f_s\ce)_{s\in\Sigma})$. In fact, the definitions of $f\ce$ and $f\ced$ make sense not just for operators, but more generally for any map which is isotone.  However, for a completely general function $f: A^n\to A$, or even for an antitone map, the definitions Eqs. (\ref{ch2:eq:mapext11}) and (\ref{ch2:eq:mapext21}) do not make sense. We therefore need a more powerful definition of map extension. Such a definition was developed for the study of canonical extensions of various classes of lattices, see e.g. \cite{1994:GehrkeJonsson,2001:GehrkeHarding,2004:GehrkeJonsson,2006:VenemaAC}. Given an object in $\A$ with underlying set $A$, the $\sigma$- and $\pi$- extensions of a map $f: A^n\to A$ are the maps $(A\ce)^n\to A\ce$ defined by:
\begin{align}
f\ce(x)=\bigvee\{\bigwedge f[d,u]\mid K^n\ni d\leq x\leq u\in  O^n\}\label{ch2:eq:mapext12} \\
f\ced(x)=\bigwedge\{\bigvee f[d,u]\mid K^n\ni d\leq x\leq u\in  O^n\}\label{ch2:eq:mapext22} 
\end{align}
where $A\ce$ is the set underlying $\ba\ce$ and $f[d,u]=\{f(a)\mid a\in A^n, d\leq a\leq u\}$. Note that $f[d,u]$ is always non-empty by compactness, which is the reason for choosing this type of `intervals'. It is straightforward to check that if $f$ is isotone Eqs. (\ref{ch2:eq:mapext11}) and (\ref{ch2:eq:mapext12}) coincide, and similarly for the dual definitions of Eqs. (\ref{ch2:eq:mapext21}) and (\ref{ch2:eq:mapext22}). We can now define the canonical extension of a DLE (resp.  BAE) $\mathcal{A}=(\mathbb{A},(f_s)_{s\in\Sigma})$ as the DLE (resp. BAE) $\mathcal{A}^\sigma=(\mathbb{A}\ce, (f_s\ce)_{s\in\Sigma})$. This choice is purely arbitrary since using the $(\cdot)\ced$ extension would work too, but it is conventional. In fact the choice of which extension to use can be guided, as we shall see later, by the algebraic properties of the maps which we want to extend, since some properties are preserved by the $(\cdot)\ce$ extension only, and others by the $(\cdot)\ced$ extension, so having two possible definitions can be turned into an advantage. Let us start by examining some of the basic properties of these extensions.

\begin{proposition}\label{ch2:prop:anymap}
Let $\ba$ be an object in $\A$ with underlying set $A$, and let $f:A^n\to A$ be \emph{any} map, then 
\begin{enumerate}[(i)]
\item $f\ce\upharpoonright A^n=f\ced\upharpoonright A^n=f$
\item $f\ce\leq f\ced$ under the pointwise order
\end{enumerate}
\end{proposition}
\begin{proof}
The proof can be found in for example \cite{2001:GehrkeHarding} or \cite{2006:VenemaAC} for the case of unary operators and is readily adapted to the $n$-ary case by considering tuples of arguments.
\end{proof}

\noindent If $f$ is monotone we can say a bit more.

\begin{proposition}\label{ch2:prop:monotonemap}
Let $\ba$ be an object in $\A$ and let $f:\mathbb{A}^n\to\mathbb{A}$ be a monotone map, then \[f\ce \upharpoonright (K\cup O)^n=f\ced \upharpoonright(K \cup O)^n\]
\end{proposition}
\begin{proof}
The case where $f$ is isotone is shown in for example \cite{2001:GehrkeHarding,2006:VenemaAC}, so we just show the antitone case. Let us start by assuming that $f$ is unary. We know from Proposition \ref{ch2:prop:anymap} that $f\ce(x)\leq f\ced(x)$ for any $x\in A\ce$, so we only need to show the opposite inequality. Start with $p\in K$, then for all $p\leq a\in A$ we have by antitonicity $f(a)=f\ce(a)\leq f\ce(p)$ and since this holds for any such $a$ we have \[\bigvee\{f(a)\mid p\leq a\in A\}\leq f\ce(p)\]
i.e. $f\ced(p)\leq f\ce(p)$ by Eq. (\ref{ch2:eq:mapext22}) applied to an antitone map. Similarly, for $u\in O$ we have for any $u\geq a\in A$ that $f\ced(u)\leq f\ced(a)=f(a)$ and thus \[f\ced(u)\leq \bigwedge\{f(a)\mid u\geq a\in A\}\]
i.e. $f\ced(u)\leq f\ce(u)$  by Eq. (\ref{ch2:eq:mapext12}) applied to an antitone map.

The case where $f$ is $n$-ary is treated by first considering tuples consisting of open elements for the isotone arguments and closed elements for the antitone arguments. From there the general case is simply computed. We show how this is done in the case where $f: A^2\to A$ is binary, isotone in the first argument and antitone in the second. By Proposition \ref{ch2:prop:anymap}, we know that for any $x,y\in A\ce, f\ce(x,y)\leq f\ced(x,y)$. We briefly show the opposite inequality for $u\in O, p\in K$. Let $a_1,a_2\in A$ such that $a_1\leq u, p\leq a_2$, then we have
\[
f(a_1,a_2)=f\ce(a_1,a_2)\leq f\ce(u,p)
\]
and since this holds for any such $a_1,a_2$ we get
\[
f\ced(u,p)=\bigvee_{a_1\leq u, p\leq a_2}f(a_1,a_2)\leq f\ce(u,p)
\]
A completely analogous dual proof shows that $f\ced(p,u)\leq f\ce(p,u)$. Consider now, $p,q\in K$, from what we have just shown and the definitions of Eqs. \ref{ch2:eq:mapext12} and \ref{ch2:eq:mapext22}, we get
\begin{align*}
f\ced(p,q)& =\bigwedge_{p\leq u_1\in O}\bigvee_{\substack{u\geq a_1\in A \\ q\leq a_2\in A}}f(a_1,a_2)\\
&=\bigwedge_{p\leq u\in O} f\ced(u,q)\\
&=\bigwedge_{p\leq u\in O} f\ce(u,q)\\
&=\bigwedge_{p\leq u\in O} \bigvee_{\substack{u\geq d\in K \\ q\leq v\in O}} \bigwedge_{\substack{d\leq a_1\in A \\ v\geq a_2\in A}} f(a_1,a_2)\\
&=\bigvee_{q\leq v\in O}\bigwedge_{\substack{p\leq a_1\in A \\ v\geq a_2\in A}}f(a_1,a_2)\\
&=f\ce(p,q)
\end{align*}
where the penultimate step follows from the fact that taking the meet over all $u\in O$ such that $p\leq u$, of the join over all $d_1\in K$ such that $d_1\leq u$, amounts to doing nothing since every element of the canonical extension can be written in this way. An analogous dual proof shows that $f\ced=f\ce$ for pairs of open elements too.
\end{proof}

\noindent The natural question is now: when do $f\ce$ and $f\ced$ agree on \emph{any} $x\in A\ce$, i.e. when is $f$ smooth? We will present a necessary and sufficient condition for a map to be smooth in Section \ref{ch2:sec:topol}, but we can already show that some frequent algebraic conditions on $f\ce$ and $f\ced$ (which can themselves be inherited from conditions on $f$ as we shall see in a short while) guarantee smoothness. We will say that a map $f:A^n\to A$ is \textbf{smooth in its $i^{th}$ argument}\index{Smooth map} $(1\leq i\leq n)$ if for any $n$-tuple $(x_1,\ldots,x_i,\ldots,x_n)$ such that $x_i\in A\ce$ and $x_j\in K\cup O$ for every $1\leq j\leq n, j\neq i$ we have $$f\ce(x_1,\ldots,x_i,\ldots,x_n)=f\ced(x_1,\ldots,x_i,\ldots,x_n)$$  

\begin{proposition}\label{ch2:prop:smooth}
Let $\ba$ be an object in $\A$ with underlying set $A$ and let $f:A^n\to A$ be an isotone map. 
\begin{enumerate}[(i)]
\item If $f\ce$ preserves down-directed meets or non-empty joins in an argument, then $f$ is smooth in that argument.
\item If $f\ced$ preserves upwardly directed joins or non-empty meets in an argument, then $f$ is smooth in that argument.
\end{enumerate}
\end{proposition}
\begin{proof}
The proof for $n$-ary functions is exactly the same as the proof for unary functions, so for notational clarity we stick to unary functions. If $f\ce$ preserves down-directed meets then the proof is essentially in \cite{2001:GehrkeHarding} and goes as follows: let $x\in A\ce$, then by definition of the canonical extension we can write $$x=\bigwedge\{u\mid x\leq u\in O\}$$
Since the set $\{u\mid x\leq u\in O\}$ is down-directed we have by our hypothesis that $f\ce(x)=\bigwedge\{f\ce(u)\mid x\leq u\in O\}$ but since $f\ce=f\ced$ on $O$ we have $$f\ce(x)=\bigwedge\{f\ced(u)\mid x\leq u\in O\}=f\ced(x)$$
since $f\ce$ and $f\ced$ agree on $K\cup O$. The proof for non-empty join preservation can be found in \cite{2001:GehrkeHarding}. The proof for $f\ced$ follows by duality. 
\end{proof}

\noindent We will say that $f:A\to A$ \textbf{anti-preserves (down-directed) meets}\index{Anti-preservation! of meets}\index{Anti-preservation! of joins}\index{Anti-preservation! of down-directed meets}\index{Anti-preservation! of up-directed joins} if for any (down-directed) subset $X\subseteq A$ $$f(\bigwedge X)=\bigvee f[X]$$ 
and dually for joins. Note that if $f$ anti-preserves meets $a\leq b$ in $A$, i.e. $a\wedge b=a$ then $f(a)=f(a\wedge b)=f(a)\vee f(b)$ and $f$ is automatically antitone, and dually for joins. Similarly, by considering the down-directed closure of $\{a,b\}$, i.e. $\{a, b, a\wedge b\}$ the anti-preservation of down-directed meets gives 
\begin{align*}
f(a) & =f(\bigwedge\{a, b, a\wedge b\}) \\
& = \bigvee f[\{a, b, a\wedge b\}] \\
&= f(a) \vee f(b) \vee f(a) = f(a)\vee f(b)
\end{align*}
and we again automatically get that $f$ is antitone. This of course dually holds for the anti-preservation of up-directed joins too.

\begin{corollary}\label{ch2:cor:smoothcor}
Let $\ba$ be an object in $\A$ with underlying set $A$ and let $f:A^n\to A$.
\begin{enumerate}[(i)]
\item If $f\ce$ anti-preserves up-directed joins or non-empty meets in one argument, then $f$ is smooth in that argument.
\item  If $f\ced$ anti-preserves down-directed meets or non-empty joins in one argument, then $f$ is smooth in that argument.
\end{enumerate}
\end{corollary}
\begin{proof}
The proof is very similar to \ref{ch2:prop:smooth}, but we show (i) as an illustration of working with unary antitone maps. The case of arbitrary $n$-ary maps follows easily from the unary cases by working component-wise. The proof of (ii) follows by duality.

\noindent Let us first assume that $f\ce$ anti-preserves up-directed joins. As we just saw this means that $f\ce$ is antitone, and since $f\ce=f$ on $A$, this means that $f$ must be antitone too. We now write $x\in A\ce$ as $$x=\bigvee\{p\mid x\geq p\in K\}$$
which is an up-directed join. We thus have
\begin{align*}
f\ce(x)&=f\ce(\bigvee\{p\mid x\geq p\in K\})\\
&=\bigwedge\{f\ce(p)\mid x\geq p\in K\}\\
&=\bigwedge\{f\ced(p)\mid x\geq p\in K\}=f\ced(x)
\end{align*}
by the fact that $f$ is antitone, $f\ce$ anti-preserves down-directed meets, Proposition \ref{ch2:prop:monotonemap}, and the definition of $f\ced$.

\noindent Let us now assume that $f\ce$ anti-preserves non-empty meets. In particular this means that $f$ must be antitone, and thus for any $x\in A\ce$  we have
\begin{align*}
f\ce(x)& =\bigvee\{\bigwedge\{f(a)\mid u\geq a\in A\} \mid x\leq u\in O\} \\
&=\bigwedge\{\bigvee \{f(a)\mid a \in \im(\gamma)\}\mid \gamma: \up x\cap O\to A \text{ s.th. } \gamma(u)\leq u\} \\
&=\bigwedge\{\bigvee \{f\ce(a)\mid a \in \im(\gamma)\}\mid \gamma: \up x\cap O\to A \text{ s.th. } \gamma(u)\leq u\} \\
&=\bigwedge\{f\ce(\bigwedge \im(\gamma)) \mid \gamma: \up x\cap O\to A \text{ s.th. } \gamma(u)\leq u\} \\
&=\bigwedge\{f\ced(\bigwedge \im(\gamma)) \mid \gamma: \up x\cap O\to A \text{ s.th. } \gamma(u)\leq u\} \\
& \geq \bigwedge\{f\ced(p)\mid x\geq p\in K\}=f\ced(x)
\end{align*}
where the first step uses general distributivity (Theorem \ref{ch2:thm:GD}) and the second step follows from the anti-preservation property of $f\ce$. The second to last step comes from $\bigwedge\im(\gamma)\in K$ and $f\ce=f\ced$ on $K\cup O$. The last step of the derivation follows from the fact that each $\bigwedge\im(\gamma)\leq x$ since each $x$ is the meet of all the opens $u$ above it and each $\gamma$ picks elements of $A$ below each $u$.
\end{proof}

\section{Algebraic Propeties of the Canonical Extension}\label{ch2:sec:algprop}
In this section we will examine how the canonical extension construction interacts with the algebraic properties of maps. Assume a map $f: A^n\to A$ satisfies a certain algebraic property, for example meet-preservation in its $i^{th}$ argument, does $f\ce:(A\ce)^n\to A\ce$ satisfy a similar property? We will examine several such properties and see that the canonical extension of a map often satisfies a `complete version' of the algebraic properties of the map it extends. We will show how this allows for the integration of the notion of canonical extension of DLEs and BAEs with their abstract presentation as $L$-algebras (see Section \ref{ch1:sec:boolstruct}). Finally, we will examine how canonical extensions interact with function composition. 

\subsection{Properties preserved and strengthened by canonical extension}

We start with an algebraic property which is not often considered but which will be crucial for the applications at the end of this Chapter. The following concepts are due to Henkin in \cite{1970:Henkin}.

\begin{definition}
Let $\ba$ be an object in $\A$ with underlying set $A$ and let $f:A\to A$. We will say that $f$ is \textbf{$k$-additive}\index{$k$-additive} if for any finite set $U\subseteq A$
\[
f(\bigvee U)=\bigvee \{f(\bigvee V)\mid V\in \pow_k(U)\} 
\]
where $\pow_k(U)$ is the set of subsets of $U$ with at most $k$ elements. A map $f:A\to A$ will be called \textbf{completely $k$-additive}\index{$k$-additive!Completely} if for any set $U\subseteq A$, the join $\bigvee \{f(\bigvee V)\mid V\in \pow_k(U)\}$ exists and equals $f(\bigvee U)$. We will say that a map is (completely) $\omega$-additive if it is (completely) $k$-additive for some $k<\omega$.
\end{definition}
The following two remarks illustrate the importance of the concept of $k$-additivity:
\begin{enumerate}
\item $1$-additivity just means additivity, i.e. distribution over joins.
\item A completely $k$-additive map on a complete lattice preserves up-directed joins: indeed if $U\subseteq A$ is up-directed, then for any $V\in \pow_k(U)$, $\bigvee V\in U$ and thus 
\[
\bigvee \{f(\bigvee V)\mid V\in \pow_k(U)\}=\bigvee f[U]=f(\bigvee U)
\]
\end{enumerate}

It is natural to generalize the concept of $k$-additivity as follows.
\begin{definition}
Let $\ba$ be an object in $\A$ with underlying set $A$ and let $f:A\to A$. We will say that $f$ is \textbf{$k$-multiplicative}\index{k-multiplicative} if for any finite set $U\subseteq A$
\[
f(\bigwedge U)=\bigwedge \{f(\bigwedge V)\mid V\in \pow_k(U)\} 
\]
A map $f:A\to A$ will be called \textbf{completely $k$-multiplicative}\index{$k$-multiplicative!Completely} if for any set $U\subseteq A$, the meet $\bigwedge \{f(\bigwedge V)\mid V\in \pow_k(U)\}$ exists and equals $f(\bigwedge U)$. We will say that a map is (completely) $\omega$-multiplicative if it is (completely) $k$-multiplicative for some $k<\omega$. We will say that a map $f:A\to A$ is \textbf{anti-$k$-additive}\index{$k$-additive!Anti-} if 
\[
f(\bigvee U)=\bigwedge \{f(\bigwedge V)\mid V\in \pow_k(U)\} 
\]
and similarly that it is \textbf{anti-$k$-multiplicative}\index{$k$-multiplicative!Anti-} if 
\[
f(\bigwedge U)=\bigvee \{f(\bigvee V)\mid V\in \pow_k(U)\} 
\]
Complete anti-$k$-additivity and multiplicativity are defined analogously.
\end{definition}

The following is an easy consequence of the definitions.

\begin{lemma}\label{ch2:lem:kaddisotone}
For any $k>0$, $k$-additive and $k$-multiplicative operators are isotone. Anti-$k$-additive and anti-$k$-multiplicative operators are antitone.
\end{lemma}
\begin{proof}
Let $f: A\to A$ be a $k$-additive operator and let $a,b\in A$ s.th. $a\leq b$, i.e. $a\vee b=b$. If $k=1$ we obviously get $f(a\vee b)=f(b)=f(a)\vee f(b)$ whence $f(a)\leq f(b)$. If $k\geq 2$ we get $f(a\vee b)=f(b)=f(a)\vee f(b)\vee f(a\vee b)= f(a)\vee f(b)\vee f(b)=f(a)\vee f(b)$ whence $f(a)\leq f(b)$ also. The proof is similar for the other types of operators.
\end{proof}
The following theorem will be used heavily at the end of this chapter, and since we will only need it for unary operators, this is how we will present the result. 

\begin{theorem}\label{ch2:thm:complkadd}
Let $\ba$ be an object in $\A$ with underlying set $A$, and let $f:A\to A$ be a function.
\begin{enumerate}[(i)]
\item If $f$ is $k$-additive, then $f\ce:A\ce\to A\ce$ is completely $k$-additive. In particular, $f\ce$ preserves up-directed joins.
\item If $f$ is $k$-multiplicative, then $f\ced:A\ce\to A\ce$ is completely $k$-multiplicative. In particular, $f\ced$ preserves down-directed meets.
\item If $f$ is anti-$k$-additive, then $f\ced:A\ce\to A\ce$ is completely anti-$k$-additive. In particular, $f\ced$ anti-preserves up-directed joins.
\item If $f$ is anti-$k$-multiplicative, then $f\ce:A\ce\to A\ce$ is completely anti-$k$-multiplicative. In particular, $f\ce$ anti-preserves down-directed meets.
\end{enumerate}
\end{theorem}
\begin{proof}
We show (i), (ii) then follows by duality, (iii) is shown in a very similar way to (i) and (iv) follows from (iii) by duality. Let $f: A\to A$ be $k$-additive, and let us first show that $f\ce$ is completely $k$-additive on sets of closed elements. Let $P\subseteq K$, we need to show $f\ce(\bigvee P)=\bigvee\{f(\bigvee V)\mid V\in \pow_k(P)\}$. It follows immediately from Lemma \ref{ch2:lem:kaddisotone} that 
\[
\bigvee\{f\ce(\bigvee V)\mid V\in \pow_k(P)\}\leq f\ce(\bigvee P)
\]
So we need only show
\begin{equation}\label{ch2:thm:complkadd:eq1}
f\ce(\bigvee P)\leq \bigvee \{f\ce(\bigvee V)\mid V\in \pow_k(P)\}
\end{equation}
We will now rewrite this inequation in a $\bigvee\leq\bigwedge$ format. For the left hand-side we simply use the definition of canonical extension:
\[
f\ce(\bigvee P)=\bigvee \{f\ce(q)\mid \bigvee P\geq q\in K\}
\]
For the left-hand side, we use the definition of canonical extension, the fact that a finite join of closed sets is closed, and the complete distributivity property of Theorem \ref{ch2:thm:GD} to get:
\begin{align*}
&\bigvee\{f\ce(\bigvee V)\mid V\in \pow_k(P) \}\\
=&\bigvee\{\bigwedge \{f(a)\mid A\ni a\geq \bigvee V\}\mid V\in \pow_k(P)\}\}\\
=&\bigwedge\{\bigvee f[\im \gamma]\mid \gamma: \pow_k(P)\to A\text{ s.th. }\gamma(V)\geq \bigvee V\}
\end{align*}
Thus to show (\ref{ch2:thm:complkadd:eq1}), we need to show that for any choice of closed element $q\leq \bigvee P$, and any choice function $\gamma$ as above, $f\ce(q)\leq \bigvee f[\im\gamma]$. To see this, note first that by definition $q\leq\bigvee P\leq \bigvee \im\gamma $, since $\gamma$ picks for a set of at most $k$ elements of $P$ a clopen $a$ above their join. Then note that $\bigvee\im\gamma$ is open (as a join of elements of $A$) and that $q$ is assumed to be closed. So by compactness, there exists a finite subset $F$ of the domain of $\gamma$ (i.e. $\pow_k(P)$) such that 
\[
q\leq \bigvee \im(\gamma\restrict F)
\]
We now use a trick which we shall use again in Theorem \ref{ch2:thm:kaddSmooth} which is to make $\gamma$ `locally join preserving'. We can assume, modulo the addition of a finite number of elements to $F$, that $F=\pow_K(I_q)$ for a finite set $I_q\subseteq P$ such that $q\in I_q$. We now define the `locally join preserving' version of $\gamma$ as the map $\gamma_q: \pow_k(I_q)\to A$ defined on singletons as
\[
\gamma_q(\{p\})=\bigwedge \{\gamma(U)\mid p\in U\in \pow_k(I_q)\}
\]
and one the other subsets $U\subseteq I_q$ as
\[
\gamma_q(U)=\bigvee\{\gamma_q(p)\mid p\in U\}
\]
Since $I_q$ is finite, all meets and all joins are finite, and $\gamma_q$ does indeed take its values in $A$. For notational clarity we also define the map $\psi_q: I_q\to A, p\mapsto \gamma_q(\{p\})$. We now have
\begin{align*}
\bigvee f[\im\gamma_q]&=\bigvee \{f(\gamma_q(U))\mid U\in \pow_k(I_q)\}\\
&=\bigvee \{f(\bigvee\{\gamma_q(p)\mid p\in U\})\mid U\in \pow_k(I_q)\}\\
&=\bigvee \{f (\bigvee V)\mid V\in\pow_k(\im\psi_q)\}\\
&=f(\bigvee\im\psi_q)\\
&=f(\bigvee \{\gamma_q(\{p\})\mid p\in I_q\})\\
&=f(\bigvee \im \gamma_q)
\end{align*}
where we have used the $k$-additivity of $f$ at the fourth step. Moreover, since $q\leq \bigvee U\leq \gamma(U)$ for each $q\in U\in\pow_k(I_q)$ we have $q\leq\gamma_q(\{q\})\leq \bigvee\im\gamma_q$. We also have $\gamma_q(\{p\})\leq \gamma(U)$ for each $p\in U$ and thus $\gamma_q(U)\leq \gamma(U)$. It now follows that
\[
f\ce(q)\leq f\ce(\bigvee\im\gamma_q)=\bigvee f\ce[\im\gamma_q]\leq f\ce[\im\gamma\restrict\pow_k(I_q)]\leq \bigvee f\ce[\im\gamma]
\]

For the case of a general set $U\subseteq A\ce$, we will show that 
\[
f\ce(\bigvee U)\leq \bigvee\{f(\bigvee V)\mid V\in \pow_k(\overline{U})\}
\]
where $\overline{U}$ is the set of closed elements below an element of $U$. The result will then follow by the monotonicity result of Lemma \ref{ch2:lem:kaddisotone} and the definition of canonical extension. By definition we have
\[
f\ce(\bigvee U)= \bigvee\{f\ce(p)\mid \bigvee U\geq p\in K\}
\]
and any such $p$ has the property that $p\leq \bigvee\overline{U}$ be definition of $\overline{U}$. By using monotonicity and the fact that we have established the result for sets of closed elements we can now conclude that for any $p\leq \bigvee U$ we have
\begin{align*}
f\ce(p)&\leq f\ce(\bigvee\overline{U})\\
&\leq \bigvee\{ f\ce(\bigvee V)\mid V\in\pow_k(\overline{U})\}\\
&\leq \bigvee\{f\ce(\bigvee V)\mid V\in\pow_k(U)\}
\end{align*}
which concludes the proof.
\end{proof}

We now focus on meet and join preservation. We prove the following for binary monotone operators, but the results clearly generalises to arbitrary finite arities.

\begin{lemma}\label{ch2:lem:pres}
Let $\ba$ be an object in $\A$  with underlying set $A$ and let $f:A\times A\to A$ be monotone in both arguments.
\begin{enumerate}[(i)]
\item If $f$ preserves finite joins in one argument then $f\ce$ preserves non-empty joins in that argument
\item If $f$ anti-preserves finite meets in one argument then $f\ce$ anti-preserves arbitrary meets in that argument
\item If $f$ preserves finite meets in one argument then $f\ced$ preserves non-empty meets in that argument
\item If $f$ anti-preserves finite joins in one argument then $f\ced$ anti-preserves arbitrary joins in that argument
\end{enumerate}
\end{lemma}
\begin{proof}
We will assume without loss of generality that $f$ has the preservation property in the first argument. 
\noindent The proofs of (i) can be found e.g. \cite{2001:GehrkeHarding} and \cite{2006:VenemaAC} in the case where $f$ is isotone in both arguments, our proof of (ii) will show how to generalise this proof to maps which have both isotone and antitone arguments. Note also that the proof of \ref{ch2:thm:complkadd} also implies (i) (in fact it is an adaptation of the proof from \cite{2001:GehrkeHarding}).

\noindent Let us now show (ii) when $f$ is antitone in the first and isotone in the second argument. We start with showing $f\ce(\bigwedge U, p)\leq \bigvee \{f\ce(u,p)\mid u\in U\}$ where $U\subseteq O$ and $p\in K$ (since $\bigwedge U\leq u$ for all $u\in U$ and $f$ is antitone in the first argument $f(\bigwedge U, p) \geq f(u,p)$ for each $u\in U$ and the opposite inequality follows). Again by definition $$f\ce(\bigwedge U, p)=\bigvee\{f\ce(v,p)\mid  \bigwedge U\leq v\in O\}$$ We also have by generalised distributivity that:
\begin{align*}
& \bigvee \{f\ce(u,p)\mid u\in U\} \\
 = &\bigvee \{\bigwedge\{f(a,b) \mid u\geq a\in A, p\leq b\in A\} \mid u\in U\} \\
= & \bigwedge\{\bigvee\{f(a,b) \mid (a,b)\in\im(\gamma)\}\mid \gamma: U\to A\times (p\uparrow\cap A) \text{ s.th. } \pi_1(\gamma(u))\leq u \}
\end{align*}
So we need to show that for any $v\in O, \bigwedge U\leq v$ and any choice function $\gamma$ as defined above $f\ce(v,p)\leq \bigvee f[\im(\gamma)]$. For this note first that by definition of $O$
\begin{align*}
\bigwedge U & =\bigwedge \{\bigvee \{a\mid u\geq a\in A\} \mid u\in U\} \\
& = \bigvee \{\bigwedge \im(\delta)\mid \delta:U\to A\text{ s.th. } \delta(u)\leq u\}
\end{align*} 
Thus if $\bigwedge U\leq v$, then for any $\delta$ as above $\bigwedge \im(\delta)\leq v$, and since $\bigwedge(\im(\delta))$ is closed and $v$ is open, there must exist by compactness a finite set $V_\delta\subseteq_\omega A$ such that $\bigwedge V_\delta\leq v$. Notice now that for any of the functions $\gamma$ defined above, $\pi_1\circ \gamma$ is one of these maps $\delta$. Thus for any $\gamma$ we have a finite set $V_\gamma\subseteq \im(\pi_1\circ \gamma)$ such that $\bigwedge U\leq \bigwedge V_\gamma\leq v$. This finite set also determines a finite subset $B_\gamma\subseteq \im(\pi_2\circ \gamma)$ by simply taking for each element of $V_\gamma$ its $\pi_2$ correspondent. Each element of $B_\gamma$ is, by definition of $\gamma$, a clopen above $p$. But since $f$ anti-preserves finite meets and is antitone in the first argument we then have 
\begin{align*}
f\ce(v,p) & \leq f\ce(\bigwedge V_\gamma,p)\\
& =f\ce (\bigwedge V_\gamma, \bigwedge B_\gamma ) \\
& =f (\bigwedge V_\gamma, \bigwedge B_\gamma ) \\
& =\bigvee\{f(a,\bigwedge B_\gamma)\mid a\in V_\gamma\} \\
& \leq \bigvee \{f\ce(a,b)\mid (a,b)\in \im(\gamma)\}
\end{align*}
We now need to show that the inequality holds for an arbitrary set $X\subseteq A\ce$ and element $y\in A\ce$. Again, we proceed similarly to (i) by showing that $f\ce(\bigwedge X, y)\leq \bigvee \{f\ce(\tilde{x}, y)\mid \tilde{x}\in \tilde{X}\}$ where $\tilde{X}$ is the set of open elements above an element of $X$. By definition $$f\ce(\bigwedge X, y)=\bigvee\{f\ce(u,p)\mid \bigwedge X\leq u\in O, y\geq p\in K\}$$
Since for any such $(u,p)$ we have $\bigwedge \tilde{X}\leq u$
\begin{align*}
f\ce(u,p) & \leq  f\ce(\bigwedge\tilde{X}, p) \\
& = \bigvee \{f(\tilde{x},y)\mid \tilde{x}\in\tilde{X}\}
\end{align*}
which completes the proof of this case. The case for (ii) when $f$ is antitone in the second argument is easy to prove by combining the two cases we have already shown.

\vspace{10pt} \noindent The proof for (iii) and (iv) follow from (i) and (ii) by duality.
\end{proof}

\begin{corollary}\label{ch2:cor:pres}
Let $\ba$ be an object in $\A$ with underlying set $A$ and let $f:A\times A\to A$.
\begin{enumerate}[(i)]
\item If $f$ preserves finite joins in one argument then $f\ced$ preserves non-empty joins in that argument.
\item If $f$ preserves finite meets in one argument, then $f\ce$ preserves non-empty meets in that argument.
\item If $f$ anti-preserves finite meets in one argument then $f\ced$ anti-preserves non-empty meets in that argument.
\item If $f$ anti-preserves finite joins in one argument then $f\ce$ anti-preserves non-empty joins in that argument.
\end{enumerate}
\end{corollary}
\begin{proof}
For (i) we know by Lemma \ref{ch2:lem:pres} that if $f$ preserves finite joins, then $f\ce$ preserves non-empty-joins. Moreover, we know from Proposition \ref{ch2:prop:smooth} that if $f\ce$ preserves non-empty joins, then $f\ce=f\ced$ and thus $f\ced$ preserves non-empty joins, and dually for (ii).

\noindent Similarly for (iii), by  Lemma \ref{ch2:lem:pres} if $f$ anti-preserves finite meets, then $f\ce$ anti-preserves non-empty meets and by Corollary \ref{ch2:cor:smoothcor} this means that $f\ce=f\ced$, thus $f\ced$ anti-preserves non-empty meets, and dually for (iv).
\end{proof}

The list of algebraic properties which we may wish to prove to be preserved by canonical extension is clearly infinite, and the choice will greatly depend on the application, i.e. the logic, at hand. We can show some simple preservation results for frequent algebraic properties such as being a expanding, contracting, or idempotent.

\begin{proposition}\label{ch2:prop:expcont}
Let $\ba$ be an object in $\A$  with underlying set $A$ and let $f:A\to A$ be expanding, then $f\ce$ and $f\ced$ are expanding. Similarly, if $f$ is contracting, then so are $f\ce$ and $f\ced$.
\end{proposition}
\begin{proof} \textbf{(1) Expanding maps.} We start by showing the property on closed elements $p\in K$, i.e. we want to show that $$p\leq f\ce(p)=\bigwedge\{f(a)\mid p\leq a\in A\}$$
But recall that $p$ is closed precisely when $p=\bigwedge\{a\mid p\leq a\in A\}$, thus we need to show that $$\bigwedge\{a\mid p\leq a\in A\}\leq \bigwedge\{f(a)\mid p\leq a\in A\} $$ which is immediate from the assumption since for every term in the RHS meet there exist a smaller term on the LHS meet. Similarly, for a general element $x\in A\ce$ we have
\[
x=\bigvee\{p\mid x\geq p\in K\}\leq \bigvee\{f\ce(p)\mid x\geq p\in K\}=f\ce(x)
\]
The proof for $f\ced$ uses similar ideas: starting with elements $u\in O$ we get
\[
u=\bigvee\{a\mid u\geq a\in A\}\leq \bigvee\{f(a)\mid u\geq a\in A\}=f\ced(u)
\]
And for a general $x\in A\ce$ we thus have
\[
x=\bigwedge\{u\mid x\leq u\in O\}\leq \bigwedge\{f\ce(u)\mid x\leq u\in O\}=f\ced(x)
\]
\textbf{(2) Contracting maps.} For the contracting case, consider again $p\in K$, then by definition and by assumption it is easy to see that 
\[
f\ce(p)=\bigwedge\{f(a)\mid p\leq a\in A\}\leq \bigwedge\{a\mid p\leq a\in A\}=p
\]
And for a general $x\in A\ce$ we then have
\[
f\ce(x)=\bigvee\{f\ce(p)\mid x\geq p\in K\}\leq \bigvee\{p\mid x\geq p\in K\}=x
\]
The proof for $f\ced$ follows exactly the same pattern.
\end{proof}

\begin{proposition}\label{ch2:prop:idem}
Let $\ba$ be an object in $\A$ with underlying set $A$ and let $f:A\to A$ be isotone and idempotent, then $f\ce$ is idempotent.
\end{proposition}
\begin{proof}
Since $f$ is assumed to be monotone, we have by Lemma \ref{ch2:lem:ineq} that 
\[
f\ce=(ff)\ce\leq f\ce f\ce
\]
So we need only to show $f\ce f\ce\leq f\ce$. We start by showing that this property holds on closed elements $p\in K$. By definition and the fact that $f\ce$ sends closed elements to closed elements, we must show that for any $p\in K$
\[
\bigwedge\{f(a)\mid f\ce(p)\leq a\in A\}\leq \bigwedge\{f(b)\mid p\leq b\in A\}
\]
i.e. we must show that for any $b\in A$ such that $p\leq b$
\[
\bigwedge\{f(a)\mid f\ce(p)\leq a\in A\}\leq f(b)
\]
This can be shown by finding an $a\in A$ such that $f\ce(p)\leq a$ and $f(a)\leq f(b)$. We get such an $a$ by putting $a=f(b)$. Since $p\leq b$ we then indeed have $f\ce(p)\leq f\ce(b)=f(b)=a$ and by idempotency we have $f(a)=f(f(b))=f(b)$ and in particular $f(a)\leq f(b)$. Thus $f\ce f\ce=f\ce$ on $K$.

\noindent It is now immediate that for an arbitrary $x\in A\ce$ we have
$$(f\ce f\ce)(x)=\bigvee\{(f\ce f\ce)(p)\mid x\geq p\in K\}=\{\bigvee (f\ce)(p)\mid x\geq p\in K\}=f\ce(x)$$
\end{proof}

\noindent It is easy to check that an antitone idempotent map is constant, so this case need not be considered.

\subsection{Canonical extension and $L$-algebras.}\label{ch2:subsec:CanExtLAlg}

Let us briefly explore what the preservation results of this section mean in terms of the $L$-algebras in $\A$ described in Section \ref{ch1:subsec:LAlg}. As we have seen in Section \ref{ch1:sec:coalglog} and as we will see in Chapter 5, $L$-algebras of interest will often be defined by functors of the type
\[
L\ba=(\Free\polyFunc\Forg \ba)/\simeq
\]
where $\polyFunc$ is a polynomial $\Set$-functor and $\simeq$ an equivalence relation generated by some axioms. These axioms are typically used to enforce algebraic conditions on the distributive lattice or boolean algebra expansions. The functor $L:\BA\to\BA$ used to define classical modal logic for example is defined by 
\[
L\ba=\Free\{\dia a\mid a\in A\}/(\dia(a\vee b)=\dia a\vee \dia b, \dia \bot=\bot)
\]
The axioms ensure that the structure map of an $L$-algebra $L\ba\to \ba$ is equivalent to a map on $A$ which preserves binary joins and the bottom element. More generally, if $\simeq$ defines some preservation property, then we have the following adjunction:

\vspace{2ex}
\begin{tabular}{c}
$\A$-morphisms $\left(\Free\polyFunc\Forg \ba)/\simeq\right)\to \ba$\\
\hline\hline
$\Set$-morphisms $\polyFunc A\to A$ with the algebraic properties defined by $\simeq$
\end{tabular}
\vspace{2ex}

The preservation results developed in this section can now be understood in a new light. Let us consider a functor $L\ba=\Free(\Forg \ba)^n/\simeq$ where $\simeq$ describes some algebraic property of an $n$-ary expansion $f: (A)^n\to A$ (for example distribution over joins in its first argument). If this algebraic property is preserved under canonical extension, i.e. $f\ce: (A\ce)^n\to A\ce$ also satisfies this property, then by the general adjunction sketched above, the canonical extension construction can be lifted from $\A$ to $\Alg_{\A}(L)$, i.e. we have a map $f\ce: L(A\ce)\to A\ce$. Moreover, since $f\ce\restrict A^n=f$ we clearly have
\[
\xymatrix
{
L\ba\ar[r]^{L\eta_A}\ar[d]_{f} & L(\ba\ce)\ar[d]^{f\ce}\\
\ba\ar[r]_{\eta_A} & \ba\ce
}
\]
This is a version of the J\'{o}nsson-Tarski theorem to which we will return in Chapter 5. Note however that for a general varietor $L$, there is no reason for the canonical extension construction to lift from $\A$ to $\Alg_{\A}(L)$, but this section has only shown a set of situations, which occur very frequently in practise, in which such a lifting is possible. 

\subsection{Canonical extension and function composition}

We now move on to studying how the canonical extension procedure interacts with the composition of maps. Understanding this interaction will be crucial to understanding the canonicity of terms built by nesting modal operators, since this essentially amounts to functional composition. For clarity's sake we will denote the functional composition of $f,g: A\to A$ as $fg$ rather than the more conventional $f\circ g$. The following result will be crucial and dates back to \cite{Jonsson51} and \cite{Ribeiro52} in the case of purely isotone maps. We extend it to monotone functions.

\begin{lemma}\label{ch2:lem:ineq} 
Let $\ba$ be an object in $\A$ with underlying set $A$, $f_i: A^{n_i}\to A, 1\leq i\leq m$ be $n_i$-ary monotone maps and $g: A^m\to A$ be an $m$-ary monotone map, then
\[
\left(g(f_1,\ldots,f_m)\right)\ce\leq g\ce(f_1\ce,\ldots,f_m\ce)
\]
\end{lemma}
\begin{proof} 
We can assume w.l.o.g. that the arguments of the functions are gathered into isotone and antitone groups, starting with isotone arguments. For notational convenience, we show the case where all maps are binary with the first argument isotone and the second argument antitone, but the proof shown below is readily generalized to the arbitrary case. So let us consider $g, f_1, f_2: A^2\to A$ isotone in the first and antitone in the second argument. We start by showing the result for tuples of arguments $(p,u,v,q)\in K\times O\times O\times K$. By (\ref{ch2:eq:mapext12}) we have
\begin{equation}\label{ch2:lem:ineq:eq:f1}
f_1\ce(p,u)=\bigwedge\{f_1(a_1,a_2)\mid p\leq a_1\in A\ni a_2\leq u\}
\end{equation}
and by Proposition \ref{ch2:prop:monotonemap} and (\ref{ch2:eq:mapext22})  we know that
\begin{equation}\label{ch2:lem:ineq:eq:f2}
f_2\ce(v,q)=f_2\ced(v,q)=\bigvee\{f_2(b_1,b_2)\mid q\leq b_2\in A\ni b_1\leq v\}
\end{equation}
Note in particular that $f_1\ce(p,u)$ is thus closed, whilst $f_2\ce(v,q)$ is open. From this it follows that
\[
g\ce(f_1\ce(p,u),f_2\ce(v,q))=\bigwedge\{g(c_1,c_2)\mid f_1\ce(p,u)\leq c_1\in A\ni c_2\leq f_2\ce(v,q)\}
\]
Take any $c_1, c_2$ as above, i.e. such that
\[
\bigwedge\{f_1(a_1,a_2)\mid p\leq a_1\text{ and } a_2\leq u\}\leq c_1\hspace{2ex}c_2\leq \bigvee\{f_2(b_1,b_2)\mid q\leq b_2\text{ and } b_1\leq v\}
\] 
since $c_1$ (resp. $c_2$) is clopen, it is open (resp. closed). By compactness and the fact that the sets above are down-directed and up-directed respectively, there must exist $\tilde{a}_1, \tilde{a}_2$ and $\tilde{b}_1,\tilde{b}_2$ in $A$ such that $p\leq \tilde{a}_1, \tilde{a}_2\leq u$, $ \tilde{b}_1\leq v, q\leq \tilde{b}_2$ and
\[
f_1(\tilde{a}_1,\tilde{a}_2)\leq c_1\hspace{3ex}c_2\leq f_2(\tilde{b}_1,\tilde{b}_2)
\]
Since $g(f_1,f_2)$ is isotone in its first and last parameters and antitone in its second and third parameters, we get that
\begin{align*}
(g(f_1,f_2))\ce(p,u,v,q)& \leq (g(f_1,f_2))\ce(\tilde{a}_1,\tilde{a}_2,\tilde{b}_1,\tilde{b}_2)\\
&=(g(f_1,f_2))(\tilde{a}_1,\tilde{a}_2,\tilde{b}_1,\tilde{b}_2)\\
&\leq g(c_1,c_2)
\end{align*}
Since this holds for any $c_1,c_2$ such that $f_1\ce(p,u)\leq c_1$ and $c_2\leq f_2\ce(v,q)$, we must have
\begin{align*}
(g(f_1,f_2))\ce(p,u,v,q)&\leq \bigwedge \{g(c_1,c_2)\mid f_1\ce(p,u)\leq c_1\in A\ni c_2\leq f_2\ce(v,q) \}\\
&=g\ce(f_1\ce(p,u),f_2\ce(v,q))
\end{align*}
From this we recover the general case as follows. Let $x_1,x_2,y_1,y_2\in A\ce$, then we can re-write Eq. (\ref{ch2:eq:mapext12}) as
\begin{align}
&(g(f_1,f_2))\ce(x_1,x_2,y_1,y_2)=\nonumber \\
&\bigvee\{(g(f_1,f_2))\ce(p,u,v,q)\mid x_1\geq p\in K, x_2\leq u\in O, y_1\leq v\in O, y_2\geq q\in K\}\label{ch2:lem:ineq:eq:gf1f2}
\end{align}
and
\begin{align}
&g\ce(f_1\ce(x_1,x_2), f_2\ce(y_1,y_2))= \nonumber \\
&\bigvee\{g\ce(r,w)\mid f_1\ce(x_1,x_2)\geq r\in K, f_2\ce(y_1,y_2)\leq w\in O\}\label{ch2:lem:ineq:eq:gf1f2ce}
\end{align}
To prove the inequality we show that for every $(g(f_1,f_2))\ce(p,u,v,q)$ in the set of Eq. (\ref{ch2:lem:ineq:eq:gf1f2}), there exist a $g\ce(r,w)$ in the set of Eq. (\ref{ch2:lem:ineq:eq:gf1f2ce}) such that 
\[
(g(f_1,f_2))\ce(p,u,v,q)\leq g\ce(r,w)
\]
But this follows immediately from what we have shown above, we just need to take $r=f_1\ce(p,u)$ which is indeed closed as was shown earlier, and $w=f_2\ce(v,q)$ which is open. The result then follows from the special case we started off with.
\end{proof}

\begin{corollary}\label{ch2:cor:ineq} 
Let $\ba$ be an object in $\A$ with underlying set $A$, $f_i: A^{n_i}\to A, 1\leq i\leq m$ be $n_i$-ary monotone maps and $g: A^m\to A$ be an $m$-ary monotone map, then
\[
(g(f_1,\ldots, f_m)\ce \upharpoonright (K\cup O)^N= g\ce (f_1\ce,\ldots,f_m\ce)\upharpoonright (K\cup O)^N
\]
where $N=\sum_i n_i$
\end{corollary}
\begin{proof}
The proof can be found in \cite{2001:GehrkeHarding} for unary isotone maps. For general monotone maps, we get by duality with the preceding Lemma \ref{ch2:lem:ineq} that 
\[
g\ced (f_1\ced,\ldots,f_m\ced)\leq (g(f_1,\ldots,f_m))\ced
\]
By Proposition \ref{ch2:prop:monotonemap} we know that for every $1\leq i\leq m$ we have $f_i\ce\upharpoonright (K\cup O)^{n_i}=f_i\ced\upharpoonright (K\cup O)^{n_i}$, and from Proposition \ref{ch2:prop:anymap} we know that $g\ce\leq g\ced$, we therefore get :
\begin{align*}
(g(f_1,\ldots,g_m))\ce & \leq g\ce(f_1\ce,\ldots,f_m\ce)\\
& = g\ce(f_1\ced,\ldots,f_m\ced)\\
&\leq g\ced (f_1\ced,\ldots,f_m\ced)\\
&\leq (g(f_1,\ldots,f_m))\ced
\end{align*}
on $(K\cup O)^N$. By applying Proposition \ref{ch2:prop:monotonemap} one more time, we see that 
\[
(g(f_1,\ldots,g_m))\ce \upharpoonright (K\cup O)^N= (g(f_1,\ldots,f_m))\ced\upharpoonright (K\cup O)^N
\]
and as a consequence, all the terms in the above sequence of inequalities are equal on $(K\cup O)^N$, and the result follows.
\end{proof}

Ideally, we would like to strengthen the inequality which we have just proved in Lemma \ref{ch2:lem:ineq} to a full equality, i.e. we would like 
\[
g\ce(f_1\ce,\ldots,f_m\ce)=(g(f_1,\ldots,f_m))\ce
\]
As it turns out, several of the algebraic properties which we have come across guarantee this equality. For example, in the case of unary operators, if $g$ preserves joins, then $g\ce f\ce=(gf)\ce$ for any monotone $g$. If $f$ preserves meets, then $g\ce f\ce=(gf)\ce$ for any isotone $f$. These results can be proven directly by standard canonical extension algebra (see e.g. \cite{2001:GehrkeHarding} for several such results). However, there is a more compact and modular method for proving this type of results. This method was developed in \cite{2004:GehrkeJonsson} and \cite{2006:VenemaAC} and relies on some topology.

\section{Topology to the rescue}\label{ch2:sec:topol}

In this section we will make heavy use of the notation and presentation of \cite{2006:VenemaAC} and in particular of the \emph{principle of matching continuities}. 
\begin{propdef}
Let $\ba$ be an object in $\A$ with canonical extension $\mathbb{A}\ce$ and let us define the collections of subsets of $A\ce$. 
\begin{itemize}
\item[-] $\sigma\eup=\{\up p\mid p\in K\}$
\item[-] $\sigma\edown=\{\down u\mid u\in O\}$
\item[-] $\sigma=\{\up p\hspace{2pt}\cap\hspace{-2pt}\down u\mid K\ni p\leq u\in O\}$
\end{itemize}
then $\sigma\eup,\sigma\edown$ and $\sigma$ (together with the empty set) constitute topologies on $A\ce$ and $\sigma$ is the join of $\sigma\eup$ and $\sigma\edown$ in the lattice of topologies on $A\ce$.
\end{propdef} 
\begin{proof}
Let's first show that $\sigma\eup$ forms a topology. Clearly $A\ce\in \sigma\eup$ as $A\ce=\up\bot$ and $\bot\in K$ (since it is clopen). If $p_1,\ldots,p_n\in K$, then \begin{align*}
\bigcap_{i=1}^n (\up p_i)&=\{x\in A\ce\mid p_i\leq x, 1\leq i\leq n\}\\
&=\{x\in A\ce\mid \bigvee_{i=1}^n p_i\leq x\}=\up(\bigvee_{i=1}^n p_i)
\end{align*}
and $(\bigvee_{i=1}^n p_i)\in K$. Similarly, if $(p_i)_{i\in I}\subseteq K$, then
\begin{align*}
\bigcup_{i\in I} (\up p_i)&=\{x\in A\ce\mid \exists i\in I: p_i\leq x\}\\
&=\{x\in A\ce\mid \bigwedge_{i\in I} p_i\leq x\}=\up(\bigwedge_{i\in I} p_i)
\end{align*}
and by definition of $K$, $\bigwedge_{i\in I} p_i\in K$.

The proof that $\sigma\edown$ is a topology is dual to that of $\sigma\eup$. 

The join of $\sigma\eup$ and $\sigma\edown$ in the lattice of topologies over $A\ce$ is the topology generated by taking $\sigma\eup \cup \sigma\edown$ as a sub-base, i.e. it is the topology whose opens are unions of finite intersections of elements of  $\sigma\eup \cup \sigma\edown$. These finite intersections are of the form $\up p \hspace{2pt}\cap \hspace{-2pt}\down u$ for some $K\ni p\leq u\in O$, and it is easy to verify from the case of $\sigma\eup$ and $\sigma\edown$ that the set of all such finite intersections - i.e. $\sigma$ - is closed under arbitrary unions.
\end{proof}

The next three topologies are well known to domain theorists and are defined as follows. A \textbf{Scott open}\index{Scott topology} subset of $A\ce$ is defined as a subset $U\subseteq A\ce$ which has the properties that it is (1) an up-set, and (2) $U\cap D\neq\emptyset$ for any up-directed set $D$ such that $\bigvee D\in U$. The \textbf{Scott closed} sets corresponding to the Scott opens can be defined as the down-sets $C$ such that for any up-directed set $D\subseteq C$, $\bigvee D\in U$. The \textbf{Scott topology} on $A\ce$ is defined as the collection of all Scott open subsets of $A\ce$ and is denoted by $\gamma\eup$. By order duality, one can also define a collection $\gamma\edown$ of dual Scott opens, i.e. down-sets $U$ with the property that for any down-directed sets $D$ such that $\bigwedge D\in U$, $U\cap D\neq \emptyset$. Finally as in the previous case we can define $\gamma=\{U\cap V\mid U\in \gamma\eup, V\in\gamma\edown\}$. This set of topologies is related to the previous one in the lattice of topologies on $A\ce$ by:

\begin{proposition}Using the above notation, $\gamma\eup\subseteq\sigma\eup$, $\gamma\edown\subseteq\sigma\edown$, $\gamma\subseteq\sigma$
\end{proposition}
\begin{proof}
We show that $\gamma\eup \subseteq \sigma\eup$. Take an Scott-open set $U\in\gamma\eup$. By definition of the canonical extension operation, any $x\in U$ can be written as $x=\bigvee\{p\mid x\geq p\in K\}$, and since $\{p\mid x\geq p\in K\}$ is clearly up-directed, by definition of the Scott-open we must have $U\cap \{p\mid  x\geq p\in K\}\neq\emptyset$. This means that there exist at least one $p\in K$ such that $x\in \up p$. We therefore have
$$U\subseteq \bigcup\{\up p\mid p\in U\cap K\}$$
Conversely, if $x\in \bigcup\{\up p\mid p\in U\cap K\}$, there must exist a $p\in U\cap K$ such that $p\leq x$ and since $U$ is an up-set this means that $x\in U$. Thus $$U=\bigcup\{\up p\mid p\in U\cap K\}\in \sigma\eup$$
\end{proof}

Note that whilst the topologies $\sigma\eup,\sigma\edown$ and $\sigma$ only make sense for objects in $\A$ which are canonical extensions (or at least which have a dense sub-object), the topologies $\gamma\eup,\gamma\edown$ and $\gamma$ are legitimate topologies on any object in $\A$.

\begin{lemma}[\cite{2004:GehrkeJonsson,1980:gierzcompendium}]\label{ch2:lem:ScottTopolBasis}
In the canonical extension of a distributive lattice $A$, the Scott topology has a basis of sets $\uparrow j$ where $j\in J_\omega(A\ce)$, the set of finite joins of completely join irreducible elements.
\end{lemma}
\begin{proof}
Let $U\in\gamma\eup$ and $x\in U$. By Lemma \ref{ch2:lem:joinIrredPrimeFilters} we can write $x=\bigvee\{j\in J(A\ce)\mid p\leq x\}$ which is the join of an up-directed set. It follows immediately from the definition of $\gamma\eup$ that there exists $j\in J(A\ce)$ such that $\up p\subseteq U$ and $x\in \up j$. Taking finite joins of elements of $J(A\ce)$ allows the collection $\{\up j\mid j\in J_\omega(A\ce)\}$ to be a basis (since it is then closed under finite intersections).
\end{proof}

With these topologies in place we can talk about continuous maps from $A\ce$ to itself, but there are of course as many notions of continuity as there are choices of topologies for the domain and codomain. Formally, we will say that $f:A\ce\to A\ce$ is $(\tau_1,\tau_2)$-continuous if $f$ is a continuous map between the topological spaces $(A\ce,\tau_1)$ and $(A\ce,\tau_2)$. For $n$-ary maps $f: (A\ce)^n$, given a topology $\tau$ on $A\ce$, we will also denote by $\tau$ the $n$-fold product topology on $A\ce$, i.e. the topology defined by the subbase of elements $\pi_i\inv(U), U\in\tau, 1\leq i\leq n$. 


The following Theorem highlights the importance of the topological approach. It shows that the topologies can be used to characterise the $(.)\ce$ operation on maps. 

\begin{theorem}\label{ch2:thm:topolChar}
Let $\ba$ be an object in $\A$ with underlying set $A$ and let $f:A^n\to A$, be a monotone map which for notational convenience we assume to be isotone in its first $m$ arguments ($m\leq n$) and antitone in the subsequent $m'=n-m$ arguments. Then $f\ce: (A\ce)^n\to A\ce$ is the largest $((\sigma\eup)^m\times(\sigma\edown)^{m'},\gamma\eup)$-continuous extension of $f$. 

\noindent Moreover, for any $f:A^n\to A$ (i.e. not necessarily monotone), $f\ce$ is the largest $(\sigma^n,\gamma\eup)$-continuous extension of $f$. 
\end{theorem}
\begin{proof}
The proof is adapted from Proposition 7.14 of \cite{2006:VenemaAC}. To see that $f\ce$ is $((\sigma\eup)^m\times(\sigma\edown)^{m'},\gamma\eup)$-continuous, let $U$ be a Scott-open set and let $(x_1,\ldots,x_n)\in (A\ce)^n$ be such that $f\ce(x_1,\ldots,x_n)\in U$. By definition of $(\cdot)\ce$ we know that
\begin{align*}
f\ce(x_1,\ldots,x_n)=\bigvee\{f\ce(p_1,\ldots,p_m,u_1,\ldots,u_{m'})\mid & K\ni p_i\leq x_i, 1\leq i\leq m, \\
& x_{m+i}\leq u_i\in O, 1\leq i\leq m'\}
\end{align*}
It is easy to see that that this join is taken over an up-directed set of tuples in $K^m\times O^{m'}$ (by taking joins of closed sets and meets of open sets). By the definition of Scott opens, since $f\ce(x_1,\ldots,x_n)\in U$ there must exist an $n$-tuple $(p_1,\ldots,p_m,u_1,\ldots, u_{m'})\in K^m\times O^{m'}$ such that $p_i\leq x_i, 1\leq i\leq m$, $x_{m+i}\leq u_i, 1\leq i\leq m'$ and $f\ce(p_1,\ldots,p_m,u_1,\ldots,u_{m'})\in U$, and thus
\[
(x_1,\ldots,x_n)\in \bigcap_{1\leq i\leq m}\pi_i\inv(\up p_i)\cap \bigcap_{1\leq i\leq m'}\pi_{m+i}\inv(\down u_i)
\]
where $\pi_i: (A\ce)^n\to A\ce$ is the usual projection map. Thus $(x_1,\ldots,x_n)\in (\sigma\eup)^m\times(\sigma\edown)^{m'}$ by definition of the product topology.

Let us now show that $f\ce$ is the largest such extension. Assume that $g:(A\ce)^n\to A$ is another $((\sigma\eup)^m\times(\sigma\edown)^{m'},\gamma\eup)$-continuous extension of $f$. To show that $g(x)\leq f\ce(x)$ for every $x\in (A\ce)^n$, we will use Lemma \ref{ch2:lem:joinIrredPrimeFilters} and show that if $p$ is completely join irreducible element of $A\ce$ and $p\leq g(x)$, then $p\leq f\ce(x)$ too. Note that if $p\in J(A\ce)$ then $\up p\in\gamma\eup$. Indeed, if $D$ is a directed set and $\bigvee D\in \up p$, then $p\leq \bigvee D$, and thus $p\leq d$ for some $d\in D$ by Lemma \ref{ch2:lem:compJoinPrime}. It then follows that $g(x)\in \up p\in \gamma\eup$, and since $g$ is assumed to be $((\sigma\eup)^m\times(\sigma\edown)^{m'},\gamma\eup)$-continuous there must exist $(p_1,\ldots,p_m,u_1,\ldots,u_{m'})\in K^m\times O^{m'}$ such that $p_i\leq x_i, 1\leq i\leq m$, $x_{m+i}\leq u_i, 1\leq i\leq m'$ and $p\leq g(p_1,\ldots,p_m,u_1,\ldots,u_{m'})$. It then follows that for any $(a_1,\ldots,a_n)\in A^n$ such that $a_i\geq p_i, 1\leq i\leq m$ and $a_{m+i}\leq u_i, 1\leq i\leq m'$
\begin{align*}
p\leq g(p_1,\ldots,p_m,u_1,\ldots,u_{m'})&\leq g(a_1,\ldots a_n)=f(a_1,\ldots a_n)
\end{align*}
Since this holds for any such $n$-tuples of elements of $A$ it follows that
\begin{align*}
p&\leq \bigvee\{f(a_1,\ldots a_n)\mid a_i\geq p_i, 1\leq i\leq m, a_{m+i}\leq u_i, 1\leq i\leq m'\}\\
&=f\ce(p_1,\ldots,p_m,u_1,\ldots,u_{m'})\leq f\ce(x)
\end{align*}
as desired.

The proof for arbitrary maps is similar. If $f\ce(x_1,\ldots,x_n)\in U\in\gamma\eup$, then by Eq.(\ref{ch2:eq:mapext12}), 
\[
f\ce(x_1,\ldots,x_n)=\bigvee\{\bigwedge f[d,u]\mid K^n\ni d\leq x\leq u\in O^n\}
\]
and it is not hard to see that the join is over an up-directed set: if $\bigwedge f[d_1,u_1]$ and $\bigwedge f[d_2,u_2]$ are in this set, then $\bigwedge f[d_1\vee d_2,u_1\wedge u_2]$ also lies in this set and $\bigwedge f[d_1,u_1]\leq \bigwedge f[d_1\vee d_2,u_1\wedge u_2]$ and $\bigwedge f[d_2,u_2]\leq \bigwedge f[d_1\vee d_2,u_1\wedge u_2]$. Thus there exists an interval $[d,u]$ such that $d\leq x\leq u$ and $\bigwedge f[d,u]\in U$, and therefore $f[d,u]\subseteq U$ since $U$ is an up-set, and $(\sigma,\gamma\eup)$-continuity then follows from the definition of $\sigma$. The proof that $f\ce$ is the greatest such extension is identical to the monotone case by using meets of intervals rather than closed and open elements.
\end{proof}

By duality we then immediately get the following two corollaries.

\begin{corollary}\label{ch2:cor:topolCharDual}
Let $\ba$ be an object in $\A$ with underlying set $A$ and let $f:A^n\to A$, be a monotone map which for notational convenience we assume to be isotone in its first $m$ arguments ($m\leq n$) and antitone in the subsequent $m'=n-m$ arguments. Then $f\ced: (A\ce)^n\to A\ce$ is the smallest $(\sigma\edown)^m\times(\sigma\eup)^{m'},\gamma\edown)$-continuous extension of $f$.

\noindent Moreover, for any $f:A^n\to A$ (i.e. not necessarily monotone), $f\ced$ is the smallest $(\sigma^n,\gamma\edown)$-continuous extension of $f$. 
\end{corollary}

\begin{corollary}\label{ch2:cor:smoothUnique}Let $\ba$ be an object in $\A$ with underlying set $A$ and let $f:A^n\to A$. Then $f$ is smooth iff it has a unique $(\sigma,\gamma)$ extension.
\end{corollary}
\begin{proof}
If $f$ is smooth, then $f\ce=f\ced$ is both $(\sigma\eup,\gamma\eup)$- and $(\sigma\edown, \gamma\edown)$- continuous by Proposition \ref{ch2:thm:topolChar} and Corollary \ref{ch2:cor:topolCharDual} and thus $(\sigma,\gamma)$-continuous by definition of $\gamma$. Any other $(\sigma,\gamma)$ extension will be both smaller and greater, and thus equal to $f\ce$, proving unicity.

Conversely, if $f$ has a $(\sigma,\gamma)$-extension $g$, then by the previous results we know that $g\leq f\ce$ and $f\ced\leq g$. Moreover, by Proposition \ref{ch2:prop:anymap}, $f\ce\leq f\ced$, and smoothness follows.
\end{proof}


As the proof of Proposition \ref{ch2:thm:topolChar} shows there is no difficultly in dealing with $n$-ary maps, one just needs to consider $n$-tuples, pointwise operations, and use the product topology. For notational convenience, we will therefore proceed with unary maps only, with the understanding that every result in the remainder of this section can easily be extended to the $n$-ary case.

\begin{proposition}\label{ch2:prop:directedcont}
Let $\ba$ be an object in $\A$ with underlying set $A$, then $f:A\ce\to A\ce$ is:
\begin{enumerate}[(i)]
\item $(\gamma\eup,\gamma\eup)$-continuous iff it preserves up-directed joins, 
\item $(\gamma\edown,\gamma\edown)$-continuous iff it preserves down-directed meets, 
\item $(\gamma\edown,\gamma\eup)$-continuous iff it anti-preserves down-directed meets,
\item $(\gamma\eup,\gamma\edown)$-continuous  iff it anti-preserves  up-directed joins
\end{enumerate}
\end{proposition}
\begin{proof}
The proof of (i) is a standard domain theoretical result and (ii) is  the dual result. (iii) and (iv) follow the same lines and we will show (iii) as an illustration. Let us first assume that $f$ anti-preserves down-directed meets, and let $U$ be a Scott-open set in $\gamma\eup$. We need to show that $f\inv(U)\in \gamma\edown$. Note first that if $x\in f\inv(U)$, and $z\leq x$, since $f$ is antitone we have $f(z)\geq f(x)\in U$ and since $U$ is an up-set we have $f(z)\in U$, i.e. $z\in f\inv(U)$. So $f\inv(U)$ is a down set as required. Let us now take $D$ a down-directed set such that $\bigwedge D\in f\inv(U)$, we need to show that $D\cap f\inv(U)\neq \emptyset$. Since $f$ anti-preserves down-directed meets we have $f(\bigwedge D)=\bigvee f[D]\in U$ and since $f$ is antitone we have that $f[D]$ is up-directed. Indeed if $d_1, d_2\in f[D]$, then there exist $c_1,c_2\in D$ such that $f(c_1)=d_1$ and $f(c_2)=d_2$ and since $D$ is down-directed, there exist $r\in D, r\leq c_1$ and $r\leq c_2$ and thus $d_1\leq f(r), d_2\leq f(r)$ with $f(r)\in f[D]$. But if $f[D]$ is directed and $\bigvee f[D]\in U$, then $f[D]\cap U\neq \emptyset$, since $U$ is a Scott-open, and thus there exists $d\in D$ such that $f(d)\in U$, i.e. $D\cap f\inv(U)\neq\emptyset$ as required.

For the opposite direction, assume that $f$ is $(\gamma\edown, \gamma\eup)$-continuous and let $D$ be a down-directed set. We need to show that $f(\bigwedge D)=\bigvee f[D]$. Let us first show that $f$ is antitone. For any $x,y\in A\ce$ such that $x\leq y$, consider the set $\down f(x)$. It is straightforward to check that $\down f(x)$ is a Scott closed set and thus $f\inv(\down f(x))$ is a dual Scott closed set, and in particular it is an up-set, hence $y\in f\inv(\down f(x))$, i.e. $f(y)\leq f(x)$. By antitonicity we immediately get $$\bigvee f[D]\leq f(\bigwedge D)$$
Now, consider the Scott closed set $\down (\bigvee f[D])$. By continuity $f\inv(\down (\bigvee f[D]))$ is a dual Scott closed set and in particular it has the property that it must be closed under taking meets of down-directed subsets. Since for any $d\in D$. $f(d)\leq \bigvee f[D]$ we have $D\subseteq f\inv(\down (\bigvee f[D]))$, and since $D$ is assumed to be down-directed we must have $\bigwedge D\in  f\inv(\down (\bigvee f[D]))$, i.e. $$f(\bigwedge D)\leq \bigvee f[D]$$ which concludes the proof.
\end{proof}

The following two seemingly technical lemmas have in fact very important topological consequences which will be detailed in Proposition \ref{ch2:prop:cont}.

\begin{lemma}\label{ch2:lem:interp1} 
Let $\ba$ be an object in $\A$ with underlying set $A$ and let $f:A\to A$ preserve finite meets. For any $p\in K$ and $x\in A\ce$, if $p\leq f\ce(x)$, there exists $p'\in K$ such that $p'\leq x$ and $p\leq f\ce(p')\leq f\ce(x)$.
\end{lemma}
\begin{proof}
If $f$ preserves finite meets then by Lemma \ref{ch2:lem:pres} $f\ced$ preserves non-empty meets and by Proposition \ref{ch2:prop:smooth} $f$ is then smooth. Hence 
\[
p\leq f\ce(x)=f\ced(x)=\bigwedge\{f\ced(u)\mid x\leq u\in O\}\]
i.e. for any $u\geq x$ we must have $p\leq f\ced(u)$ and since \[
f\ced(u)=\bigvee\{f(a)\mid u\geq a\in A\}
\]
and $\{f(a)\mid u\geq a\in A\}$ is up-directed, there must exist by compactness an element $a_u\in A$ with $a_u\leq u$ such that $p\leq f(a_u)$. Consider now the closed element defined as the meet of all these elements of $A$, namely $p'=\bigwedge\{a_u\mid x\leq u, a_u\in A\}$. Since $x$ is the meet of all opens above it and since each $a_u$ lies below one of these opens we must have $p'\leq x$. Moreover, since $f\ce$ preserve meets we have that $f\ce(p')=\bigwedge\{f(a_u)\mid x\leq u, a_u\in A\}$ and thus $p\leq f\ce(p')$ by construction of the elements $a_u$.
\end{proof}

\begin{lemma}\label{ch2:lem:interp2} Let $\ba$ be an object in $\A$ with underlying set $A$ and let $f\ce:A\ce\to A\ce$ anti-preserve joins then for any $p\in K$ and $x\in A\ce$, if $p\leq f\ce(x)$, there exists $u\in O$ such that $u\geq x$ and $p\leq f\ce(u)\leq f\ce(x)$.
\end{lemma}
\begin{proof}
The proof is an antitone version of the preceding Lemma \ref{ch2:lem:interp1}. By Lemma \ref{ch2:lem:pres} $f\ced$ must anti-preserve non-empty joins and by Corollary \ref{ch2:cor:smoothcor} $f$ must be smooth. Hence
\[p\leq f\ce(x)=f\ced (x)=\bigwedge\{f\ced(q)\mid x\geq q\in K\}\]
i.e. for any $q\in K, q\leq x$ we must have $p\leq f\ced(q)$. By definition this means
\[p\leq \bigvee\{f(a)\mid q\leq a\in A\}\]
and again by compactness and the fact that $\{f(a)\mid q\leq a\in A\}$ is up-directed (by antitonicity of $f$) this means that there exist an element $a_q\in A$ such that $p\leq f(a_q)$. Consider the open element defined as the join of all these elements, i.e. $u=\bigvee\{a_q\mid q\in K, x\geq q\}$. Since $x$ is the join of all closed elements below it and since each $a_q$ lies above one such closed element $u\geq x$. It remains to check that $p\leq f\ce(u)$. By definition and our assumption on $f$ we have \[p\leq \bigwedge\{f\ce(a_q)\mid q\in K, x\geq q\}=f\ce(\bigvee\{a_q\mid q\in K, x\geq q\})=f\ce(u)\]
\end{proof}
Using the second set of topologies, we can restate and generalise the technical Lemmas \ref{ch2:lem:interp1} and \ref{ch2:lem:interp2} topologically as:

\begin{proposition}\label{ch2:prop:cont}
Let $\ba$ be an object in $\A$ with underlying set $A$, a map $f:A\ce\to A\ce$ is
\begin{enumerate}[(i)]
\item $(\sigma\eup,\sigma\eup)$-continuous if it preserves non-empty meets 
\item $(\sigma\edown,\sigma\edown)$-continuous if it preserves non-empty joins
\item $(\sigma\edown,\sigma\eup)$-continuous if it anti-preserves non-empty joins
\item $(\sigma\eup,\sigma\edown)$-continuous if it anti-preserves non-empty meets
\end{enumerate}
\end{proposition}
\begin{proof}
We will show (i) and (iii), (ii) and (iv) follow by duality. For (i) we need to show that for any $p\in K, f\inv(\up p)\in\sigma\eup$. Take $x\in f\inv(\up p)$, by lemma \ref{ch2:lem:interp1} we know that there exists $p'(x)\in K$ such that $p'(x)\leq x$ and $p\leq f(p'(x))$. We now define the closed element $q=\bigwedge\{p'(x)\mid x\in f\inv(\up p) \}$ and claim that $\up q=f\inv(\up p)$. Since $f$ preserves non-empty meets we have $f(q)=\bigwedge \{f(p'(x))\mid x\in f\inv(\up p)\}$ and $p\leq f(q)$, thus $$\up q\subseteq  f\inv(\up p)$$
Conversely, since for any $x\in  f\inv(\up p)$ there exist $p'(x)\leq x$ we have by definition of $q$ that $x\in \up q$ and thus
$$f\inv(\up p)\subseteq \up q$$

For (iii) we proceed in a similar way: take $x\in f\inv(\up p)$, by lemma \ref{ch2:lem:interp2} we know that there exists $u(x)\in O$ such that $x\leq u(x)$ and $p\leq f(u)$. Define an open element $v=\bigvee\{u(x)\mid x\in f\inv(\up p) \}$, then by assumption on $f$ we have $f(v)=\bigwedge\{f(u(x))\mid x\in f\inv(\up p)\}$ and thus $\down v\subseteq f\inv(\up p)$ and conversely if $x\in f\inv(\up p)$, there exists $x\leq u(x)$ and $x\leq v$ and thus $f\inv(\up p) \subseteq \down v$.
\end{proof}

Since preservation of all non-empty joins in particular implies that of directed ones, it is clear that join preserving maps are also $(\gamma\eup,\gamma\eup)$-continuous, and similarly for the other combinations of Propositions \ref{ch2:prop:directedcont} and \ref{ch2:prop:cont}.

We can also use the topological formalism to show an important property of (anti-)$k$-additive and (anti-)$k$-multiplicative maps\index{$k$-additive,$k$-multiplicative}.
\begin{theorem}\label{ch2:thm:kaddSmooth}
Let $\ba$ be an object in $\A$  with underlying set $A$ and if $f:A\to A$ is $k$-additive, $k$-multiplicative, anti-$k$-additive, or anti-$k$-multiplicative, then its extension is $(\sigma,\gamma)$-continuous and thus smooth.
\end{theorem}
\begin{proof}
We show the result for $k$-additive maps, the other cases are treated similarly/dually. Recall first that by Theorem \ref{ch2:thm:complkadd} if $f$ is $k$-additive, then $f\ce$ is completely $k$-additive, and in particular it preserves up-directed joins. It follows from Theorem \ref{ch2:prop:directedcont} that it is $(\gamma\eup,\gamma\eup)$-continuous. We now show that it is $(\sigma\edown,\gamma\edown)$-continuous too. Let $\Gamma\in\gamma\edown$, i.e. $\Gamma$ is a down-set such that if a down directed set $D$ is such that $\bigwedge D\in \Gamma$, then $\Gamma\cap D\neq\emptyset$. Assume now that $f\ce(x)\in \Gamma$ for some $x$, we will show that there exist $x\leq u\in O$ such that $f\ce(u)\in \Gamma$. From Lemma \ref{ch2:lem:kaddisotone}, we know that $f$ is isotone and by Theorem \ref{ch2:thm:complkadd} we know that it is completely $k$-additive and we thus have:
\begin{align*}
f\ce(x)&=f\ce\left(\bigvee K\cap\hspace{-1ex}\down x \right)\\
&\stackrel{(1)}{=}\bigvee\{f\ce(\bigvee U)\mid U\in\pow_k(K\cap\hspace{-1ex}\down x)\}\\
&\stackrel{(2)}{=}\bigvee\{\bigwedge \{f\ce(a)\mid \bigvee U\leq a\in A\}\mid U\in\pow_k(K\cap\hspace{-1ex}\down x)\}\\
&\stackrel{(3)}{=}\bigwedge\{\bigvee f\ce[\im\phi]\mid \phi: \pow_k(K\cap\hspace{-1ex}\down x)\to A, \bigvee U\leq\phi(U)\}
\end{align*}
where $(1)$ follows by definition of complete $k$-additivity, $(2)$ follows from the fact that each $\bigvee U$ being a finite join of closed elements is closed, and $(3)$ follows from the complete distributivity property of Theorem \ref{ch2:thm:GD}. It is easy to see by taking pointwise meets of choice maps that $\{\bigvee f\ce[\im\phi]\mid \phi: \pow_k(K\cap\hspace{-1ex}\down x)\to A, \bigvee U\leq\phi(U\}$ is down-directed. Since $\Gamma\in\gamma\edown$, there must exist $\phi_0: \pow_k(K\cap\hspace{-1ex}\down x)\to A, \bigvee U\leq\phi_0(U)$ such that $\bigvee f\ce[\im\phi_0]\in\Gamma$. Recall that $f\ce(x)=\bigvee \{f\ce(p)\mid x\geq p\in K\}$, and for the equality (3) above to hold, it must be the case that  for every closed element $p\leq x$ we must have
\[
f\ce(p)\leq \bigvee f\ce[\im\phi]
\]
for any choice function $\phi$, and a fortiori for $\phi_0$. So let $p\leq x$ be closed, then $f\ce(p)$ is closed as well, and by compactness there must exist a finite subset $F\subseteq_{\omega}\pow_k(K\cap \hspace{-1ex}\down x)$ such that
\[
f\ce(p)\leq \bigvee f\ce[\im(\phi_0\restrict F)]
\]
Modulo the addition of a finite number of elements to $F$ we can assume that $F=\pow_k(I_p)$ for $I_p$ a finite subset of $K\cap\hspace{-1ex}\down x$ such that $p\in I_p$. We now define a map $\phi_p: \pow_k(I_p)\to A$ by first defining it on singletons $\{q\}, q\in I_p$
\[
\phi_p(\{q\})=\bigwedge\{\phi_0(U)\mid q\in U\in F\}
\]
which is an element of $A$ since it is a finite meet of elements of $A$. We then extend $\phi_p$ to the whole of $\pow_k(I_p)$ by
\[
\phi_p(U)=\bigvee\{\phi_p(\{q\})\mid q\in U\}
\]
It is clear that $\phi_p$ preserves all joins of at most $k$-singletons by construction. It follows that if we define $\psi_p: I_p\to A$ by $\psi(q)=\phi_p\{q\})$ for each $q\in I_p$, we have
\begin{align*}
\bigvee f\ce[\im\phi_p]&=\bigvee \{f\ce(\phi_p(U))\mid U\in \pow_k(I_p)\}\\
&=\bigvee \{f\ce(\bigvee\{\phi_p(\{q\})\mid q\in U\})\mid U\in \pow_k(I_p)\}\\
&=\bigvee\{f\ce(\bigvee V)\mid V\in \pow_k(\im\psi_p)\}\\
&=f\ce(\bigvee \im\psi_p)\\
&=f\ce(\bigvee \{\phi_p(\{q\})\mid q\in I_p\})\\
&=f\ce(\bigvee \im\phi_p)
\end{align*}
where we have used the $k$-additivity of $f\ce$ at the fourth step. Moreover, given $U\in \pow_k(I_p)$ we have by construction that $\phi_p(q)\leq \phi(U)$ for every $q\in U$ and thus $\phi_p(U)\leq \phi_0(U)$. In particular
\[
f\ce(\bigvee \im\phi_p)=\bigvee f\ce[\im\phi_p]\leq \bigvee f\ce(\im\phi_0\restrict F)\leq \bigvee f\ce(\im\phi_0)
\]
and thus $\bigvee f\ce(\im\phi_p)\in \Gamma$. Note also that since $q\leq \bigvee U\leq \phi_0(U)$ whenever $q\in U$ we have $q\leq\phi_p(U)$ whenever $q\in U$ and in particular since we're assuming $p\in I_p$ we have
\[
p\leq \bigvee \im\phi_p
\]
and $\im\phi_p$ is an element of $A$. Let us write $a_p=\im\phi_p$. For every $p\leq x$ we can thus find a clopen element $a_p$ such that $p\leq a_p$ and $f\ce(a_p)\in \Gamma$. We claim that if we define $u=\bigvee_p a_p$, then $u$ is the open element we are looking for, i.e. $x\leq u$ and $f\ce(u)\in\Gamma$. The fact that $x\leq u$ is easy: $x$ is the join of all closed elements $p$ below it, and for each such closed element $p\leq a_p$. Now let us show that $f\ce(u)\in\Gamma$. By complete $k$-additivity we have
\[
f\ce(u)=f\ce(\bigvee_p a_p)=\bigvee \{f\ce(\bigvee U)\mid U\in \pow_k\{a_p\mid p\in K\cap\hspace{-1ex}\down x\}\}
\]
The result will follow immediately if we can show that each $f\ce(\bigvee U)\leq \bigvee f\ce[\im \phi]$, since $\bigvee f\ce[\im \phi]\in\Gamma$. So let us consider $U\in \pow_k\{a_p\mid p\in K\cap\hspace{-1ex}\down x\}$, by definition it is of the form $\bigvee_i \bigvee\im\phi_{p_i}$ where each $\phi_{p_i}$ is constructed as above. Based on these maps we define $I_U=\bigcup_i I_{p_i}$ and $\phi_U: \pow_k(I_U)\to A$ in the same way as we defined $\phi_{p_i}$: we start on singleton sets
\[
\phi_U(\{q\})=\bigvee\{\phi_{p_i}(q)\mid q\in I_{p_i}\}
\]
and then on the entire domain:
\[
\phi_U(V)=\bigvee\{\phi_U(\{q\})\mid q\in V\}
\]
Note that since all joins are finite, the map $\phi_U$ does indeed take its values in $A$. We now proceed as above: we define $\psi_U: I_U\to A, q\mapsto \phi_U(\{q\})$ and compute
\begin{align*}
\bigvee f\ce[\im\phi_U]&=\bigvee \{f\ce(\phi_U(V))\mid V\in \pow_k(I_U)\}\\
&=\bigvee \{f\ce(\bigvee\{\phi_U(\{q\})\mid q\in U\})\mid U\in \pow_k(I_U)\}\\
&=\bigvee\{f\ce(\bigvee V)\mid V\in \pow_k(\im\psi_U)\}\\
&=f\ce(\bigvee \im\psi_U)\\
&=f\ce(\bigvee \{\phi_U(\{q\})\mid q\in I_U\})\\
&=f\ce(\bigvee \im\phi_U)
\end{align*}
For an given $q\in I_{p_i}$, we already know that $\phi_{p_i}(q)\leq \phi_0(q)$, and it follows that $\phi_U(q)\leq \phi_0(p)$ too. Similarly, given $V\in \pow_k(I_U)$, each $\phi_{p_i}(q)\leq \phi_0(V)$ if $q\in V$ and $q\in I_{p_i}$, and thus $\phi_U(q)\leq \phi_0(V)$, whence $\phi_U(V)\leq \phi_0(V)$. It follows that
\[
f\ce(\bigvee \im\phi_U)=\bigvee f\ce[\im\phi_U]\leq \bigvee f\ce(\im\phi_0\restrict \pow_k(I_U))\leq \bigvee f\ce(\im\phi_0)
\]
which concludes the proof.
\end{proof}


The following Theorem shows why these topological results are useful to study the interaction of canonical extension and functional composition. The result is essentially Theorem 2.30. of \cite{2004:GehrkeJonsson}.

\begin{theorem}[Principle of matching topologies]\label{ch2:thm:MatchingTopologies}
Let $\ba$ be an object in $\A$ with underlying set $A$, and $f: A^n\to A$ and $g_i:A^{m_i}\to A, 1\leq i\leq n$ be arbitrary maps, then
\begin{enumerate}[(i)]
\item If there exist topologies $\tau_i$ on $A$, $i\leq i\leq n$ such that each $g_i\ce$ is $(\sigma^{m_i},\tau_i)$-continuous and $f\ce$ is $(\tau_1\times\ldots\times\tau_n,\gamma\eup)$-continuous, then
\[
f\ce(g_1\ce,\ldots,g_n\ce)\leq (f(g_1,\ldots,g_n))\ce
\]
\item If there exist topologies $\tau_i$ on $A$, $i\leq i\leq n$ such that each $g_i\ce$ is $(\sigma^{m_i},\tau_i)$-continuous and $f\ce$ is $(\tau_1\times\ldots\times\tau_n,\gamma\edown)$-continuous, then
\[
f\ce(g_1\ce,\ldots,g_n\ce)\geq (f(g_1,\ldots,g_n))\ce
\]
\item If there exist topologies $\tau_i$ on $A$, $i\leq i\leq n$ such that each $g_i\ce$ is $(\sigma^{m_i},\tau_i)$-continuous and $f\ce$ is $(\tau_1\times\ldots\times\tau_n,\gamma)$-continuous, then
\[
f\ce(g_1\ce,\ldots,g_n\ce)=(f(g_1,\ldots,g_n))\ce
\]
\end{enumerate}
\end{theorem}
\begin{proof}
For (i) we use Theorem \ref{ch2:thm:topolChar}: the existence of the matching set of topologies $\tau_1,\ldots,\tau_n$ means that $f\ce(g_1\ce,\ldots,g_n\ce)$ is $(\sigma^{m_1}\times\ldots\times \sigma^{m_n}, \gamma\eup)$-continuous, and since $(f(g_1,\ldots,g_n))\ce$ is the largest such map the result follows. For (ii), we use Corollary \ref{ch2:cor:topolCharDual} and the existence of the set of matching topologies to conclude that $f\ce(g_1\ce,\ldots,g_n\ce)\geq (f(g_1,\ldots,g_n))\ced$ and the result then follows from Theorem \ref{ch2:prop:anymap}. Claim (iii) is simply the conjunction of (i) and (ii).
\end{proof}

This \emph{principle of matching topologies} will be pivotal in our understanding of canonicity in the next sections.

\section{Canonicity}\label{ch2:sec:can}
We are now ready to define and study canonicity algebraically. Recall from Eq. (\ref{ch1:eq:free}) that given a signature $\Sigma$ we can define the free $\mathsf{S}_\Sigma$-algebra over $\mathsf{F}(V)$, i.e. $\mathsf{G}\circ \Free(V)$. This $\Sigma$-AE will be denoted henceforth by $\mathsf{T}(V)$ and is the set of terms in the language of $\Sigma$-AEs modulo equivalence in $\A$. From Propositions \ref{ch1:prop:catequ}, \ref{ch1:prop:adj} and the universal property of free objects, it is clear that for every $\mathcal{A}=(\mathbb{A}, (f_s)_{s\in\Sigma})$ in $\AEs(\Sigma)$ and every valuation $v:V\to A$, there exist a unique interpretation map $\lsem \cdot\rsem_v\eA: \mathsf{T}(V)\to \mathcal{A}$ such that the following triangle commutes (in $\Set$):

\[
\xymatrix@R=10pt@C=36pt
{
& \mathsf{T}(V) \ar@{-->}[dd]^{\lsem \cdot\rsem_v\eA} \\
V \ar[ur]^{\eta_V} \ar[dr]_\pi & \\
& \mathcal{A}
}
\]

An equation between two terms $s,t\in \mathsf{T}(V)$ will be called \textbf{canonical}\index{Canonical!equation} if $\lsem s\rsem_v\eA=\lsem t\rsem _v\eA$ for any valuation $v: V\to A$ implies that $\lsem s\rsem_v\eAs=\lsem t\rsem _v\eAs$ for any valuation $v: V\to A\ce$. Using the notation introduced in Section \ref{ch1:subsec:algsem}, $s=t$ is canonical if
\[
\mathcal{A}\models s=t  \hspace{2ex}\Rightarrow\hspace{2ex} \mathcal{A}\ce\models s=t
\]
Recall from the previous Chapter that equations between terms of the free algebra define varieties. From this perspective, an equation is canonical if the variety it defines is closed under taking canonical extensions. Similarly, we will call an inequation $s\leq t$ canonical \index{Canonical!inequation} if 
\[
\mathcal{A}\models s\leq t  \hspace{2ex}\Rightarrow\hspace{2ex} \mathcal{A}\ce\models s\leq t
\] 

In order to be able to say anything about the canonicity of equations, we somehow need to compare the interpretations in $\mathcal{A}$ with interpretations in $\mathcal{A}\ce$. It is natural to try to use the map $(\cdot)\ce$ to mediate between these interpretations, but $(\cdot)\ce$ is defined on maps, not on terms. Moreover, not every valuation on $\mathcal{A}\ce$ originates from valuation on $\mathcal{A}$. We would therefore like to recast the problem in such a way that (1) terms are viewed as maps, and (2) we do not need to worry about valuations. 
 
We will therefore adopt the language of \emph{term functions}\index{Term function}, i.e. we will view each term $t\in\mathsf{T}(V)$ built using $n$ variables of $V$ as defining for each AE $\mathcal{A}$, a map $t\eA: A^n\to A$. This will allow us to consider its canonical extension $(t\eA)\ce$, and it will also allow us to reason without having to worry about specifying valuations. Formally, given a signature $\Sigma$ and a set $V$ a propositional variables, we inductively define the term function associated with an element $t\in\mathsf{T}(V)$ built from variables $x_1,\ldots,x_n\in V$ as follows:
\begin{itemize}
\item $x_i\eA=\pi_i^n:A^n\to A, 1\leq i\leq n$
\item $(f(t_1,\ldots, t_m))\eA=f\eA\circ \langle t_1\eA, \ldots, t_m\eA\rangle$
\end{itemize}
where $\pi_i$ is the usual projection on the $i^{th}$ component, $f\eA$ is the interpretation of the symbol $f$ in $\mathcal{A}$ and $\langle t_1\eA,\ldots,t_m\eA\rangle$ is usual the product of $m$ maps. Note that in this definition we work in $\Set$, and the building blocks of term functions are thus the variables in $V$ (interpreted as projections) and all operation symbols, including boolean operations. However, we could equally well work directly in $\A$ in which case we would consider $\Free V$ as the set of variables and only interpret the symbols defined by the signature $\Sigma$. 

We now show how term functions allow us to reason without explicit reference to valuations. Note first that when we enforce an equation $s=t$, the set of variables $\mathrm{Var}(s)$ and $\mathrm{Var}(t)$ of $s$ and $t$ need not be equal. However for the equation $s\eA=t\eA$ to be well-typed, the term functions must have the same domain, i.e. be built on the same number of variables. This presents no problem, since we can always consider a term on $m$ variables to be a term on $n>m$ variables with $n-m$ `silent' variables which are simply ignored. We can thus assume without loss of generality that $s\eA=t\eA$ is well-typed by simply adding silent variables where needed. In fact we can assume that $\mathrm{Var}(s)=\mathrm{Var}(t)$ by using this procedure.

\begin{proposition}\label{ch2:prop:termfct}
Let $s,t\in\mathsf{T}(V)$ and $\mathcal{A}$ be an object in $\AEs(\Sigma)$ then
\[
\mathcal{A}\models s=t\text{ iff }s\eA=t\eA
\]
\end{proposition}
\begin{proof}
We show that if $\mathrm{Var}(t)=\{x_1,\ldots x_n\}\subseteq V$, and $v: V\to \mathcal{A}$ is a valuation 
\[
\lsem t\rsem_v\eA=t\eA(v(x_1), \ldots, v(x_n))
\]
The result then follows: since we can assume $\mathrm{Var}(s)=\mathrm{Var}(t)$ whenever $s\eA=t\eA$, then for any valuation
\[
\lsem s\rsem_v\eA=s\eA(v(x_1), \ldots, v(x_n))=t\eA(v(x_1), \ldots, v(x_n))=\lsem t\rsem_v\eA
\]
and thus $\mathcal{A}\models s=t$. Conversely, to show $s\eA=t\eA$ we need to show that $s\eA(a_1,\ldots,a_n)=t\eA(a_1,\ldots,a_n)$ for any $(a_1,\ldots,a_n)\in A^n$. But any such choice can be used to build a valuation $v: V\to A$ such that $v(x_i)=a_i, 1\leq i\leq n$. If $\mathcal{A}\models s=t$, then in particular 
\begin{align*}
s\eA(a_1,\ldots,a_n)& =s\eA(v(x_1), \ldots, v(x_n))\\
&=\lsem s\rsem_v\eA=\lsem t\rsem_v\eA\\
&=t\eA(v(x_1), \ldots, v(x_n))\\
&=t\eA(a_1,\ldots,a_n)
\end{align*}
To show that $\lsem t\rsem_v\eA=t\eA(v(x_1), \ldots, v(x_n))$ we proceed by induction on the complexity of $t$. If $t=x$ is just a propositional variable, then 
\[
\lsem t\rsem_v=v(x)=\pi_1^1(v(x))=x\eA(v(x))=t\eA(v(x))
\]
If $t=f(t_1,\ldots,t_n)$ where each $t_i$ is $m_i$-ary and such that $\bigcup_i \mathrm{Var}(t_i)=\mathrm{Var}(t)$, then
\begin{align*}
\lsem t\rsem_v\eA &=f\eA(\lsem t_1\rsem_v\eA,\ldots,\lsem t_n\rsem_v\eA)\\
&=f\eA(t_1\eA(v(x_1^1),\ldots,v(x_{m_1}^1)),\ldots,t_n\eA(v(x_1^n),\ldots,v(x_{m_n}^n)))\\
&=t\eA(v(x_1),\ldots,v(x_n))
\end{align*}
\end{proof}
A completely analogous result holds for inequations.

We now consider the canonical extensions $(t\eA)\ce$ of the term functions $(t)\eA$. For example, for a signature containing a single unary operator $\dia$, the term $t=\neg\dia p\in \mathsf{T}(V)$ defines the maps $\lambda p.(\neg\dia p)\eA: A\to A$ and $\lambda p.(\neg\dia p)\eAs: A\ce\to A\ce$. Note that unlike \cite{Jonsson94}, the terms $t\in\mathsf{T}(V)$ need not define an isotone map since the definition of canonical extension of Eq. (\ref{ch2:eq:mapext12}) is suitable for any map. Note also that the idea of interpreting all propositional variables as the identity function makes sense from the point of view of quantification over valuations: a propositional variable can be mapped to any element of $A$ and in this sense propositional variables are indistinguishable.

Following \cite{Jonsson94} we say for any $t\in\mathsf{T}(V)$ that:
\begin{itemize}
\item[-] $t$ is \textbf{stable} if $(t\eA)\ce=t\eAs$
\item[-] $t$ is \textbf{expanding} if $(t\eA)\ce\leq t\eAs$
\item[-] $t$ is \textbf{contracting} if $(t\eA)\ce\geq t\eAs$
\end{itemize}
for any $\mathcal{A}$ (and valuation $\pi:V\to\mathbb{A}$). The inequality between maps is taken pointwise. The following trivial propositions illustrate the usefulness of these notions.
\begin{proposition}\label{ch2:prop:caninequ}
If $s,t\in\mathsf{T}(V)$ are stable then the equation $s=t$ is canonical. Similarly, if $s$ is contracting and $t$ is expanding, then the inequality $s\leq t$ is canonical.
\end{proposition}
\begin{proof}
Let $\mathcal{A}$ be an arbitrary object in $\AEs(\Sigma)$. If $\mathcal{A}\models s=t$ then $s\eA=t\eA$ by Proposition \ref{ch2:prop:termfct}, therefore $(s\eA)\ce=(t\eA)\ce$ and thus $s\eAs=t\eAs$ by stability, and it follows that $\mathcal{A}\ce\models s=t$ by Proposition \ref{ch2:prop:termfct}. 

Similarly, if $\mathcal{A}\models s\leq t$ then $s\eA\leq t\eA$  by Proposition \ref{ch2:prop:termfct} and thus $(s\eA)\ce\leq (t\eA)\ce$. By the assumptions on $s$ and $t$, this means that we also have $s\eAs\leq t\eAs$, and thus $\mathcal{A}\ce\models s\leq t$ by Proposition \ref{ch2:prop:termfct}. 
\end{proof}

If we can determine if the terms of an (in)equation are stable, expanding or contracting, we can thus also determine if it canonical. It is easy to see that stable terms always exist.

\begin{lemma}\label{ch2:lem:meetjoinStab} The meet and join operations are stable.
\end{lemma}
\begin{proof}
Take $t=p\wedge q\in\mathsf{T}(V)$, then consider $t\eA=\lambda p\lambda q. (p\wedge q)\eA$. Since $\wedge\eA$ is isotone in both argument, we have by definition 
\begin{align*}
(t\eA)\ce(x,y)& =\bigvee\{\bigwedge\{ t\eA(a,b)\mid p\leq a\in A, q\leq b\in A\}\mid x\geq p\in K, y\geq q\in K\} \\
&=\bigvee\{\bigwedge\{a\wedge\eA b\}\mid p\leq a\in A, q\leq b\in A\}\mid x\geq p\in K, y\geq q\in K\} \\
&=\bigvee\{\bigwedge\{a\wedge\eAs b\}\mid p\leq a\in A, q\leq b\in A\}\mid x\geq p\in K, y\geq q\in K\} \\
&=\bigvee\{\bigwedge\{a\mid p\leq a\in A\} \wedge\eAs \bigwedge\{b\mid q\leq b\in A\}\mid x\geq p\in K, y\geq q\in K\}\\
&=\bigvee\{p\wedge\eAs q\mid x\geq p\in K, y\geq q\in K\}\\
&=\bigvee\{p\mid x\geq p\in K\}\wedge\eAs\bigvee\{q\mid x\geq q\in K\}\\
&=x\wedge\eAs y=t\eAs(x,y)
\end{align*}
where we've used infinite distributivity (which holds for all canonical extension as was noted in Section \ref{ch2:sec:canext}) and the fact that $\ba$ is a sub-object of $\ba\ce$ and thus $\wedge\eA=\wedge\eAs$ on elements of $A$.
The proof for joins is almost identical.
\end{proof}

By using Definition (\ref{ch2:eq:mapext12}) for the extension of a map we can also show:
\begin{lemma}\label{ch2:lem:negStab}
In the case of boolean algebras, the complementation operation is stable.
\end{lemma}
\begin{proof}
Let $t=\neg p\in\mathsf{T}(V)$, then $t\eA=\lambda p.(\neg p)\eA$ and since $\neg\eA$ is antitone we have
\begin{align*}
(t\eA)\ce(x)&=(\neg\eA)\ce(x) \\
& =\bigvee\{\bigwedge\{\neg\eA(a)\mid u\geq a\in A\}\mid x\leq u\in O\}\\
&= \bigvee\{\bigwedge\{\neg\eAs(a)\mid u\geq a\in A\}\mid x\leq u\in O\}\\
& = \bigvee\{\neg\eAs\bigvee\{a\mid u\geq a\in A\}\mid x\leq u\in O\}\\
& = \bigvee\{\neg\eAs u\mid x\leq u\in O\}\\
&= \neg\eAs \bigwedge\{u\mid x\leq u\in O\} \\
&=\neg\eAs x=(t)\eAs(x)
\end{align*}
where we have used the infinite version of the de Morgan laws in a CABA and the fact that since $\ba$ is a sub boolean algebra of $\ba\ce$, $\neg\eAs=\neg\eA$ on $\ba$.
\end{proof}

Moreover, for any signature $\Sigma$, all the functions $f_s, s\in\Sigma$ define stable terms in $\mathsf{T}(V)$. Indeed, let $f_s$ be $n$-ary and $t=f_s(p_1, \ldots, p_n)\in \mathsf{T}(V)$, then if we consider $t\eA=\lambda p_1\ldots\lambda p_n. f_s(p_1, \ldots, p_n): A^n\to A$ we have 
\[
(t\eA)\ce(x_1,\ldots,x_n)=f_s\ce(x_1,\ldots,x_n)=t\eAs(x_1,\ldots,x_n)
\]
by definition of the canonical extension of a AE $\mathcal{A}=(\mathbb{A}, (f_s)_{s\in\Sigma})$ as the AE $\mathcal{A}\ce=(\mathbb{A},(f_s\ce)_{s\in\Sigma})$. We have thus shown that all the building blocks of a modal language are stable. The key to studying canonicity is thus to understand how strings of these building blocks behave, i.e. to understand the interaction between function composition and canonical extension. This explains why we dedicated a large part of the previous sections to this very topic, and our main tool to establish canonicity will be the \emph{Principle of Matching Topologies} of Theorem \ref{ch2:thm:MatchingTopologies}.

Armed with this principle we build Table \ref{ch2:table} which gives a list of sufficient conditions on two unary terms to guarantee the preservation of stability under functional composition. Note that two columns are particularly well-behaved: if $f\ce$ preserves non-empty or just up-directed joins, then canonical extension always commutes with functional composition if $g\ce$ is monotone. Terms with this property (i.e. whose composition with any monotone term is stable) have been called \textbf{conservative} in \cite{Jonsson94}. We will rely on the results summarized in this table to define a notion of Sahlqvist formulas. Table \ref{ch2:table2} summarizes how the preservation properties we have examined combine under functional composition, together with Table \ref{ch2:table} it will allows us to build well-behaved terms in the coming section.

\begin{table}
\centering
\caption{Stability inheritance of unary terms under composition ($f\ce \circ g\ce$)}
\renewcommand{\arraystretch}{1.8}
\begin{tabular}{| l | l  *{10}{| c } |}
\hline\hline
\multicolumn{2}{| r | }{$f\ce$} & \multicolumn{5}{c}{Preservation} & \multicolumn{5}{| c |}{Anti-preservation} \\
\cline{3-12}
\multicolumn{2}{| l | }{$g\ce$} & $\leq$  & $\bigwedge$ & $\bigvee$ & d.d. $\bigwedge$ & u.d. $\bigvee$ & $\leq$ & $\bigwedge$ & $\bigvee$ & d.d. $\bigwedge$ & u.d. $\bigvee$ \\
\hline
\multirow{5}{*}{\begin{turn}{90}Preservation\end{turn}} & $\leq $ & \xmark  & \xmark  & \cmark  & \xmark & \cmark  & \xmark & \xmark & \xmark & \xmark & \xmark\\
\cline{2-12}
& $\bigwedge$ & \cmark & \cmark & \cmark & \cmark & \cmark & \xmark  & \xmark & \xmark & \xmark & \xmark\\
\cline{2-12}
& $\bigvee$ & \xmark & \xmark & \cmark & \xmark & \cmark & \cmark & \cmark & \cmark & \cmark & \cmark\\
\cline{2-12}
 & d.d. $\bigwedge$ & \xmark & \xmark & \cmark & \xmark & \cmark & \xmark & \xmark & \xmark & \xmark & \xmark\\
\cline{2-12}
& u.d. $\bigvee$ & \xmark & \xmark & \cmark & \xmark & \cmark & \xmark & \xmark & \xmark & \xmark & \xmark\\
\hline
\multirow{5}{*}{\begin{turn}{90}Anti-preservation\end{turn}} & $\leq $ & \xmark & \xmark & \cmark & \xmark & \cmark & \xmark & \xmark & \xmark & \xmark & \xmark\\
\cline{2-12}
& $\bigwedge$ & \xmark & \xmark   & \cmark & \xmark & \cmark & \cmark & \cmark & \cmark & \cmark & \cmark\\
\cline{2-12}
& $\bigvee$ & \cmark & \cmark  & \cmark  & \cmark & \cmark & \xmark & \xmark & \xmark & \xmark & \xmark\\
\cline{2-12}
& d.d. $\bigwedge$ & \xmark & \xmark  & \cmark & \xmark & \cmark &  \xmark & \xmark & \xmark & \xmark & \xmark \\
\cline{2-12}
& u.d. $\bigvee$ & \xmark & \xmark  & \cmark & \xmark & \cmark & \xmark & \xmark & \xmark & \xmark & \xmark\\
\hline
\end{tabular}
\label{ch2:table}
\end{table}

\begin{sidewaystable}
\renewcommand{\arraystretch}{1.8}

\caption{Property inheritance under composition ($f\ce \circ g\ce$)}
\centering
\begin{tabular}{| l | l  *{10}{| c } |}
\hline\hline
\multicolumn{2}{| r | }{$f\ce$} & \multicolumn{5}{c}{Preservation} & \multicolumn{5}{| c |}{Anti-preservation} \\
\cline{3-12}
\multicolumn{2}{| l | }{$g\ce$} & $\leq$  & $\bigwedge$ & $\bigvee$ & d.d. $\bigwedge$ & u.d. $\bigvee$ & $\leq$ & $\bigwedge$ & $\bigvee$ & d.d. $\bigwedge$ & u.d. $\bigvee$ \\
\hline
\multirow{5}{*}{\begin{turn}{90}Preservation\end{turn}} & $\leq $ & $\leq$  & $\leq$   & $\leq$   & $\leq$  & $\leq$   & $\geq$ &$\geq$ & $\geq$ & $\geq$ & $\geq$\\
\cline{2-12}
& $\bigwedge$ & $\leq$ & $\bigwedge$ & $\leq$ & d.d.$\bigwedge$ & $\leq$ & $\geq$  & a-$\bigwedge$ & $\geq$ & a-d.d. $\bigwedge$ & $\geq$\\
\cline{2-12}
& $\bigvee$ & $\leq$ &$\leq$ & $\bigvee$ & $\leq$ & u.d. $\bigvee$ & $\geq$ & $\geq$ & a-$\bigvee$ & $\geq$ & a-u.d. $\bigvee$\\
\cline{2-12}
 & d.d. $\bigwedge$ & $\leq$ & d.d. $\bigwedge$ & $\leq$ & d.d. $\bigwedge$ & $\leq$ & $\geq$ & a-d.d. $\bigwedge$ & $\geq$ & a-d.d. $\bigwedge$ & $\geq$\\
\cline{2-12}
& u.d. $\bigvee$ & $\leq$ & $\leq$ & u.d. $\bigvee$ & $\leq$ & u.d. $\bigvee$ & $\geq$ & $\geq$ & a-u.d. $\bigvee$ & $\leq$ & a-u.d. $\bigvee$\\
\hline
\multirow{5}{*}{\begin{turn}{90}Anti-preservation\end{turn}} & $\leq $ & $\geq$ & $\geq$ & $\geq$ & $\geq$ &$\geq$ & $\leq$ & $\leq$ & $\leq$ & $\leq$ & $\leq$\\
\cline{2-12}
& $\bigwedge$ & $\geq$ & $\geq$  & a-$\bigwedge$ & $\geq$ & a-d.d. $\bigwedge$ & $\leq$ & $\leq$ & $\bigvee$ & $\leq$ & d.d. $\bigwedge$\\
\cline{2-12}
& $\bigvee$ & $\geq$ & a-$\bigvee$  & $\geq$  & a-u.d. $\bigvee$ & $\geq$ & $\leq$ & $\bigvee$ & $\leq$ & u.d. $\bigvee$ & $\leq$\\
\cline{2-12}
& d.d. $\bigwedge$ & $\geq$ & $\geq$  & a-d.d.$\bigwedge$ & $\geq$ & a-d.d.$\bigwedge$ &  $\leq$ & $\leq$ & d.d.$\bigwedge$ & $\leq$ & d.d.$\bigwedge$ \\
\cline{2-12}
& u.d. $\bigvee$ & $\geq$ & a-u.d. $\bigvee$  & $\geq$ & a-u.d. $\bigvee$ & $\geq$ & $\leq$ & u.d. $\bigvee$ & $\leq$ &u.d. $\bigvee$ & $\leq$\\
\hline
\end{tabular}
\caption*{$\leq$ stands for isotonicity, $\geq$ for antitonicity, `u.d.' for up-directed, `d.d.' for down-directed and `a-' for anti-preservation.}
\label{ch2:table2}

\end{sidewaystable}

\section{Sahlqvist identities}\label{ch2:sec:Sahl}

We've seen in the previous section how to build stable terms and thus canonical equations. However, there exists a more general technique to build canonical equations, where not all terms need to be stable: the theory of Sahlqvist formulas. This theory was first laid out in 1975 by Henrik Sahlqvist in the now legendary \cite{1975:Sahlqvist}. This theory was developed in the context of classical `box-diamond style' modal logic. The theory was then generalised to the context of arbitrary boolean Algebras with Operators in \cite{Jonsson94} and \cite{deRijkeVenema95}, and subsequently to more general structures \cite{2005:GehrkeVenema, 2008:Teheux} and to other completions of boolean algebras \cite{1999:GivantVenema}. Here our aim is to generalize the theory to boolean Algebras Expansions, with the hope of dealing with some non-normal modal logics. Our path is straightforward: we will follow the algebraic treatment of \cite{Jonsson94,2006:VenemaAC} but without assuming that the maps in the signature are operators.

The algebraic treatment of Sahlqvist formulas hinges on the notion of quasi-equation and how they can be reduced to equations in the context of AEs. Let $\mathsf{T}(V)$ be the free AE for a signature $\Sigma$. A \textbf{quasi-equation} \index{Quasi-equation} is a set of $n+1$ pairs of terms $\{(s_1,t_1),\ldots,(s_n,t_n),(u,v)\}\subset \mathsf{T}(V)\times\mathsf{T}(V)$, denoted by $s_1=t_1, \ldots, s_n=t_n \Rightarrow u=v$. A quasi-equation is interpreted as being valid in a AE $\mathcal{A}$ if for any interpretation map $(\cdot)\eA: \mathsf{T}(V)\to\mathcal{A}$, $$s_1\eA=t_1\eA,\ldots, s_n\eA=t_n\eA \text{ implies } u\eA=v\eA$$ in which case we write $s_1\eA=t_1\eA,\ldots, s_n\eA=t_n\eA\Rightarrow u\eA=v\eA$. Just as an equation, a quasi-equations is canonical if $$s_1\eAs=t_1\eAs,\ldots, s_n\eAs=t_n\eAs\Rightarrow u\eAs=v\eAs$$
whenever $$s_1\eA=t_1\eA,\ldots, s_n\eA=t_n\eA\Rightarrow u\eA=v\eA$$
We will only need quasi-equations of the shape $s=\bot\Rightarrow t=u$.

We now show how the validity of a quasi-equation over some variety is equivalent to the validity of a certain equation over a certain collection of AEs. This makes precise Proposition 1.2 of \cite{Jonsson94} which also appears, in a slightly different format, in \cite{2006:VenemaAC}. Let $\Sigma$ be a signature and $\mathsf{T}(V)$ be the corresponding free $\Sigma$-AE and let $s,t,u\in\mathsf{T}(V)$. We are concerned with the validity over a variety $\mathbb{V}$ of $\Sigma$-AEs of the quasi-equation $s=\bot\Rightarrow t=u$. Consider now the signature $\Sigma\cup\{\mathsf{g}\}$ with $\mathsf{g}$ unary and the corresponding free $\Sigma\cup\{\mathsf{g}\}$-AE $\mathsf{T}'(V)$. We first build a variety of $\Sigma\cup\{\mathsf{g}\}$-AE by quotienting $\mathsf{T}'(V)$ with all the equations defining the variety $\mathbb{V}$, let us call this variety $\mathbb{V}'$. We then consider the collection of AEs in $\mathbb{V}'$ which satisfy: 
\[
\mathsf{g}(\bot)=\bot\text{ and } \mathsf{g}(x)=\top\text{ where } \bot\neq x\in\mathsf{T'}(V)
\]
Note that this collection is \emph{not} a variety, since $x\neq\bot$ cannot be expressed equationally. We will write $\mathcal{C}$ for this collection. Note also that every $\Sigma$-AE $\mathcal{A}\in\mathbb{V}$ trivially gives rise to an element $\mathcal{A}_{\mathsf{g}}\in\mathcal{C}$ by simply appending the operator $\mathsf{g}$ to $\mathcal{A}$ and declaring that it satisfies the condition above. Similarly any valuation $\pi: V\to \mathcal{A}$ trivially gives rise to an interpretation map 
\[
(\cdot)^{\mathcal{A}_{\mathsf{g}}}:\mathsf{T}'(V)\to\mathcal{A}_{\mathsf{g}}
\] 
by using $(\cdot)\eA$ and defining $(\mathsf{g})^{\mathcal{A}_{\mathsf{g}}}$ in the obvious way. It is now straightforward to check from the construction that for any $\mathcal{A}\in \mathbb{V}$ we have 
\[
s\eA=\bot \implies t\eA=u\eA\hspace{2ex}\text{ iff }\hspace{2ex} t^{\mathcal{A}_{\mathsf{g}}}\vee (\mathsf{g}(s))^{\mathcal{A}_{\mathsf{g}}}=u^{\mathcal{A}_{\mathsf{g}}}\vee (\mathsf{g}(s))^{\mathcal{A}_{\mathsf{g}}} 
\]
We can now use this relation between quasi-equation and equation to investigate the canonicity of quasi-equations.
\begin{lemma}\label{ch2:lem:Sahl}
Let $s\in\mathsf{T}(V)$ be an expanding term, and let $t\in\mathsf{T}(V)$ be a stable term, then the quasi-equation $s=\bot\Rightarrow t=\bot $ is canonical.
\end{lemma}
\begin{proof}
Assume that $s\eA=\bot\Rightarrow t\eA = \bot$. By the preceding construction this is equivalent to $$t^{\mathcal{A}_{\mathsf{g}}} \vee (\mathsf{g}(s))^{\mathcal{A}_{\mathsf{g}}}= (\mathsf{g}(s))^{\mathcal{A}_{\mathsf{g}}}$$
or in other words,
\[
t^{\mathcal{A}_{\mathsf{g}}}\leq  (\mathsf{g}(s))^{\mathcal{A}_{\mathsf{g}}}
\]
We need to check that  $t^{\mathcal{A}_{\mathsf{g}}\ce}\leq  (\mathsf{g}(s))^{\mathcal{A}_{\mathsf{g}}\ce}$. Since $t$ is stable we have $(t^{\mathcal{A}_{\mathsf{g}}})\ce=t^{\mathcal{A}_{\mathsf{g}}\ce}$ and thus
$$t^{\mathcal{A}_{\mathsf{g}}\ce}\leq  ((\mathsf{g}(s))^{\mathcal{A}_{\mathsf{g}}})\ce$$
Notice now that $\mathsf{g}$ trivially preserves finite joins, and thus by the results of the previous section, and Table \ref{ch2:table} in particular, it is clear that irrespective of $s$,  
\[
((\mathsf{g}(s))^{\mathcal{A}_{\mathsf{g}}})\ce=(\mathsf{g}^{\mathcal{A}_{\mathsf{g}}})
\ce(s^{\mathcal{A}_{\mathsf{g}}})\ce
\]
Since $\mathsf{g}$ is one of the extension maps, it is stable, and since $s$ is assumed to be expanding we thus get
\[
t^{\mathcal{A}_{\mathsf{g}}\ce}\leq ((\mathsf{g}(s))^{\mathcal{A}_{\mathsf{g}}})\ce \leq \mathsf{g}^{\mathcal{A}_{\mathsf{g}}\ce}(s^{\mathcal{A}_{\mathsf{g}}})\ce
\leq \mathsf{g}^{\mathcal{A}_{\mathsf{g}}\ce}(s^{\mathcal{A}_{\mathsf{g}}\ce})
\]
which is equivalent to $s\eAs=\bot\Rightarrow t\eAs = \bot$.
\end{proof}

The following result captures the key idea behind Sahlqvist identity. In a AE defined by an arbitrary signature $\Sigma$, we will take this Proposition as the definition of a Sahlqvist identity. However, since we would really like Sahlqvist identities to be syntactically defined, rather than through algebraic properties of their subterms, we will use the terminology of \textbf{abstract Sahlqvist identity} in the following result and of \textbf{concrete Sahlqvist identity} when it is instantiated purely syntactically in a specific modal language.

\begin{propdef}[\cite{Jonsson94}]\label{ch2:prop:Sahl}
Let $\mathsf{T}(V)$ be the modal language defined by a signature $\Sigma$. Every identity in $\mathsf{T}(V)$ of the type
\begin{align}\label{ch2:prop:Sahl:eq:1}
s(t_1,\ldots, t_n, \neg u_1,\ldots,\neg u_m)=\bot
\end{align}
where $t_1,\ldots, t_n$ are stable terms, $u_1,\ldots, u_m$ are expanding terms, and $s\in$ is a conservative term will be called an \textbf{abstract Sahlqvist identity}. Every abstract Sahlqvist identity is canonical.
\end{propdef}
\begin{proof}
Let us define
\begin{align*}
v&=(u_1\wedge p_1) \vee \ldots \vee (u_n\wedge p_m) \\
w&=s(t_1,\ldots, t_n, p_1,\ldots, p_m)
\end{align*} 
where $p_1,\ldots, p_m\in V$ are fresh variables not occurring in any of the terms. Then the identity (\ref{ch2:prop:Sahl:eq:1}) is equivalent to the quasi-identity $v=\bot\Rightarrow w=\bot$. Theorem 5.5 in \cite{Jonsson94} shows that $v$ must be expanding and, by definition of stability and of being conservative, $w$ must be stable. The results then follows from Lemma \ref{ch2:lem:Sahl}.
\end{proof}

The key to a practical use of Proposition \ref{ch2:prop:Sahl} is to isolate, for a given logic, syntactically defined classes of stable, expanding and conservative terms. Of course, going from abstract to concrete Sahlqvist identities will always be specific to the language at hand. Here we have gathered results for languages whose operations satisfy some of the preservation properties listed in Table \ref{ch2:table}, with the case of Graded Modal Logic and Probability Logic, detailed later in this Chapter, in mind. As we shall detail shortly, Tables \ref{ch2:table} and \ref{ch2:table2} provide a tool for building stable and conservative terms in such languages. However, we do not yet have a general technique for building expanding terms. For this we will need the following nomenclature. Let us restrict our attention to the class of AE whose endo-maps $\{f_s\}_{s\in\Sigma}$ are all monotone. We introduce the following dual notions: let $t\in\mathsf{T}(V)$ and $p\in V$ be a (propositional) variable, we will say that a term $t$ in the language associated with the signature $\Sigma$ is \textbf{positive in $p$}\index{Positive term} if every occurrence of $p$ in $t$ is in the scope of an even number of antitone operations (including the complementation). Dually, a term $t\in T(V)$ will be called \textbf{negative in $p$}\index{Negative term} if every occurrence of $p$ in $t$ is in the scope an odd number of antitone operations. For example, if $f:A^2\to A$ is isotone in the first and antitone in the second parameter, then the term $f(p,p)$ is neither positive nor negative in $p$ since the first occurrence of $p$ is under the scope of no antitone operations, whereas the second occurrence is under the scope of one antitone operation. The terms $f(p,\neg p)$ or $f(p,g(p))$ where $g$ is unary and antitone, are both positive in $p$. A term is \textbf{positive} if it is positive in all its propositional variables, and dually for \textbf{negative} terms. Positivity and negativity as we have just defined it are interesting concepts because they provide a purely syntactic description of isotone and antitone terms as the following well-known result shows.

\begin{proposition}\label{ch2:prop:PosIsoNegAnti}
Let $\Sigma$ be a signature containing only functions which are monotone in each of their arguments, let $\mathsf{T}(V)$ be the free $\Sigma$-AE and let $t\in \mathsf{T}V$. If $t$ is positive in $p$, then it is isotone in $p$, if it is negative in $p$, then it is antitone in $p$.
\end{proposition}
\begin{proof}
We prove both parts simultaneously by induction on the complexity of $t$. The base case for terms positive in $p$ is easy since $t$ is then the propositional variable $p$, which is trivially isotone since the identity map is isotone. The base case for terms negative in $p$ is either $\neg p$ or a term $f_s(q_1,\ldots, q_{i-1}, p, q_{i+1},\ldots, q_n)$ where $f_s, s\in\Sigma$ is antitone in its $i^{th}$ argument. Both cases are clearly antitone.

Now for the inductive step. We can treat all cases in one go. Assume $t=f(s_1,\ldots,s_{n_i}, s_{n_i+1},\ldots, s_{n_i+n_a})$ where $n_i$ is the number of isotone parameters of $f$, and $n_a$ is the number of antitone parameters. For example if $f=\wedge$, then $n_i=2, n_a=0$ and if $f=\neg$, then $n_i=0, n_a=1$, and similarly for the monotone maps $f_s, s\in\Sigma$.

If $t$ is positive in $p$, then each $s_i$, $1\leq i\leq n_i$ must be positive in $p$, and thus by induction isotone in $p$. Dually each $s_i$, $n_i+1\leq i\leq n_i+n_a$ must be negative in $p$ and thus antitone in $p$ by the induction hypothesis. The term $t$ is then clearly isotone in $p$. The case where $t$ is negative in $p$ works in exactly the same way.
\end{proof}

The following key result is Theorem 5.5 of \cite{Jonsson94} (Theorem 7.20 of \cite{2006:VenemaAC}).

\begin{theorem}\label{ch2:thm:posexp}
If a signature $\Sigma$ contains only isotone operations, then every positive term is expanding.
\end{theorem}

In this instance, we were not able to generalise this important theorem to include antitone operations. The reason is that Lemma \ref{ch2:lem:ineq} induces a profound asymmetry in the algebraic theory of canonicity. By working uniformly for all monotone maps, it gives a preferred direction to all inequalities which quickly becomes incompatible with order reversal, i.e. with antitone maps. We illustrate this phenomenon with a simple example: let a signature contain a binary map $g$ which is isotone in its first and antitone in its second argument, unary isotone maps $f_1, f_2$ and a unary antitone map $h$. Then the term
\[
g(f_1(p), f_2(h(q)))
\]
is positive but if we try to apply Lemma \ref{ch2:lem:ineq} repeatedly, we get
\[
(g(f_1, f_2(h)))\ce\leq g\ce(f_1\ce,(f_2(h))\ce)\geq g\ce(f_1\ce,(f_2\ce(h\ce)))
\]
and it is thus impossible to use the expanding/contracting properties of the subterms to infer this property for the entire term, despite the term being positive. Because of its full generality, Lemma \ref{ch2:lem:ineq} cannot be used for nested sequences of mixed or antitone terms. In consequence, for a signature $\Sigma=\Sigma_{\mathrm{iso}}+\Sigma_{\mathrm{mix+anti}}$ containing antitone and/or mixed operations, our only source of expanding terms will be those terms which are positive for the sub-signature $\Sigma_{\mathrm{iso}}$. 

Having built this (not entirely satisfying) syntactically defined class of expanding terms, let us now show how to build stable and conservative terms. We start with the well-known case of classical modal logic defined by the signature $\Sigma=\{\dia,\square\}$. Since all operators are either join- (the diamond operator $\dia$) or meet- (the box operator $\square$) preserving, we can use Table \ref{ch2:table} to notice that terms involving exclusively $\dia$, meets and joins are stable. These terms are usually defined as the \textbf{strictly positive} terms since $\dia$ is often considered as the fundamental operator and $\square$ the derived one. Moreover, by using the distributivity rules of boolean algebras, we have that:
\begin{lemma} In the signature $\Sigma=\{\dia\}$ where $\dia$ is unary and join preserving, the strictly positive terms preserve joins. In particular, the strictly positive terms are conservative
\end{lemma}
\begin{proof}
By induction on the complexity of the terms. For the base case, note that $\dia$ preserves joins, $\wedge$ considered as a binary operation also preserves joins by virtue of the distributivity of $\wedge$ over $\vee$, finally $\vee$ considered as binary operation trivially preserves $\vee$. The inductive step follows for the same reason. The fact that terms preserving joins are conservative follows from Lemma \ref{ch2:lem:pres} and Table \ref{ch2:table}.
\end{proof}
Another conclusion of Table \ref{ch2:table} is that nested sequences of meet-preserving operators, i.e. nested sequences of $\square$, are also stable, but are not conservative, so they can play the role of the terms $t_1,\ldots,t_n$ in Proposition \ref{ch2:prop:Sahl}. This provides us with the classical definition of Sahlqvist identities.
\begin{propdef}
A \textbf{classical Sahlqvist identity} is an equation of the form 
\[
s(t_1,\ldots,t_n,\neg u_1,\ldots,\neg u_m)=\bot
\]
where $s$ is a strictly positive term, $t_1,\ldots,t_n$ are nested sequences of $\square$, and the terms $u_1,\ldots,u_m$ are positive terms. Classical Sahlqvist identities are canonical.
\end{propdef}
\begin{proof}
This is just a special case of Proposition \ref{ch2:prop:Sahl}.
\end{proof}

Note that we now have a purely syntactic implementation of the abstract notion of Sahlqvist identities, i.e. we now have what we have defined as `concrete Sahlqvist identities. Note also that other stable combinations suggests themselves from Tables \ref{ch2:table} and \ref{ch2:table2}. In fact we can use the two tables as the basis of an algorithm to generate stable terms: pick two stable terms $s$ and $t$ with one of the preservation or anti-preservation properties, then from Table \ref{ch2:table} check whether or not the composition $st$ is also stable. If it is, use Table \ref{ch2:table2} update the preservation or anti-preservation property of $st$, select a new term $r$ and repeat the process. This leads us to the following revised definition of Sahlqvist identities:

\begin{propdef}\label{ch2:def:genSahlqvist}
Given a signature $\Sigma$ whose operations satisfy any of the preservation properties of Table \ref{ch2:table}, a \textbf{general Sahlqvist identity} is an equation between terms in the language defined by $\Sigma$ of the form 
\[
s(t_1,\ldots,t_n,\neg u_1,\ldots,\neg u_m)=\bot
\]
where $s$ is a conservative term built from the Table \ref{ch2:table}, $t_1,\ldots,t_n$ are stable terms built from the Table \ref{ch2:table}, and the terms $u_1,\ldots,u_m$ are terms which are positive for the sub-signature of $\Sigma$ consisting of isotone operations only.
\end{propdef}

\begin{example}
Let us illustrate this definition in the case of BAOs defined by the signature $\{\dia\}$. 
\begin{itemize}
\item As an example of stable terms which are not nested boxes, consider the following formulas: $\dia\square p$  and $\dia\square (p\wedge \neg q)$. From table \ref{ch2:table} it is immediate that $\dia\square p$ is stable. It is clear by Table \ref{ch2:table} that since $\neg$ anti-preserves joins and $\wedge$ trivially preserves meets in both arguments that $\wedge(\cdot,\neg(\cdot))$ is a stable term which anti-preserves joins in its second argument. Thus by Table \ref{ch2:table} $\square(\wedge(\cdot,\neg(\cdot)))$ is also stable and by Table \ref{ch2:table2} it preserves meets in its first argument and anti-preserves joins in its second. Finally, since $\dia$ is conservative and stable, we get that $\dia\square (p\wedge \neg q)$ is stable.
\item As an example of general but not classical Sahlqvist formula, consider the term 
\[
t'=\dia p\wedge \dia (q\wedge\neg p)\wedge\dia(r\wedge\neg p\wedge\neg q)
\] 
The equation $t'=\top$ clearly enforces the existence of at least three distinct successors in a Kripke frame. In the formalism of Definition \ref{ch2:def:genSahlqvist}, this equation can be re-written as
\[
t=\neg t'=\square \neg p\vee \square(\neg q\vee p)\vee \square(\neg r\vee p\vee q)=\bot
\]
It cannot be a classical Sahlqvist term since $\square(\neg q\vee p)$ for example is neither a nested box nor the negation of a positive formula. However, by using De Morgan's laws as a proof that $\neg$ anti-preserves joins, it is easy to check directly from Table \ref{ch2:table} that each of the terms $t_1=\square \neg p, t_2=\square(\neg q\vee p)$ and $t_3=\square(\neg r\vee p\vee q)$ is stable, and  $t=\vee(t_1,t_2,t_3)$ with each $t_i$ stable, is thus a general Sahlqvist formula. Thus the logic $K+(t'=\top)$ is sound and strongly complete w.r.t. Kripke frames whose elements all have at least three distinct successors. This example clearly generalizes to $n$-successors for any $n$.
\item As an example of non general Sahlqvist and non-canonical formula, consider McKinsey's axiom 
\[
\square\dia p\to\dia\square p
\]
It can be re-written as $\vee(\square\dia p, \neg\dia\square p)$ and we immediately see that one subterm is positive in $p$ and the other negative, so the only way this formula could be a general Sahlqvist is if one of the two terms was stable. From Table \ref{ch2:table} we can easily see that $\square\dia$ is a problematic combination which is not stable, and whilst $\dia\square p$ is stable, it has no preservation property (apart from being isotone) and thus $\neg\dia\square p$ is not in general stable either.
\end{itemize}
\end{example}

For non-normal logics, we propose using the other rows and columns of Table \ref{ch2:table} to isolate classes of stable and conservative terms. The next section will provide us with concrete case studies involving most of the types of extension functions studied so far.

\section{Applications}\label{ch2:sec:Ex}
\subsection{Graded Modal Logic}\label{ch2:subsec:GML}

In this section we consider Graded Modal Logic (see e.g. \cite{1972:Fine, 1985:GML}), or GML for short,  as an example of a logic whose modal operators are not operators in the technical sense, i.e. do not preserve joins. This logic is based on the choice of $\BA$ as a fundamental reasoning structure, and on the signature $\Sigma=\{\kop\mid k\in\mathbb{N}\}$ where all the function symbols are unary. In the BNF formalism, GML formulas are thus given by:
\[
\phi::=\bot\mid p\mid \neg \phi\mid \phi\wedge\phi\mid \kop\phi, k\in\mathbb{N}
\]
The intended interpretation of $\kop p$ in a Kripke frame is that it should be true at a point $w$ if $w$ has \emph{at least} $k$ successors where $p$ holds. The function symbol $\kop[1]$ is therefore equivalent to the $\dia$ operator of classical modal logic. We will not use the dual operators, but it is frequent to define $[k] =\neg\kop\neg a$ whose interpretation is thus `there are \emph{strictly fewer} than $k$ successors where $a$ does not hold'. These operators can thus be understood as counting `exceptions'. Clearly $[1]$ is equivalent to $\square$ in classical modal logic.

Formally the semantics of GML can be given either in terms of Kripke frames or in terms of coalgebras for the so-called `bag' or multiset functor $\Bag:\Set\to\Set$. For a Kripke frame $(W,R)$, a valuation $\pi:V\to W$ and a point $w\in W$, the semantics of the terms $\mathsf{T(V)}$ is defined inductively in the usual way for propositional variables and boolean connectives and by:
\[
(W,R,w)\models \kop p \text{ iff } \mid\hspace{-1pt}\{w'\in W \mid wRw'\text{ and } w'\models p\}\hspace{-1pt}\mid \geq k
\]
The coalgebraic semantics of GML is given by the bag functor: 
\[
\Bag(W)=\{f: W\to\mathbb{N}\mid \supp(f)\text{ is finite}\}
\]
The modal operators $\kop$ are then interpreted as the predicate liftings 
\[
\lsem\kop\rsem: \pow(W)\to \pow(\Bag W), U\mapsto \{f\in \Bag(W) \mid \sum_{x\in U} f(x)\geq k\}
\] 
and the coalgebraic semantics of GML terms is then defined in the usual inductive way.

GML can be axiomatized in several ways, but we follow the axiomatization of \cite{1972:Fine} which provides an elegant strong completeness proof. 
\begin{enumerate}[GML1]
\item \label{ch2:ax:GML1} $\kop[1](a\vee b)=\kop[1](a)\vee\kop[1](b)$
\item \label{ch2:ax:GML2} $\kop[1]\bot=\bot$
\item \label{ch2:ax:GML3} $\kop a\to\kop[l] a,\hspace{1ex} l<k$
\item \label{ch2:ax:GML4} $\kop a \leftrightarrow \bigvee_{i=0}^k \kop[i](a\wedge b)\wedge\kop[k-i](a\wedge \neg b)$
\item \label{ch2:ax:GML5} $\neg\kop[1](a\wedge\neg b)\to(\kop a\to\kop b)$
\end{enumerate}

The axiomatization of GML defines a quotient of the free $\Sigma$-BAE $\mathsf{T}(V)$, and thus a variety of $\Sigma$-BAEs, which we denote by $\mathsf{Q}_{GML}$.

\begin{theorem}[\cite{1972:Fine,1985:GML}]\label{ch2:thm:GMLsoundcomp}
GML is sound and strongly complete w.r.t. to its Kripke frame semantics, in other words for any term $t\in\mathsf{T}(V)$, $(t)^{\mathsf{Q}_{GML}}=\top$ iff  $(W,R)\models t$ for any Kripke frame $(W,R)$
\end{theorem}

It is easy to verify from the above theorem that the expansions $\kop$ are isotone for any $k\in\mathbb{N}$. Moreover, the following proposition captures a fundamental distributivity rule satisfied by each $\kop$. This distributivity rule will be, via Theorem \ref{ch2:thm:complkadd}, key to building Sahlqvist GML terms via Table \ref{ch2:table}.

\begin{proposition}[GML distributivity]\label{ch2:prop:gmldistrib}
For each $k$, $\kop$ is $k$-additive.
\end{proposition}
\begin{proof}
We use a semantic argument in conjunction with the soundness and completeness result of Theorem \ref{ch2:thm:GMLsoundcomp}. To show 
\[
\kop\bigvee X\leq \bigvee\{\kop\bigvee U\mid U\in\pow_k(X)\}
\]
is easy: let $(W,R)$ be a Kripke frame and $w\in W$, then if $(w, W, R)\models \kop\bigvee X$, then there are at least $k$ successors of $w$ which land in $\bigvee X$. Pick $k$ of them, each of them must land in at least one of the elements $a\in X$. By choosing one such element $a(i)$ for each $1 \leq i\leq k$ we define subset $U$ of $X$ of cardinality at most $k$ and it is clear that $x\models \kop\bigvee U$. The opposite direction follows from the monotonicity of $\kop$ and the observation that $\bigvee U\leq \bigvee X$ for any $U\in\pow_k(X)$.
\end{proof}

Note that for $k=1$, the distributivity law of Proposition \ref{ch2:prop:gmldistrib} is the usual join-preservation property of $\dia$. Note also that if $k=|X|$, we would just have the triviality $\kop\bigvee X=\kop\bigvee X$ since one of the functions $\phi$ can then pick every element of $X$.

Recall from the comments made before Theorem \ref{ch2:thm:complkadd} that if $\kop\ce$ is completely $k$-additive, then in particular it preserves up-directed joins, i.e. it is $(\gamma\eup,\gamma\eup)$-continuous. From Table \ref{ch2:table}, it is clear that, for all intent and purposes, functions preserving up-directed joins are just as good as functions preserve non-empty joins. In particular terms built from meets, joins and up-directed join preserving functions are stable and conservative. Note that GML has modal operators satisfying each of the preservation properties detailed in Table \ref{ch2:table}: $\kop[1]$ preserves joins, $\kop, k>1$ preserve up-directed joins, $[0]$ preserves meets and $[k], k>1$ preserves down-directed meets. We can now define Sahlqvist formulas for GML.

\begin{propdef} A \textbf{general Sahlqvist identity}\label{ch2:propdef:SahlqvistGML} for GML is an equation of the form
\[
s[t_1,\ldots,t_n,\neg u_1,\ldots,\neg u_m]=\bot
\]
where $s$ is strictly positive (i.e. it is built from $\vee,\wedge,\kop$ and propositional  variables), $t_1,\ldots,t_n$ are stable terms obtained from Table \ref{ch2:table} and  $u_1,\ldots,u_m$ are positive terms. Sahlqvist identities for GML are canonical.
\end{propdef}
\begin{proof}
The proof follows immediately from Proposition \ref{ch2:prop:Sahl} and the fact that since the functions $\kop\ce$ preserve up-directed joins by Theorem \ref{ch2:thm:complkadd}, strictly positive terms are both stable and conservative. 
\end{proof}

\begin{example}\label{ch2:ex:GML}
Let us consider a scenario with a Description Logic flavour. Assume we have a GML with several classes of graded operators intended to represent genealogical relations with number restrictions. Concretely we have a $\kop^\mathtt{Child}$ operator, with the intended interpretation of $\kop^\mathtt{Child} p$ being `has at most $k$ children satisfying p', and an operator $\kop^{\mathtt{G-Child}}$ whose interpretation is  similar but with grand-children instead. Now it would be natural that any model for such a logic should have cardinality restrictions of the type: 
\[
\kop[k_1]^\mathtt{Child}\kop[k_2]^\mathtt{Child}p\to \kop[k_1\cdot k_2]^{\mathtt{G-Child}}p
\] 
Such a formula is easily seen to belong to the class of general Sahlqvist formula for GML defined above since $\kop[k_1]^\mathtt{Child}\kop[k_2]^\mathtt{Child}p$ is stable and $\neg\kop[k_1\cdot k_2]^{\mathtt{G-Child}}p$ is negative in $p$.
\end{example}

\subsection{Intuitionistic Logic and Distributive Substructural Logics}\label{ch2:subsec:intuitionistic}

In this section we will provide examples of logics build on $\BDL$ and $\DL$ and containing non-isotone operators. We will also use these as examples of logics presented in the abstract flavour of coalgebraic logic.

We start with the well-know \textbf{intuitionistic logic}. Algebraically, intuitionistic logic over a set $V$ of propositional variables is the free \textbf{Heyting algebra}\index{Intuitionistic logic}\index{Heyting algebra} over $V$. However, it can also be viewed as a positive modal logic with a binary operator $\to$, i.e. of modal logic based on the free distributive lattice over $V$ (since Heyting algebras are necessarily distributive). Algebraically, this amounts to viewing a Heyting algebra as a DLE of a particular kind. This is the point of view which we adopt here and we therefore define the following syntax constructor for any $f;A\to B$ in $\BDL$:
\[
\LHey: \BDL\to\BDL, \begin{cases}A\mapsto \Free\{a\to b\mid a,b\in \Forg A\}/\equiv\\
\LRL f: \LHey A\to \LHey B, [a]_{\equiv}\mapsto [f(a)]_{\equiv}
\end{cases}
\]
where $\equiv$ is the fully invariant equivalence relation in $\BDL$ (see Section \ref{ch1:subsec:fullyinvariant}) generated by the following Heyting Distribution Laws:
\begin{enumerate}[HDL1]
\item $a\to(b\wedge c)=(a\to b)\wedge (a\to c)$
\label{ch2:ax:HDL1}
\item $(a\vee b)\to c=(a\to c)\wedge(b\to c)$
\label{ch2:ax:HDL2}
\end{enumerate}
The language defined by $\LHey$ for a set $V$ of propositional variables is the free $\LHey$-algebra over $\Free V$, i.e. the language of intuitionistic propositional logic quotiented by the axioms of bounded distributive lattices and HDL\ref{ch2:ax:HDL1}-\ref{ch2:ax:HDL2}. Note that an $\LHey$-algebra is \emph{not} a Heyting algebra, the axioms HDL\ref{ch2:ax:HDL1}-\ref{ch2:ax:HDL2} only capture some of the Heyting algebra structure. Instead, an $\LHey$-algebra is simply a distributive lattice with a binary map satisfying the distribution laws above (which happen to be valid in HAs), i.e. a DLE. The remaining features of HAs will be captured in a second stage in Chapter 5 via \emph{canonical frame conditions}, for reason which will become clear. The reason for choosing $\BDL$ rather than $\DL$ is that one of the additional axioms needed to fully capture intuitionistic logic involve $\top$.

Since the (modal) operator $\to$ preserves meet in one argument and anti-preserves joins in the other, it is isotone in its first argument and antitone in its second, and $\LHey$-algebras fall within the class of DLEs for which we can use the results of Sections \ref{ch2:sec:canext}-\ref{ch2:sec:Sahl} to define a notion of canonical (in)equation. Indeed this is the reason we have chosen to define $\LHey$ as we did. On the other hand, the notion of Sahlqvist identity for $\LHey$-algebra is problematic. By the comment following Theorem \ref{ch2:thm:posexp}, \emph{positive terms} cannot be built from non-isotone expansions. In particular no $\LHey$-term is positive. Moreover, no $\LHey$-term can be strictly positive (even though we are working without negation!), since we do not have any conservative operation. Thus a general Sahlqvist identity in $\LHey$ simply amount to an equation
\[
t=\bot
\]
where $t$ is stable. However, it is perfectly possible to define useful and interesting \emph{canonical (in)equalities} in the language of $\LHey$-algebra, as we will see in Chapter 5. The format of Sahlqvist identities is simply not very well-suited to $\LHey$-algebras.

We proceed in an analogous way to define a basic language for \textbf{distributive substructural logics}\index{Substructural logics} (see \cite{2003:onoresiduated} for an overview of the topic). In a nutshell, substructural logic disallow some or all of the structural rule of propositional logic, viz. weakening, contraction and exchange. In this setting an additional operator, which we denote by $\ast$, is added to the language as a kind of $\wedge$ which does not satisfy the structural rule. This operator is required to be left and right-residuated, i.e. there must exist $\lRes$ and $\rRes$ operators such that $a\ast b\leq c$ iff $b\leq a\lRes c$ iff $a\leq c\rRes a$ and posses a unit $I$. Algebraically, such a structure corresponds to a \textbf{residuated lattice}, i.e. a lattice with a monoidal operation which is left and right-residuated (see \cite{2003:onoresiduated,2007:galatosresiduated}\index{Residuated lattice}. Just as Heyting algebra can be seen as a DLE, a residuated lattice can be seen as a lattice with expansions. We will in fact restrict ourselves to distributive residuated lattices in order to make use of the theory of canonicity developed in the previous Sections, which means that we will view distributive residuated lattices as DLEs. We define the 
following syntax constructor:
\[
\LRL: \DL\to\DL, \begin{cases}
\LRL A =\Free\{I, a\ast b, a\lRes b, a\rRes b\mid a,b\in \Forg A\}/\equiv\\
\LRL f: \LRL A\to \LRL B, [a]_{\equiv}\mapsto [f(a)]_{\equiv}
\end{cases}
\]
where $\equiv$ is the fully invariant equivalence relation in $\DL$ (see Section \ref{ch1:subsec:fullyinvariant}) generated by the Distribution Laws:

\begin{enumerate}[DL1]
\item$(a\vee b)\ast c=(a\ast c)\vee(b\ast c)\label{ch2:ax:DL1} $ 
\item$a\ast (b\vee c)=(a\ast b)\vee (a\ast c)$
\label{ch2:ax:DL2} 
\item$a\lRes(b\wedge c)=(a\lRes b)\wedge (a\lRes c)$
\label{ch2:ax:DL3} 
\item$(a\vee b)\lRes c=(a\lRes c)\wedge (b\ast c)$
\label{ch2:ax:DL4} 
\item$(a\wedge b)\rRes c=(a\rRes c)\wedge (b\rRes c)$
\label{ch2:ax:DL5} 
\item$a\rRes (b\vee c)=(a\rRes b) \wedge (a\rRes c)$
\label{ch2:ax:DL6}
\end{enumerate}

The language defined by $\LRL$ is the free $\LRL$-algebra over $\Free V$, which is the language of the full Lambek calculus (or residuated lattices) quotiented under the axioms of $\DL$ and DL\ref{ch2:ax:DL1}-\ref{ch2:ax:DL6}. An $\LRL$-algebra is simply an object of $\DL$ endowed with a nullary operation $I$ and binary operations $\ast,\lRes$ and $\rRes$ satisfing the distribution laws above, i.e. a DLE whose expansions are amenable to the techniques developed earlier in the Chapter. Again, note that an $\LRL$-algebra is \emph{not} a distributive residuated lattice. Only some features of this structure have been captured by the axioms above. But several are still missing, and will be added in Chapter 5 as \emph{canonical frame conditions}. Note again that $\lRes$ and $\rRes$ are isotone in one argument but antitone in the other. 

We now proceed to define a notion of general Sahlqvist identity for the language $\LRL$-algebras. The positive $\LRL$-terms are those built using $I,\wedge,\vee$ and $\ast$ only, since $\lRes$ and $\rRes$ are not isotone. But since $\ast$ is like a binary $\dia$ operator (it preserves joins in each argument), and negation is not part of the language, the positive terms are also the strictly positive ones. Consequently we have
\begin{propdef}
A \textbf{general Sahlqvist formula} for distributive substructural logics is an equation of the form
\[
s[t_1,\ldots, t_n,u_1,\ldots, u_m]=\bot
\]
where $s,u_1,\ldots,u_n$ are positive terms, and $t_1,\ldots,t_n$ are stable terms obtained from Table \ref{ch2:table2}. Sahlqvist identities for distributive substructural logics are canonical.
\end{propdef}
As we will see in Chapter 5, the notions of canonical equation and inequation are in fact more useful and better suited to the language of $\LRL$-algebras.

\chapter{Functor Presentations}
The power of coalgebraic logic comes from its high level of abstraction, both at the syntactic and at the semantic level. In its most abstract formalism (promoted by Alexander Kurz, see e.g. \cite{2005:UfExtCoalg,2007:GoldThomCoalg,DirkOverview}) a coalgebraic logic and its interpretation are determined by an endofunctor $L:\BA\to\BA$, an endofunctor $T:\Set\to\Set$ and a natural transformation $\lambda: L\pow\to\pow T$. In itself a functor is a very abstract entity, with very simple specifications (viz. it needs to preserve identities and commute with functional composition). Faced with such abstraction the logician may yearn for something concrete to manipulate, guide his/her intuition and prove things with. This is where the idea of representing abstract functors using simple, well known functors often proves useful. This idea of representing abstract entities by using simpler and more well-known ones is pervasive in much of mathematics and it is therefore not surprising that the theory of functor representation should follow very similar lines. We will return to this point in the third section of this chapter, but for now let us point out that whilst we use the term `representation' loosely and intuitively, the main type of functor representation we will study is the notion of functor \emph{presentation} which has a strict technical meaning.

For functors $F:\cat\to\Set$, a natural candidate for the role of simple functors which can be used to represent arbitrary ones is the class of covariant $\hom$ functors $\hom(A,-), A$ in $\cat$. More generally for a category $\cat$ enriched over a category $\mathbf{V}$, it is natural to try to represent functors $F:\cat\to\mathbf{V}$ in terms of the `enriched $\hom$ functors' in $\cat[V]$. The theory of enriched functors representation has been studied in e.g. \cite{1993:KellyPower,2011:KurzEqPresMon,2012:finfuncpres}, although from a slightly different perspective to the one developed in what follows. For our purpose we only need enrichment over $\Set$, i.e. no enrichment at all, but the enriched reader might find comfort in knowing that most of what follows can be extended to the enriched setting. 

Underlying every type of functor representations we will define and study, is the ability to express a functor as a certain colimit of $\hom$ functors. A representation of a functor $T$ in these terms will be called a \emph{presentation} of $T$. When $\cat$ is small, the basic construction is a textbook category theory construction (although it is usually carried out for presheaves, see e.g. \cite{SheavesTopos}), and is essentially a consequence of Yoneda's lemma. For sufficiently well-behaved functors $F:\cat\to\Set$, this construction can be generalized to a large class of categories called accessible categories which are characterised by the fact that they contain a small subcategory which can be used to approximate all its objects. In this case, the notion of functor presentation is also a standard construction which is detailed for example in \cite{AAC,LPAC,setFuncPres}. A general theory of $\Set$-endofunctor presentations which includes a characterization of properties of functors in terms of properties of the presentations can be found in \cite{setFuncPres}. 

Our small contribution to this already well-developed theory is twofold. 

\begin{itemize}
\item We clarify the notions of functor presentation which exist in the literature (e.g. \cite{1971:Freyd,AAC,setFuncPres,2005:cococo,2011:KurzEqPresMon,2012:finfuncpres})
and relate them to standard notions from the theory of locally presentable and accessible categories (see \cite{LPAC,1989:MakkaiPare}) by considering functors as objects in functor categories. From this, we extract the notions of \emph{functor presentation}, $\lambda$-\emph{presentable} functors, $\lambda$-\emph{presented} functor, $\lambda$-\emph{generatable} functor and $\lambda$-\emph{generated} functor. We highlight and detail a striking similarity between $\Set$-functors and algebraic varieties (already mentioned in passing in \cite{1971:Freyd,KellyEnriched}). 
\item We isolate and characterize a `minimal presentation' (already introduced, although less systematically, in \cite{AAC}) and thereby study the interaction between presentations and the notion of \emph{base} which is introduced in the first section of this chapter. This notion is key to the formulation of the nabla-style of coalgebraic logic, but is interesting in its own right to understand $\Set$-functors.
\end{itemize}

\section{The base transformation.}\label{ch3:sec:base}

We start this chapter with a slight detour by looking at the notion of base. For $\Set$-endofunctors this notion is well understood and underpins the entire edifice of coalgebraic logic in the nabla-style. Definitions and basic properties can thus be found in treatments of the nabla-logics, for example the excellent \cite{KKV:2012:Journal}. The notion of base for a $\Set$-endofunctor is also discussed in some details, although not under the name `base', in \cite{Gumm05} which has inspired this section significantly. Intuitively, given a functor $T:\Set\to\Set$, a set $X$ and an element $\alpha\in TX$, the base of $\alpha$, which we will denote as $\Base_X(\alpha)$, is the smallest subset of $X$ that is required to `support' $\alpha$, i.e. the collection of elements of $X$ that are `ingredients' in the construction of $\alpha$. In this section we will formalize and discuss this intuition in as general a context as possible. 

More generally, given an arbitrary category $\cat$ and a functor $T:\cat\to\Set$, what do we need in order be able to talk about the `support' of $\alpha\in TC$? Well, we clearly need to be able to talk about subobjects of $C$, since $\Base(\alpha)$ is going to be such a subobject. We will therefore require that $\cat$ be \textbf{well-powered}\index{Well-powered}, i.e. that every object $C$ of $\cat$ should posses a small (i.e. set-sized) poset of subobjects denoted by $\Sub(C)$. We refer the reader to e.g. \cite{MacLane} or \cite{Goldblatt} for more details on the precise construction of the poset $\Sub(C)$. We will denote a subobject $U$ of $C$ either by the object $U$ or by the injection map $i_U^C: U\mono C$ depending on the context and we will denote the partial order on $\Sub(C)$ by $\subseteq$. The fact that a subobject $i_U^C$ `supports' $\alpha$,  can now be formalized as 
\[
\alpha\in Ti_U^C[TU]
\]
where $Ti_U^C[TU]$ is the usual direct image (in $\Set$) of $TU$ under $Ti_U^C$.

Secondly, we would like to identify the subobject of $C$ which \textit{only} contains the ingredients that are \textit{strictly} necessary to build $\alpha$, in other words we want the smallest possible subobject supporting $\alpha$. We therefore require that $\cat$ should have have pullbacks of monos (i.e. intersections when $\cat=\Set$). In fact, requiring $\cat$ to have (all) pullbacks is probably a more `natural' requirement, and in fact the categories we will be looking at in what follows (viz. locally presentable categories) will even be complete. The definition of the base of $\alpha$ is then given by:
\begin{definition}\index{Base}
Let $\cat$ be a well-powered category with pullbacks of monos, let $T:\cat\to \Set$ and let $C$ be an object of $\cat$, then the base of $\alpha\in TC$ is the subobject of $C$ defined as
\[
\Base_C(\alpha)=\bigcap \{ U\stackrel{i}{\hookrightarrow} C\mid \alpha\in Ti[TU] \}
\]
where the $\bigcap$ symbol signifies the (limiting object of the) pullback of all the monos in the set, which of course reduces to the usual intersection when $\cat=\Set$.
\end{definition}

Whilst the object $\Base_C(\alpha)$ is well-defined in $\cat$ as soon as $\cat$ has small intersections, it does not necessarily mean that it constitutes a meaningful representation of the aforementioned vision of a smallest subobject from which $\alpha$ can be built, as the following important example shows. 

\begin{example}\label{ch3:ex:filter}\index{Filter}\index{Filter functor}
Let $\powc:\Set\to\Set\op$ be the contravariant powerset functor, then $\powc^2:\Set\to\Set$ is a covariant functor sometimes called the neighbourhood functor\index{Neighbourhood functor}. Let us now defined the subfunctor of $\powc^2$ defined on sets $X$ as
\[
\Filt X=\{\Filt\subseteq \pow X\mid \Filt\text{ is a filter}\}
\]
and on functions $f: X\to Y$ as 
\[
\Filt f=(f\inv)\inv: \Filt X\to\Filt Y, \Filt\mapsto\{U\subseteq Y\mid f\inv(U)\in\Filt\}
\]
It is easy to verify that $\Filt f(\Filt)$ is indeed a filter on $\pow Y$ and that $\Filt$ is thus a functor. It is interesting for our purpose to understand how $\Filt$ acts on inclusion maps. Let $i_U^X: U\inc X$ and consider the trivial filter $\Filt_U=\{\{U\}\}\in \Filt U$, it is easy to see that 
$$\Filt i_U^X(\Filt_U)=\{V\subseteq X\mid X\cap U\in\Filt_U\}=\up U$$
Thus if we apply our definition of $\Base$ to $\uparrow U$ we get $$\Base(\up U)=\bigcap\{V\subseteq X\mid \up U\in \Filt i_V^X[\Filt V]\}=U$$
which is in line with the intuition of what the base should be, namely the least amount of information required to build the element $\uparrow U$; and in particular $\uparrow U\in \Filt i^X_U[\Filt U]$. However, assume now that $X$ is infinite, and consider the cofinal (or Fr\'{e}chet) filter $\Filt^c_X\in\Filt X$, and let $x\in X$ and $X_x=X\setminus \{x\}$. It is easy to check that if $\Filt^c_{X_x}$ is the cofinal filter on $X_x$, then
\[
\Filt i^X_{X_x}(\Filt^c_{X_x})=\{V\subseteq X\mid V\cap X_x\in \Filt^c_{X_x}\}=\Filt^c_X
\]
Note that the same holds for any subset of $X$ to which finitely many elements have been removed. In particular, there is no smallest such subset. Note also that
\[
\Base_X(\Filt^c_X)=\bigcap\{U\subseteq X\mid \Filt^c_X\in \Filt i_U^X[\Filt U] \}\subseteq \bigcap_{x\in X} X_x=\emptyset
\]
This clearly violates the intuition of what the base of an element should be, in particular we have the problematic situation that $\Filt^c_X\notin \Filt i^X_{\Base_X(\Filt^c_X)}[\Filt \Base_X(\Filt^c_X)]$.
\end{example}

To remedy the pathology illustrated by the previous example we need to add some criterion to the functors for which we want to define a notion of base. The criterion we are interested is very natural in light of the definition of the base, namely the preservation of small intersections. However, let us take this opportunity to define, more generally, the notion of preservation of classes of limits.

\begin{definition}\index{Preservation!of limits}\index{Preservation!of colimits}\index{Weak preservation!of limits} \index{Weak preservation!of colimits}\index{Continuous functor}\index{Co-continuous functor}
Let $\cat[J]$ be a small category and $D:\cat[J]\to \cat$ be a diagram in $\cat$.  We will say that a functor $T:\cat\to\Set$ \textbf{preserves $D$-limits} if every limiting cone of $D$ in $\cat$ gets mapped by $T$ to a limiting cone of $T\circ D$ in $\Set$. A functor preserving all small limits is called continuous.

\noindent Similarly, we will say that a functor $T:\cat\to\Set$ \textbf{weakly preserves $D$-limits} if every limiting cone of $D$ in $\cat$ gets mapped by $T$ to a weakly limiting cone of $T\circ D$ in $\Set$. Preservation of $D$-colimits, co-continuity and weak-preservation of $D$-colimits are defined similarly. 
\end{definition}

For our purpose, pullbacks will be the most important type of limits, and in particular we will be most interested in the (weak) preservation pullbacks involving monic arrows, viz. inverse images and small intersections. We will say that a functor $T:\cat\to\Set$ \textbf{preserves monos} if for any monic arrow $m:U\mono A$, $Tm: TU\to TA$ is also monic, and dually, $T$ \textbf{preserves epis} if for any epic arrow: $e:B\epi A$, $Te: TB\to TA$ is also epi. The following two sets of dual criteria can be used to determine when a functor preserves monos and or epis.

\begin{proposition}
If $\cat$ is a category where all monos are sections (resp. all epis are retractions), then all functors $T:\cat\to\cat'$ preserve monos (resp. epis).
\end{proposition}
\begin{proof}
Every section is a mono (and every retraction is an epi) and being a section (or a retraction) is clearly preserved by functoriality.
\end{proof}

\begin{proposition}\label{ch3:prop:MonoPres}
If $\cat$ is a category that has all kernel pairs (resp. co-kernel pairs) and $T:\cat\to\Set$ is a functor weakly preserving these pairs, then $T$ preserves monos (resp. epis).
\end{proposition}
\begin{proof}
It is easy to check that a morphism $f: A\to B$ is a monomorphism iff the following commutative square is a pullback square
\[
\xymatrix
{
A\ar[d]_{\Id_A}\ar[r]^{\Id_A} & A\ar[d]^{f} \\
A\ar[r]_{f} & B 
}
\]
(and dually for epimorphisms), and the result follows (almost) immediately from the assumption on $T$.
\end{proof}
Note in particular that weak-pullback preserving functors which are the bread and butter of coalgebraic logic must therefore preserve monos.

Given an arrow $f:A\to B$ in $\cat$ and an element $i_V^B\in\Sub(B)$, the \textbf{inverse image}\index{Inverse image} of $i_V^B$ along $f$ is defined as the pullback:
\[
\xymatrix
{
f\inv(V)\ar@{^{(}->}[d]_{i_*^V} \ar[r]^-{f_*^V} & V\ar@{^{(}->}[d]^{i_V^B}\\
A\ar[r]_f & B
}
\]
A mono-preserving functor $T:\cat\to\Set$ will thus be said to \textbf{(weakly) preserve inverse images} if applying $T$ to the diagram above yields a (weak) pullback in $\Set$. As we have already seen, an \textbf{intersection}\index{Intersection} is a (small) pullback of monomorphisms. Thus, we will say that $T:\cat\to\Set$ \textbf{(weakly) preserves  intersections} if for any set of monos $\{i_x: U_x\mono A\mid x\in X\}$ the cone of the diagram $\{Ti_x: U_x\mono A\mid x\in X\}$ given by
$$\left(T \bigcap\{i_x\mid x\in X\},\{Tj_x: T\bigcap\{i_x\mid x\in X\}\mono TU_x\}_{x\in X} \right)$$
is a (weak) pullback. If $T:\cat\to\Set$ preserves intersections we have the following very natural property.

\begin{lemma}\label{ch3:lem:interbase}
Let $\cat$ be well-powered with pullbacks of monos let $T:\cat\to\Set$ preserve intersections and monos, then for any $A$ in $\cat$ and any $\alpha\in TA$: $$\alpha\in Ti_{\Base_A(\alpha)}^A[T\Base_A(\alpha)]$$
\end{lemma}
\begin{proof}
Let us write $$\mathscr{B}(\alpha)=\{Ti_U^A[TU]\mid i_U^A\in\Sub(A), \alpha\in Ti_U^A[TU]\}$$
By definition, for each $Ti^A_U[TU]\in\mathscr{B}(\alpha)$ there exists an element $\beta_{TU}\in Ti_U^A[TU]$ which gets mapped to $\alpha$ along the corresponding maps $Ti^A_U$. Since $T$ preserve monos, each $Ti^A_U$ is a mono, and they therefore define an intersection $\bigcap\mathscr{B}(\alpha)$. From the fact that this intersection $\bigcap\mathscr{B}(\alpha)$ is a pullback, any collection of elements $\{\beta_{TU}\}_{TU\in \mathscr{B}(\alpha)}$ which all get mapped to $\alpha$ (and there exists at least one) determines a unique element 
\[
\beta_0\in \bigcap \mathscr{B}(\alpha)
\]
which gets mapped to the various $\beta_{TU}\in TU$ and then to $\alpha$  along the various monos defining the pullback. Since $T$ preserves intersections, we have
\[
\beta_0\in T\left(\bigcap\{i^A_U\in\Sub(A)\mid \alpha\in Ti_U^A[TU]\}\right)=T\Base_A(\alpha)
\]
and thus $Ti_{\Base_A(\alpha)}^A(\beta_0)=\alpha$.
\end{proof}

\noindent Note that weak-preservation of intersections is actually sufficient.

We will spend the rest of this section showing that, for a large class of functors, the base construction actually defines a natural transformation between $T$ and the $\Sub$ functor. To show this we must first turn $\Sub$ into a (covariant) functor. This is done by specifying a generalized notion of `direct image' for the category $\cat$. Given a morphism $f: A\to B$ in $\cat$, we will define the \textbf{direct image}\index{Direct image} $f[A]$ as the smallest subobject of $B$ through which $f$ factors. Since we're assuming that $\cat$ has pullbacks of monos to define bases, we can build $f[A]$ explicitly as 
$$f[A]=\bigcap\{i_V^B: V\inc B\mid f\text{ factors through } i_V^B\}$$
Note that the preservation of intersections does not imply the preservation of direct images, since for $f:A\to B$, there are generally subsets of $TB$ through which $Tf$ factors but which are not of the form $TU$ for $U\in\Sub(B)$. Note also that whilst the direct image $f[A]$ is, strictly speaking, a subobject of $B$, i.e. an equivalence class of monomorphisms into $B$, for notational convenience we will often denote is as an object, i.e. $f[A]$, with the understanding that we are really looking at (the equivalence class of) $i_{f[A]}^A$. In order for $f$ to factor through $f[A]$, we need a map $A\to f[A]$ which we will denote $f^*$, i.e. $f^*$ will from now on denote the restriction of a morphism to its (direct) image.  This construction endows $\cat$ with an \textbf{orthogonal factorization system}\index{Factorization system}

\begin{proposition}Let $\cat$ be well-powered and with small pullbacks and let $f: A\to B$ be a $\cat$-morphism. The decomposition $$A\stackrel{f}{\to} B=A\stackrel{f^*}{\to} f[A]\stackrel{i^*}{\inc}B$$ has the property that $f^*$ is orthogonal to the class of all monomorphism in $\cat$, i.e. for any monomorphism $m: C\mono D$ and arrows $g: A\to C, h: f[A]\to D$ such that the following square commutes
\[
\xymatrix
{
A\ar[r]^{f^*}\ar[d]_g & f[A]\ar[d]^h \\
C\ar@{>->}[r]_m & D
}
\]
there exist a unique diagonal fill-in morphism $e: f[A]\to C$ such that $g=e\circ f^*$ and $h=m\circ e$.
\end{proposition}\label{ch3:prop:ortho}
\begin{proof}
Recall we have defined a decomposition $f=i^*\circ f^*$. Let us first show that if $f^*=n\circ k$ for a certain monomorphism $n$ and a certain morphism $k$, then $n$ is an isomorphism. Note first that $i^*\circ n\circ k=i^*\circ f^*=f$ and since $i^*\circ n$ is the composition of two monos, it is itself a mono through which $f$ factors. But $i^*$ is the smallest such mono, so $i^*$ must factor through $i^*\circ n$, i.e. there must exist $p$ such that $i^*\circ n\circ p=i^*$, and since $i^*$ is mono, this means $n\circ p=\id$. As a consequence, we also have that $(i^*\circ n)\circ (p\circ n)=i^*\circ(n\circ p)\circ n=i^*\circ n$ and thus $(p\circ n)=\id$ too, by virtue of $i^*\circ n$ being mono. Thus $n$ is iso.

Now staring from the commutative square above, since $\cat$ has pullbacks we can build the pullback of $h$ along $m$:
\[
\xymatrix
{
A \ar@/_1pc/[ddr]_g \ar@/^1pc/[rrd]^{f^*} \ar@{-->}[rd]\\
& h\inv(C)\ar@{>->}[r]^n\ar[d] & f[A]\ar[d]^h \\
& C\ar@{>->}[r]_m & D
}
\]

By the universal property of the pullback there exist a unique morphism $A\to h\inv(C)$, and since pulling back along a mono produces a mono, we have that $f^*$ factors through a mono $n$ and, by our previous point, this means that this mono $n$ is actually an iso. It is then easy to see that we have an arrow from $f[A]\to C$. The unicity follows from the fact that inverses are unique up-to-isomorphism and so are pullbacks.
\end{proof}

Note that algebraic theories, and in particular $\Set$ and $\BA$, have a regular epi-mono factorisation. More generally, every locally presentable category (see next section) has strong-epi mono factorisation. Both are orthogonal factorization systems.

If $\cat$ is well-powered and has pullbacks we can therefore turn $\Sub$ into a covariant functor $\Sub: \cat\to\Set$ by defining it as usual on objects and by setting for any\index{Subobject functor} $f:A\to B$: 
\[
\Sub(f):\Sub(A)\to\Sub(B), \quad i^A_U\mapsto \bigcap\{i_V^B\mid f\circ i_U^A\text{ factors through } i_V^B\}
\]
For clarity's sake we need to make the following notational point precise. It is customary mathematical notation to write $\Sub(f)(i_U^A)$ as $f[U]$, but in our very general categorical context this notation can be misleading. Consider again the $\cat$-morphism $f: A\to B$ and the subobject $i_U^A: U\inc A$. Whilst $f[A]$ is the direct image of $f$, $f[U]$ is the direct image of $f\circ i_U^A$. Thus $f[U]$ is an abuse of notation since $U$ is not the domain of $f$. In fact we can make this notation both convenient and precise as follows: since $\Sub(f)(i_U^A)$ is the smallest subobject of $B$ through which $f\circ i_U^A$ factors, there must exist a morphism $f^*_U$ such that  $ \Sub(f)(i_U^A)\circ f^*_U=f\circ i^A_U$. For brevity's sake we will often denote $\Sub(f)(i^A_U)$ by $i^*_U$. Notice that we are using a notation for the maps defining the limiting cone which is dual to that in the inverse image case, e.g. $f_*^U$ versus $f^*_U$. With this notation in place, it is clear that $f^*_U=(f\circ i_U^A)^*$, i.e. in $\Set$-language $f^*_U$ is the image restriction of $f$ restricted to $U$. 

Let us now check that $\Sub$ indeed defines a functor. Consider first the diagram:
\[
\xymatrix
{
U\ar@{^{(}->}[d]_i\ar@{^{(}->}[r] & V\ar@{^{(}->}[d] \\
A\ar[r]_{\id_A} & A
}
\]
Clearly, if $V$ is a subobject of $A$ through which $\id_A\circ i=i$ must factor, then $U\inc V$, and the smallest such $V$ is $U$ itself. Thus $\Sub(\id_A)=\id_{\Sub(A)}$. Now for composition consider the following diagram:
\[
\xymatrix@C=12ex
{
U\ar[r]^{f^*_U}\ar@{^{(}->}[d]_{i^A_U} & f^*_U[U] \ar[r]^{g^*_{f^*_U[U] }}\ar@{^{(}->}[d]_{i^*_U} & g^*_{f^*_U[U]}[ f^*_U[U]] \ar@{^{(}->}[d]^{i^*_{f^*_U[U]}}  \\
A\ar[r]_f & B\ar[r]_g & C
}
\]
In light of the above diagram, the abuse of notation consisting in writing $f[U]$ for $f^*_U[U]$ is here fully justified, and we shall thus use this convention for what follows. By definition, $(g\circ f)[U]$ is the smallest subobject of $C$ through which $g\circ f\circ i^A_U$ factors. On the other hand $g[f[U]]$ is the smallest subobject of $C$ through which $g\circ i^B_{f[U]}$ factors. But that means that $g\circ i^B_{f[U]}\circ f^*$ also factors through $g[f[U]]$, and by commutativity of the above diagram this means that $g\circ f\circ i_U^A$ factors through $g[f[U]]$, and thus $ (g\circ f)[U]\inc g[f[U]]$. For the converse direction, note that from the diagram above and the definition of $(g\circ f)[U]$ we can extract the following diagram:
\[
\xymatrix
{
U \ar[d]_{(g\circ f)^*_U}\ar[r]^{f^*_U} & f[U] \ar[d]^{g\circ i^*_U} \\
(g\circ f)[U]\ar@{^{(}->}[r] & C
}
\]
By Proposition \ref{ch3:prop:ortho} we know that $f^*$ is orthogonal to the class of all monomorphisms and thus there must exist a unique diagonal fill-in morphism $f[U]\to (g\circ f)[U]$. This in turns means that $g\circ i_{f[U]}^B$ factors through $(g\circ f)[U]$ and since $g[f[U]]$ is the smallest subobject of $C$ with this property we have $g[f[U]]\inc (g\circ f)[U]$. \medbreak

The notion of base can now be understood as a transformation $\Base: T\to\Sub$ between the functors $T:\cat\to\Set$ and $\Sub:\cat\to\Set$ defined at each stage $A$ in $\cat$ by:
$$\Base_A: TA\to\Sub(A), \quad \alpha\mapsto i_{\Base_A(\alpha)}^A=\bigcap \{ i_U^A: U\inc A\mid \alpha\in Ti_U^A[TU] \}$$ 
This transformation is in general not natural, but we will now isolate preservation properties for $T$ which guarantee the naturality of $\Base$. We first need the following lemmas.

\begin{lemma}\label{ch3:lem:subofmono}
Consider $U\stackrel{i_U^V}{\mono}V \stackrel{i_V^A}{\mono} A$, then $\Sub(i^A_V)(i_U^V)=i^A_U$.
\end{lemma}
\begin{proof}
If $i_W^A\in\Sub(A)$ is such that $i^A_V\circ i^V_U$ factors through $i^A_W$, then since $i_U^A=i^A_V\circ i^V_U$, it is trivially the case that $i_U^A$ factors through $i_W^A$ and thus $i_U^A\leq i_W^A$ which shows that $i_U^A$ is the smallest such object and the result follows.

\end{proof}

\begin{lemma}\label{ch3:lem:submono}
The functor $\Sub$ preserves monos.
\end{lemma}
\begin{proof}
Let $i:A\mono B$ be a $\cat$-monomorphism. Since $\Sub(i):\Sub(A)\to\Sub(B)$ is an arrow in $\Set$, it will be equivalent to show that it is injective. By the preceding Lemma \ref{ch3:lem:subofmono}, if $i_U^A:U\inc A$ is a subobject of $A$, then $\Sub(i)(i\circ i_U^A)=i\circ i_U^A$. Thus if we have two subobjects $i_U^A, i_V^A\in\Sub(A)$ such that $\Sub(i)(i_U^A)=\Sub(i)(i_V^A)$, then clearly $i\circ i_U^A=i\circ i_V^A$, and since $i$ is mono, $i_U^A=i_V^A$.
\end{proof}

\begin{lemma}\label{ch3:lem:subinter}
The functor $\Sub$ preserves intersections.
\end{lemma}
\begin{proof}
Let us consider an arbitrary collection of monos $\{i_x: A_x\mono B\mid x\in X\}$ in $\cat$. By definition, any collection of monos $\{j_x: U_x\mono A_x \in \Sub(A_x)\mid x\in X\}$ such that $\Sub(i_x)(j_x)=k\in \Sub(B)$ for all $x\in X$ defines an element of $\bigcap\{\Sub(i_x)\mid x\in X\}$, we will show that it also defines a unique element of $\Sub\left(\bigcap\{i_x\mid x\in X\}\right)$, providing us with a map
\[
\bigcap\{\Sub(i_x)\mid x\in X\}\to \Sub\left(\bigcap\{i_x\mid x\in X\}\right)
\]
From which the result follows immediately. So let us consider such a collection of monos $j_x\in \Sub(A_x), x\in X$, from Lemma \ref{ch3:lem:subofmono} we know that $\Sub(i_x)(j_x)=i_x\circ j_x$, which is a mono into $B$. Since $\cat$ has intersections we can form the intersection $\bigcap\{i_x\circ j_x: U_x\inc B\mid x\in X\}$. Since this intersection is a cone for $\{i_x: A_x\to B\mid x\in X\}$, there must exist a unique monomorphism 
\[
\bigcap\{i_x\circ j_x: U_x\inc B\mid x\in X\}\mono \bigcap\{i_x: A_x\to B\}
\]
But this precisely defines an element of $\Sub\left(\bigcap\{i_x\mid x\in X\}\right)$, as desired.
\end{proof}

\begin{lemma}\label{ch3:lem:subinverse}
The functor $\Sub$ preserves inverse images.
\end{lemma}
\begin{proof}
Let $f:A\to B$ and $i_V^B: V\mono B$ in $\cat$. We need to show that $\Sub(f\inv(V))$ together with the arrows $\Sub(i^V_*)$ and $\Sub(f^V_*)$ is a pullback for $\Sub(A)\stackrel{\Sub(f)}{\longrightarrow}\Sub(B)\stackrel{\Sub(i_V^B)}{\longleftarrow}\Sub(V)$. As usual, we pick $m\in \Sub(A)$ and $j_W^V\in \Sub(V)$ such that $\Sub(f)(m)=\Sub(i_V^B)(j_W^V)=i_V^B\circ j_W^V$ and we need to find a unique element of $\Sub(f\inv(V))$ which gets mapped to $m$ and $j_W^V$. Consider the following commutative diagram:
\[
\xymatrix@C=10ex@R=8ex
{
U\ar@/_2.5pc/@{>->}[ddr]_{m} \ar@{-->}[dr] \ar[rr]^{(f\circ m)^*_U} & & W=(f\circ m)^*_U[U]\ar@{>->}[d]^{j_W^V}\\
& f\inv(V)\ar@{>->}[d]_{i_*^V}\ar[r]^{f_*^V} & V\ar@{>->}[d]^{i_V^B} \\
& A\ar[r]_f & B
}
\]
Since the bottom square is a pullback and since $\left(U, \{j^V_W\circ (f\circ m)^*_U, m\}\right)$ forms a cone for the same diagram, there must exist a unique arrow $n: U\to f\inv(V)$. Moreover, this arrow needs to be a monomorphism since $i^V_*$ and $m$ are monomorphisms. We therefore have a unique element of $n\in \Sub(f\inv(V))$. It remains to check that it gets mapped to $m$ and $j^V_W$ by $\Sub(i^V_*)$ and $\Sub(f^V_*)$ respectively. By Lemma \ref{ch3:lem:subofmono} we know that $\Sub(i^V_*)(n)=i^V_*\circ n=m$. Moreover, if $\Sub(f_*^V)(n)$ was a subobject of $W$, then $f\circ m$ would factor through it and since $W$ is the smallest subobject of $B$ with this property we must have $\Sub(f_*^V)(n)=j^V_W$.
\end{proof}

\begin{theorem}\label{ch3:thm:baseNat}
Let $\cat$ be well-powered with pullbacks and direct images, then $T:\cat\to\Set$ preserves monos, intersections and inverse images iff $\Base: T\to\Sub$ is natural and sub-cartesian.
\end{theorem}
\begin{proof}
Let us start by assuming that $T$ preserves monos, intersections and inverse images. We must then show that for any $f:A\to B$ in $\cat$ the following diagram commutes.
\[
\xymatrix@C=12ex
{
TA\ar[d]_{\Base_A}\ar[r]^{Tf} & TB\ar[d]^{\Base_B}\\
\Sub(A)\ar[r]_{\Sub(f)} & \Sub(B)
}
\]
Let us pick an $\alpha\in TA$. In order to show that $\Sub(f)(\Base_A(\alpha))\subseteq\Base_B(Tf(\alpha))$. First note that by Lemma \ref{ch3:lem:subinter}, $\Sub$ preserves intersections and thus
\begin{align*}
\Sub(f)(\Base_A(\alpha))&=\Sub(f)\left(\bigcap\{i_U^A\mid \alpha\in Ti_U^A[TU]\}\right)\\
&=\bigcap\{\Sub(f)(i_U^A)\mid \alpha\in Ti_U^A[TU]\}
\end{align*}
Now starting with a mono $i_V^B:V\mono B$ such that $Tf(\alpha)\in Ti_V^B[TV]$ let us build the preimage of $TV$ under $Tf$. Since $T$ preserves preimages we have the following pullback square:
\[
\xymatrix@C=24ex
{
Tf\inv(TV)=T(f\inv(V)) \ar[d]_{Ti_{f\inv(V)}^A} \ar[r]^-{Tf_{Tf\inv(TV)}=T(f_{f\inv(V)})} & TV\ar[d]^{Ti_V^B} \\
TA\ar[r]_{Tf} & TB
}
\]
And since this square is a pullback, the fact that there exist $\beta\in TV$ such that  $Ti_V^B(\beta)=Tf(\alpha)$ means that there must exist a $\beta'\in T(f\inv(V))$ such that $T(f_{f\inv(V)})(\beta')=\beta$ and $Ti_{f\inv(V)^A}(\beta')=\alpha$. If we now set $U=f\inv(V)$, then $\alpha\in Ti_U^A[TU]$ and $\Sub(f)(i_U^A)$ is the smallest subobject of $B$ through which $f\circ i_U^A$ factors. Clearly $V$ is such a subobject since $i_V^B\circ f_{f\inv(V)}=f\circ i_U^A$, so $\Sub(f)(i_A^U)\inc V$. 
Since we can find such an $U$ for any $V$ with $Tf(\alpha)\in Ti_V^B[TV]$ we must have 
$$\bigcap\{\Sub(f)(i_U^A)\mid \alpha\in Ti_U^A[TU]\}\subseteq \bigcap\{i_V^B\mid Tf(\alpha)\in Ti_V^B[TV]\}$$
i.e. $\Sub(f)(\Base_A(\alpha))\subseteq\Base_B(Tf(\alpha))$.

\medskip
Now for the opposite direction, i.e. $\Base_B(Tf(\alpha))\subseteq\Sub(f)(\Base_A(\alpha))$. For notational convenience let us write $i^*$ for the mono $i^*:\Sub(f)(\Base_A(\alpha))\mono B$. It is enough to show that $Tf(\alpha)\in Ti^*[T(\Sub(f)(\Base_A(\alpha)))]$. But this follows immediately from Lemma \ref{ch3:lem:interbase}: since $\alpha\in Ti_{\Base_A(\alpha)}^A[T\Base_A(\alpha)]$, then there must exist $\beta\in T\Base_A(\alpha)$ with $Ti_{\Base_A(\alpha)}^A(\beta)=\alpha$. Since $\Sub(f)(\Base_A(\alpha))$ is the smallest subobject of $B$ through which $f\circ i^A_{\Base_A(\alpha)}$ factors, $Tf\circ Ti^A_{\Base_A(\alpha)}$ factors through $T(\Sub(f)(\Base_A(\alpha)))$ and so there must exist a $\beta'\in T(\Sub(f)(\Base_A(\alpha)))$ such that $Ti^*(\beta')=Tf(\alpha)$, which is what we needed to show.

\medskip
Let us conclude the `if' direction by showing that $\Base$ is sub-cartesian. Let $m: A\mono B$ be a monic arrow in $\cat$, we need to show that the following square is a pullback:
$$
\xymatrix@C=12ex
{
TA\ar@{>->}[r]^{Tm}\ar[d]_{\Base_A} & TB\ar[d]^{\Base_B}\\
\Sub(A)\ar@{>->}[r]_{\Sub(m)} & \Sub(B)
}
$$
Pick an element $\beta\in TB$ and an element $i_U^A\in\Sub(A)$ such that $\Sub(m)(i_U^A)=\Base_B(\beta)$. From Lemma \ref{ch3:lem:submono} we know that $\Sub(m)$ is injective and from Lemma \ref{ch3:lem:subofmono} we know that $\Sub(m)(i_U^A)=m\circ i_U^A$. Since $T$ preserves intersections we know from Lemma \ref{ch3:lem:interbase} that $\beta\in Ti_{\Base_B(\beta)}^B[T\Base_B(\beta)]=T(m\circ i_U^A)[TU]$. Thus there must exist $\alpha\in TU$ such that $Tm\circ Ti_U^A(\alpha)=\beta$, and since $T$ preserves monos, this $\alpha$ must be unique. If we now choose $\alpha'=Ti_U^A(\alpha)$ we get a unique element of $TA$ such that $Tm(\alpha')=\beta$ and $\Base_A(\alpha')=U$.

\medskip
For the `only if' direction: assume that $\Base:T\to\Sub$ is natural and subcartesian. The fact that $T$ preserves monos is a simple consequence of $\Base$ being subcartesian: for any $m: A\mono B$, if $f,g: TX\to TA$ are such that $Tm\circ f=Tm\circ g$, then by forming the naturality square for $m$, which must be a pullback, we immediately get that $f=g$ and thus $Tn$ is a mono.  Let us now show that $T$ must preserve inverse images. Consider the following commutative diagram:
\[
\xymatrix@C=8ex
{
X\ar@/_1.5pc/[ddddr] \ar@/^3pc/[rrrdd] \ar@{-->}[drr] \ar@{-->}[ddr]\\ 
& &  \Sub(f\inv(V)) \ar[rr]^{\Sub(f_*^V)}\ar[dd]_>>>>>>{\Sub(i^V_*)} &  & \Sub(V)\ar[dd]^{\Sub(i_V^B)} \\
& T(f\inv(V))\ar[rr]^>>>>>>>>>{T(f_*^V)}\ar[dd]_{T(i_*^V)}\ar[ur]^{\Base_{f\inv(V)}} & & TV\ar[dd]^<<<<<<<<<{T(i_V^B)} \ar[ur]^{\Base_V}\\
& &  \Sub(A)\ar[rr]_<<<<<<<{\Sub(f)} & & \Sub(B) \\
& TA\ar[rr]_{Tf}\ar[ur]^{\Base_A} & & TB\ar[ur]^{\Base_B}
}
\]
To show that $T(f\inv(V))$ is a pullback square, take a competitor $X$, and let's show that there exist a unique morphism $X\to T(f\inv(V))$. First note that by post-composition with the component $\Base_A$ and $\Base_V$, $X$ becomes a competitor to $\Sub(f\inv(V))$ which we know by Lemma \ref{ch3:lem:subinverse} to be a pullback. So there must exist a unique morphism $X\to \Sub(f\inv(V))$. But since $i^V_*$ is a mono and that both $\Sub$ and $T$ preserve monos, the naturality square for $i^V_*$ must be a pullback since $\Base$ is sub-cartesian. So there must exist a unique function $X\to T(f\inv(V))$.

The proof that $T$ preserves intersections is almost identical since $\Sub$ has been shown to preserve intersections in Lemma \ref{ch3:lem:subinter}.
\end{proof}

\section{The concept of functor presentation}\label{ch3:sec:funcPresIntro}

The purpose of this section is to show how functors $T:\cat\to\Set$ from a small category $\cat$ can be written in terms of elementary functors, namely the $\hom$ functors. `Written in terms of' will formally mean `expressed as colimit of' $\hom$ functors. We can immediately make the following remarks: first, a colimit of $\hom$ functors will by definition inhabit a functor category where objects are functors and arrows are natural transformations, and as is customary we will denote by $\Set^{\cat}$ the category of functors $T: \cat\to\Set$.  Secondly, since we are only concerned with covariant functors we will only consider the covariant $\hom$ functors.

\subsection{The category $\el$.}

The Yoneda lemma has been described as the deepest triviality known to humanity. The following subsection adds further weight to this observation by showing how it underlies the theory of functor presentation. We start by recalling the Lemma.
\begin{lemma}[Yoneda Lemma]\index{Yoneda Lemma}
Let $\mathbf{C}$ be a locally small category and $T: \mathbf{C}\to\Set$ be a functor, then for any object $A$ of $\mathbf{C}$, there exist a bijection
\begin{equation}
\Nat(\hom(A,-),T)\simeq TA \nonumber
\end{equation}
\end{lemma}
\begin{proof}
See for example \cite{MacLane}.
\end{proof}
\noindent In the special case where $T$ is also a $\hom$ functor, say $\hom(B,-)$, then the Yoneda lemma tells us that 
\begin{equation}\label{ch3:eq:YonEmb}
(\hom(A,-),\hom(B,-))\simeq\hom(B,A)
\end{equation}
i.e. to each $\cat$-arrow $f: B\to A$ corresponds a $\Set^{\cat}$-arrow (i.e. a natural transformation) $\eta_f: \hom(A,-)\to \hom(B,-)$. Moreover this correspondence is well-behaved with respect to arrow composition and identities. This suggests that the assignment $A\mapsto \hom(A,-)$ in fact defines a contravariant functor. This is indeed the case and this functor is known as the Yoneda embedding. It is defined formally as the functor $\Yon:\cat\op\to\Set^{\cat}$ defined on objects as $\Yon(A)=\hom(A,-)$ and on $\cat\op$-morphisms $f:B\to A$ by the natural transformation $\Yon f: \hom(A,-)\to \hom(B,-)$ defined by:
\[
\Yon f_C: h\in \hom(A,C) \mapsto h\circ f \in \hom(B,C)
\]

\begin{corollary}The Yoneda embedding $\Yon$ defined above is a fully faithful functor.
\end{corollary}
\begin{proof}
This is the content of the isomorphism of Eq. (\ref{ch3:eq:YonEmb}).
\end{proof}

Given a functor $T:\cat\to \Set$ and an object $A$ of $\cat$, the Yoneda lemma suggests viewing elements of $TA$ as natural transformations. Moreover, the existence of the Yoneda embedding suggests that since we have morphisms between $\hom$ functors, it may also be possible to define morphisms between these natural transformations. As we shall shortly show, this is indeed the case and allows us to view the \textit{elements} of $TA$ as \textit{objects} in a category which is called the \textbf{category of elements}\index{Category of elements} of $T$ and denoted $\el$ or $\int T$ (we will use the former notation throughout). The formal construction is as follows: the objects of $\el$ are the natural transformations $\eta: \hom(A,-)\to T$ for some $A$ in $\cat$ and there is a morphism between $\eta: \hom(A,-)\to T$ and $\eta':\hom(B,-)\to T$ whenever there is a $\cat$-morphism $f:B\to A$, such that the following diagram commute:

\[
\xymatrix
{
\hom(A,-) \ar[rr]^{\Yon f}\ar[dr] _{\eta} & & \hom(B,-)\ar[dl]^{\eta'} \\
& T &
}
\]

\noindent Thus $\el$ is the comma category $\Yon \downarrow T$. There is an obvious forgetful functor $\Forg_T:\el\to \mathbf{C}\op$ which is defined on objects by: $$\left(\hom(A,-)\stackrel{\eta}{\longrightarrow}T\right)\mapsto A$$ and on arrows by $(\eta\to\eta')\mapsto f$ where $f:B\to A$ is the $\cat$-arrow making $\eta'\circ \Yon(f)=\eta$. 

We can use Yoneda's lemma to give a more intuitive description of the category $\el$ which is often easier to work with. Formally, $\el$ can equivalently be defined as follows: to each $\eta: \hom(A,-)\to T$ we associate the pair $(A,\alpha)$ where $\alpha\in TA$ is the element corresponding to $\eta$ by Yoneda, i.e. $\eta_A(\id_A)$. This assignment is clearly bijective. For arrows, we need to define an arrow $(A,\alpha)\to(B,\beta)$ for $\beta\in TB$ whenever the natural transformation $\eta': \hom(B,-)\to T$ satisfies a commutative triangle as above, i.e. $\eta'\circ \Yon(f)=\eta$. In other words we want 
\[
\alpha=(\eta'\circ \Yon f)_A(\id_A)=\eta'_A (f)=\eta'_A\circ \hom(f)(\id_B)=Tf\circ\eta'_B(\id_B)=Tf(\beta)
\]
It is then clear that the category whose objects are pairs $(A,\alpha)$ where $A$ is in $\cat$ and $\alpha\in TA$ and where there exists an arrow $(A,\alpha)\to (B,\beta)$ whenever there exists a map $f:B\to A$ such that $\alpha=Tf(\beta)$ is another implementation of $\el$. Note that the reversal of arrows from $\mathbf{C}$ to $\el$ is conventional and is not universal in the literature, but is convenient for our purpose, since we are interested in covariant functors. But in order to reason about the category $\el$ itself it often is conceptually and notationally convenient to forget about the arrow reversal and consider $\el\op$ instead, since we then have an arrow $(A,\alpha)\to (B,\beta)$ whenever there exists a map $f:A\to B$ such that $\beta=Tf(\alpha)$, and then proceed by duality. The category $\el\op$ is also sometimes called the Grothendieck construction\index{Grothendieck construction}, although this terminology usually has a broader meaning in the context of fibrations. As the following examples will illustrate, the category $\el$ can often be seen as a generalisation of the concept of `pointed space'.

\begin{example}
\begin{enumerate}
\item If $T$ is taken to be the identity functor on $\Set$, then $\el[\id]\op$ is just the category whose objects are pairs $(X,x)$ with a distinguished $x\in X$ and morphisms in $\el\op$ send distinguished elements to distinguished elements. In other words, $\el$ is the category of pointed sets.\index{Pointed set}
\item Consider the finitary covariant powerset functor $\powf:\Set\to\Set$. The objects of $\el[\powf]\op$ are pairs $(X, \{x_1,\ldots,x_n\})$ where $x_i\in X, 1\leq i\leq n$ and there is a morphisms between $(X, \{x_1,\ldots,x_n\})$ and $(Y, \{y_1,\ldots,y_m\})$ only if there exists a function $f: X\to Y$ such that $\powf f(\{x_,\ldots, x_n\})=f[\{x_,\ldots, x_n\}]=\{y_1,\ldots,y_m\}$. This is a generalisation of the previous example in the sense that a set is equipped with a distinguished element, but this time the set of of the form $TX=\powf X$.
\item Next, consider the forgetful functor $\Forg: \BA\to\Set$. An object of $\el[\Forg]\op$ is a pair $(A,a)$ where $A$ is a boolean algebra and $a\in A$. There will exist a morphism between $(A,a)$ and $(B,b)$ in $\el[\Forg]\op$ if there exists a boolean homomorphism $f:A\to B$ such that $f(a)=b$. So $\el[\Forg]\op$ is the category of `pointed boolean algebras'.
\item For a slightly different type of example, recall that a group $G$ can be seen as a category with a single object and arrows for every element $g\in G$ (encoding `multiplication' by $g\in G$). A functor $T: G\to\Set$ defines a set $S$ with a group action, or $G$-set. In this case, $\el\simeq S$.
\end{enumerate}
\end{example}

As we shall now see, the category $\el\op$ (and therefore $\el$ too) is not very well behaved when it comes to the existence of limits and co-limits.

\begin{proposition}\label{ch3:prop:CatElHom}
The category $\el\op$ has an initial object iff $T=\hom(A,-)$ for some $A$ in $\cat$.
\end{proposition}
\begin{proof}
If $T=\hom(A,-)$, then $(A,\id_A)$ is the initial object of $\el\op$. Indeed, take any $(B,f)$ in $\el\op$ (i.e. $f\in \hom(A,B)$), we need to show that there exists a unique arrow $(A,\id_A)\to (B,f)$, i.e. that there must exist a unique arrow $!: A\to B$ such that $\hom(A,-)(!)(\id_A)=f$, where $\hom(A,-)(!): \hom(A,A)\to \hom(A,B)$ is the usual application of the covariant $\hom(A,-)$ functor to $!$, i.e. $\hom(A,-)(!)(\id_A)=!\circ \id_A$. Clearly, taking $!=f$ does the job.

Conversely, assume that $(A,\alpha)$ is a initial object in $\el\op$, then we show that this provides a bijection $\phi:\hom(A,B)\to TB$ for any object $B$ in $\cat$. Start with an element of $f\in\hom(A,B)$ and define $\phi(f)=Tf(\alpha)\in TB$. This assignment is injective because it defines an arrow $(A,\alpha)\to(B,Tf(\alpha))$ in $\el\op$ which must be unique since $(A,\alpha)$ is assumed to be initial. To see that $\phi$ is surjective, start with an element $\beta\in TB$, then $(B,\beta)$ is an object of $\el\op$ and since $(A,\alpha)$ is initial, there must exist a unique arrow $(A,\alpha)\to(B,\beta)$, and thus an arrow $f: A\to B$ in $\cat$ such that $Tf(\alpha)=\beta$, i.e. an element of $\hom(A,B)$ such that $\phi(f)=\beta$.
\end{proof}

Thus $\el$ has an terminal object iff $T$ is a $\hom$ functor. As we shall see later, having a terminal object means that $\el$ is filtered which will be an important feature of a category in what follows. More generally we have the following result. 

\begin{proposition}\label{ch3:prop:Kelly} If $\cat$ has limits of some class and $T$ preserves them, then $\el\op$ has these limits too. In particular, if $\cat$ is complete and $T$ is continuous, then $\el\op$ is complete.
\end{proposition}
\begin{proof}
See Proposition 4.87 in \cite{KellyEnriched}.
\end{proof}

We conclude this brief description of the category $\el$ with the following simple observation.

\begin{lemma}\label{ch3:lem:Elsmall}
If $\cat$ is small, then for any $T:\cat\to \Set$,  $\el$ is small.
\end{lemma}
\begin{proof}
By smallness of $\cat$ both $\cat$ and $\bigcup_{A\in \cat} TA$ are sets, and thus the objects $(A,\alpha), \alpha\in TA$ of $\el$ all lie in a set.
\end{proof}

\subsection{Colimits of representables}

\noindent We now use what is sometimes known as the co-Yoneda lemma (particularly in the context of enriched category theory) to build a representation of functors in $\Set^{\mathbf{C}}$ when $\mathbf{C}$ is small. This construction is a classic construction and is detailed e.g. in Proposition 1 of \cite{SheavesTopos} for the case of contravariant functors.

\begin{theorem} \label{ch3:thm:fstpresthm}
Let $\mathbf{C}$ be a small category and $T: \mathbf{C}\to\Set$ be a functor, then $T$ is isomorphic to a colimit of $\hom$-functors.
\end{theorem}
\begin{proof}
We need to define a diagram $\Diag_T: \mathbf{D}\to \Set^{\mathbf{C}}$ such that $T\simeq \colim \Diag_T$. We take as index category $\mathbf{D}$, the category of elements of $T$, and we use the Yoneda embedding $\Yon: \mathbf{C}\op\to \Set^{\mathbf{C}}$ together with the forgetful functor $\Forg_T$ to define a diagram $\Diag_T$ $$\Diag_T:\el \stackrel{\Forg_T}{\longrightarrow}\mathbf{C}\op\stackrel{\Yon}{\longrightarrow}\Set^{\mathbf{C}}$$ 

\noindent We claim that $T$ is its colimit, i.e.
$$T\simeq \colim \Diag_T $$
To see that $\colim \Diag_T$ is well-defined, note that since $\Set$ is co-complete, so is $\Set^{\cat}$, and thus since $\cat$ is small, $ \colim \Diag_T $ exists by Lemma \ref{ch3:lem:Elsmall}. To check that $T$ is the colimit, note first that by construction $T$ is a co-cone of the diagram, the arrows defining this co-cone being precisely those forgotten by $\Forg_T$ (when $\el$ is viewed in terms of natural transformations). Let's now check that it is a colimit, i.e. we consider any other co-cone $G$ of $\Diag_T$ and show that there exists a unique natural transformation $\theta: T\to G$. Consider the $\el$-morphism $\eta\to\eta'$ where $\eta:\Hom(A,-)\to T$ and $\eta':\Hom(B,-)$ are such that $\eta=\eta'\circ \Yon(f)$ for an $f: B\to A$ in $\cat$. 

$$
\xymatrix
{
\Diag_T(\eta)=\Hom(A,-)\ar[rr]^{\Yon(f)} \ar[dr]_\eta \ar@/_1pc/[ddr]_{\zeta} & & \Diag_T(\eta')=\Hom(B,-)\ar[dl]^{\eta'}\ar@/^1pc/[ddl]^{\zeta'} \\
& T\ar@{-->}[d]^\theta \\
& G
}
$$
Following our alternative characterisation of $\el$ and Yoneda, $\eta,\eta',\zeta,\zeta'$ correspond to elements $x(\eta)\in TA, x'(\eta')\in TB, y(\zeta)\in GA$ and $y'(\zeta')\in GB$ respectively which are related by $x(\eta)=Tf(x'(\eta'))$ and $y(\zeta)=Gf(y'(\zeta'))$. Define the transformation $\theta: T\to G$ by 
\begin{align*}
\theta_A &: TA\to GA, x(\eta)\mapsto y(\zeta)
\end{align*}
If $\theta$ is well-defined, then the above diagram is commutative since $\theta\circ \eta=\zeta$ iff they correspond by Yoneda to the same element of $y(\zeta)\in GA$. To see that this transformation is well-defined note since \emph{every} element of $TA$, for an $A$ in $\cat$, corresponds to a natural transformation $\eta$, and since these are the objects of $\el$ and $G$ is a co-cone there must exist a $\zeta$ corresponding to every $\eta$. Thus $\theta_A$ is defined everywhere on $TA$. The fact that $\theta_A$ is a functional follows from the fact that $G$ is a co-cone of $\Diag_T$.

\smallskip
Now we need to check that the transformation is natural, i.e. that for $f: B\to A$, $\theta_A\circ Tf=Gf\circ \theta_B$. Staring with $y(\eta')\in TB$, let $x(\eta)=Tf(y)$. This means that we have an arrow $\eta\to\eta'$ in $\el$ as in the diagram above and we therefore have 
\begin{align*}
\theta_A\circ Tf(y(\eta'))&=\theta_A(x(\eta)) \\
&=y(\zeta)\\
&=Gf(y'(\zeta'))\\
&=Gf\circ \theta_B(\eta')
\end{align*}
\end{proof}

Thus any functor $T:\cat\to\Set$ can be expressed as a colimit of representables when $\cat$ is small. More generally, we propose the following definition.

\begin{definition}[Functor presentation]\index{Presentation}
A \textbf{presentation} for a functor $T:\cat\to\Set$, $\cat$ small, is a diagram $\Diag:\cat[J]\to\Set^{\cat}$ which factors through the Yoneda embedding $\Yon:\cat\to\Set^{\cat}$ such that $T=\colim\Diag$.
\end{definition}

\begin{remark}\label{ch3:rem:analogy1}
This is a good place to start describing an analogy between $\Set$-functors on a small category and algebraic varieties. For a presentation $\Diag:\cat[J]\to\Set^{\cat}$ of a functor $T$, we can think of the objects of $\cat[J]$ as \emph{generators} and the morphisms of $\cat[J]$ as \emph{relations}. In particular, $\el$ can be seen as a canonical category of generators and relations for $T$. We will make this analogy more precise further on.
\end{remark}

\section{Some essential tools and concepts}\label{ch3:sec:tools}

Before we move on to describe special types of functor presentations, we need some essential concepts from the theory of locally presentable and accessible categories, and some essential tools from general category theory.

\subsection{Locally presentable and accessible categories}

\noindent The constraint that $\cat$ be small is too strong for many practical applications, not least the study of $\Set$ endofunctors. However, the construction of Theorem \ref{ch3:thm:fstpresthm} seems to require \textit{all} the objects of $\cat$. So what are we to do? A natural solution is to consider categories which are `spanned' by a small subcategory.  Given a functor $T:\cat\to\Set$ we can then apply Theorem \ref{ch3:thm:fstpresthm} to the restriction of $T$ to this small subcategory and find criteria under which this is enough to present the full functor $T$. To make this strategy precise we must first settle on a notion of `spanning a category'. Such a notion is provided by the theory of accessible and locally presentable categories, developed in great detail in  \cite{1989:MakkaiPare} and \cite{LPAC} which we will follow in this subsection. We will only mention the essential definitions here and will discuss the salient facts as they become necessary in the remained of the chapter. We refer the reader to \cite{LPAC} for further details. We assume throughout that $\lambda$ is a regular cardinal and that all categories are locally small.

\begin{definition} A non-empty partially ordered set $J$ is $\lambda$-\textbf{directed}\index{Directed set} if every subset $J_0\subseteq J$ such that $|J_0|\leq\lambda$ has an upper bound. If $J$ is $\lambda$-directed, the colimit of a diagram $D: J\to \cat$ is called a $\lambda$-\textbf{directed colimit}\index{Directed colimit}. An object $A$ of a locally small category $\cat$ is said to be $\lambda$-\textbf{presentable} if $\hom(A,-)$ preserves all $\lambda$-directed colimits in $\cat$. The $\aleph_0$-presentable objects are called \textbf{ finitely presentable}\index{Presentable} and an object will be simply called \textbf{presentable} if it is $\lambda$-presentable for some $\lambda$.
\end{definition}

An equivalent characterisation of $\lambda$-presentable objects is as follows. Assume $\Diag: J\to \cat$ is a $\lambda$-directed diagram and $B=\colim \Diag=\colim_{j\in J} B_j$ is its $\lambda$-directed colimit. If there exists a morphism $f:A\to B$, then $A$
is $\lambda$-presentable iff $f$ factorises essentially uniquely through one of the morphisms $c_j: B_j\to B$ of the colimiting co-cone, where 'essentially uniquely' means that if $f$ factors through through $c_j$ in two different ways, i.e. $f=c_j\circ g=c_j\circ g'$, then we can find an $i\geq j$ where these two different ways are unified, i.e. such that $\Diag(j\to i)\circ g=\Diag(j\to i)\circ g'$. The $\lambda$-presentable objects will form the basis or generators of the notion of `spanning a category' which we will use. We now need to specify how to assemble these basic objects.

\begin{definition}
A locally small category $\cat$ will be called $\lambda$-\textbf{accessible}\index{Accessible category} if it has $\lambda$-directed colimits and a small subcategory $\cat_\lambda$ of $\lambda$-presentable objects such that every object in $\cat$ is a $\lambda$-directed colimit of objects in $\cat_\lambda$. If in addition $\cat$ is cocomplete (i.e. $\cat$ has \emph{all} colimits, not just $\lambda$-directed ones), then $\cat$ is said to be \textbf{locally} $\lambda$-\textbf{presentable}\index{Locally presentable category}. Again, for $\lambda=\aleph_0$ we will say finitely accessible and locally finitely presentable, whilst accessible and locally presentable will mean $\lambda$-accessible and $\lambda$-locally presentable for some $\lambda$.
\end{definition}

\begin{example}
\begin{enumerate}
\item The standard example is that of the category $\Set$ which is locally finitely presentable. A set $X$ is finitely presentable iff it is finite, and since any set is the directed colimit (i.e. union in this instance) of its finite subsets, it is finitely accessible. Since $\Set$ is also cocomplete, it is locally finitely presentable.
\item The category $\BA$ is also locally finitely presentable. A boolean algebra $A$ is finitely presentable iff it can be presented by finitely many generators and finitely many relations. This is in fact a general property of all varieties with a finitary signature. For details of why this is the case and why $\BA$ and all finitary varieties are locally finitely presentable, we refer the reader to chapter 3 of \cite{LPAC}.
\item As we will prove later, the category of functors $\Set^{\cat}$ is also locally finitely presentable when $\cat$ is small.
\end{enumerate}
\end{example}

In light of these examples, let us clarify the status of $\cat_\lambda$. There are two ways of looking at the smallness of $\cat_\lambda$. The first is to consider $\cat_\lambda$ as the entire subcategory of $\lambda$-presentable objects. The correct notion of smallness is then to say that $\cat_\lambda$ is \emph{essentially small}, i.e. that $\cat_\lambda$ has a small skeleton, i.e. that it is locally small and has a small number of isomorphism classes of objects. This is the approach taken in \cite{1989:MakkaiPare}. The other approach, taken in \cite{LPAC}, is to assume that a choice of representative of each isomorphisms class of $\lambda$-presentable object has been made, such as taking the finite ordinals in the case of $\Set$, in which case $\cat_\lambda$ is small in the usual sense. This is the approach we will choose throughout. 

It is often convenient to work with a class of colimits which is equivalent to directed colimits. A category $\cat[J]$ is called $\lambda$-\textbf{filtered}\index{Filtered category} if it satisfies the following conditions:
\begin{enumerate}
\item it is non-empty
\item if $I$ is a set of cardinality strictly smaller than $\lambda$ and $\{A_i\}_{i\in I}$ are objects in $\cat[J]$, then there exists an object $B$ and $\cat[J]$-morphisms $f_i: A_i\to B$
\item if $I$ is a set of cardinality strictly smaller than $\lambda$ and $\{f_i:A\to B\}_{i\in I}$ are $\cat[J]$-morphisms, there exists a $\cat[J]$-morphism $h: B\to C$ which simultaneously equalizes all $f_i, i\in I$ 
\end{enumerate} 
We will say `filtered' for $\aleph_0$-filtered. A $\lambda$-filtered diagram is a diagram $D:\cat[J]\to\cat$ where $\cat[J]$ is $\lambda$-filtered. A $\lambda$-\textbf{filtered colimit}\index{Filtered colimit} is the colimit of such a diagram. We cite the following two useful results without proof.

\begin{lemma}
A category has $\lambda$-filtered colimits iff it has $\lambda$-directed colimits. A functor $T:\cat\to \Set$ from such a category preserves $\lambda$-filtered colimits iff it preserves $\lambda$-directed colimits
\end{lemma} 
\begin{proof}
See \cite{LPAC} 1.7.
\end{proof}

\begin{lemma}\label{ch3:lem:colimofpres}
A colimit of a $\lambda$-small diagram of $\lambda$-presentable objects is $\lambda$-presentable.
\end{lemma}
\begin{proof}
See \cite{LPAC} 1.16.
\end{proof}

This result generalises as follows.

\begin{lemma}
Let $\lambda, \mu$ be two regular cardinals, then $\lambda$ is said to be \emph{sharply smaller}\index{Sharply smaller} than $\mu$, written $\lambda\triangleleft \mu$, if $\lambda<\mu$ and for any set $X$ such that $|X|<\mu$, $\pow_\lambda(X)$ has a cofinal set of cardinality strictly smaller than $\mu$. Assume now that $\cat$ is a $\lambda$-accessible category and $\lambda\triangleleft\mu$, then an object $A$ in $\cat$ is $\mu$-presentable iff it is a split subobject of a $\mu$-small $\lambda$-directed colimit of $\lambda$-presentable objects.
\end{lemma}
\begin{proof}
See \cite{1989:MakkaiPare} 2.3.11.
\end{proof}

The relation $\triangleleft$ turns out to be very important in the study of accessible categories, for example if a category is $\lambda$-accessible, and $\lambda\triangleleft\mu$, then it is also $\mu$-accessible. As a special example of such a $\lambda\triangleleft\mu$ relation, note that $\lambda\triangleleft\lambda^+$, where $\lambda^+$ is the successor cardinal of $\lambda$. It is also worth mentioning that $\triangleleft$ is transitive.

To conclude this brief introduction to locally presentable and accessible categories we need to define the following crucial class of functors.

\begin{definition}\index{Accessible functor}
Let $\cat$ be a category with $\lambda$-directed colimits and let $\cat[D]$ be a $\lambda$-accessible category. We will say that a functor $T:\cat\to\cat[D]$ is $\lambda$-\textbf{accessible} if it preserves $\lambda$-directed colimits.\end{definition}

Given an accessible category $\cat$, this property on a functor $\cat\to\Set$ will be the key to lifting Theorem \ref{ch3:thm:fstpresthm} from the small subcategory of presentable objects of $\cat$ to the full category $\cat$.

\subsection{Kan extensions and cofinal diagrams} 

We now present two extremely useful technical concepts on which we will rely in what follows. The notion of Kan extension is very useful in the context of extending  a functor defined on a subcategory to the entire category. Here we will merely recall the definition and a few easy but salient facts, for more details the reader is referred to Chapter X of \cite{MacLane}. Since we are dealing with colimits in this Chapter, the useful notion is that of \emph{left} Kan extension (in Section \ref{ch5:sec:semcomp} we will deal with limits and use right Kan extensions).

\begin{definition}\index{Kan extension}
Given two functors $\cat[B]\stackrel{F}{\longleftarrow}\cat[A]\stackrel{G}{\longrightarrow}\cat$, a left Kan extension $\Lan_G F$ of $F$ along $G$ is a functor $\Lan_G F: \cat\to\cat[B]$ together with a natural transformation $\eta: F\to \Lan_G F\circ G$. The pair $(\Lan_G F, \eta)$ is universal in the sense that for any other pair $(L',\eta')$ with $L': \cat\to\cat[B]$ and $\eta': F\to L'\circ G$, there must exist a natural transformation $\zeta: \Lan_G F\to L'$ such that $\eta'=\zeta_G\circ\eta $ where $\zeta_G$ is the obvious natural transformation $\zeta_G: \Lan_G F\circ G\to L'\circ G$.
\end{definition}

\noindent In light of the definition of accessible and presentable categories, let $$\Inc_\lambda: \cat_\lambda\to \cat$$ be the inclusion functor of the subcategory of $\lambda$-presentable objects of $\cat$. Let also $T:\cat\to \Set$ be an arbitrary functor and $T_\lambda=T\Inc_\lambda: \cat_\lambda\to\cat$ be its restriction to the subcategory $\cat_\lambda$. We will examine the left Kan extension $\Lan_{\Inc_\lambda}T_\lambda$. First recall,

\begin{proposition}\label{ch3:prop:KanExists}
Let $\cat[B]\stackrel{F}{\longleftarrow}\cat[A]\stackrel{G}{\longrightarrow}\cat$ such that $\cat[A]$ is small and $\cat[B]$ is cocomplete, then $F$ has a left Kan extension $\Lan_G F$ along any $G$.
\end{proposition}
\begin{proof}
See \cite{MacLane} X.3.2.
\end{proof}

Thus $\Lan_{\Inc_\lambda}T_\lambda$ always exists. Its explicit construction is given by
\[
\Lan_{\Inc_\lambda}T_\lambda A=\colim \left((\cat_\lambda\downarrow A)\stackrel{\Diag_c^a}{\longrightarrow}\cat_\lambda\stackrel{T_\lambda}{\longrightarrow}\Set\right)
\]
where $\Diag^A_c$ is the functor which projects an object of $\cat_\lambda\downarrow A$ to its domain (and is called the canonical diagram, see below)\index{Canonical diagram}. Moreover, we have the following:

\begin{proposition}\label{ch3:prop:KanInclusion}
Let $\cat[B]\stackrel{F_0}{\longleftarrow}\cat[A_0]\stackrel{I}{\longrightarrow}\cat[A]$ such that $\cat[A_0]$ is a small full subcategory of $\cat[A]$ and $\cat[B]$ is cocomplete, then $F_0$ has a left Kan extension $\Lan_I F_0$ whose associated natural transformation $\eta: F_0\to \Lan_I F_0\circ I$ is the identity natural transformation.
\end{proposition}
\begin{proof}
See \cite{MacLane} X.3.3.
\end{proof}

The fact that $\Lan_{\Inc_\lambda}T_\lambda$ can be written as a colimit suggests that if this colimit is filtered and if the functor $F$ is accessible we should be able to recover $T$ from $T_\lambda$. This is indeed the case, but to show it we will need the following concept, which will prove very useful later on in this chapter. 

\begin{definition}
A functor $G: \cat[A]\to \cat[B]$ is called \textbf{cofinal}\index{Cofinal functor} (or sometimes `final', see e.g. \cite{MacLane}) if for any object $B$ of $\cat[B]$ the comma category $B\downarrow G$ is no-empty and connected, i.e. if the following two conditions hold for any $B$
\begin{enumerate}[(i)]
\item there exists an object $A$ of $\cat[A]$ and a morphism $B\to GA$
\item any two objects in $B\downarrow G$, i.e. any two morphisms $B\to GA$ and $B\to GA'$,  are related by a zigzag of morphisms:
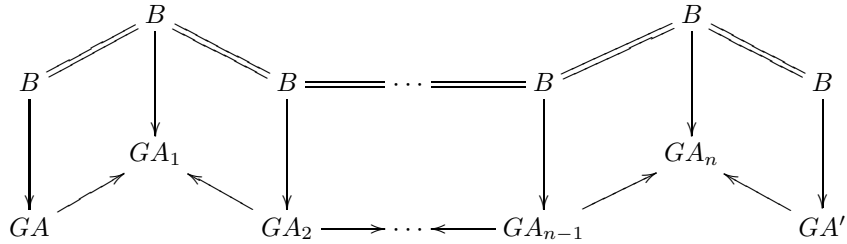
\[
\xymatrix@R=3ex
{
& B\ar[dd]\ar@{=}[dl]\ar@{=}[dr] & & & & B\ar[dd]\ar@{=}[dl]\ar@{=}[dr] \\
B\ar[dd] & & B\ar[dd] & \cdots\ar@{=}[l] \ar@{=}[r] & B\ar[dd] & & B\ar[dd] \\
& GA_1  & & & & GA_n \\
GA\ar[ur] & & GA_2\ar[ul]\ar[r]  & \cdots  & GA_{n-1}\ar[ur]\ar[l] & & GA' \ar[ul] 
}
\]
(or the same pattern with the arrows between $G$-terms reversed).
\end{enumerate}
\end{definition}

Cofinal functors are very useful for computing colimits. Indeed, if a functor $G:\cat[A]\to\cat[B]$ is cofinal, then we can restrict diagrams $\Diag$ on $\cat[B]$ to diagrams on $\cat[A]$ along $G$ without altering their colimits. In this context we will speak of a \textbf{cofinal diagram} rather than cofinal functor.

\begin{theorem}\label{ch3:thm:cofinal}
If $G: \cat[A]\to\cat[B]$ is cofinal and $\Diag: \cat[B]\to \cat[C]$ is a diagram such that $\colim \Diag G$ exists, then $\colim \Diag\simeq\colim \Diag G$.
\end{theorem}
\begin{proof}
See \cite{MacLane}, Chapter IX, Section 3, Theorem 1.
\end{proof}

We will be particularly interested in the case where $G$ is an inclusion functor and $\cat[A]$ and $\cat[B]$ are directed or filtered categories. Note that if $G$ is full and $\cat[A]$ is filtered, then the existence of a zigzag of morphisms between any two objects $B\to GA$ and $B\to GA'$ of $B\downarrow G$ implies that the span $GA\leftarrow B\to GA'$ can be completed into a commutative square
$$
\xymatrix
{
B\ar[r]\ar[d] & GA\ar[d] \\
GA'\ar[r] & GA''
}
$$
which, together with the non-emptiness of $B\downarrow G$, is the - slightly non-standard but perhaps more \emph{cofinal} - definition of cofinality given in 0.11 of \cite{LPAC}. 

Let us now return to the question of determining whether the restriction $T_\lambda$ of a functor $T$ whose domain is accessible determines $T$. To answer the question positively we will need the following important lemma.

\begin{lemma}\label{ch3:lem:candiag} Let $\cat$ be a $\lambda$-accessible category whose subcategory of $\lambda$-presentable objects is $\cat_\lambda$, and let $A$ be an object of $\cat$. Define the \textbf{canonical diagram} to be the projection functor $\Diag_c^A: \cat_\lambda\downarrow A\to\cat$ defined above, then
\begin{enumerate}[(i)]
\item $\cat_\lambda\downarrow A $ is a $\lambda$-filtered category
\item $A=\colim\Diag_c^A$
\end{enumerate}
\end{lemma}
\begin{proof}
We start by showing (i). By definition of a $\lambda$-accessible category, $A$ must be expressible as a $\lambda$-directed colimit $A=\colim_{j\in J} A_j$ of $\lambda$-presentable objects. We will use this colimit to show that $\cat_\lambda\downarrow A$ is $\lambda$-filtered. Note first that $\cat_\lambda\downarrow A$ is clearly non-empty, or else $J$ would have to be empty which would contradict the fact that it is $\lambda$-filtered. Note also that if we take any object of $\cat_\lambda\downarrow A$, i.e. any arrow $B\to A$ with $B$ $\lambda$-presentable, then by definition of $\lambda$-presentability $B\to A=\colim_{j\in J}A_j$ must factor through an essentially unique $c_j: A_j\to A$. If we select strictly fewer than $\lambda$ such objects $f_\kappa: B_\kappa\to A, \kappa<\lambda$, then each $f_\kappa$ must factor through some map $c_{j_\kappa}:A_{j_\kappa}\to A$ and since $J$ is $\lambda$-directed, there must exist $c_j: A_j\to A$ such that $c_j\circ(A_{j_\kappa}\to A_j)=c_{j_\kappa}$ for each $\kappa$ and thus such that $f_\kappa=c_{j}\circ (B_\kappa\to A_{j_\kappa}\to A_j)$ for each $\kappa$, i.e. we have arrows in $\cat_\lambda\downarrow A$ from every $B_\kappa\to A$ to a common codomain. The situation is illustrated by the following commutative diagram.
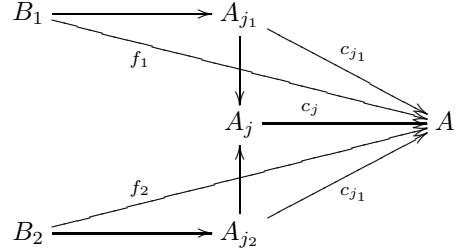
\[
\xymatrix@R=6ex@C=14ex
{
B_1 \ar[drr]_<<<<<<<<<<<<<{f_1}\ar[r] & A_{j_1}\ar[dr]^{c_{j_1}}\ar[d]\\
& A_j\ar[r]^<<<<<<<{c_j} & A \\
B_2\ar[urr]^<<<<<<<<<<<<<{f_2}\ar[r] & A_{j_2}\ar[u]\ar[ur]_{c_{j_1}}
}
\]

Similarly, given strictly fewer than $\lambda$ $\cat_\lambda\downarrow A$-morphisms $f_\kappa:(B_1\to A)\to(B_2\to A), \kappa<\lambda$ we need to find a morphism $g: (B_2\to A)\to (C\to A)$ equalizing all $f_\kappa$s simultaneously. By the above comment, there must exist $c_j:A_j\to A$ such that $B_2\to A$ factors through $A_j$. But this means that $B_1\to A$ also factors through $c_j$, and it does so in strictly fewer than $\lambda$ different ways (i.e. $c_j\circ(B_2\to A_j)\circ f_\kappa, 1\leq \kappa<\lambda$). Since any such factorization must be essentially unique for $B_1$ to be $\lambda$-presentable, there must exist $j\leq k \in J$ such that $(B_2\to A_j\to A_k)$ equalizes every $f_\kappa$. The morphisms $A_k\to A$ and $(B_2\to A_j\to A_k)$ thus give us the $\cat_0\downarrow A$-morphism we were looking for, as can be seen from the commutativity of the following diagram:
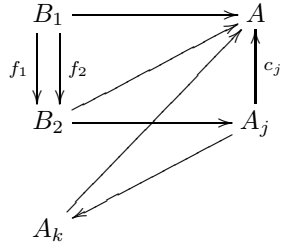
\[
\xymatrix@R=6ex@C=14ex
{
B_1 \ar[r]\ar@<1ex>[d]^{f_2} \ar@<-1ex>[d] _{f_1} & A \\
B_2\ar[r]\ar[ur] & A_j\ar[u]_{c_j}\ar[dl] \\
A_k\ar[uur]
}
\]

To show (ii) we will use the idea of a cofinal diagram developed earlier.  By definition we know that $A$ can be expressed as a certain $\lambda$-directed diagram $\Diag^A: \cat[J]\to \cat$ of $\lambda$-presentable objects. In other words, for any $j\in \cat[J]$ (since $\cat[J]$ is a set), $\Diag^Aj $ is by definition an object of $\cat_\lambda$ from which there exists a morphism $c_j$ (part of the cocone formed by $A$) whose codomain is $A$, i.e. an object of $\cat_\lambda\downarrow A$. We can therefore define the functor $I: \cat[J]\to \cat_\lambda\downarrow A$ by $I(j)=c_j$ and $I(i\to j)$ as the obvious morphism between $c_j$ and $c_i$ defined by the fact that $A$ is a cocone of $\Diag^A$. Clearly, $\Diag_c^A\circ I=\Diag^A$, and so by Theorem \ref{ch3:thm:cofinal}, if we can show that $I$ is cofinal, then we will have shown that $$A=\colim \Diag^A=\colim \Diag_c^A\circ I=\colim \Diag_c^A$$
Let us pick any $f: B\to A$ in $\cat_\lambda\downarrow A$. Since $B$ is presentable and $A=\colim \Diag^A=\colim_{j\in \cat[J]}A_j$, there must exist a presentable object $A_j$ such that $f$ factors through $c_j: A_j\to A$. This provides us with an arrow $(B\to A)\to I(j)$ as required. Now, for the zigzag condition, assume that we have morphisms $(B\to A)\to (A_i\to A)$ and $(B\to A)\to (A_j\to A)$ in $\cat_\lambda\downarrow A$ where $(A_i\to A)=I(i)$ and similarly for $j$. Since $\cat[J]$ is directed there must exist $k$ such that $i,j\leq k$ and consequently we must have
$$
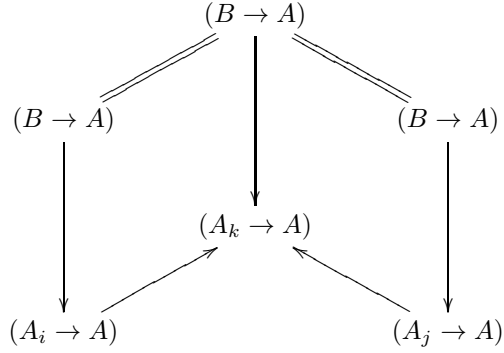
\xymatrix
{
& (B\to A)\ar[dd]\\
(B\to A)\ar@{=}[ur]\ar[dd]& & (B\to A)\ar@{=}[ul] \ar[dd]\\
& (A_k\to A) \\
(A_i\to A) \ar[ur]& & (A_j\to A)\ar[ul]
}
$$
which provides us with a zigzag of morphisms as required.
\end{proof}

\begin{corollary}\label{ch3:cor:pressizemin}
Let $\cat$ be a non-small $\lambda$-accessible category, then the set of objects in $\cat_\lambda$ has cardinality at least $\lambda$.
\end{corollary}
\begin{proof}
Assume, for the sake of contradiction that $|(\cat_\lambda)\ob|<\lambda$. Take any object $A$ in $\cat$. From Lemma \ref{ch3:lem:candiag} we know that $A=\colim\Diag_c^A$. But since $|(\cat_\lambda)\ob|<\lambda$ and a $\lambda$-small colimit of $\lambda$-presentable objects is $\lambda$-presentable by Lemma \ref{ch3:lem:colimofpres}, this means that $A$ is $\lambda$-presentable, and thus $\cat=\cat_\lambda$ which is a contradiction since we're assuming that $\cat$ is not small.
\end{proof}

We can now state the main result of this subsection.
\begin{proposition}\label{ch3:prop:LanF0}  Let $\cat$ be a $\lambda$-accessible category whose subcategory of $\lambda$-presentable objects is $\cat_\lambda$ and let $\Inc_\lambda:\cat_\lambda\to\cat$ be the injection functor. For any $\lambda
$-accessible functor $T:\cat\to\Set$, with restriction $T_\lambda:\cat_\lambda\to\Set$, it is the case that $$\Lan_{\Inc_\lambda} T_\lambda= T$$
\end{proposition}
\begin{proof} So far we have used the notation $\Diag_c^A$ to denote the projection functors with codomain $\cat_\lambda$ or $\cat$ depending on the context. Here we will need to draw the distinction, and the two variants are introduced in the following trivially commutative diagram:
\[
\xymatrix
{
& \cat_\lambda\downarrow A\ar[dl]_{\Diag_{c_\lambda}^A}\ar[dr]^{\Diag_c^A} \\
\cat_\lambda\ar@{^{(}->}^{\Inc_\lambda}[rr]\ar[dr]_{T_\lambda} & & \cat\ar[dl]^{T} \\
&\Set
}
\]
By construction we have 
\begin{align*}
\Lan_{\Inc_\lambda} T_\lambda A &=\colim \left((\cat_\lambda\downarrow A)\stackrel{\Diag_{c_\lambda}^A}{\longrightarrow}\cat_\lambda\stackrel{T_\lambda}{\longrightarrow}\Set\right) \\
&\stackrel{1}{=}\colim \left((\cat_0\downarrow A)\stackrel{\Diag_c^A}{\longrightarrow}\cat\stackrel{T}{\longrightarrow}\Set\right) \\
&\stackrel{2}{=}T\colim \Diag_c^A \\
&\stackrel{3}{=}TA
\end{align*}
where step 1 is by definition of $T_\lambda$ and the commutativity of the diagram above, step 2 is by the fact that $\Diag_c^A$ is a $\lambda$-filtered diagram (by Lemma \ref{ch3:lem:candiag}) and that $T$ is $\lambda$-accessible, and step 3 is by the fact that $\colim\Diag_c^A=A$ by Lemma \ref{ch3:lem:candiag}.
\end{proof}

\section{Size-bound functor presentations}\label{ch3:sec:sizeBoundPres}

We now return to the functors $T:\cat\to\Set$ where $\cat$ is small. As was shown in Theorem \ref{ch3:thm:fstpresthm}, any such functor has a presentation arising from its category of elements $\el$. In this section we will consider cases where simpler presentations of $T$ are possible, and relate the existence of such presentations to the status of $T$ in $\Set^{\cat}$.

\subsection{$\lambda$-presentable and $\lambda$-presented functors}\label{ch3:sec:presCsmall}

We start with the following obvious definition.

\begin{definition}[Presentable functor]\index{Presentable functor} We will say that a functor $T:\cat\to\Set$ ($\cat$ arbitrary) is $\lambda$-\textbf{presentable} if it is a $\lambda$-presentable object in $\Set^{\cat}$, i.e. if $\Nat(T,-)$ preserves $\lambda$-directed colimits. As usual if $\lambda=\aleph_0$ we will say \emph{finitely} presentable\index{Finitely presentable!functor}.
\end{definition}

We have the following easy example of presentable functors.

\begin{proposition}\label{ch3:prop:hompres}
If $\cat$ is a small category, then all $\hom$ functors $\hom(A,-):\cat\to\Set$ are $\lambda$-presentable.
\end{proposition}
\begin{proof}
In fact we will show a stronger result, namely that $\hom$ functors are \emph{absolutely presentable} (see \cite{2005:cococo})\index{Absolutely presentable} in $\Set^{\cat}$, i.e. that for any object $A$ of $\cat$, $\Nat(\hom(A,-),-)$ preserves \emph{all} colimits, and a fortiori the $\lambda$-directed ones. Moreover, the smallness of $\cat$ is not necessary, and the same proof works for $\cat$ locally small. This latter requirement cannot be relaxed as the Yoneda lemma is key.

Let $A$ be an object of $\cat$, and let $\Diag: \cat[J]\to \Set^{\cat}$ be a diagram whose colimit is $G$. We need to show that if there exists a natural transformation $\eta:\hom(A,-)\to G$, then it must factor through an essentially unique object $\Diag(j)$ for a certain object $j$ of $\cat[J]$. For notational clarity, let us write $G_j=\Diag(j)$, i.e. $G=\colim_j G_j$. By Yoneda's lemma  any morphism $\eta:\hom(A,-)\to G$ corresponds to a unique element $\alpha\in GA$ where $\alpha=\eta_A(\id_A)$. Since colimits in $\Set^{\cat}$ are computed pointwise, $GA$ is the colimit of the diagram  $\{G_j A\}_{j\in J}$ in $\Set$ and thus, by the general construction of colimits in $\Set$, $\alpha\in GA$ must originate from an essentially unique $\beta\in G_j A$. By using Yoneda in the opposite direction, it is clear that this $\beta$ corresponds to a natural transformation $\theta: \hom(A,-)\to G_j$ defined by $\theta_A(\id_A)=\beta$, and for any $f:A\to B$ by $\theta_B(f)=Tf(\beta)$. If we denote by $c_j$ the natural transformation $c_j: G_j\to G$, we can then check that $\eta=c_j\circ \theta$. For every $B$ in $\cat$ we have:
\begin{align*}
\eta_B(f)&=\eta_B\circ \hom(A,-)(f)(\id_A) \\
&=Gf\circ \eta_A(\id_A) & \eta\text{ natural}\\
&= Gf(\alpha) & \text{Definition of }\alpha\\
&=Gf((c_j)_A(\beta)) & \text{Definition of }\beta\\ 
&= Gf\circ (c_j)_A\circ \theta_A(\id_A) & \text{Definition of }\theta\\
&=(c_j)_B\circ \theta_B\circ \hom(A,-)(f)(\id_A) & c_j,\theta\text{ natural}\\
&=(c_j)_B\circ \theta_B(f)
\end{align*}
i.e. $\eta$ factors through $\theta$ as desired. The essential unicity of the factorisation is a consequence of the essential unicity of $\beta$ and of the corresponding $\theta$ by Yoneda's lemma.
\end{proof}

We can in fact strengthen the previous Proposition \ref{ch3:prop:hompres} as follows:

\begin{proposition}\label{ch3:prop:homifflexpres}
Let $\cat$ be a non-empty small finitely complete category and $T:\cat:\to\Set$. T is left-exact (i.e. preserves finite limits) and finitely presentable in $\Set^{\cat}$ iff it is a $\hom$ functor.
\end{proposition}
\begin{proof}
The `if' direction follows from the previous Proposition \ref{ch3:prop:hompres} and the fact that (covariant) $\hom$ functors preserve limits. For the `only if' direction we will first show that $\el$ is filtered. Since $\cat$ is non empty, $\el$ must be non-empty too. Moreover, since $\cat$ has finite limits it must have products and equalizers (in fact these two requirements are equivalent). Thus for any $(A,\alpha), (B,\beta)$ in $\el$ we can build the object $(A\times B, (\alpha,\beta))$ in $\el$ and since $T$ is left-exact the projections give us arrows $T\pi_1: T(A\times B)=TA\times TB\to TA$  and $T\pi_1: T(A\times B)=TA\times TB\to TB$ such that $T\pi_1(\alpha,\beta)=\alpha$ and $T\pi_2(\alpha,\beta)=\beta$, i.e. arrows in $(A,\alpha)\to (A\times B,(\alpha,\beta))$ and $(B,\beta)\to (A\times B,(\alpha,\beta))$ in $\el$. Similarly, given two arrows $(A,\alpha)\rightrightarrows (B,\beta)$, i.e. arrows $f,g: B\to A$ such that $Tf(\beta)=Tg(\beta)=\alpha$, we can build the equalizer $e: E\to B$ of $f,g$ in $\cat$. Since $T$ preserves equalizers, we know that $Te$ equalizes $Tf$ and $Tg$ and thus since $Tf(\beta)=Tg(\beta)$, there must exist an element $\gamma$ in the equalizer $TE$ such that $Te(\gamma)=\beta$. We therefore have an $\el$-arrow $(B,\beta)\to(E,\gamma)$ which composes with the two arrows  $(A,\alpha)\rightrightarrows (B,\beta)$ to give a unique arrow $(A,\alpha)\to(E,\gamma)$.

Since $\el$ is filtered, it is obvious that if we write $T$ as a colimit of representables following the construction of Theorem \ref{ch3:thm:fstpresthm}, $T$ can in fact be written as a \emph{filtered} colimit of representables. Since it is assumed to be finitely presentable in $\Set^{\cat}$ it means that the identity natural transformation $\Id_T: T\to T$ must factor uniquely through one of the $\hom$ functors of the filtered colimit representing $T$. This in turn, means that $T$ must be equivalent to this $\hom$ functor.
\end{proof}

In fact, we can demand slightly less that left-exactness. If we instead require that $T$ \emph{weakly} preserve limits, then the proof goes through unchanged. Note also that weak preservation of limits is equivalent to weak preservation of pullbacks and terminal objects. The proof can be generalised to higher cardinalities, in which case left exactness must be generalised to $\lambda$-left exactness, i.e. the preservation of $\lambda$-small limits. We can also characterize left-exact functors.

\begin{proposition}\label{ch3:prop:lexiffelfiltered}
Let $\cat$ be a small category, then $T:\cat\to\Set$ is left exact iff $\el$ is filtered.
\end{proposition}
\begin{proof}
Assume first that $T$ is left exact, then as we saw in the proof of the previous Proposition \ref{ch3:prop:homifflexpres}, $\el$ is filtered. Let is now assume that $\el$ is filtered, and let $\Diag:\cat[J]\to\cat$ be a finite diagram, we then have
\begin{align*}
T(\lim \Diag)&=\colim(\Yon\Forg_T)(\lim \Diag)\\
&\stackrel{1}{=}\colim(ev_{\lim\Diag}\Yon\Forg_T)\\
&\stackrel{2}{=}\colim(\lim ((ev_{(-)}\Diag)\Yon\Forg_T))\\
&\stackrel{3}{=}\lim(\colim ((ev_{(-)}\Diag)\Yon\Forg_T))\\
&\stackrel{4}{=}\lim(\colim(\Yon\Forg_T)\Diag)\\
&=\lim(T\Diag)
\end{align*}
where (1) is because colimits are evaluated pointwise in $\Set^{\cat}$, (2) follows from the fact that both $\Yon$ and $\Forg_T$ preserves limits, (3) is the well-known fact that finite limits commute with filtered colimits (see \cite{MacLane} IX.2.), and (4) is because each evaluation functor $ev_A:\Set^{\cat}\to\Set$ preserves colimits (see \cite{MacLane} V.3.).
\end{proof}

One of the motivations behind the concept of finitely present\emph{able} object in a category, is that in the case of algebraic varieties, it provides and abstract characterisation of the objects which can be given a finite presentation by generators and relations, i.e. finitely present\emph{ed} objects. And in fact, as shown in Theorem 3.12. of \cite{LPAC}, the two concepts are equivalent. We would like to show the same sort of correspondence for functors $\Set\to\cat$ when $\cat$ is small. The purpose of this section is to isolate classes of functors for which the cardinality of the set of relations and/or generators is bounded. There are two natural ways to specify the cardinality of generators and relations, and two associated notions of finitely-presented and finitely-generated functors. Recall from Remark \ref{ch3:rem:analogy1} that for a presentation $\Diag:\cat[J]\to\Set^{\cat}$ of a functor $T$, we drew a parallel between the objects of $\cat[J]$ and generators and between the morphisms of $\cat[J]$ and relations.

\begin{definition}[Strongly $\lambda$-presented functor]\index{Strongly presented functor}
A functor $T:\cat\to\Set$ is \textbf{strongly $\lambda$-presented} if there exists a $\lambda$-small presentation $\Diag:\cat[J]\to \Set^{\cat}$ of $T$, i.e. a presentation such that $\cat[J]$ has fewer than $\lambda$ objects and fewer than $\lambda$ morphisms.\index{Finitely strongly presented!functor} As usual, we will say that $T$ is \textbf{finitely strongly presented} if it is strongly $\aleph_0$-presentable, i.e. if it can be written as a finite colimit of $\hom$ functors.
\end{definition}

The intuition behind the alternative notion of finite presentation is the following. We have seen that we can write any functor $T:\cat\to\Set$ as a colimit of representable functors. By the general construction of colimits as regular quotients of coproducts, this means that any functor $T$ can be viewed as a regular quotient of a coproduct of $\hom$ functors. From this perspective it is natural to consider the $\hom$ functors as `generators', and the following definition is then natural.

\begin{definition}[$\lambda$-presented functor]\index{Presented functor}
A functor $T:\cat\to\Set$ is $\lambda$-\textbf{presented} if there exist a set $I$ of cardinality $|I|<\lambda$, and a regular epimorphism (coequalizer) $e$
\[
\xymatrix
{
S\ar@<3pt>[r]\ar@<-2pt>[r] & \coprod\limits_{i\in I}\ar@{->>}[r]^>>>>>>{e} \hom(A_i,-) & T
}
\]
such that $S$ can itself be expressed as a regular quotient of a coproduct of fewer than $\lambda$ representable functors.
\end{definition}

In this definition, the only measure of `size' is the cardinality of the sets indexing coproducts of $\hom$s. The number of relations necessary to encode $T$ is therefore measured by the size of the functor encoding the relations, i.e. by the `size' of the domain of the pair of arrows coequalized by $e$. As suggested by the terminology, we have

\begin{proposition}\label{ch3:prop:strongPresPres}
If a functor $T:\cat\to\Set$ is strongly $\lambda$-presented, then it is $\lambda$-presented.
\end{proposition}
\begin{proof}
If $T$ is strongly finitely presented, then $T=\colim \Yon\circ\Diag$ with $\Diag:\cat[J]\to \cat\op$ such that $\cat[J]$ has  strictly fewer than $\lambda$ objects and fewer than $\lambda$ morphisms. The result follows immediately by expressing the colimit as a regular quotient of a coproduct in the canonical way, i.e.
\[
\xymatrix
{
\coprod\limits_{f\in \cat[J]\mor}\hom(\cod\Diag( f),-)\ar@<3pt>[r]^{\phi}\ar@<-2pt>[r]_{\psi} & \coprod\limits_{A\in \cat[J]\ob}\hom(\Diag(A),-)\ar@{->>}[r]^>>>>>>{e} & T
}
\]
where $\phi\circ \mathrm{in}_f=\mathrm{in}_{\cod f}$, $\psi\circ \mathrm{in}_f=\mathrm{in}_{\dom f}\circ ((-)\circ \Diag(f))$ and $\lambda<|\cat[J]\mor|,|\cat[J]\ob|$. We can then take $S=\coprod_{f\in \cat[J]\mor}\hom(\cod\Diag( f),-)$ and the identity as a trivial regular quotient in the definition of $\lambda$-presented.
\end{proof}

In fact, we can also show the converse.
\begin{proposition}\label{ch3:prop:StrongPresIffPres}
If a functor $T:\cat\to\Set$ is $\lambda$-presented, then it is also strongly $\lambda$-presented.
\end{proposition}
\begin{proof}
By unravelling the definition of being $\lambda$-presented, it is immediate that there exist sets $I,J$ of cardinality strictly less than $\lambda$ such that
\[
\xymatrix
{
\coprod\limits_{j\in J}\hom(B_j,-)\ar@{->>}[r]^>>>>>>{q} & S \ar@<3pt>[r]^>>>>>{\phi_1}\ar@<-2pt>[r]_>>>>>{\phi_2} & \coprod\limits_{i\in I}\hom(A_i,-)\ar@{->>}[r]^>>>>>>{e} & T 
}
\]
It is easy to check that since $q$ is epi, $e$ is also the coequalizer of $\phi_1\circ q,\phi_2\circ q$. So we can assume w.l.o.g. that we in fact have a coequalizer 
\[
\xymatrix
{
\coprod\limits_{j\in J}\hom(B_j,-) \ar@<3pt>[r]^>>>>>{\phi_1}\ar@<-2pt>[r]_>>>>>{\phi_2} & \coprod\limits_{i\in I}\hom(A_i,-)\ar@{->>}[r]^>>>>>>{e} & T 
}
\]
By the Yoneda lemma, the transformations $\phi_1,\phi_2$ are equivalent to the data for each $j\in J$ of pair maps $f_j^1: A_i\to B_j, f_j^2: A_k\to B_j$ for some indices $i,k$. We now define the category $\cat[J]$ as the subcategory of $\cat$ such that
\[
\cat[J]\ob=\{A_i\mid i\in I\}\cup\{B_j\mid j\in J\}
\]
and
\[
\cat[J]\mor=\text{closure of }\{f^1_j,f^2_j\mid j\in J\}\text{ under composition and identities}
\]
We claim that if we write $\Yon:\cat[J]\op\to\Set^{\cat}$ for the Yoneda transformation, then $T=\colim \Yon$. Since $\lambda$ is regular, we have that $|\cat[J]\ob|,|\cat[J]\mor|< \lambda$, and if $T=\colim \Yon$ then it is strongly $\lambda$-presentable, as desired. 

Let $T'=\colim \Yon$, we now work pointwise to establish $T'=T$. Let $X$ be any object of $\cat$, by assumption and the description of $\phi_1,\phi_2$ from the Yoneda lemma, $TX$ is the quotient of $\coprod_{i\in I}\hom(A_i,X)$ under the smallest equivalence relation generated by
\[
h_i: A_i\to X \sim_X h_k: A_k\to X
\]
if there exist $j\in J$ and $g: B_j\to X$ such that
\begin{equation}\label{ch3:prop:StrongPresIffPres:d1}
\xymatrix@R=2ex
{
A_i\ar[drr]^{h_i}\ar[dr]_<<<{f^1_j}\\
& B_j\ar[r]^<{g} & X\\
A_k\ar[urr]_{h_k}\ar[ur]^<<<{f^2_j}
}
\end{equation}
commutes. If we build $T'=\colim \Yon$ as a regular quotient of a coproduct we get:
\[
\xymatrix
{
\coprod\limits_{f\in \cat[J]\mor}\hom(\cod f,X)\ar@<3pt>[r]^>>>>>{\phi}\ar@<-2pt>[r]_>>>>>{\psi} & \coprod\limits_{C\in \cat[J]\ob}\hom(C,X)\ar@{->>}[r] & T'X
}
\]
where $\phi\circ\mathrm{in}_f=\mathrm{in}_{\cod f}, \psi\circ \mathrm{in}_f=\mathrm{in}_{\dom f}\circ ((-)\circ f)$. $T'X$ is the quotient of $\coprod_{C\in \cat[J]\ob}\hom(C,X)$ under the smallest equivalence relation generated by
\[
h_j: B_j\to X \approx_X h_k: A_k\to X
\]
if there exists $f_j: A_k\to B_j\in \cat[J]\mor$ such that
\begin{equation}\label{ch3:prop:StrongPresIffPres:d2}
\xymatrix@R=2ex
{
B_j\ar[r]^{h_j} & X\\
A_k\ar[ur]_{h_k}\ar[u]_{f_j}
}
\end{equation}
commutes. We can now see that $T'X\simeq TX$ by defining a bijection between the two sets. We simply put
\[
\theta_X: TX\to T'X, [A_i\xto{h_i} X]_{\sim_X}\mapsto [A_i\xto{h_i} X]_{\approx_X}
\]
where the square brackets simply denote the equivalence classes under the respective equivalence relations. To see that the map $\theta_X$ is well-defined notice that from diagram (\ref{ch3:prop:StrongPresIffPres:d1}) we get two copies of diagram (\ref{ch3:prop:StrongPresIffPres:d2}), i.e. 
\begin{align*}
& \text{ if }A\xto{h_k}X\sim_X A\xto{h_i}X \text{ then }\\ 
& A_i\xto{h_i}X\approx_X  B_j\xto{g} X \text{ and } B_j\xto{g} X\approx_X  A_i\xto{h_i}X
\end{align*}
From this it is clear that if we pick two representatives $A_i\xto{h_i} X, A_j\xto{h_j}X$ of the same equivalence class under $\sim_X$, then they will belong to the same equivalence class under $\approx_X$. This also shows that $\theta_X$ is injective: if $\theta_X([A_i\xto{h_i} X]_{\sim_X})=\theta_X([A_k\xto{h_k} X]_{\sim_X})$, i.e. $[A_i\xto{h_i} X]_{\approx_X}=[A_k\xto{h_k} X]_{\approx_X}$, then we must have $[A_i\xto{h_i} X]_{\sim_X}=[A_k\xto{h_k} X]_{\sim_X}$. Finally to see that $\theta_X$ is surjective, we only need to check that an equivalence class with representative $[B_j\xto{g} X]_{\approx_X}$ has a per-image under $\theta_X$. This follows from the fact that the we have morphisms of the type $f_j: A_i\to B_j$ by construction of $\cat[J]$, and thus $[B_j\xto{g} X]_{\approx_X}=[A_i\xto{g f_j} X]_{\approx_X}$ which has pre-image $[A_i\xto{g f_j} X]_{\sim_X}$ under $\theta_X$.
\end{proof}

The reader would be justified in asking why we have defined two notions of $\lambda$-presentability if they are equivalent. The reason will become clear when we consider the associated notions of $\lambda$-generated and strongly $\lambda$-generated functors which are, in general, not equivalent. First though, let us prove the promised results that just as in the case of algebraic varieties, being finitely (strongly) present\emph{able} is equivalent to being finitely present\emph{ed}. We start with the following technical lemma.

\begin{lemma}\label{ch3:lem:subfuncpres}
Let $\cat$ be a category and let $\Inc:\cat_0\inc\cat$ be the embedding of a full subcategory. Assume that $T:\cat\to\Set$ is a functor and that $T_0$ is its restriction to $\cat_0$, assume further that $T$ and $T_0$ have the property that $T=\Lan_{\Inc} T_0$, then if $T$ is a $\lambda$-presentable object in $\Set^{\cat}$ so is $T_0$ in $\Set^{\cat_0}$.
\end{lemma}
\begin{proof}
Assume that $\Diag_0:\cat[J]\to\Set^{\cat_0}$ is a $\lambda$-filtered diagram and that there exists a natural transformation $\sigma_0:T_0\to\colim\Diag_0$. We need to show that $\sigma_0$ factors through an essentially unique object in the diagram $\Diag_0$. To show this recall first that we can turn $\Diag_0$ into a $\lambda$-filtered diagram in $\Set^{\cat}$ by post-composing with the functor $\Lan_{\Inc}(-):\Set^{\cat_0}\to\Set^{\cat}$, i.e. we define $\Diag=\Lan_I(-)\circ \Diag_0$. Since the left Kan extension construction is a left adjoint, i.e.
\[
\Lan_{\Inc}(-) \dashv (-)\circ \Inc
\]
it preserves all colimits, and thus 
\[
\colim \Diag=\colim \Lan_{\Inc}(-)\circ \Diag_0=\Lan_{\Inc}(\colim\Diag_0)
\]
Since $T=\Lan_{\Inc}T_0$, by functoriality of $\Lan_{\Inc}(-)$ there must exist a natural transformation $\sigma=\Lan_{\Inc}(\sigma_0):T\to\Lan_{\Inc}(\colim\Diag_0)$, i.e. a natural transformation $\sigma:T\to\colim \Diag$ where $\Diag$ is $\lambda$-filtered. Since $T$ is $\lambda$-presentable $\sigma$ must therefore factor essentially uniquely through an object $\Lan_{\Inc}(\Diag_0(j))$ for some $j$ of $\cat[J]$. Finally, since we are taking left Kan extensions along a full inclusion functor, we are under the conditions of Proposition \ref{ch3:prop:KanInclusion}, i.e. $\Lan_{\Inc}(F) \circ \Inc= F$ for any $F$ in $\Set^{\cat_0}$, and thus by applying the right adjoint $(-)\circ\Inc$ of $\Lan_{\Inc}(-)$ to the commutative triangle
\[
\xymatrix
{
& \Lan_{\Inc} (\Diag_0(j)) \ar[dr] \\
T=\Lan_{\Inc}(T_0)\ar[ur] \ar[rr]_{\sigma} & & \colim \Diag=\Lan_{\Inc}(\colim\Diag_0)
}
\]
we get that $T_0$ factors through $\Diag_0(j)$. The essential unicity of this factorisation is likewise inherited from the essential unicity of the factorisation for $T$. 
\end{proof}

We can now prove that finitely presentable and finitely presented functors coincide. This proof generalises, and is to some extent adapted from, Theorem 2.16 of \cite{2011:superfinitary}.

\begin{proposition}\label{ch3:prop:presfunc}
Let $\cat$ be a small category and let $T: \cat\to\Set$, then $T$ is finitely presentable in $\Set^{\cat}$ iff it is strongly finitely presented.
\end{proposition}
\begin{proof}
The `if' direction follows from the fact that $\hom$ functors are absolutely presentable in $\Set^{\cat}$ (see the proof of Lemma \ref{ch3:prop:hompres}) and thus finitely presentable objects of $\Set^{\cat}$. Therefore, by Lemma \ref{ch3:lem:colimofpres}, so is a finite colimit of $\hom$ functors.

For the `only if' direction, assume that $T$ is finitely presentable in $\Set^{\cat}$. We need to show that $T$ is a finite colimit of $\hom$ functors. In order to show this we need to prove two things: (1) that $T$ can be written as a the colimit of a diagram defined on a category with finitely many \emph{objects}, and (2) that this category also has finitely many \emph{arrows}.

Let us first show (1). We use the fact that the objects of $\cat$ form a set, and that all sets can be expressed as the colimit of their finite subsets, in order to express $T$ as a filtered colimit. Let $\cat[J]$ denote the category whose objects are the full subcategories of $\cat$ with finitely many objects and whose arrows are the inclusion functors between these categories. For any $\cat_0$ of $\cat[J]$, let $T_0:\cat_0\to\Set$ denote the restriction of $T$ to $\cat_0$ and let $\Inc_0$ denote the inclusion functor $\cat_0\inc\cat$. We now define a diagram $\Diag:\cat[J]\to\Set^{\cat}$ whose colimit will be $T$: for any $\cat_0$ in $\cat[J]$ let $$\Diag(\cat_0)=\Lan_{\Inc_0} T_0$$
The definition of $\Diag$ on inclusion functors follows automatically: let $\Inc_0^1:\cat_0\inc\cat_1$ in $\cat[J]$, then by construction of the left Kan extension we have for any object $A$ of $\cat$ that
\[
\Lan_{\Inc_0} T_0 A=\colim\left(\cat_0\downarrow A\stackrel{P_0}{\longrightarrow}\cat_0\stackrel{T_0}{\longrightarrow}\Set\right)
\]
and since $\cat_0\downarrow A$ is a subcategory of $\cat_1\downarrow A$, the colimit defining $\Lan_{\Inc_1} T_1 A$ is a cocone for the diagram defining $\Lan_{\Inc_0} T_0 A$ and so there must exist a function
\[
\Diag(\Inc_0^1)_A: \Lan_{\Inc_0} T_0 A\to \Lan_{\Inc_1} T_1 A
\]
These functions define a natural transformation $\Diag(\Inc_0^1):\Lan_{\Inc_1} T_0 \to \Lan_{\Inc_1} T_1$. Note that whilst all the morphisms in $\cat[J]$ are monos, their images under $\Diag$ typically are not, as the following example shows: let $\cat_0\ob=\{A_1,A_2\}$ and $\cat_0\mor=\emptyset$ with unique morphisms $f:A_1\to B$ and $g: A_2\to B$ in $\cat$, then $\Diag(\cat_0)B=\Lan_{\Inc_0} T_0 B=TA_1+TA_2$. Let now $\cat_1\ob=\{A_1,A_2,B\}$ and $\cat_1\mor=\{f,g\}$, then $\Diag(\cat_1) B=TB$ which in general is not a subset of $TA_1+TA_2$. 

Let us now show that $T=\colim\Diag$. Since colimits of functors are computed pointwise, it is enough to show that $TA=\colim (ev_A\circ \Diag)$. To see that $TA$ is a cocone, it is enough to see that we have maps $\Diag(\cat_0)(A)=\Lan_{\Inc_0}T_0(A)\to TA$ for every $\cat_0$. From the construction of the left Kan extension as a colimit in $\Set$, an element of $\Lan_{\Inc_0}(A)$ is an equivalence class of elements of the type $\alpha_0\in TA_0$ where $f: A_0\to A$ is in $\cat_0\downarrow A$. So let us pick $\alpha_0$ as a representative of this equivalence class, then we can associate with it $$(d_{\cat_0})_A: \Lan_{\Inc_0}T_0(A)\to TA, [\alpha_0]\mapsto Tf(\alpha_0)$$ To check that this assignment is well-defined, let us choose another representative of the equivalence class of $\alpha_0$, say $\alpha_1\in TA_1$ with $g:A_1\to A$ in $\cat_0\downarrow A$. Because $\alpha_0$ and $\alpha_1$ are in the same equivalence class, there must exist a zigzag of morphisms connecting $\alpha_0$ and $\alpha_1$. By definition of the morphisms in $\cat_0\downarrow A$, it is easy to see that this implies $Tg(\alpha_1)=Tf(\alpha_0)$, and thus the choice of representative does not matter, and our assignment is well-defined. This shows $TA$ is a cocone for $ev_A\circ \Diag$ and thus $T$ is a cocone for $\Diag$. We now check that $TA=\colim(ev_A\circ\Diag)$, i.e. that for every $\alpha\in TA$, there exist an object $\cat_0$ of $\cat[J]$ and an element $\alpha_0\in \Lan_{\Inc_0}T_0 (A)$ such that $(d_{\cat_0})_A(\alpha_0)=\alpha$ and that any other choice is related to this one by a zigzag. Clearly, we can always choose $\cat_0$ as the full subcategory of $\cat$ with only one object: $A$, in which case we get $\Lan_{\Inc_0}T_0(A)=TA$ and we can take $\alpha_0=\alpha$ since $(d_{\cat_0})_A=\id_{TA}$ in this case. For the essential unicity, assume that there exist another object $\cat_1$ in $\cat[J]$ and another element $[\alpha_1]\in\Lan_{\Inc_1}T_1 A$ with $(d_{\cat_1})_A([\alpha_1])=\alpha$. By definition of $d_{\cat_1}$, this means that there exists an object $f: A_1\to A$ in $\cat_1\downarrow A$ such that $Tf(\alpha_1)=\alpha$. If $A$ is an object of $\cat_1$ we can take $\alpha_1=\alpha$ and $\Diag(\Inc_0^1)_A=\id_{TA}$ gives us a (very simple) zigzag. If $A$ is not in $\cat_1$ we can embed $\cat_1$ and $\cat_0$ in the full subcategory $\cat_2$ whose objects are those of $\cat_1$ plus $A$. It is easy to see that $\Diag(\Inc_0^2)_A=\id_{TA}$ and $\Diag(\Inc_1^2)_A(\alpha_1)$ is precisely $Tf(\alpha_1)=\alpha$, and this provides us with a zigzag 
\[
\Lan_{\Inc_0}T_0 A=TA\stackrel{\Diag(\Inc_0^2)_A}{\longrightarrow}\Lan_{\Inc_2}T_2 A=TA\stackrel{\Diag(\Inc_1^2)_A}{\longleftarrow}\Lan_{\Inc_1}T_1 A
\]
connecting $\alpha_0$ and $\alpha_1$ which concludes the proof that $T=\colim\Diag$.

Since $\Diag$ is directed, and $T$ is assumed to be finitely presentable, $\Id_T$ must factor through a unique object $\Diag(\cat_0)$, i.e. 
\[
T=\Lan_{\Inc_0}T_0
\]
for a certain full subcategory $\cat_0$ with finitely many object. We are now going to show that (a) $T$ can be written as a certain colimit of $\hom$ functors (b) that this colimit arises from a diagram over a category with only finitely many objects. This will conclude the proof of (1).

For (a), recall by Theorem \ref{ch3:thm:fstpresthm} that $T_0=\colim(\Yon_0 \Forg_{T_0})$ where $\Yon_0:\cat_0\to\Set^{\cat_0}$ is the Yoneda transformation. We will now show that we can extend this result and that 
\[
T=\colim\left(\el[T_0]\stackrel{\Inc_0 \Forg_{T_0}}{\longrightarrow}\cat\op\stackrel{\Yon}{\longrightarrow}\Set^{\cat}\right)
\] 
To see that $T$ is a cocone of the diagram $\Diag_0=\Yon\Inc_0\Forg_{T_0}$, we use the fact that by Yoneda's lemma for any $(A,\alpha)$ in $\el[T_0]$, $\alpha$ determines a unique natural transformation $c^\alpha:\hom(A,-)\to T$, all these transformations are compatible by definition of $\el[T_0]$, and $T$ is thus indeed a cocone. To see that $T=\colim\Diag_0$ we can again check that $TB=\colim(ev_B\circ\Diag_0)$ at every stage $B$ of $\cat$. So let $\beta\in TB$, since $TB=\Lan_{\Inc_0} T_0 B$ there must exist $f:A\to B$ in $\cat_0\downarrow B$ and $\alpha\in TA$ such that $Tf(\alpha)=\beta$, but this data provides us precisely with an $f\in\Diag_0(A,\alpha)B=\hom(A,B)$ such that $c_B^\alpha(f)=\beta$. Now we just need to check that it is essentially unique. Assume that there exists $f'\in\Diag_0(A',\alpha')$ such that $c_B^\alpha(f')=Tf'(\alpha')=\beta $, i.e. there exists another object of $\cat_0\downarrow B$ and another element $\alpha'\in TA'$ which gets mapped to $\beta\in TB=\Lan_{\Inc_0}T_0 B$. But by construction of the left Kan extension as a colimit, this means that there must exist a zigzag in $\cat_0\downarrow B$ connecting $f$ and $f'$, and this immediately provides us with a zigzag (identical but with all the arrows reversed) connecting $\hom(A,B)$ and $\hom(A',B)$. Thus $T=\colim \Diag_0$.

For (b), recall that the objects of $\el[T_0]$ are given by the pairs $(A,\alpha)$ with $A$ an object of $\cat_0$ and $\alpha\in T_0 A$. We know that $\cat_0$ has finitely many objects, thus to show that $\el[T_0]$ has finitely many objects we need to show that $T_0 A$ is always finite. To achieve this let us write $\cat_0\ob=\{A_1,\ldots,A_n\}$ and consider the collection
$$\mathscr{U}(\cat_0)=\{(U_1,\ldots,U_n)\mid U_i\subseteq T_0A_i, |U_i|<\omega, 1\leq i\leq n\}$$
i.e. the $n$-tuples of finite subsets of the images of $\cat_0$-objects under $T_0$. We now define for any $\vect[U]\in\mathscr{U}(\cat_0)$ the functor
$T_0^{\vect[U]}:\cat_0\to\Set$ on objects by
\[
T_0^{\vect[U]} A=\bigcup_{f\in(\cat_0\downarrow A)\ob}Tf[U_{\dom f}]
\]
where $U_{\dom f}$ is the component of $\vect[U]$ corresponding to $\dom f$. On morphisms $g:A\to B$ the functor is defined as
$$T_0^{\vect[U]}g=T_0 g\restrict T_0^{\vect[U]} A$$
Note the following: (i) $\mathscr{U}$ can be partially ordered by taking the usual set inclusion component-wise and if $\vect[U],\vect[V]\in \mathscr{U}$ are such that $\vect[U]\subseteq\vect[V]$, then $T_0^{\vect[U]}\inc T_0^{\vect[V]}$; (ii) $\mathscr{U}(\cat_0)$ is directed (by taking the union component-wise); and (iii) if we define $\Diag_\mathscr{U}: \mathscr{U}\to\Set^{\cat_0}$ (which defines a diagram of monos by (i)) it is clear that $T_0=\colim\Diag_{\mathscr{U}}$. We know from Lemma \ref{ch3:lem:subfuncpres} that since $T=\Lan_{\Inc_0} T_0$, $\cat_0$ is a full subcategory, and $T$ is finitely presentable, $T_0$ must also be finitely presentable. Therefore, $T_0$ must be isomorphic to one of the $T_0^{\vect[U]}$, $\vect[U]\in\mathscr{U}$, i.e. image finite (or `super finitary', see \cite{2011:superfinitary}). Therefore $\el[T_0]\ob$ is indeed finite and $T$ is a colimit of $\hom$ functors whose diagram is defined on a category with finitely many objects. This concludes the proof of (1).

Now we need to show (2), i.e. that $\el[T_0]$ also has finitely many \emph{morphisms}. To prove this, let us consider the set of all finite subcategories of $\cat_0$, i.e. the subcategories of $\cat_0$ with only finitely many morphisms. These form a set $\mathscr{V}$ which is partially ordered by the usual inclusion. Note also that by taking the union of two subcategories (i.e. the union of their sets of objects and morphisms),we easily get that $\mathscr{V}$ is directed. We now define the diagram $\Diag_m: \mathscr{V}\to\Set^{\cat_0}$ on objects by
\[
\Diag_m(\cat_i)=\Lan_{\Inc_i} (T_0\circ\Inc_i): \cat_0\to\Set
\] 
where $\Inc_i:\cat_i\inc\cat_0$. The definition of $\Diag_m$ on morphisms follows automatically: suppose $\Inc_{ik}:\cat_i \inc \cat_k$ is an inclusion between two finite subcategories of $\cat_0$, and let us write $T_i$ for $T_0\circ\Inc_i$; then at any stage $A$ of $\cat_0$ we have by construction of the left Kan extension that
\[
\Lan_{\Inc_i}T_i A =\colim\left(\cat_i\downarrow A\stackrel{P_i}{\longrightarrow}\cat_i\stackrel{T_i}{\longrightarrow}\Set\right)
\]
where $P_i(f:\Inc_i(B)\to A)=B$ . If $\cat_i\inc\cat_k$, the colimit defining $\Lan_{\Inc_k} T_k A $ is a cocone for the diagram $T_i P_i$ and therefore there must exist a unique function 
\[
\Diag_m(\Inc_{ik})_A:\Lan_{\Inc_i} T_i A\to \Lan_{\Inc_k} T_k A
\]
These functions define a natural transformation $\Diag_m(\Inc_{ik}):\Diag_m(\cat_i)\to\Diag_m(\cat_k)$. 


The proof that $T_0=\colim \Diag_m$ is exactly the same as the proof above that $T$ was colimit for $\Diag$; the fact that the subcategories $\cat_i$ are not full does not play any role in the construction (however it does imply that the restriction functors $T_i$ are \emph{not} finitely presentable). Therefore, since $T_0$ is finitely presentable we must have $T_0=\Lan_{I_i} T_i$ for a subcategory $\cat_i$ of $\cat_0$ with finitely many morphisms. As was shown in Proposition \ref{ch1:prop:adj}, the composition of two adjunctions is an adjunction and thus we have that $\Lan_{\Inc_0}(-)\circ \Lan_{\Inc_i}(-)\dashv ((-)\circ\Inc_i) \circ ((-)\circ \Inc_0)$, i.e. $$T=\Lan_{\Inc_0}(\Lan{\Inc_i} T_i)=\Lan_{\Inc_0\Inc_i} T_i$$
We can then use this fact to show that 
\[
T=\colim(\el[T_i]\stackrel{\Inc_0\Inc_i\Forg_{T_i}}{\longrightarrow}\cat\op\stackrel{\Yon}{\longrightarrow}\Set\cat)
\]
in exactly the same way as we showed that $T=\colim\Yon\Inc_0\Forg_{T_0}$ earlier, and $\el[T_i]$ is a category with finitely many objects and finitely many morphisms.

\end{proof}

\begin{remark}
We have voluntarily  decomposed the proof above into two steps: first show that $T$ can we written as a colimit of a diagram of $\hom$ functors defined on an index category with finitely many \emph{objects}, and second that this index category can also be assumed to have finitely many \emph{morphisms}. The reader might wonder if this decomposition is necessary, and why we didn't immediately write $T$ as a filtered colimit of left Kan extensions of restrictions of $T$ to finite categories. The problem with this approach is that by taking non full subcategories, the restriction functors do not inherit presentability (i.e. we cannot use Lemma \ref{ch3:lem:subfuncpres}), which means that we cannot prove image finiteness. Thus, though it looks like we are performing the same trick twice, there is a good reason to proceed in this way. Moreover, as we shall see when we reach the notion of $\lambda$-generated functor, this distinction between the number of objects and the number of morphisms in a diagram will become very important. 
\end{remark}

\begin{corollary}\label{ch3:cor:funcfinpres}
For any small category $\cat$, the category $\Set^{\cat}$ is locally finitely presentable.
\end{corollary}
\begin{proof}
Using exactly the same construction as in the previous proof of Proposition \ref{ch3:prop:presfunc}, it is clear that any $T$ in $\Set^{\cat}$ can be expressed as a filtered colimit of functors of the type $T_{\cat_0}$ where $\cat_0$ is a finite subcategory of $\cat$. Since by the same proof each of these $T_{\cat_0}$ is a  finitely presentable object of $\Set^{\cat}$, it means that every object in $\Set^{\cat}$ is a filtered colimit of presentable objects. And since $\Set^{\cat}$ is also cocomplete, it is locally finitely presentable. 
\end{proof}

\begin{corollary}\label{ch3:cor:lambdapresfunc}
Let $\cat$ be a small category, and let $\lambda$ be a regular cardinal, then an object $T$ in $\Set^{\cat}$ is $\lambda$-presentable iff it is a $\lambda$-small colimit of $\hom$ functors. 
\end{corollary}
\begin{proof}
For the `if' direction, notice first that since $\hom$ functors are absolutely presentable, they are $\lambda$-presentable, and thus so is a $\lambda$-small colimit of them by Lemma \ref{ch3:lem:colimofpres}.
For the `only if' direction we proceed as in the proof of Proposition \ref{ch3:prop:presfunc}, but this time we use the fact that the set $\cat\ob$ can be expressed as the colimit of its subsets of cardinality smaller than $\lambda$. Since $\lambda$ is regular, this colimit is $\lambda$-directed. In consequence, we can write $T$ as a $\lambda$-directed colimit of subfunctors in the same fashion as in the proof of Proposition \ref{ch3:prop:presfunc} and the rest of the proof follows.
\end{proof}

\begin{remark}\label{ch3:rem:analogy2} Similarly to Theorem 3.12 of $\cite{LPAC}$, Proposition \ref{ch3:prop:presfunc} is not trivial, but it tightens the analogy, started in Remark \ref{ch3:rem:analogy1}, between functors $\cat\to\Set$, $\cat$ small, and algebraic varieties. A presentation defined on a small index category $\cat[J]$ is like the presentation of a variety, with $\cat[J]\ob$ as its set of generators and $\cat[J]\mor$ as its sets of relations. If $\cat[J]$ is a $\lambda$-small category, then the functor is $\lambda$-presented, i.e. presented by fewer than $\lambda$ `generators' and fewer than $\lambda$ `relations'. In particular coproducts of $\hom$ functors (i.e. when $\cat[J]$ is discrete) can be thought of as the analogue of free algebraic structures.
\end{remark}

\subsection{$\lambda$-generatable and $\lambda$-generated functors}

The standard notion of functor presentation in the literature, is to express a functor $T$ as an epimorphic image of a coproduct of representables whose size is typically bounded by some cardinal (see e.g. \cite{setFuncPres}). This notion probably dates back to \cite{1971:Freyd} where a functor with such a representation is called `petty'. For reasons that will soon be clear we prefer the terminology of \cite{2005:cococo}:

\begin{definition}[$\lambda$-generated functor, \cite{2005:cococo}]\index{Generated functor}
A functor $T:\cat\to\Set$ will be called \textbf{$\lambda$-generated} if there exists a set $I$ of cardinality $|I|< \lambda$, and regular epimorphism 
\begin{align*}
e: \coprod_{i\in I}\Hom(A_i,-)\epi T
\end{align*}
\end{definition}

The associated `strong' notion is defined as in the case of strongly presentable functor:

\begin{definition}[Strongly $\lambda$-generated functor]\index{Strongly generated functor}
A functor $T:\cat\to\Set$ will be called \textbf{strongly $\lambda$-generated}, if there exists a presentation $\Diag:\cat[J]\to\Set^{\cat}$ of $T$ (i.e. $T=\colim\Diag$) such that $\cat[J]$ has fewer than $\lambda$ objects, there are no restrictions on the number of morphisms in $\cat[J]$.
\end{definition}

\begin{proposition}
A functor $T:\cat\to\Set$ is $\lambda$-generated (in the sense of \cite{2005:cococo}) if it is strongly $\lambda$-generated.
\end{proposition}
\begin{proof}
The proof is the same as the proof of Proposition \ref{ch3:prop:strongPresPres}.
\end{proof}

Unlike Proposition \ref{ch3:prop:StrongPresIffPres}, we cannot in general show the converse implication, i.e. that $\lambda$-generated functors are strongly $\lambda$-generated. However, in the case where the base category $\cat$ is sufficiently well-behaved we do have a converse result which we will return to when discussing the canonical presentation in Section \ref{ch3:sec:canPres}. A necessary technical condition on the base category for the following Proposition is that the cardinality of the set of quotients of any object should be bounded. We will call a category \textbf{$\lambda$-co-well powered} if for every object $A$ of $\cat$, the collection of (equivalence classes of) epimorphisms with domain $A$ is a set of cardinality strictly less than $\lambda$.

\begin{proposition}\label{ch3:prop:StrongGenIffGen}
Let $\cat$ be a $\lambda$-co-well powered small category with an epi-split-mono factorisation system, then every $\lambda$-generated functor $T:\cat\to\Set$ is strongly $\lambda$-generated. 
\end{proposition}
\begin{proof}
By assumption, we have a set $I$ of cardinality less than $\lambda$, a functor $S$ and a natural transformation $e$
\[
\xymatrix
{
S \ar@<3pt>[r]^-{\phi_1}\ar@<-2pt>[r]_-{\phi_2} & \coprod\limits_{i\in I}\hom(A_i,-)\ar@{->>}[r]^-{e} & T
}
\]
which is the coequalizer of $\phi_1,\phi_2$. By Yoneda, the transformation $e$ uniquely determines (and is uniquely determined by) a set $\{x^{A_i}\in TA_i\mid i\in I\}$, and for any object $X$ in $\cat$, two morphisms $f_i: A_i\to X$ and $f_j: A_j\to X$ in the coproduct $\coprod_i \hom(A_i,X)$ are identified precisely when
\[
Tf_i(x^{A_i})=Tf_j(x^{A_j})
\]
We use this characterization of $e$ to define the following category $\cat[J]$ in two steps. In the first step, the objects of $\cat[J]$ will be all the pairs $(A_i, x^{A_i}), i\in I$, and there will be morphism $(A_i, x^{A^i}) \to (A_j, x^{A_j})$ whenever there exist $f: A_j\to A_i$ such that $Tf(x^{A_j})=x^{A_i}$. In a second step we complete $\cat[J]$ by adding for every morphism as defined above its epi-split mono factorisation (which exists by assumption on $\cat$), i.e. for every $f: A_j\to A_i$ such that $Tf(x^{A_j})=x^{A_i}$ we consider the decomposition $f=f^m\circ f^e$ where $f^e$ is epi and $f^m$ is split mono. We then add $f[A_j]$ to $\cat[J]\ob$ and we add the morphisms induced by $f^e$ and $f^m$ to $\cat[J]\mor$: $f^e$ defines a morphism $(f[A_j],Tf^e(x^{A_j}))\to (A_j,x^{A_j})$, and $f^m$ defines a morphism $(A_i, x^{A_i})\to(f[A_j], Tf^e(x^{A_j}))$. Finally, we close this collection of morphisms under composition and identity maps. Note that since $\cat$ is assumed to be $\lambda$-co-well powered and since $\lambda$ is regular $|\cat[J]\ob|<\lambda$.
 
We now define 
\[
T'=\colim(\cat[J]\stackrel{\Forg}{\longrightarrow}\cat\op\stackrel{\Yon}{\longrightarrow}\Set^{\cat})
\]
where $\Forg$ is the obvious forgetful functor. We want to show that $T'=T$. By working pointwise, this means that for any $X$ in $\cat$, the following coequalizers 
\[
\xymatrix
{
SX \ar@<3pt>[r]\ar@<-2pt>[r] & \coprod\limits_{i\in I}\hom(A_i,X)\ar@{->>}[r]^<<<<<{e_X} & TX
}
\]
and
\[
\xymatrix
{
\coprod\limits_{f\in \cat[J]\mor} \hom(\dom f, X) \ar@<3pt>[r]^>>>>>>{\phi}\ar@<-2pt>[r]_>>>>>>{\psi} & \coprod\limits_{A\in\cat[J]\ob}\hom(A_i,X)\ar@{->>}[r]^>>>>>>{q_X} & T'X
}
\] 
where $\mathrm{in}_f\circ \phi=\mathrm{in}_{\dom f}$ and $\mathrm{in}_f\circ \psi=\mathrm{in}_{\cod f}\circ ((-)\circ f)$, should be the same object. To check that this is the case, note first that we can characterise $T'X$ as the quotient of the coproduct under the smallest equivalence class generated by
\[
f: A\to X \approx_X f': A'\to X 
\]
if there exist $h: A'\to A$ such that $T h(x^{A})=x^{A'}$, where $x^{A}$ is the element of $TA$ associated with $A$ in $\cat[J]$, and similarly for $x^{A'}$. Note that by construction, the objects $A$ of $\cat[J]$ are either of the shape $A_i, i\in I$ or $f[A_i]$ for an $f\in\cat[J]\mor$. If $A$ is of the shape $f[A_i]$, we can immediately see that $h: A\to X$ will be in the same equivalence class under $\approx_X$ as $A_i\xto{f^e}f[A_i]\xto{h} X$, by construction of $\cat[J]$. Thus we can always reason w.l.o.g. by using representatives of the shape $f: A_i\to X$ for an $i\in I$. From this it follows that we can define a map
\[
\theta_X:  T'X\to TX, [A_i\xto{f} X]_{\approx}\mapsto[A_i\xto{f} X]_{e_X}
\]
where $[A_i\xto{f} X]_{\approx}$ is the equivalence class of $f$ under $\approx_X$, and $[A_i\xto{f} X]_{e_X}$ is the equivalence class of maps identified with $f$ by $e_X$. It is easy to check that $\theta_X$ is well-defined: if we start with maps $f_i: A_i\to X$ and $f_j: A_j\to X$ such that $f_i\approx_ X f_j$, then there exists $h: A_j\to A_i$ with $Th(x^{A_j})=x^{A_i}$, and thus $Tf_j(x^{A_j})=Tf_i(Th (x^{A_j})=Tf_i(x^{A_i})$, i.e. $e_X(f_i)=e_X(f_j)$. Moreover, $\theta_X$ is trivially surjective. It remains to show that $\theta_X$ is injective. For this assume that $e_X(f_i)=e_X(f_j)$, and build the epi-split mono decomposition $f_i=f_i^m\circ f_i^e$. We then construct the following diagram
\[
\xymatrix@C=10ex
{
& &  TA_j\ar[d]^{f_j} \\
T A_i\ar@{->>}[r]_{Tf^e_i} & T f[A_i]\hspace{1ex}\ar@{>->}@<-2pt>[r]_{Tf^m_i} & TX\ar@<-3pt>[l]_{r}
}
\]
where $r$ is a retraction of $f_i^m$, i.e. $r\circ f^m_i=\id_{f[A_i]}$. Since $Tf_j(x^{A_j})=Tf_i(x^{A_i})$ by assumption, it follows that $Tr\circ Tf_j(x^{A_j})=Tr\circ Tf^m_i\circ Tf^e_i(x^{A_i})=Tf^e_i(x^{A_i})$, and thus
\[
[A_j\xto{f_j} X]_{\approx_X}=[f[A_i]\xto{f^m_i} X]_{\approx_X}
\]
and since we have $[f[A_i]\xto{f^m_i} X]_{\approx_X}=[A_i\xto{f_i} X]_{\approx_X}$ by construction, we can conclude by transitivity that 
\[
[A_j\xto{f_j} X]_{\approx_X}=[A_i\xto{f_i} X]_{\approx_X}
\]
and thus $\theta_X$ is also injective, which concludes the proof.
\end{proof}

The assumptions on $\cat$ under which strongly $\lambda$-generated and $\lambda$-generated functors coincide are identical (modulo the $\lambda$ bound) to the conditions isolated in Proposition 4.6 of \cite{1971:Freyd}. This Proposition establishes under which conditions functors called `lucid' are `petty', i.e. generated. In the analogy between functors and elements of a variety, `$\lambda$-lucid' functors are $\lambda$-coherent elements, i.e. $\lambda$-generated functor whose $\lambda$-generated subfunctors are $\lambda$-presented (they are therefore $\lambda$-presented themselves).   

We would now like  to have an abstract characterization of functors which are $\lambda$-generated, in the same way that being $\lambda$-presentable is an abstract characterization of being $\lambda$-presented. As it turns out, this characterization exists, and is detailed in section 1.E. of \cite{LPAC}, but unfortunately, an object satisfying this characterization is also called $\lambda$-generated. In order to both avoid confusion and make a parallel with presentable and presented objects we will use the following terminology.

\begin{definition}[$\lambda$-generatable object]\index{Generatable}
An object $A$ in a category will be called $\lambda$-\textbf{generatable} if $\hom(A,-)$ preserves $\lambda$-directed colimits of monomorphisms. In particular a functor $T:\cat\to\Set$ will be called $\lambda$-generatable if it is a $\lambda$-generatable object in $\Set^{\cat}$, i.e. if $\Nat(T,-)$ preserves $\lambda$-directed colimits of monomorphisms.
\end{definition}

We cite without proof the following relevant results from \cite{LPAC}.

\begin{proposition}[\cite{LPAC} 1.69]\label{ch3:prop:geniffquotpres}
In a locally $\lambda$-presentable category, an object is $\lambda$-generatable iff it is a strong quotient of a $\lambda$-presentable object.
\end{proposition}

\begin{theorem}[\cite{LPAC} 1.70] A category $\cat$ is locally-presentable iff it is cocomplete and there exist a regular cardinal $\lambda$ and a small subcategory $\cat[G]_\lambda$ of $\lambda$-generatable objects such that every object of $\cat$ is a $\lambda$-directed 
colimit of its subobjects from $\cat[G]_\lambda$.
\end{theorem}

Let us now show that as expected, $\lambda$-generat\emph{ed} functors $\cat\to\Set$, with $\cat$ small, are precisely the $\lambda$-generat\emph{able} objects in $\Set^{\cat}$. We start with the analogue of Lemma \ref{ch3:lem:subfuncpres} for $\lambda$-generated functors.

\begin{lemma}\label{ch3:lem:subfuncgen}
Let $\cat$ be a category and let $\Inc:\cat_0\inc\cat$ be the embedding of a full subcategory. Assume that $T:\cat\to\Set$ is a functor and that $T_0$ is its restriction to $\cat_0$, assume further that $T$ and $T_0$ have the property that $T=\Lan_{\Inc} T_0$, then if $T$ is a $\lambda$-generatable object in $\Set^{\cat}$ so is $T_0$ in $\Set^{\cat_0}$.
\end{lemma}
\begin{proof}
The proof is identical to that of  Lemma \ref{ch3:lem:subfuncpres}  once it is established that if $\Diag_0: \cat_0\to\Set^{\cat_0}$ is a diagram of monomorphisms, then so is $\Lan_{\Inc}(-)\circ \Diag_0$, i.e. $\Lan_{\Inc}(-)$ preserves monos. This is rather obvious from the construction of $\Lan_{\Inc}$: if $F_i\inc F_j$ are two functors in $\Set^{\cat_0}$ then clearly for all $A$ in $\cat_0$
$$\colim\left(\cat_0\downarrow A\stackrel{P}{\longrightarrow}\cat_0\stackrel{F_i}{\longrightarrow}\Set\right)\inc \colim\left(\cat_0\downarrow A\stackrel{P}{\longrightarrow}\cat_0\stackrel{F_j}{\longrightarrow}\Set\right)$$
i.e. $\Lan_{\Inc}F_i A\inc \Lan_{\Inc}F_j A$.
\end{proof}

\begin{proposition}\label{ch3:prop:geniffgen}
A functor $T:\cat\to\Set$, $\cat$ small, is $\lambda$-generated iff it is a $\lambda$-generatable object of $\Set^{\cat}$.
\end{proposition}
\begin{proof}
For the `if' direction we proceed as in Proposition \ref{ch3:prop:presfunc} and Corollary \ref{ch3:cor:lambdapresfunc} and we write $T$ as the following $\lambda$-filtered colimit of functors. We define $\cat[J]$ to be the category whose objects are the full subcategories of $\cat$ with strictly fewer than $\lambda$ objects and whose arrows are the inclusions between these subcategories. Then we can write define $\Diag:\cat[J]\to\Set^{\cat}$ by $\Diag(\cat_0)=\Lan_{\Inc_0}T_0$ where $T_0$ is the restriction of $T$ to the full subcategory $\cat_0$. The definition of $\Diag$ on inclusions follows automatically as in Proposition  \ref{ch3:prop:presfunc} and we once again get that 
\[
T=\colim\Diag
\]
Note once again that whilst $\cat[J]$ is a $\lambda$-directed category all of whose arrows are monos, the colimit is \emph{not} a colimit of monos. To remedy this we need to take an epi-mono factorisation of all the morphisms $d_{\cat_0}: \Lan_{\Inc_0}T_0\to T$ making $T$ a cocone for $\Diag$ which is defined for every $A$ of $\cat$ as
\[
(d_{\cat_0})_A:\Lan_{\Inc_0}T_0 A\to TA, [\alpha_0]\to Tf(\alpha_0)
\]
where $[\alpha_0]$ is an equivalence class with representative $\alpha_0\in TA_0$ for a certain $f:A_0\to A$ in $\cat_0\downarrow A$.

Taking the epi-mono factorization of $d_{\cat_0}$ is always possible since $\Set^{\cat}$ is a topos. We can therefore define the following diagram $$\Diag_m:\cat[J]\to\Set^{\cat}, \Diag_m(\cat_0)=d_{\cat_0}[\Lan_{\Inc_0} T_0]$$ For any inclusion $\Inc_0^1:\cat_0\inc\cat_1$ in $\cat[J]$, $\Diag_m(\Inc_0^1)$ is the obvious inclusion of $d_{\cat_0}[\Lan_{\Inc_0} T_0]$ into $d_{\cat_1}[\Lan_{\Inc_1} T_1]$. For $A$ in $\cat$ we then have that $T$ is the colimit over all $\cat_0$ in $\cat[J]$ of the direct images $$(d_{\cat_0})_A[\Lan_{\Inc_0} T_0 A]=\bigcup_{f\in(\cat_0\downarrow A)\ob} Tf[T\dom f]$$
It is clear that $T=\colim \Diag_m$ and that $\Diag_m$ is a $\lambda$-filtered diagram of monos since $(d_{\cat_0})_A[\Lan_{\Inc_0} T_0 A]\subseteq TA$ and for an inclusion $\Inc_0^1:\cat_0\inc\cat_1$ in $\cat[J]$, we clearly have $(d_{\cat_0})_A[\Lan_{\Inc_0} T_0 A]\subseteq (d_{\cat_0})_A[\Lan_{\Inc_1} T_0 A]$. Note that this colimit of monos is the `$\lambda$-generalization' of the colimit of monos in Theorem 2.16 of \cite{2011:superfinitary} which is the basis of the notion of \emph{superfinitary functor}.

Since $T$ is $\lambda$-generatable, $\Id_T$ must factor through one of these monomorphisms and thus $T=d_{\cat_0}[\Lan_{\Inc_0} T_0]$ for a full subcategory $\cat_0$ with less than $\lambda$-objects. From Proposition \ref{ch3:prop:presfunc} we know that 
\[
\Lan_{\Inc}T_0=\colim\left(\el[T_0]\stackrel{\Inc\Forg_{T_0}}{\longrightarrow}\cat_0\op\stackrel{\Yon}{\longrightarrow}\Set^{\cat}\right)
\]
By the previous Lemma \ref{ch3:lem:subfuncgen} we know that $T_0$ itself is $\lambda$-generatable. Therefore, if we proceed as in Proposition 
\ref{ch3:prop:presfunc}, and express $T_0$ as a colimit of monos by restricting the cardinality of $T_0 A$ for all $A_0$ in $\cat_0$, we can show that $|T_0 A|< \lambda$ for all $A$ in $\cat_0$, and thus $\el[T_0]$   
has less that $\lambda$ objects. In other words $\Lan_{\Inc}T_0$ can be expressed a quotient of a coproduct of less than $\lambda$ $\hom$ functors, and since $T$ is a quotient of $\Lan_{\Inc}T_0$, we immediately get that $T$ is a quotient of a coproduct of less than $\lambda$ $\hom$ functors (all epimorphisms in $\Set^{\cat}$ are regular).

For the `only if' direction, we assume that $T$ can be written as $$e: \coprod_{i\in I}\Hom(A_i,-)\epi T$$where $|I|<\lambda$. Note that $\coprod_{i\in I}\Hom(A_i,-)$ is a $\lambda$-small colimit of $\hom$ functors for the diagram $\Diag=\Yon\circ A_{(-)}: I\to\Set^{\cat}$ where $A_{(-)}(i)=A_i$ and $I$ is viewed as a discrete category. By Corollary \ref{ch3:cor:lambdapresfunc}, this means that $\coprod_{i\in I}\Hom(A_i,-)$ is a $\lambda$-presentable object in $\Set^{\cat}$. Since $T$ is a regular (and therefore strong) quotient of a $\lambda$-presentable object, it must be a $\lambda$-generated object of $\Set^{\cat}$ by Proposition \ref{ch3:prop:geniffquotpres}.
\end{proof}

\begin{corollary}
Let $\cat$ be a small category. Every $\lambda$-presentable functor in $\Set^{\cat}$ is $\lambda$-generated.
\end{corollary}

\begin{remark}We can now extend the analogy of Remarks \ref{ch3:rem:analogy1} and \ref{ch3:rem:analogy2} further by noting the similarity between $\lambda$-generated functors and $\lambda$-generated algebraic varieties. The analogy is detailed in Table \ref{ch3:table1}. As indicated in this table, we do not know if there exist an abstract characterisation of strongly finitely generated functors of the type `preservation of a subclass of all cofiltered limits'. We conjecture that it may be possible to take this analogy further, that a theory of `varieties of functor' might exists in which classes of functors (e.g. left-exact functors) may be described by small categories of generators and relations with a specified form, and that a Birkhoff style theorem for $\Set$-valued functors might also be found for an appropriate set of operation on functors.
\end{remark}

\begin{sidewaystable}
\caption{Analogy between varieties and functors}
\centering
\begin{tabular}{|p{4cm} || p{5.5cm} || p{5.5cm} |}
\hline
Algebraic variety $\mathbf{V}$ & \multicolumn{2}{c|}{$\Set$-functors on small category $\cat$ }\\
\hline
 & Strong version & Normal version \\
\hline\hline
Presentation $\langle G, R\rangle$ by sets of generators $G$ and relations  $R$ & Diagram $\Diag:\cat[J]\to\Set^{\cat}$ factoring through $\Yon:\cat\op\to\Set^{\cat}$ s. th. \[T=\colim\Diag\] & Regular epimorphism: \[S\rightrightarrows \coprod\limits_{i\in I}\hom(A_i,-)\epi T\]\\
\hline
Generators in $G$ & $\cat[J]\ob$ & $\{\hom(A_i,-)\mid i\in I\}$\\
\hline
Relations in $R$ & $\cat[J]\mor$ & $S$\\
\hline
Finitely presented objects: $G$ and $R$ finite & Strongly finitely presented functor: $\cat[J]$ a finite category & Finitely presented functor $I$ finite, $S$ finitely generated\\
\hline
Finitely presentable object in $\cat[V]$ & \multicolumn{2}{c|}{Finitely presentable object in $\Set^{\cat}$}\\
\hline
Free object in $\cat[V]$ & \multicolumn{2}{c|}{$\cat[J]$ discrete, i.e. coproduct of $\hom$} \\
\hline 
Finitely generated objects: $G$  finite & Strongly finitely generated functor: $\cat[J]\ob$ finite & Finitely generated functor: $I$ finite \\
\hline
Finitely generatable object in $\cat[V]$ & ? & Finitely generatable object in $\Set^{\cat}$\\
\hline
\end{tabular}
\label{ch3:table1}
\end{sidewaystable}

\subsection{Free cocompletions}

We end this section with a different way of looking at the functor presentations. As we have just seen, when $\cat$ is small, the finite colimits of representables are precisely the finitely presentable objects in $\Set^{\cat}$ (Proposition \ref{ch3:prop:presfunc}). The filtered colimits of representable functors are precisely the objects of $\Set^{\cat}$ which are left exact functors \ref{ch3:prop:lexiffelfiltered}. Finally, by Theorem \ref{ch3:thm:fstpresthm}, general colimits of representable functors are precisely the functors in $\Set^{\cat}$. These three facts can be interpreted as three different `free cocompletions' of the small category $\cat$.
 
\begin{definition}\index{Free cocompletion!finite}
Let $\cat$ be a category. A \textbf{free finite cocompletion} of $\cat$ is a full embedding $\Inc_r:\cat\to\rex(\cat)$ such that 
\begin{enumerate}[(i)]
\item $\rex(\cat)$ is finitely cocomplete (in other words right exact or `rex')
\item $\Inc_r$ is universal amongst such embeddings in the sense that for any other functor $F: \cat\to\cat[A]$ with $\cat[A]$ finitely cocomplete, there must exist a unique (up to isomorphism) right exact functor $F^*:\rex(\cat)\to\cat[A]$ such that $F^* \Inc_r= F$.
\end{enumerate}
A \textbf{free filtered cocompletion}\index{Free cocompletion!filtered} of $\cat$ is a full embedding $\Inc_f: \cat\to\filtC(\cat)$ such that 
\begin{enumerate}
\item $\filtC(\cat)$ has all filtered colimits (and a small subcategory generating $\filtC$ by taking filtered colimits, i.e.  $\filtC(\cat)$ is finitely accessible)
\item $\Inc_f$ is universal amongst such embeddings,  i.e. for any other functor $F:\cat\to\cat[A]$ where $\cat[A]$ has all filtered colimits, there exists an accessible functor $\tilde{F}:\filtC(\cat)\to\cat[A]$ such that $\tilde{F} \Inc_f=F$
\end{enumerate}
A \textbf{free cocompletion}\index{Free cocompletion}
of $\cat$ is a full embedding $\Inc_c: \cat\to\CC(\cat)$ such that 
\begin{enumerate}
\item $\CC(\cat)$ has all colimits
\item $\Inc_c$ is universal amongst such embeddings,  i.e. for any other functor $F:\cat\to\cat[A]$ where $\cat[A]$ has all colimits, there exists an accessible functor $\tilde{F}:\CC(\cat)\to\cat[A]$ such that $\tilde{F} \Inc_c=F$
\end{enumerate}
\end{definition}

As it turns out, each embedding can be taken to be the Yoneda embedding, but we must be careful in our choice of Yoneda embedding. In what preceded, we were interested in studying the structure of the category $\Set^{\cat}$, and to this end we relied heavily on the Yoneda transformation $\Yon: \cat\op\to\Set^{\cat}, A\mapsto\hom(A,-)$. In the present context, we are interested in freely completing $\cat$, so the source of the functor must be $\cat$, not $\cat\op$, i.e. we consider the Yoneda embedding $\Yon: \cat\to\Set^{\cat\op}, A\mapsto\hom(-,A)$, i.e. the embedding into the category $\hcat=\Set^{\cat\op}$ of presheaves. This change of convention does not modify anything: everything we have done so far can be re-phrased using this embedding. For example, all presheaves can be written as a colimit of \emph{contravariant} $\hom$ functors, the proof is exactly the same as that of Theorem \ref{ch3:thm:fstpresthm} with the following small difference. If $T:\cat\to\Set$ is contravariant, then there exists a morphism between $(A,\alpha)$ and $(B,\beta)$ in $\el$ if there exists a morphism $f$ in $\cat$ such that $Tf(\beta)=\alpha$. Since $T$ is contravariant, this means that $f$ must be a morphism from $A\to B$. The forgetful functor therefore has type $\el\to\cat$.

\begin{proposition}\label{ch3:prop:finitecocomp}
For a small category $\cat$, a free finite cocompletion is given by the category $\mathrm{f.p.}(\hcat)$ of finitely presentable objects in $\hcat$, together with the Yoneda embedding $\Yon: \cat\to\mathrm{f.p.}(\hcat)$.
\end{proposition}
\begin{proof}
First, we need to show that $\Yon$ indeed maps $\cat$ into $\mathrm{f.p.}(\hcat)$, but this follows immediately by Proposition \ref{ch3:prop:hompres}. Moreover, $\mathrm{f.p.}(\hcat)$ is right exact by Lemma \ref{ch3:lem:colimofpres}, and $\Yon$ is a full embedding, so we need only check the universal property. This is where Kan extensions come into play. Let $F:\cat\to\cat[A]$ be a functor and $\cat[A]$ be finitely cocomplete, then the span
$$\cat[A]\stackrel{F}{\longleftarrow}\cat\stackrel{\Yon}{\longrightarrow}\mathrm{f.p.}(\hcat)$$ satisfies the conditions of Proposition \ref{ch3:prop:KanExists} and there must therefore exist a functor $F^*=\Lan_{\Yon}F:\mathrm{f.p.}(\hcat)\to \cat[A]$ such that $F^* \Yon=F$. So all we need to check is that $F^*=\Lan_{\Yon}F$ is right extact. The functor $F^*$ is sometimes known as the `Yoneda extension' of $F$ and is known to be a left adjoint, and thus cocontinuous, and a fortiori right exact (see for example \cite{LPAC} 1.45. (i)).
\end{proof}

\begin{proposition}
For a small category $\cat$, a free filtered cocompletion is given by the category $\widetilde{\cat}$ of presheaves whose category of elements is filtered, i.e. by Proposition \ref{ch3:prop:lexiffelfiltered} the category of left-exact functors, together with the Yoneda embedding $\Yon: \cat\to\widetilde{\cat}$.
\end{proposition}
\begin{proof}
Note first that since the category of elements of a $\hom$ functor has a terminal element by Proposition \ref{ch3:prop:CatElHom}, it is trivially filtered, and the Yoneda embedding is indeed defined into $\widetilde{\cat}$. The rest of the proof is exactly as in the proof of Proposition \ref{ch3:prop:finitecocomp}, with Yoneda extentions preserving all colimit, and thus in particular filtered ones.
\end{proof}

\begin{proposition}
For a small category $\cat$, a free cocompletion is given by the category $\Set^{\cat\op}$ of presheaves together with the Yoneda embedding.
\end{proposition}
\begin{proof}
The only thing to show is the universality of $\Inc_c$, and the proof is identical to Proposition \ref{ch3:prop:finitecocomp}.
\end{proof}

Note that the embedding $\Inc_r$ into $\rex(\cat)$ does not in general preserve \emph{existing} colimits in $\cat$, i.e. $\Inc_r$ is not cocontinuous. However, this can be remedied by restricting the embedding to the subcategory of left exact presheaves on $\cat$ (which makes sense since the covariant $\hom$ preserves limits, see Proposition 1.45 (ii) of \cite{LPAC}). This gives rise to the stronger notion of free cocompletion, which is to the free finite completion described above as the Dedekind MacNeille completion is to the canonical extension. Note also that the notion of free finite cocompletion above can be generalized to higher cardinalities by essentially appending a $\lambda$ in front of all the relevant concepts, i.e. the $\lambda$-presentable functors provide a free $\lambda$-small cocompletion.

\section{Presentations of accessible functors}\label{ch3:sec:presaccessfunc}

In this section we will extend the results of the previous section to the case where $\cat$ is no longer small, but can be approximated by a small subcategory, i.e. we will consider $\Set$-functors on accessible categories. As we will see, all the results obtained so far can essentially be transferred to this case, provided that the functor is accessible.

\subsection{General presentations}

We will start by generalizing the construction of Theorem \ref{ch3:thm:fstpresthm} to $\lambda$-accessible functors. Before we can present the main construction we need a few technical lemmas.

%

\begin{lemma}\label{ch3:lem:LanHom}
Let $\cat$ be a $\lambda$-accessible category, let $\cat_\lambda$ be its subcategory of $\lambda$-presentable objects and let $A$ be such a $\lambda$-presentable object. If we denote by $\Inc$ the inclusion functor $\cat_\lambda\inc \cat$, then 
\[
\Lan_{\Inc}(\hom_{\cat_\lambda}(A,-))=\hom_{\cat}(A,-)
\]
\end{lemma}
\begin{proof}
By construction of the left Kan extension, and using the notation of Proposition \ref{ch3:prop:LanF0}, we have for any object $B$ of $\cat$
\begin{align*}
\Lan_{\Inc}(\hom_{\cat_\lambda}(A,B))&=\colim (\hom_{\cat_\lambda}(A,-)\circ \Diag_{c_\lambda}^B)\\
&\stackrel{1}{=}\colim(\hom_{\cat}(A,-)\circ \Diag_c^B)\\
&\stackrel{2}{=}\hom_{\cat}(A,\colim\Diag_c^B)\\
&\stackrel{3}{=}\hom_{\cat}(A,B)
\end{align*}
where (1) follows from the trivial commutative diagram of Proposition \ref{ch3:prop:LanF0}, (2) follows from the fact that $\Diag_c^B$ is $\lambda$-filtered (see (i) of Lemma \ref{ch3:lem:candiag}) and that $A$ is $\lambda$-presentable, and (3) follows from (ii) of Lemma \ref{ch3:lem:candiag}.
\end{proof}

Lemma \ref{ch3:lem:LanHom} establishes the commutativity of the following diagram
\begin{equation}\label{ch3:eq:LanSquare}
\xymatrix
{
\cat_\lambda\op\ar[r]^{\Yon_\lambda}\ar[d]_{\Inc\op} & \Set^{\cat_\lambda}\ar[d]^{\Lan_{\Inc}(-)} \\
\cat\op\ar[r]_{\Yon} & \Set^{\cat}
}
\end{equation} 
where $\Yon_\lambda$ is the Yoneda transformation $\cat_\lambda\to\Set^{\cat_\lambda}$ and $\Yon$ is the Yoneda transformation from the entire category $\cat$. Note also that the $\lambda$-presentability of the object defining the $\hom$ functor is essential, for a general $A$ in $\cat$ we would \emph{not} have $\Lan_{\Inc}\hom_{\cat_\lambda}(A,-)=\hom_{\cat}(A,-)$.

\begin{lemma}\label{ch3:lem:colimHom}
Let $\cat$ be a $\lambda$-accessible category and let $\Diag: \cat[J]\to\Set^{\cat}$ be a small diagram such that $\Diag(j)$ is a functor of the form $\hom(A,-)$ where $A$ is $\lambda$-presentable for each object $j$ of $\cat[J]$, then $\colim \Diag$ preserves $\lambda$-filtered colimits.
\end{lemma}
\begin{proof}
Let $\DiagF: \cat[K]\to \cat$ be a $\lambda$-filtered diagram in $\cat$. Note first that
\begin{align*}
\colim\Diag(\colim \DiagF)&=ev_{\colim\DiagF}\colim\Diag\\
&=\colim ev_{\colim \DiagF}\circ \Diag
\end{align*}
where $ev_A: \Set^{\cat}\to\Set$ is the `evaluation at $A$' functor which preserves colimits (see \cite{MacLane} V.3.). Consider now the functor $ ev_{\colim \DiagF}\circ \Diag$: for any $j$ in $\cat[J]$ we have by assumption that $\Diag(j)$ is of the form $\hom(A_j,-)$ with $A_j$ presentable. Thus $$ev_{\colim \DiagF}\circ \Diag(j)= \Diag(j)(\colim \DiagF)=\colim(\Diag(j)\circ \DiagF)$$
i.e. viewed as functors $\cat[J]\to\Set$ we have $ev_{\colim\DiagF}\circ\Diag=\colim (\Diag(-)\circ \DiagF)$. Finally, the formula of interchange of colimits (see \cite{MacLane} IX.2.) gives us 
\[
\colim(\colim(\Diag(-)\circ \DiagF))=\colim((\colim\Diag)\circ \DiagF)
\]
and we can conclude 
\[
\colim\Diag(\colim\DiagF)=\colim((\colim\Diag)\circ \DiagF))\]
\end{proof}

\begin{lemma}\label{ch3:lem:LanPres}
Let $\cat$ be a $\lambda$-accessible category, $\Inc: \cat_\lambda\inc\cat$ be the inclusion of the subcategory of $\lambda$-presentable objects, and let $T:\cat\to\Set$ be a $\lambda$-accessible functor and $T_\lambda=T\circ\Inc$, under those assumptions, $T$ inherits any presentation of $T_\lambda$. Formally, if 
\[
T_\lambda=\colim \Yon_\lambda\circ\Diag
\] 
for a functor $\Diag:\cat[J]\to\cat_\lambda\op$, then
\[
T=\colim \Yon\circ \Inc\op\circ \Diag
\]
\end{lemma}
\begin{proof}
For every $A$ in $\cat$ we have
\begin{align*}
TA&=\Lan_{\Inc}T_\lambda A & \text{Proposition \ref{ch3:prop:LanF0}}\\
&=\Lan_{\Inc}\colim (\Yon_\lambda\circ \Diag) (A) \\
&=\colim \Lan_{\Inc}(\Yon_\lambda\circ\Diag) (A)& \Lan_{\Inc}(-)\text{ is a left adjoint}\\
&=\colim (\Yon\circ\Inc\op\circ \Diag) (A) & \text{ Diagram } (\ref{ch3:eq:LanSquare})
\end{align*}
\end{proof}

We are now ready to present the main theorem on functor presentation.

\begin{theorem}\label{ch3:thm:scdpresthm}
Let $\cat$ be a $\lambda$-accessible category and $T:\cat\to \Set$. $T$ is accessible iff it is a small colimit of hom-functors $\hom(A,-)$ where $A$ is $\lambda$-presentable.
\end{theorem}
\begin{proof}
The `if' direction was shown in the previous Lemma \ref{ch3:lem:colimHom}.

For the `only if direction', we once again write $\Inc:\cat_\lambda\inc\cat$ and we start by considering the restriction $T_\lambda=T\circ\Inc$. Since this $\cat_\lambda$ is small we can perform the construction of Theorem \ref{ch3:thm:fstpresthm} and  express $T_\lambda$ as 
\[
T_\lambda=\colim\left(\el[T_\lambda]\stackrel{\Forg_{T_\lambda}}{\longrightarrow} \cat_\lambda\op\stackrel{\Yon_\lambda}{\longrightarrow}\Set^{\cat_\lambda}\right)
\]
and where $\Yon_\lambda$ is used as above. It follows that $TA=\colim (\Yon_\lambda\circ\Forg_{T_\lambda}) (A)$ by Lemma \ref{ch3:lem:LanPres}.

\end{proof}

\subsection{Size-bound presentations}

The proof above hints at the fact that accessible functors are those for which the results developed in the case where $\cat$ was small, can be extended to the case where $\cat$ is accessible. As the following results will illustrate, this is indeed the case. We start with the accessible version of Proposition \ref{ch3:prop:presfunc} and Corollary \ref{ch3:cor:lambdapresfunc} wrapped together. The general strategy of all these proofs is to reduce the study of $T$ to that of $T_\lambda$, its restriction to the small subcategory of $\lambda$-presentable objects. We start with the following useful lemma.


\begin{proposition}\label{ch3:prop:accesspresfunc}
Let $\cat$ a $\lambda$-accessible category and let $\mu\leq\lambda$ be a regular cardinal, then an object $T$ in the category $\mathbf{Acc}_{\lambda}$ of $\lambda$-accessible functors $\cat\to\Set$ is $\mu$-presentable iff it is a $\mu$-small colimit of functors of the shape $\hom(A,-)$, $A$ $\lambda$-presentable.
\end{proposition}
\begin{proof}
The `if' direction is clear: if $A$ is $\lambda$-presentable, then $\hom(A,-)$ is accessible. From the proof of Proposition \ref{ch3:prop:hompres} it is clear that all $\hom$ functors are $\lambda$-presentable  since they are absolutely presentable and by Lemma \ref{ch3:lem:colimofpres}, a $\lambda$-small colimit of $\lambda$-presentable objects is $\lambda$-presentable and $\mu\leq \lambda$. 

For the `only if' direction, note first that since $T$ is $\lambda$-accessible, we have by Proposition \ref{ch3:prop:LanF0} that $T=\Lan_{\Inc} T_\lambda$ where $\Inc:\cat_\lambda\to\cat$ is the inclusion of the small subcategory $\lambda$-presentable objects and $T_\lambda$ is the restriction of $T$ to $\cat_\lambda$. By Lemma \ref{ch3:lem:subfuncpres} we therefore know that $T_\lambda$ is $\mu$-presentable because $T$ is. We now apply Proposition \ref{ch3:prop:presfunc} to $T_\lambda$, and we get that $T_\lambda$ is a $\mu$-small colimit of $\hom$ functors in $\Set^{\cat_\lambda}$. We can now follow the same derivation as in the proof of Theorem \ref{ch3:thm:scdpresthm} with the difference that the colimit of the diagram $\Diag_{T_\lambda}: \el[T_\lambda]\to\Set^{\cat_\lambda}$ expressing $T_\lambda$ is now a $\mu$-small colimit of $\hom$ functors, and thus we finally get that $T$ itself is a $\mu$-small colimit of $\hom(A,-)$ functors with all the $A$s $\lambda$-presentable as desired.

\end{proof}

Note that the `if' direction of the previous Proposition can be strengthened in the following way: from Proposition \ref{ch3:prop:hompres} we know that \emph{any} $\hom$ functor is $\lambda$-presentable in $\Set^{\cat}$ for any $\lambda$, and thus a $\mu$-small colimit of $\hom$ functors is $\mu$-presentable in $\Set^{\cat}$. However, $\hom(A,-)$ can only be $\mu$-presentable in $\mathbf{Acc}_\lambda$ if $A$ is $\lambda$-presentable.  

\begin{proposition}\label{ch3:prop:accfuncLocFinPres}
Let $\cat$ be a $\lambda$-accessible category and let $\mathbf{Acc}_\lambda$ be the subcategory of $\Set^{\cat}$ consisting of $\lambda$-accessible functors. $\mathbf{Acc}_\lambda$ is locally finitely presentable.
\end{proposition}
\begin{proof}
Let us first show that $\mathbf{Acc}_\lambda$ is cocomplete. Let us consider a diagram $\Diag_F:\cat[I]\to\mathbf{Acc}_\lambda$ and let $F=\colim\Diag_F$. Of course, since $\mathbf{Acc}_\lambda$ is a subcategory of $\Set^{\cat}$ which is cocomplete, we can build $F$ in $\Set^{\cat}$, but to check that $F$ is an object of $\mathbf{Acc}_\lambda$ we need to check that it preserves $\lambda$-filtered colimits. So let $\DiagF: \cat[J]\to\cat$ be a $\lambda$-filtered diagram, by using the interchange of colimits we have:
\begin{align*}
F(\colim\DiagF)&=\colim\Diag_F(\colim\DiagF) \\
&=\colim(ev_{\colim\DiagF} \circ\Diag_F)\\
&=\colim(\colim(\Diag_F(-)\circ \DiagF)) \\
&=\colim((\colim\Diag_F)\circ\DiagF)\\
&=\colim (F\circ\DiagF)
\end{align*}
and thus $\mathbf{Acc}_\lambda$ is cocomplete.

Let us now show that all $\lambda$-accessible functors are filtered colimits of finitely presentable objects in $\mathbf{Acc}_\lambda$. Again, we use the fact that $T=\Lan_{\Inc} T_\lambda$, and apply known results to $T_\lambda$. In this instance we apply Corollary \ref{ch3:cor:funcfinpres} and get that $T_\lambda$ is a filtered colimit of finitely presentable objects in $\Set^{\cat_\lambda}$, i.e. a filtered colimit of finite colimits of functors of the shape $\hom(A,-)$ with $A$ $\lambda$-presentable (in $\cat$). Lemma \ref{ch3:lem:LanPres} then allows us to extend this result to $T$ itself, i.e. $T$ is  also a filtered colimit of finite colimits of functors of the shape $\hom(A,-)$ with $A$ $\lambda$-presentable. By Proposition \ref{ch3:prop:accesspresfunc} this means that $T$ is a filtered colimit of finitely presentable objects in $\mathbf{Acc}_\lambda$, which concludes the proof.
%
\end{proof}

In is interesting to note from the previous Proposition that the `$\lambda$' of `$\lambda$-accessible' controls the type of $\hom$ functors used to present a $\lambda$-accessible functor, i.e. $\hom$s of the shape $\hom(A,-)$ for $A$ $\lambda$-presentable in the base category, but is independent of the `size' of the presentation which is controlled by size of the small category $\cat_\lambda$. For a $\lambda$-accessible functor $T:\cat\to\Set$, any bounds on the size of the presentation are therefore also inherited from bounds on the size of the presentation of its restriction $T_\lambda$. The notion of strongly $\lambda$-presented and strongly $\lambda$-generated functors, can therefore be lifted directly to accessible functor.

\begin{definition}A $\lambda$-accessible functor on a $\lambda$-accessible category $\cat$ will be called \textbf{strongly $\mu$-presented}\index{Strongly generated functor}\index{Strongly presented functor} (resp. generated) if its restriction $T_\lambda$ to the subcategory of $\lambda$-presentable objects is strongly $\mu$-presented (resp. generated). 
\end{definition}

Similarly, we can use the left Kan extension functor to lift the notion of presented and generated functors to the accessible case. Using the same notation as above, if we assume that $T_\lambda$ is $\mu$-generated, then $T_\lambda$ can be expressed as a regular epi $\coprod_{i\in I}\Hom(A_i,-)\stackrel{e}{\epi T_\lambda}$ for a set $|I|<\mu$. We now apply the left Kan extension functor $\Lan_{\Inc}(-)$ to get 
\[
\Lan_{\Inc}(\coprod\limits_{i\in I}\Hom(A_i,-))=\coprod\limits_{i\in I}\Lan_{\Inc}\Hom(A_i,-)=\coprod\limits_{i\in I}\Hom(A_i,-)\epi \Lan_{\Inc}(T_\lambda)=T
\]
i.e. $\Lan_{\Inc}(-)(e): \coprod_{i\in I}\Hom(A_i,-)\epi T$, which is a regular epi too since $\Lan_{\Inc}(-)$ being a left adjoint preserves all colimits. The same construction clearly defines the notion of being $\mu$-presented for a $\lambda$-accessible functor. \index{Generated functor}\index{Presented functor} Finally, the lifting by $\Lan_{\Inc}(-)$, together with Proposition \ref{ch3:prop:accesspresfunc} also allows us to transfer Propositions \ref{ch3:prop:presfunc} and \ref{ch3:prop:geniffgen} to accessible functors, i.e. a  an $\lambda$-accessible functor is $\mu$-presented (resp. generated) iff it is a $\mu$-presentable (resp. $\mu$-generatable) object in $\mathbf{Acc}_\lambda$.

The following proposition describes a class of base categories for which finitely-generated and finitely-presented functors coincide. This class includes the categories of distributive lattices and of boolean algebras.

\begin{proposition}\label{ch3:prop:locfinvar}
Let $\cat$ be a locally finite variety (i.e. a variety whose finitely generated objects are finite, see \cite{2001:NicksBro}) and let $T:\cat\to\Set$ be a finitary functor preserving monos. Then $T$ is finitely presentable iff it is finitely generatable.
\end{proposition}
\begin{proof}
We only need to show that finitely generatable functors are finitely presentable. So let $T$ be finitely generatable and let $T_\omega$ be its restriction to the subcategory $\cat_\omega\stackrel{\Inc_\omega}{\inc}\cat$ of finitely presented objects. Since $T$ is finitary, $T=\Lan_{\Inc_\omega} T_\omega$ by \ref{ch3:prop:LanF0}, and thus if we can express $T_\omega$ as a finite colimit of $\hom$ functors, we are done by Lemma \ref{ch3:lem:LanPres}.

The idea of the proof is as follows. Just as we did in the proof of Propositions \ref{ch3:prop:presfunc} and \ref{ch3:prop:geniffgen}, we start by considering the category whose objects are the subcategories $\cat_0$ of $\cat_\omega$ with finitely many objects and whose arrows are the obvious inclusions. We can immediately note that since $\cat$ is a locally finite variety, (i.e. finitely generated objects are finite), each such subcategory has finitely many morphisms. We will now modify this category slightly: for every finite subcategory $\cat_0$ of $\cat_\omega$, we define the category $\overline{\cat}_0$ as the full subcategory of $\cat$ whose objects are those of $\cat_0$ plus all their strong quotients. Since the strong quotient of a finitely presented variety is a finitely generated one, and since those are finite by assumption, $\overline{\cat}_0$ also has finitely many objects and finitely many morphisms. We now consider the category $\cat[J]$ whose objects are all the completions $\overline{\cat}_0$ of finite subcategories $\cat_0$, and whose morphisms are the inclusions, and we define a diagram $\Diag:\cat[J]\to \Set^{\cat_\omega}$ by $\Diag(\overline{\cat}_0)=\Lan_{\Inc_0}(T_\omega\circ \Inc_0)$ where $\Inc_0$ is the inclusion $\Inc_0: \overline{\cat_0}\inc \cat$. It is easy to see that $\cat[J]$ is directed and that
\[
T_\omega=\colim\Diag
\]
So far, the construction looks a lot like that of Propositions \ref{ch3:prop:presfunc} and \ref{ch3:prop:geniffgen}, but the key difference is that each $\overline{\cat}_0$ has all quotients. This means that transformations $d_{\overline{\cat}_0}: \Lan_{\Inc_0}(T_\omega\circ \Inc_0)\to T_\omega$ are monos. Indeed, by construction of the left Kan extension as a colimit, for every $A\in\cat_\omega, \Lan_{\Inc_0}(T_\omega\circ \Inc_0) A$ is taken as a colimit over the objects of $\overline{\cat}_0 \downarrow A$, but any such $f: C\to A$ for $C$ in $\overline{\cat}_0$ can be factored as a strong epi (which lives in $\overline{\cat}_0$) followed by a mono, and since $T$ is assumed to preserve monos, the colimit is in fact taken over a collection of monos pointing into $TA$. Formally, 
\[
\colim\left(\overline{\cat}_0 \downarrow A\stackrel{\Forg_0}{\longrightarrow}\cat_\omega\stackrel{T_\omega}{\longrightarrow}\Set\right)=\colim\left((\overline{\cat}_0 \mono A)\stackrel{\mathsf{V}_0}{\longrightarrow}\cat_\omega\stackrel{T_\omega}{\longrightarrow}\Set\right)
\]
where $(\overline{\cat}_0 \mono A)$ is the category whose objects are all the monos from $\overline{\cat}_0$ into $A$, and whose morphisms are defined in the obvious fashion. The functors $\Forg_0,\mathsf{V}_0$ are the obvious forgetful functors. From this it follows that for every $\overline{\cat_0}$ the map $d_{\overline{\cat_0}}:\Lan_{\Inc_0}(T_\omega\circ \Inc_0)\mono T_\omega$ is a mono, and thus $\Diag$ in fact defines a filtered diagram of monos. Since $T$ is finitely generatable, $T_\omega$ is also finitely generatable by Lemma \ref{ch3:lem:subfuncgen}, and it must therefore be equal to one of the elements of this diagram, i.e. 
\begin{equation}\label{ch3:prop:locfinvar:eq}
T_\omega=\Lan_{\Inc_0}(T_\omega\circ\Inc_0)
\end{equation}
for a certain finite full subcategory $\overline{\cat}_0\stackrel{\Inc_0}{\inc}\cat_\omega$. 

Now we can use the exact same technique as in the proof of Proposition \ref{ch3:prop:presfunc} and show that $T_0=T_\omega\circ\Inc_0$ is image finite. The key is that this technique allows us to write $T_0$ as a filtered colimit of monos (all of which are image finite), and we can thus use the fact that $T_\omega$ is finitely generatable together with Lemma \ref{ch3:lem:subfuncgen} and Eq. (\ref{ch3:prop:locfinvar:eq}) to conclude that $T_0$ is finitely generatable and must therefore be equal to one of the image finite functors. From there it follows easily that the category $\el[T_0]$ has finitely many objects and finitely many morphisms (since these arise from the finitely many morphisms of $\overline{\cat}_0$). Two applications of \ref{ch3:lem:LanPres} then allow us to conclude that $T$ is strongly finitely presented, and thus finitely presented and thus finitely presentable (by Proposition \ref{ch3:prop:presfunc}).
\end{proof}

\begin{remark}
The construction above (viz. extending the subcategory defining a $\lambda$-generated functor by adding all the quotients) can be carried out in general, but the result is that the enlarged subcategory contains objects which are no longer $\lambda$-presentable, but $\lambda$-generated. So we can express a $\lambda$-generated functor either as a direct image of a left Kan extension over a subcategory of $\lambda$-\emph{presentable} objects, \emph{or}, as a Kan extension over a subcategory of $\lambda$-\emph{generated} objects, in which case the direct image is taken in the base category, and we need preservation of monos.
\end{remark}

\subsection{The canonical presentation.}\label{ch3:sec:canPres}

Let again $\cat$ be a $\lambda$-accessible category, $\Inc:\cat_\lambda\inc\cat$ be the inclusion of the subcategory of $\lambda$-presentable objects, $T:\cat\to\Set$ be a $\lambda$-accessible functor, and $T_\lambda=T\circ\Inc$. By Theorem \ref{ch3:thm:scdpresthm} we have
\[
T=\colim(\el[T_\lambda]\stackrel{\Forg}{\longrightarrow}\cat_\lambda\op\stackrel{\Yon\circ \Inc}{\longrightarrow}\Set^{\cat})
\]
This presentation of $T$ is what we will call the \textbf{canonical presentation}\index{Canonical presentation}. The purpose of this section is to show what this colimit actually looks like. Re-writing this colimit as a coequalizer of a coproduct gives us
\[
\xymatrix
{
\coprod\limits_{f\in\el[T_\lambda]\mor}\Yon\circ I\circ\Forg(\dom(f))\ar@<1ex>[r]^\phi \ar@<-1ex>[r]_\psi & \coprod\limits_{(A,\alpha)\in\el[T_\lambda]\ob}\Yon\circ I\circ\Forg((A,\alpha)) \ar@{->>}[r] & T
}
\]
where $\phi$ and $\psi$ are defined uniquely by their action on the summands of the coproduct (by the universal property of the coproduct) as follows: $\phi\circ \mathrm{in}_f =\mathrm{in}_{\dom(f)}$ and $\psi\circ \mathrm{in}_f=\mathrm{in}_{\cod(f)}\circ \Yon I\Forg(f)$. By using the definitions and Yoneda's lemma, this can be re-written as
\[
\xymatrix
{
\coprod\limits_{f\in \cat_\lambda\downarrow A, A\in\cat_\lambda\ob}T\dom(f)\times\hom(A,-)\ar@<1ex>[r]^<<<<{\phi} \ar@<-1ex>[r]_<<<<{\psi} & \coprod\limits_{A\in\cat_\lambda\ob}TA\times \hom(A,-)\ar@{->>}[r] & T
}
\]
since by definition of $\el[T_\lambda]$ there exists a morphism in $\el[T_\lambda]$ for each choice of an object $A$ in $\cat_\lambda$, a morphism $f: B\to A$ and an element $\beta\in TB=T\dom(f)$. The natural transformations $\phi$ and $\psi$ are now defined as follows: $\phi\circ \mathrm{in}_f=\mathrm{in}_{\cod(f)}\circ \langle Tf\times Id\rangle$ and $\psi\circ \mathrm{in}_f=\mathrm{in}_{dom(f)}\circ\langle Id \times ((-)\circ f)\rangle$. For any $C$ of $\cat$, we get that
\[
\xymatrix
{
\coprod\limits_{f\in \cat_\lambda\downarrow A, A\in\cat_\lambda\ob}T\dom(f)\times\hom(A,C)\ar@<1ex>[r]^<<<<{\phi} \ar@<-1ex>[r]_<<<<{\psi} & \coprod\limits_{A\in\cat_\lambda\ob}TA\times \hom(A,C)\ar@{->>}[r] & TC
}
\]
i.e. $TC$ is the quotient of the set $\coprod_{A\in\cat_\lambda\ob}TA\times \hom(A,C)$ by the smallest equivalence relation generated for the relation $\sim_C$ defined for any $(\alpha, g)\in TA\times\hom(A,C)$ and $(\alpha', A')\in TA'\times\hom(A',C)$ by 
\begin{align*}
(\alpha, g)\sim_C(\alpha',g')&\text{ if there exists }f: B\to B'\text{ and }(\beta, f')\in TB\times\hom(B',C)\\
&\text{ such that } \phi(\beta,f')=(\alpha,g), \psi(\beta,f)=(\alpha', g')
\end{align*}
i.e. if there exist $f: B\to B'$ and $(\beta,f')\in TB\times\hom(B',C)$ such that
\[
Tf(\beta)=\alpha, f'=g  \text{ (and thus in particular }A=B'\text{)}
\]
and 
\[
\beta=\alpha', g'=f'\circ f \text{ (and thus in particular }A'=B\text{)}
\]
We thus get the following characterisation:
\[
(\alpha, g)\sim_C(\alpha',g')\text{ if there exists }f: A'\to A\text{ such that }g'=g\circ f\text{ and }\alpha=Tf(\alpha')
\] 
The \textbf{canonical presentation}\index{Canonical presentation} can thus equivalently be defined as the regular epi transformation 
\[
q^c: \coprod_{A\in\cat_\lambda\ob}TA\times \hom(A,-)\epi T
\] 
obtained by quotienting the functor $\coprod_{A\in\cat_\lambda\ob}TA\times \hom(A,-)$ at each stage $C$ of $\cat$ by the smallest equivalence relation generated by $\sim_C$. 

In the general case where $\cat$ is an arbitrary $\lambda$-accessible category, we know that the cardinality of $\cat\ob_\lambda$ must be at least $\lambda$ by Corollary \ref{ch3:cor:pressizemin}, and thus the size of the coproduct must also be at least $\lambda$. However, the precise cardinality of $\cat\ob_\lambda$ seems to be a largely open question. In the case where $\cat$ is the locally finitely presentable category $\Grp$, it is known that the set of finitely presentable objects is countable (see \cite{1997:countfinpresgrp} Chapter 9) but non-recursively enumerable (see \cite{2003:nonrecenumgrp}). 
In this latter case the canonical presentation $q^c$ of a finitary functor takes the shape $T$
\[
q^c_X: \coprod_{n\in\mathbb{N}} Tn\times \hom(n,X)\epi TX
\]
Our notion of canonical presentation (i.e. via the equivalence relation generated by $\sim_X$) can also be found in e.g. \cite{2011:KurzEqPresMon} but is not the `standard' notion found in most of the literature (e.g. \cite{setFuncPres,AAC, 2012:finfuncpres, Thesis:Leal, KurzLeal}) where the focus is to present functors as quotients of polynomial functors rather than as a colimit of representables. Nevertheless, in the case of accessible $\Set$-endofunctors the two notions coincide, although this is not immediately obvious.

\begin{proposition}\label{ch3:prop:twoequiv}
Let $\sim_X$ be the relation on $\coprod_{n\in\mathbb{N}} Tn\times \hom(n,X)$ defined above and let $\approx_X$ be the relation on $\coprod_{n\in\mathbb{N}} Tn\times \hom(n,X)$ defined for any $(\sigma,f)\in Tm\times\hom(m,X)$ and $(\tau, g)\in Tn\times\hom(n,X)$ as in 3.9. of \cite{setFuncPres}, namely
\[
(\sigma,f)\approx_X(\tau,g)\text{ if }Tf(\sigma)=Tg(\tau)
\]
The relations $\sim_X$ and $\approx_X$ generate the same equivalence relation on $X$.
\end{proposition}
\begin{proof}
The reader might recognise the relations $\sim_X$ and $\approx_X$, since they are the same relations as in the proof of Proposition \ref{ch3:prop:StrongGenIffGen}. In fact the proof that $\sim_X$ and $\approx_X$ define the same quotient on a set $X$ is identical to the proof that the quotients defined by the corresponding relations in Proposition \ref{ch3:prop:StrongGenIffGen} coincide. 
\end{proof}

As we have seen in Section \ref{ch1:subsec:algsem}, a regular epimorphism can be seen as a quotient under some equations. In particular a regular epi from a polynomial functor can alternatively be seen as an pair $(\Sigma, E)$ where $\Sigma$ is the signature defining the polynomial functor - in the case of a $\lambda$-accessible functor this would be the set of $\lambda$-presentable objects - and $E$ is a set of equations - i.e. the set of pairs $((\sigma,f),(\tau, g))$ originating from a common element through the maps $\phi,\psi$ defined above. It is usually in terms of such a pair $(\Sigma, E)$ that the notion of \emph{presentation of a functor} is defined (see for example \cite{setFuncPres}).

\subsection{Atomic objects of $\el$.}

In this section we will show that the canonical presentation which tends to have a lot a redundant information about the functor, can often be simplified. To show this we will need the notion of atomic object in a category. Let $\cat$ be a well-powered category, and let us firstly assume that $\cat$ has not got an initial object, we will then say that an object $A$ in $\cat$ is \textbf{atomic}\index{Atomic object} if it has no proper subobjects, i.e. if all the monos with codomain $C$ are isos, or using the $\Sub$ functor if $\Sub(A)=\{\id_A\}$. If $\cat$ has an initial object, we will define its atomic objects as the atomic objects of $\cat_0$, the subcategory of $\cat$ where the initial objects and all arrows from it have been removed. The atomic elements of $\mathbf{C}$ form a fully faithful subcategory $\atob[\cat]$ of $\cat$.

\begin{example}
\begin{enumerate}
\item If we consider a poset $\cat[P]$ as a category, then since all arrows are monic, an element of $\cat[P]$ is atomic precisely when it's an atom in the order-theoretic sense.
\item In $\Set$, an object is atomic iff it is a singleton set, so there really is only one atom in $\Set$, and it is $1$.
\item The atomic objects in $\Grp$, i.e. the groups without proper subgroups are the cyclic groups $\mathbb{Z}_n$ with $n$ prime. Indeed, if $G$ is an atomic object in $\Grp$, then for any $g\in G, g\neq e$ the subgroup $\langle g\rangle$ generated by $g$ must be the entire group $G$. Thus $G$ must be cyclic. If $G$ is infinite, then it must be isomorphic to $\mathbb{Z}$, but $\mathbb{Z}$ has plenty of proper subgroups (e.g. the even integers), so $G$ must be finite and cyclic. Since the subgroups of cyclic groups are cyclic and since for a finite cyclic group of order $n$ its subgroups' order must be a divisor of $n$, it is clear that $\mathbb{Z}_n$ cannot have a proper subgroup if $n$ is prime. 
\item A similar argument as in the previous case, or a duality argument from example 2, shows that the only atomic object in $\BA$ is the free boolean algebra on one element, due to the idempotency of all the operators defining a BA.
\item The category which will interest us is $\el\op$, for a functor $T:\cat\to\Set$ where $\cat$ is small or $\lambda$-accessible. Intuitively, an object $(A,\alpha)$ of $\el\op$ is atomic if $\alpha$ is built using the entirety of the object $A$, i.e. there cannot exist $i:A'\inc A$ and $\alpha'\in TA'$ such that $\alpha= Ti(\alpha')$. For example, if $T=\pow$ and $A=3=\{0,1,2\}$ then $(3,\{0,1\})$ is not atomic since it comes from the injection of $(2,\{0,1\})$, but $(2,\{0,1\})$ itself is atomic. We will denote the subcategory of atomic objects of $\el\op$ by $\atob\op$.
\end{enumerate}
\end{example}

From the example above it is clear that atomicity in $\el\op$ is related to the notion of base. 

\begin{proposition}
Let $\cat$ be well-powered and have pullbacks of monos, and let $T:\cat\to\Set$ preserve intersections, then $(A,\alpha)$ is atomic in $\el\op$ iff $A=\Base(\alpha)$.
\end{proposition}
\begin{proof}
If $A=\Base(\alpha)$, then by definition of $\el$ we have $\alpha\in T\Base(\alpha)$. Clearly if there existed a strict mono $i:A'\inc A$ and an element $\alpha'\in TA'$ such that $Ti_{A'}^A(\alpha')=\alpha$, then $A'$ would be a strict subobject of the intersection of all such subobjects which is a contradiction, so $(A,\alpha)$ is indeed atomic.

For the converse, assume that $(A,\alpha)$ is atomic. Since $i^A_{\Base(\alpha)}:\Base(\alpha)\inc A$, if we can show that $\alpha\in Ti^A_{\Base(\alpha)}[T\Base(\alpha)]$, then we're done since $(A,\alpha)$ is atomic. But this is exactly the content of Lemma \ref{ch3:lem:interbase}.
\end{proof}

\begin{corollary}\label{ch3:cor:atombelow}
Let $\cat$ be well-powered and have pullbacks of monos, and let $T:\cat\to\Set$ preserve intersections, then for every object $(A,\alpha)$ in $\el\op$, there exists an atomic object $(B,\beta)$ and a monomorphism $i: B\to A$ such that $Ti(\beta)=\alpha$.
\end{corollary}

Note that since all locally presentable categories are complete and well-powered (see \cite{LPAC} 1.56),  Corollary \ref{ch3:cor:atombelow} applies to all $\Set$-valued functors on locally presentable categories which preserve intersections, in particular all functors defined on sets or algebraic varieties which preserve intersections. 

\subsection{The minimal presentation}

\begin{theorem}\label{ch3:thm:atomcofinal}
Let $\cat$ be a small category with pullbacks of monos and let $T:\cat\to\Set$ preserve intersections, then $\Inc:\atob\inc \el$ is cofinal.
\end{theorem}
\begin{proof}
Let $(A,\alpha)$ be an object in $\el$, we need to find an atomic object $(B,\beta)$ and an $\el$-morphism $(A,\alpha)\to(B,\beta)$. The existence of such an object and such a morphism is precisely the content of Corollary \ref{ch3:cor:atombelow} (since small categories are automatically well-powered).

Next consider two $\el$-morphisms $f_1:(A,\alpha)\to (B_1,\beta_1)$ and $f_2:(A,\alpha)\to (B_2,\beta_2)$ with $(B_1,\beta_1)$ and $(B_2,\beta_2)$ atomic. Then we need to find a zigzag of morphisms in $(A,\alpha)\downarrow\Inc$ between $f_1$ and $f_2$. Since $\cat$ has all intersections, we can factorize $f_1, f_2$ through their direct images $f_1[B_1]$ and $f_2[B_2]$ and we get the following commutative diagram
\[
\xymatrix
{
B_1\ar[dr]_{f_1^*} & & \Base(\alpha)\ar@{^{(}->}[dl]^{i_1}\ar@{_{(}->}[dr]_{i_1} & & B_2\ar[dl]^{f_2^*} \\
& f_1[B_1]\ar@{_{(}->}[dr]_{m_1} & & f_2[B_2]\ar@{^{(}->}[dl]^{m_2} \\
& & A
}
\]
by definition of the base of $\alpha$ and the fact that $m_1$ and $m_2$ are monomorphisms. It is then easy to check that we have the following zigzag of morphisms in $(A,\alpha)\downarrow\Inc$:
\[
\xymatrix@C=4ex
{
& (A,\alpha)\ar[dd]_{m_1}  & & (A,\alpha)\ar@{=}[dr] \ar[dd]^{m_2}\\
(A,\alpha)\ar@{=}[ur]\ar[dd]_{f_1} & & (A,\alpha)\ar@{=}[ul]\ar@{=}[ur]\ar[dd]^{i^A_{\Base(\alpha)}} & & (A,\alpha)\ar[dd]^{f_2} \\
& (f_1[B_1], Tf^*_1(\beta_1))\ar[dl]^{f^*_1}\ar[dr]_{i_1} & & (f_2[B_2], Tf^*_2(\beta_2))\ar[dl]^{i_2}\ar[dr]_{f_2^*}\\
(B_1,\beta_1) & & \Base(\alpha,\alpha')  & & (B_2,\beta_2) \\
}
\]
Notice the reversal of arrows between the two diagrams which is due to the definition of $\el$.
\end{proof}

As an immediate corollary of the previous result we get the following representation theorem.
\begin{theorem}\label{ch3:thm:smallatomrep}
Let $\cat$ be a small category closed under intersections and let $T:\cat\to\Set$ preserve intersections, then $$T=\colim\left(\atob\stackrel{\Inc}{\longrightarrow}\el\stackrel{\Forg_T}{\longrightarrow}\cat\op\stackrel{\Yon}{\longrightarrow}\Set^{\cat}\right)$$ 
\end{theorem}
\begin{proof}
Immediate by Theorems \ref{ch3:thm:atomcofinal} and \ref{ch3:thm:cofinal}.
\end{proof}

We would now like to proceed as we did for the previous representation theorem, namely by extending the result for small category to a result for $\lambda$-accessible categories and $\lambda$-accessible functors. However, compared with Theorem \ref{ch3:thm:fstpresthm}, Theorem \ref{ch3:thm:smallatomrep} has an additional condition placed on the small category, and this condition might not in general hold in the small subcategory $\cat_\lambda$ of $\lambda$-presentable objects in a $\lambda$-accessible category. The following example from \cite{1968:interfingengrp} shows that this is indeed the case.

\begin{example}
Let $\Free:\Set\to\Grp$ be the free group functor, and consider the direct product $\Free 2\times \Free 1$, i.e. $\Free 2\times\mathbb{Z}$. This group can be finitely presented by $\langle a,b,c\mid ac=ca, bc=cb\rangle$, in particular it is finitely presented. Now consider the finitely presented subgroups $G=\Free 2=\langle a,b\rangle $ and $H=\langle a,bc\rangle$. As the following argument will show, the intersection $G\cap H=\langle b^ia b^{-i}, i\in \mathbb{N}\rangle$, which is clearly \emph{not} finitely presentable. Using the commutation rules of the presentation of $\Free 2\times\mathbb{Z}$, it is clear that for any $i\in \mathbb{N}$ we have 
$$b^ia b^{-i}=b^ic^ic^{-i}ab^{-i}=b^ic^iac^{-i}b^{-i}=(bc)^i a(bc)^{-i}$$
and thus $ \langle b^ia b^{-i}, i\in \mathbb{N}\rangle$ is clearly a subgroup both of $G$ and $H$. Conversely, let $w(a,bc)$ be a word in $\Free 2\times\mathbb{Z}$ built from $a$ and $bc$ only, then using the commutation rules we can re-group all the $c$'s and re-write $w(a,bc)=w'(a,b)c^k$ where $k$ is the sum of the exponents of the $b$'s in $w'(a,b)$. If $w(a,bc)$ lies in the intersection of $G$ and $H$, we must clearly have $k=0$, i.e. $w'(a,b)\in \langle b^ia b^{-i}, i\in \mathbb{N}\rangle$.
\end{example}

To our knowledge, no characterisation of the $\lambda$-accessible categories whose $\lambda$-presentable objects are closed under intersections exists. Indeed, as the previous example has shown us, even finite intersections are in general problematic. So the best we can hope for is to isolate some useful classes of accessible categories that do exhibit this property. The following Proposition aims at doing just that.

\begin{proposition}\label{ch3:prop:locallyfinite}
Let $\cat$ be a locally finite variety, then the class of finitely presentable objects of $\cat$ is closed under intersections.
\end{proposition}
\begin{proof}
Let $\{A_i\inc B\}_{i\in I}$ be a set of finitely presented subobjects of an algebra $B$. Since each $A_i$ is finite, and since limits of algebras are created in $\Set$, $A=\bigcap_i A_i$ is also finite. To show that it is finitely generated, let $\Free\dashv\Forg:\Set\to\cat$ denote the usual free/forgetful adjunction. The counit of the adjunction $\epsilon_A$ provides a regular epi $\epsilon_A: \Free\Forg A\epi A$ since the adjunction $\Free\dashv\Forg$ is monadic (see \cite{1993:KellyPower}). We thus have a presentation of $A$ using finitely many generators given by $\langle \Forg A\mid \Forg\ker(\epsilon_A)\rangle$.
\end{proof}

Examples of locally finite varieties include $\DL,\BA$ and $\Set$ which should convince the reader that requiring the closure of the set of presentable objects under intersection is not too strict a requirement in the context of coalgebraic logics. Moreover, for $\Set$-endofunctors we have the following useful fact.

\begin{proposition}[\cite{KKV:2012:Journal}]\label{ch3:prop:weakPBinter}
Let $T:\Set\to\Set$ be finitary, then $T$ preserves intersections.
\end{proposition}

\begin{theorem}\label{ch3:thm:accessatomrep}
Let $\cat$ be a $\lambda$-accessible category whose small subcategory $\Inc_\lambda:\cat_\lambda\inc\cat$ of $\lambda$-presented objects is closed under intersections. For any intersection-preserving $\lambda$-accessible functor $T:\cat\to\Set$ let $T_\lambda$ be its restriction to $\cat_\lambda$, we then have
$$T=\colim\left(\atob[T_\lambda]\stackrel{\Inc}{\longrightarrow}\el[T_\lambda]\stackrel{\Inc_\lambda\op\Forg_{T_\lambda}}{\longrightarrow}\cat\op\stackrel{\Yon}{\longrightarrow}\Set^{\cat}\right)$$ 
\end{theorem}
\begin{proof}
The proof is identical to that of Theorem \ref{ch3:thm:scdpresthm}, with the difference that Theorem \ref{ch3:thm:smallatomrep} allows us to represent $T_\lambda$ as the colimit $$T_\lambda=\colim\left(\atob[T_\lambda]\stackrel{\Inc}{\longrightarrow}\el[T_\lambda]\stackrel{\Forg_{T_\lambda}}{\longrightarrow}\cat_\lambda\op\stackrel{\Yon}{\longrightarrow}\Set^{\cat_\lambda}\right)$$
and the conclusion follows from Lemma \ref{ch3:lem:LanPres}.
\end{proof}

As we saw earlier, colimits can be expressed as coequalizers whose domains are coproducts. Under the hypotheses of Theorem \ref{ch3:thm:accessatomrep}, a functor $T$ can be represented via the regular epi-transformation:
\[
q^m: \coprod_{A\in\cat_\lambda\ob} \Base_A\inv\times\Hom(A,-)\epi T
\]
where $\Base_A\inv=\{\alpha\in TA\mid \Base_A(\alpha)=A\}$ and where $q^m_C$ identifies $(\sigma,f)\in \Base_A\inv\times\Hom(A,C)$ and $(\tau, g)\in \Base_B\inv\times\Hom(B,C)$ if there exists $h: A\to B$ such that $f=g\circ h$ and $\tau=Th(\sigma)$. The transformation $q^m$ will be called the \textbf{minimal presentation} of $T$.

\begin{example}
We illustrate the construction in the case of the finite powerset functor $\powf$. The canonical presentation of $\powf$ is given by 
\[
q^c_X: \coprod_{n\in\mathbb{N}} \powf n\times \hom(n,X)\epi \powf X
\]
where $q^c_X$ identifies $(U, f)$ and $(V, g)$ where $U\subset \{1,\ldots, m\}$ and $V\subset \{1,\ldots, n\}$ when there exist a map $h: m\to n$ such that $f=g\circ h$ and $\powf h(U)=h[U]=V$, or, equivalently by Proposition \ref{ch3:prop:twoequiv}, when $\powf f (U)=f[U]=\powf g(V)=g[V]$. The minimal presentation is given by
\[
q^m_X: \coprod_{n\in\mathbb{N}} n\times \hom(n,X)\epi \powf X
\]
since for each $n$ there is only one atomic element, namely $(n, \{n\})$. The regular epi $q^m_X$ identifies $(U, f)$ and $(V, g)$ under the same circumstances as for $q^m_C$. Notice how much redundant information was removed when going from the canonical to the minimal presentation.
\end{example}

\section{Lifting presentations using adjunctions}\label{ch3:sec:liftpres}

Let $\cat$ be a $\lambda$-accessible category for which there exists an adjunction $\Free\dashv\Forg$ where $F:\Set\to\cat$ is a `free' functor and $\Forg:\cat\to\Set$ is a forgetful functor. In this section we will show how we can use the results developed so far together with the adjunction $\Free\dashv\Forg$ to give a representation of a large class of functors $T:\cat\to\cat$. We restrict ourselves to finitary functors, but all the results that follow can be adapted easily to the general case of $\lambda$-accessible functors.

One way to represent $T$ might be the following: we know how to represent accessible functors $\cat\to\Set$ as colimits of representable functors, and from there as regular quotients of coproducts of representables. So, for an endofunctor $T$ on $\cat$ we could try to represent $\Forg T:\cat\to\Set$. However, as we have illustrated earlier in the case of $\cat=\Grp$, representing a $\lambda$-accessible functor $\cat\to\Set$ can be difficult practically since it will involve a coproduct over all $\lambda$-presentable objects, which is already an incredibly complicated set for the case of finitely presented groups. The situation is simpler in the case of $\BA$, where the finitely presented boolean algebras are those of the form $\pow n$, for all $n$ finite. However, apart from the complexity of listing all presentable objects in a category, there is another reason for which we choose another path, namely that in the case of $\cat=\BA$ we would like BAEs to emerge from our presentations, i.e. we would like $\BA$-morphisms between freely generated BAs, or equivalently $\Set$-morphisms between their sets of generators. Ideally, we would therefore like to involve the free functor $\Free$.

With these two objections to representing $\Forg T$ in mind, a better way to proceed is to find a representation of the functor $\Forg T\Free: \Set\to\Set$, since presentable objects in $\Set$ are easy to deal with, and with the free functor in the mix we will hopefully get to BAEs in the case of $\cat=\BA$.

The first requirement to build a representation of a finitary $T:\cat\to\cat$ by using a representation of $\Forg T\Free$ is naturally that $\Forg T\Free$ should be finitary. Since $\Free$ is cocontinuous and $T$ is finitary, this means that $\Forg T\Free$ is finitary iff $\Forg$ is finitary. A adjunction $\Free\dashv\Forg$ with this property is called a \textbf{finitary adjunction}, and the associated monad is called a \textbf{finitary monad}\index{Finitary monad}\index{Finitary adjunction}. These concepts are readily extended to $\lambda$-accessible adjunctions and monads. A well-known class of finitary adjunctions is given by the following well-known result:

\begin{lemma}
If $\cat$ is a finitary algebraic variety, then $\Forg: \cat\to\Set$ \emph{creates} filtered colimits
\end{lemma}
\begin{proof}
See e.g. \cite{MacLane} IX. 2, or \cite{1986:Johnstone:Stone}.
\end{proof}

Assuming $\Forg$ and $T$ finitary, we can represent $\Forg T\Free$ by using Theorem \ref{ch3:thm:scdpresthm}:
\[
\Forg T\Free=\colim\left(\el[(\Forg T\Free)_\omega]\stackrel{\Inc\circ\mathsf{V}}{\longrightarrow}\Set\stackrel{\Yon}{\longrightarrow}\Set^{\Set}\right)
\]
where $\Inc:\Set_f\inc\Set$ is the inclusion of the full subcategory of finite sets, $(\Forg T\Free)_\omega=\Forg T\Free\Inc$ and $\mathsf{V}:\el[(\Forg T\Free)_\omega]\to \Set_f$ is the obvious forgetful functor. This yields the canonical presentation
\begin{equation}\label{ch3:eq:lift1}
q^m: \coprod_{n\in\mathbb{N}}\Forg T\Free(n)\times \hom(n,-)\epi \Forg T\Free
\end{equation}

However, it is ultimately $T$ we want a representation for, and to go from a representation of $\Forg T\Free$ to a representation of $T$ we use the adjunction as follows. We pre-compose with $\Forg$ and get a presentation
\begin{equation}\label{ch3:eq:lift2}
q_{\Forg}^m: \coprod_{n\in\mathbb{N}} \Forg T\Free(n)\times \hom(n,\Forg(-))\epi \Forg T\Free\Forg 
\end{equation}
which is regular and has a polynomial functor as its domain.
We will now `kill' the $\Free\Forg$ term by using the counit of the adjunction, or more precisely $\Forg T\epsilon$. This will give us a regular transformation onto $\Forg T$, which by the adjunction will give us a transformation onto $T$. 
Since we want a presentation, i.e. a regular epi from a coproduct of $\hom$ functors, we want $\Forg T\epsilon$ to be a regular epi. For a general adjunction the counit is in general \emph{not} a regular epimorphism. The class of adjunctions for which the counit is a regular epi is in fact well-known to category theorists.

\begin{definition}[\cite{1993:KellyPower}]
An adjunction $\Free\dashv \Forg: \Set\to\cat$ is said to be of \textbf{descent type} if any of the following equivalent conditions is satisfied:
\begin{enumerate}
\item the comparison functor $\mathsf{K}: \cat\to\Set^{\Forg\Free}$ is full and faithful
\item $\epsilon$ is a regular epi-transformation
\end{enumerate}
\end{definition}

This class of adjunctions has been used in \cite{1993:KellyPower} and \cite{2011:KurzEqPresMon} to prove a slightly different type of presentation results (i.e. not directly related to the canonical colimit description of accessible functors) for finitary functors between enriched categories. Note that the more frequently studied class of \emph{monadic} adjunctions, i.e. those for which the comparison functor is \emph{an equivalence} of categories, is a subclass of the class of adjunctions of descent type. 

\begin{lemma}The counit of an adjunction $\Free\dashv\Forg:\cat\to\cat[D]$ of monadic type is a regular epi. 
\end{lemma}
\begin{proof}
Consider the following fork:
\[
\xymatrix
{
\Forg\Free\Forg\Free\Forg A\ar@<3pt>[r]^{\Forg\Free\Forg \epsilon_A}\ar@<-2pt>[r]_{\Forg\epsilon_{\Free\Forg A}} & \Forg\Free\Forg A\ar[r]^{\Forg\epsilon_A} & \Forg A
}
\]
and let us show that it is a split coequalizer, the result will then follow from Beck's Monadicity Theorem. First, notice that $\eta_{\Forg A}$ is a section of $\Forg\epsilon_A$ by definition of the interaction between the unit and the counit in an adjunction. Similarly, $\eta_{\Forg\Free\Forg A}$ is a section of $\Forg\epsilon_{\Free\Forg A}$. So we just need to show that 
\[
\eta_{\Forg A}\circ \Forg \epsilon_{A}=\Forg\Free\Forg\epsilon_A\circ \eta_{\Forg\Free\Forg A}
\]
but this is just a consequence of the naturality of $\eta$.
\end{proof}

For $T\epsilon$ to be regular, there is unfortunately no choice: we need to impose that $T$ preserve regular epimorphisms. This requirement was already shown to be particularly helpful in Sections \ref{ch1:sec:boolstruct} and \ref{ch1:sec:rel}, and also, albeit in a slightly different context, in Chapter 3 of \cite{Thesis:Rob}. As it turns out, this is all we need to ask from our functors as the following result indicates.

\begin{proposition}[\cite{Borceux}]\label{ch3:prop:regepiAdj}
Let $\Free\dashv\Forg:\cat\to\cat[D]$ be an adjunction, then $\Forg$ preserves regular epimorphisms.
\end{proposition}
\begin{proof}
Proposition 4.3.9. of \cite{Borceux}.
\end{proof}

In particular if $\epsilon$ is a regular epi-transformation and $T$ preserves regular epis, then $\Forg T\epsilon$ is a regular epi. There is one last step before we have shown that $\Forg T$ can be written as a regular quotient of a polynomial functor. Recall the regular epi $q^m_{\Forg}$ of Eq. (\ref{ch3:eq:lift2}), we need to combine it with $\Forg T\epsilon$ to get a presentation
\[
\Forg T\epsilon\circ q^m_{\Forg}: \coprod_{n\in\mathbb{N}} \Forg T\Free(n)\times \hom(n,\Forg(-))\epi \Forg T\Free\Forg\epi \Forg T 
\]
To combine these two regular epis we need regular epis to compose. This is not always the case as is shown in \cite{2000:quotLocales}, but we have discussed a natural framework in which this is the case in Section \ref{ch1:sec:rel}, namely regular categories. Since the categories we are interested in are all regular, this will not cause us any restriction. We summarize our results as follows.
\begin{theorem}
Let $\cat$ be an accessible category in which the composition of two regular epis is a regular epi, $\Free\dashv\Forg: \cat\to\Set$ be a finitary adjunction of descent type, and let $T:\cat\to\cat$ be a finitary functor which preserves regular epis, then $T$ has a canonical presentation by the regular epi:
\[
\Free\left(\coprod_{n\in\mathbb{N}} \Forg T\Free(n)\times \hom(n,\Forg(-))\right)\epi T 
\] 
\end{theorem}
\begin{proof}
From the presentation $\Forg T\epsilon\circ q^m_{\Forg}$ defined above, we get the adjoint transpose transformation described in the statement, and we just need to show that it is a regular epi. This follows immediately from the fact that $\Free$ is cocontinuous and that $\epsilon$ is a regular epi.
\end{proof}

In particular any finitary endofunctor $T:\BA\to\BA$ or $T:\DL\to\DL$ which preserves regular epis can be given a presentation of the type described in the Theorem above. As we will now see in Chapter 4, such a transformation allows us to present $T$-algebras on $\BA$ as quotiented $\Free(\coprod_{n\in\mathbb{N}} \Forg T\Free(n)\times \hom(n,\Forg(-)))$-algebras, i.e. a BAE as we have shown in Chapter 1.

A very similar presentation of endofunctors on $\BA$ can be found in \cite{2006:bonsanguepresenting} and in \cite{2011:KurzEqPresMon}. In \cite{2006:bonsanguepresenting} the notion of presentation is introduced in an \textit{ad-hoc} fashion, motivated by the example of modal algebras. Our presentation here is closer to that of  \cite{2011:KurzEqPresMon}, although their starting point lies with adjunctions $\Free\vdash\Forg$ of descent type (in an enriched setting), whilst our presentation starts with the notion of presentable object in $\Set^{\cat}$, with the adjunction used to lift presentations to $\cat^{\cat}$.
\chapter{Translations}

This chapter will be of a relatively technical nature. Its primary use will be to develop the tools necessary to formulate a theory of completeness-via-canonicity for the $\nabla$ flavour of coalgebraic logic. This version of coalgebraic logic is based on the category $\BA$, its positive fragment (which would be defined over $\DL$) has not yet been studied. This chapter will thus be concerned entirely with algebras in $\BA$ on the syntax side and coalgebras in $\Set$ on the semantics side. We note also that Theorem \ref{ch4:thm:semthm} makes essential use of a property of $\Set$, and it is not clear how this could be adapted to a more general category such as $\Pos$.

In the previous chapter we have seen how well-behaved $\BA$-endofunctors can be given a presentation in terms of polynomial functors.  A natural question to ask in this context is: given two such functors $S,T$ and a natural transformation $q: S\to T$ (where $S$ might define a presentation of $T$), what can we say about the relationship between the coalgebraic logics associated with $S$ and $T$? Is there a syntactic relationship? And what happens at the semantic level? These questions seem natural but, as far as we know, have not really been studied systematically in the literature.

\section{Syntax translations}\label{ch4:sec:syntrans}

Recall from Chapter 1, that we can consider coalgebraic logics as coming in three flavours. In the predicate lifting style, the syntax is determined by a signature $\Sigma$ of operators and associated arities. The result is that we can view the language of such a logic as the initial algebra for the functor 
\[
S:\BA\to\BA, A\mapsto \Free \left(\coprod_{\sigma\in\Sigma} (\Forg A)^{\ari(\sigma)}\right)=\Free\polyFunc\Forg
\]
where $\polyFunc$ is the polynomial $\Set$ functor associated with the signature $\Sigma$. In light of the previous chapter, we can consider this functor as being trivially presented by itself, and we can leave it at that. It is clear that the structure map of any $S$-algebra $s:SA\to A$ is in bijective correspondence with a $\Set$-morphism  $\coprod_{\sigma\in\Sigma}(\Forg A)^{\ari(\sigma)}\to \Forg A$, i.e. with the data of a $\Sigma$-BAE with reduct $A$.

In the nabla formalism, the syntax is determined by the $\Set$-endofunctor $T$ defining the semantics, and the language $\lang_T$ defined in Chapter 1 can be viewed as the initial algebra for the functor 
\[
\Free T\Forg: \BA\to\BA
\]
In this case, the functor defining the syntax is therefore not presented. To get a representation of $\Free T\Forg$, note that by the adjunction $\Free\dashv\Forg$, any presentation of $T\Forg$ will lead to a presentation of $\Free T\Forg$, and any presentation of $T$ trivially gives a presentation of $T\Forg$. Thus we need to present $T$. Since $T$ is assumed to be finitary and weak pullback preserving, we have two canonical choices: we can use the canonical or the minimal presentation, since, by Proposition \ref{ch3:prop:weakPBinter}, $T$ will then preserve all intersections. As it turns out, the minimal presentation has some very nice properties, and we will therefore often choose this presentation for $T$, i.e. we have a regular epi
\[
SX=\coprod_{n\in\mathbb{N}}\hom(n,X)\times \Base_{n}\inv\epi T
\]
The structure map $s: \Free S\Forg A\to A$ of any $\Free S\Forg$-algebra is therefore once again in bijective correspondence with a $\Set$-morphism  $\coprod_{n\in\mathbb{N}}\hom(n,\Forg A)\times \Base_{n}\inv\to\Forg A$, i.e. with the data of a $(\Base_{n}\inv)_{n\in \mathbb{N}}$-BAE with reduct $A$.

Finally in the abstract formalism, the syntax is determined by a $\BA$-functor $L$. As was mentioned in Chapter 1, we need to make some assumptions on $L$, namely that it preserves regular epis and weak pullbacks and that it is a varietor. Again, the language defined by $L$ can be thought of as the initial $L$ algebra which exists by \ref{ch1:cor:initAlgFin}. If $L$ is finitary (or just accessible), we can use the results from the last section of the previous chapter to give a presentation to $L$.

In the last two cases (and trivially in the first), the presentation of the syntax functor provides us with a new syntax functor whose algebras can be viewed as BAEs, which is precisely the reason for which we consider these presentations in the first place. The task is now to understand how we can go from the language of the presentation to the original language.

Let $K,L$ be two regular epi preserving varietors, and let $q: K\epi L$ be a regular epi-transformation. Let $\mathsf{Q}:\Alg_{\BA}(L)\to\Alg_{\BA}(K)$ denote the functor defined by $\mathsf{Q}(A,\alpha)=(A,\alpha\circ q_A)$ and for $f:(A,\alpha)\to (B,\beta)$ (i.e. $f\circ \alpha=\beta\circ Lf$) by $\mathsf{Q}f=f$ which defines a $K$-algebra morphism by naturality of $q$.

\begin{theorem}\label{ch4:thm:syntaxthm}\index{Translation! syntax}
Let $K,L:\BA\to\BA$ be two finitary functors such that $K$ preserves regular epimorphisms, and let $q: K\epi T$ be a  natural transformation. Let $\Free_K\dashv \Forg_K: \Set\to\Alg_{\BA}(K)$ denote the obvious free/forgetful adjunction associated with $K$, and similarly $\Free_L\dashv \Forg_L$ for $L$. There exist a unique natural transformation 
\[
\xi: \Free_K\to\mathsf{Q}\Free_L
\]
Moreover, $\xi$ is a regular epi-transformation whenever $q$ is a regular epi-transformation.
\end{theorem}
\begin{proof}
The existence of $\xi$ is not difficult to show. Since the free $K$-algebra over an object $A$ of $\BA$ is the initial $K(-)+A$ algebra, we get by initiality the existence of a unique morphism
\begin{equation}\label{ch4:thm:syntaxthm:diag}
\xymatrix@C=12ex
{
K\Free_K (A)+A\ar[dd]\ar@{-->}[r]^{K\xi_A+\id_A} & K \Free_L (A)+A\ar@{->>}[d]^{q_{\Free_L (A)}+\id_A} \\
& L\Free_L (A) +A\ar[d] \\
\Free_K (A)\ar@{-->}[r]_{\xi_A} & \Free_L (A)
}
\end{equation}
It is not difficult to see that $\xi$ is natural, and the first statement follows.

Now, let us show that $\xi_A$ is a regular epi. Let us write $K'=K(-)+A$ and $L'=L(-)+A$. By construction of the initial algebras, we have the following diagram where $\Forg: \BA\to\Set$ is the forgetful functor and $0$ is the initial object of $\BA$
\[
\xymatrix@R=4ex
{
& \Forg K' 0\ar[r]^{\Forg K'!}\ar@{->>}[dd]^{\Forg q_0}\ar[dl] & \Forg K'^2 0\ar@{->>}[dd]^{\Forg(q_{L'0}\circ K' q_0)}\ar[dll]\ar[r]^{\Forg K'^2!} & \cdots \\
\colim_i \Forg K'^i 0\ar@{-->}[dd]_{\phi} & & \\
& \Forg L'0\ar[r]^{\Forg L'!}\ar[dl] & \Forg L'^2 0\ar[r]^{\Forg L'^2 !}\ar[dll] & \cdots \\
\colim_j \Forg L'^j 0
}
\]
where $\phi$ is the unique arrow arising from the fact that $\colim_j \Forg L'^j 0$ is a cocone for the diagram defining $\colim_i \Forg K'^i 0$. It is easy to check that this is a commutative diagram. Note also that since we're assuming that $q$ is a regular epi-transformation, then $\Forg q$ is also a regular epi by Proposition \ref{ch3:prop:regepiAdj}. Moreover, since $K$ is assumed to preserve regular epis, we in fact have that every vertical arrow in the diagram is epi in $\Set$. By construction of colimits in $\Set$, we know that any element $x\in \colim_j \Forg L'^j 0$ can be traced back to an element $y\in \Forg L'^n 0$ for a certain $n$, and thus by the fact that the vertical arrow landing in $\Forg L'^n0$ is surjective, there must exist an element $y'\in \Forg K'^n 0$ which gets mapped to an element $x'\in\colim_i \Forg K'^i 0$, and by commutativity of the diagram $\phi(x')=x$. Thus $\phi$ is surjective.

Next, notice that since $\Forg$ preserves (and creates) filtered colimits, we actually have $$\colim_i\Forg K'^i 0=\Forg \colim_i K'^i 0=\Forg\init[K']$$ and similarly for $L'$. By unicity of the the arrow $\phi$ we must also have $\phi=\Forg\xi_A$. So we have that $\Forg\xi_A$ is regular epi, and in particular it is the coequalizer of its kernel pair. Since $\Forg$ is monadic over $\Set$, it reflects exact sequences (see Proposition \ref{ch1:prop:MonadicExactOverSet}), and $\xi_A$ must be therefore be a regular epi.

\end{proof}

The following trivial fact will be useful: it is easy to see that
\[
\Forg_K\mathsf{Q}=\Forg_L
\]
This means in particular that 
\[
\Forg_K\xi_{\Free V}: \Forg_K \Free_K\epi\Forg_K\mathsf{Q}\Free_L=\Forg_L\Free_L
\]
It is clear that since $\Forg_K\xi_{\Free V}$ is just the underlying $\Set$ map of $\xi_{\Free V}$, it is (regular) epi. We will sometimes write $\Forg_K\xi_{\Free V}$ as $(-)^q: \Forg_K \Free_K\to\Forg_L \Free_L$ and call it the \textbf{syntax translation} or \textbf{translation map} between the languages.

The natural transformation $\xi$ also interacts in an interesting and useful way with the counits of the adjunctions $\Free_K\dashv \Forg_K$ and $\Free_L\dashv\Forg_L$ as the following Proposition shows:

\begin{proposition}\label{ch4:prop:xicounits}
Let $K,L,q,\mathsf{Q}$ and $\xi$ be as above, and let $\epsilon^K$ and $\epsilon^L$ be the counits of the adjunctions $\Free_K\dashv \Forg_K$ and $\Free_L\dashv\Forg_L$ respectively, we then have for any $A$ in $\Alg_{\BA}(L)$ that
\begin{equation}\label{ch4:eq:1}
\epsilon^K_{\mathsf{Q}A}=\mathsf{Q}\epsilon^L_{A}\circ \xi_{\Forg_LA}
\end{equation}
\end{proposition} 
\begin{proof}
The proof is a consequence of the construction of free algebras. Recall that the carrier $\init[(L(-)+\Forg_L A)]$ of $\Free_L\Forg_L A$ is the colimit of the initial sequence (in $\BA$):
\[
0\xto{!}L(0)+\Forg_L A \xto{L!+\id_{\Forg_L A}} L(L(0)+\Forg_L A)+\Forg_L A\to\ldots
\] 
It is easy to see that $\Forg_L A$ is a cocone for this sequence and there must therefore exist a unique morphism 
\[
\epsilon_{\Forg_L A}^L: \init[L(-)+\Forg_L A]\to \Forg_L A
\]
Our choice of notation is not innocent, since $\epsilon_{\Forg_L A}$ is indeed the counit of $\Free_L\dashv\Forg_L$ (on carriers). The same construction goes for $\epsilon_{\Forg_L A}^K$. Since the initial sequences for $L$ and $K$ are connected by the natural transformation $q$ we get the following commutative diagram
\[
\xymatrix
{
\init[(K(-)+\Forg_L A)]\ar@{-->}[dd]^{\epsilon_{\Forg_L A}^K}\ar@/_2pc/@{-->}[dddd]_{\Forg_K\xi_{\Forg_L A}}\\
& 0\ar[dl]\ar[dd]^{\id_0}\ar[ul]\ar[r] & K(0)+\Forg_L A \ar[dll]\ar[ull]\ar[r]\ar[dd]^{u_0=q_0+\id_{\Forg_L A}}& K(K(0)+\Forg_L A)+\Forg_L A\ar[dlll]\ar[ulll]\ar[dd]^{q_{L(0)+\Forg_L A}\circ Ku_0+\id_{\Forg_L A}}\ar[r] & \ldots\\
\Forg_L A\\
& 0\ar[dl]\ar[ul]\ar[r] & L(0)+\Forg_L A\ar[r]\ar[ull]\ar[dll] & L(L(0)+\Forg_L A)+\Forg_L A \ar[r]\ar[ulll]\ar[dlll]& \ldots\\
\init[(L(-)+\Forg_L A)]\ar@{-->}[uu]_{\epsilon_{\Forg_L A}^L}
}
\]
The result follows by the unicity of the three dashed arrows and the naturality of $q$.
\end{proof}

Theorem \ref{ch4:thm:syntaxthm} shows us how a natural transformation induces a translation between the languages associated with two functors. As an illustration, we show how this works in the case of the nabla flavour of coalgebraic logic. Since the nabla languages are defined by $\Set$-endofunctors, let $S, T:\Set\to\Set$ be two weak-pullback preserving functors (note that $\Set$ functors always preserve regular epis) and let $q: S\to T$ be any natural transformation. We can read off the effect of the translation transformation 
\[
\xi_{\Free V}: \init[(\Free S\Forg(-)+\Free V)]\to \init[(\Free T \Forg(-)+ \Free V)]
\]
from Diagram \ref{ch4:thm:syntaxthm:diag} of Theorem \ref{ch4:thm:syntaxthm}. By defining $(-)^q=\Forg_{\Free S\Forg}\xi_{\Free V}$ as above, we get the recursive definition for every $\alpha\in \init[(\Free S\Forg(-)+\Free V)]$:
\[
(\nabla\alpha)^q=\nabla (q_{\init[(\Free T\Forg(-)+\Free V)]}\circ S(-)^q)(\alpha)
\]
If we use the nabla languages $\lngS$ and $\lngT$ (see Section \ref{ch1:sec:coalgLang}) rather than the abstract language associated with $\Free S\Forg$ or $\Free T\Forg$, we can rewrite this translation map a bit more pithily for every $\alpha\in\lngS$ as
\[
(\nabla\alpha)^q=\nabla (q_{\lngT}\circ S(-)^q)(\alpha)
\]
 
\section{Models and semantics translations}\label{ch4:sec:semtrans}

This section starts with the dual counterpart of the previous section. We will then examine how the connections between syntax and models, i.e. the semantics, interacts with natural transformations, and in particular functor presentations. We end with an application to the case of the nabla flavour of coalgebraic logics.

\subsection{Semantic translations}

Since models are described in exactly the same way for the predicate-lifting, nabla and abstract descriptions of coalgebraic logic, i.e. by coalgebras, we just need to look at the impact of natural transformations on coalgebras. Note first that a natural transformation $q: S \to T$ between two $\Set$-endofunctors induces a functor $Q: \Coalg(S) \to \Coalg(T)$ on the corresponding categories of coalgebras, given for any $(W,\gamma)$ in $\Coalg(S)$ and any $f:(W,\gamma)\to (W',\delta)$ (i.e. $Tf\circ \gamma=\delta\circ  f$) by
\[
Q(W, \gamma) = (W,q_W \circ \gamma), \hspace{3ex}Qf=f
\]
and $Qf$ is a coalgebra morphism by naturality of $q$. We will now show how we can use $Q$ to define a translation from $S$-behaviours to $T$-behaviours, in a way that is dual to the syntactic translation of Theorem \ref{ch4:thm:syntaxthm}. First recall the classic result.

\begin{theorem}[Ad\`{a}mek]\label{ch4:thm:TermCoalgEx}
Let $T:\cat\to\cat$ be an endofunctor on a category with a terminal object 1. Assume further, that the limit of the following diagram exists in $\cat$
\[
\cdots\longrightarrow T^3 1\stackrel{T^2 !}{\longrightarrow}T^2 1\stackrel{T !}{\longrightarrow} T 1\stackrel{!}{\longrightarrow} 1
\]
and that $T$ preserves it, then $T$ has a terminal coalgebra.
\end{theorem}

Note that if $\cat$ is locally presentable, then it is complete and cocomplete and in particular $\cat$ has a terminal object and the limit above always exists, so the only requirement left is that $T$ should preserve it. This is always the case for polynomial functors:

\begin{proposition}Coproducts of $\hom: \cat\to\Set$ functors preserve $\omega\op$-limits.
\end{proposition}
\begin{proof}
This is a consequence of the fact that $\hom$ functors preserve limits and that small coproducts in $\Set$ commute with connected limits, and in particular with $\omega\op$-limits.
\end{proof}

In particular polynomial functors always have terminal coalgebras. So if $T$ is a finitary $\Set$-endofunctor and $S$ is a polynomial functor arising from a presentation of $T$, i.e. $q:S\epi T$, then we can always build the terminal coalgebra of $S$. Intuitively, the terminal coalgebra of $T$ should be a quotient of the terminal coalgebra for $S$, since the latter is a more fine-grained specification of behaviours. This is indeed the case and is at the core of the method developed by Ad{\'a}mek  and Milius in \cite{2006:AdamekTermCoalg} to compute terminal coalgebras for functors which do not preserve the $\omega\op$-limit of Theorem \ref{ch4:thm:TermCoalgEx}. The procedure is in two steps. First, they show that if $\term[S]$ is the terminal coalgebra of $S$, then the coalgebra \[
\term[S]\longrightarrow S\term[S]\stackrel{q_{\term[S]}}{\longrightarrow}T\term[S]
\]
is a \emph{weakly} terminal $T$-coalgebra. They then quotient this coalgebra to turn it into the terminal $T$-coalgebra by identifying behaviours that $T$ cannot tell apart. In this way it can be shown that every finitary functor on $\Set$ has a terminal coalgebra. In fact it is the case that every accessible functor on a locally finitely presentable category has a terminal coalgebra (see Theorem 4.2.12.  \cite{2010:AdamekSurvey}).  With this fact established, we can easily quotient $S$-behaviours onto $T$-behaviours as follows.

\begin{theorem}\label{ch4:thm:semthm}\index{Translation! semantics}
Let $S,T:\Set\to\Set$ be two finitary functors, and let $q: S\epi T$ be an epi-transformation. Let $\Cofree_S:\Set\to\Coalg(S)$ denote the obvious cofree functor associated with $S$, and similarly with $\Cofree_T$ for $L$. There exist a natural transformation 
\[
\chi: Q\circ \Cofree_S\to \Cofree_T
\]
which is an epi-transformation.
\end{theorem}
\begin{proof}
The existence proof is dual to that of Theorem \ref{ch4:thm:syntaxthm}. Let $V$ be any set, by using the fact that $\Cofree_T(V)$ is the terminal $T(-)\times V$-coalgebra, we get: 
\begin{equation}\label{ch4:thm:semthm:Diag}
\xymatrix@C=12ex
{
T\Cofree_S V\times V\ar@{-->}[r]^{T\chi_V\times\id_V} & T\Cofree_T V\times V\\
S\Cofree_S V\times V\ar@{->>}[u]^{q_{\Cofree_S V}} \\
\Cofree_S V\ar[u]\ar@{-->}[r]_{\chi_V} & \Cofree_T V\ar[uu]
}
\end{equation}
To prove that $\chi_V$ is surjective when $q$ is an epi-transformation, we use the fact, in $\Set$, every epimorphism has a section. In particular, there exist a section
\[
s: T\Cofree_T V\to S\Cofree_T V, \text { such that }q_{\Cofree_T V}\circ s=\id_{T\Cofree_T V}
\]
Using $s$ we can build for every set $V$ the following commutative diagram (where we drop all the $-\times V$ terms and $-\times \id_V$ maps for legibility):
\[
\xymatrix@C=12ex
{
T\Cofree_T V\ar@<-2pt>[r]_{T\sigma} & T\Cofree_S V\ar@<-3pt>@{-->}[l]_{T\chi_V}\\
S\Cofree_T V\ar@<-2pt>@{-->}[r]_{S\sigma}\ar[u]_{q_{\Cofree_T V}} & S\Cofree_S V\ar[u]_{q_{\Cofree_S V}}\ar@<-3pt>[l]_{S\chi_V}\\
T\Cofree_T V\ar[u]^{s}\ar@/^2pc/[uu]^{\id_{T\Cofree_T V}} \\
\Cofree_T V\ar@<-2pt>@{-->}[r]_{\sigma} \ar[u]& \Cofree_S V\ar[uu]\ar@<-3pt>@{-->}[l]_{\chi_V}
}
\]
Where $\sigma$ exists by terminality of $\Cofree_S V$ and $\chi$ exists by terminality of $\Cofree_T V$. It follows that $\chi_V\circ\sigma=\id_{\Cofree_T V}$, i.e. $\chi_V$ is a split epi, and in particular a surjection.
\end{proof}

If there is no ambiguity about the set $V$ of colours, we will denote the $\Set$-morphisms $\chi_V: \Cofree_T V\to \Cofree_S V$ by $(-)_q$, in duality to the syntax translation map $(-)^q$ defined in the previous section. We will call this map the \textbf{semantic translation}.


\subsection{Compatible semantics}

We now know how functor presentations (or more generally natural transformations) give rise to translations between languages and between models. The next natural question is to investigate how presentations interact with the relationship between language and model, i.e. the semantics.

Recall that for the abstract description of coalgebraic logic given by the diagram
\[
\xymatrix@C=12ex
{
\BA\ar@/^1pc/[r]^{\uf} \ar@(l,u)^{L}& \Set\op\ar@/^1pc/[l]^{\pow}\ar@(r,u)_{T\op}
}
\]
the semantics is given by a natural transformation $\delta: L\pow\to\pow T\op$. Now, let $q:K\epi L$ be a transformation arising from a presentation of $L$ (recall that $K$ is then of the form $\Free P \Forg$, where $P$ is polynomial in $\Set$) and $r:S\epi T$ be a transformation arising from a presentation of $T$ (e.g. the minimal presentation). Since $\pow$ is contravariant, we have a canonical choice of natural transformation $\lambda:K\pow \to \pow S$ given by:
\begin{equation}\label{ch4:diag:badDiag}
\xymatrix@C=12ex
{
K\pow\ar@{->>}[d]_{q\pow}\ar[r]^{\lambda} & \pow S\\
L\pow\ar[r]_{\delta} & \pow T\ar@{>->}[u]_{\pow r}
}
\end{equation}
which is clearly natural. However, $\lambda$ defined in this way has some unpleasant properties. As was mentioned in Chapter 1, the properties of $\delta$ (or its adjoint transpose) reflect important properties of the logic described by the diagram above, for example expressivity and completeness. If we choose $\lambda$ as above, then $\lambda$ will be neither mono (since $q\pow$ is epi) nor epi (since $\pow r$ generally isn't). So no matter how well-behaved $\delta$ is, it is all lost by building $\lambda$ in this way. As will soon become clear, we need a well behaved logic at the polynomial level, in particular we will need completeness in order to build models. We therefore need to proceed differently, and the case of the nabla logic will guide us in finding a good notion of `lifted semantics' for our purpose. Recall that in the case of the nabla style of coalgebraic logic, the abstract semantics takes the form:
\[
\delta: \Free T\Forg \pow\to \pow T
\]
which is freely generated by the (adjoint) maps
\[
\tilde{\delta}_X: T\Forg \pow X\to \Forg\pow TX,\hspace{6pt} A\mapsto\{\alpha\in TX\mid \alpha\Tmem A\}
\]
Clearly, this assignment is fully generic in the functor $T$, and we can similarly define for a polynomial functor $S$ such that $q:S\epi T$, the natural transformation
\[
\lambda: \Free S\Forg \pow\to \pow S
\]
which is freely generated by the (adjoint) maps
\[
\tilde{\lambda}_X: S\Forg \pow X\to \Forg\pow SX,\hspace{6pt} A\mapsto\{\alpha\in SX\mid \alpha\Smem A\}
\]
The crucial observation to be made about this choice of semantics is that if $\alpha\Smem A$, then there must exist a witness $w\in S(\in_X)$ which projects to $\alpha$ and $A$. By naturality of $q$ we therefore get a witness $q_{\in_X}(w)\in T(\in_X)$ which projects to $q_X(\alpha)$ and $q_{\Forg\pow X}(A)$. This means that we have the following weaker condition than Diagram (\ref{ch4:diag:badDiag}):
\[
\hat{\lambda}_X(A)\subseteq \pow q_{\Forg\pow X}\circ \hat{\delta}_X\circ q_{\Forg\pow X}(A)
\]
This relation can be understood as the statement that the $\lambda$-interpretation of $A$ agrees with the $\delta$-interpretation of $q_{\Forg\pow X}(A)$. We formalize and generalize this notion from the nabla style of coalgebraic logic in the following definition.
\begin{definition}[Compatible semantics]\index{Compatible semantics}\label{ch4:def:compSem}
Let $(K,S,\lambda: K\pow\to\pow S)$ and $(L,T,\delta: L\pow\to \pow T)$ define two abstract coalgebraic logics and their semantics, and let $q: K\to L$ and $r: S\to T$ be two natural transformations. We will say that $\lambda$ and $\delta$ are \textbf{compatible semantics} w.r.t. to $q,r$  whenever for every set $X$, if we view $K\pow X$ and $\pow SX$ as skeletal categories, then there exists a transformation
\[
\mu_X: \lambda_X\to \pow r_X\circ \delta_X\circ q_{\pow X} 
\]
Or using `evil' (i.e. mentioning elements) terminology, for every $a\in K\pow X$
\[
\lambda_X(a)\subseteq \pow r_X\circ \delta_X\circ q_{\pow X}(a)
\]
\end{definition}

Let us now show that the term `compatible' is indeed appropriate.

\begin{theorem}\label{ch4:thm:compSem}
Let $(K,S,\lambda: K\pow\to\pow S)$ and $(L,T,\delta: L\pow\to \pow T)$ define two abstract coalgebraic logics and their semantics, and let $q: K\to L$ and $r: S\to T$ be two natural transformations making $\lambda$ compatible with $\delta$. Let also $V$ be a set (of propositional variables) and let $a\in \Free_K V$, if for some coalgebra $W\xto{\gamma} TW$, some valuation $v: V\to\mathcal{Q} W$ and some $w\in W$
\[
w,\gamma,v\models a
\]
then
\[
w,r_W\circ\gamma,v\models (a)^q
\]
Moreover, if $w\in \Cofree_S \mathcal{Q}V$ and
\[
w\models a
\]
then
\[
(w)_q\models (a)^q
\]
\end{theorem}
\begin{proof}
The interpretation maps are defined by initiality via the following commutative diagram (where we have dropped all the $(-)+\Free V$ and $(-)+\id_V$ terms for clarity's sake)
\[
\xymatrix
{
K\Free_K V\ar[ddd]\ar@{-->}[rr]^<<<<<<<<<{K\lsem-\rsem_\gamma^K}\ar[dr]_{K(-)^q}  & &  K\pow W\ar[d]^{\lambda_W}\ar@{=}[dr]\\
& K\Free_L V\ar[d]_{q_{\Free_L V}}\ar@/_1pc/[rr]_>>>>{K\lsem-\rsem^L_{r_W\circ\gamma}} & \pow S W\ar[dd]^>>>>>{\pow \gamma} & K\pow W\ar[d]^{q_{\pow W}}\\
& L\Free_L V\ar@{-->}[rr]^<<<<<<<{L\lsem-\rsem^L_{r_W\circ\gamma}}\ar[ddd] & & L\pow W\ar[d]^{\delta_W}\\
\Free_K V\ar@{-->}[rr]_<<<<<<<<<{\lsem-\rsem_\gamma^K}\ar[ddr]_{(-)^q} & & \pow W & \pow TW\ar[d]^{\pow r_W}\\
& & & \pow S W\ar[d]^{\pow \gamma}\\
 & \Free_LV\ar@{-->}[rr]^<<<<<<<<<{\lsem-\rsem^L_{r_W\circ\gamma}} & & \pow W
}
\]
Staring from $a=\langle\alpha\rangle_K, \alpha\in K\Free_K V$, we know from the compatibility assumption that 
\[
\lambda_W(K\lsem -\rsem_\gamma^K(\alpha))\subseteq \pow r_W(\delta_W(q_{\pow W}( K\lsem -\rsem_\gamma^K(\alpha))))
\]
from which it follows by commutativity of the diagram that
\[
\lsem a\rsem_\gamma^K\subseteq \lsem (a)^q\rsem_{r_W\circ\gamma}^L
\]
which is what we needed to show. In terms of the terminal coalgebra semantics, we adapt the diagram above as follows
\[
\xymatrix
{
K\Free_K V\ar[ddd]\ar@{-->}[rr]^<<<<<<<<<{K\lsem-\rsem^K}\ar[dr]_{K(-)^q}  & &  K\pow \Cofree_S\mathcal{Q}V\ar[d]^{\lambda_{\Cofree_S\mathcal{Q}V}}\ar@{=}[rr] & & K\pow \Cofree_S\mathcal{Q}V\ar[d]^{q_{\pow \Cofree_S\mathcal{Q}V}}\\
& K\Free_L V\ar[d]_{q_{\Free_L V}}\ar@/_1pc/[rr]_>>>>>>>{K\lsem-\rsem^L} & \pow S \Cofree_S\mathcal{Q}V\ar[dd] & K\pow \Cofree_T\mathcal{Q}V \ar[d]^{q_{\pow \Cofree_T\mathcal{Q}V}}\ar[ur]^{K\pow (-)_q}& L\pow \Cofree_S\mathcal{Q}V\ar[d]^{\delta_{\Cofree_S\mathcal{Q}V}} \\
& L\Free_L V\ar@{-->}[rr]^<<<<<<<<<<{L\lsem-\rsem^L}\ar[ddd] & & L\pow \Cofree_T\mathcal{Q}V\ar[d]_{\delta_{\Cofree_T\mathcal{Q}V}} \ar[ur]_{L\pow (-)_q}& \pow T\Cofree_S\mathcal{Q}V\ar[d]^{\pow r_{\Cofree_S\mathcal{Q}V}}\\
\Free_K V\ar@{-->}[rr]_<<<<<<<<<{\lsem-\rsem^K}\ar[ddr]_{(-)^q} & & \pow \Cofree_S\mathcal{Q}V & \pow T\Cofree_T\mathcal{Q}V\ar[dd]\ar[ur]_{\pow T(-)_q} & \pow S\Cofree_S\mathcal{Q}V\ar[d]\\
& & & & \pow \Cofree_S\mathcal{Q}V \\
 & \Free_LV\ar@{-->}[rr]^<<<<<<<<<<<<{\lsem-\rsem^L} & & \pow \Cofree_T\mathcal{Q}V\ar[ur]_{\pow (-)_q}
}
\]
Staring from $a=\langle\alpha\rangle_K, \alpha\in K\Free_K V$, the compatibility assumption once again entails that
\[
\lambda_{\Cofree_S\mathcal{Q}V}(K\lsem -\rsem^K(\alpha))\subseteq \pow r_{\Cofree_S\mathcal{Q}V}(\delta_{\Cofree_S\mathcal{Q}V}(q_{\pow \Cofree_S\mathcal{Q}V}( K\lsem -\rsem^K(\alpha))))
\]
from which it follows by commutativity of the diagram that
\[
\lsem a\rsem^K\subseteq \pow(-)_q(\lsem (a)^q\rsem^L)
\]
i.e.
\[
\text{if }w\models a\text{ then }(w)_q\models (a)^q
\]
as desired.
\end{proof}

\subsection{The case of nabla logics}\label{ch4:subsec:nabla}
We illustrate the notion of compatible semantics with the case that inspired its definition, namely the nabla flavour of coalgebraic logic. However, we will adopt an abstract perspective on nabla logic as follows. Let $T:\Set\to\Set$ be a weak pullback preserving finitary functor, we saw in Section \ref{ch1:sec:coalglog} that we can view the nabla language modulo boolean equivalence as the free $L$-algebra for $L=\Free T\Forg$, that is we can use $L$ to define a simple logic where the only inference rules are those of propositional calculus. Here we add more structure to $L$ by defining for any finitary $\Set$-functor $T$, any boolean algebra $A$ and morphism $f: A\to B$
\[
\begin{cases}
L^T A=\Free(\{\nabla\alpha\mid \alpha\in T\Forg A\})/\simeq_{\KKV(T)}\\
L^T f: L^T A\to L^T B, [\nabla \alpha]\mapsto [\nabla Tf \alpha]
\end{cases}
\]
where the quotient is taken under provability in $\KKV(T)$, i.e. $a\simeq_{\KKV(T)} b$ iff $\KKV(T)\vdash a\leq b$ and $\KKV(T)\vdash b\leq a$.

We now define the semantic transformation $\delta: L\pow \to\pow T$. The kind of definition of $L$ given above will occur repeatedly in all the Examples we will examine, particularly in Chapter 5. In every case $L$ will be of the form $L=(\Free T\Forg)/\simeq$, so it is worth detailing exactly how we define a semantic natural transformation on such an $L$. The procedure will always be the same: for every set $X$, we will define the map $\hat{\delta}_X: T\Forg \pow X\to \Forg\pow TX$, which uniquely determines a $\BA$-morphism $\delta_X:\Free T\Forg \pow X\to \pow TX$ (via the adjunction). If $\delta_X$ can be shown to respect the equivalence relation, i.e. if $\delta_X(a)=\delta_X(b)$ whenever $a\simeq b$, then $\delta_X$ defines a $\BA$-morphism $(\Free T\Forg \pow X)/\simeq\to\pow TX$. We usually commit the abuse of terminology consisting in saying that we define $\delta_X$ on generators, rather than saying we define its adjoint transpose. For notational clarity we will also drop the $[-]$ brackets and simply remember that elements of $LA$ are equivalence classes, of which we pick representatives. 

In the case at hand we proceed as follows: we define $\delta_X$ on generators $\nabla\alpha, \alpha \in T\Forg\pow X$ by
\[
\delta_X( \nabla \alpha)= \{t\in TX\mid t\Tmem \alpha\}
\]
\begin{lemma}
$\delta_X$ respects the equivalence relation $\simeq_{\KKV(T)}$. 
\end{lemma}
\begin{proof}
We proceed by induction on the depth of the proof of $\nabla \Phi\leq \nabla \Psi$. The base case if trivial: if $\nabla\Phi\leq \nabla\Phi$, then $\delta_X(\nabla\Phi)\subseteq \delta_X(\nabla\Phi)$. For the inductive case we need only check the case where the last step of the proof is $(\nabla 1), (\nabla 2), (\nabla 3)$, the propositional rules being automatically taken care of by being in $\BA$. 
\begin{itemize}
\item[$(\nabla 1)$]Assume that the last applied rule was
\begin{prooftree}
\AxiomC{$\{U\subseteq V\mid (U,V)\in R\}$}
\LeftLabel{$(\nabla 1)$}
\RightLabel{$(\alpha,\beta)\in \Tlift[R]$}
\UnaryInfC{$\nabla \alpha\leq \nabla\beta$}
\end{prooftree}
From the premise we can conclude that $R\subseteq \subseteq_{\pow X}$, and thus $\Tlift[R]\subseteq (\Tlift[\subseteq_{\pow X}])$. Moreover, it is clear that $(\in_X;\subseteq_{\pow X})\subseteq \in_X$, and thus since $T$ preserves weak pullbacks we get $(\Tmem[X];\Tlift[\subseteq_X])\subseteq \Tmem[X]$. We can then conclude that
\begin{align*}
t\Tmem[X]\alpha\Tlift[R]\beta &\Rightarrow t\Tmem[X]\alpha\Tlift[\subseteq_X]\beta\\
&\Rightarrow t\Tmem[X]\beta
\end{align*} 
i.e. $\delta_X(\nabla \alpha)\subseteq \delta_X(\nabla\beta)$ as desired.
\item[$(\nabla 2)$] Assume that the last applied rule was
\begin{prooftree}
\AxiomC{$\{\nabla (T\bigcap)\Phi\leq b\mid\Phi\in SRD(A)\}$}
\LeftLabel{$(\nabla 2)$}
\UnaryInfC{$\bigcap\{\nabla\alpha\mid \alpha\in A\}\leq b$}
\end{prooftree}
We need to show that if for all $\alpha\in A$, $t\Tmem\alpha$, then there exists a $\Phi\in SRD(A)$ such that $t\Tmem(T\bigcap)\Phi$, and we can then use the Induction Hypothesis on shorter proofs. We proceed as follows: $A$ is a finite set, say of cardinality $n$, i.e. $A=\{\alpha_1,\ldots,\alpha_n\}$, and we can therefore consider the kernel $n$-tuple $\ker_n(\pi_1)$ of the map $\pi_1: \in_X\to X$, i.e. the pullback of $n$ copies of $\pi_1$. Intuitively an element of $\ker_n(\pi_1)$ is an $n$-tuple of subsets of $X$ which all contain a common element.
Since $T$ weakly preserves pullbacks, it weakly preserves kernel $n$-tuples, and the data of $t\Tmem\alpha$ for all $\alpha\in A$ defines an element of $x\in T\ker_n(\pi_1)$.

Next, note that $\ker_n$ defines a relation $K_n$ on $X\times\powf\pow X$ defined by  
\[
x K_n \mathcal{V} \text{ if }|\mathcal{V}|=n\text{ and }x\in U\text{ for all }U\in\mathcal{V}
\]
and that there exist a map $ h:\ker_n(\pi_1)\to K_n$ defined by
\[
((x, U_1), \ldots,(x, U_n))\mapsto(x, \{U_1,\ldots, U_n\})
\] 
If we denote by $p_i, 1\leq i\leq n$ the maps from the kernel $n$-tuple to the various copies of $\in_X$, it is clear that by construction 
\[
p_i;\pi_2;\in_{\pow X}=h;\pi_2
\] 
i.e. the $\pow X$-component of the $i^{th}$ projection of an $n$-tuple belongs to a finite collection of subsets iff this finite collection of subsets is the one built by $h$. Note also that $K_n \subseteq (\in_X;\bigcap)$, and that since $T$ preserves weak pullbacks we therefore have $\Tlift[K_n] \subseteq \Tmem;T\bigcap$. 

We can now gather our results and see that (1) by the action of the map $h$, our witness $x$ gives us an element $Th(x)\in TK_n$, (2) this element also belongs to $\Tmem[X];T\bigcap$, i.e. $t\Tmem[X] T\bigcap \Phi$ for some $\Phi\in T\powf\pow X$, and finally (3) this $\Phi$ is in $SRD(A)$ since $T\pi_2\circ T\pi_i(x)=\alpha_i$ and $T\pi_2\circ Th(x)=\Phi$, thus $\alpha_i\Tmem[\pow X]\Phi$ for $1\leq i\leq n$ as desired.
\item[$(\nabla 3)$] Assume that the last applied rule was
\begin{prooftree}
\AxiomC{$\{\nabla\alpha\leq b\mid \alpha\Tmem\Phi\}$}
\LeftLabel{$(\nabla 3)$}
\UnaryInfC{$\nabla (T\bigcup)\Phi\leq b$}
\end{prooftree}
We need to show that if $t\Tmem (T\bigcup\Phi)$ then there exists $\alpha\Tmem[\pow X]\Phi$ such that $t\Tmem[X]\alpha$, and we can then use the I.H. on a shorter proof. To show this we first notice that if $x\in \bigcup_i U_i$ for a certain collection $(U_i)_{i\in I}$ of subsets of $X$, then $x$ must belong to at least one $U_i$. We choose a selection map for any such pair $(x,\bigcup_i U_i)$, i.e. we define a map $h: (\in_X; \bigcup)\to (\in_{X})$ which chooses a subset to which a given element of a union of subsets belongs. Note that this does not require the axiom of choice since all unions are over finite sets. This map has the pleasant property that $\pi_1\circ h=\pi_1$ (i.e. we keep the element), and that $h;\pi_2;\in_{\pow X}=\pi_2$ (i.e. the subset we choose is an element of the collection of subset we started off with). From this it follows that since $T$ weakly preserve pullbacks, we can send the witness $w$ of $t\Tmem[X] (T\bigcup) \Phi$ to a witness $Th(w)\in T\hspace{-4pt}\in_X$ of $t\Tmem \alpha$ where $\alpha=T \pi_2Th(w)$ is such that $\alpha\Tmem[\pow X]\Phi$ as desired.
\end{itemize}
\end{proof}
Having shown that $\delta_X$ defines an appropriate semantic transformation for $L^T$ we conclude by showing that for every epi-transformation $r$ between two weak pullback preserving functors $S, T$ if $\lambda: L^S\to \pow S$ and $\delta: L^T\to \pow T$ are the semantic transformations we have just defined, then there exists a natural transformation $q: L^S\epi L^T$ making the semantics compatible in the sense of definition \ref{ch4:def:compSem}. This transformation is uniquely determined by $r: S\to T$, but is a little messy to formulate.

We temporarily define $q: L^S\to L^T$ by its action on generators
\[
q_A: (\Free S\Forg A)/\simeq_{\KKV(S)}\hspace{2pt}\to (\Free T\Forg A)/\simeq_{\KKV(T)}, \nabla\alpha\mapsto\nabla(r_{\Forg A}(\alpha))
\]
For $q_A$ to be well-defined in must be compatible with the $\KKV(T)$ axiomatization encoded in the target functor $L^T$. Specifically, for any $a,b\in L^S A$ if $\KKV(S)\vdash a\leq b$, then we need $\KKV(T)\vdash q_A(a)\leq q_A(b)$. As it turns out $q$ defined as above is \emph{not} well-defined. Let us try to show why this is, and how we can fix the problem. We proceed by induction on depth of the proof of $\KKV(S)\vdash a\leq b$, as in the Lemma above, then the base case is again trivial. For the inductive step let us first consider the $(\nabla 1)$ rule i.e.
\begin{prooftree}
\AxiomC{$\{a\leq b\mid (a,b)\in R\}$}
\LeftLabel{$(\nabla 1)$}
\RightLabel{$(\alpha,\beta)\in \Slift[R]$}
\UnaryInfC{$\nabla \alpha\leq \nabla\beta$}
\end{prooftree}
By definition, if $(\alpha,\beta)\in \Slift[R]$, then there exists a witness $w\in SR$ which we can turn into a witness $r_R(w)\in TW$, from which it follows immediately that $\KKV(T)\vdash \nabla(\Free r_{\Forg A}( \alpha))\leq \nabla(\Free r_{\Forg A}( \beta))$ by the induction hypothesis and an application of the $(\nabla 1)$ rule. 

However, the two other rules do not behave so well, in particular the induction hypothesis does \emph{not} allow us to turn $\KKV(S)$-proof into a $\KKV(T)$-proof.  Let us see what happens with the $(\nabla 2)$ rule. Assume that we have a proof $\KKV(S)\vdash a\leq b$ whose last step is
\begin{prooftree}
\AxiomC{$\{\nabla (S\bigwedge)\Phi\leq b\mid \Phi\in SRD(U)\}$}
\LeftLabel{$\nabla 2$}
\UnaryInfC{$\bigwedge\{\nabla \alpha\mid \alpha\in U\}\leq b$}
\end{prooftree}
for some $U\in \powf S\Forg A$. If we wanted to show $\bigwedge\{\nabla\alpha\mid \alpha\in q_A[A]\}$ (i.e. the image of the conclusion under the transformation $q_A$ defined above) by using the $(\nabla 2)$ rule of $\KKV(T)$, then the following lemma shows us what would be needed in the premise.

\begin{lemma}\label{ch4:lem:SRD}
Let $S,T$ be two weak-pullback preserving functors on $\Set$, let $q: S\epi T$ be a epi natural transformation. For any boolean algebra $A$ and $U\in \powf S\Forg A$, the two following conditions are equivalent
\begin{enumerate}[(1)]
\item $\Phi\in SRD(q_{\Forg A}[U])$
\item there exist $\Phi'\in S\powf\Forg A$ such that $q_{\powf \Forg A}\Phi'=\Phi$ and for all $\alpha\in U$ there exist $\alpha'$ such that $\alpha'\Smem[\Forg A]\Phi'$ and $q_{\Forg A}(\alpha')=q_{\Forg A}(\alpha)$
\end{enumerate}
\end{lemma}
\begin{proof} For notational convenience we will use the following abbreviations:
\begin{itemize}
\item $(-)^q: S\Forg A\to S\Forg A, \alpha\mapsto q_{\Forg A}(\alpha)$
\item $[-]^q: \powf S\Forg A\to \powf T\Forg A, U\mapsto q_{\Forg A}[U]$
\item $\{-\}^q: S\powf A\to T\powf\Forg A, \Phi\mapsto q_{\powf\Forg A}(\Phi)$
\end{itemize}

\noindent \textbf{From (1) to (2).} By definition of redistributions, we know that all elements $\alpha$ of $[U]^q$ are $T$-lifted members of $\Phi$, i.e. we can always find a witness $w$ in $T\hspace{-4pt}\in_{\Forg A}$ such that $T\pi_1(w)=\alpha$ and $T\pi_2(w)=\Phi$ where $\pi_1: \in_{\Forg A}\fto\Forg A, \pi_2: \in_{\Forg A}\fto\powf\Forg A$. Since $[U]^q$ is finite, we can encode this property categorically by considering the kernel $n$-tuple of $T\pi_2$ where $n=|[U]^q|$, i.e. the limit of diagram 
\[
\xymatrix@C=5ex@R=6ex{
T\hspace{-4pt}\in_{\Forg A} \ar[drr]_{T\pi_2} & T\hspace{-4pt}\in_{\Forg A}\ar[dr]|-{T\pi_2} & \ldots \ar[d]|-{T\pi_2} & T\hspace{-4pt}\in_{\Forg A}\ar[dl]|-{T\pi_2} & T\hspace{-4pt}\in_{\Forg A}\ar[dll]^{T\pi_2} \\
 &  & T\powf\Forg A
}
\]
with one arrow $T\pi_2$ for each element $\alpha\in[U]^q$. Let us denote the limit of this diagram by $\ker_n(T\pi_2)$. This set contains an $n$-tuple of witnesses each of whom witnesses a relation $\alpha\Tmem[\Forg A]\Phi$ for all $\alpha\in[U]^q$.

\noindent Since $T$ preserves weak pullbacks, it also weakly preserves kernel pairs. Moreover, it is not difficult to see that the weak preservation of kernel pairs implies the weak preservation of kernel $n$-tuples when $n$ is finite. We therefore have that the set of witnesses $w_\alpha$ of each relation $\alpha\Tmem[]\Phi$ determines (non-uniquely) a `global witness' $w_{\Phi}\in T\ker_n(\pi_2)$. Since we have an epi natural transformation $q: S\fto T$, we have a surjective map 
\[
q_{\ker_n(\pi_2)}: S\ker_n(\pi_2)\epi T\ker_n(\pi_2)
\]
and in particular our `global' witness $w_{\Phi}$ has a pre-image $w'_{\Phi}\in S\ker_n(\pi_2)$ under this map. It is straightforward to check that by naturality of $q$, this $w'_{\Phi}$ is an $n$-tuple of witnesses, all of whom witness a relation $\alpha'\Smem[\Forg A]\Phi'$ for a common element $\Phi'$ such that $\{\Phi'\}^q=\Phi$. Moreover, each $\alpha'$ gets mapped to a different element $(\alpha')^q$ of $[U]^q$.

\noindent \textbf{From (2) to (1).} If for any $\alpha\in U$ we have an $\alpha'$ such that $(\alpha)^q=(\alpha')^q$ and $\alpha' \Smem[\Forg A]\Phi'$ then there must exist a witness $w_{\alpha'}\in S\hspace{-4pt}\in_{\Forg A}$ such that $S\pi_1(w_{\alpha'})=\alpha', S\pi_2(w_{\alpha'})=\Phi'$. Using the map 
\[
q_{S\hspace{-1pt}\in_{\Forg A}}:S\in_{\Forg A}\fto T\hspace{-4pt}\in_{\Forg A}
\]
we can turn $w_{\alpha'}$ into an element $w_{(\alpha')^q}\in T\hspace{-4pt}\in_{\Forg A}$ witnessing $(\alpha')^q\Tmem[\Forg A]\Phi$ by naturality of $q$. Since this holds for any $\alpha\in U$ and since $(\alpha)^q=(\alpha')^q$ we clearly have that every element of $[U]^q$ is a $T$-lifted member of $\Phi$, i.e. $\Phi\in SRD([U]^q)$.
\end{proof}

Using the result and the notation of the lemma, it is clear that if we wanted to conclude that $\bigwedge\{\nabla\alpha\mid \alpha\in [U]^r\}$ by using the $(\nabla 2)$ rule, we would need proofs of $\{\nabla(S\bigwedge) \Phi\leq b\mid \Phi\in SRD(U')\}$ for \emph{every} $U'$ such that $[U']^r=[U]^r$. To remedy this problem, we will need $q$ to identify more formulas than those identified by $r_{\Forg A}$. The easiest solution is to identify formulas $\bigwedge\{\nabla \alpha\mid\alpha \in  U\}$ and $\bigwedge\{\nabla \alpha\mid \alpha\in V\}$ whenever $[U]^r=[V]^r$. Since the converse of the $(\nabla 2)$ rule is derivable using the $(\nabla 1)$ rule (see Remark 6.4 of \cite{KKV:2012:Journal}), identifying all these meets will automatically guarantee that the corresponding slim redistributions will also be identified. Formally we define the set of equations as follows. Let $\mathscr{S}_r(X)$ denote the set of all sections of the epi $r_X: SX\epi TX$, using this notation we define
\[
E_A^\wedge=\mathscr{S}_r(\Forg A))^2\times \powf T\Forg A
\]
and the maps
\[
e_i^\wedge E_A^\wedge\to \Forg L^S A,\hspace{2ex} (s_1,s_2, A)\mapsto \bigwedge s_i[A], \hspace{2ex}i=1,2
\]
this set $E_A^\wedge$ together with the maps $e_1^\wedge,e_2^\wedge$ define the equations under which we will quotient $L^S A$ before applying the transformation $r_{\Forg A}$ in order to be compatible with $(\nabla 2)$.

Let us now turn our attention to the $(\nabla 3)$ rule by assuming that we have a $\KKV(S)$ proof of $a\leq b$ whose last step is
\begin{prooftree}
\AxiomC{$\{\nabla\alpha\leq b\mid \alpha\Smem[\Forg A]\Phi\}$}
\LeftLabel{$(\nabla 3)$}
\UnaryInfC{$\nabla (S\bigvee)\Phi\leq b$}
\end{prooftree}
The following lemma shows what would be needed as a premise of a $(\nabla 3)$ rule to conclude $\nabla  (T\bigvee)\{\Phi\}^r \leq b$. 

\begin{lemma}\label{ch4:lem:liftmem} Let $S,T$ be two weak-pullback preserving functors, let $q: S\epi T$ be an epi natural transformation. For any boolean algebra $A$ and $\Phi\in T\powf\Forg A$ and using the same conventions as in Lemma \ref{ch4:lem:SRD}, the following two conditions are equivalent
\begin{enumerate}[(1)]
\item $\alpha\Tmem[\Forg A]\Phi$
\item there exist $\Phi'\in S\powf\Forg A$ such that $\{\Phi'\}^q=\Phi$ and $\alpha'\in S\Forg A$ such that $(\alpha')^q=\alpha$ and $\alpha'\Smem[\Forg A]\Phi'$
\end{enumerate} 
\end{lemma}
\begin{proof}\textbf{From (1) to (2).} Assume $\alpha\Tmem[\Forg A]\Phi$, by definition, there must exist a witness $w\in T(\in_{\Forg A})$ such that $T\pi_1(w)=\alpha, T\pi_2(w)=\Phi$. Since $q$ is epi, there must exist a pre-image $w'\in S(\in_{\Forg A})$ such that $q_{\in_{\Forg A}}(w')=w$ which witnesses that $\alpha'\Smem[\Forg A]\Phi'$ for a certain $\alpha'\in S\Forg A$ and $\Phi\in S\powf\Forg A$. By naturality of $q$ we have $(\alpha')^q=\alpha$ and $\{\Phi'\}^q=\Phi$. 

\noindent{\textbf{From (2) to (1).}} Immediate by pushing the witness of $\alpha'\Smem\Phi'$ down to $T(\in_{\Forg A})$ by the maps defined above.
\end{proof}

From this lemma we can see that if we wanted to conclude $\nabla (T\bigvee) \{\Phi\}^r \leq b$ by using the $(\nabla 3)$ rule, we would need to have a proof of $\nabla\alpha'\leq b$ for every lifted member $\alpha'$ of every $\Phi'$ which gets mapped to $\{\Phi\}^q$. Again, we have no choice but to identify more formulas than those identified by $r_{\Forg A}$. We proceed as in the $(\nabla 2)$ case and identify $\nabla (S\bigvee)\Phi$ and $\nabla (S\bigvee)\Psi$ whenever $\{\Phi\}^r=\{\Psi\}^r$. Since the converse of rule $(\nabla 3)$ is derivable using $(\nabla 1)$ (see Remark 6.6 of \cite{KKV:2012:Journal}) we will then automatically get the relevant premise for the $(\nabla 3)$ rule. We proceed formally as in the case of the $(\nabla 2)$ rule and define for every boolean algebra $A$ the set (of equations)
\[
E^\vee_A: \mathscr{S}(\powf\Forg A)^2\times T\powf\Forg A
\]
and the maps
\[
e^\vee_i: E^\vee_A \to \Forg L^S A, \hspace{2ex}(s_1,s_2,\Phi)\mapsto s_i(\Phi), \hspace{2ex}i=1,2
\]
We can finally define $q: L^S\to L^T$. We proceed in two steps. Let $A$ be any boolean algebra, we first construct the coequalizer of the pair of morphisms
\[
\xymatrix@C=12ex
{
\Free (E^\wedge_A+E^\vee_A)\ar@<-2pt>[r]_<<<<<<<<<<{\Free(e^\wedge_2+e^\vee_2)} \ar@<3pt>[r]^<<<<<<<<<<{\Free(e^\wedge_1+e^\vee_1) }& L^S A\ar@{->>}[r]^{q_{E_A}} & Q_{E_A}
}
\]
The second step is defined on `generators' of $Q_{E_A}$ as
\[
[\nabla\alpha]\mapsto[\nabla r_{\Forg A}(\alpha)]
\]
By combining the two steps we get an epi-transformation
\[
q_A: L^S A\to L^T A
\]
which is a well-defined boolean algebra morphism. It is easy to see from the definition of the first quotient that $r_{\Forg A}$ is compatible with it. For example since we have identified formulas $\nabla (S\bigvee)\Phi$ and $\nabla (S\bigvee)\Psi$ precisely when $\{\Phi\}^r=\{\Psi\}^r$, this means that if $\nabla (S\bigvee)\Phi$ and $\nabla (S\bigvee)\Psi$ are two representatives of an equivalence class under the equations of $E^\wedge_A+E^\vee_A$, then they are mapped to the same formula $\nabla T\bigvee\{\Phi\}^r=\nabla T\bigvee\{\Psi\}^r$. From this perspective, it looks like the quotient under the equations of $E^\wedge_A+E^\vee_A$ is not doing anything since we end up with the same elements as if we had only used $(-)^r$. This is true on formulas, but what the quotient achieves is to ensure that the transformation $q_A$ preserve the $\leq$ relation. Put differently, let $\Phi,\Psi$ such that $\{\Phi\}^r=\{\Psi\}^r$ and assume that $\nabla(S\bigvee)\Phi\leq b$ in $L^S A$ and the proof of this ends in $(\nabla 3)$. For $\Psi\neq\Phi$, there is no reason to have $\nabla(S\bigvee)\Psi\leq b'$ for a $b'$ such that $(b')^r=(b)^r$, i.e. there is no reason for $(-)^r$ to be compatible with the $\leq$ structure on $L^S A$. By quotienting under $E^\wedge_A+E^\vee_A$ we force $\nabla(S\bigvee)\Phi$ and $\nabla(S\bigvee)\Psi$ to be the same object, and in particular we then have $[\nabla\alpha]\leq [\nabla(S\bigvee)\Phi]=[\nabla(S\bigvee)\Psi]\leq b$ for any lifted member of $\Phi$ or $\Psi$. In this way we make it possible to prove by induction that if $\nabla(S\bigvee \Phi)\leq b$ and the proof of this ends in $(\nabla 3)$, then $q_A(\nabla (T\bigvee)\{\Phi\}))\leq q_A(b^q)$ can be shown in $\KKV(T)$ using a proof ending in $(\nabla 3)$.

Let us finally show that with $q$ thus defined, $\lambda: L^S\pow\to\pow S$ and $\delta: L^T\pow\to\pow T$ are compatible semantics in the sense of Definition \ref{ch4:def:compSem}. We only need to show the inequality defining compatible semantics on generators, i.e. for $\alpha\in S\Forg \pow X$ we need to check that
\[
\lambda_X(\nabla\alpha)\subseteq \pow r_X\circ \delta_X\circ q_{\pow X}(\nabla\alpha)
\]
But this follows trivially by definition and naturality of $r$, i.e. if $t\Smem\alpha$, then by pushing the witness from $S\hspace{-4pt}\in_X$ to $T\hspace{-4pt}\in_X$ with $r_{\hspace{-1pt}\in_X}$ we get a witness of $r_X(t)\Tmem r_{\Forg \pow X}(\alpha)$ as desired.

\section{Proof-theoretic translations }\label{ch4:sec:prooftrans}

In this section we will deal with purely syntactic questions, i.e. no models will be involved. We will study the link between provability in a logic for a functor $K:\BA\to\BA$ and a functor $L:\BA\to\BA$ when they are related by a regular epi-transformation $q: K\epi L$. 

We will use the notation and constructions from our discussion on the algebraic semantics of coalgebraic logics in Section \ref{ch1:subsec:algsem}. Let $K,L:\BA\to\BA$ be weak-pullback and regular epi preserving finitary functors, in particular $K,L$ are varietors. Let $V$ be a set of proposition variables, let $\Free_L\dashv \Forg_L: \Set\to\Alg_{\BA}(L)$ be the obvious free and forgetful functors and similarly for $\Free_K\dashv \Forg_K: \Set\to\Alg_{\BA}(K)$. By definition, the abstract languages over a set of propositional variables $V$, associated with $L$ and $K$, are simply $\Free_L V$ and $\Free_K V$. Let also $e_1,e_2: E \rightrightarrows \Forg_L\Free_L V$ define a set $E$ of equations in the abstract coalgebraic language $\Free_L V$. These equations define a variety $\mathbb{V}_{E}$ of $L(-)+\Free V$-algebras defined by the coequalizer of the adjoint transpose morphisms $\hat{e}_1,\hat{e}_2: \Free_L E \rightrightarrows \Free_L V$. Given a natural transformation $q: K\epi L$, we will now canonically lift these equations to define a variety of $K$-algebras.  
We follow the technique developed in Section 1.5.4, i.e. we start by building the smallest equivalence relation generated by the equations defined by $e_1,e_2$, i.e. we take the kernel pair of the coequalizer of the adjoint transpose morphisms $\hat{e}_1,\hat{e}_2: \Free_L E\to \Free_L V$. For notational simplicity, and because this kernel pair can still be thought of as `equations' we will denote the kernel pair as $E$ and the projections to $\Free_L V$ as $e_1, e_2$ since these also pick the left and right-hand side of an `equation'. We denote the equalizer as $q_E: \Free_L V\to Q_E$, and we thus have the following exact sequence:
\begin{equation}\label{ch4:diag:exactsequ1}
\xymatrix
{
E\ar@<1ex>[r]^{e_1}\ar@<-1ex>[r]_{e_2} & \Free_L V\ar@{->>}[r]^{q_E} & Q_E
}
\end{equation}
We now lift this exact sequence to a $K$-algebra by using the transformation $\mathsf{Q}: \Alg_{\BA}(L)\to\Alg_{\BA}(K)$ defined at the beginning of this chapter as well the following easy Proposition:
\begin{proposition}\label{ch4:prop:QpreservesCoeq}
Let $K,L$ be two functors on $\BA$ such that $K$ is a varietor preserving regular epis, let $q: K\to L$ be a natural transformation and let $\mathsf{Q}:\Alg_{\BA}(L)\to\Alg_{\BA}(K)$ be the associated functor, then $\mathsf{Q}$ preserves coequalizers.
\end{proposition}
\begin{proof}
The proof is exactly the same as Proposition \ref{ch1:prop:AlgRegCat} (ii) and Proposition \ref{ch1:prop:AlgExact}, and hinges on $K$ preserving regular epis.
\end{proof}

\begin{proposition}\label{ch4:prop:QpreservesKernelPairs}
Let $K,L$ be two regular epi-preserving functors on $\BA$, let $q: K\to L$ be a natural transformation and let $\mathsf{Q}:\Alg_{\BA}(L)\to\Alg_{\BA}(K)$ be the associated functor, then $\mathsf{Q}$ preserves kernel pairs and exact sequences.
\end{proposition}
\begin{proof}
Recall from Section 1.4 that a kernel pair $p_1,p_2: (\ker f, \phi)\rightrightarrows (A,\alpha)$ of an $L$-algebra morphism $f: (A,\alpha)\to (B,\beta)$ defines a relation, i.e. a monomorphism $p_1\times p_2: (\ker f,\phi)\to (A,\alpha)\times (A,\alpha)$ in $\Alg_{\BA}(L)$. Let $r_1,r_2:(C,\gamma)\rightrightarrows\mathsf{Q}(A,\alpha)$ be a cone for the diagram $\mathsf{Q}(A,\alpha)\stackrel{\mathsf{Q}f}{\to} \mathsf{Q}(B,\beta)\stackrel{\mathsf{Q}f}{\leftarrow}(A,\alpha)$, we need to show that there exists unique $K$-algebra morphism $(C,\gamma)\to \mathsf{Q}(\ker f,\delta)$. Since $L$ preserves regular epis, $\Forg_L:\Alg_{\BA}(L)\to \BA$ is monadic by Proposition \ref{ch1:prop:Lvarietal}, and thus creates all limits in $\Alg_{\BA}(L)$. In particular, $\ker f$ is the kernel pair of $f:A\to B$ in $\BA$. There must therefore exist a unique $\BA$-morphism $u: C\to \ker f$ such that $p_1\circ u=r_1$ and $p_2\circ u=r_2$. Let us show that $u$ defines a $K$-algebra morphism, i.e. that $u\circ\gamma=\delta\circ Ku$. The cone $(C,\gamma)$ defines a morphism $r_1\times r_2: (C,\gamma)\to \mathsf{Q}((A,\alpha)\times(A,\alpha))$ in $\Alg_{\BA}(K)$, and have the following commutative diagram:
\[
\xymatrix@C=12ex
{
& KC\ar[dd]^(0.3){\gamma}\ar[dr]^{K(r_1\times r_2)}\ar[dl]_{Ku} \\
K\ker f\ar[dd]_{\phi\circ q_{\ker f}}\ar[rr]_(0.35){K(p_1\times p_2)} & & K(A\times A)\ar[dd]^{(L\pi_1\circ \alpha\times L\pi_2\times \alpha)\circ q_{A\times A}}\\
& C\ar[dl]_{u}\ar[dr]^{r_1\times r_2} \\
\ker f\hspace{1ex}\ar@{>->}[rr]_(0.4){(p_1\times p_2)} & & A
}
\]
from which it is clear that
\begin{align*}
r_1\times r_2\circ \gamma &=(p_1\times p_2)\circ u\circ \gamma\\
&= (L\pi_1\circ \alpha\times L\pi_2\circ\alpha)\circ q_{A\times A}\circ K(r_1\times r_2)\\
&= (L\pi_1\circ \alpha\times L\pi_2\circ\alpha)\circ q_{A\times A}\circ K(p_1\times p_2)\circ Ku\\
&=(p_1\times p_2)\circ \phi\circ q_{\ker f}\circ Ku
\end{align*}
and thus $\phi\circ q_{\ker f}\circ Ku=u\circ \gamma$ as required since $p_1\times p_2$ is a monomorphism. The preservation of exact sequences is now a consequence of Proposition \ref{ch4:prop:QpreservesCoeq}.
\end{proof}

Thus, as long as $K, L$ preserve regular epis, the functor $\mathsf{Q}$ associated with the natural transformation $q$ turns the exact sequence (\ref{ch4:diag:exactsequ1}) in $\Alg_{\BA}(L)$ into an exact sequence
\[
\xymatrix
{
\mathsf{Q}E\ar@<1ex>[r]^(0.45){\mathsf{Q}e_1}\ar@<-1ex>[r]_(0.45){\mathsf{Q}e_2} & \mathsf{Q}\Free_L V\ar@{->>}[r]^{\mathsf{Q}q_E} & \mathsf{Q}Q_E
}
\]
in $\Alg_{\BA}(K)$. We are now ready to define the lifting of the equations defined by $E,e_1,e_2$ into the syntax functor defined by $K$: we take the pullback of $e_1\times e_2$ along the translation transformation $\xi_V: \Free_K V\to\mathsf{Q}\Free_L V$ and we call this $K$-algebra of equations $E^*$:
\begin{equation}\label{ch4:diag:E*def}
\xymatrix@C=12ex
{
E^*\hspace{1ex}\ar@{>->}[r]^(0.4){e^*_1\times e^*_2}\ar@{->>}[d]_{r} & \Free_K V\times \Free_K V\ar[d]^{\xi_V\times\xi_V} \\
\mathsf{Q} E\hspace{1ex}\ar@{>->}[r]_(0.4){\mathsf{Q}e_1\times \mathsf{Q}e_2} & \mathsf{Q}\Free_L V\times \mathsf{Q}\Free_L V
}
\end{equation}
Note that pulling back a mono always yields a mono, and thus $e^*_1\times e^*_2$ is a mono, and as was shown in Section 1.4 this mono defines a pair a jointly monic morphisms $e_1^*, e_2^*: E^*\to \Free_K V$, which is what we expect from a pair of morphisms defining `equations'. Moreover since we are assuming $K$ to be a varietor preserving regular epis, $\Alg_{\BA}(K)$ is a regular category by Proposition \ref{ch1:prop:AlgRegCat}. By Theorem \ref{ch4:thm:syntaxthm}, $\xi_V$ is a regular epi, and since pulling back a regular epi gives a regular epi in a regular category, $r$ must be a regular epi.

We will now show that the maps $e_1^*, e_2^*$ form the kernel pair of their coequalizer, this will in particular show that $E^*$ is an equivalence relation, and intuitively shows that the procedure above gives us the `tightest' lifting of the equations of $E$ into the $K$-language, i.e. that we in fact have lifted the entire exact sequence of Diagram (\ref{ch4:diag:exactsequ1}). So let $q_{E^*}: \Free_K V\to Q_{E^*}$ be the coequalizer of $e_1^*, e_2^*$, and notice first that there exist a unique morphism between $Q_{E^*}$ and $\mathsf{Q}Q_E$, i.e. we have the following commutative diagram:
\begin{equation}\label{ch4:diag:equationLift}
\xymatrix@C=12ex
{
E^*\ar@<1ex>[r]^{e_1^*}\ar@<-1ex>[r]_{e^*_2} \ar[d]_{r} & \Free_K V \ar@{->>}[r]^{q_{E^*}}\ar[d]^{\xi_V} & Q_{E^*}\ar@{-->>}[d]^{q^{E^*}_E}\\
\mathsf{Q}E\ar@<1ex>[r]^{\mathsf{Q}e_1}\ar@<-1ex>[r]_{\mathsf{Q}e_2} & \mathsf{Q}\Free_L V\ar@{->>}[r]^{\mathsf{Q}q_E} & \mathsf{Q}Q_E
}
\end{equation}
The reason is simply that by commutativity of the left-hand side square, $\mathsf{Q}q_{E}\circ \xi_V$ also coequalizes $e_1^*$ and $e_2^*$. This being established, let $p_1,p_2: C\to\Free_K V$  be a cone for the diagram $\Free_K V\stackrel{q_{E^*}}{\to} Q_{E^*}\stackrel{q_{E^*}}{\leftarrow}\Free_K V$, if $E^*$ is the kernel pair of $q_{E^*}$, there must exist a unique $K$-algebra morphism $u:C\to E^*$ such that $e_1^*\circ u=p_1$ and $e^*_2\circ u=p_2$. By the same argument as the one leading to Diagram (\ref{ch4:diag:equationLift}), if $q_{E^*}$ coequalizes $p_1,p_2$, then so does $\mathsf{Q}q_E\circ \xi_V$, and therefore $C, \xi_V\circ p_1, \xi_V\circ p_2$ is a cone for the diagram $\mathsf{Q}\Free_L V\stackrel{\mathsf{Q}q_E}{\to}\mathsf{Q}Q_E\stackrel{\mathsf{Q}q_E}{\leftarrow}\mathsf{Q}\Free_L V$, whose limit is $\mathsf{Q}E$. There must therefore exist a unique $K$-algebra morphism $v:C\to\mathsf{Q}E$ such that $e_1\circ v=\xi_V\circ p_1$ and $e_2\circ v=\xi_V\circ p_2$. Consequently, 
\begin{align*}
e_1\times e_2\circ v&=e_1\circ v\times e_2\circ v\\
&=\xi_V\circ p_1\times \xi_V\circ p_2\\
&=\xi_V\times\xi_V\circ p_1\times p_2
\end{align*}
and $C$ is a cone for Diagram (\ref{ch4:diag:E*def}) which defines $E^*$ as a pullback, so there must exist a unique morphism $u: C\to E^*$ such that $e_1^*\times e_2^*\circ u=p_1\times p_2$, and the result follows.

We have thus shown three things: (1) we can lift equations in the logic defined by the functor $L$ to equations in the logic defined by the functor $K$ in a canonical manner, (2) the two algebras of equations and the two quotients obtained in this way are related by the nice commutative Diagram (\ref{ch4:diag:equationLift}), (3) both original algebra of equations $E$ and its lifted counterpart are equivalence relations on their respective free algebras. 
We can now state and prove the key result of this Section. Recall from Section \ref{ch1:subsec:algsem}, that an object $A$ in a category is said to be \emph{orthogonal} to a morphism $q$ (notation $A\perp q$) is for every morphism $f:\dom q\to A$ there exists a unique morphism $h: \cod q\to A$ such that $f=h\circ q$.

\begin{theorem}\label{ch4:thm:prooftrans}
For any $L$-algebra $A$, $A\perp q_E$ iff $\mathsf{Q}A\perp q_{E^*}$.
\end{theorem}
\begin{proof}
Let us first assume that $A\perp q_{E}$. To show that that $\mathsf{Q}A\perp q_{E^*}$, we take any $K$-algebra  morphism $h: \Free_K A\to \mathsf{Q}A$ and we will show that $h$ must factor through $\mathsf{Q}\Free_L V$. We proceed by using the usual adjunctions: $h: \Free_K V\to \mathsf{Q} A$ defines a unique morphism $\tilde{h}: V\to\Forg_K\mathsf{Q}A=\Forg_L A$ which in turns define a unique morphism $\hat{h}: \Free_L V\to A$. Let us now show that $\mathsf{Q}\hat{h}\circ \xi_V=h$, we do this by unravelling the construction of the two adjoint transpose maps and we get that $\hat{h}=\epsilon^L_A\circ \Free_L\Forg_K h\circ \Free_L\eta^K_V$. For $\mathsf{Q}\hat{h}$ we have:
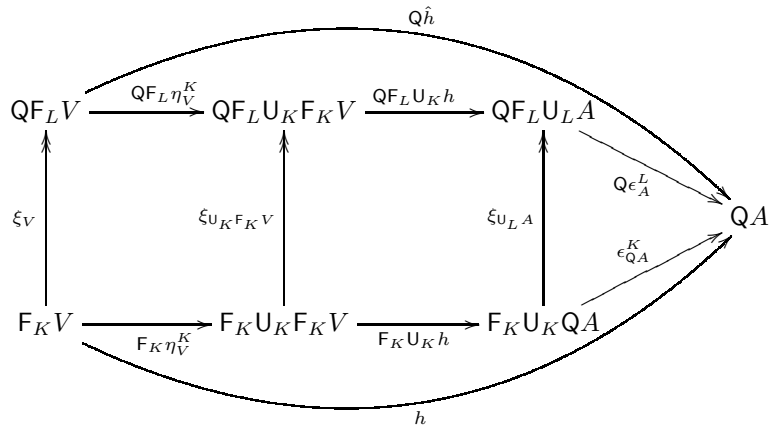
\[
\xymatrix@C=10ex@R=6ex
{
\mathsf{Q}\Free_L V\ar@/^4.2pc/[rrrd]^{\mathsf{Q}\hat{h}}\ar[r]^{\mathsf{Q}\Free_L\eta^K_V} & \mathsf{Q}\Free_L \Forg_K\Free_K V \ar[r]^{\mathsf{Q}\Free_L\Forg_K h}& \mathsf{Q}\Free_L\Forg_L A\ar[dr]_{\mathsf{Q}\epsilon^L_A}\\
& & & \mathsf{Q}A\\
\Free_K V\ar@{->>}[uu]^{\xi_V}\ar[r]_{\Free_K \eta^K_V}\ar@/_4.2pc/[rrru]_{h} & \Free_K\Forg_K\Free_K V\ar@{->>}[uu]^{\xi_{\Forg_K\Free_K V}}\ar[r]_{\Free_K\Forg_K h} & \Free_K\Forg_K\mathsf{Q}A\ar@{->>}[uu]^{\xi_{\Forg_LA}}\ar[ur]^{\epsilon^K_{\mathsf{Q}A}}
}
\]
All the squares commute by naturality of $\xi$, so all we need to check is the commutativity $\mathsf{Q}\epsilon^L_A\circ \xi_{\Forg_L A}=\epsilon^K_{\mathsf{Q}A}$, but this precisely what we have shown in Proposition \ref{ch4:prop:xicounits}. Thus we have shown that $h=\mathsf{Q}\hat{h}\circ \xi_V$. But since $A\perp q_E$, there must exist a unique morphism $\hat{u}: Q_E\to A$ such that $\hat{u}\circ q_E=\hat{h}$. If we now define $u=\mathsf{Q}\hat{u}\circ q^{E^*}_E$, where $q^{E^*}_E:Q_{E^*}\to \mathsf{Q}Q_E$ is the unique morphism define in Diagram (\ref{ch4:diag:equationLift}), we get a unique (since $\hat{u}$ and $q^{E^*}_E$ are unique) $K$-algebra morphism such that 
\[
u\circ q_{E^*}=\mathsf{Q}\hat{u}\circ \mathsf{Q}q_{E}\circ \xi_V=\mathsf{Q}\hat{h}\circ \xi_V=h
\]
i.e. $\mathsf{Q}A\perp q_{E^*}$ as desired.

For the converse, we start in a completely analogous fashion. Assume $\mathsf{Q}A \perp q_{E^*}$, and let $h: \Free_L V\to A$. By using the adjunctions, $h$ defines a function $\hat{h}:V\to \Forg L A=\Forg_K\mathsf{Q}A$, which in turns defines a map $\tilde{h}: \Free_K V\to \mathsf{Q}A$. By using the same argument as above and Proposition \ref{ch4:prop:xicounits}, it is easy to show that $\tilde{h}=h\circ \xi_V$. Since we're assuming that $\mathsf{Q}A\perp q_{E^*}$, there must exist a unique $K$-algebra morphism $u: Q_{E^*}\to \mathsf{Q}A$ such that $u\circ q_{E^*}=\tilde{h}$. By using the notation of Diagram (\ref{ch4:diag:equationLift}) we then get that
\[
\tilde{h}\circ e^*_1=u\circ q_{E^*}\circ e^*_1=u\circ q_{E^*}\circ e^*_2=\tilde{h}\circ e^*_2
\]
From this it follows that 
\[
\mathsf{Q}h\circ \xi_V\circ e^*_1=\mathsf{Q}h\circ \xi_V\circ e^*_2
\]
and thus
\[
\mathsf{Q}h\circ \mathsf{Q}e_1\circ r=\mathsf{Q}h\circ \mathsf{Q}e_2\circ r
\]
i.e. $\mathsf{Q}h\circ \mathsf{Q}e_1=\mathsf{Q}h\circ \mathsf{Q}e_2$ since $r$ is epi. But this means that $h$ coequalizes $e_1,e_2$, and since $q_E$ is the coequalizer of $e_1,e_2$, there must exist a unique map $v: Q_E\to A$ such that $v\circ q_E=h$, i.e. $A\perp q_E$ as desired.
\end{proof}

We have thus show that not only can we lift equations from the $L$-language to the $K$-language, but that by doing so we define varieties which are in extremely tight correspondence. From our soundness and completeness results for the abstract algebraic semantic of coalgebraic logics developed in Section \ref{ch1:subsec:algsem} (in particular Theorem \ref{ch1:thm:algsem}), this means that the derivability of a formula $a$ in the abstract Hilbert system associated with the set of equations $E$, is equivalent to the derivability of some pre-image $a'$ such that $(a')^q=a$ in the abstract Hilbert system associated with $E^*$. And similarly for equations and equational systems. In particular, if a formula $a$ in the logic defined by $L$ and the axioms in $E$ is consistent, then it must have an inverse image $a'$ under the syntax translation which is consistent in the logic defined by $K$ and the axioms of $E^*$.

\chapter{Completeness-via-canonicity}

In this Chapter we will weave together the strands developed in all the previous Chapters. The first section is an overview of known results on weak-completeness (see for example \cite{2005:UfExtCoalg,2010:AlexDaniela}). The second section will develop the notion of strong completeness via canonicity in the absence of frame conditions (i.e. of axioms) which is key to developing a theory of strong completeness in the presence of canonical axioms. This section uses the work of \cite{2005:UfExtCoalg,2012:KurzStrongComp,2009:DirkStrongComp}, although our presentation is slightly different. Finally the third section collects all the results of the thesis into completeness-via-canonicity theorems for coalgebraic logics both in their abstract and in their nabla formulation.

\section{Weak completeness}\label{ch5:sec:weakComp}

There are two `natural' candidates for building coalgebraic models and proving things like completeness. The first is the terminal coalgebra (if it exists), or finitary version thereof (if it does not). It is the key to understanding weak completeness and is intimately connected with the properties of the semantic transformation $\delta$. However, this type of construction is not amenable to techniques exploiting canonicity, and for this we must use the second natural candidate which is the notion of `canonical model'. The idea there is, in a nutshell, to put a coalgebraic structure on the set of ultrafilters of the initial $L$-algebra defining the logic. It is intimately related to the \emph{inverse} of the transpose of the semantic transformation ($\hat{\delta}$), if it exists. Thus the two techniques work, in a certain sense, in opposite directions.

\subsection{Completeness theorem}

The following theorem captures the essential features of terminal coalgebra semantics and of weak completeness, it is similar in some respects to Theorem 6.15 of \cite{2010:AlexDaniela} and many ideas date back to \cite{AlgSem2004}. We place ourselves once again in the fundamental situation of Diagram (\ref{ch1:diag:fundamentalSituation}), i.e.
\[
\xymatrix
{
\cat\ar@/^1pc/[rr]^{F} \ar@(l,u)^{L} \ar@<5pt>[d]^{\Forg} & \perp&\cat[D]\op\ar@/^1pc/[ll]^{G}\ar@(r,u)_{T\op}\\
\Set\ar@<5pt>[u]^{\Free}_{\dashv}
}
\]
Recall that $\cat$ is assumed to be a finitary variety. In particular $\Forg$ creates limits and directed colimits.

\begin{theorem}\label{ch5:thm:weakcomp1}
In the fundamental situation described above, let $L:\cat\to\cat$ be a finitary varietor, let $T:\cat[D]\to\cat[D]$ have a terminal coalgebra $\nu T$, and let $\delta: LG\to G T$ be a semantic transformation. For any set $V$, if $L_V=L(-)+\Free V$ and $T_V=T(-)\times F\Free V$, then:
\begin{itemize}
\item There exists a unique interpretation morphism $\lsem-\rsem_V: \init[L_V]\to G\nu T_V$.
\item This map is injective whenever (i) there morphism $0_{\cat}\to G1_{\cat[D]}$ is injective, (ii) $\delta$ is injective at every object and (iii) $L$ preserves injective maps. In particular any two formulas $a,b\in \init[L_V]$ such that $a\neq b$ have different denotation, i.e. $L$ is weakly complete w.r.t. $\Coalg(T)$
\item This map is surjective whenever (i) there exist an regular epi $0_{\cat}\to G1_{\cat[D]}$ (ii) $\delta$ is a regular epi-transformation and (iii) $L$ preserve regular epis.
\end{itemize}
\end{theorem}
\begin{proof}
\textbf{Existence of $\lsem-\rsem_V$:} The structure map of $\nu T_V$ is of the form:
\[
\gamma\times v: \nu T_V\to T(\nu T_V)\times F\Free V
\]

By the fact that $F\dashv G$, $v$ has an adjoint transpose $\hat{v}: \Free V\to G\nu T_V$, and thus the structure map of the cofree coalgebra $\nu T_V$ determines uniquely the map:
\[
G \gamma+\hat{v}: G T\nu T_v+ \Free V\to G \nu T_V
\]

This having been established, we proceed as usual and use the natural transformation $\delta$ and the initiality of $\init[L_V]$ to define the interpretation map as the catamorphism:
\[
\xymatrix@C=12ex
{
L\init[L_V]+\Free V\ar@{-->}[r]^{L\lsem-\rsem_V+\id_{\Free V}}\ar[dd]& L G\nu T_V+\Free V\ar[d]^{\delta_{\nu T_V}+\id_{\Free V}}\\
&G T \nu T_V+\Free V\ar[d]^{G\gamma+\hat{v}}\\
\init[L_V]\ar@{-->}[r]_{\lsem - \rsem_V} & G \nu T_V
}
\]
Note that we do not need to specify any valuation, instead the map $\hat{v}$ is purely endogenous to the cofree coalgebra, in fact it is merely a rewriting of its canonical colouring map.

\textbf{Construction of $\lsem-\rsem_V$ from the initial sequence.} Let us now give an alternative description of $\lsem-\rsem_V$ which will be essential in what follows. We use both the initial sequence of $L_V$ and the terminal sequence of $T_V$ and relate them as follows. 
\[
\xymatrix@C=11ex
{
& & \init[L_V]\\
0\ar[r]^{i}\ar[d]_{\lsem-\rsem_0}\ar[urr] & L_V0\ar[r]^{L_V i}\ar[d]_{\lsem-\rsem_1=\delta_1\circ L_V \lsem-\rsem_0}\ar[ur] & L_V^20\ar[r]^{L^2_V i}\ar[d]_{\lsem-\rsem_2=\delta_{T_V1}\circ L_V \lsem-\rsem_1}\ar[u] & \ldots\ar[r] & L^n_V0\hspace{2ex}\ldots\ar[d]_{\lsem-\rsem_n=\delta_{T_V^{n-1}1}\circ L_V \lsem-\rsem_{n-1}}\ar[ull] \\
G 1\ar[r]_{G t}\ar[drr] & G T_V 1\ar[r]_{G T_V t}\ar[dr] & G T_V^2 1\ar[r]_{G T^2_V t}\ar[d] & \ldots\ar[r] & G T_V^n 1\hspace{2ex\ldots}\ar[dll]\\
& & G \nu T_V
}
\]
where $0$ is the initial object in $\cat$ (which always exists since $\cat$ is a finitary variety) and $i:0\to L_V0$ is the unique morphism arising from initiality of $0$. Dually, 1 is the terminal object in $\cat[D]$ and $t: T_V 1\to 1$ is the unique morphism arising from the fact that 1 is terminal. Since $G \lim_i T_V^i 1=G \nu T_V$ forms a cocone for the cochain $G 1\to G T_V1\to \ldots$, it clearly also forms a cocone for the initial sequence $0\to L_V0\to\ldots$, and thus there must exist a unique morphism 
\[
\lsem-\rsem: \init[L_V]=\colim_i L_V^i0 \to G \nu T_V
\]
It is quite clear by construction that in fact $\lsem-\rsem=\lsem-\rsem_V$. In order to determine the set theoretic properties of $\lsem-\rsem_V$ (viz. injectivity and surjectivity), we now apply the forgetful functor $\Forg:\cat\to\Set$ to the above diagram. Recall that $\Forg$ creates filtered colimits (and a fortiori $\omega$-cochains) and thus preserves them. Hence we have that:
\[
\Forg\init[L_V]=\Forg \colim_i L_V^i0=\colim_i \Forg L_V^i0
\]
Moreover, since $UG\nu T_V$ forms a cocone for the cochain $\Forg G 1\to\Forg G T_V1\to \ldots$ - and thus for the cochain $\Forg 0\to \Forg L_V 0\to\ldots$ - there must exist a unique morphism $\Forg\init[L_V]\to UG\nu T_V$, which by unicity must be $\Forg\lsem-\rsem_V$ itself. 

\textbf{Injectivity of $\lsem-\rsem_V$.} Let $a,b\in \Forg\init[L_V]$ such that $\Forg \lsem a\rsem_V=\Forg \lsem b\rsem_V$. By construction of colimits in $\Set$ as coproducts quotiented by an equivalence relation, we know that there must exist $i,j$ and pre-images $a'\in L^i_V0,b'\in \Forg L^j_V0$ which map to $a,b$ respectively. By setting $k=\max\{i,j\}$ and choosing the appropriate representative in the equivalence class we can in fact choose $a',b'\in \Forg L^k_V0$. By the fact that $\Forg$ creates limits and the construction of limits in $\Set$, any element of $\Forg G \nu T_V=\Forg G\lim_i T^i_V 1$ is a set of coherent sequences $(u_0, u_1, u_2,\ldots)\in \prod_i T_V^i 1$, where coherent means that $u_i=T_V^it(u_{i+1})$. In particular $X=\Forg \lsem a\rsem_V=\Forg \lsem b\rsem_V$ is such a set.

Consider now $X_a= \Forg \lsem a'\rsem_k$ and $X_b=\Forg \lsem b'\rsem_k$, by commutativity of the diagram, they must get mapped to $X$ by the inverse image of the projection map $\pi_k: \lim_i T^i_V 1\to T^k_V 1$. In other words a sequence $(x_i)\in X$ `goes through' $X_a$ iff it `goes through' $X_b$. Thus $X_a=X_b$. It is easy to check by induction that each $\Forg \lsem-\rsem_i$ is injective: $\Forg \lsem-\rsem_0$ is injective by assumption and since $\lsem-\rsem_{n+1}=\delta_{T^n_V 1}\circ L_v \lsem-\rsem_n$, $\lsem-\rsem_{n+1}$ is injective since $\delta_{T^n_V 1}$ is injective by assumption and $L_v$ preserves injective maps. Thus $X_a=X_b$ implies that $a'=b'$, which in turns implies that $a=b$.

\textbf{Surjectivity of $\lsem-\rsem_V$} The proof is very similar to the one for injectivity. Since $X\in\Forg G \nu T_V$ can be represented as a set of coherent sequences, we can take its inverse image under the various projections maps and get a coherent sequence of elements $X_i=\pi_i\inv(X)\in \Forg G  T_V^i 1$, where coherent now means that $G T_V^i(X_i)=(T_V^i)\inv(X_i)=X_{i+1}$. Since $\lsem-\rsem_0$ is trivially a regular epi, since $\delta$ is assumed to be a regular epi, and since $L$ is assumed to preserve them, it is clear that by induction each $c_n$ is a regular epi, and thus each $\Forg c_n$ is surjective. Any set $X_i$ therefore has a preimage $a_i\in \Forg L_V^i0$ which determines an element of $\Forg \init[L_V]$.
\end{proof}
\begin{remark} The condition that the initial object $0_{\cat}$ of $\cat$ should map injectively (resp. surjectively) into (resp. onto) $G1_{\cat[D]}$ (where $1_{\cat[D]}$ is the terminal object of $\cat[D]$) may at first seem ad hoc and unnatural. However, in most practical cases the categories $\cat$ and $\cat[D]$ are connected in a very special way, namely the subcategories $\cat_f$ and $\cat[D]_f$ of finite objects are dually equivalent and $\cat=\mathrm{Ind}(\cat_f), \cat[D]=\mathrm{Pro}(\cat[D]_f)$ (see \cite{2012:KurzStrongComp}). Since initial and terminal objects tend to be finite, it is actually generally the case that $G$ sends $1_{\cat[D]}$ to $0_{\cat}$, i.e. we have an isomorphism between  $0_{\cat}$ and $G1_{\cat[D]}$ and both conditions are satisfied. This is what happens in the case of $\pf\dashv\ups: \DL\to\Pos\op$ and in the case of $\uf\dashv\pow:\BA\to\Set\op$.
\end{remark}
\begin{remark}
Note that if $\cat=\BA$, then any finitary functor preserves injective maps as is shown in Lemma 6.14. of \cite{2010:AlexDaniela}. We do not know if this holds for $\DL$, but the proof in \cite{2010:AlexDaniela} does not seem to be directly adaptable to $\DL$ as it relies on a very special property of $\BA$. 
\end{remark}
In the cases of interest to us, i.e. when $\cat=\DL$ or $\BA$, we have the following representation of the `colours' $F \Free V$ of the cofree coalgebra. 

\begin{lemma}\label{ch5:lem:ufF=UP}
There is a natural isomorphism $\pf\Free\simeq\mathcal{Q}$, where $\mathcal{Q}$ is the contravariant powerset functor $\Set\to\Pos\op$. Similarly, $\uf\Free\simeq\mathcal{Q}$ where $\mathcal{Q}$ is the contravariant powerset functor $\Set\to\Set\op$.
\end{lemma}
\begin{proof}
We use the the well-known fact that for any $A$ in $\DL$ a prime filter $u\in \pf A$ is equivalent to a $\DL$-morphism $\chi_u: A\to \mathbbm{2}$, where $\mathbbm{2}$ is the two element distributive lattice. In particular, prime filters $u\in \pf\Free V$ are in one-to-one correspondence with $\DL$-morphisms $\Free V\to \mathbbm{2}$, which are themselves, by adjunction, in one-to-one correspondence with maps $V\to \Forg \mathbbm{2}=2$ where $2$ is the set with two elements. In other words, prime filters $u\in \pf\Free V$ are in one-to-one correspondence with subsets of $V$.
\end{proof}
The terminal coalgebra semantics in theses cases is thus given by the terminal coalgebra for the functor $T(-)\times\mathcal{Q} V$, i.e. `$T$-behaviours' are simply coloured by sets of propositional variables.
Using terminal coalgebra semantics we can show weak completeness of the abstract Hilbert system developed in Section \ref{ch1:subsec:Hilbert}. We show weak completeness in the case of $\cat=\DL$ since the case of $\BDL$ and $\BA$ are clearly special sub-cases.
\begin{corollary}
Let $L:\DL\to\DL$ be a varietor which preserves injective maps, $T:\Pos\to\Pos$ and that $\delta: L\ups\to\ups T$ be injective at every stage, then the Hilbert system $\deriv[ML]$ defined by $L$ for a set $V$ of propositional variables is weakly complete w.r.t. to $\Coalg(T)$.
\end{corollary}
\begin{proof}
Given a formulas $a, b\in \init[(L(-)+\Free V)]$ such that $a\not\vdash_{\mathrm{ML}} b$, we need to find a $T$-model which satisfies $a$ but not $b$. Recall from Proposition \ref{ch1:prop:MLEQconnection} that $a\deriv[ML]b$ iff $\deriv a\wedge b=a$. Our staring assumption is thus that $\not\vdash_{\mathrm{EL}} a\wedge b=a$, i.e. $a\wedge b\neq a$ in $\init[(L(-)+\Free V)]$ by Theorem \ref{ch1:thm:algsem}. We can now use Theorem \ref{ch5:thm:weakcomp1} and the assumption that $\delta$ is injective to get that the interpretations of $a\wedge b$ and $a$ in the terminal coalgebra do not coincide, i.e.
\[
\lsem a\wedge b\rsem_V\neq\lsem a\rsem_V
\]
In particular, there exists $t\in \lsem a\rsem_V$ such that $t\notin \lsem a\wedge b\rsem_V$, which by definition of the coalgebraic semantics in the abstract style means exactly that
\[
t\models a\text{ and }t\not\models a\wedge b
\]
and thus $t\not\models b$ as required.
\end{proof}
If the terminal coalgebra does not exist we have the following result from \cite{2010:AlexDaniela}, whose proof is very similar the proof of Theorem \ref{ch5:thm:weakcomp1}, but where the coalgebraic semantics is not given by the terminal coalgebra (which may not exist), but rather built from an arbitrary section $s: 1\to T_V1\times F\Free V$ of the terminal morphism $T_V1\to 1$ (which is always possible when $\cat[D]=\Pos$ or $\Set$ since it simply amounts to choosing an element in $T_V1\times F\Free V$).

\begin{theorem}\label{ch5:thm:weakcomp2}
In the fundamental situation described above, let $L:\cat\to\cat$ be a finitary varietor, let $T: \cat[D]\to\cat[D]$ be such that $\cat[D]$ has a terminal object $1_{\cat[D]}$ and that there exists a right-inverse $s:1_{\cat[D]}\to T 1_{\cat[D]}\times F\Free V$ of the unique morphism to the terminal object. For any set $V$, if $L_V=L(-)+\Free V$ and $T_V=T(-)\times F\Free V$ then:
\begin{itemize}
\item For any $n$, there exist a unique morphism $\lsem -\rsem_n: L_V^n(0_{\cat})\to GT^n1_{\cat[D]}$ interpreting formulas of modal depth at most $n$
\item This map is injective if (i) the morphism $0_{\cat}\to G1_{\cat[D]}$ is injective (ii) $\delta$ is injective, and (iii) $L$ preserves injective maps. In particular for any two formulas $a,b\in \init[L_V]$ such that $a\neq b$ there exist an $n$ which provide $a,b$ with different denotations, i.e. $L$ is weakly complete w.r.t. $\Coalg$.
\item This map is surjective if (i) there exists an regular epi $0_{\cat}\to G1_{\cat[D]}$ (ii) $\delta$ is a regular epi-transformation, and (iii) $L$ preserves regular epis.
\end{itemize}
\end{theorem}
\begin{proof}
See Theorem 6.15 of \cite{2010:AlexDaniela}.
\end{proof}

In many, but not all, practical cases it is possible to prove the injectivity of $\delta$ from its injectivity on finite objects. The following result bears some similarity to the construction of Section 6 of \cite{2012:KurzStrongComp}, but the strict connection between $L$ and $T$ is abandoned for the assumption that $L$ be finitary.

\begin{proposition}\label{ch5:prop:approxdelta}
Let the adjunction $F\dashv G:\cat\to\cat[D]\op$ be either the adjunction $\pf\dashv\ups: \DL\to\Pos\op$ or $\uf\dashv\pow: \BA\to\Set\op$, let $L:\cat\to\cat$ be finitary, and let $\delta:LG\to GT$. If $\delta_X$ is injective for any finite $\cat[D]$-object $X$, then $\delta$ is injective.
\end{proposition}
\begin{proof}
This is an approximation result using the facts that (1) every object of $\DL$ or $\BA$ can be represented as a filtered colimit of finitely presentable objects, (2) that these are all finite, i.e. $\DL$ and $\BA$ are locally finite, and (3) that by Birkhoff's representation theorem every finite distributive lattice can be represented as the downsets of its poset of join irreducibles, and similarly every finite boolean algebra can be represented as the powerset of its atoms. We show the result for the adjunction $\uf\dashv\pow: \BA\to\Set\op$, the proof is exactly the same in the distributive lattice case.

So let $X$ be any set, $\pow X$ is a boolean algebra which can be written as 
\[
\pow X=\colim_j A_j
\]
where each $A_j$ is a finitely presentable boolean algebra. Since such algebras are finite and since finite boolean algebras are isomorphic to the powerset algebra on their atoms, we can assume that each $A_j$ is of the form $A_j=\pow X_j$ for a finite set $X_j$, i.e. $\pow X=\colim_j \pow X_j$. From this, it is intuitively quite clear that $X=\colim_j X_j$. Formally, by the fact that the diagram $X_j$ is a diagram of monos and by the construction of colimits in $\Set$, it is clear that the maps $\mathrm{in}_j:\pow X_j\to \pow X$ give, by restriction to singletons, a collection of maps $\widehat{\mathrm{in}}_j: X_j\to X$, and $X$ is thus a cocone for the diagram $X_j$. To see that it is a colimiting cocone, let $(Y, h_j: X_j\to Y)$ be another cocone on the same diagram, then the direct images $h_j[-]: \pow X_j\to\pow Y$ turn $\pow Y$ into a cocone (in $\Set$!) for the diagram $\pow X_j$ whose colimit is $\pow X$. There must therefore exist a unique morphism $u: \pow X\to \pow Y$ making the obvious diagram commute. Since $\mathrm{in}_j$ and $h_j[-]$ send singletons to singletons, we can also restrict $u$ to singletons and get a unique map $\hat{u}: Y\to X$, proving that $X$ is the claimed colimit. Let us denote by $i_j$ the injection $X_j\to X$ and by $\tilde{i}_j $ the injection $\pow X_j\to \pow X$. It is not difficult to see that $\pow i_j\circ \tilde{i}_j=\id_{\pow X_j}$. By naturality of $\delta$ we have the following commutative diagrams:
\[
\xymatrix
{
L\pow X\ar[r]^{\delta_{X}}\ar[d]_{L\pow i_j} & \pow TX\ar[d]^{\pow T i_j}\\
L\pow X_j\ar[r]_{\delta_{X_j}} & \pow TX_j
}
\]
Now let $x,y\in L\pow X$ and assume that $\delta_X(x)=\delta_X(y)$. Since $L$ is finitary we have
\[
L\pow X=L\colim _j\pow X_j=\colim_j L\pow X_j
\]
and by construction of colimits in $\Set$, there must exist $x_j\in L\pow X_j, y_k\in L\pow X_k$ which map to $x,y$ by $L\tilde{i}_j,L\tilde{i}_k$. Since $\pow X_j$ is a filtered diagram, we can assume w.l.o.g. that $x_j$ and $y_k$ actually belong to the same set (by pushing them up to a common superset if necessary), i.e. we take $x_j, y_j\in L\pow X_j$. We now have
\begin{align*}
\pow T i_j\circ \delta_X (x) & = \pow T i_j\circ \delta_X(y) \hspace{1em}\Leftrightarrow\\
 \pow T i_j\circ \delta_X\circ L\tilde{i}_j(x_j) & = \pow T i_j\circ \delta_X\circ L\tilde{i}_j(y_j)\hspace{1em}\Leftrightarrow\\
 \delta_{X_j}\circ L\pow i_j\circ L\tilde{i}_j(x_j) & = \delta_{X_j}\circ L\pow i_j\circ L\tilde{i}_j(y_j)\hspace{1em}\Leftrightarrow\\
\delta_{X_j}(x_j) & = \delta_{X_j}(y_j)
\end{align*}
and thus $x_j=y_j$ since $\delta_{X_j}$ is assumed to be injective. It follows immediately that $x=y$, and thus $\delta_X$ is injective too.
\end{proof}

It is clear from the proof above that if $T$ preserves finite objects (which is quite often the case in practise), then for $\delta_X$ to be injective on finite objects, it must also be the case that $L$ should preserve finite objects. This can never be the case if $L$ represents a logic with infinitely many modal operators such as GML (see Section \ref{ch2:subsec:GML}). This problem is addressed in the next section.

\subsection{Completeness by filtration}\label{ch5:sub:filtration}

When a language includes infinitely many operators (and thus $L$ does not preserve finite sets), it is often easier to prove completeness (or, the injectivity of the semantic transformation) on finite sublanguages. This is the idea behind the filtration method which is well-known to modal logicians. Here we follow the presentation of this method by Kurz and Petri\c{s}an in \cite{2010:AlexDaniela}.

The idea is to express the logic's functor $L$ by a cofiltered limit of functors which preserve finite boolean algebras. As we shall see in the next section, the preservation of finite algebras is a particularly useful property for reasoning about the semantic transformation and its transpose. For now however, we will state the following more general approximation result.

\begin{theorem}[Weak completeness by filtration]\label{ch5:thm:filt}
In the fundamental situation of Diagram (\ref{ch1:diag:fundamentalSituation}), let $L:\cat\to\cat$ be a finitary functor, let $T:\cat[D]\to\cat[D]$, and assume that 
\[
L=\colim_k L_k
\]
where the index $k$ runs over a cofiltered category. Assume further that for each $k$ there exists a natural transformation
\[
\delta_k: L_k G\to G T
\]
such that for every natural transformation $\zeta_{k,k'}: L_k\to L_{k'}$ in the diagram defining $L$, $\delta_k=\delta_{k'}\circ \zeta_{k,k'}$, then there exists a unique natural transformation $\delta: L G \to G T$ which agrees with $\delta_k$ for each $k$ and is injective if each $\delta_k$ is injective.
\end{theorem}
\begin{proof}
Since $\cat$ is a finitary variety, $\Forg:\cat\to\Set$ creates filtered colimits, and in particular
\[
\Forg LA=\Forg \colim_k L_k A=\colim_k \Forg L_k A
\]
By construction of colimits in $\Set$, if $a\in  LG X$ for some set $X$, then there must exist $k$ in the index category and an element $a'\in L_k G X$ such that $\xi_{k,X}(a')=a$, where $\xi_{k}: L_kG\to LG $ is the obvious natural transformation. We now define 
\[\delta_X (x): LG X\to G TX, a\mapsto \delta_{k,X}(a')\]
To see that this is well-defined, assume that there exist  $k'$ and $a''$ such that $\xi_{k',X}(a'')=a$. By construction of colimits in $\Set$, this means that there must exist a zigzag of natural transformations between $L_k$ and $L_{k'}$ which connects the elements $a'$ and $a''$. By the assumption that $\delta_k=\delta_{k'}\circ \zeta_{k,k'}$ for any natural transformation $\zeta_{k,k'}:L_k\to L_{k'}$ in the diagram, it follows that $\delta_{k',X}(a'')=\delta_{k,X}(a')$, and $\delta$ is thus well-defined. If each $\delta_k$ is injective, it is clear from the construction we have just described that so is $\delta$.
\end{proof}

Weak completeness by filtration is then an immediate consequence of Theorems \ref{ch5:thm:filt} and \ref{ch5:thm:weakcomp2}.
\subsection{A brief note on expressivity}

Given a $T$-coalgebra $\gamma: W\to TW$ and a valuation $v: V\to G W$, the interpretation map $\lsem-\rsem_W: \init[L_V]\to G W$ has an adjoint transpose:
\[
\theta_W: W\to F\init[L]
\]
called the theory map and which associates to each point $w\in W$ the ultrafilter of formulas which are satisfied at $w$. The following result was elegantly shown in \cite{JacobsExpressivity}:

\begin{theorem}[\cite{JacobsExpressivity}]
Let $L:\cat\to\cat$ be a finitary functor, let $T:\cat[D]to\cat[D]$ where $\cat[D]$ has an strong-epi mono factorisation system, and let $\delta:LG \to GT$ be a semantic transformation. If the adjoint transpose $\hat{\delta}: TF\to  FL$ is injective then logically indistinguishable points are behaviourally equivalent, i.e. if two points have the same theory, then there exists a coalgebra homomorphism identifying them.   
\end{theorem}
\begin{proof}
Let $\gamma: W\to TW$ be a coalgebra, and let $v: V\to G W$ be a valuation. This data defines the interpretation map $\lsem-\rsem: \init[L_V]\to G W$, where $L_V=L(-)+\Free V$. The idea is to factor the adjoint transpose of $\lsem-\rsem$, i.e. the theory map $\theta: W\to F\init[L_V]$, into a strong-epi mono factorisation, which is always possible in $\cat[D]$ by assumption. We have
\[
\xymatrix@C=10ex
{
W\ar[d]_{\gamma}\ar@{->>}[r]^{q}\ar@/^2pc/[rrr]^{\theta} & Q\hspace{1ex}\ar@{-->}[d] \ar@{>->}[rr]^{m} & & F\init[L_V]\ar[d]^{\simeq}\\
TW\ar@{->>}[r]_{Tq} & TQ\hspace{1ex}\ar@{>->}[r]_{Tm} & TF\init[L_V]\hspace{1ex}\ar@{>->}[r]_{\hat{\delta}_{\init[L_V]}} & F L\init[L_V]
}
\]
where the unique diagonal fill-in morphism exists by virtue of $q$ being strong. If $\theta(x)=\theta(y)$, then since $m$ is injective, we must have $q(x)=q(y)$, and the two states are thus behaviourally equivalent.
\end{proof}

\section{Strong completeness}\label{ch5:sec:strongComp}

In this section we will present the basic set-up for the study of coalgebraic completeness-via-canonicity. In the language of modal logic we will investigate `canonical' models in the coalgebraic framework, i.e. coalgebraic models built from maximally consistent sets of formulas.

\subsection{The coalgebraic J\'{o}nsson-Tarski theorem}
The construction of canonical models for coalgebraic logics was first studied in \cite{2005:UfExtCoalg} via the notion of ultrafilter extension, and subsequently in \cite{2012:KurzStrongComp} and \cite{2009:DirkStrongComp} where general conditions are presented for the existence of `canonical' models (more on this terminology shortly). The idea of building models from formulas is pervasive in logics, and is a fundamental tool in modal logic, but the J\'{o}nsson-Tarski theorem actually shows more than this. Although it is not usually stated explicitly, the classical J\'{o}nsson-Tarski theorem can in fact be split into two independent components: 
\begin{enumerate}
\item the first shows how to build Kripke models of normal modal logics from the algebra of formulas using the usual notion of maximally consistent sets (also called theory), 
\item the second shows how validity in this Kripke model is equivalent to (algebraic) validity in the \emph{canonical extension} of the BAO defined by the logic.
\end{enumerate}
The first component of the theorem is used to show strong completeness of normal modal logics, whilst the second is used to show strong completeness in the presence of canonical equations. This section and \cite{2005:UfExtCoalg,2012:KurzStrongComp,2009:DirkStrongComp} focus solely on the first part of the theorem, we will examine the second part in the next section. 

We start by making an important comment on the term `canonical', which is related to the dual role of the J\'{o}nsson-Tarski theorem. In its first role, it builds `canonical models', which in the coalgebraic framework correspond to $T$-coalgebra over $F\init[(L(-)+\Free V)]$. However, as pointed out in \cite{2009:DirkStrongComp,2012:KurzStrongComp} for example, these `canonical' models are usually not defined uniquely. In fact an element of choice is always required, for example when a transformation $FL\to TF$ is used to define them (see \cite{2012:KurzStrongComp}), it is in general not natural as we shall soon see, making canonical models incompatible with one another. The term `canonical' is thus rather imprecise in this context, prompting Pattinson and Schr\"{o}der to suggest the term `quasi-canonical' in \cite{2009:DirkStrongComp}. In the context of models we will adopt the following convention: we will use the term `canonical' as meaning `governed by a law' and thus call a model canonical if it can be explicitly constructed. Models which can merely be shown to exist will be call `quasi-canonical'.

In its second role, the J\'{o}nsson-Tarski theorem connects the building of models from formulas to the notion of \emph{canonical} extension. As we have seen in Chapter 2, every DLE can be embedded into a better behaved DLE, called its canonical extension, whose underlying distributive lattice structure is complete and whose join irreducibles are join-dense. Moreover, the canonical extension is \emph{uniquely defined} (up to isomorphism, see e.g. \cite{GivantHalmos}) by the requirement that that $\ba$ must be dense and compact in $\ba\ce$.  The term `canonical' is thus completely unambiguous in the context of DLEs (and BAEs). There is therefore a fundamental contradiction between the two usages of the term `canonical', but as we shall see in the next section, this contradiction often resolve itself spontaneously thanks to the notion of \emph{smoothness} introduced in Chapter 2, viz. we can get a uniquely defined canonical extension through the use of a non-uniquely defined canonical model. 

Our criterion for the J\'{o}nsson-Tarski theorem is essentially (6.11) of \cite{2012:KurzStrongComp}, which is slightly different from that of \cite{2005:UfExtCoalg,2009:DirkStrongComp}, but is easier to state. We will show how it is related to the conditions of Theorem 3 of \cite{2005:UfExtCoalg}, and Theorem 2.5 of \cite{2009:DirkStrongComp}. We place ourselves once again in the following situation
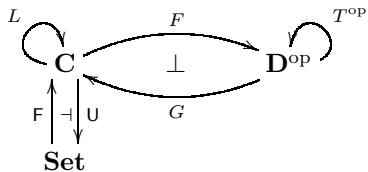
\[
\xymatrix
{
\cat\ar@/^1pc/[rr]^{F} \ar@(l,u)^{L} \ar@<5pt>[d]^{\Forg} & \perp&\cat[D]\op\ar@/^1pc/[ll]^{G}\ar@(r,u)_{T\op}\\
\Set\ar@<5pt>[u]^{\Free}_{\dashv}
}
\]
where $\cat$ is a finitary algebraic variety.
\begin{theorem}[Coalgebraic J\'{o}nsson-Tarski Theorem]\label{ch5:thm:jontarski}\index{J\`{o}nsson-Tarski theorem}
Let $\delta: L G\to G T$ be a semantic natural transformation and let $\eta,\epsilon$ denote the unit and counit of the adjunction $F\dashv G$. If the adjoint semantic transformation $\hat{\delta}: TF \to F L$ given by $\hat{\delta}=F L\eta\circ F\delta_{F}\circ \epsilon_{TF}$ has (not necessarily natural) right inverse $\hat{\delta}\inv:F L\to TF$, then for any $A$ in $\Alg_{\cat}(L)$, the morphism $\eta_A: A\to GF A$ lifts to an $L$-algebra morphism.
\end{theorem}
\begin{proof}
Starting with an algebra $\alpha: LA\to A$, we show that the following diagram commutes:
\begin{equation}\label{ch5:diag:jontar}
\xymatrix
{
LA\ar[r]^{L\eta_A}\ar[ddd]_{\alpha}\ar[ddr]_{\eta_{LA}} & L GF A \ar[d]^{\delta_{F A}}\\
& GT F A \ar[d]^{G(\hat{\delta}_A\inv)}\\
& GF L A\ar[d]^{GF\alpha} \\
A\ar[r]_{\eta_A} & GF A
}
\end{equation}
To see that it commutes, notice first that the trapezium at the bottom of the diagram defined by $\eta_{LA}$ commutes by naturality of $\eta$. So we need only to show that the triangle of the top of the diagram commutes. To see this, consider the following diagram:
\[
\xymatrix
{
LA\ar[r]^{L\eta_A}\ar[d]_{\eta_{LA}} & LGF A\ar[r]^{\delta_{F A}} & G T F A \ar[d]^{\eta_{G TF A}}\\
GF LA\ar[r]_{GF L\eta_A} & GF LGF A\ar[r]_{GF \delta_{F A}} & GF G T F A\ar@/_4pc/[u]_{G\epsilon_{TF A}}
}
\]
The naturality square clearly commutes, and moreover, by the fact that $F\dashv G$ it is also the case that $G\epsilon_{TF A}\circ \eta_{G TF A}=\id_{G TF A}$, and thus 
\begin{align*}
G\hat{\delta}_A\circ \eta_{LA}& = G\epsilon_{TF A}\circ GF \delta_{F A}\circ GF L\eta_A\circ \eta_{LA} \\
& = G\epsilon_{TF A}\circ \eta_{G T F A}\circ \delta_{F A}\circ L\eta_A \\
& = \delta_{G A}\circ  L\eta_A
\end{align*}
From this we get:
\[
G(\hat{\delta}_A\inv)\circ \delta_{F A}\circ L\eta_A=G (\hat{\delta}_A\inv)\circ G\hat{\delta}_A\circ \eta_{LA}=G( \hat{\delta}_A\circ \hat{\delta}_A\inv)\circ \eta_{LA}=\eta_{LA}
\]
By definition of $\hat{\delta}_A\inv$ as a right inverse.
\end{proof}

In the case where $A=\mathsf{G}\Free V$ is the free $L$-algebra over $V$, i.e. the abstract language, we will call the coalgebra 
\[
\hat{\delta}_{\mathsf{G}\Free V}\inv\circ F\langle-\rangle_V: F(\mathsf{G}\Free V)\to TF(\mathsf{G}\Free V)
\]
(where $\langle-\rangle_V$ is the structure map of the language), the \textbf{(pseudo)-canonical $T$-model}\index{Canonical model} (depending on whether $\hat{\delta}\inv$ can be given explicitly) associated with the semantic transformation $\delta$. More generally, for any object $\alpha: LA\to A$ in $\Alg_{\cat}(L),$ we will call a coalgebra $\hat{\delta}_{A}\inv\circ F\alpha: FA\to TFA$ a \textbf{(pseudo)canonical model on $A$}\index{Canonical model!on an algebra}. We will call the $L$-algebra $GF\alpha\circ G\hat{\delta}_A\inv\circ \delta_{FA}: LGF A\to GFA$ the \textbf{J\'{o}nsson-Tarski extension} of the $L$-algebra $\alpha: LA\to A$\index{J\'{o}nsson-Tarski extension}. Will will show in that $\hat{\delta}\inv$ can in practise often be given explicitly, modulo a non-constructive choice principle.

\begin{remark}
We could recast the coalgebraic J\'{o}nsson-Tarski Theorem without any reference to the semantics of the language defined by $L$ in terms of the \textbf{distributive law of a monad over a functor}\index{Distributive law of a monad}. As is well-known, every adjunction gives rise to a monad, and in particular $GF:\cat\to\cat$ is a monad. A distributive law of $GF$ over a functor $L:\cat\to\cat$ is a natural transformation $\lambda: L GF\to GF L$ which interacts well with the unit and multiplication of the monad, in particular the following triangle must commute
\[
\xymatrix
{
LA\ar[r]^(0.4){L\eta_A}\ar[dr]_{\eta_{LA}} & LGF A\ar[d]^{\lambda_A}\\
& GF L A
}
\]
But this is exactly the triangle from the proof of Theorem \ref{ch5:thm:jontarski}. In other words, given a distributive law of the monad $GF$ over $L$, we immediately get an injective embedding of an $L$ algebra into its canonical extension. Notwithstanding the elegance of this approach, our purpose is, at the end of the day, to build models, and factoring through $G TF A$ is therefore essential.
\end{remark}

\subsection{Strong completeness for positive coalgebraic logics}\label{ch5:subsec:DL}
In this section we will apply our coalgebraic J\'{on}sson-Tarski Theorem \ref{ch5:thm:jontarski} to $L$-algebras in $\DL$ interpreted by $T$-coalgebras in $\Pos$, i.e. to positive coalgebraic logics (see \cite{1995:DunnPositiveML,2012:PositiveCoalgExpress,2013:PositiveCoalg}). Clearly, weak completeness follows from the J\'{on}sson-Tarski theorem if we can show that the unit $\eta_{\mathsf{G}\Free V}$ of the adjunction $F\dashv G$, evaluated at the `language object' $\mathsf{G}\Free V$, is injective; in which case different formulas will have different denotations. In the cases which interest us here, i.e. when $\cat=\DL$ and $\cat[D]=\Pos$, $\eta$ is given by
\[
\eta_A: A\to \ups\pf A, a\mapsto\{p\in \pf A\mid a\in p\}
\] 
and we must distinguish two problems: (1) proving that $\eta_{\mathsf{G}\Free V}$ is injective and (2) proving that $\eta$ is injective in general. The problems are different because $V$ is traditionally assumed to be countable, and thus so is the logic $\mathsf{G}\Free V$. In this case it can easily be shown that $\eta_{\mathsf{G}\Free V}$ is injective via a Lindenbaum lemma argument.

\begin{lemma}[Lindenbaum's Lemma, \cite{1995:DunnPositiveML}]\label{ch5:lem:Lindenbaum}
If $L:\DL\to\DL$ preserves countable sets and $V$ is countable, then $\eta_{\mathsf{G}\Free V}$ is injective.
\end{lemma}
\begin{proof}
Let $a,b\in \mathsf{G}\Free V$ such that $a\neq b$, we need to build a prime filter which contains one formula and not the other. If $a\neq b$ then either $a\nleq b$ or $b\nleq a$, we assume the former and it follows easily that $\uparrow a\cap \downarrow b=\emptyset$. Since $V$ is countable and $L$ preserves countable sets, so is $\mathsf{G}\Free V$ and we can enumerate all the formulas $a_0,a_1,\ldots\in \mathsf{G}\Free V$. The strategy is to add $a_0$ to $\uparrow a$ if $\uparrow (a\wedge a_0)\cap\downarrow b=\emptyset$ or else add it to $\downarrow b$, and then iterate the construction. It can be shown that each formula can in this way be added to the left or to the right whilst preserving disjointness (see Lemma 3.5 of \cite{1995:DunnPositiveML}) and that the limit of the procedure provides a prime filter which contains $a$ but not $b$.
\end{proof}

In fact this kind of procedure gives more than weak completeness since it can be carried out for arbitrary pairs of sets forming a non-intersecting ideal-filter pair. This provides us with a strong completeness result for the abstract Hilbert system of Section \ref{ch1:subsec:Hilbert}.

\begin{theorem}[Strong completeness for countable $L$-algebras over $\DL$]\label{ch5:them:strongComplCountableDL}
If $L:\DL\to\DL$ preserves countable sets and satisfies the conditions of the J\'{o}nsson-Tarski Theorem \ref{ch5:thm:jontarski} for a choice of semantics $\delta:L\ups\to\ups T$, and if $V$ is countable, then the Hilbert system $\deriv[ML]$ associated with $L$ is strongly complete w.r.t. $\Coalg(T)$
\end{theorem}
\begin{proof}
Given arbitrary sets $\Phi,\Psi\subseteq \mathsf{G}\Free V$ of formulas such that $\Phi\not\vdash_{\mathrm{ML}}\Psi$ we must show that there exists a model satisfying all the formulas of $\Phi$ and none of the formulas of $\Psi$. It is not difficult to show that $\Phi\not\vdash_{\mathrm{ML}}\Psi$ iff for every \emph{finite} sets of formulas $\phi_1,\ldots,\phi_m\in \Phi$ and $\psi_1,\ldots,\psi_n\in\Psi$ 
\[
\phi_1\wedge\ldots\wedge\phi_m\nleq \psi_1\vee\ldots\psi_n
\]
i.e. the filter $\langle\Phi\rangle\eup$ generated by $\Phi$ has an empty intersection with the ideal $\langle\Psi\rangle\edown$ generated by $\Psi$ (in particular, if we work in $\BDL$ then no meet of elements of $\Phi$ can equal $\bot$, i.e. we have a `consistent' set of formulas). By the same procedure as above, we can enumerate all formulas in $\mathsf{G}\Free V$ and successively add them either to $\langle\Phi\rangle\eup$ or to $\langle\Psi\rangle\edown$ whilst maintaining the intersection between the sets empty. The process yields a prime filter $p_\Phi$ containing every formula in $\Phi$, and thus
\[
p_\Phi\in\eta_{\mathsf{G}\Free V}(\phi),\text{ i.e. }p_\Phi\in\lsem \phi\rsem_{\pf \mathsf{G}\Free V},\text{ i.e. } p_\Phi\models \phi, \text{ for each }\phi\in\Phi
\]
and since $p_\Phi$ is disjoint from $\Psi$ it follows that
\[
p_\Phi\notin\eta_{\mathsf{G}\Free V}(\psi),\text{ i.e. }p_\Phi\notin\lsem \psi\rsem_{\pf \mathsf{G}\Free V},\text{ i.e. }p_\Phi\not\models \psi, \text{ for each }\psi\in\Psi
\]
which is what we wanted to show.
\end{proof}

If we wish a more categorically pleasing result, and in particular include cases where $V$ is not countable, then we must prove the injectivity of $\eta$ in general. This requires an additional ingredient, namely the Prime Filter Theorem (see Theorem \ref{ch2:thm:PrimeFilterThm}). Indeed, to extend a filter $\up a$ such that $\up a\hspace{2pt}\cap\hspace{-2pt}\down b=\emptyset$ to a prime filter $F$ such that $F\cap\hspace{-2pt} \down b\in\emptyset$ (as in the proof of Lemma \ref{ch5:lem:Lindenbaum}) in a general distributive lattice requires precisely the Prime Filter Theorem. Note that this theorem is independent of ZF, but is strictly weaker than ZFC, and we must thus assume that we are doing mathematics in which the Prime Filter Theorem holds.  Modulo this hypothesis, we can adapt Theorem \ref{ch5:them:strongComplCountableDL} and show the following result:

\begin{theorem}[Strong completeness for arbitrary $L$-algebras over $\DL$]\label{ch5:thm:strongComplDL}
Let $L:\DL\to\DL$, $T:\Pos\to\Pos$ and $\delta:L\ups\to\ups T$ satisfy the conditions of the J\'{o}nsson-Tarski Theorem \ref{ch5:thm:jontarski}. For any distributive lattice $A$ and any sets $\Phi,\Psi\subseteq A$ such that no finite meet of elements of $\Phi$ lies below a finite join of element of $\Phi$, there exist a canonical $T$-model of $A$ satisfying all formulas of $\Phi$ and no formula of $\Psi$.
\end{theorem}

For an important class of problems it is in fact possible to explicitly build
a right inverse $\hat{\delta}\inv$. This class of problem includes positive modal logic (see \cite{1995:DunnPositiveML}) and our description of intuitionistic and sub-structural logics of Section \ref{ch2:subsec:intuitionistic} (see \cite{2015:self}). It is characterised by $n$-ary modalities which either (1) preserve joins or anti-preserve meets in each of their arguments, or (2) preserve meets or anti-preserves joins in each of their arguments. Formally, for a given finitary signature $\Sigma$ we define the syntax building functor:
\[
L_\Sigma:\DL\to\DL\begin{cases}
L_\Sigma A=\Free\polyFunc\Forg A/\equiv\\
L_\Sigma f: L_\Sigma A\to L_\Sigma B, [a]_\equiv\mapsto [\polyFunc\Forg f(a)]_\equiv
\end{cases}
\]
where $f:A\to B$ is a $\DL$-morphism and $\equiv$ is the fully invariant relation in $\DL$ generated by equations of the type
\begin{equation}\label{ch5:eq:DL1}
\heartsuit(a_1\vee b_1,\ldots,a_k\vee b_k, a_{k+1}\wedge b_{k+1},\ldots,a_n\wedge b_n)=\heartsuit(a_1,\ldots,a_n)\vee\heartsuit(b_1,\ldots,b_n)
\end{equation}
or 
\begin{equation}\label{ch5:eq:DL2}
\heartsuit(a_1\wedge b_1,\ldots,a_k\wedge b_k, a_{k+1}\vee b_{k+1},\ldots,a_n\vee b_n)=\heartsuit(a_1,\ldots,a_n)\wedge\heartsuit(b_1,\ldots,b_n)
\end{equation}
for every $n$-ary symbol $\heartsuit$ in the signature $\Sigma$. We will always group the arguments of each $n$-ary symbol according to whether they are isotone or antitone, starting with $k$ isotone arguments and followed $n-k$ antitone arguments. Note that the syntax builders $\LHey$ and $\LRL$ defined in Section \ref{ch2:subsec:intuitionistic} do indeed fall into this category. We will call logics defined by this kind of functor \textbf{positive relational logics}\index{Relational logic!positive}. The reason for this terminology is that we are going to interpret these logics in coalgebras defined by the $\Pos$ equivalent of the powerset functor. Following \cite{2011:KurzEqPresMon}, we define the convex powerset functor $\cpow_c:\Pos\to\Pos$ as the functor associating with each poset $X$ its set of convex subsets (i.e. subsets $Y$ with the property that if $x,z\in Y$ then $y\in Y$ whenever $x\leq y\leq z$) ordered by the Egli-Milner order\index{Egli-Milner order}, i.e. $Y\leq Y'$ in $\cpow_c$ if
\begin{align*}
&(\forall y\in Y)(\exists y'\in Y')\text{ s.th. } y\leq y', \text{ and }\\
&(\forall y'\in Y')(\exists y\in Y)\text{ s.th. } y\leq y'
\end{align*}
Note that upsets are particular examples of convex sets. The convex powerset functor associates with a monotone map the convex closure of its direct image, i.e. if $f: X\to Y$ is a monotone map, then $\cpow_c f: \cpow_c X\to\cpow_c Y, U\mapsto \overline{f[U]}$, where $\overline{(-)}$ denotes the convex closure. We now define for a signature $\Sigma$ the functor:
\[
T_\Sigma: \Pos\to\Pos\begin{cases}
T_\Sigma X=\prod_{\heartsuit\in\Sigma} \cpow_c(X^k\times (X\op)^{n-k})\\
T_\Sigma f= \prod_{\heartsuit\in\Sigma} \cpow_c f^n
\end{cases}
\]
where $k$ is the number of isotone arguments of $\heartsuit$, $n$ its total arity, $X\op$ is the dual poset of $X$ (i.e. same carrier, order reversed), and $f^n$ is the usual $n$-ary product of maps (it is easy to see that $\Pos$ has products which are created in $\Set$). 

For better readability we will follow the following convention. For any symbol $\heartsuit$ in the signature $\Sigma$ such that $n=\ari(\heartsuit)$ and $k$ denotes the number of isotone arguments (regrouped as the first arguments), $i$ will always denote an index value between 1 and $k$, and such an index will be called an isotone index. Similarly, we assume from now on that $j$ will always denote an index value between $k+1$ and $n$, and such an index will be called an antitone index. Using this convention, we interpret the logic defined by $L_\Sigma$ in $T_\Sigma$-coalgebras via the following semantic transformation $\delta_X^\Sigma: L_\Sigma \ups X\to\ups T_\Sigma X$:
\begin{align*}
&[\heartsuit(u_1,\ldots,u_k,u_{k+1},\ldots,u_n)]\mapsto \\
& \{t\in T_\Sigma X \mid \exists (a_1,\ldots,a_n)\in \pi_\heartsuit(t), (a_i\in u_i\text{ for all }i)\text{ and }(a_j\notin u_j\text{ for all }j)\}\\
&\text{if }\heartsuit\text{ satisfies Eq. (\ref{ch5:eq:DL1})}\\
&[\heartsuit(u_1,\ldots,u_k,u_{k+1},\ldots,u_n)]\mapsto \\
&\{t\in T_\Sigma X\mid \forall (a_1,\ldots,a_n)\in \pi_\heartsuit(t), (a_j\in u_j \text{ for all } j)\Rightarrow (a_i\in u_i \text{ for all }i)\}\\
&\text{if }\heartsuit\text{ satisfies Eq. (\ref{ch5:eq:DL2})}
\end{align*}
where $\pi_\heartsuit$ is the projection on the $\heartsuit$-component of the product defining $T_\Sigma$.
\begin{proposition}\label{ch5:prop:deltaWellDefined}
$\delta^\Sigma$ is well-defined.
\end{proposition}  
\begin{proof}
For $\delta^\Sigma$ to be well defined, the image of each `generator' must form an upset, and  $\delta^\Sigma$ must preserve the equations (\ref{ch5:eq:DL1}) and (\ref{ch5:eq:DL2}). Let us fix an arbitrary poset $X$, a symbol $\heartsuit\in\Sigma$ with $n=\ari(\heartsuit)$, and let $(u_1,\ldots, u_n), (v_1,\ldots, v_n)\in (\ups X)^n$. We first show that if $t\in \delta^\Sigma(\heartsuit(u_1,\ldots,u_n))$ and $t\leq t'\in T_\Sigma X$, then $t'\in \delta^\Sigma(\heartsuit(u_1,\ldots,u_n))$ too. By definition of the partial order on $T_\Sigma X$, this means that $\pi_\heartsuit(t)\leq \pi_\heartsuit(t')$ for the (component-wise) Egli-Milner order, i.e. for each $(a_1,\ldots,a_k,a_{k+1},\ldots,a_n)\in \pi_\heartsuit(t)$ there exist $(a_1',\ldots,a_k',a_{k+1}',\ldots,a_n')\in \pi_\heartsuit(t')$ such that $a_i\leq a_i'$ and $a_j'\leq a_j$ (note the reversal of direction due to the presence of $(-)\op$ in the definition of $T_\Sigma$), and conversely. We need to distinguish two cases:
\begin{enumerate}
\item $\heartsuit$ satisfies (\ref{ch5:eq:DL1}). By definition there exists $(a_1,\ldots,a_n)\in\pi_\heartsuit(t)$ such that $a_i\in u_i$ and $a_j\notin u_j$. We want to show that there exist $(a_1',\ldots,a_n')$ in $\pi_\heartsuit(t')$ with the same property. By definition of the Egli-Milner order  we know that there exist $(a_1'\ldots,a_n')\in \pi_\heartsuit(t')$ s.th. $a_i\leq a_i'$, and $a_j'\leq a_j$. Since $u_i$ is an upset, $a_i'\in u_i$ and since $u_j$ is an upset it follows that $a_j'\notin u_j$.
\item $\heartsuit$ satisfies (\ref{ch5:eq:DL2}). By definition, for all $(a_1,\ldots,a_n)\in\pi_\heartsuit(t)$, if $a_j\in u_j$ for every antitone index $j$ then $a_i\in u_i$ for every isotone index $i$. We want to show that the same holds for every $(a_1',\ldots,a_n')\in\pi_\heartsuit(t')$. So let us assume that each $a_j'\in u_j$. By definition of the Egli-Milner order, we know that there exists $(a_1,\ldots,a_n)$ in $\pi_\heartsuit(t)$ s.th. $a_i\leq a_i'$ and $a_j'\leq a_j$. Since since $u_j$ is an upset it follows that $a_j\in u_j$ for each antitone $j$, and thus $a_i\in u_i$ for every isotone $i$, and thus $a_i'\in u_i$ since $u_i$ is an upset.
\end{enumerate} 
Finally, we need to show that $\delta^\Sigma$ preserves equations (\ref{ch5:eq:DL1}) and (\ref{ch5:eq:DL2}). We have (dropping the square brackets for notational clarity):
\begin{align*}
&\delta^\Sigma_X(\heartsuit(u_1\vee v_1,\ldots,u_k\vee v_k,u_{k+1}\wedge v_{k+1},\ldots,u_n\wedge v_n))\\
&=\{t\in T_\Sigma X \mid \exists (a_1,\ldots,a_n)\in \pi_\heartsuit(t), (a_i\in u_i\vee v_i)\text{ and }(a_j\notin u_j\wedge v_j)\}\\
&=\{t\in T_\Sigma X \mid \exists (a_1,\ldots,a_n)\in \pi_\heartsuit(t), (a_i\in u_i)\text{ and }(a_j\notin u_j)\}\cup\\
&\hspace{2em}\{t\in T_\Sigma X \mid \exists (a_1,\ldots,a_n)\in \pi_\heartsuit(t), (a_i\in v_i)\text{ and }(a_j\notin v_j)\}\\
&=\delta^\Sigma_X(\heartsuit(u_1,\ldots, u_n))\vee \delta^\Sigma_X(\heartsuit(v_1,\ldots, v_n))
\end{align*}
A similar proof shows that $\delta^\Sigma$ preserves equation (\ref{ch5:eq:DL2}).
\end{proof}

We will now show how a right inverse to $\hat{\delta}^\Sigma$ can be defined explicitly, providing strong completeness of the logic defined by $L_\Sigma$ w.r.t. $T_\Sigma$-coalgebras. To keep the proof readable we will need the following additional notation: if $\vect[x]$ is a $k-1$-tuple of isotone arguments (resp. an $n-k-1$-tuple of antitone arguments) and $y_i$ (resp. $y_j$) is a particular isotone (resp. antitone) argument, then $\vect[x/y_i]$ (resp. $\vect[x/y_j]$) denotes the $k$-tuple (resp. $n-k$-tuple) formed by inserting $y_i$ (resp. $y_j$) in $i^{th}$ (resp. $j^{th}$) position in $\vect[x]$.

\begin{theorem}\label{ch5:thm:strongcomplRelational}
Any positive relational logic defined by a signature $\Sigma$ is strongly complete with respect to $T_\Sigma$-coalgebras.
\end{theorem}
\begin{proof}
By the coalgebraic J\'{o}nsson-Tarski Theorem \ref{ch5:thm:jontarski} and Theorem \ref{ch5:thm:strongComplDL} it suffices to define a right-inverse to the transformation $\hat{\delta}: T_\Sigma \pf\to\pf L_\Sigma$. For the purpose of finding a right inverse to $\hat{\delta}$, we only need to specify when a generator of $L_\Sigma A$ belongs to $\hat{\delta}_A(t)$ for some $t\in T_\Sigma \pf A$.
\begin{align*}
&\text{if }\heartsuit\text{ satisfies Eq. (\ref{ch5:eq:DL1}):}\\
&\heartsuit(a_1,\ldots,a_n)\in \hat{\delta}_A(t)\text{ iff }\\
&\exists (F_1,\ldots,F_n)\in \pi_\heartsuit(t)\text{ s.th. }(a_i\in F_i\text{ for all }i) \text{ and }  (a_j\notin F_j \text{ for all } j)\\
&\text{if }\heartsuit\text{ satisfies Eq. (\ref{ch5:eq:DL2}):}\\
&\heartsuit(a_1,\ldots,a_n)\in \hat{\delta}_A(t)\text{ iff }\\
&\forall (F_1,\ldots,F_n)\in \pi_\heartsuit(t)(a_j\in F_j\text{ for all }j) \Rightarrow  (a_i\in F_i \text{ for all } i)
\end{align*}
We now define a morphism $\gamma_A: \pf L_\Sigma A\to T_\Sigma \pf A$ which we will show is a right inverse. It maps $F\in \pf L_\Sigma A$ to the $|\Sigma|-$tuple of sets of $n$-tuples $(F_1,\ldots,F_n), n=\ari(\heartsuit), \heartsuit\in\Sigma$ of prime filters of $A$ defined by:
\begin{align*}
\pi_\heartsuit(\gamma_A(F))=&\{(F_1,\ldots,F_n)\mid \\
&(a_i\in F_i)\text{ and }(a_j\notin F_j)\Rightarrow\heartsuit(a_1,\ldots,a_n)\in F \text{ if }\heartsuit\text{ satisfies (\ref{ch5:eq:DL1})} \\
&\heartsuit(a_1,\ldots,a_n)\in F\Rightarrow \text{ if }(a_j\in F_j)\text{ then } (a_i\in F_i)\text{ if }\heartsuit\text{ satisfies (\ref{ch5:eq:DL2})}\}
\end{align*}

To see that $\gamma_A$ is well-defined, we need to check that for each symbol $\heartsuit\in\Sigma$, $\pi_\heartsuit(\gamma_A(F))$ is a convex subset of $\cpow_c(\pf A^k\times (\pf A\op)^{n-k})$. So let $(F_1,\ldots,F_n)\leq (F'_1,\ldots,F'_n)\leq (F''_1,\ldots,F''_n)$ with $(F_1,\ldots,F_n),(F''_1,\ldots,F''_n)\in \pi_\heartsuit(\gamma_A(F))$. We need to show that $(F'_1,\ldots,F'_n)\in \pi_\heartsuit(\gamma_A(F))$. From the presence of $(-)\op$ in the definition, we get that $F_i\subseteq F_i'\subseteq F_i''$ for isotone indices and $F_j''\subseteq F_j'\subseteq F_j$ for antitone indicex. Again we distinguish two cases
\begin{enumerate}
\item $\heartsuit$ satisfies (\ref{ch5:eq:DL1}). Assume $(a_i\in F_i')$ and $(a_j\notin F_j')$. From the inequalities above, it follows that $a_i\in F_i''$ and $a_j\notin F_j''$ and thus $\heartsuit(a_1,\ldots,a_n)\in F$ since $(F_1'',\ldots,F_n'')\in \pi_\heartsuit(\gamma_A(F))$.
\item $\heartsuit$ satisfies (\ref{ch5:eq:DL2}). Assume that $\heartsuit(a_1,\ldots,a_n)\in F$ and that $a_j\in F_j'$. It follows that $a_j\in F_j$ and thus $a_i\in F_i$ and thus $a_i\in F'_i$ as desired.
\end{enumerate}

With all these checks in place, we now move to the difficult part of the proof: showing that $\gamma_A$ is a right inverse of $\hat{\delta}_A$. For this it must be the case that $\heartsuit(a_1,\ldots,a_n)\in F\Leftrightarrow\heartsuit(a_1,\ldots,a_n)\in \hat{\delta}_A(\gamma_A(F))$, i.e. by unravelling the definitions:
\begin{align}
\heartsuit(a_1,\ldots,a_n)\in F \Leftrightarrow &\exists (F_1,\ldots,F_n)\in \pi_\heartsuit(\gamma_A(F)) \text{ s.th. } \label{ch5:eq:iffDL1} \\
&(a_i\in F_i\text{ for all i})\text{ and } (a_j\notin F_j, \text{ for all j} )\nonumber \\
&\text{if }\heartsuit\text{ satisfies Eq. (\ref{ch5:eq:DL1})} \nonumber \\
\heartsuit(a_1,\ldots,a_n)\in F \Leftrightarrow &\forall (F_1,\ldots,F_n)\in \pi_\heartsuit(\gamma_A(F)) \text{ s.th. }  \label{ch5:eq:iffDL2}\\
&(a_j\in F_j, \text{ for all j})\Rightarrow (a_i\in F_i\text{ for all i}) \nonumber\\
&\text{if }\heartsuit\text{ satisfies Eq. (\ref{ch5:eq:DL2})} \nonumber
\end{align}
It is easy to see that if $\heartsuit$ satisfies Eq. (\ref{ch5:eq:DL1}), then the right-to-left direction of Eq. (\ref{ch5:eq:iffDL1}) follows immediately from the definition of $\gamma_A$. Similarly, if  $\heartsuit$ satisfies Eq. (\ref{ch5:eq:DL2}), then the left-to-right direction of Eq. (\ref{ch5:eq:iffDL2}) follows from the definition. The difficult part of the proof are the converse implications. 

\noindent \textbf{Left-to-right of Eq. (\ref{ch5:eq:iffDL1})}  Assume that $\heartsuit(a_1,\ldots,a_n)\in F$, we must build an $n$-tuple $(F_1,\ldots,F_n)\in\pi_\heartsuit(\gamma_A(F))$ of prime filters such that $a_i\in F_i$ for every isotone $i$, and $a_j\notin F_j$ for every antitone $j$. Since this tuple must belong to $\pi_\heartsuit(\gamma_A(F))$ it should also have the property that if $(a_i'\in F_i)$ for all $i$ and $(a_j'\notin F_j)$ for all $j$ then $\heartsuit(a_1',\ldots,a_n')\in F$, or equivalently that if $\heartsuit(a_1',\ldots,a_n')\notin F$ then there must exist $i$ with $a'_i\notin F_i$ or $j$ with $a'_j\in F_j$. To achieve this we use a Prime Filter Theorem argument on a set of ideal-filter pairs $((F_1,I_1),\ldots,(F_n,I_n))$ which we will call $\mathscr{P}(a_1,\ldots,a_n)$ and define as follows. For $1\leq l\leq k$ we put:
\begin{enumerate}[({iso} 1)]
\item $\uparrow a_l\subseteq F_l\subseteq \{c_l\mid (\forall\vect[c]\in\prod_{i\neq l} F_i)(\forall\vect[d]\in\prod_j I_j), \heartsuit(\vect[c/c_l],\vect[d])\in F\}$
\item $I_l=\{c_l\mid (\exists\vect[c]\in \prod_{i\neq l} F_i) (\exists \vect[d]\in \prod_j I_j)\text{ s.th. }\heartsuit(\vect[c/c_l],\vect[d])\notin F\}$
\end{enumerate}

\noindent For $k+1\leq l\leq n$ we put:
\begin{enumerate}[({anti} 1)]
\item $\downarrow a_j\subseteq I_j\subseteq\{d_l\mid (\forall\vect[c]\in\prod_i F_i)(\forall \vect[d]\in\prod_{j\neq l} I_j), \heartsuit(\vect[c],\vect[d/d_l])\in F\}$
\item $F_l=\{d_l\mid (\exists\vect[c]\in\prod_i F_i)(\exists \vect[d]\in\prod_{j\neq l} I_j) \text{ s.th. }\heartsuit(\vect[c],\vect[d/d_l])\notin F\}$
\end{enumerate}
We show the following facts about $\mathscr{P}(a_1,\ldots,a_n)$
\begin{itemize}
\item $((\uparrow a_1,I_1),\ldots,(F_{k+1},\downarrow a_{k+1}),\ldots)$ belongs to  $\mathscr{P}(a_1,\ldots,a_n)$, which is therefore not empty. Since the $I_i$s and $F_j$s are defined from the filters $\uparrow a_i$ and the ideals $\downarrow a_j$, we need only check the inclusions in (iso 1) and (anti 1). For this let $1\leq l\leq k$, $a_l\leq c_l$ and assume for the sake of contradiction that there exists $\vect[c]\in\prod_{i\neq l} \downarrow a_i$ and $\vect[d]\in \prod_j \downarrow a_j$ such that $\heartsuit{\vect[c/c_l],\vect[d]}\notin F$, but then $\heartsuit(\vect[c/c_l],\vect[d])\geq \heartsuit(a_1,\ldots,a_n)\notin F$, a contradiction with the starting assumption. Thus (iso 1) holds, and the proof that (anti 1) holds is dual.
\item $\mathscr{P}(a_1,\ldots,a_n)$ forms a poset under component-wise set inclusion
\item Each $I_l, 1\leq l\leq k$ is an ideal. Assume $c_l\in I_l$ and $c_l'\leq c_l$. By definition, there exists $\vect[c]\in \prod_{i\neq l} F_i$ and $\vect[d]\in \prod_j I_j$ s.th. $\heartsuit(\vect[c/c_l],\vect[d])\notin F$. It follows immediately from the fact that $l$ is an isotone argument and that $F$ is a filter that $\heartsuit(\vect[c/c_l'],\vect[d])\notin F$ and thus $c_{l'}\in I_l$. Now assume that $c_l,c_l'\in I_l$. By definition there exist $\vect[c],\vect[c']\in \prod_{i\neq l} F_i$ and $\vect[d],\vect[d']\in \prod_j I_j$ s.th. $\heartsuit(\vect[c/c_l],\vect[d]), \heartsuit(\vect[c'/c_l'],\vect[d'])\notin F$. It follows that $\heartsuit(\vect[c\wedge c'/c_l],\vect[d]\vee\vect[d']), \heartsuit(\vect[c\wedge c'/c_l'],\vect[d]\vee\vect[d'])\notin F$. Since $F$ is prime and $\heartsuit$ satisfies (\ref{ch5:eq:iffDL1}) it follows that
\begin{align*}
&\heartsuit(\vect[c\wedge c'/c_l],\vect[d]\vee\vect[d'])\vee \heartsuit(\vect[c\wedge c'/c_l'],\vect[d]\vee\vect[d'])\\
=&\heartsuit(\vect[c\wedge c'/c_l\vee c_l'],\vect[d]\vee\vect[d'])\notin F
\end{align*}
and since $F_i$ is a filter and each $I_j$ is an ideal, it follows that $\vect[c\wedge c']\in\prod_{i\neq l}F_i$ and $\vect[d]\vee\vect[d']\in \prod_j I_j$ witness that $c_l\wedge c_l'\in I_l$.
\item For totally dual reasons, $F_l, k+1\leq l\leq n$ is a filter.
\item $F_i\cap I_i=\emptyset=F_j\cap I_j$. For an isotone index $l$ assume that there exist $x_l\in F_l, y_l\in I_l$ such that $x_l\leq y_l$. By definition of $I_l$ there exists $\vect[c]\in \prod_{i\neq l} F_i$ and $\vect[d]\in \prod_j I_j$ s.th. $\heartsuit(\vect[c/y_l],\vect[d])\notin F$. It follows that $\heartsuit(\vect[c/x_l],\vect[d])\notin F$, contradicting the property (iso 1) of $F_l$. The proof for antitone arguments is identical.
\end{itemize}

Let us now check that $\mathscr{P}(a_1,\ldots,a_n)$ has upper bound of chains. Assume $((F_1^k,I_i^k),\ldots,(F_n^k,I_n^k))_{k\in \omega}$ is a chain of elements of $\mathscr{P}(a_1,\ldots,a_n)$, and define
\[
F_l^\infty=\bigcup_k F_l^k\hspace{2em}I_l^\infty=\bigcup_k I_l^k,\hspace{2em}k=1,\ldots,n
\]
It is well-known and easy to check that the union of a chain of filters (resp. ideals) is a filter (esp. ideal). Let us now check that conditions (iso 1)-(iso 2) and (anti 1)-(anti 2) are satisfied.  The first inclusions of (iso 1) and (anti 1) are trivially satisfied. For the second inclusion, let $1\leq l\leq k$, $c_l\in F_l^\infty,$ and let $\vect[c]\in \prod_{i\neq l} F_i^\infty$ and $\vect[d]\in \prod_j I_j^\infty$. By definition each component $c_i$ of $\vect[c]$ can be traced back to $F_i^k$ for a certain $k$, and similarly for each component $d_j$ of $\vect[d]$. By taking the maximum of these $k$s we get an integer $\tilde{k}$ s.th. $\vect[c]\in \prod_{i\neq l} F_i^{\tilde{k}}, \vect[d]\in \prod_j I_j^{\tilde{k}}$. It follows immediately by the condition (iso 1) on $F_l^{\tilde{k}}$ that $\heartsuit(\vect[c/c_l],\vect[d])\in F$ as desired. The proof for $l$ antitone is identical. Finally we need to show that (iso 2) and (anti 2) are satisfied. Let $c_l\in I_l^\infty$ for $l$ isotone. By construction, there must exist $k$ such that $c_l\in I_l^k$, from which it follows that there exist $\vect[c]\in \prod_{i\neq l} F_i^k$ and $\vect[d]\in \prod_j I_j^k$ s.th. $\heartsuit(\vect[c/c_l],\vect[d])\notin F$. By definition it follows immediately that there exist $\vect[c]\in \prod_{i\neq l} F_i^\infty$ and $\vect[d]\in \prod_j I_j^\infty$ s.th. $\heartsuit(\vect[c/c_l],\vect[d])\notin F$, i.e. we have shown the inclusion from left to right in (iso 2). For the opposite direction, assume that there exist $\vect[c]\in \prod_{i\neq l} F_i^\infty$ and $\vect[d]\in \prod_j I_j^\infty$ s.th. $\heartsuit(\vect[c/c_l],\vect[d])\notin F$. Once again, each component of $\vect[c]$ and of $\vect[d]$ can be traced back to a filter or an ideal indexed by a finite integer $k$. By taking $\tilde{k}$ to be the maximum over all these integers, we can assume that $\vect[c]\in \prod_{i\neq l} F_i^{\tilde{k}}$ and $\vect[d]\in \prod_j I_j^{\tilde{k}}$, and thus $c_l\in I^{\tilde{k}}_l$, and thus $c_l\in I^\infty_l$, which concludes the proof that (iso 2) holds. The proof that (anti 2) holds is identical.

Since $\mathscr{P}(a_1,\ldots,a_n)$ is closed upper bound of chains, we can apply Zorn's lemma and get the existence of a maximal element of $\mathscr{P}(a_1,\ldots,a_n)$ which we call
\[
((G_1,J_1),\ldots,(G_n,J_n))
\]
We claim that $G_1,\ldots,G_n$ are the prime filters we are looking for. Let us first check that they are prime. For isotone arguments, let assume that $c_l\vee c_l'\in G_l$ but that $c_l, c_l'\notin G_l$. It follows that
\begin{align*}
&((G_1,J_1),\ldots, (G_l, J_l), \ldots,(G_n,J_n))\subsetneq\\
&((G_1,J_1'),\ldots,(\langle G_l\cup\{c_l\}\rangle, J_l),(G_{l+1},J'_{l+1}),\ldots,(G_{k+1}',J_{k+1}),\ldots,(G_n',J_n))
\end{align*}  
where $\langle G_l\cup\{c_l\}\rangle$, the filter generated by $G_l\cup\{c_l\}$, and for each isotone index $i'$, $J'_{i'}$ is defined by (iso 2) by allowing components in this filter. Explicitly:
\[
J'_{i'}=\{c_{i'}\mid \hspace{-1pt}(\exists \vect[c]\hspace{-2pt}\in\hspace{-1em} \prod_{i<l, i\neq i'}\hspace{-1em} G_i\times \langle G_l\cup\{c_l\}\rangle\times\hspace{-1em}\prod_{i>l, i\neq i'}\hspace{-10pt} G_i)(\exists\vect[d]\hspace{-2pt}\in\hspace{-1pt}\prod_j J_j)\text{ s.th. }\heartsuit(\vect[c/c_{i'}],\vect[d])\notin F\}
\]
Similarly, for $j'$ antitone, we make sure that (anti 2) is respected and define:
\[
G_{j'}=\{d_{j'}\mid \hspace{-1pt}\exists \vect[c]\hspace{-1pt}\in\hspace{-1pt}\prod_{i< l}\hspace{-2pt}G_i\times\langle G_l\cup\{c_l\}\rangle\times \hspace{-1pt}\prod_{i>l} G_i)(\exists \hspace{-1pt}\vect[d]\hspace{-2pt}\in\hspace{-4pt} \prod_{j\neq j'} \hspace{-1pt}J_j)\text{ s.th. }\heartsuit(\vect[c],\vect[d/d_{j'}])\notin F\}
\]
Since $(G_1,J_1,\ldots,(G_n,J_n))$ is maximal in $\mathscr{P}(a_1,\ldots,a_n)$, the new $n$-tuple of filter-ideal pairs cannot be in $\mathscr{P}(a_1,\ldots,a_n)$. Since (iso 2) and (anti 2) are satisfied by definition, and since the first inclusion of (iso 1) and (anti 1) are also necessarily satisfied, for the new $n$-tuple not to be in $\mathscr{P}(a_1,\ldots,a_n)$, it must be the case that the second inclusion of (iso 1) or of (anti 1) must be violated. Since all $G_i$s are unchanged apart from $G_l$, and since every $J_j$ is unchanged, it is not too hard to see that the violations of either inclusions are equivalent and happen precisely if there exists $(x_1,\ldots, x_l\wedge c_l, \ldots, x_k)\in \prod_{i<l}G_i\times\langle G_l\cup\{c_l\}\times \prod_{i>l} G_i$ and a $\vect[d]\in\prod_j J_j$ such that
\[
\heartsuit(x_1,\ldots,x_l\wedge c_l,\ldots,x_k,\vect[d])\notin F
\]  
By definition, this means that $x_l\wedge c_l\in J'_{l}$, i.e. there exist $y_l\in J'_{l}$ s.th. $x_l\wedge c_l\leq y_l$. A completely analogous argument shows that there must exist $(x_1',\ldots, x_l'\wedge c_l', \ldots, x_k')\in \prod_{i<l}G_i\times\langle G_l\cup\{c_l\}\times \prod_{i>l} G_i$ and $y_l'\in J'_l$ s.th. $x_l'\wedge c_l'\leq y_l'$. By combining the two inequalities we get
\[
(x_l\wedge c_l)\vee (x_l'\wedge c_l')\leq y_l\vee y_l'
\]
i.e.
\[
(x_l\vee x_l')\wedge(x_l\vee c_l')\wedge(c_l\vee x_l')\wedge(c_l\vee c_l')\leq y_l\vee y_l'
\] 
since $J'_l$ is an ideal, $G_l$ is a filter, $x_l,x_l'\in G_l$ and $c_l\vee c_l'\in G_l$ is assumed, we get $G_l\cap J'_l\neq\emptyset$ which is easily seen to be a contradiction by construction of $J'_l$. Thus we cannot have both $c_l\notin G_l$ and $c_l'\notin G_l$, and $G_l$ is thus prime. A completely dual proof shows that each $J_j$ is prime. It follows that $J_i$ and $G_j$ are prime too. We show this for $J_l$, $l$ isotone, and a completely dual argument shows that each $G_j$ is prime too. If $c_l\wedge c_l'\in J_l$, then by definition there exists $\vect[c]\in\prod_{i\neq l} G_i, \vect[d]\in\prod_j J_j$ s.th. $\heartsuit(\vect[c/c_l\wedge c_l'],\vect[d])\notin F$. Assume now for the sake of contradiction that $c_l,c_l'\notin J_l$. Then they must both belong to $G_l$ since $G_l$ is maximal, and not being in $J_l$ precisely ensures that the second inequality of (iso 1) is respected. But if $c_l,c_l'\in G_l$, then $c_l\wedge c_l'\in G_l$ since it's a filter, which contradicts the property established above that $G_l\cap J_l=\emptyset$. Thus either $c_l$ or $c_l'$ belongs to $J_l$ which is thus prime.

Finally, we need to check that $((G_1,J_1),\ldots,(G_n,J_n))\in \pi_\heartsuit(\gamma_A(F)))$, i.e. that if $\heartsuit(a'_1,\ldots,a'_n)\notin F$ then there must exist either an isotone $i$
with $a_i'\notin G_i$ or an antitone $j$ with $a_j'\in G_j$. Assume that $a_i'\in G_i$ apart from a single isotone index $l$ and that $a'_j\notin G_j$ for every antitone index $j$, we must show that $a_l\notin G_l$. By definition (iso 2), the $k$-tuple $(a_i')_{i\neq l}$ and the $n-k$-tuple $(a_j)$ clearly witness the fact that $a_l\in J_l$, and since $G_l\cap J_l=\emptyset$ it follows that $a_l\notin G_l$ as desired. Similarly, if $a_i'\in G_i$ for each isotone index and $a_j'\notin G_j$ apart from a single antitone index $l$, then $a_l\in G_l$ by definition (anti 2).

\noindent \textbf{Right-to-left of Eq. (\ref{ch5:eq:iffDL2})}. Is shown in a very similar way as the left-to-right direction of  Eq. (\ref{ch5:eq:iffDL1}) above. We show the contrapositive: if $\heartsuit(a_1,\ldots,a_n)\notin F$, then there must exist $(F_1,\ldots,F_n)\in \pi_\heartsuit(\gamma_A(F))$ s.th. $a_j\in F_j$ for all antitone $j$ but $a_i\notin F_i$ for all isotone $i$ (and thus for a particular $i$). Since $(F_1,\ldots,F_n)\in \pi_\heartsuit(\gamma_A(F))$ we must ensure that if there exist $a'_1,\ldots,a'_n$ s.th. $a_j'\in F_j$ for each $j$ but $a'_i\notin F_i$ for a certain $i$, then $\heartsuit(a_1',\ldots,a_n')\notin F$.  We proceed as above and define a set $\mathscr{P}(a_1,\ldots,a_n)$ of filter-ideal pairs $((F_1,I_1),\ldots,(F_n,I_n))$ as follows. For $1\leq i\leq k$ we put:

\begin{enumerate}
\item $F_i=\{c_i\mid \exists \vect[d]\in \prod_j F_j\text{ s.th. }\heartsuit(c_i,\vect[d])\in F\}$
\item $\downarrow a_i\subseteq I_i\subseteq \{c_i\mid \forall \vect[d]\in \prod_j F_j, \heartsuit(c_i,\vect[d])\notin F\}$
\end{enumerate}

\noindent For $k+1\leq j\leq n$ we put:

\begin{enumerate}
\item $\uparrow a_j\subseteq F_j\subseteq\{d_j\mid \forall \vect[c]\in \prod_i I_i, \heartsuit(\vect[c],d_j)\notin F\}$
\item $I_j=\{d_j\mid \exists\vect[c]\in \prod_i I_i\text{ s.th. } \heartsuit(\vect[c], d_j)\in F\}$ 
\end{enumerate}

The desired tuple of prime filters is obtained in the same way as above by dualization, i.e. by showing that the collection $\mathscr{P}(a_1,\ldots,a_n)$ forms a non-empty poset, that each $F_i$ (resp. $I_j$) is a filter (resp. an ideal), that the intersection of each filter-ideal pair is empty, and that $\mathscr{P}(a_1,\ldots,a_n)$ is closed under union of chains. By applying Zorn's lemma we once again find a maximal element consisting of prime filter-prime ideal pairs which fulfill the desired requirement.

\end{proof}

\subsection{Strong completeness for boolean coalgebraic logics}\label{ch5:subsec:strngcomplBA}

We now turn our attention to the case where $\cat=\BA$ and $\cat[D]=\Set$. The category of sets has a very remarkable property which allows for a reformulation of the coalgebraic J\'{o}nsson-Tarski theorem.
In Theorem 3 of \cite{2005:UfExtCoalg}, the criterion for the existence of a  J\'{o}nsson-Tarski embedding an $L$-algebra $A$ is essentially that of our Theorem \ref{ch5:thm:jontarski}, but is formulated in the following way: there must exist a map $h_A: \uf LA\to T\uf A$ such that for any $u\in \uf LA$, 
\begin{equation}\label{ch5:eq:h}
b\in u\Leftrightarrow h_A(u)\in \delta_{\uf A}\circ L\eta_A \circ h_A(u)
\end{equation}
If such an $h_A$ exists then it is easy to see that $\hat{\delta}_A\inv$ can be substituted by $h_A\inv$ in the Diagram (\ref{ch5:diag:jontar}) of the proof above and preserve commutativity, i.e. $h_A$ is a right inverse to $\hat{\delta}_A$: if we unravel the definition of $\hat{\delta}$ we get for any $u\in\uf LA$:
\begin{align*}
\hat{\delta}_A\circ h_A(u)&=\uf L\eta_A\circ \uf\delta_{\uf A}\circ \epsilon_{T\uf A}\circ h_A(u)\\
&=(\delta_{\uf A}\circ L\eta_A)\inv\circ \epsilon_{T\uf A}(h_A(u))\\
&=(\delta_{\uf A}\circ L\eta_A)\inv(\{U\in \pow T\uf A\mid h_A(u)\in V\})
\end{align*}
If $b\in u$, then by (\ref{ch5:eq:h}) we know that $h_A(u)\in \delta_{\uf A}\circ L\eta_A$, and thus $\delta_{\uf A}\circ L\eta_A(b)$ is one of these $U$s and in particular $b\in \hat{\delta}_A\circ h_A(u)$. Conversely, if $b\in \hat{\delta}_A\circ h_A(u)$, then $\delta_{\uf A}\circ L\eta_A(b)\in \epsilon_{T\uf A}(h(u))$, i.e. $h_A(u)\in \delta_{\uf A}\circ L\eta_A(b)$ and thus $b\in u$ by (\ref{ch5:eq:h}). So we have established that $\hat{\delta}_A\circ h_A=\id_{\uf LA}$, i.e. that $h_A$ is a section of $\hat{\delta}_A$, which is thus a split epi. 

The interesting property of $\Set$ which we will now exploit is that all epimorphisms are split (modulo the axiom of choice). In other words, if $\hat{\delta}_A$ is an epimorphism, then it must have a section, i.e. a right-inverse, and \emph{any} section of $\hat{\delta}_A$ will act as a good $h_A$ and define a J\'{o}nsson-Tarski embedding. Indeed, assume that $s_A: \uf LA\to T\uf A$ is such a section, by reusing the last steps of the proof of the coalgebraic J\'{o}nsson-Tarski Theorem we get:
\begin{align*}
\pow s_A \circ\delta_{\uf A}\circ L\eta_A&=\pow s_A\circ \pow \hat{\delta}_A\circ\eta_{LA}\\
&=\pow(\hat{\delta}_A\circ s_A)\circ \eta_{LA}=\eta_{LA}
\end{align*}
as was shown above and as required for Diagram (\ref{ch5:diag:jontar}) to commute. 

It is thus clear that the requirement that $\hat{\delta}$ be an epi-transformation is enough to embed an $L$-algebra $A$ into its canonical extension $\pow \uf A$ equipped with an $L$-algebra structure. However, this embedding is not uniquely defined and relies completely on the axiom of choice. We therefore call a model obtained in this way \textbf{`quasi-canonical'}\index{Quasi-canonical!embedding}, following the terminology of \cite{2009:DirkStrongComp}, and reserve the term `canonical' for the case where the right inverse to $\hat{\delta}$ can be built explicitly, such as in the case of (positive) relational logics described above. Moreover, the surjectivity of $\hat{\delta}$ only guarantees the existence of pseudo-canonical models for the $\uf\dashv\pow$ adjunction, i.e. for boolean coalgebraic logic. In general not all surjective morphisms are split. Indeed, even in the case of $\Pos$ it is easy to find surjective morphisms which do not split. Thus whilst useful and simple in the boolean case, showing that $\hat{\delta}$ is surjective does not generalize, even to the case of coalgebraic logics based on distributive lattices.

Following the discussion above and the prevalence of boolean coalgebraic logic, we will spend the remainder of this Section investigating  conditions on $L,T$ and $\delta$ which guarantee that $\hat{\delta}$ is an epi-transformation in the context of the adjunction $\uf\dashv\pow:\BA\to\Set$. As we will also show, we can in many cases explicitly build the right inverse of $\hat{\delta}$ when it is also assumed to be a mono-transformation, i.e. when the logic is assumed to be expressive. We first need a few easy technical results.

\begin{lemma}\label{ch5:lem:epitomono}
The functors $\uf$ and $\pow$ turn monos into epis and vice-versa.
\end{lemma}
\begin{proof}
To see that $\uf$ turns epi into monos is easy: since it is a left adjoint it preserves epis, i.e. it sends an epi in $\BA$ to an epi in $\Set\op$, i.e. to a mono in $\Set$. The dual argument shows that $\pow$ sends epis to monos.

To see that $\pow$ sends monos to epis, let $f: X\to Y$ be a mono in $\Set$, and let $U\in\pow A$, we need to find a $V\in\pow B$ such that $U=\pow f V=f\inv V$. Since $f$ is injective we can simply take $V=f[U]$ and we then get $f\inv f[U]=U$ by injectivity.

Let us finally check that $\uf$ also sends monos to epis. Let $f: A\to B$ be a monomorphism in $\BA$, then $f$ is injective on the carriers. To check that $\uf f$ is epi, we need to show that for every $u\in\uf A$, there exists a $v\in \uf B$ such that $u=\uf f(v)=f\inv (v)$. We choose $v$ as the ultrafilter generated by $f[u]$ (this means that we need to assume the axiom of choice) and show that $u=f\inv (v)$. Clearly, if $a\in u$ then $f(a)\in f[u]\subseteq v$, i.e. $u\subseteq f\inv (v)$. Conversely, if we choose $a\in f\inv (v)$, then either $a$ of $\neg a$ must belong to $u$. If $\neg a\in u$, then $f(\neg a)\in v$, i.e. $\neg f(a)\in v$. But since $f(a)$ is in $v$ by assumption that $a\in f\inv (v)$, and since $v$ is an ultrafilter, we must have $f(a)=f(\neg a)$ to avoid a contradiction. But since $f$ is injective this means that $a=\neg a$ which is impossible. Thus it must be the case that $a\in u$ and we thus have $f\inv (v)\subseteq u$, and therefore $u=f\inv (v)$.
\end{proof}

\begin{lemma}\label{ch5:lem:unitcounit}
The unit and counit $\eta,\epsilon$ and of the adjunction $\uf\dashv\pow: \BA\to\Set$ are isomorphisms on finite boolean algebras and finite sets respectively.
\end{lemma}
\begin{proof}
Let $A$ be a finite boolean algebra, then $\pow\uf A$ is also finite. Recall that $\eta_A: A\to \pow\uf A, a\mapsto\{u\in\uf A\mid a\in u\}$ is a monomorphism and $\Forg\eta_A$ is injective. Since every ultrafilter in a finite boolean algebra is principal, an element $U$ of $\pow\uf A$ is a finite set of principal ultrafilter and is thus in one-to-one correspondence with a finite set $\{a_1,\ldots,a_n\}$ of elements of $A$. We can therefore define for any finite $A$
\[
h_A: \pow\uf A\to A, \{\up a_1,\ldots,\up a_n\}\mapsto a_1\vee\ldots\vee a_n
\]
It is then easy to see that 
\[
\eta_A\circ h_A(U)=\eta_A(a_1\vee\ldots\vee a_n)=\bigcup_{i=1}^n\eta_A(a_i)=U
\]
and $h_A$ is thus a right inverse of $\eta_A$. Similarly, it is easy to check that $h_A\circ \eta_A(a)=a$ for any $a\in A$, and thus $h_A$ is also a left inverse and $\eta_A$ is therefore iso.

For the counit, let $X$  be a finite set, and recall that $\epsilon_X: X\to\uf\pow X, x\mapsto\{U\subseteq X\mid x\in U\}=\uparrow\{x\}$. This map is clearly injective. To see that it is surjective, pick $u\in \uf\pow X$, and consider the set $U=\bigcap u$. By the properties of ultrafilters we automatically get that $U\neq\emptyset$ and $u=\uparrow U$. To see that $U$ is in fact a singleton, assume $\{x,y\}\subseteq U, x\neq y$. Since $u$ is an ultrafilter, either $\{x\}\in u$ or $\{x\}^c\in u$. If $\{x\}^c\in u$, then clearly $x$ cannot be an element of $U$, so it must in fact be the case that $\{x\}\in u$, but then we clearly cannot have $\{x,y\}\subseteq U$ and thus it must be the case that $U=\{x\}$. This shows that $u=\uparrow\{x\}=\epsilon_X(x)$ for some $x$.
\end{proof}

From these lemmas, we can immediately show that if $\hat{\delta}$ is surjective at each stage $A$, then $\delta$ is injective and the logic is (weakly) complete as we have seen in Section \ref{ch5:sec:weakComp}.

\begin{proposition}
If $L:\BA\to\BA$ preserves monos (in particular if it preserves weak pullbacks) and $\hat{\delta}:T\uf\to\uf L$ is surjective, then $\delta: L\pow \to\pow T$ is injective.
\end{proposition}
\begin{proof}
By using the same argument as in the proof of Theorem \ref{ch5:thm:jontarski}, we know that the following diagram commutes:
\[
\xymatrix
{
LA\ar[r]^{L\eta_A}\ar[d]_{\eta_{LA}} & L\pow\uf A\ar[r]^{\delta_{\uf A}} & \pow T \uf A \\
\pow\uf LA\ar[r]_{\pow \uf L\eta_A}\ar[urr]^{\pow\hat{\delta}_A} & \pow \uf L\pow \uf A\ar[r]_{\pow \uf \delta_{\uf A}} & \pow \uf \pow T \uf A\ar[u]_{\pow\epsilon_{T\uf A}}
}
\]
By Lemma \ref{ch5:lem:epitomono}, since $\hat{\delta}_A$ is epi (in $\Set$) $\pow\hat{\delta}_A$ must be mono (in $\BA$). Moreover, since the unit $\eta:\Id\to\pow\uf$ is a mono-transformation and $L$ preserves monos, we must have $\delta_{\uf A}$ mono.
\end{proof}

\begin{lemma}\label{ch5:lem:delatdeltahatfin}
Let $L:\BA\to\BA$ and let $T:\Set\to\Set$ be two functors related by a semantic transformation $\delta:L\pow\to\pow T$, and assume further that $T$ preserve finite sets. The adjoint transformation $\hat{\delta}: T\uf\to \uf L$ is injective (resp. surjective) on finite objects iff $\delta$ is surjective (resp. injective) on finite objects.
\end{lemma}
\begin{proof}
This follows by the definition of the adjoint transformation by
\[
\hat{\delta}_A=\uf L\eta_A\circ \uf\delta_{\uf A}\circ \epsilon_{T\uf A}
\]
Clearly, if $A$ is a finite boolean algebra, then by Lemma \ref{ch5:lem:unitcounit} and the assumption on $T$, $\uf L\eta_A$ and $\epsilon_{T\uf A}$ are isomorphisms. The result then follows easily from Lemma \ref{ch5:lem:epitomono}.
\end{proof}

The following technical lemma is not strictly necessary, but it makes the proof of Theorem \ref{ch5:thm:deltahatsurj} more symmetric, and it also provides an important example of functor weakly preserving cofiltered limits.

\begin{lemma} \label{ch5:lem:powcofiltered}
The covariant powerset functor $\mathsf{P}:\Set\to\Set$ weakly preserves cofiltered limits.
\end{lemma}
\begin{proof}
Let $(X_i)_{i\in I}$ be a cofiltered diagram in $\Set$, let $X=\lim_i X_i$ and let $\pi_i: X\to X_i$. The limit $X$ is the set of coherent collection of elements $(x_i)_{i\in I}$, where coherent means that if $x_i\in X_i$ and $f_{ij}: X_i\to X_j$, then $x_j=f_{ij}(x_i)$. 

We need to show that $\mathsf{P} \lim_i X_i$ is a weak limit for the diagram $(\mathsf{P}X_i)_{i\in I}$. It is clear that by functoriality of $\mathcal{Q}$, it is a cone for this diagram. Given a coherent collection of subsets $U_i\in\mathsf{P} X_i$, i.e. a collection of subsets such that $f_{ij}[U_i]=U_j$ for any $f_{ij}: X_i\to X_j$, we need to find an element of $\mathsf{P} X=\mathsf{P}\lim_i X_i$ mapping onto the various elements of the collection. For this we can simply take $U=\{(x_i)_{i\in I}\in \lim_i X_i\mid x_i\in U_i\}$. For $U$ to be a good choice, it must be the case that $\mathsf{P}\pi_i(U)=\pi_i[U]=U_i$. Consider any $y_i\in U_i$, then we must find in any $U_j\in\mathsf{P}X_j$, an element $y_j$ belonging to a coherent collection containing $y_i$. Since the diagram is cofiltered, there exist an $X_k$, a $U_k\in\mathsf{P}X_k$, and morphisms $f_{ki}: X_k\to X_i$  and $f_{kj}: X_k\to X_j$ such that $f_{ki}[U_k]=U_i$ and $f_{kj}[U_k]=U_j$. In particular, there must exist $y_k$ such that $f_{ki}(y_k)=y_i$ and $f_{kj}(y_k)=y_j$, and since this holds for any $U_j$, $y_i$ must belong to a coherent family in $U$, i.e. $\pi_i[U]=U_i$ as required.
\end{proof}

Another important example, which will be crucial in the next subsection will be the following (the proof is very similar to the one just given):

\begin{lemma}\label{ch5:lem:homcofiltered}
Let $n$ be a finite set, then contravariant $\hom$ functor $\hom(-,n):\Set\op\to\Set$ weakly preserves cofiltered limits (i.e. turns cofiltered limits into weak filtered colimits).
\end{lemma}
\begin{proof}
Let $X=\lim_i X_i$ be the limit of a cofiltered diagram, and let us consider a collection of maps $f_i\in\hom(X_i,n)$ which is coherent in the sense that if $h_{ij}: X_i\to X_j$ is the image of a morphism from the index category, then 
\[
\hom(-, n)(f_j)=f_j\circ h_{ij}=f_i
\]
We need to show that we can find a (not necessarily unique) element $f\in \hom(X, n)$ such that $\hom(-,n)(\pi_i)(f_i)=f$ for all the projection maps $\pi_i: X\to X_i$. Let us define
\[
f: X\to n, x\mapsto \begin{cases}y & \text{ if }\forall i, f_i(\pi_i(x))=y\\
0 & \text{ else}
\end{cases}
\]
Let us show that for any $x\in X$ we do have $f(x)=f_i(\pi_i(x))$. For this we need to show that if $f_i(\pi_i(x))=y$ for a particular $i$, then $f_j(\pi_j(x))=y$ for any other $j$ of the index category. For this we use the fact that the diagram is cofiltered, i.e. given $i,j$, there must exist a $k$ and morphisms $h_{ki}: X_k\to X_i$ and $h_{kj}: X_k\to X_j$. By a proof completely analogous to that of Proposition \ref{ch5:prop:surjenough}, we can assume without loss of generality that all morphisms $h_{ki}$ are surjective, i.e. $x_i\in X_i$ must have a pre-image $x_k$ under the map $h_{ki}$. By the fact that $f_k=f_i\circ h_{ki}$ we have
\[
f_k(x_k)=f_i\circ h_{ki}(x_k)=f_i(x_i)=y
\]
and thus
\[
y=f_k(x_k)=f_j\circ h_{kj}(x_k)=f_j(x_j)
\]
Thus we indeed have that if $f_i(\pi_i(x))=y$ then $f_j(\pi_j(x))=y$ too, for any index object $j$. This establishes $f(x)=f_i(\pi_i(x))$, and proves that $x$ defined as above does indeed define a weak colimit.
\end{proof}

Note the duality with the notion of finitely presentable object: in $\Set$ an object $X$ is finitely presentable iff $\hom(X,-)$ preserves filtered colimits iff $X$ is finite. Here we have that finite sets are also `finitely co-presentable'.

\begin{corollary}\label{ch5:cor:contravPow}
The contravariant powerset functors $\mathcal{Q}:\Set\op\to\Set$ weakly preserve cofiltered limits.
\end{corollary}
\begin{proof}
This is a simple consequence of the fact that $\mathcal{Q}=\hom(-,2)$.
\end{proof}

The following characterisation of situations where completeness implies strong completeness is due to \cite{2012:KurzStrongComp}, but here we decouple $L$ from $T$, whereas \textit{op.cit.} considers the case where the syntax functor $L$ is defined as $L_T=\pow T\uf$ (which is similar in spirit to the nabla logic case where $L_T=\Free T\Forg$). We start by focusing on the the epi nature of $\hat{\delta}$. Recall from the remark above that this will imply the existence of \emph{quasi}-canonical models.

\begin{theorem}\label{ch5:thm:deltahatsurj}
Let $L:\BA\to\BA$ and let $T:\Set\to\Set$ preserve finite sets (i.e. if $X$ is finite, so is $TX$), and assume further that $L$ is finitary and that $T$ weakly preserves cofiltered limits. If the semantic transformation $\delta: L\pow \to\pow T$ is a mono-transformation, then its transpose $\hat{\delta}:T\uf \to \uf L$ is an epi-transformation.
\end{theorem}
\begin{proof}
The idea of the proof comes from Section 6 of \cite{2012:KurzStrongComp}. Since $\BA$ is locally finitely presentable, we can write any object $A$ of $\BA$ as a filtered colimit $\colim_{i\in I} A_i$ where each $A_i$ is finitely presentable. Moreover, since $\BA$ is a locally finite variety and finitely generated boolean algebras are finite, each $A_i$ is \emph{finite}.

Recall that $\hat{\delta}= \uf L\eta\circ \uf\delta_{\uf}\circ \epsilon_{ T\uf}$. We examine each step of this composition starting from the last, in order to find an inverse image to an arbitrary element $u\in \uf LA$. Consider first:
\begin{equation}\label{ch5:diag:step3}
\xymatrix
{
& \uf L\pow \uf A_i\ar[dd]\ar[rr]^{\uf L \eta_{A_i}}_{\simeq} & & \uf L A_i\ar[dd]\\
\uf L\pow \uf A\ar[rr]^(0.3){\uf L \eta_A}\ar[dr]\ar[ur] & & \uf L A\ar[ur]\ar[dr]\\
& \uf L\pow \uf A_j\ar[rr]^{\uf L \eta_{A_j}}_{\simeq}  & & \uf L A_j
}
\end{equation}
Since $A=\lim_i A_i$, $L$ is finitary and $\uf$ if left adjoint, we have $\uf LA=\uf L\colim_i A_i=\lim_i \uf LA_i$. Thus $u\in \uf LA$ is uniquely associated with a coherent collection $u_i\in \uf LA_i, i\in I$. Since $L$ and $\uf$ preserve finite objects, it follows from Lemma \ref{ch5:lem:unitcounit} that each $\uf L\eta_{A_i}$ is an iso. We can therefore map the collection $(u_i)_{i\in I}$ to a coherent collection $(v_i)_{i\in I}, v_i\in \uf L\pow\uf A_i$. By using the aforementioned properties of $\uf$ and $L$ and Corollary \ref{ch5:cor:contravPow}, we get that $\uf L\pow\uf A$ is a weak limit of the diagram $(\uf L\pow\uf A_i)_{i\in I}$, and thus the collection $(v_i)_{i\in I}$ defines a (not necessarily unique) element $v\in \uf L\pow\uf A$. Since each diagram of the shape of (\ref{ch5:diag:step3}) commutes it is clear that $\uf L\eta_A(v)=u$.

\noindent For the second step we have:
\begin{equation}\label{ch5:diag:step2}
\xymatrix
{
 & \uf\pow T\uf A_i\ar[dd]\ar@{->>}[rr]^{\uf\delta_{\uf A_i}} & & \uf L\pow \uf A_i\ar[dd]\\
\uf\pow T\uf A\ar@{->>}[rr]^(0.3){\uf\delta_{\uf A}}\ar[dr]\ar[ur] & & \uf L\pow \uf A\ar[dr]\ar[ur]\\
 & \uf\pow T\uf A_i\ar@{->>}[rr]^{\uf\delta_{\uf A_i}}  & & \uf L\pow \uf A_j
}
\end{equation}
By Lemma \ref{ch5:lem:epitomono}, since $\delta_{\uf A}$ is assumed to be mono,  $\uf\delta_{\uf A}$ is epi for any $A$, and we can thus find an inverse image $w$ such that $\uf\delta_{\uf A}(w)=v$. By projecting onto the various $\uf \pow T\uf A_i$ this $w$ defines a coherent family $(w_i)_{i\in I}, w_i\in \uf\pow T\uf A_i$. 

\noindent Finally, at the last step we have:
\begin{equation}\label{ch5:diag:step1}
\xymatrix
{
& T\uf A_i\ar[dd]\ar[rr]^{\epsilon_{T\uf A_i}}_{\simeq} & & \uf\pow T\uf A_i\ar[dd] \\
T\uf A\ar[rr]^(0.35){\epsilon_{T\uf A}}\ar[ur]\ar[dr] & & \uf\pow T\uf A\ar[dr]\ar[ur]\\
& T\uf A_j\ar[rr]^{\epsilon_{T\uf A_i}}_{\simeq} & & \uf\pow T\uf A_i
}
\end{equation}
Since $\uf$ and $T$ preserve finite objects, each $T\uf A_i$ is finite, and by Lemma \ref{ch5:lem:unitcounit}, each $\epsilon_{T\uf A_i}$ is an isomorphism. The coherent family $(w_i)_{i\in I}$ thus defines a coherent family $x_i\in T\uf A_i, i\in I$. Since $T$ weakly preserves cofiltered limits and $\uf$ preserves filtered colimits, $T\uf A$ is a weak limit of the diagram $(T\uf A_i)_{i\in I}$, and the coherent collection $(x_i)_{i\in I}$ therefore defines a (not necessarily unique) element $x\in \uf TA$.  Since the each diagram of the shape of (\ref{ch5:diag:step3}) commutes it is clear that $\epsilon_{T\uf A}(x)=w$. By combining our results, it is clear that 
\[
\hat{\delta}_A(x)=\uf L\eta_A(\uf\delta_{\uf A}(\epsilon_{ T\uf A}(x)))=u
\]
i.e. we have shown that $\hat{\delta}_A$ is surjective.
\end{proof}

We combine the previous Proposition \ref{ch5:thm:deltahatsurj} with the assumption of expressivity to build a natural right inverse to $\hat{\delta}$, and get the J\'{o}nsson-Tarski theorem \ref{ch5:thm:jontarski} (as well as strong completeness).

\begin{theorem}\label{ch5:thm:strgcomp1}
Let $L:\BA\to\BA$ be finitary, let $T:\Set\to\Set$ preserve finite sets, and assume further that $T$ weakly preserves cofiltered limits. If the semantic transformation $\delta: L\pow \to\pow T$ is injective, and the logic defined by $L$ is expressive w.r.t. $\Coalg(T)$, then the J\'{o}nsson-Tarski Theorem \ref{ch5:thm:jontarski} holds.
\end{theorem}
\begin{proof}
We want to show that the construction detailed in the proof of Proposition \ref{ch5:thm:deltahatsurj} defines a right inverse to $\hat{\delta}$.

We use the same idea as in the previous proof, but we choose a particular representation of boolean algebras. For this let $f: A\to B$ be a morphism in $\BA$, and let $\Diag_c^A$ and $\Diag_c^B$ be the canonical diagrams of $A$ and $B$ respectively. Recall from Chapter 3, that the canonical diagram for an object $A$ in a locally presentable category (such as $\BA$) is the diagram defined by the projection functor $\Diag_c^A:\cat_{\lambda}\downarrow A\to C$ (where $\cat_\lambda$ is the subcategory of $\lambda$-presentable objects), recall also from Lemma \ref{ch3:lem:candiag} that $A=\colim \Diag_c^A$. Note that if $f:A\to B$, then any object $A_i$ of the canonical diagram $\Diag_c^A$ is also an object in the canonical diagram $\Diag_c^B$ satisfying
\[
\xymatrix
{
A_i\ar[d]_{h_i}\ar[dr]^{g_i=f\circ h_i}\\
A\ar[r]_{f} & B 
}
\]
In consequence, by applying $\uf L$, we have commutative diagrams between the corresponding cofiltered limits of the type:
\[
\xymatrix
{
\uf LA\ar[d]_{\uf Lh_i}\ar[dr]_{\uf L h_j} & & \uf LB\ar[d]^{\uf L g_k}\ar[ll]_{\uf L f}\ar[dl]^{\uf L g_j}\ar[dll]_{\uf L g_i}\\
\uf LA_i & \uf LA_j & \uf LA_k
}
\]
and it is clear that every coherent family in $\uf LB=\lim \uf L\Diag_c^B$ gets mapped to a coherent family in $\uf LA=\lim \uf L\Diag_c^A$.

We have shown in Theorem \ref{ch5:thm:deltahatsurj} that by lifting each element of a coherent family $v=(v_i)_{i\in \{\BA_\omega\downarrow B\}\ob}$ (with $v_i\in \uf LA_i$) in $\uf LB$ against the corresponding maps $\uf L\eta_{A_i}\circ \uf\delta_{\uf A_i}\circ \epsilon_{T\uf A_i}$, we can find a coherent family $(t_i)_{i\in \{\BA_\omega\downarrow B\}\ob}$ with $t_i\in T\uf A_i$ and thus an element $t_v\in T\uf B$ such that $\hat{\delta}_B(t_v)=v$. Since we are assuming that $\hat{\delta}$ is injective, this preimage is unique. The naturality of the maps $v\mapsto t_v$ then follows from the connection between $\Diag_c^A$ and $\Diag_c^B$ and thus between the unique coherent families representing the preimages.
\end{proof}

The same proof can be used when the logic is not necessarily expressive, i.e. when $\hat{\delta}$ is not necessarily injective, but we must then require the surjectvity of $\delta$ on finite structures.

\begin{theorem}\label{ch5:thm:strgcomp2}
Let $L:\BA\to\BA$ be finitary, let $T:\Set\to\Set$ preserve finite sets, and assume further that $T$ weakly preserves cofiltered limits. If the semantic transformation $\delta: L\pow \to\pow T$ is injective and surjective on finite sets, then the J\'{o}nsson-Tarski Theorem \ref{ch5:thm:jontarski} holds.
\end{theorem}
\begin{proof}
The proof is similar to the proof of the preceding Theorem \ref{ch5:thm:strgcomp1}. Since $T$ preserves finite sets, and since $\delta, \eta$ and $\epsilon$ are iso on finite sets, it is easy to see that $\hat{\delta}$ is iso on finite sets. By fixing a filtered diagram $(A_i)_{i\in I}$ representing $A$ and using the fact that $L$ is finitary and that $\BA$ is locally finite, we construct the following diagrams, simplifying the diagrams (\ref{ch5:diag:step1})-(\ref{ch5:diag:step3}):
\[
\xymatrix
{
& T\uf A_i\ar[r]^{\hat{\delta}_{A_i}}_\simeq\ar[dd] & \uf L A_i\ar[dd]\\
T\uf A\hspace{1ex}\ar[rrr]^{\hat{\delta}_A}\ar[ur]\ar[dr] & & & \uf L A\ar[dl]\ar[ul]\\
& T\uf A_j\ar[r]_{\hat{\delta}_{A_j}}^\simeq & \uf L A_j
}
\]
with each $\hat{\delta}_{A_i}$ an isomorphism. Starting from an element $t$ of $\uf LA$, which is the limit of the cofiltered diagram $\uf LA_i\to\uf LA_j$,  and using the fact that $T$ weakly preserves cofiltered limits, it is clear that we can find an inverse image under $\hat{\delta}_A$ of $t$.
\end{proof}

Unfortunately, there are many interesting examples of functors $T$ which do \emph{not} preserve finite sets, such as the bag functor or the finite probability distribution functor. This question is addressed in \cite{2009:DirkStrongComp} via the notion of \textbf{strong completeness over finite algebras}\index{Strong completeness!over finite algebras}. Formally, given a semantic transformation $\delta:L\pow\to\pow T$, the logic defined by $L$ is strongly complete over finite algebras if $\hat{\delta}_A:T\uf A\to\uf LA$ is an isomorphism whenever $A$ is finite. Note that for functors $T$ \emph{not} preserving finite sets, this is not equivalent to asking for $\delta$ to be an iso on finite \emph{sets} (which is what we required in the previous Theorem). Intuitively, strong completeness over finite algebras allows to approximate the full logic by considering its finite fragments.

\begin{theorem}\label{ch5:thm:strgcomp3}
Let $L:\BA\to\BA$ be finitary, let $T:\Set\to\Set$ weakly preserve cofiltered limits, and let $\delta:L\pow\to\pow T$ be a semantic transformation such $\hat{\delta}$ is an iso on finite boolean algebras, i.e. the logic defined by $L$ is strongly complete over finite algebras then the J\'{o}nsson-Tarski \ref{ch5:thm:jontarski} Theorem holds.
\end{theorem}
\begin{proof}
We again proceed as in Proposition \ref{ch5:thm:deltahatsurj}, i.e. we use the fact that $L$ is finitary and that $\BA$ is locally finitely presentable and write any $A$ in $\BA$ as a filtered colimit $A=\colim_{i\in I} A_i$. We then have the following diagram, simplifying the diagrams (\ref{ch5:diag:step1})-(\ref{ch5:diag:step3}), for any $A_i\to A_j$ in the filtered diagram representing $A$:
\[
\xymatrix
{
& T\uf A_i\ar[r]^{\hat{\delta}_{A_i}}_\simeq\ar[dd] & \uf L A_i\ar[dd]\\
T\uf A\hspace{1ex}\ar[rrr]^{\hat{\delta}_A}\ar[ur]\ar[dr] & & & \uf L A\ar[dl]\ar[ul]\\
& T\uf A_j\ar[r]_{\hat{\delta}_{A_j}}^\simeq & \uf L A_j
}
\]
Since $\uf LA=\lim_i\uf LA_i$, every element of $u\in \uf LA$ defines a coherent family $(u_i)_{i\in I}, u_i\in \uf L A_i$. By the fact that $\hat{\delta}_{A_i}$ is iso (since each $A_i$ is finite) and that the above diagram commutes, this coherent family defines a coherent family $(t_i)_{i\in I}$ with $t_i\in T\uf A_i$, and thence an element $t\in \uf TA$ as it is a weak limit of the diagram $(\uf TA_i)_{i\in I}$. By commutativity of the diagram $\hat{\delta}(t)=u$, and thus $\hat{\delta}$ is surjective.
\end{proof}

Let us show a few examples where the Theorems developed above prove strong completeness. 

\begin{example}[A modal logic for trees with at most $n$ successors]\label{ch5:ex:treef}
Consider the functor $\polyFunc[n]:\Set\to\Set$ defined for any $f: X\to Y$ in $\Set$ by
\begin{align*}
\polyFunc[n] X & =\coprod_{i=0}^n X^i\\
\polyFunc[n] f & =[!+f+\ldots+f^n]
\end{align*}
where $X^0=1$ the terminal object in $\Set$, $!: 1\to 1$, $f^k$ is the $k$-fold product of $f$, and $[.+\ldots+.]$ denotes the obvious coproduct of maps. We will write $\mathrm{in}_k$ for the injections $X^k\to \polyFunc[n] X, 1\leq k\leq n$.

We consider the nabla language $\lang_{\polyFunc[n]}$ for this functor which is the initial algebra for the functor $\Free\polyFunc[n]\Forg(-)+\Free V$ (see Section 1.5.1 and \cite{2013:self} for more details), and the language is thus equivalent to the free BAE defined by a single $k$-ary operator $\langle k\rangle$ for each $1\leq k\leq n$. Due to the particularly simple shape of the functor, the axioms of the nabla logic take the form:
\begin{enumerate}[(Tree 1)]
\item $\bigwedge_i\{\langle k_i\rangle(a_1^i,\ldots,a_{k_i}^i)\}=\begin{cases}
\langle k\rangle(\bigwedge_i a_1^i, \ldots,\bigwedge_i a_k^i)&\text{ if }k_i=k\text{ for all }i\\
\bot&\text{ else}
\end{cases}$
\item $\langle k \rangle(\bigvee \phi_1,\ldots,\bigvee \phi_k)=\bigvee\{\langle k\rangle(a_1,\ldots,a_k)\mid a_i\in \phi_i, 1\leq i\leq k\}$
\end{enumerate}
where $a_j^i\in \lang_{\polyFunc[n]}$, $\phi_i\in \powf\lang_{\polyFunc[n]}, 1\leq i\leq k$. In other words, operator $\langle k \rangle$ distributes over meets and joins in each of its arguments.  This last feature is very desirable as we have seen in Chapter 2, and we will return to the significance of this later on. Using the discussion on negation from \cite{KKV:2012:Journal} (Section 6), we can also write down the following axiom which is derivable but which is convenient for the calculations that will follow.
\[
\text{(Tree 3) }\neg\langle k\rangle(a_1,\ldots,a_k)=\hspace{-2ex}\bigvee\limits_{1\leq j\neq k\leq n}\langle j \rangle(\top, \ldots, \top)\vee \hspace{-1ex}\bigvee\limits_{1\leq i\leq k}\langle k\rangle(\epsilon_{1i}a_1,\ldots,\epsilon_{ki}a_k)
\]
where $\epsilon_{ij}=\neg$ if $i=j$ and is the empty symbol otherwise. Using these axioms we can define the functor $L$ defining our logic of trees of degree at most $n$ as the functor $\BA\to\BA$ defined for any object $A$ and morphism $f: A\to B$ as
\begin{align}
LA &=(\Free\polyFunc[n]\Forg))/(\text{Tree 1-3}) \label{ch5:eq:treef}\\
Lf &: LA\to LB, [\langle k\rangle(a_1,\ldots,a_k)]\mapsto [\langle k\rangle(f(a_1),\ldots,f(a_k))] \nonumber
\end{align}
where the quotient is taken under the smallest equivalence relation (in $\BA$) containing the axioms Tree 1-3. This equivalence relation can be explicitly constructed using the method described in Section \ref{ch1:subsec:algsem} via the kernel pair of the coequalizer associated with the set of axioms. For notational clarity we drop the square brackets $[-]$ and simply remember that we are dealing with equivalence classes under the axioms Tree 1-3. 

Before moving on to the semantics, we make the following important observation about terms in the language: every term can be expressed in the shape $\bigvee_i\langle k_i\rangle(a^i_1,\ldots,a^i_{k_i})$ where each $k_i$ is different (i.e. $k_i\neq k_j$ when $i\neq j$). To see that this is the case, start with an arbitrary term and re-write it in conjunctive (or disjunctive) normal form. By (Tree 3), we can remove all the innermost negation, modulo the addition of some joins. Now by distributivity, we can re-write the term thus obtained as a join of meets. But by (Tree 1), we can also see that each meet can only be a meet of operators of the same arity, or else we have a bottom element which we can get rid of. Finally, we can use (Tree 1) to push the meets of subterms of the same arity inside the operator, leaving us with a join of the shape $\bigvee_i\langle k_i\rangle(a^i_1,\ldots,a^i_{k_i})$. Finally, we can use (Tree 2) to push the joins of identical $k_i$s inside the modal operator, leaving us with a join over different $k_i$s.

The semantic transformation $\delta: L\pow \to \pow \polyFunc[n] $ is given by the transpose maps:
\begin{align*}
& \tilde{\delta}_X: \polyFunc[n]\Forg\pow X\to \Forg\pow \polyFunc[n] X, \\
&(U_1,\ldots, U_k)\mapsto \{(x_1,\ldots,x_k)\in \mathrm{in}_k[X]\mid x_i\in U_i, 1\leq i\leq k\}
\end{align*}
We can easily check that $\delta_X$ is well-defined, i.e. that it is independent of the choice of representative. As an illustration, let us show that it is the case for (Tree 1), i.e. the distribution of meets.
\begin{align*}
\delta_X(\bigwedge_i\langle k\rangle(U_1^i,\ldots,U_k^i)) &= \bigcap_i \delta_X(\langle k\rangle(U_1^i,\ldots,U_k^i))\\
&= \bigcap_i\{\langle k \rangle(x_1^i, \ldots, x_n^i)\mid x^i_j\in U^i_j\}\\
&=\{\langle k \rangle(x_1, \ldots, x_n)\mid x_j\in \bigcap_i U^i_j\}\\
&=\delta_X(\langle k \rangle(\bigwedge_i U^i_1, \ldots, \bigwedge_i U^i_n))
\end{align*}

By using the general result that nabla logics are weakly complete, we get that this logic is sound and weakly complete with respect to infinite $n$-ary trees. But we can show this directly. We introduce a technique which we will use repeatedly in the following examples. The idea is to use the fact that $L$ is finitary and Proposition \ref{ch5:prop:approxdelta} and show that $\delta$ is injective on \emph{finite} sets by exhibiting a retraction for $\delta_X$ with $X$ finite. We define $r_X: \pow \polyFunc[n] X\to L\pow X$ by
\[
r_X(\{\langle k_i\rangle(x_1^i,\ldots,x_{k_i}^i)\}_{i\in I})=\bigvee_{k}\langle k_i \rangle(\{x^i_1\}_{i\mid k_i=k}, \ldots, \{x^i_k\}_{i\mid k_i=k})
\]
where the join is taken over all the $k$s that occur in $\{\langle k_i\rangle(x_1^i,\ldots,x_{k_i}^i)\}_{i\in I}$, this set is finite since $\polyFunc[n] X$ is finite.

It is easy to check that $r_X$ is a right inverse of $\delta_X$. By the comment above, every term in $LX=\Free\polyFunc[n]\Forg \pow X$ can be written as $\bigvee_i\langle k_i\rangle(U^i_1,\ldots,U^i_{k_i})$ with every $k_i$ different and with each $U_j^i\in \pow X$. It then follows that 
\begin{align*}
r_X(\delta_X(\bigvee_i\langle k_i\rangle(U^i_1,\ldots,U^i_{k_i})))&=r_X(\bigcup_i \delta_X(\langle k_i\rangle(U^i_1,\ldots,U^i_{k_i})))\\
&=r_X(\bigcup_i\{(x_1,\ldots,x_{k_i})\mid x_j\in U^i_j, 1\leq j\leq k_i\})\\
&=\bigvee_i\langle k_i\rangle(\{x_1\mid x_1\in U_1^i\}, \ldots,\{x_{k_i}\mid x_{k_i}\in U_{k_i}^i\})\\
&=\bigvee_i\langle k_i\rangle(U^i_1,\ldots,U^i_{k_i})
\end{align*}
Thus $r_X\circ \delta_X=\id_{L\pow X}$, and we can conclude that $\delta_X$ is injective on finite sets, and thus everywhere from Proposition \ref{ch5:prop:approxdelta} since $L$ is finitary. Weak completeness follows by Theorem \ref{ch5:thm:weakcomp2}. The adjoint transformation $\hat{\delta}:\polyFunc[n]\uf \to \uf L$ is given by:
\[
\hat{\delta}_A((u_1,\ldots, u_k))=\{\langle k\rangle(a_1,\ldots,a_k)\mid a_i\in u_i, 1\leq i\leq k\} 
\]
It is easy to check that $\hat{\delta}_A((u_1,\ldots, u_k))$ is an ultrafilter, and that $\hat{\delta}_A$ injective. The logic is thus expressive. 

Let us check that we can use our theorems. The functor $\Free\polyFunc[n]\Forg$ is finitary since $\Forg$ is monadic, and thus preserves filtered colimits, $\polyFunc[n]$ is finitary, and $\Free$ being a left adjoint preserve all colimits, and in particular filtered ones. To see that $\polyFunc[n]$ preserves cofiltered limits, note first that each summand in the coproduct is essentially a hom functor $\hom(k,-)$ which preserves all limits, and thus cofiltered ones. As for the coproduct, it is a standard result that coproducts commute with connected limits in $\Set$, and since cofiltered limits are connected, we can conclude that $\polyFunc[n]$ preserves cofiltered limits. 
Finally, since $\delta$ is injective, $\hat{\delta}$ is surjective by Theorem \ref{ch5:thm:deltahatsurj}. We can thus use either Theorem \ref{ch5:thm:strgcomp1} or Theorem \ref{ch5:thm:strgcomp2} since we have both expressivity and a cofiltered limit preserving functor to conclude that the logic is strongly complete with respect to its semantics. This example exhibits the best possible case: we have an expressive and strongly complete logic. As we shall see, combining these two features is rarely possible.
\end{example}

\begin{example}[Classical modal logic]\label{ch5:ex:mod} Consider the functor $LA:\BA\to\BA$ defined by
\begin{align*}
LA &=\Free(\{\dia a\mid a\in A\})/(\dia\bot=\bot, \dia(a\vee b)=\dia a\vee \dia b)\\
Lf &: LA\to LB, [\dia a]\mapsto [\dia f(a)]
\end{align*}
where the quotient is under the smallest equivalence relation (in $\BA$) generated by the two equations. The map $Lf$ is well defined by virtue of $f$ being a $\BA$-morphism. Note that this is a particular instance of a relational logic, as defined in the previous section, with $\dia$ being a unary operator satisfying (Eq. \ref{ch5:eq:DL1}). For notational clarity, we will from now on drop the square brackets of the equivalence class $[\dia a]$ and simply write $\dia a$. The functor $L$ defines \textbf{classical modal logic}\index{Classical modal logic}, and is interpreted in $\cpow$-coalgebras, i.e. Kripke frames, via the the semantic transformation $\delta: L\pow\to \pow\cpow$ defined inductively via it action on the generators $\dia U$:
\[
\delta_X(\dia U)=\{V\subseteq X\mid V\cap U\neq \emptyset\}
\]
The fact that $\delta$ is well-defined is a direct application of Proposition \ref{ch5:prop:deltaWellDefined}. We will now show that classical modal logic is strongly complete with respect to its Kripke frame semantics. Of course, we already know this to be true from Theorem \ref{ch5:thm:strongcomplRelational}, but the purpose of this example is to illustrate how the theorems developed in this section can be used in practise.
An important requirement in the application of Theorems \ref{ch5:thm:deltahatsurj}, \ref{ch5:thm:strgcomp1}, \ref{ch5:thm:strgcomp2} and \ref{ch5:thm:strgcomp3} proving strong completeness is that $\delta$ should be injective, and in particular the logic should be \emph{weakly} complete (Theorems \ref{ch5:thm:weakcomp1} or \ref{ch5:thm:weakcomp2}).
To show that $\delta$ is injective, we show that it is injective on \emph{finite} sets, and the general case then follow from Proposition \ref{ch5:prop:approxdelta} since $L$ is finitary and the fundamental adjunction is $\uf\dashv\pow:\BA\to\Set\op$. To show that $\delta_X$ is injective when $X$ is finite, we will exhibit a retraction $r_X$, i.e. a left inverse, to $\delta_X$. To define this retraction we follow the following steps.

\begin{enumerate}[(Step 1)]
\item For every finite set $X$, any element of $L\pow X$ can be expressed as 
\[
\bigvee_i\left(\bigwedge_j \dia\{x_j^i\}\wedge\neg\dia V_i\right)
\]
where each $V_i\subseteq X$ and each $x_j^i\in X$ is such that $x^i_j\notin V_i$. To see that this is the case, first note that for any finite set $U$ the term $\dia U$ can be written as $\bigvee_{x\in U}\dia \{x\}$ by using the distributivity over joins and the finiteness of $U$. Thus starting with a term $t\in L\pow X$, we can rewrite it in such a way that every $\dia$ operator is applied to a singleton set. Next, we put this re-written term in disjunctive normal form. If in a clause (i.e. a meet) we find a literal and its negation, then this clause evaluates to $\bot$ and we can get rid of it. Finally, by using the de Morgan law and distributivity once more, we can regroup the negative literals to form the sets $V_i$ and obtain a term of the form above. The fact that $x_j^i\notin V_i$ follows from the previous step.
\item Let us now consider the image under $\delta_X$ of a term without joins, i.e. a set of the form $\delta_X(\bigwedge_{x\in U} \dia\{x\}\wedge \neg \dia V)$. It is the collection of subsets of $X$ which include $U$ but avoids $V$, i.e. 
\[
\delta_X(\bigwedge_{x\in U} \dia\{x\}\wedge \neg \dia V)=\{W\subseteq X\mid U\subseteq W\subseteq V^c\}
\]
The two inclusions are compatible since, as was shown in the previous step, we can assume $U\cap V=\emptyset$.
\item To recover $U$ and $V$ from the set above we define for any element $\mathcal{V}\in\pow\cpow X$ the following set of pairs of subsets:
\[
\mathcal{I}(\mathcal{V})=\{(Y,Z)\mid Y\subseteq W\subseteq Z^c \Rightarrow W\in \mathcal{V}\}
\]
This set forms a poset under component-wise inclusion, and it is easy to check that $(\top,\top)\in\mathcal{I}(\mathcal{V})$ trivially.
\item Let us write for notational convenience $t=\bigwedge_{x\in U} \dia\{x\}\wedge \neg \dia V$. We now show that $\mathcal{I}(\delta_X(t))$ has $(U,V)$ as a minimal element. It follows immediately from the definitions that $(U,V)\in \mathcal{I}(\delta_X(t))$. Now assume $(Y,Z)\in \mathcal{I}(\delta_X(t))$, we need to show that $U\subseteq Y$ and $V\subseteq Z$. Assume for the sake of contradiction that $U\nsubseteq Y$ and pick $W=Y$. It is clear that $Y\subseteq W\subseteq Z^c$, but $W\notin \delta_X(t)$ since $U\nsubseteq Y$. Similarly, if we assume that $V\nsubseteq Z$ i.e. $Z^c\nsubseteq V^c$, and pick $W=Z^c$, it is clear that $Y\subseteq W\subseteq Z^c$, but that $W\notin\delta_X(t)$ since $W\nsubseteq V^c$. We can thus recover a term $t=\bigwedge_{x\in U} \dia\{x\}\wedge \neg \dia V$ from its interpretation $\delta_X(t)$ by taking the minimal element of $\mathcal{I}(\delta_X(t))$. 
\item Let us now show that we can recover a general term of the form detailed above by finding all the minima of $\mathcal{I}(\delta_X(t))$. Assume 
\[
t=\bigvee_i\left(\bigwedge_{x\in U_i} \dia\{x\}\wedge\neg\dia V_i\right)
\]
We claim that the minima of $\mathcal{I}(\delta_X(t))$ are precisely the pairs $(U_i,V_i)$. First note that each $(U_i,V_i)\in \mathcal{I}(\delta_X(t))$ by definition. Now let us take an arbitrary $(Y,Z)\in \mathcal{I}(\delta_X(t))$, we need to show that $U_i\subseteq Y$ and $V_i\subseteq Z$ for some $i$. If $(Y,Z)\in \mathcal{I}(\delta_X(t))$, then in particular we must have $Y\in\delta_X(t)$, and thus there must exist $i$ such that $U_i\subseteq Y\subseteq V_i^c$. Similarly, we must have $Z^c\in\delta_X(t)$, and thus there must exist $j$ such that $U_j\subseteq Z^c\subseteq V_j^c$. If $i=j$ we are done, so let us assume that $i\neq j$, i.e. that we cannot find $i$ such that $U_i\subseteq Y$ and $Z^c\subseteq V_i^c$. In particular this means that we are assuming that $U_j\nsubseteq Y$, i.e. there exists $x\in U_j, x\notin Y$, and $Z^c\nsubseteq V_i^c$, i.e. there exists $y\in Z^c, y\notin V_i^c$. We now have three possibilities:
\begin{itemize}
\item $y\in U_j^c$: We then consider $W=Y\cup\{y\}$: it is obvious that $Y\subseteq W$, but since $y\in Z^c$ it is also clear that $W\subseteq Z^c$, since $Y\subseteq Z^c$. Let us now show that $W\notin \delta_X(t)$, which will contradict the assumption that $(Y,Z)\in \mathcal{I}(\delta_X(t))$. It is enough to show that $U_j\nsubseteq W$ and $W\nsubseteq V_i^c$. Since we are assuming both $U_j\nsubseteq Y$ and $y\in U_j^c$ it is clear that $U_j\nsubseteq W$ holds. Similarly, since $y\notin V_i^c$, and $y\in W$, it follows that $W\nsubseteq V_i^c$. Thus $W\notin\delta_X(t)$ and we get the desired contradiction.
\item $y\in U_j$ and $x\in V_i^c$: We once again put $W=Y\cup\{y\}$ and show the same contradiction as above. We again have $Y\subseteq W\subseteq Z^c$. Since $y\notin V_i^c$, and $y\in W$, it follows that $W\nsubseteq V_i^c$. To show that $U_j\nsubseteq W$, note first that since $x\in V_i^c$ and $y\notin V_i^c$, we must have $x\neq y$. Since $x\notin Y$ and $x\neq y$ it follows that $x\notin W$, and  since $x\in U_j$ it follows that $U_j\nsubseteq W$.
\item $y\in U_j$ and $x\in V_i$: This is the difficult case because we cannot immediately discount $x=y$. If we can find two distinct points $x\neq y$ satisfying the conditions above then $W=Y\cup\{y\}$ will lead to the same contradiction as above. For there to be no two distinct points $y,x$ such that $y\in U_j$ and $x\in V_i$ it must be the case that $U_j=V_i=\{x\}$. We now show that this situation can be discounted. For notational convenience we will write $\bigwedge \dia U$ for $\bigwedge_{u\in U}\dia\{u\}$ and $\dia x$ for $\dia\{x\}$. We have the following join of clauses in the expression of $t$:
\begin{align*}
&(\bigwedge \dia U_i\wedge\neg\dia x)\vee (\dia x\wedge\neg\dia V_j)=\\
&(\bigwedge\dia U_i\vee\dia x)\wedge(\bigwedge\dia U_i\vee\neg\dia V_j)\wedge(\neg \dia x\vee\neg\dia V_j)=\\
&(\bigwedge\dia U_i\wedge\neg \dia x)\vee(\bigwedge\dia U_i\wedge \neg\dia V_j)\vee (\dia x\wedge \neg\dia V_j)
\end{align*}
where the first step is by distributing meets over the join and the second step is by distributing the joins over the meets. Recall that we had assumed that $U_i\subseteq Y\subseteq V_i^c$, we show that $U_i\subseteq Y\subseteq V_j^c$. Assume that there exist $y\in Y,y\notin V_j^c$, i.e. $y\in V_j$. Since we have also assumed that $V_j\subseteq Z$ and $Z$ is disjoint from $Y$, we get a contradiction. Thus $U_i\subseteq Y\subseteq V_j^c$, which means that modulo the re-writing of $t$ described above we can find pairs $(\tilde{U}_i,\tilde{V}_i^c)$ and $(\tilde{U}_j,\tilde{V}_j^c)$ sandwiching $Y$ and $Z^c$ respectively but without the problematic property that $\tilde{V}_i=\tilde{U}_j=\{x\}$, and we can then proceed as above to find a contradictory $W$.
\end{itemize}
\item We have thus shown that by taking the minimum elements of $\mathcal{I}(\delta_X(t))$ we can (modulo re-writing, which makes no difference since $L\pow X$ is in $\BA$) always recover $t$,  via the map $r_X: \pow\cpow X\to L\pow X$
\[
\mathcal{V}\mapsto \bigvee_{\text{ minima }(U,V)\text { of }\mathcal{I}(\mathcal{V})}\left (\bigwedge_{x\in U}\dia\{x\}\wedge \neg\dia V\right)
\]
which as we have shown is a retraction of $\delta_X$, i.e. $\delta_X$ is injective on finite sets.
\end{enumerate}

Since $L$ is finitary can now apply Proposition \ref{ch5:prop:approxdelta} to conclude that $\delta$ is actually a mono-transformation, and modal logic is thus weakly complete with respect to its $\pow$-coalgebra, i.e. Kripke frame semantics by Theorem \ref{ch5:thm:weakcomp2}. Alternatively, we could use the fact that $\cpowf$ has a terminal coalgebra which in particular is a $\cpow$-coalgebra, and use Theorem \ref{ch5:thm:weakcomp1}. Note that unlike the usual weak completeness proofs (e.g. via tableaux systems, see for example \cite{1972:fittingtableau}), we have in effect shown `categorical weak completeness', i.e. a result which is independent of the choice of a base object in $\Set$, i.e. of a model carrier. To show strong completeness of modal logic we can use the fact that $\cpow$ preserves finite sets, that $L$ is finitary and that $\cpow$ weakly preserves cofiltered limits (see Lemma \ref{ch5:lem:powcofiltered}) to conclude that modal logic is strongly complete with respect to $\cpow$-coalgebra by Theorem \ref{ch5:thm:deltahatsurj}. Note that this only ensures the existence of \emph{quasi}-canonical models. 

We can also show that $\delta$ is surjective on finite sets. Let $X$ be a finite set and let $\{U_i\}_{1\leq i\leq n}$ be a collection of subsets of $X$. For each $i$, since $X$ is finite, we can write
\[
\{U_i\}=\delta_X\left(\bigwedge_{x_i\in U_i}[\dia \{x_i\}]\wedge \neg \bigwedge_{y_i\in \neg U_i}[\dia\{y_i\}]\right)
\]
i.e. each $U_i$ is the intersection of the collection of subsets which intersect each of its elements and none of the elements of its complement. It follows that
\[
\{U_i\}_{1\leq i\leq n}=\delta_X\left(\bigvee_{i\leq i\leq n}\left[\bigwedge_{x_i\in U_i}[\dia \{x_i\}]\wedge \neg \bigwedge_{y_i\in \neg U_i}[\dia\{y_i\}]\right]\right)
\]
and thus $\delta_X$ is surjective. Since it is always injective, we have that $\delta$ is bijective on finite sets. We can therefore also prove strong completeness of classical modal logic via Theorem \ref{ch5:thm:strgcomp2}. Moreover, since $\cpow$ preserves finite sets it follows from \ref{ch5:lem:delatdeltahatfin} that $\hat{\delta}$ is an iso on finite algebras, and the logic is thus strongly complete over finite algebras and strong completeness can be obtained from Theorem \ref{ch5:thm:strgcomp3} as well. 

It is difficult to give an explicit definition of the adjoint transformation, but the following observation provides us with a way around this difficulty. By definition, $LA$ is the quotient of $\Free(\{\dia a\mid a\in A\})$ under an equivalence relation, i.e. there exist a (regular) epi $q_A:\Free(\{\dia a\mid a\in A\})\epi LA$. There must therefore exist a mono transformation $\uf q_A: \uf LA\mono \uf \Free(\{\dia a\mid a\in A\})$. But by Lemma \ref{ch5:lem:ufF=UP}, this is to say that we have a monomorphism $\uf LA\mono \Forg\pow (\{\dia a\mid a\in A\})$, i.e. we can view ultrafilters on $LA$ as collections of elements $\dia a, a\in A$. Note that not all sets of generators define valid ultrafilters, since the axioms encoded in $L$ must be satisfied. For example, if $\dia a$ and $\dia b$ belong to an ultrafilter, then so does $\dia a\vee\dia b$ and thus $\dia(a\vee b)$ by the axioms defining $L$. Modulo this remark, viewing an ultrafilter in terms of the generators it contains allow us to write:
\begin{align*}
\hat{\delta}_A (U)&\simeq\{\dia a\in LA \mid U\in \delta_{\uf A}(L\eta_A)(\dia a)\}\\
&= \{\dia a\in LA\mid U\in\delta_{\uf A}(\dia \eta_A(a))\}\\
&=\{\dia a\in LA\mid \exists u\in U. a\in u\}
\end{align*}
and it is easy to check that this collection of generators does indeed satisfy the axioms defining $L$, and does therefore define an element of $\uf LA$. 

The adjoint transformation $\hat{\delta}$ is not injective in general, reflecting the well-known fact that classical modal logic is not expressive (i.e. has not got the Hennessy-Milner property) for arbitrary Kripke frames (see \cite{2001:ModalLogic} Section 2.2). However, $\hat{\delta}$ is injective on finite subsets. This can be shown from the surjectivity of $\delta$ on finite sets and Lemma \ref{ch5:lem:delatdeltahatfin}, or directly by using an argument dual to the proof of Theorem 9 in \cite{JacobsExpressivity}. Let $U$ and $V$ be finite sets of ultrafilters of $A$ such that $U\neq V$. Since $V$ is assumed to be finite, we can write it as $V=\{v_1,\ldots,v_n\}$ with $v_i\in\uf A, 1\leq i\leq n$, and since $U\neq V$, there must exist an $u\in U$ such that $u\notin V$, i.e. $\forall i \exists b_i\in u$ such that $b_i\notin v_i$. Consider now the element $a=\dia\bigwedge b_i\in LA$. It is easy to check that:
\begin{align*}
a\in \hat{\delta}_A(W) & \Leftrightarrow W\in \delta(L\eta_A(\dia\bigwedge b_i))\\
&\Leftrightarrow W\cap L\eta_A(\dia\bigwedge b_i)\neq \emptyset\\
&\Leftrightarrow \exists w\in W \forall i, b_i\in w
\end{align*}
and thus we have $a\in \hat{\delta}_A(U)$ and $a\notin\hat{\delta}_A(V)$ by construction of $a$. The fact that the logic is not expressive means that we cannot use Theorem \ref{ch5:thm:strgcomp1}.

Of course we can also define a right inverse to $\hat{\delta}$ as we did in Theorem \ref{ch5:thm:strongcomplRelational} (this is essentially the content of the traditional canonical model construction) by:
\[
\hat{\delta}\inv_A: \uf LA\to \cpow\uf A, u\mapsto\{v\in \uf A\mid a\in v \text{ whenever }\dia a\in u\}
\]
Strong completeness then follows immediately from the  J\'{o}nsson-Tarski Theorem \ref{ch5:thm:jontarski}. 
\end{example}

\subsection{Strong completeness by semantic completion}\label{ch5:sec:semcomp}

We start this subsection by presenting three important cases for which the theorems developed in the previous section do not hold. We will then suggest a general recipe for recovering strong completeness in such cases. The examples are taken from \cite{2009:DirkStrongComp}, but the general method to recover strong completeness is, to the best of our knowledge, new.
\begin{example}[Modal logic for trees with unbounded branching degree]\label{ch5:ex:treeOmega}
Consider the functor $\polyFunc[\omega]:\Set\to\Set$ defined for any $f: X\to Y$ in $\Set$ by
\begin{align*}
\polyFunc[\omega]X & =\coprod_{i\in\omega} X^i\\
\polyFunc[\omega] f & =[!+ f + \ldots]
\end{align*}
where $X^0=1$ is the terminal object in $\Set$ and $!: 1\to 1$. We consider the unbounded version of the language detailed in Example \ref{ch5:ex:treef} , i.e. the free BAE defined by $k$-ary expansions $\kop$ for each $k\in \omega$. The axioms and the semantics are given in the same way as in Example \ref{ch5:ex:treef}, but the formula (Tree 3) does not hold since it would involve an infinitary disjunction. The semantic transformation $\delta$ is shown to be injective in exactly the same way as in the bounded branching case of Example \ref{ch5:ex:treef}, the only subtlety in defining the retraction $r_X$ is to define it on the image of $\delta_X$, since $\pow \polyFunc[\omega] X$ is not finite anymore. Elements in $\delta_X[L\pow X]$ for $X$ finite are however finite.

The adjoint transpose of the semantic transformation $\delta$ is given as in Example \ref{ch5:ex:treef} by:
\[
\hat{\delta}_A: \polyFunc[\omega]\uf A\to \uf L A, (u_1,\ldots,u_n)\mapsto\{\kop(a_1,\ldots,a_k)\mid a_i\in u_i, 1\leq i\leq k\}
\]
To see that $\hat{\delta}_A$ cannot be surjective, consider the filter generated by
\[
\{\neg\kop(\top,\ldots,\top)\mid k\in\omega\}
\]
This set is consistent, and thus generates an ultrafilter in $\uf LA$, but there can clearly not be any tuple $(u_1,\ldots,u_k)$ mapped to it by $\hat{\delta}_A$.
\end{example}
\begin{example}[Image finite semantic of classical modal logic]\label{ch5:ex:modf}
We proceed exactly as in Example \ref{ch5:ex:mod}, but we replace the semantic functor $\cpow$ by its finitary version $\cpowf$ and the semantic transformation is then defined on generators by the maps:
\[
\delta_X:L\pow X\to \pow\cpowf X, [\dia U]\mapsto\{V\subseteq_\omega X\mid V\cap U\neq \emptyset\}
\]
The properties of $\delta_X$ which we showed in Example \ref{ch5:ex:mod} remain true, i.e. $\delta_X$ remains injective for each $X$, and it remains surjective on finite sets. Moreover, as is well known, the logic is now expressive, i.e. $\hat{\delta}$ is injective (see \cite{JacobsExpressivity}). However, $\cpowf$ does not weakly preserve cofiltered limits, as the following simple example immediately shows. Consider the sets $X_n=\{1,\ldots, n\}$ together with the surjective maps 
\[
p_n:X_{n+1}\to X_n, i\mapsto\begin{cases} i &\text{ if }i<n+1\\
n &\text{else}\end{cases}
\]
It is clear that the $X_n$ and $p_n$ form a cofiltered diagram and that $\lim_n X_n=\mathbb{N}$. A coherent family $(U_n)_{n\in\mathbb{N}}$ of elements $U_n\in\cpowf X_n$ does not in general define an element of $\cpowf \lim_n X_n=\cpowf \mathbb{N}$ since the size of the $U_n$ need not be bounded. 

From this failure to preserve cofiltered limits we can show that $\hat{\delta}_A$ cannot be surjective for a general algebra $A$. Indeed, from the proof of Theorem \ref{ch5:thm:strgcomp3}, it is clear that if $\hat{\delta}_A$ was surjective, we could find for every coherent family of the cofiltered diagram $(\cpowf\uf A_i)_{i\in I}$ (where $\colim_i A_i=A$ is a representation of $A$ as a filtered colimit of finitely presentable boolean algebras) an element of $\cpowf\uf A$ projecting onto its components in $\cpowf\uf A_i$, i.e. $\cpowf$ would weakly preserve cofiltered limits. Thus the failure of $\cpowf$ to weakly preserve cofiltered limits in facts forbids any Tarski-J\'{o}nsson embedding, and therefore strong completeness by canonicity or quasi-canonicity. Thus while we recover expressivity (which we did not have in Example \ref{ch5:ex:mod}), we loose strong completeness. We will return to this tension between the two concepts.
\end{example}

\begin{example}[Graded modal logic]\label{ch5:ex:GMLf}
We have introduced graded modal logic in Chapter 2, and explained its semantics in terms of the bag functor. Graded modal logic can also be treated in its nabla flavour, as was done in \cite{2013:self}. In the abstract style, the syntax is given by the functor $L: \BA\to\BA$ defined by
\begin{align*}
& LA=\Free(\{\langle k\rangle a\mid k\in\mathbb{N}, a\in A\})/\Ax_{\mathrm{GML}}\\
& Lf: LA\to LB, [\langle k\rangle a]\mapsto [\langle k\rangle f(a)]
\end{align*}
The quotient in the definition of $L$ is taken under the smallest equivalence relation (in $\BA$) containing the axioms $\Ax_{\mathrm{GML}}$ of graded modal logic. Recall from Chapter 2 that these are given by:
\begin{enumerate}[GML1]
\item \label{ch5:ax:GML1} $\langle 1\rangle(a\vee b)=\langle 1\rangle(a)\vee\langle 1\rangle (b)$
\item \label{ch5:ax:GML2} $\langle 1\rangle\bot=\bot$
\item \label{ch5:ax:GML3} $\dia_k a\to\dia_l a,\hspace{1ex} l<k$
\item \label{ch5:ax:GML4} $\dia_k a \leftrightarrow \bigvee_{i=0}^k \dia_i(a\wedge b)\wedge\dia_{k-i}(a\wedge \neg b)$
\item \label{ch5:ax:GML5} $\neg\dia_1(a\wedge\neg b)\to(\dia_k a\to\dia_k b)$
\end{enumerate}

Once again we will drop the square brackets and simply write the equivalence classes of the generators under $\Ax_{\mathrm{GML}}$ as $\langle k\rangle a$. The semantic transformation $\delta: L \pow\to \pow \Bag$ is given by its action on generators:
\[
\delta_X: L\pow X\to\pow\Bag X, \langle k\rangle U\mapsto\{f\in\Bag(X)\mid \sum_{x\in U} f(x)\geq k \}
\]
which is precisely the definition of the predicate lifting we defined in Chapter 2. It is not very hard to check that $\delta_X$ preserves the axioms (GML \ref{ch5:ax:GML1}-GML \ref{ch5:ax:GML5}). To see that $\delta_X$ is injective, we proceed as in Example \ref{ch5:ex:mod} and show that $\delta_X$ is injective when $X$ is finite. Since $L$ is finitary and we are working with the adjunction $\uf\dashv\pow: \BA\to\Set\op$, it will follow that $\delta_X$ is injective in general.  We exhibit a retraction $r_X$ for $X$ finite, just as we did in Example \ref{ch5:ex:mod}, which we construct in several steps:
\begin{enumerate}[(Step 1)]
\item For every finite set $X$, any element $t\in L\pow X$ can be expressed as:
\[
t=\bigvee_i \left(\bigwedge_{j_i} \langle k_{j_i}\rangle \{x_{j_i}\} \wedge \bigwedge_{j_i'} \neg \langle k_{j_i'}\rangle \{y_{j_i'}\}\right)
\]
where for a given $i$ we can only have $\langle k_{j_i}\rangle \{x\}$ and $\neg \langle k_{j_i'}\rangle \{x\}$ in the corresponding meets if $k_{j_i}<k_{j_i'}$. To see that this is the case, consider first a term $\langle k\rangle U$ for $U\subseteq X$. Since $X$ is finite, we can write $U$ as $\{x_1\}\vee\ldots\vee\{x_n\}$, which for notational clarity we will abbreviate as $x_1\vee\ldots\vee x_n$. We will now repeatedly use the axiom (GML 4) as follows:
\begin{align*}
\langle k\rangle U &=\langle k\rangle(x_1\vee\ldots\vee x_n)\\
=&\bigvee_{i=0}^k \left(\langle i\rangle((x_1\vee\ldots\vee x_n)\wedge x_1^c) \wedge \langle k-i\rangle((x_1\vee\ldots\vee x_n)\wedge x_1)\right)\\
=&\bigvee_{i=0}^k \left(\langle i\rangle(x_2\vee\ldots\vee x_n) \wedge \langle k-i\rangle(x_1)\right)\\
=&\bigvee_{i=0}^k \left(\bigvee_{j=0}^{i}\langle j\rangle((x_2\vee\ldots\vee x_n)\wedge x_2^c) \wedge \langle i-j\rangle((x_2\vee\ldots\vee x_n)\wedge x_2)\right) \\
& \wedge \langle k-i\rangle(x_1)\\
=&\bigvee_{i=0}^k \left(\bigvee_{j=0}^{i}\langle j\rangle(x_3\vee\ldots\vee x_n) \wedge \langle i-j\rangle(x_2)\right)\wedge \langle k-i\rangle(x_1)\\
=&\ldots
\end{align*}
We keep on proceeding in this way for modalities $\langle k\rangle$ with $k>1$. As soon as we reach a term $\langle 1\rangle (x_1,\ldots,x_m)$ or $\langle 1 \rangle\emptyset$ we use (GML 1) and (GML 2) to rewrite it as a join a modalities applied to singletons, or as $\bot$. It is clear that in this way we can re-write $\langle k\rangle U$ as a (much more complicated) term involving only singleton sets after the graded modalities. Thus we can assume without loss of generality that an arbitrary term has graded modalities applied to singleton sets only. We then put such a term in disjunctive normal form to get a term of the form announced above. The fact that for a given $i$ we can only have $\langle k_{j_i}\rangle \{x\}$ and $\neg \langle k_{j_i'}\rangle \{x\}$ in the corresponding meets if $k_{j_i}<k_{j_i'}$ is an easy consequence of axiom (GML 3).
\item We can make an extra assumption on the form of the terms above which will prove useful in the last step of this proof. For a given $x\in X$, we can assume that all the modal operators appearing positively (resp. negatively) in front of $x$ are identical. It is in fact a simple consequence of distributivity. Assume that we have a clause $\langle k_1\rangle x\wedge\neg\langle k_2\rangle x\wedge S$ where $S$ stands for the sub-clause not involving $x$, and that we have another clause $\langle k_1'\rangle x\wedge\neg\langle k_2'\rangle x\wedge S'$. Part of the join defining $t$ will thus be
\[
(\langle k_1\rangle x\wedge\neg\langle k_2\rangle x\wedge S)\vee \langle k_1'\rangle x\wedge\neg\langle k_2'\rangle x\wedge S'
\]
By applying distributivity back and forth using (GML 3) it is not difficult to see that this join can be re-written as
\begin{align*}
&\langle (\min(k_1,k_1')\rangle x\wedge\neg \langle\max(k_2,k_2')\rangle x\wedge S)\vee \\
&\langle (\min(k_1,k_1')\rangle x\wedge\neg \langle\max(k_2,k_2')\rangle x\wedge S')
\end{align*}
i.e. we do have the same modal operators in front of $x$ in both clauses. This clearly generalizes to all clauses, by each time widening the interval of values which $x$ is allowed to take.
\item  Assume a term $t=\bigwedge_{j} \langle k_{j}\rangle \{x_{j}\} \wedge \bigwedge_{j'} \neg \langle k_{j}\rangle \{y_{j}\}$. The set $\delta_X(t)$ is the collection of maps $f: X\to\mathbb{N}$ such that $f(x_j)\geq k_j$ and $f(y_{j'})< k_{j'}$. To recover the sets $\{x_j\}$ and $\{y_{j'}\}$ and all the integers $k_j$ and $k_{j'}$ for $\delta_X(t)$, we define for any element $\mathcal{V}\in\pow\Bag(X)$, the following set of pairs of maps $X\to \mathbb{N}$:
\begin{align*}
\mathcal{I}(\mathcal{V})=\{(f,g)\mid & f\restrict(\supp(f)\cap\supp(g))<g\restrict(\supp(f)\cap\supp(g))\text{ and }\\
& \text{ if }f\leq h\restrict \supp(f)\text{ and } h\restrict\supp(g)\leq g\text{ then } h\in \mathcal{V} \}
\end{align*}
where the inequalities are of course defined pointwise. The first condition in the definition ensures that the second is not always vacuously satisfied. We define a partial order on $\mathcal{I}(\mathcal{V})$ as follows:
\[
(f_1,g_1)\leq (f_2,g_2)\text{ iff } f_1\leq f_2\restrict\supp(f_1)\text{ and }g_1\geq g_2\restrict \supp (g_1)
\] 
Note the reversal of order and the presence of the support in the second argument.
\item Let us define the maps $f_t, g_t: X\to\mathbb{N}$ by
\begin{align*}
f_t(x)=\begin{cases}k_j & \text{if }\exists j\text{ s.th. }x=x_j\\
0& \text{else}\end{cases} & \hspace{1em} g_t(x)=\begin{cases}k_{j'} & \text{if }\exists j'\text{ s.th. }x=y_{j'}\\
0& \text{else}\end{cases}
\end{align*}
We show that $\mathcal{I}(\delta_X(t))$ has $(f_t,g_t)$ as minimum. Clearly $(f_t,g_t)\in \mathcal{I}(\delta_X(t))$ by definition. Assume that $(f,g)\in\mathcal{I}(\delta_X(t))$ and define
\[
h: X\to\mathbb{N}, x\mapsto\begin{cases}f(x)&\text{if }x\in \supp(f)\\
g(x) & \text{if }x\in\supp (g)\cap\supp(f)^c\end{cases}
\]
and
\[
\tilde{h}: X\to\mathbb{N}, x\mapsto\begin{cases}f(x)&\text{if }x\in \supp(f)\cap\supp(g)^c\\
g(x) & \text{if }x\in\supp (g)\end{cases}
\]
Note that the conditions $x\in\supp(f)\cap\supp(g)^c$ and $x\in\supp (g)\cap\supp(f)^c$ ensures that $h$ and $\tilde{h}$ are well defined, since the two supports might have a non-empty intersection. It also follows that $h$ coincides with $f$ on $\supp(f)$, whilst $\tilde{h}$ coincides with $g$ on $\supp(g)$. It follows from theses definitions that $f\leq h\restrict \supp(f)$, and $h\restrict\supp(g)\leq g$, and similarly $f\leq \tilde{h}\restrict \supp(f)$, and $\tilde{h}\restrict\supp(g)\leq g$, since $f$ is strictly below $g$ on $\supp(f) \cap \supp (g)$. Since we're assuming that $(f,g)\in\mathcal{I}(\delta_X(t))$, it must be the case that $h,\tilde{h}\in\delta_X(t)$, which by definition means in particular that $f_t\leq h\restrict\supp (f)= f$, and that $g_t\geq \tilde{h}\restrict \supp (g)=g$. We have thus shown that $(f_t,g_t)\leq (f,g)$ for the partial order defined above, and it follows that $(f_t,g_t)$ is indeed the minimum element of $\mathcal{I}(\delta_X(t))$ as claimed.
\item Let us now consider a general term $t$ as described above, and for each $i$ define $f_i$ and $g_i$ as above, i.e.
\begin{align*}
f_i(x)=\begin{cases}k_j^i & \text{if }\exists j\text{ s.th. }x=x_j^i\\
0& \text{else}\end{cases} & \hspace{1em} g_i(x)=\begin{cases}k_{j'}^i & \text{if }\exists j'\text{ s.th. }x=y_{j'}^i\\
0& \text{else}\end{cases}
\end{align*}
We claim that the minima of $\mathcal{I}(\delta_X(t))$ are the pairs $(f_i,g_i)$. Clearly, each such pair belongs to $\mathcal{I}(\delta_X(t))$ by definition. Now, assume $(f,g)\in \mathcal{I}(\delta_X(t))$ and define $h^{(0)}$ as we defined $h$ in the previous step. By definition $f\leq h^{(0)}\restrict \supp(f)$ and $h^{(0)}\restrict\supp(g)\leq g$, and thus we must have $h^{(0)}\in\delta_X(t)$, i.e. there must exist $i_0$ s.th. $f_{i_0}\leq h^{(0)}\restrict\supp(f_{i_0})$ and $h^{(0)}\restrict\supp(g_{i_0})\leq g_{i_0}$. If we also have $g\restrict \supp(g_{i_0})\leq g_{i_0}$, then we are done, since we will have shown $(f_{i_0},g_{i_0})\leq (f,g)$. If $g\restrict\supp(g_{i_0})\nleq g_{i_0}$, then there must exist $x_{i_0}\in \supp(f)\cap \supp(g)$ such that $g_{i_0}(x_{i_0})< g(x_{i_0})$ ($x_{i_0}$ cannot belong to $\supp(f)^c\cap \supp(g)$, or we would contradict that $h^{(0)}\restrict\supp(g_{i_0})\leq g_{i_0}$). Now define
\[
h^{(1)}(x)=\begin{cases} h^{(0)}(x)& \text{if }x\neq x_{i_0}\\
g(x_{i_0})&\text{else}
\end{cases}
\]
By construction $h^{(1)}\restrict\supp (g_{i_0})\nleq g_{i_0}$, and moreover $f\leq h^{(1)}\restrict \supp(f)$ and $h^{(1)}\restrict \supp(g)\leq g$ (since the only change from $h^{(0)}$ is $g(x_{i_0})$). Thus if $(f,g)\in\mathcal{I}(\delta_X(t))$ we must have $h^{(1)}\in\delta_X(t)$, and thus there must exist $i_1$ such that $f_{i_1}\leq h^{(1)}\restrict\supp(f_{i_1})$ and $h^{(1)}\restrict\supp(g_{i_1})\leq g_{i_1}$. We have established earlier that we can assume without loss of generality that $f_{i_0}$ and $f_{i_1}$ coincide on the intersection of their supports, and similarly for $g_{i_0}$ and $g_{i_1}$. It follows that $f_{i_1}\leq f\restrict\supp(f_{i_1})$, indeed on $\supp(f)$ $h^{(1)}$ only differs from $f$ at $x_{i_0}$ and if $x_{i_0}\in\supp(f_{i_1})$ then 
\[
f_{i_1}(x_{i_0})=f_{i_0}(x_{i_0})\leq h^{(0)}(x_{i_0})=f(x_{i_0})
\]
Thus if $g\restrict\supp(g_{i_1})\leq g_{i_1}$ we are done, and we will have shown $(f_{i_1},g_{i_1})\leq (f,g)$. Otherwise, we iterate our construction: assume that $g\restrict\supp(g_{i_1})\nleq g_{i_1}$, then there exist $x_{i_1}\in \supp(f)\cap \supp(g)$ such that $g_{i_1}(x_{i_1})<g(x_{i_1})$. We now define 
\[
h^{(2)}(x)=\begin{cases} h^{(1})(x)& \text{if }x\neq x_{i_1}\\
g_(x_{i_1})&\text{else}
\end{cases}
\]
This map has the property that $h^{(2)}\restrict \supp(g_{i_0})\nleq g_{i_0}$ and $h^{(2)}\restrict \supp(g_{i_1})\nleq g_{i_1}$ (since we include two counter-example points in its image), but also $f\leq h^{(2)}\restrict \supp(f)$ and $h^{(2)}\restrict \supp(g)\leq g$. Thus if $(f,g)\in\mathcal{I}(\delta_X(t))$ we must have $h^{(2)}\in\delta_X(t)$. We then repeat the procedure detailed above. It is clear that at each step $n$ we either find a pair $(f_{i_n},g_{i_n})\leq (f,g)$ and we have shown what we wanted, or else we construct a new map $h^{(n)}$ with the property that $h^{(k)}\restrict\supp(g_{i_k})\nleq g_{i_k}$, for every $1\leq k\leq n$, i.e. $(f_{i_k},g_{i_k})\nleq (f,g)$ for every $1\leq k\leq n$. Since there is a finite number $p$ of clauses in our term, this procedure must stop with either a pair $(f_{i_k},g_{i_k})\leq (f,g)$ (where $0\leq k\leq p$) or with a map $h^{(p)}$ partially sandwiched between $f$ and $g$ with the property that $(f_{i_k},g_{i_k})\nleq (f,g)$ for every $1\leq k\leq p$, and thus $h^{(p)}\notin\delta_X(t)$, which contradicts $(f,g)\in \mathcal{I}(\delta_X(t))$.
\item We have thus exhibited the following retraction $r_X: \pow\Bag (X)\to L\pow X$, defined by:
\[
\mathcal{V}\mapsto \bigvee_{(f,g)\text{ minima of }\mathcal{I}(\mathcal{V})}\left(\bigwedge_{x\in \supp f}\langle f(x)\rangle\{x\}\wedge\bigwedge_{y\in\supp g} \neg\langle g(y)\rangle\{y\}\right)
\]
and thus shown that $\delta_X$ is injective on finite sets, and thus everywhere by \ref{ch5:prop:approxdelta}.
\end{enumerate}
Since $\Bag$ is finitary and $\Set$ is locally presentable, we know from (see Theorem 4.2.12. of \cite{2010:AdamekSurvey}) that it has a terminal coalgebra, and  we thus get weak completeness of GML with respect to its multiset semantics via Theorem \ref{ch5:thm:weakcomp1}. However, since $\Bag$ does \emph{not} preserve finite sets, we cannot hope to prove strong completeness from weak-completeness via Theorems \ref{ch5:thm:deltahatsurj}-\ref{ch5:thm:strgcomp2}. Let us now see if Theorem \ref{ch5:thm:strgcomp3} can be used.
To explicitly characterise the adjoint transformation $\hat{\delta}: \Bag\uf\to \uf L$, we use the same observation as in Example \ref{ch5:ex:mod}, i.e. that $LA$ is a quotient of a free algebra, and thus ultrafilters on $LA$ can be seen as sets of generators $\langle k\rangle a, a\in A$ (again not all sets of generators define an ultrafilter since the axioms encoded by $L$ must hold). We can then write
\begin{align*}
\hat{\delta}(f)&\simeq\{\langle k\rangle a\in LA\mid f\in \delta_{\uf A}((L \eta_A)(\langle k\rangle a))\}\\
&=\{\langle k\rangle a\in LA\mid f\in \delta_{\uf A}(\langle k\rangle \eta_A(a))\}\\
&=\{\langle k\rangle a\in LA\mid \sum_{u\mid a\in u}f(u)\geq k\}
\end{align*}
Note that since $\Bag$ does \emph{not} preserve finite sets, we cannot use Lemma \ref{ch5:lem:epitomono} to infer properties of $\hat{\delta}$ on finite algebras from properties of $\delta$ on finite sets. In particular, the fact that $\delta$ is injective, does \emph{not} imply that $\hat{\delta}$ should be surjective on finite algebras. We can in fact explicitly check that $\hat{\delta}$ is \emph{not} surjective, even on finite algebras. To see this, let $A$ be a finite boolean algebra and consider any ultrafilter $u$ containing the set of generators $\{\langle k\rangle \top\mid k\in\mathbb{N}\}$. It is clear from the above characterisation that any such $u$ cannot have a pre-image since such a map $f: \Bag\uf A\to\mathbb{N}$ would then have to satisfy $\sum_{u\in \supp(f)}f(u)\geq k$ for each $k$. Since the support of each $f$ is bounded we get a contradiction, thus $\hat{\delta}$ fails to be surjective, and in particular we cannot use Theorem \ref{ch5:thm:strgcomp3} to strengthen weak completeness to strong completeness.

Note also $\Bag$ does not preserve cofiltered limits. Consider once again the sets $X_n$ and the maps $p_n$ defined in the previous Example \ref{ch5:ex:modf}. A coherent family $f_n\in \Bag(X_n)$ does not in general define an element of $\Bag(\lim_n X_n)=\Bag(\mathbb{N})$ for the same reason as above: whilst every element of $\Bag(\mathbb{N})$ must have a finite support, the family $f_n$ might have unbounded supports.
\end{example}

What characterises these three examples is that the functor $T$ providing the semantic domain is `too small' to accommodate the logic. Alternatively and equivalently, the functor $L$ defining the logics is too large in that it allows total logical descriptions of states (i.e. ultrafilters) which cannot be realized by $T$-models. We present a generic technique to try to solve this problem which essentially consists of enlarging the semantic domain by `completing' the functor $T$ in an appropriate fashion. We therefore call this procedure \textbf{semantic completion}. This technique seems to work well in cases where the only obstacle to applying Theorem \ref{ch5:thm:strgcomp3} is the non weak preservation of cofiltered limits by the semantic functor.

\paragraph{The finitary version of a functor, revisited.} Recall that the finitary version of a set functor $T:\Set\to\Set$ was defined as
\[
T_\omega: \Set\to\Set, X\mapsto\bigcup_{U\subseteq_\omega X} TU
\]
Following our work in Chapter 3, we know from Proposition \ref{ch3:prop:LanF0}, that if a functor $T$ is finitary and if we denote by $I$ the inclusion functor $\Inc: \Set_f\to\Set$, we have
\[
T=\Lan_I (T\circ \Inc)
\] 
More generally, as was shown in Propositions \ref{ch3:prop:presfunc} and \ref{ch3:prop:accesspresfunc}, if $T$ is any functor then $T_\omega$ is just the left Kan extension of $T\circ I$ w.r.t. the inclusion functor $\Inc: \Set_f\to\Set$ of the category of finite sets into the category of sets. Formally
\[
T_\omega=\Lan_I (T\circ \Inc)
\]
The functors $\Bag$ and $\cpowf$ which lead to the failures of strong completeness in the examples above can be seen either as finitary functors or as finitary versions of more general functors. The finitary version of a functor clearly preserves filtered colimits and we present here a very simple proof.

\begin{proposition}\label{ch5:prop:finitaryversion}
Let $\cat$ be a locally $\lambda$-presentable category, let $\Inc:\cat_\lambda\to \cat$ be the inclusion of the subcategory of $\lambda$-presentable objects, and let $T:\cat\to\cat$ be any functor. The $\lambda$-ary version of the functor given by
\[
T_\lambda=\Lan_{\Inc}(T\circ\Inc)
\]
preserves $\lambda$-filtered colimits.
\end{proposition}
\begin{proof}
Since $\cat$ is locally $\lambda$-presentable, it is in particular cocomplete, and the left Kan extension does indeed exist. Moreover it can be computed using the following coend formula (see \cite{MacLane} X.4):
\[
\Lan_{\Inc}(T\circ \Inc)A=\int^B \hom(\Inc B, A)\cdot (T\circ \Inc)B
\]
where $-\cdot (T\circ \Inc)B$ denotes the copower functor. Since a coend is a colimit, it will preserve the colimits preserved by the composition of $\hom(\Inc B, -)$ and $-\cdot(T\circ \Inc) B$. From the fact that 
\[
-\cdot(T\circ \Inc) B\dashv\hom((T\circ \Inc) B, -)
\]
it is clear that the copower functor preserves all colimits, and in particular cofiltered ones. So we need only show that $\hom(\Inc B, -)$ preserves $\lambda$-filtered colimits, but this follows immediately from the fact that $B$ is a $\lambda$-presentable object.
\end{proof}

\paragraph{The co-finitary version of a functor.}

By dualising the above procedure, we can create a version of any $\Set$-endofunctor which weakly preserves cofiltered limits, which we will call the \textbf{co-finitary version}\index{Co-finitary version} of the functor, for lack of a better name. Note that what we present now is \emph{not} the dual version of the finitary functor, it simply uses dual tools, namely the right Kan extension. The best property we will be able to show for the co-finitary version of a functor is \emph{weak} preservation of cofiltered limits in the category of $\Set$, which is a much more restricted result than Proposition \ref{ch5:prop:finitaryversion}, but sufficient for our purpose. 

\begin{definition}
A functor $G: \cat[A]\to \cat[B]$ is called \textbf{initial}\index{Initial functor} if for any object $B$ of $\cat[B]$ the comma category $G\downarrow B$ is non-empty and connected, i.e. if the following two conditions holds for any $B$
\begin{enumerate}[(i)]
\item there exists an object $A$ of $\cat[A]$ and a morphism $GA\to B$
\item any two objects in $B\downarrow G$, i.e. any two morphisms $GA\to B$ and $ GA'\to B$,  are related by a zigzag of morphisms.
\end{enumerate}
\end{definition}

Initial functors are used to compute limits as cofinal functors are used to compute colimits.

\begin{theorem}\label{ch5:thm:initial}
If $G: \cat[A]\to\cat[B]$ is initial and $\Diag: \cat[B]\to \cat[C]$ is a diagram such that $\lim \Diag G$ exists, then $\lim \Diag\simeq\lim \Diag G$.
\end{theorem}
\begin{proof}
Dual to the result on cofinal functors (see \cite{MacLane}, Chapter IX, section 3, Theorem 1).
\end{proof}

The notion of \textbf{right Kan extension}\index{Kan extension} is defined dually to that of left Kan extension and we refer the reader to Section \ref{ch3:sec:tools} or to see \cite{MacLane} X.3. for all the definitions. For our purpose, the following result will be sufficient.

\begin{proposition}\label{ch3:prop:KanExists2}
Let $\cat[B]\stackrel{F}{\longleftarrow}\cat[A]\stackrel{G}{\longrightarrow}\cat$ such that $\cat[A]$ is small and $\cat[B]$ is complete, then $F$ has a right Kan extension $\Ran_G F$ along any $G$.
\end{proposition}
\begin{proof}
See \cite{MacLane} X.3.2.
\end{proof}

Thus if we consider the inclusion $\Inc:\Set_f\to \Set$, the right Kan extension $\Ran_{\Inc}(T\circ \Inc)$ for any $T:\Set\to\Set$ always exists. Its explicit construction is given by
\[
\Ran_{\Inc}(T\circ \Inc) X=\lim \left((X\downarrow \Inc)\stackrel{Q}{\longrightarrow}\Set_f\stackrel{T\circ\Inc}{\longrightarrow}\Set\right)
\]
where $Q$ is the obvious projection functor. We will call $\Ran_{\Inc}(T\circ \Inc)$ the co-finitary version of $T$ and denote it by $T^\omega$. We can in fact restrict the index category of this limit as the following Proposition shows.

\begin{proposition}\label{ch5:prop:surjenough}
Let $X$ be a set, and let us consider the subcategory of $X\downarrow \Set_f$ whose objects consists of surjections $X\epi n, n\in\mathbb{N}$ and whose morphisms between $g: X\epi n, h: X\epi p$ are surjections $f:n\epi p$ such that $h=f\circ g$. Let us call this category $\mathbf{J}$, and let $I$ denote the obvious inclusion $\mathbf{J}\to X\downarrow \Set_f$, then $I$ is initial.
\end{proposition} 
\begin{proof}
We need to show that for any $g: X\to n$ there exist an arrow $I(h)\to g$ (in $X\downarrow\Set_f$) for some object $h$ of $\mathbf{J}$. For this, we consider the strong epi-mono factorisation of $g$ as $g=m\circ e$ and get the morphism $m: I(e)\to g$ in $X\downarrow\Set_f$:
\[
\xymatrix
{
& X\ar[dr]^{g}\ar@{->>}[dl]_{e}\\
n'\hspace{1ex}\ar@{>->}[rr]_{m} &  & n
}
\]
For the second condition, we need to show that given $e: X\epi p$ and $e': X\epi p'$ such that there exist $g:X\to n$ and morphisms $h: p\to n,  h': p'\to n$ such that $h\circ e=h'\circ e'=g$, we have a zigzag of morphism connecting $e$ and $e'$. For this we take an strong epi-mono factorisation of $g$ as $g=m\circ q$. Using the fact that the epimorphisms $e$ and $e'$ are strong we get fill-in morphisms $u,u'$ which make the following diagram commute and provide us with the zigzag of morphisms:
\[
\xymatrix
{
& X\ar@{->>}[dl]_{e}\ar@{->>}[dr]^{e'} \ar@{->>}[d]^{q} \\
p\ar[dr]_{h}\ar@{-->}[r]^{u} & n'\ar@{>->}[d]^{m} & p'\ar@{-->}[l]_{u'}\ar[dl]^{h'}\\
& n
}
\]
Since strong epimorphisms have the property that if $u\circ e$ is a strong epimorphism, then $u$ must be a strong epimorphism, $u,u'$ indeed provide the zigzag of morphism in $\mathbf{J}$ connecting $e$ and $e'$.
\end{proof}

Let us now show that the co-finitary version of a functor does what we want it to do, namely weakly preserve cofiltered limits.

\begin{proposition}\label{ch5:prop:cofinitaryversion}
Let $T:\Set\to\Set$ be an endofunctor, let $\Inc: \Set_f\to\Set$ be the inclusion functor of the subcategory of finite sets, and let $T^\omega$ be defined as $\Ran_{\Inc}(T\circ \Inc)$, then $T^\omega$ weakly preserves cofiltered limits. 
\end{proposition}
\begin{proof}
We write $\Ran_{\Inc}(T\circ \Inc)A$ as the end (see \cite{MacLane} X.4)
\[
\Ran_{\Inc}(T\circ \Inc)X=\int_n (T\circ\Inc)n^{\hom(X, \Inc n)}
\]
and use an argument totally analogous to Proposition \ref{ch5:prop:finitaryversion}: since an end is a limit, it weakly preserves the limits preserved by the composite of the contravariant functors $\hom(-, \Inc n)$ and $(T\circ\Inc) n^{(-)}$ where the latter is the power functor. Since
\[
(T\circ\Inc) n^{(-)}\vdash \hom(-,(T\circ\Inc) n)
\]
the power functor (which is contravariant) turns all colimits into limits, and in particular it turns (weak) filtered colimits into (weak) cofiltered limits. Thus we need to show that $\hom(-, \Inc n)$ weakly preserves cofiltered limits, i.e. that it turns cofiltered limits into weak filtered colimits. This is precisely what we have shown in Lemma \ref{ch5:lem:homcofiltered}.
\end{proof}

We now have a uniform way to turn a $\Set$-endofunctor into one that weakly preserves cofiltered limits, and is thus amenable to Theorems \ref{ch5:thm:strgcomp1}, \ref{ch5:thm:strgcomp2} and \ref{ch5:thm:strgcomp3}. Note that in general, this will result in loosing expressivity.
\begin{example}[Modal logic for trees with unbounded branching degree]\label{ch5:ex:treeOmegaStrComp}
It is easy to see that the functor $\polyFunc[\omega]$ does not weakly preserve cofiltered limits by considering a sequence of increasingly `large' $n$-tuples of elements of $X$, whose limit cannot be an $n$-tuple, but a map $\mathbb{N}\to X$. Indeed this is what makes $\hat{\delta}$ non-surjective, even on finite algebras. We apply the construction above to remedy this situation:
\begin{align*}
\polyFunc[\omega]^\omega (X)&=\Ran_{\Inc}(\polyFunc[\omega])(X)\\
&=\lim \left((X\downarrow \Inc)\stackrel{Q}{\longrightarrow}\Set_f\stackrel{\polyFunc[\omega]}{\longrightarrow}\Set\right)\\
& \simeq \{f: \mathbb{N}\to X\}=\hom(\mathbb{N},X)
\end{align*}
where finite sequences of elements of $X$ are identified with functions $f:\mathbb{N}\to X$ which eventually become constant. We can interpret the language defined in Example \ref{ch5:ex:treeOmega} in $\polyFunc[\omega]^\omega-$coalgebras in exactly the same way as we did in $\polyFunc[\omega]-$coalgebras, and the semantic transformation remains injective. The key difference is that the adjoint transpose $\hat{\delta}: \polyFunc[\omega]^\omega\uf\to\uf L$ is now surjective. For any $A$ in $\BA$, define the following section, $s_A: \uf LA\to \polyFunc[\omega]^\omega\uf A$
\[
s_A(t)=\begin{cases} \kop(u_1,\ldots,u_k)\text{ s.th. } a_i\in u_i, 1\leq i\leq k\text{ iff }\kop(a_1,\ldots,a_k)\in t \\
\hspace{20em} \text{if }\exists k, \kop(\top,\ldots,\top)\in t\\
(\langle\{\neg\kop(\top,\ldots,\top)\mid k\in\omega\}\rangle\eup,\ldots) \hspace{55pt}\text{else}
\end{cases}
\]
where $\langle-\rangle\eup$ denotes the ultrafilter generated by the set. Note that from the axiom (Tree 1), it is clear that if there exist a $k$ with $\kop(\top,\ldots,\top)\in t$, then it is unique. It is easy to see that $s_A$ is indeed a section of $\hat{\delta}_A$ (see Example \ref{ch5:ex:treef} for a definition of $\hat{\delta}$), which is thus surjective. Since it is surjective on any boolean algebra, we can apply the J\'{o}nsson-Tarski Theorem \ref{ch5:thm:jontarski} directly to conclude that the logic defined by $\Free\polyFunc[\omega]\Forg$ and the axioms (Tree 1), (Tree 2) is strongly complete with respect to $\polyFunc[\omega]^\omega-$coalgebras. 

However, the definition of $s_A$ above suggests that $\polyFunc[\omega]^\omega-$ is `too big' a functor. Indeed, most of the elements gained by enlarging $\polyFunc[\omega]$ to $\polyFunc[\omega]^\omega$ are not really necessary. In fact a much more economical alternative (and much more fruitful as we shall see at the end of the Chapter) is to extend $\polyFunc[\omega]$ by a singleton, i.e. we define
\[
\polyFunc[\omega]^*: \Set\to\Set, \polyFunc[\omega]^*X=\polyFunc[\omega] X+1 
\]
We can interpret $L$-algebras in $\polyFunc[\omega]^*$-coalgebras exactly as in Examples \ref{ch5:ex:treef} and \ref{ch5:ex:treeOmega}, and $\delta$ remains injective. Moreover, by mimicking the definition above, we can show that the adjoint transpose $\hat{\delta}: \polyFunc[\omega]^*\uf\to\uf L$ is surjective by defining the section
\[
s_A(t)=\begin{cases} \kop(u_1,\ldots,u_k)\text{ s.th. } a_i\in u_i, 1\leq i\leq k\text{ iff }\kop(a_1,\ldots,a_k)\in t \\
\hspace{19em} \text{if }\exists k, \kop(\top,\ldots,\top)\in t\\
^* \hspace{185pt}\text{else}
\end{cases}
\]
where $*$ is the new singleton in the coproduct defining $\polyFunc[\omega]^*$. Since there exists only one problematic ultrafilter, namely the one generated by $\{\neg \kop(\top,\ldots,\top)\mid k\in\omega\}$, it is clear that $\hat{\delta}_A\circ s_A=\id_{\uf LA}$. It follows from the J\'{o}nsson-Tarski Theorem \ref{ch5:thm:jontarski} that the logic defined by $L$ is strongly complete with respect to $\polyFunc[\omega]^*-$coalgebras. 
\end{example}

\begin{example}(Countable semantics of classical modal logic)\label{ch5:ex:modc}
Recall from Examples \ref{ch5:ex:mod} and \ref{ch5:ex:modf} that we have the traditional semantic of classical modal logic which is strongly complete but not expressive and given in terms of the covariant powerset functor $\cpow$, and the image finite semantics which is expressive but not strongly complete and given in terms of the finite powerset functor $\cpow_\omega$. If we apply the construction above to this functor we get (by using Proposition \ref{ch5:prop:surjenough}) that
\begin{align*}
\cpow^\omega(X)&=\Ran_{\Inc}(\cpow_\omega)(X)\\
&=\lim \left((X\downarrow \Inc)\stackrel{Q}{\longrightarrow}\Set_f\stackrel{\cpow_\omega}{\longrightarrow}\Set\right)\\
& =\cpow_c (X)
\end{align*}
where $\cpow_c(X)$ is the set of countable subsets of $X$. We can keep the definition of the semantic transformation $\delta$ from Example \ref{ch5:ex:mod} since $\cpow_\omega(X)\subseteq \cpow_c(X)$ for all sets $X$. Moreover, since $\cpow_c$ preserves cofiltered limits by Proposition \ref{ch5:prop:cofinitaryversion}, and since the properties of the semantic transformation and its adjoint are the same as in Example \ref{ch5:ex:mod}, we can also use Theorem \ref{ch5:thm:strgcomp3} to conclude that $\cpow_c$ provide a strongly complete semantic to classical modal logic. Again, the logic fails to be expressive.
\end{example}
\begin{example}(Graded Modal Logic)\label{ch5:ex:GMLomega}
We apply the co-finitary construction to the finitary bag functor which has provided the semantic for Graded Modal Logic in Example \ref{ch5:ex:GMLf}. There were two obstacles to strong completeness in this example: (1) the finitary bag functor $\Bag$ does not preserve cofiltered limits, and (2) the adjoint semantic transformation $\hat{\delta}$ is not surjective on finite sets. We will now show that by fixing the first problem via the co-finitary construction, we also fix the second. Let us first compute $\Bag^\omega$:
\begin{align*}
\Bag^\omega(X)&=\Ran_{\Inc}(\Bag)(X)\\
&=\lim \left((X\downarrow \Inc)\stackrel{Q}{\longrightarrow}\Set_f\stackrel{\Bag}{\longrightarrow}\Set\right)\\
& =\{f: X\to\mathbb{N}\cup\{\infty\}\mid \supp(f)\subseteq_c X\}
\end{align*}
where $\subseteq_c$ means `is a countable subset of'. Since $\Bag(X)\subseteq \Bag^\omega(X)$ for any set $X$, we can keep the definition of the semantic transformation $\delta$ from Example \ref{ch5:ex:GMLf}. To see that $\Bag^\omega$ addresses the second problem raised above, we show that $\hat{\delta}$ is an isomorphism on finite boolean algebras. Since $\delta$ is essentially unchanged, we can recycle $\hat{\delta}:\Bag^\omega\uf A\to\uf LA$ too, which as we have shown above is defined by:
\[
\hat{\delta}(f)=\{\kop b\mid \sum_{u\in \eta_A(b)} f(u)\geq k\}
\]
where $\eta$ is the unit of the adjunction $\uf\dashv\pow$. For a finite boolean algebra $A$, every ultrafilter on $A$ is principal, so let us denote by $a_u$ the elements of $A$ which define an ultrafilter $u=\up a_u$. A map $f: \uf A\to\mathbb{N}\cup\{\infty\}$ can be seen as a map $f:\{a_u\}_{u\in\uf A}\to\mathbb{N}\cup\{\infty\}$ and we then have:
\[
\hat{\delta}_A(f)=\{\langle k \rangle b\mid \sum_{a_u\leq b}f(a_u)\geq k \}
\]
It is straightforward to check that this map is injective. To see that it is surjective, we define a section $s: \uf LA \to\Bag \uf A$ of $\hat{\delta}_A$ as follows. Recall that an ultrafilter of $LA$ is determined uniquely by the generators $\kop b, b\in A$ it contains. Thus we will alternatively write $w\in \uf LA$ as a set $\{\langle k_i \rangle b_i\mid i\in I\}$ of generators of $LA$. We now define
\begin{align*} s(w)=\lambda a_u. 
\begin{cases}
\max\left(\{k_i\mid \exists i\in I. b_i=a_u\}\cup\{0\}\right) & \text{if it exists}\\
\infty & \text{else}
\end{cases}
\end{align*}
Note that since the $k_i$ need not be bounded, we do need the $\{\infty\}$ option provided by $\Bag^\omega$ for this map to be well defined. Let us check that it is indeed a section. We start with a collection $\{\kop[k_i]a_i\}_{i\in I}$ defining an ultrafilter in $LA$ and for notational clarity we write $f_w=s(w)$ with $w$ described by the set $\{\langle k_i\rangle a_i\mid i\in I\}$. We need to show that $\hat{\delta}_A(f_w)=w$ i.e. that the sets of generators describing $\hat{\delta}_A(f_w)$ and $w$ are equal, in other words that
\[
\{\kop[l] b\mid \sum_{a_u\leq b}f_w(a_u)\geq l\}=\{\kop[k_i]a_i\}_{i\in I}
\]
Let us first show the inclusion from right to left, i.e. that each $\kop[k_i]a_i\in\hat{\delta}_A(f_w)$. We need the following lemma.
\begin{lemma}\label{lem:GMLhelp}
In the situation described above, if $\langle k\rangle a$ belongs to an ultrafilter of $\uf LA$, then this ultrafilter must also contain a finite set of terms $\{\langle k_j\rangle a_{u_j}\}_{1\leq j\leq m}$, $u_j\in \uf A$, such that 
\[
\sum_{j=1}^m k_j\geq k
\]
\end{lemma}
\begin{proof}
The first step is to show that there exist elements of the type $a_u$ below $a$. So let us assume that $\up a$ is not a principal ultrafilter, i.e. there must exist a finite set $\{a_i\mid 1\leq i\leq n\}$ of elements of $A$ such that neither $a_i$ nor $\neg a_i$ belong to $\up a$. Each selection function
\[
h: n\mono \{a_i\}_{1\leq i\leq n}\cup\{\neg a_i\}_{1\leq i\leq n}
\] 
of either $a_i$ or $\neg a_i$ for each $1\leq i\leq n$ completes the proper filter  $\up a$ into an ultrafilter $u_h=\up (a\wedge \bigwedge_i h(i))=\up a_{u_h}$, and $a_{u_h}\leq a$. Let us write $H$ for the set of all such selection functions. We have, 
\begin{align*}
& \bigvee_{h\in H} a_{u_h}= \bigvee_{h\in H} (a\wedge \bigwedge_i h(i))=a\wedge\bigvee_{h\in H}\bigwedge_i h(i)=a\wedge \top=a
\end{align*}
Let us now focus on the operators. There are two possibilities, either $\kop a$ is the `largest' such formula in the ultrafilter, or there is a larger integer $k'>k$ such that $\kop[k']a$ is in the ultrafilter. In the first case, since $a=\bigvee_h a_{u_h}$ and since $a_{u_h}\wedge a_{u_{h'}}=\bot$ whenever $h\neq h'$ we can use an argument totally analogous to Step 1 of Example \ref{ch5:ex:GMLf} (using axioms GML 1-2, 4) to show that there must exist a partition of $k$ into a collection $\{k_h\}_{h\in H_0}$ (i.e. the $k_h$s add up to $k$) where $H_0\subseteq H$ such that for every $h\in H_0$, $\kop[k_h]a_{u_h}$ is in the ultrafilter, and the result follows.

If there exists $k'>k$ such that $\kop[k'] a$ is in the ultrafilter then there are two possibilities: either there exists a largest such $k'$, in which case we can apply the same reasoning as above to $\kop[k']a$, or there is no largest such $k'$ in which case there are also terms of the type $\kop[k_h]a_{u_h}$ for arbitrary large $k_h$s by the same argument as above instantiated at each $k'$, in which case we will have $\infty\geq k$.
\end{proof}
From this Lemma it follows that to any $\kop[k_i]a_i$ there must exist a collection of formulas $\{\langle k_{i_j}\rangle a_{u_j}\}_{1\leq j\leq m}$ in the ultrafilter, where each $a_{u_j}$ defines a principal ultrafilter. From the definition of $f_w$, it then follows that
\[
\sum_{a_u\leq a_i} f_w(a_u)\geq\sum_{j=1}^m k_{i_j}\geq k_i
\]
We therefore have $\kop[k_i] a_i\in \hat{\delta}_A(f_w)$. 

Let us now show the opposite inclusion. If $\kop[l] b\in \hat{\delta}_A(f_w)$, then by definition there exist a (finite) collection of elements $a_{u_k}\leq b, 1\leq k\leq p$ such that $f_w(a_{u_k})> 0$ and $\sum_{k}f_w(a_{u_k})\geq l$. By definition of $f_w$, this means that for each $k$, $a_{u_k}=a_i$ for some $i\in I$, and since $f_w$ picks the largest integer $l_k$ associated with $a_{u_k}$, this means that $\kop[l_k] a_{u_k}$ is in the ultrafilter. The condition that $\kop[l] b\in \hat{\delta}_A(f_w)$ takes the form $\sum_k l_k\geq l$. It is easy to see that for any ultrafilter $u,v\in \uf A$, $a_u\wedge a_v=\bot$. We use this fact and the axiom (GML \ref{ch5:ax:GML4}) to show that $\kop[l]b\in w$ as follows. To fix ideas the let consider three elements $\kop[l_1]a_{u_1},\kop[l_2]a_{u_2},\kop[l_3]a_{u_3}$, the result can straightforwardly be generalized to any number $n$ of such elements. We have
\begin{align*}
&\kop[l_1+l_2+l_3](a_{u_1}\vee a_{u_2}\vee a_{u_3})\\
=&\bigvee_{j=0}^{l_1+l_2+l_3}\hspace{-1ex}\kop[j]((a_{u_1}\vee a_{u_2}\vee a_{u_3})\wedge a_{u_1})\wedge\kop[l_1+l_2+l_3-j]((a_{u_1}\vee a_{u_2}\vee a_{u_3})\wedge a_{u_1}^c)\\
=&\bigvee_{j=0}^{l_1+l_2+l_3}\hspace{-1ex}\kop[j]a_{u_1}\wedge\kop[l_1+l_2+l_3-j](a_{u_2}\vee a_{u_3})\\
=&\bigvee_{j=0}^{l_1+l_2+l_3}\hspace{-1ex}\kop[j]a_{u_1}\wedge\left(\bigvee_{j'=0}^{l_1+l_2+l_3-j}\kop[j']a_{u_2}\wedge\kop[l_1+l_2+l_3-j-j']a_{u_3}\right)\\
=&\bigvee_{j=0}^{l_1+l_2+l_3}\bigvee_{j'=0}^{l_1+l_2+l_3-j}(\kop[j]a_{u_1}\wedge\kop[j']a_{u_2}\wedge\kop[l_1+l_2+l_3-j-j']a_{u_3})
\end{align*}
In particular, we therefore have
\[
\kop[l_1]a_{u_1}\wedge\kop[l_2]a_{u_2}\wedge\kop[l_3]a_{u_3}\leq \kop[l_1+l_2+l_3](a_{u_1}+a_{u_2}+a_{u_3})
\]
More generally, using exactly the same argument we have
\[
\bigwedge_k \kop[l_k]a_{u_k}\leq \left\langle\sum_k l_k\right\rangle(\bigvee_k a_{u_k})
\]
Since each $\kop[l_k]a_{u_k}$ is in the ultrafilter and that there are finitely many of them, their meet is in the ultrafilter and we then have that
\[
\bigwedge_k \kop[l_k]a_{u_k}\leq \left\langle\sum_k l_k\right\rangle(\bigvee_k a_{u_k})\leq \langle l\rangle (\bigvee_k a_{u_k})\leq \langle l\rangle b
\]
and thus $\langle l\rangle b$ belongs to the ultrafilter $w$ defined by $\{\langle k_i\rangle a_i \mid i\in I\}$. We have thus proved that 
\[
\hat{\delta}_A(s(w)=w
\]
i.e. that $s$ is indeed a section and that in consequence $\hat{\delta}$ is surjective on finite boolean algebras.

We have now shown that $\hat{\delta}$ is an iso on finite algebra, i.e. that GML is strongly complete over finite algebras. Moreover, since $\Bag^\omega$ weakly preserves filtered colimits by construction, it follows that we can now apply Theorem \ref{ch5:thm:strgcomp3} and conclude that GML is strongly complete with respect to $\Coalg(\Bag^\omega)$.
\end{example}

\section{Completeness-via-canonicity}\label{ch5:sec:compCanon}

\subsection{J\'{o}nsson-Tarski vs canonical extensions}

Strong completeness via the coalgebraic J\'{o}nsson-Tarski Theorem \ref{ch5:thm:jontarski} is the bedrock of coalgebraic completeness-via-canonicity, but so far we have made no use of canonicity as it was developed in Chapter 2. To do this, we need first of all to restrict our attention to coalgebraic logics over $\DL$ or $\BA$, which we will now assume until the end of this Chapter. Secondly, if we are to make use of the algebraic theory of canonicity developed in Chapter 2 in conjunction with the theory of strong completeness developed in the previous Section \ref{ch5:sec:strongComp}, we need to relate two constructions which are a priori unrelated: (1) the canonical extension of a DLE (or BAE) and (2) the J\'{o}nsson-Tarski extension defined by Theorem \ref{ch5:thm:jontarski}. Whilst these two entities share the same carrier, namely the canonical extension of a distributive lattice or of a boolean algebra, there is no a priori reason for their structure maps to be the same. Formally, given a functor $L:\DL\to\DL$ defining DLEs, if the logic defined by $L$ is interpreted in $T$-coalgebras for $T:\Pos\to\Pos$ via a semantic transformation $\delta: L\ups\to\ups T$ whose adjoint transpose has right inverses, then for a given $\alpha:LA\to A$ in $\Alg_{\DL}(L)$ what is the relationship between 
\[
LA\ce\xto{\alpha\ce}A\ce
\]
the canonical extension of the DLE $(A,\alpha)$ (defined in Section \ref{ch2:subsec:CanExtLAlg}), and the J\'{o}nsson-Tarski extension given by
\[
LA\ce=L\ups\pf A\xto{\delta_{\pf A}}\ups T\pf A\xto{\ups \hat{\delta}_A\inv}\ups\pf LA\xto{\ups \pf \alpha}\ups \pf A=A\ce
\]
In particular, when can we say that these two structure maps define the same $L$-algebra? Answering this question is the purpose of this section and the last technical obstacle to a coalgebraic theory of completeness-via-canonicity. For notational convenience we will denote the J\'{o}nsson-Tarski extension of a structure map $\alpha: LA\to A$ by $\tilde{\alpha}\ce$, i.e. $\tilde{\alpha}\ce=\ups \pf \alpha\circ \ups \hat{\delta}_A\inv\circ \delta_{\pf A}$ and we want to understand the relationship between $\alpha\ce$ and $\tilde{\alpha}\ce$.

We can answer this question in two different ways depending on whether we are dealing with DLEs or BAEs. Let us first make the nomenclature absolutely precise. By definition, if $L$ defines a variety of DLEs or BAEs whose expansion symbols belong to a signature $\Sigma$, then a structure morphism (in $\DL$ of $\BA$) $\alpha: LA\to A$ can be seen as a collection of \emph{functions} $\alpha_\heartsuit: A^n\to A$ for each $\heartsuit\in\Sigma$ of arity $n$. Thus $\alpha\ce$ defines a collection of maps $\alpha\ce_\heartsuit: (A\ce)^n\to A\ce$ whose are by definition the canonical extensions of each $\alpha_\heartsuit$. We will say that $\alpha$ is \textbf{smooth}\index{Smooth} if each $\alpha_\heartsuit,\heartsuit\in\Sigma$ is smooth. Similarly, we will say that the map $\alpha\ce$ or the map $\tilde{\alpha}\ce$ are $(\tau_1,\tau_2)$-continuous if each $\alpha\ce_\heartsuit$ or $\tilde{\alpha}\ce_\heartsuit$ is $(\tau_1,\tau_2)$-continuous in each argument.

In the case of positive logics (i.e. over $\DL$) we will use the following result.

\begin{lemma}\label{ch5:lem:kaddsmoothdelta}
Let $L:\DL\to\DL$ define a variety of DLEs whose expansions are (anti-)$k$-additive or (anti-)$k$-multiplicative, and assume that $L$-algebras can be interpreted in $T$-coalgebras for $T:\Pos\to\Pos$ via a semantic transformation $\delta: L\ups \to \ups T$ whose adjoint transpose has right inverses. If $\delta$ defines maps which are completely (anti-)$k$-additive or completely (anti-)$k$-multiplicative in every argument, then for any $L$-algebra $(A,\alpha)$, $\tilde{\alpha}\ce$ is $(\sigma,\gamma)$-continuous.
\end{lemma}
\begin{proof}
By definition of the J\'{o}nsson-Tarski extension, since $\ups \pf \alpha$ and $\ups \hat{\delta}_A\inv$ are just inverse images and therefore preserve all meets and all joins, the preservation properties of $\tilde{\alpha}_\heartsuit\ce$ are determined by $\delta_{\pf A}^\heartsuit$. For the sake of simplicity we assume that the extension symbol is unary, but the proof clearly extends to $n$-ary operators in a straightforward way. Assume that $\delta_{\pf A}^\heartsuit$ is completely $k$-additive, then by the comment we just made it follows that $\tilde{\alpha}\ce_\heartsuit$ is also completely $k$-additive. It follows immediately that $\tilde{\alpha}\ce_\heartsuit$ preserves up-directed joins, and is therefore $(\gamma\eup,\gamma\eup)$-continuous. 

Let us now show that $\tilde{\alpha}\ce_\heartsuit$ is also $(\sigma\edown,\gamma\edown)$-continuous. Since it is completely $k$-additive we have for every $u\in O(A)$
\begin{align*}
\tilde{\alpha}\ce_\heartsuit(u)&\stackrel{1}{=}\tilde{\alpha}\ce_\heartsuit(\bigvee A\cap\hspace{-1ex}\down u)\\
&\stackrel{2}{=}\bigvee\{\tilde{\alpha}\ce_\heartsuit(\bigvee U)\mid U\in \pow_k(A\cap \hspace{-1ex}\down u)\}\\
&\stackrel{3}{=}\bigvee\{\tilde{\alpha}\ce_\heartsuit a\mid u\leq a\in A\}\\
&\stackrel{4}{=}\bigvee\{\alpha\ce_\heartsuit a\mid u\leq a\in A\}\\
&=\alpha\ce_\heartsuit(u)
\end{align*}
where (1) is by definition of open elements, (2) is by definition of $k$-additivity, (3) is because the set $A\cap\hspace{-1ex}\down u$ is up-directed, and (4) is because both the J\'{o}nsson-Tarski and the canonical extensions are extensions, i.e. they agree with $\alpha$ on $A$. Thus $\tilde{\alpha}\ce_\heartsuit$ and $\alpha\ce_\heartsuit$ agree on open elements. Now we use the fact that since $\alpha_\heartsuit$ is assumed to be $k$-additive, $\alpha_\heartsuit\ce$ is completely $k$-additive by Theorem \ref{ch2:thm:complkadd}, and thus $(\sigma\edown,\gamma\edown)$-continuous by Theorem \ref{ch2:thm:kaddSmooth}. By using the dual version of Lemma \ref{ch2:lem:ScottTopolBasis}, we find that the topology $\gamma\edown$ has a basis given by $\{\down m\mid m\in M_\omega(A\ce)\}$ where $M_\omega(A\ce)$ is the set of finite meets of completely meet irreducible elements of $A\ce$. Since $\alpha_\heartsuit\ce$ is $(\sigma\edown,\gamma\edown)$-continuous, it means that for every $m\in M_\omega(A\ce)$ there exist $u\in O(A)$ such that $\alpha_\heartsuit\ce(u)=m$. But since $\alpha\ce_\heartsuit$ and $\tilde{\alpha}\ce_\heartsuit$ agree on open elements this means that for every $m\in M_\omega(A\ce)$ there exist $u\in O(A)$ such that $\tilde{\alpha}_\heartsuit\ce(u)=m$. In other words the inverse image under $\tilde{\alpha}_\heartsuit\ce$ of any element in a basis of $\gamma\edown$ is in $\sigma\edown$, and it follows that $\tilde{\alpha}_\heartsuit\ce$ must be $(\sigma\edown,\gamma\edown)$-continuous. By gathering our results, it is clear that $\tilde{\alpha}_\heartsuit\ce$ is $(\sigma,\gamma)$-continuous. The proof for anti-$k$-additivity and (anti-)$k$-multiplicativity follow along exactly the same lines.
\end{proof}

We can now give a very general criteria for the canonical and the J\'{o}nsson-Tarski extensions to coincide:

\begin{theorem}[Completeness-via-canonicty for DLEs]\label{ch5:thm:CanExtJonTarksiExt}
Let $L:\DL\to\DL$ define a variety of DLEs whose expansions are (anti-)$k$-additive or (anti-)$k$-multiplicative, and assume that $L$-algebras can be interpreted in $T$-coalgebras for $T:\Pos\to\Pos$ via a semantic transformation $\delta: L\ups \to \ups T$ whose adjoint transpose has right inverses. If $\delta$ defines maps which are completely (anti-)$k$-additive or completely (anti-)$k$-multiplicative in every argument, then for any $L$-algebra $(A,\alpha)$:
\[
(A\ce,\alpha\ce)=(A\ce,\tilde{\alpha}\ce)
\]
\end{theorem}
\begin{proof}
The result follows from the preceding Lemma \ref{ch5:lem:kaddsmoothdelta} and Corollary \ref{ch2:cor:smoothUnique} since the assumptions of the Theorem guarantee that both $\alpha\ce$ and $\tilde{\alpha}\ce$ will be smooth, and thus equal.
\end{proof}

In the case of BAEs, i.e. if we assume that the base category on which the syntax building functors are defined is $\BA$, we can in fact relax the smoothness condition, i.e. the $(\sigma,\gamma)$-continuity of $\delta$, significantly. In fact, we only require $(\sigma,\gamma\eup)$- or $(\sigma,\gamma\edown)$-continuity.  The following theorem also shows that we do not necessarily need the canonical extension and the J\'{o}nsson-Tarski extension to coincide. 


\begin{theorem}[Completeness-via-canonicty for BAEs]\label{ch5:thm:CanExtJonTarksiExtBAE}
Let $L:\BA\to\BA$ define a variety of BAEs whose expansions are isotone, and assume that $L$-algebras can be interpreted in $T$-coalgebras for $T:\Set\to\Set$ via a semantic transformation $\delta: L\pow\to\pow T$ whose adjoint transpose has right inverses. If $\delta$ defines maps which are all $(\sigma,\gamma\eup)-$continuous or all $(\sigma,\gamma\edown)-$continuous, then if the canonical extension of $(A,\alpha)$ belongs to a variety, so does its J\'{o}nsson-Tarski extension.
\end{theorem}
\begin{proof}
 
Let $V$ be a set of variables, and $q:\Free_L\Free V=\init[(L(-)+\Free V)]\epi Q$ be a regular epi (in $\Alg_{\BA}(L)$) which defines a variety by orthogonality (see Definition \ref{ch1:def:variety}). By definition, it coequalizes a pair $e_1,e_2: E\to \Free_L\Free V$  of `equations'. If the canonical extension $(A\ce,\alpha\ce)$ belongs to this variety, then it too must coequalize $e_1,e_2$, i.e. $\lsem e_1(x)\rsem_{\alpha\ce}=\lsem e_2(x)\rsem_{\alpha\ce}$ for every `equation' $x\in E$, where $\lsem-\rsem_{\alpha\ce}$ denotes the catamorphism into the $(A,\alpha)$. Since we're in $\BA$ this can be equivalently restated as
\[
\lsem \neg(e_1(x)\leftrightarrow e_2(x))\rsem_{\alpha\ce}=\bot
\]
From the fact $\delta$ defines $(\sigma,\gamma\eup)$ maps and that $\pow\uf \alpha$ and $\pow\hat{\delta}_A\inv$ are $(\gamma\eup,\gamma\eup)$-continuous, it follows that we $\tilde{\alpha}\ce$ is $(\sigma,\gamma\eup)$-continuous and from Theorem \ref{ch2:thm:topolChar} it follows that $ \tilde{\alpha}\ce\leq\alpha\ce$ pointwise, since $\alpha\ce$ defines the largest such extension of $\alpha$. It follows by a simple induction on the modal depth of terms in $\Free_L\Free V$ that
\[
\lsem-\rsem_{\tilde{\alpha}\ce}\leq \lsem-\rsem_{\alpha\ce}
\]
pointwise, and in particular we have
\[
\lsem \neg(e_1(x)\leftrightarrow e_2(x))\rsem_{\tilde{\alpha}\ce}\leq \lsem \neg(e_1(x)\leftrightarrow e_2(x))\rsem_{\alpha\ce}=\bot
\]
Hence $\lsem -\rsem_{\tilde{\alpha}\ce}$ also coequalizes $e_1,e_2$ and thus $(A\ce,\tilde{\alpha}\ce)$ also belongs to the variety defined by $q$.

The case of $(\sigma,\gamma\edown)-$continuous maps is treated dually by using Corollary \ref{ch2:cor:topolCharDual} and $e_1(x)=e_2(x)$ iff $e_1(x)\leftrightarrow e_2(x)=\top$.
\end{proof}

\begin{remark}
Whilst the previous Theorem \ref{ch5:thm:CanExtJonTarksiExtBAE} is much stronger topologically than Theorem \ref{ch5:thm:CanExtJonTarksiExt}, we have not found any natural preservation property on $\delta$ which implies $(\sigma,\gamma\eup)$ or $(\sigma,\gamma\edown)$-continuity. Of course, any $\delta$ which is completely (anti-)$k$-additive or completely (anti-)$k$-multiplicative maps in each argument satisfy this requirement, but it is a priori a much weaker constraint to satisfy.
\end{remark}

For $L$-algebras defining DLEs with smooth extensions or BAEs we will rely on the Theorems above to directly prove completeness-via-canonicity for coalgebraic logics. For more complicated functors, such as those defined by the nabla style of coalgebraic logics, we will proceed indirectly by using the translation techniques developed in Chapter 4. We start with the `straightforward' case.


\subsection{$L$-algebras defining DLEs with smooth extensions}

We work in $\DL$ but everything that follows can easily be specialized to the case of endofunctors on $\BA$. Let $L: \DL\to\DL$ be a functor defining a variety of DLEs with smooth extensions. In all concrete examples we will use DLEs whose extensions obey the conditions of Theorem \ref{ch2:thm:kaddSmooth} in each of their arguments. In other words for all practical purposes we can assume $L$ to define only $n$-ary modalities which are (anti-)$k$-additive or (anti-)$k$-multiplicative in each of their arguments, and thus smooth. We will write $\Free_L\dashv\Forg_L$ for the adjunction $\DL\to\Alg_{\DL}(L)$ and $\eta_L$ for the associated unit.

\begin{theorem}[Coalgebraic completeness-via-canonicity for smooth DLEs]\label{ch5:thm:strongcompcan}
Let $L:\DL\to\DL$ be a regular epi-preserving finitary functor defining a variety of DLEs with smooth extensions, $T:\Pos\to\Pos$ and $\delta: L\ups\to\ups T$ be such that the adjoint transpose $\hat{\delta}: T\pf\to \pf L$ has right-inverses, and let $\Ax\subseteq \Forg\Free_L \Free V\times \Forg\Free_L \Free V$ be any set of canonical equations defined over a set $V$ of variables. If $\delta_{\pf A}$ defines maps which are completely (anti)-$k$-additive or completely (anti-)$k$-multiplicative, then logic defined by $L$ together with the axioms of $\Ax$ is strongly complete with respect to the class of $T$-coalgebras on which the equations of $\Ax$ are valid.
\end{theorem}
\begin{proof}
Given sets of formulas $\Phi,\Psi$ such that $\Phi\nvdash_{\mathrm{ML}+\Ax}\Psi$, we need to find a model satisfying all the formulas of $\Phi$ and none of the formulas of $\Psi$ on which the equations of $\Ax$ are valid. We start the construction as in in Sections \ref{ch1:subsec:algsem} and \ref{ch4:sec:prooftrans} by taking the adjoint transpose of the maps $e_1,e_2$ associated with the set of axioms $\Ax$, and then taking their coequalizer.
\[
\xymatrix
{
\Free \Ax \ar@<3pt>[r]^{\hat{e}_1} \ar@<-2pt>_{\hat{e}_2}[r] & \Free_L \Free V  \ar@{->>}[r]^{q_{\Ax}} & Q_{\Ax}
}
\]
This defines the quotient under the smallest equivalence class (in $\DL$) generated by the equations of $\Ax$. We can explicitly recover this equivalence relation by taking the kernel pair of $q_{\Ax}$. By Proposition \ref{ch1:prop:fullInvClosure} we know that we can get the fully invariant closure of this equivalence relation. So let us assume that we have the following exact sequence
\[
\xymatrix
{
F_{\Ax} \ar@<3pt>[r]^{e'_1} \ar@<-2pt>_{e'_2}[r] & \Free_L \Free V  \ar@{->>}[r]^{q_{F_{\Ax}}} & \mathscr{L}_{\Ax}
}
\]
where $F_{\Ax}$ is the fully invariant closure of the equivalence relation defined by $q_{\Ax}$ (and the kernel pair of $q_{F_{\Ax}}$). The algebra $\mathscr{L}_{\Ax}$ is usually known as the \textbf{Lindenbaum-Tarski algebra }\index{Lindenbaum-Tarski algebra} for the logic defined by $L$ with additional axioms in $\Ax$. It is the quotient of the logic $\Free_L V$ under provable equivalence using the axioms of $\Ax$, and $F_{\Ax}$ can be thought of as the algebra of pairs of terms which are provably equivalent using equationalreasoning and $\Ax$. By construction, $\mathscr{L}_{\Ax}$ is in the variety defined by the axioms in $\Ax$ (in fact it completely characterizes this variety, see Proposition \ref{ch1:prop:Q*}), and since all the axioms are assumed to be canonical, this means that the canonical extension of $\mathscr{L}_{\Ax}$ also belongs to this variety. Since we're assuming that $\hat{\delta}_{\mathscr{L}_{\Ax}}$ has a right-inverse $h: \pf L\mathscr{L}_{\Ax}\to T\pf \mathscr{L}_{\Ax}$, the J\'{o}nsson-Tarski extension of $\mathscr{L}_{\Ax}$ exists by Theorem \ref{ch5:thm:jontarski}. Since $\mathscr{L}_{\Ax}$ inherits the canonical valuation from $\Free_L V$, we can consider the algebraic semantics of formulas in $\ups\pf \mathscr{L}_{\Ax}$ which is given by:
\[
\xymatrix@C=10ex
{
L\Free_L \Free V+\Free V\ar@/^2pc/[rr]^{K \lsem - \rsem_{\ups\pf\mathscr{L}_{\Ax}}+\id_{\Free V}} \ar[ddd]_{\langle-\rangle^L_V+\eta^L_{\Free V}}\ar[r]^{Lq_{F_{\Ax}}+\id_{\Free V}} & L\mathscr{L}_{\Ax}+\Free V\ar[r]^{L\eta_{\mathscr{L}_{\Ax}}+\id_{\Free V}}\ar[ddd]_{\alpha+\eta^L_{\Free V}} \ar[ddr]_{\eta_{L\mathscr{L}_{\Ax}}} & L \ups\pf\mathscr{L}_{\Ax}+\Free V  \ar[d]^{\delta_{\pf \mathscr{L}_{\Ax}}+\id_{\Free V}}\\
& & \ups T \pf \mathscr{L}_{\Ax} +\Free V \ar[d]^{\ups h+\eta_{\Free V}}\\
& & \ups \pf L \mathscr{L}_{\Ax} +\ups\pf \Free V\ar[d]^{\ups\pf\alpha+\ups\pf(\eta^L_{\Free V})} \\
\Free_L \Free V\ar[r]_{q_{F_{\Ax}}}\ar@/_2pc/[rr]_{\lsem - \rsem_{\ups\pf\mathscr{L}_{\Ax}}} & \mathscr{L}_{\Ax}\ar[r]_{\eta_{\mathscr{L}_{\Ax}}} & \ups \pf \mathscr{L}_{\Ax}
}
\]
where $\langle-\rangle^L_V$ and $\alpha$ are the structure maps of $\Free_L\Free V$ and $\mathscr{L}_{\Ax}$ respectively. Note that the algebraic semantic map $\lsem - \rsem_{\ups\pf\mathscr{L}_{\Ax}}$ is just an incarnation of the truth lemma, i.e. the denotation of a formula is simply the set of prime filters containing (the equivalence class under $q_{F_{\Ax}}$ of) this formula.

Since $\Phi\nvdash_{\mathrm{ML}+\Ax}\Psi$ the filter $\langle\Phi\rangle\eup$ in $\mathscr{L}_{\Ax}$  generated by $\Phi$ has an empty intersection with the ideal $\langle\Psi\rangle\edown$ generated by $\Psi$. We can thus use the prime filter theorem to extend $\langle\Phi\rangle\eup$ to a prime filter $w_{\Phi}\in\pf \mathscr{L}_{\Ax}$ which do not intersect $\langle\Psi\rangle\edown$. By definition of the coalgebraic semantics via the semantic transformation $\delta$, this means exactly that
\begin{align*}
(w_{\Phi},h\circ\pf\alpha, v)&\models a\text{ for all }a\in\Phi\\
(w_{\Phi},h\circ\pf\alpha, v)&\not\models a\text{ for all }a\in\Psi
\end{align*}
where $v: \pf\mathscr{L}_{\Ax}\to\mathcal{Q}V$ is the valuation given by the map $\Free V\to \ups\pf \mathscr{L}_{\Ax}$ by using the sequence of adjunctions described in Theorem \ref{ch5:thm:weakcomp1}.

Now we just need to check that all the axioms of $\Ax$ are valid on the coalgebra $h\circ\pf\alpha:\pf\mathscr{L}_{\Ax}\to T\pf\mathscr{L}_{\Ax}$. Clearly, since all the equations in $\Ax$ are assumed to be canonical, and since they are valid in the $L$-algebra $\mathscr{L}_{\Ax}$ by construction, they are also valid in its canonical extension $\ups\pf\mathscr{L}_{\Ax}$. From our assumption we can also know via Theorem \ref{ch5:thm:CanExtJonTarksiExt} that the J\'{o}nsson-Tarski extension of $\mathscr{L}_{\Ax}$ in the diagram above is precisely its canonical extension. It follows immediately that the equations of $\Ax$ are valid in the J\'{o}nsson-Tarski extension of $\mathscr{L}_{\Ax}$, and thus on the coalgebra $h\circ\pf\alpha:\pf\mathscr{L}_{\Ax}\to T\pf\mathscr{L}_{\Ax}$ by definition of the coalgebraic semantics.
\end{proof}

The condition on the semantic transformation $\delta$ required by the previous Theorem turns out not to be too restrictive in practise as the following result illustrates.

\begin{proposition}\label{ch5:prop:relationalDeltaOK}
For any finitary signature $\Sigma$, and any distributive lattice $A$, the semantic transformation $\delta^\Sigma_{\pf A}: L_\Sigma\ups\pf A\to \ups T_\Sigma \pf A$ interpreting the positive relational logic defined by $\Sigma$ (see Section \ref{ch5:subsec:DL}) defines maps which either (1) preserves all joins or anti-preserve all meets in each argument or (2) preserve all meets or anti-preserve all joins in each argument.
\end{proposition}
\begin{proof}
Recall that the modalities $\heartsuit$ in the signature $\Sigma$ come in two flavours, satisfying the distribution law of Eqs. (\ref{ch5:eq:DL1}) or (\ref{ch5:eq:DL2}) respectively, and that the $\delta^\Sigma$ accordingly defines the maps $\delta^\heartsuit: (\Forg \ups \pf A)^n\to \Forg \ups T_\Sigma \pf A$
\begin{align*}
&[\heartsuit(u_1,\ldots,u_k,u_{k+1},\ldots,u_n)]\mapsto \\
& \{t\in T_\Sigma X \mid \exists (a_1,\ldots,a_n)\in \pi_\heartsuit(t), (a_i\in u_i\text{ for all }i)\text{ and }(a_j\notin u_j\text{ for all }j)\}\\
&\text{if }\heartsuit\text{ satisfies Eq. (\ref{ch5:eq:DL1})}\\
&[\heartsuit(u_1,\ldots,u_k,u_{k+1},\ldots,u_n)]\mapsto \\
&\{t\in T_\Sigma X\mid \forall (a_1,\ldots,a_n)\in \pi_\heartsuit(t), (a_j\in u_j \text{ for all } j)\Rightarrow (a_i\in u_i \text{ for all }i)\}\\
&\text{if }\heartsuit\text{ satisfies Eq. (\ref{ch5:eq:DL2})}
\end{align*}
The fact that these maps preserves all joins or anti-preserve all meets in each argument for expansions satisfying Eq. (\ref{ch5:eq:DL1}) and preserve all meets or anti-preserve all joins in each argument for expansions satisfying Eq. (\ref{ch5:eq:DL2}) follows from elementary set-theoretic considerations.
\end{proof}

In the case of BAEs we can state the following result.
\begin{theorem}\label{ch5:thm:strongcompcan2}
Let $L:\BA\to\BA$ be a regular epi-preserving finitary functor defining a variety of BAEs, $T:\Set\to\Set$ and $\delta: L\pow\to\pow T$ be such that the adjoint transpose $\hat{\delta}: T\uf\to \uf L$ has right-inverses, and let $\Ax\subseteq \Forg\Free_L \Free V$ be any set of canonical formulas defined over a set $V$ of variables. If $\delta_{\uf A}$ defines maps which are all $(\sigma,\gamma\eup)$-continuous or all $(\sigma,\gamma\edown)$-continuous then the logic defined by $L$ together with the axioms of $\Ax$ is strongly complete with respect to the class of $T$-coalgebras on which the axioms of $\Ax$ are valid.
\end{theorem}
\begin{proof}
The result follows from Theorem \ref{ch5:thm:CanExtJonTarksiExtBAE} in exactly the same way as the Theorem \ref{ch5:thm:strongcompcan} above follows from Theorem \ref{ch5:thm:CanExtJonTarksiExt}.
\end{proof}

\begin{example}[Logic for trees with bounded degrees]
We return to our Example \ref{ch5:ex:treef} of a logic for trees with bounded degrees, say $n$. Recall that the operators of the logic defined by Eq. (\ref{ch5:eq:treef}) all preserve both meets and joins in each of their argument. They are thus particularly well-behaved with regards to Table \ref{ch2:table}. Crucially, recall that the logic has the following axiom
\[
\text{(Tree 3) }\neg\langle k\rangle(a_1,\ldots,a_k)=\hspace{-2ex}\bigvee\limits_{1\leq j\neq k\leq n}\langle j \rangle(\top, \ldots, \top)\vee \hspace{-1ex}\bigvee\limits_{1\leq i\leq k}\langle k\rangle(\epsilon_{1i}a_1,\ldots,\epsilon_{ni}a_1)
\]
which, by repeated use, allows us to push negations down to the propositional level. The consequence of this is particularly nice: since join-preserving operators are conservative (i.e. they make stable combinations when applied to  any monotone term), since all operators are join preserving, and since we can always push the negations in front on propositional variables, we get that \emph{every term in the language is stable, and thus canonical}.

As we saw in Example \ref{ch5:ex:treef}, the adjoint transpose $\hat{\delta}$ is surjective, and thus has right-inverses in $\Set$. Moreover, we can easily check that the semantic transformation $\delta$ defines maps which are $(\sigma\eup,\gamma\eup)$-continuous in each argument. By definition we have for each set $X$
\[
\delta_X(U_1,\ldots,U_k)=\{(x_1,\ldots,x_k)\in \mathrm{in}_k[X]\mid x_i\in U_i, 1\leq i\leq k\}
\]
Now let $(U_i^j)_{j\in J}$ be a collection of subsets of $X$. It is immediate that
\[
\delta_X(U_1,\ldots,\bigcup_{j\in J} U^j_i,\ldots,U_k)=\bigcup_{j\in J} \delta_X(U_1,\ldots,U^j_i,\ldots,U_k)
\]
i.e. $\delta_X$ is preserves all joins in each of its arguments, and thus in particular all up-directed ones. It follows that $\delta_X$ is $((\gamma\eup)^k,\gamma\eup)$-continuous, and we can therefore apply Theorem \ref{ch5:thm:strongcompcan2} to \emph{any} set of formulas in the logic and get strong completeness for any collection of frame conditions (note that if the set of frame conditions is not consistent we get strong completeness vacuously). 
\end{example}

\begin{example}[Classical modal logic]
We continue from Example \ref{ch5:ex:mod}, but we could also consider the countable version of classical modal logic developed in Example \ref{ch5:ex:modc}, since it is also strongly complete. We have described in Section \ref{ch2:sec:Sahl} a general class of Sahlqvist identities (see Definition \ref{ch2:def:genSahlqvist}). We have seen that the adjoint transpose $\hat{\delta}$ of the semantic transformation has a right inverse (essentially as a special case of Theorem \ref{ch5:thm:strongComplDL}), and using the same argument as in the previous example it is easy to see that $\delta_X$ is $(\sigma\eup,\gamma\eup)$-continuous, and we can therefore apply Theorem \ref{ch5:thm:strongcompcan2} to any set canonical axioms and get strong completeness-via-canonicity. Consider for example the general but not classical Sahlqvist formula described in Chapter 2:
\[
(3) \hspace{6ex }\dia p\wedge \dia (q\wedge\neg p)\wedge\dia(r\wedge\neg p\wedge\neg q)
\]
We can conclude that the classical modal logic $K+(3)$ is sound and strongly complete with respect to the class of frames where every point has at least three successors.
\end{example}

\begin{example}[GML]
Recall from Example \ref{ch5:ex:GMLomega} that GML is strongly complete w.r.t. the class of $\Bag^\omega$-coalgebra, where $\Bag^\omega$ is the semantic completion of the usual bag functor, as described in Section \ref{ch5:sec:semcomp}. Recall also from \ref{ch2:prop:gmldistrib} that the expansions $\kop$ of GML are all $k-$additive. From Theorem \ref{ch2:thm:kaddSmooth} we know that they are therefore smooth and the $L$-algebras of GML thus satisfies the conditions of Theorem \ref{ch5:thm:strongcompcan}. We need only check that the semantic transformation map $\delta:L\pow\to\pow\Bag^\omega$ define completely $k$-additive maps. 
\begin{align*}
\delta_X (\bigcup_i U_{i\in I})&=\{f\in\Bag^\omega(X)\mid \sum_{x\in \supp(f)\cap \bigcup_i U_i} f(x)\geq k\}\\
&=\bigcup_{\{i_1,\ldots,i_p\}\subseteq I}\{f\in\Bag^\omega(X)\mid \sum_{x\in \supp(f)\cap \bigcup_{j=1}^p U_{i_j}} f(x)\geq k\}\\
&=\bigcup_{\{i_1,\ldots,i_k\}\subseteq I}(\delta(\bigcup_{j=1}^p U_{i_j}))
\end{align*}
Where the second step arises from the fact that any map $f$ assigning a weight of at least $k$ to $\bigcup_i U_i$ it is necessary and sufficient to find a partition $k=k_1+\ldots+k_p$ (in at most $k$ integers) and a corresponding list of sets $U_{i_1},\ldots,U_{i_p}$ with $\{i_1,\ldots,i_p\}\subseteq I$ such that $\sum_{\supp (f)\cap U_{i_j}}f(x)\geq k_j, 1\leq j\leq p$, and thus
\[
\sum_{x\in \supp(f)\cap \bigcup_{j=1}^k U_{i_j}} f(x)\geq k
\]
The semantic transformation therefore also satisfies the conditions of Theorem \ref{ch5:thm:strongcompcan} and canonical equations in GML therefore define extensions of GML which are strongly complete with respect to the class of $\Bag^\omega$-coalgebras which validate these equations. 
As shown in Section \ref{ch2:subsec:GML} Sahlqvist identities can be defined for GML (see Proposition and Definition \ref{ch2:propdef:SahlqvistGML}), generating a large class of canonical formulas. Some examples of Sahlqvist frame conditions could for example be
\[
\neg\kop(\top)
\]
i.e. `there can never be $k$ or more successors', which is Sahlqvist since it is the negation of a positive term. GML plus this axiom are thus sound and strongly complete w.r.t. frames in which every point has no more than $k$ successors (including multiplicities). Similarly, we could consider the infinite set of axioms
\[
\{\kop[2k]\top\wedge\neg\kop[2k+1]\top\}_{k\in\mathbb{N}}
\] 
i.e. `every point must have an even number of successors (including multiplicities)', which are all Sahlqvist since the conjunction of a stable term and the negation of a positive term. GML plus this collection of axioms is thus sound and strongly complete w.r.t. frames in which every point has an even number of successors (including multiplicities).
\end{example}

\begin{example}[Intuitionistic Logic and Distributive Substructural Logics] We expand on the examples of Section \ref{ch2:subsec:intuitionistic}, which are also detailed in \cite{2015:self}. Recall that the language of intuitionistic logic (IL) was defined by the functor
\[
\LHey: \BDL\to\BDL, \begin{cases}A\mapsto \Free\{a\to b\mid a,b\in \Forg A\}/\equiv\\
\LRL f: \LHey A\to \LHey B, [a]_{\equiv}\mapsto [f(a)]_{\equiv}
\end{cases}
\]
where $\equiv$ is the fully invariant equivalence relation in $\BDL$ (see Section \ref{ch1:subsec:fullyinvariant}) generated by the following Heyting Distribution Laws:
\begin{enumerate}[HDL1]
\item $a\to(b\wedge c)=(a\to b)\wedge (a\to c)$
\item $(a\vee b)\to c=(a\to c)\wedge(b\to c)$
\end{enumerate}
The language of distributive substructural logics was defined by the functor\[
\LRL: \DL\to\DL, \begin{cases}
\LRL A =\Free\{I, a\ast b, a\lRes b, a\rRes b\mid a,b\in \Forg A\}/\equiv\\
\LRL f: \LRL A\to \LRL B, [a]_{\equiv}\mapsto [f(a)]_{\equiv}
\end{cases}
\]
where $\equiv$ is the fully invariant equivalence relation in $\DL$ (see Section \ref{ch1:subsec:fullyinvariant}) generated by the Distribution Laws:
\begin{enumerate}[DL1]
\item$(a\vee b)\ast c=(a\ast c)\vee(b\ast c) $ 
\item$a\ast (b\vee c)=(a\ast b)\vee (a\ast c)$
\item$a\lRes(b\wedge c)=(a\lRes b)\wedge (a\lRes c)$
\item$(a\vee b)\lRes c=(a\lRes c)\wedge (b\ast c)$
\item$(a\wedge b)\rRes c=(a\rRes c)\wedge (b\rRes c)$
\item$a\rRes (b\vee c)=(a\rRes b) \wedge (a\rRes c)$
\end{enumerate}

The functors $\LHey$ and $\LRL$ are examples of the class of functors described in Section \ref{ch5:subsec:DL} defining \emph{relational logics}. We therefore interpret $\LHey$-algebras in coalgebras for the functor:
\[
\THey: \Pos\to\Pos,\begin{cases} \THey W=\cpow_c(W\times W)\\
\THey f: \THey W\to \THey W', U\mapsto(f\times f)[U].
\end{cases}
\]
via the semantic transformation $\semTHey: \LHey\ups\to\ups \THey$
\[
\semTHey_W(U\to V)=\{(x,y)\in \THey W\mid x\in U\Rightarrow y\in Y\}
\]
Similarly we interpret $\LRL$-algebras in coalgebras for the functor:
\[
\TRL: \Pos\to\Pos,\begin{cases} \TRL W=\two\times(\cpow_c(W\times W))^3 \\
\TRL f: \TRL W\to \TRL W', U\mapsto(\Id_{\two}\times (f\times f)^3)[U].
\end{cases}
\]
via the semantic transformation $\semTRL: \LRL\ups\to\ups \TRL$ 
\begin{align*}
\semTRL_W(I)&=\{t\in \TRL W \mid \pi_1(t)=0\in\two\}\\
\semTRL_W(u\ast v)&=\{t\in \TRL W\mid \exists (x,y)\in \pi_2(t), x\in u, y\in v\}\\
\semTRL_W(u\lRes w)&=\{t\in \TRL W\mid \forall (x,y)\in \pi_3(t), x\in u\Rightarrow y\in w\}\\
\semTRL_W(w\rRes v)&=\{t\in \TRL W\mid\forall (x,y)\in \pi_4(t), x\in v\Rightarrow y\in w\} .
\end{align*}
where  $\pi_i, 1\leq i\leq 4$ are the usual projections maps. The intuition is that the first component of the structure map of a $\TRL$-coalgebra (to the (po)set $\two$) separates states into units and non-units. The second component sends each `state' $w\in W$ to the pairs of states which it `contains', the next two components are used to interpret $\lRes$ and $\rRes$, respectively, and turn out to be very closely related to the second component.

Since these two situations are particular instances of relational logics, as developed in Section \ref{ch5:subsec:DL}, we know by Theorem \ref{ch5:thm:strongcomplRelational} that both $\semTHey$ and $\semTRL$ have right inverses, and that the J\'{o}nsson-Tarski extension of any $\LHey$ or $\LRL$-algebra exists by Theorem \ref{ch5:thm:jontarski}. Moreover, we know from Proposition \ref{ch2:prop:smooth} that $\LHey$ and $\LRL$ define varieties of DLEs with smooth expansions, and from Proposition \ref{ch5:prop:relationalDeltaOK} that the semantic transformation  $\semTHey$ and $\semTRL$ either (1) preserves all joins or anti-preserves all meets in every argument, or (2) preserves all meets or anti-preserves all joins in every argument. It follows that $\LHey$-algebras and $\semTHey$ satisfy the the conditions of Theorem \ref{ch5:thm:strongcompcan}, and similarly for $\LRL-$algebras and $\semTRL$. We therefore have strong completeness of $\LHey$ or $\LRL$-algebras validating canonical axioms with respect to the class of $\THey$ or $\TRL$-coalgebras validating these axioms. We will apply this fact to get strongly complete semantics of IL and of the full distributive Lambek calculus.

Recall that the axioms HDL\ref{ch2:ax:HDL1}-HDL\ref{ch2:ax:HDL2} are not sufficient to completely capture IL. We also need to specify that $\to$ is the residuum of $\wedge$ and that $a\to a$ is a theorem of IL. This can be done by using the following axioms (Heyting Frame Conditions):
\begin{enumerate}[HFC1]
\item $a\to a=\top$\label{ch5:ax:Heyting1},
\item $a\wedge(a\to b)=a\wedge b$\label{ch5:ax:Heyting2}
\item $ (a\to b)\wedge b=b$\label{ch5:ax:Heyting3} 
\end{enumerate}
It is not difficult to check that HDL\ref{ch2:ax:HDL1}-HDL\ref{ch2:ax:HDL2} together with HFC\ref{ch5:ax:Heyting1}-HFC\ref{ch5:ax:Heyting3} axiomatize IL. 

Similarly, DL\ref{ch2:ax:DL1}-DL\ref{ch2:ax:DL6} does not fully axiomatize the full distributive Lambek calculus, i.e. distributive residuated lattices. To achieve this we need to specify that $I$ is a unit for $\ast$ and that $\lRes$ and $\rRes$ are the left and right residuals of $\ast$. This can be achieved via the following axioms (Frame Conditions):
\begin{enumerate}[FC1]
\item $a\ast I=a$, $I\ast a=a$, 
\label{ch5:ax:FrameCond1} 
\item $I\leq a\lRes a$, $I\leq a\rRes a$,
\label{ch5:ax:FrameCond2} 
\item $a\ast(b\lRes c)\leq (a\ast b)\lRes c$,\label{ch5:ax:FrameCond3} 
\item $(c\rRes b)\ast a\leq c\rRes(a\ast b)$,\label{ch5:ax:FrameCond4} 
\item $(a\rRes b)\ast b\leq a$, and 
\label{ch5:ax:FrameCond5} 
\item $b\ast(b\lRes a)\leq a$,
\label{ch5:ax:FrameCond6}
\end{enumerate}
Again, it is not difficult to check that DL\ref{ch2:ax:DL1}-DL\ref{ch2:ax:DL6} together with FC\ref{ch5:ax:FrameCond1}-FC\ref{ch5:ax:FrameCond6} axiomatize distributive residuated lattices.

\begin{proposition}\label{ch5:prop:FrameCondCan}
The axioms HFC\ref{ch5:ax:Heyting1}-\ref{ch5:ax:Heyting3} and FC\ref{ch5:ax:FrameCond1}-\ref{ch5:ax:FrameCond6} are canonical.
\end{proposition}
\begin{proof}The proof is an application the results of Chapter 2. We show the result for FC\ref{ch5:ax:FrameCond1}-\ref{ch5:ax:FrameCond6}, the result for HFC\ref{ch5:ax:Heyting1}-\ref{ch5:ax:Heyting3} follow very similar lines.

FC\ref{ch5:ax:FrameCond1}: Since $\ast$ preserves binary joins in each argument, it is smooth by Proposition \ref{ch2:prop:smooth}, 
and it follows that it is $(\sigma^2,\gamma)$-continuous by Corollary \ref{ch2:cor:smoothcor}. Since $\pi_1\ce$ and $I\ce$ are trivially $(\sigma,\sigma)$-continuous, it follows from Theorem \ref{ch2:thm:MatchingTopologies} that $(\ast \circ\langle \pi_1, I\rangle)\ce=\ast\ce\circ \langle \pi_1, 1\rangle\ce$. 
Each side of the equation is thus stable and the result follows from Proposition \ref{ch2:prop:caninequ}.

FC\ref{ch5:ax:FrameCond2}: $I$ is stable and thus contracting, and $(\lRes \circ \langle\pi_1,\pi_1\rangle)\ce=\lRes\ce\circ \langle\pi_1,\pi_1\rangle\ce$, 
since $\pi_1\ce$ is $(\sigma,\sigma)-$continuous and $\lRes\ce$ is smooth. The RHS of the 
inequality is thus stable, and a fortiori expanding, and the inequality is thus canonical.

FC\ref{ch5:ax:FrameCond3}-\ref{ch5:ax:FrameCond4}:  Since $\ast\ce$ preserve joins in each argument, it preserves up-directed ones, and is thus $((\gamma\eup)^2,\gamma\eup)$-continuous. Since $\lRes\ce$ is smooth it is in particular $(\sigma^2,\gamma\eup)$-continuous. Since $\pi_1\ce$ is $(\sigma,\gamma\eup)$-continuous, we get that $\ast\ce\circ\langle\pi_1\ce,\lRes\ce\circ\langle\pi_2\ce,\pi_3\ce\rangle\rangle$ is $(\sigma^3,\gamma\eup)$-continuous and thus contracting. For the RHS, note that since $\lRes\ce$ preserves meets in its first argument, it must 
in particular preserve down-directed ones, thus $\lRes\ce$ is $(\gamma\edown,\gamma\edown)$-continuous in its first argument. Similarly, since 
$\lRes\ce$ anti-preserve joins in its second argument, it must in particular anti-preserve up-directed ones, and is thus 
$(\gamma\eup,\gamma\edown)$-continuous in its second argument. This means that $\lRes\ce$ is $(\gamma^2,\gamma\edown)$-continuous. 
We thus have that the full term is $(\sigma^3,\gamma\edown)$ continuous, and thus expanding. The inequation is therefore canonical. 

FC\ref{ch5:ax:FrameCond5}-\ref{ch5:ax:FrameCond6}: The LHS is contracting by the same reasoning as above, and the RHS is stable and thus expanding.

\end{proof}

It follows from Proposition \ref{ch5:prop:FrameCondCan} and Theorem \ref{ch5:thm:strongcompcan} that
\begin{itemize}
\item The logic defined by $\LHey$ and the Heyting frame conditions HFC\ref{ch5:ax:Heyting1}-\ref{ch5:ax:Heyting3} is strongly complete with respect to $\THey$-coalgebras validating HFC\ref{ch5:ax:Heyting1}-\ref{ch5:ax:Heyting3}.
\item The logic defined by $\LRL$ and the frame conditions FC\ref{ch5:ax:FrameCond1}-\ref{ch5:ax:FrameCond6} is strongly complete with respect to $\TRL$-coalgebras validating FC\ref{ch5:ax:FrameCond1}-\ref{ch5:ax:FrameCond6}.
\end{itemize}
To conclude this example, let us examine what these classes of $\THey$ and $\TRL$-coalgebras look like. Let us first examine what $\THey$-coalgebras validating 
HFC\ref{ch5:ax:Heyting1}-\ref{ch5:ax:Heyting3} look like. For every $\gamma: W\to \THey W$ in this class, every $w\in W$ and 
every valuation, $w\models a\to a$. By considering a formula satisfied at a single point in the model is easy to see that $(x,y)\in \gamma(w)\Rightarrow x=y$, i.e. the structure map of the coalgebra only really defines a binary relation to interpret $\to$. Thus $\THey$-coalgebras validating 
HFC\ref{ch5:ax:Heyting1} are equivalent to $\cpow_c$-coalgebras where $w\models a\to b$ iff $\forall x\in \gamma(w), x\models a\Rightarrow x\models b$. 
The distributivity laws of $\to$ together with HFC\ref{ch5:ax:Heyting2}-\ref{ch5:ax:Heyting3} encode the well-known residuation property of $\to$ with 
respect to  $\wedge$. Combined with HFC\ref{ch5:ax:Heyting1} and the associated reformulation in terms of $\cpow_c$-coalgebra, 
the residuation property states that:
\[
w\models a\wedge b \hspace{1ex}\Rightarrow\hspace{1ex} w\models c \hspace{3ex}\text{ iff } 
	\hspace{3ex}w\models b\hspace{1ex}\Rightarrow\hspace{1ex} (\forall x\in \gamma(w)
		\hspace{1ex} (x\models a\hspace{1ex}\Rightarrow \hspace{1ex}x\models c))
\]
Assuming the left-hand side, for the right-hand side to hold it is necessary that if $w\models b\hspace{1ex}$, then $\forall x\in \gamma(w), x\models b $; 
that is, successor states satisfy the so-called `persistency' condition. Assuming the right-hand side, for the left hand side to hold in a model where $w\models a,b$ it is necessary that $x\in \gamma(x)$; that is, the relation is reflexive. Finally, from HFC\ref{ch5:ax:Heyting3} we get that $a\wedge b\leq c$ iff $b\leq a\to c$ iff $b\leq a\to (c\wedge (a\to c))$. By unravelling the interpretation of this last inequality, we get that the relation interpreting $\to$ must also be transitive. Thus we have recovered the traditional Kripke semantics of intuitionistic logic via a pre-order and persistent valuations by using the theory of canonicity for distributive lattices. Recall that $\gamma$ takes its values in the \emph{convex} powerset of the carrier $(W,\leq)$ of the model. Thus the order defined by $\gamma$ and the order of the carrier are closely related: if $y\in\gamma(x)$, then for every $z\in W$ such that $x\leq z\leq y$, $z\in\gamma(x)$ too. Indeed, since each poset is in particular a pre-order, each element of $\Pos$ comes with a `canonical' $\THey$-coalgebra structure validating 
HFC\ref{ch5:ax:Heyting1}-\ref{ch5:ax:Heyting3} defined by $\gamma(x)=\up x$ (upsets are convex). 

Let us now describe $\TRL$-coalgebras validating FC\ref{ch5:ax:FrameCond1}-\ref{ch5:ax:FrameCond6}. Axiom FC\ref{ch5:ax:FrameCond1} means that at 
every $w$ in a $\TRL$-coalgebra, amongst all the pairs of states into which $w$ can be `separated' there must exist a \emph{unit} state $i$, viz. $\pi_1(\gamma(i))=0$, such that $(w,i)\in \pi_2(\gamma(w))$. Similarly, there must exist a unit state $i'$ such that 
$(i',w)\in\pi_2(\gamma(w))$. This condition can be found in this form in, for example, \cite{2007:calcagnoPOPL}. The other axioms are simply 
designed to capture the residuation condition in such a way that canonicity can be used, so a model in which FC\ref{ch5:ax:FrameCond2}-\ref{ch5:ax:FrameCond6} 
are valid is simply a model in which the residuation conditions hold. By considering models with only three points it is easy to see that these 
conditions imply that
\begin{align*}
(y,z)\in \pi_1(\gamma(x)) \text{ iff } (x,z)\in \pi_2(\gamma(y)) \text { iff }(y,x)\in \pi_3(\gamma(z)),
\end{align*}
that is, the last three components of a $\TRL$-coalgebra's structure map are determined by any one of them. If we choose the second 
as defining the last two, a $\TRL$-coalgebra validating FC\ref{ch5:ax:FrameCond1}-\ref{ch5:ax:FrameCond6}, really is a coalgebra for the functor 
\[
\TRL': \Pos\to\Pos, \TRL' W=\two\times \cpow_c(W\times W)
\]
in which the interpretation of the operators is given by:
\begin{enumerate}
\item $w\models a\ast b \text{ iff }\exists  (x,y)\in \gamma(w) \text{ s.t. }x\models a \text{ and } y\models b$
\item $w\models a/b \text{ iff }\forall (x,y)\text{ s.th }(w,y)\in \gamma(x) \text{ if }y\models b\text{ then }x\models a$
\item $w\models b\lRes a \text{ iff }\forall (x,y)\text{ s.th }(y,w)\in \gamma(x) \text{ if }y\models b\text{ then }x\models a$.
\end{enumerate}
It is frequent to consider variations of the full distributive Lambek calculus where one or several of the structural rules are allowed. This can be achieved by adding yet more canonical frame conditions. Specifically, it is very easy to check that the following (in)equations, each corresponding to admitting a structural rule, are canonical: (1) Commutativity: $a\ast b=b\ast a$; (2) Increasing idempotence: $a\leq a\ast a$ (defines relevant logic); and (3) Integrality: $a\leq I$ (defines affine logic). 
\end{example}

\subsection{General $L$-algebras}\label{sec:transmeth}

We now define a method for proving completeness-via-canonicity in the presence of frame conditions for logics defined by a functor $L:\BA\to\BA$ which does not define a DLE. The main application of this method is the case of coalgebraic logics with the cover modality, but in principle, this method applies to any functor $L$ which is too complicated to deal with using the direct method detailed in the previous Section. Since the target usage of the method in this Section is the case of nabla logics, we place ourselves in the $\uf\dashv\pow: \BA\to\Set\op$ situation throughout.

For this method, we continue where we left off at the end of Chapter 4. We briefly recall the basic setting. We have two finitary regular epi-preserving functors  $K,L:\BA\to\BA$, two functors $S,T:\Set\to\Set$ and two epi-transformations $q: K\epi L$ and $r: S\epi T$. We also have two semantic natural transformations $\lambda: K\pow\to\pow S$ and $\delta: L\pow\to\pow T$, and these define \emph{compatible semantics}, i.e. for every set $X$ and every $a\in L\pow X$ we must have
\[
\lambda_X(a)\subseteq \pow r_X\circ \delta_X\circ q_{\pow X}(a)
\]
For the method to work, or even make sense, we need $K$ to define a BAE.  Typically, $K$ would arise from a presentation of $L$, as discussed in Section \ref{ch3:sec:liftpres}. In this case $K$ of the form $(\Free \polyFunc\Forg)/\simeq$, for a polynomial $\Set$-functor $\polyFunc$, and possibly some quotient $\simeq$. In other words, we can assume that whilst $L$ might not define a variety of BAEs, $K$ does. We fix a set of propositional variables $V$ throughout.

The problem is the following: given a set $E$ of axioms (frame conditions) in the logic defined by $L$, and a (possibly infinite) set $\Phi\subseteq \init[(L(-)+\Free V)]$ of $L$-formulas which are consistent with the axioms of $E$, we want to build an $T$-model in which every formula $a\in\Phi$ is satisfiable. We proceed in five steps. 

\subsubsection{Build the Lindenbaum-Tarski algebra.} 
Let us write $\Free_L\dashv\Forg_L$ for the adjunction $\BA\to\Alg_{\BA}(L)$, and similarly for $K$. We proceed as in the direct method by building the Lindenbaum-Tarski algebra associated with $E$. As usual, we start by taking the adjoint transpose of the maps $e_1,e_2$ associated with the set of equations $E$, and then taking their coequalizer.
\[
\xymatrix
{
\Free E \ar@<3pt>[r]^{\hat{e}_1} \ar@<-2pt>_{\hat{e}_2}[r] & \Free_L \Free V  \ar@{->>}[r]^{q_E} & Q_E
}
\]
This defines the quotient under the smallest equivalence class generated by the equations of $E$. By Proposition \ref{ch1:prop:fullInvClosure} we know that we can get the fully invariant closure of this equivalence relation. So let us assume that we have the following exact sequence
\[
\xymatrix
{
F_E \ar@<3pt>[r]^{e'_1} \ar@<-2pt>_{e'_2}[r] & \Free_L \Free V  \ar@{->>}[r]^{q_{F_E}} & \mathscr{L}_E
}
\]
where $F_E$ is the fully invariant closure of the equivalence relation defined by $q_E$ (and the kernel pair of $q_{F_E}$). The algebra $\mathscr{L}_E$ is the Lindenbaum-Tarski algebra\index{Lindenbaum-Tarski algebra} for the logic defined by $L$ with additional axioms in $E$. As we saw in Section \ref{ch1:subsec:algsem}, $\mathscr{L}_E$ has the nice property that it is itself a member of the variety it defines, or put succinctly, $\mathscr{L}_E\perp q_{F_E}$ (see Proposition \ref{ch1:prop:Q*}). Note that if $a\in\Phi$ is consistent with the axioms of $E$, then  $\lsem a\rsem_{\mathscr{L}_E}\neq \bot$. Since $\mathscr{L}_E$ inherits the canonical valuation of $\Free_L \Free V$, we can actually think of the formulas $a\in\Phi$ as elements of $\mathscr{L}_E$ which are not equal to $\bot$.

\subsubsection{Lift the axioms.} 
We now apply the construction leading to Theorem \ref{ch4:thm:prooftrans}, and lift the equations in $F_E$ to the $K$-logic.  Recall that we generated Diagram (\ref{ch4:diag:equationLift}):
\[
\xymatrix@C=12ex
{
E^*\ar@<1ex>[r]^{e_1^*}\ar@<-1ex>[r]_{e^*_2} \ar[d]_{r} & \Free_K \Free V \ar@{->>}[r]^{q_{E^*}}\ar[d]^{\xi_V} & Q_{E^*}\ar@{-->>}[d]^{q^{E^*}_{E'}}\\
\mathsf{Q}F_E\ar@<1ex>[r]^{\mathsf{Q}e'_1} \ar@<-1ex>[r]_{\mathsf{Q}e'_2} & \mathsf{Q}\Free_L \Free V\ar@{->>}[r]^{\mathsf{Q}q_{F_E}} & \mathsf{Q}\mathscr{L}_E
}
\]
where $\mathsf{Q}:\Alg(L)\to\Alg(K)$ pre-composes with $q: K\to L$. 

We can already point out that since all the vertical arrows are epis, every $a\in\Phi$ must have an inverse image $a'\neq \bot$ under the syntax translation map.

\subsubsection{Check for canonicity and that validity in the canonical extension implies validity in the J\'{o}nsson-Tarski extension.} 
Since $K$ is assumed to define a BAE, it makes sense to ask whether the variety defined by $q_{E^*}$ is canonical. This will in particular be the case if $E^*$ is generated by canonical equations, for example Sahlqvist identities. This leads us to the following natural definition.

\begin{definition} \label{ch5:def:SahlqvistGen}
Let $q: K\epi L$ be a regular epi-transformation between two regular epi-preserving finitary functors on $\BA$. We will say that $a\in\Free_L \Free V$ is a \textbf{abstract coalgebraic Sahlqvist formula}\index{Coalgebraic Sahlqvist formula!abstract} for the transformation $q$ if every pre-image of $a$ under the translation map $(-)^q:\Forg_K\Free_K \Free V\to\Forg_L\Free_L \Free V$ is an abstract Sahlqvist formula in the sense of Definition \ref{ch2:prop:Sahl}. We will say that $a$ is a \textbf{concrete coalgebraic Sahlqvist formula} if we have a notion of concrete Sahlqvist formula in the BAEs defined by $K$ and every pre-image of $a$ under the translation map $(-)^q$ is a concrete Sahlqvist formula, for example in the sense of Definition \ref{ch2:def:genSahlqvist}. \index{Coalgebraic Sahlqvist formula!concrete}
\end{definition}

If $E^*$ is generated by a set of Sahlqvist identities, then the variety defined by $q_{E^*}$ is canonical, and thus if $A\perp q_{E^*}$, then $\pow\uf A\perp q_{E^*}$. Recall from Theorem \ref{ch4:thm:prooftrans} that in the context of the lifting of  equations described by the diagram above, we have for any $L$-algebra $A$ that $A\perp q_{F_E}$ iff $\mathsf{Q}A\perp q_{E^*}$. Since the Lindenbaum-Tarski algebra $\mathscr{L}_E$ has the property that $\mathscr{L}_E\perp q_{F_E}$, Theorem  \ref{ch4:thm:prooftrans} tells us that $\mathsf{Q}\mathscr{L}_E \perp q_{E^*}$.  Thus if the variety defined by $q_{E^*}$ is canonical we will have
\[
\pow\uf \mathsf{Q}\mathscr{L}_E\perp q_{E^*}
\]

Since we are working over $\BA$, the most general result to verify that equations that hold in the canonical extension also hold in the J\'{o}nsson-Tarski extension is Theorem \ref{ch5:thm:CanExtJonTarksiExtBAE}, but in practise Theorem \ref{ch5:thm:CanExtJonTarksiExt} is more useful since one can often exhibit preservation properties of the expansions defined by the $S$-algebras. 

\subsubsection{Build an $S$-model.} 
If the adjoint transpose $\hat{\lambda}: T\uf\to\uf K$ has a right inverse $h$ at $\mathscr{L}_E$, then we can build a coalgebra
\[
\gamma: \uf \mathscr{L}_E\to\uf L\mathscr{L}_E\xto{\uf q_{\mathscr{L}_E}} \uf K\mathscr{L}_E\xto{h} S\uf \mathscr{L}_E
\]
which provides a J\'{o}nsson-Tarski embedding of $\mathscr{L}_E$ into $\pow \uf \mathscr{L}_E$ as $K$-algebras (see Theorem \ref{ch5:thm:jontarski} and remarks thereafter). Since this embedding $\mathscr{L}_E\to \pow\uf \mathscr{L}_E$ is injective and since we know that for every $a\in\Phi$, $a \neq \bot$, we get that $\lsem a \rsem_{\pow\uf \mathscr{L}_E}\neq \bot$ for the valuation on $\pow\uf \mathscr{L}_E$ inherited from the canonical valuation on $\mathscr{L}_E$ as follows
\[
\xymatrix@C=12ex
{
K\Free_K \Free V+\Free V\ar[ddd]_{\langle-\rangle_V+\eta_V^K}\ar[r]_{K (q_{F_E}\circ(-)^q)+\id_{\Free V}} \ar@/^2pc/[rr] ^{K\lsem - \rsem_{\pow\uf \mathscr{L}_E}+\id_{\Free V}} & 
K \mathscr{L}_E + \Free V \ar[ddr]_{\eta_{K\mathscr{L}_E}}\ar[ddd]_{\alpha\circ q_{\mathscr{L}_E}+\eta_V^L}  \ar[r]^{K\eta_{\mathscr{L}_E}+\id_{\Free V}}& K\pow\uf \mathscr{L}_E +\Free V\ar[d]^{\delta_{\uf \mathscr{L}_E}+\id_{\Free V}}
\\
& & \pow S \uf \mathscr{L}_E+\Free V\ar[d]^{\pow h+\id_{\Free V}}\\
& & \pow \uf K \mathscr{L}_E +\Free V\ar[d]_{\pow\uf (\alpha\circ q_{\mathscr{L}_E})+\pow\uf\eta_V^L\circ \eta_{\Free V}}\\
\Free_K\Free V\ar[r]_{q_{F_E}\circ(-)^q}\ar@/_2pc/[rr] _{\lsem - \rsem_{\pow\uf \mathscr{L}_E}}& \mathscr{L}_E\ar[r]_{\eta_{\mathscr{L}_E}} & \pow \uf \mathscr{L}_E
}
\]
where $\eta^K$ is the unit of the adjunction $\Free_K\dashv\Forg_K$, $\eta^L$ that of $\Free_L\dashv\Forg_L$, $\eta$ is the unit of the adjunction $\uf\dashv\pow$ and $\langle-\rangle_V$ and $\alpha$ are the structure morphisms of $\Free_K$ and $\mathscr{L}_E$ respectively. 

As noted above, there exists for every $a\in\Phi$ an $a'\in \Free_K \Free V$ such that $a'\neq \bot$ and $(a')^q=a\neq \bot$. Let us gather such a choice of pre-images in a set $\Phi'\subseteq\Free_K\Free V$. By assumption that $\Phi$ is consistent with the axioms of $E$ we also have that for each $a\in\Phi$, $q_{F_E}(a)\neq \bot$. Finally, since the J\'{o}nsson-Tarski embedding is injective we can conclude that $\lsem a'\rsem_{\pow\uf\mathscr{L}_E}\neq\bot$ for every $a'\in\Phi'$. Note that the semantic map $\lsem-\rsem_{\pow\uf\mathscr{L}_E}$ is explicitly given by $\eta_{\mathscr{L}_E}\circ q_{F^E}\circ(-)^q$. This just a reformulation of the usual \emph{truth lemma}: the interpretation of a formula $a'$ is simply the set of ultrafilters of $\mathscr{L}_E$ which contains its translation $a$. Since $\Phi$ is consistent, we can extend it to an ultrafilter by Lindenbaum's lemma, i.e. to a an element $u\in \pow\uf\mathscr{L}_E$, and for every $a'\in\Phi'$, $u\in\lsem a'\rsem_{\mathscr{L}_E}$. By definition of the semantic map (see Section \ref{ch1:sec:semantics}), if $u\in\lsem a' \rsem_{\pow\uf \mathscr{L}_E}$ for every $a'\in\Phi'$, then
\[
u,\gamma\models a'
\]
where the valuation is simply the valuation given by the map $\Free V\to \pow\uf \mathscr{L}_E$ defined in the diagram above. By the sequence of adjunctions described in the proof of Theorem \ref{ch5:thm:weakcomp1}, it is equivalent to a map $\uf \mathscr{L}_E\to\mathcal{Q}V$, i.e. a valuation on the coalgebra. Thus we have found an $S$-model in which every $a'\in\Phi$ is satisfiable (simultaneously). 

We now need to check that the lifted axioms of $E^*$ are valid over our $S$-model. Since the lifted axioms are assumed to be canonical and are valid in the $K$-algebra $\mathsf{Q}\mathscr{L}_E$ by construction, they are also valid over the canonical extension of $\mathsf{Q}\mathscr{L}_E$. By assumption we therefore know that they also hold on the J\'{o}nsson-Tarski extension of  $\mathsf{Q}\mathscr{L}_E$, and all the axioms of $E^*$ are automatically valid on the coalgebra $\gamma:\pow\uf \mathscr{L}_E\to S\pow\uf \mathscr{L}_E$ by definition of the coalgebraic semantics.

\subsubsection{Build a $T$-model.} 
To conclude, we use Theorem \ref{ch4:thm:compSem} which tells us that if the semantics are compatible, then if
\[
u,\gamma, v\models a'
\]
then 
\[
u,q_{\uf\mathscr{L}_E}\circ\gamma,v \models (a')^q=a
\]
Since this holds for every $a'\in\Phi'$, the model above is the model we were looking for, i.e. a model in which every formula in $\Phi$ is satisfiable. We now just need to check that the axioms of $E$ are valid on the coalgebra $q_{\uf\mathscr{L}_E}\circ\gamma: \pow\uf\mathscr{L}_E\to T\pow\uf\mathscr{L}_E$. It is quite clear that this should be the case, by construction of $E^*$. For every axiom $b=\top$ in $E$, the axiom $b'=\top$ is in $E^*$ for every pre-image $b'$ of $b$ (under $(-)^q$). Thus $\gamma\models b'=\top $, and thus by Theorem \ref{ch4:thm:compSem}, $q_{\uf\mathscr{L}_E}\circ\gamma\models (b')^q=b=\top$. Thus every axiom of $E$ holds everywhere on our coalgebra $q_{\uf\mathscr{L}_E}\circ\gamma: \pow\uf\mathscr{L}_E\to T\pow\uf\mathscr{L}_E$. as desired.

\begin{remark}
There are actually two ways to get a $K$-algebra embedding of $\mathscr{L}_E$ into $\pow\uf\mathscr{L}_E$. The strategy described above uses the assumption that $\hat{\lambda}$ has a right inverse at $\mathscr{L}_E$. This allows us to embed $\mathsf{Q}\mathscr{L}_E$ into $\pow\uf \mathsf{Q}\mathscr{L}_E$. The other strategy would be to assume that $\delta$ has a right inverse at $\mathscr{L}_E$, this would allow us to first embed $\mathscr{L}_E$ into its $L$-algebra canonical extension $\pow\uf \mathscr{L}_E$, and then use the transformation $\mathsf{Q}$ to get an embedding $\mathsf{Q}\mathscr{L}_E$ into $\mathsf{Q}\pow\uf\mathscr{L}_E$. However, we cannot use canonicity in the later strategy since $\mathsf{Q}\pow\uf\mathscr{L}_E$ is not the canonical extension of $\mathsf{Q}\mathscr{L}_E$.
\end{remark}

\begin{example}[Nabla logic for the finitary powerset functor $\cpow_f$]
Recall that the logic is defined by $L:\BA\to\BA$ defined by $L=\Free\cpow_f\Forg(-)/\simeq_{\KKV(\cpow_f)}$ and is interpreted in $\cpowf$ coalgebras via $\delta: L\pow\to\pow\cpowf$ which takes the `generators' of $L\pow X$ to the set of their $\cpowf$-lifted members at each $X$. A presentation of $\cpowf$ is given by the functor
\[
\polyFunc[\omega] X=\left(\coprod_{n\in\omega} X^n\right)
\]
and a regular quotient $q: \polyFunc[\omega]\epi \cpowf$ given by 
\[
q_X(x_1,\ldots,x_n)=q_X(y_1,\ldots,y_p) \text{ iff }\{x_1,\ldots,x_n\}=\{y_1,\ldots,y_p\}
\] 
the tuples with the same underlying set of elements are identified. the functor $\polyFunc[\omega]$ defines a nabla-style logic defined by the functor $K:\BA\to\BA$ given by $K=\Free \polyFunc[\omega]\Forg(-)/\simeq_{\KKV(\polyFunc[\omega])}$ and interpreted via a semantic transformation $\lambda: K\pow\to \pow\polyFunc[\omega]$ which takes the `generators' of $K\pow X$ to the set of their $\polyFunc[\omega]$-lifted members at each $X$. We have shown in Chapter 4 how the natural transformation $q$ can be used to construct a natural transformation $r: K\epi L$, in such a way that the semantics $\lambda$ and $\delta$ associated with $K$ and $L$ are \emph{compatible} with respect to $r$ and $q$ in the sense of Definition \ref{ch4:def:compSem}. We showed in Theorem \ref{ch4:thm:syntaxthm} that the transformation $r: K\epi L$ defines a translation morphism $\xi_V:\Free_K\Free V\epi \mathsf{Q}\Free_L\Free V$, where $\mathsf{Q}:\Alg_{\BA}(L)\to\Alg_{\BA}(K)$ is the functor defined by pre-composing with $r$.

As an illustration of the construction above, let us consider the axiom for transitivity $(4)=\nabla\{\nabla\{p\}\}\to\nabla\{p\}$, and let us consider an arbitrary set $\Phi$ of $L$-formulas which are $\KKV(\cpowf)+(4)$-consistent. We follow the steps detailed above. 
\begin{enumerate}[(Step 1)]
\item We start by building the Lindenbaum-Tarski algebra as the coequalizer of the equation $(4)=\top$, i.e.  $q_{(4)}: \Free_L\Free V\epi \mathscr{L}_4=\Free_L\Free V/(4)$. It is an $L$-algebra, and we will denote its structure map by $\alpha: L\mathscr{L}_4\to\mathscr{L}_4$. Note that since $\Phi$ is a $\KKV(\cpowf)+(4)$-consistent set of formulas, $\Phi$ can be extended to an ultrafilter $\langle\Phi\rangle\eup$ in $\mathscr{L}_4$. Note also that $\mathscr{L}_4$ comes with a canonical valuation $\eta_V^L: \Free V\to \mathscr{L}_4$ which simply injects propositional formulas into $\mathscr{L}_4$.
\item We then lift the axiom (4) to the language of the functor $K$ as was shown in in Section \ref{ch4:sec:prooftrans} by taking the pre-images under $\xi_V$ of (4) (and boolean combinations thereof), for example $\nabla(\nabla(p,p),\nabla(p))\to\nabla(p,p,p)$. Let us denote this set by
$(4)^*$. Theorem \ref{ch4:thm:prooftrans} tells us that $\mathsf{Q}\mathscr{L}_4$ belongs to the variety of defined by $(4)^*$, since $\mathscr{L}_4$ belongs to the variety defined by $(4)$.

\item Let us check that these inverse images are canonical. As was shown in \cite{2013:self} and Examples \ref{ch5:ex:treef} and \ref{ch5:ex:treeOmega}, the axioms of the logic for trees of unbounded branching degrees enforce that the operators defined by $\Free \polyFunc[\omega]\Forg$ preserve meets and joins in each argument, and are thus in particular smooth. All elements of $(4)^*$ can be written from elements of the general form:
\[
\vee(\neg \nabla(\nabla(p,\ldots,p),\ldots,\nabla(p,\ldots,p)), \nabla(p,\ldots,p))
\]
which is always Sahlqvist since $\vee$ is conservative, $\nabla(p,\ldots,p)$ is stable, and $\neg \nabla(\nabla(p,\ldots,p),\ldots,\nabla(p,\ldots,p))$ is the negation of a positive term. In other words each element of $(4)^*$ is canonical. By canonicity of the elements of $(4)^*$ is follows that since $\mathsf{Q}\mathscr{L}_4$ belongs to the variety defined by $(4)^*$ , so does its canonical extension.

Let us now check that validity on the canonical extension implies validity on the J\'{o}nsson-Tarski extension. Recall from Example \ref{ch5:ex:treeOmegaStrComp} that the logic defined by $K$ is strongly complete with respect to $\polyFunc[\omega]^*$-coalgebra. The J\'{o}nsson-Tarski extension of $\mathsf{Q}\mathscr{L}_4$ is thus given by
\[
L\pow\uf \mathscr{L}_4\xto{\delta_{\uf \mathscr{L}_4}}\pow \polyFunc[\omega]^* \uf \mathscr{L}_4\xto{\pow \hat{\delta}_{\mathscr{L}_4}\inv}\pow\uf K\mathscr{L}_4\xto{\pow\uf (\alpha\circ r_{\mathscr{L}_4)}}\pow\uf \mathscr{L}_4
\]  
together with the valuation $\tilde{v}: \pow\uf\eta_V^L\circ \eta_{\Free V}: \Free V\to \pow\uf\mathscr{L}_4$. It easy to check that $\delta_{\uf\mathscr{L}_4}$ defines maps which preserve all joins in each argument, and by Theorem \ref{ch5:thm:CanExtJonTarksiExt} it follows that formulas which are valid on the canonical extension of $\mathsf{Q}\mathscr{L}_4$ are also valid on its J\'{o}nsson-Tarski-extension. 

\item Let us gather our results in the following diagram:
\[
\xymatrix@C=12ex
{
K\Free_K \Free V+\Free V\ar[ddd]_{\langle-\rangle_V+\eta_V^K}\ar[r]_{K (q_{(4)}\circ\xi_V)+\id_{\Free V}} \ar@/^2pc/[rr] ^{K\lsem - \rsem_{\pow\uf \mathscr{L}_E}+\id_{\Free V}} & 
K \mathscr{L}_4 + \Free V \ar[ddr]_{\eta^K_{\mathscr{L}4}+\id_{\Free V}}\ar[ddd]_{\alpha\circ r_{\mathscr{L}_4}+\eta_V^L}  \ar[r]^{K\eta_{\mathscr{L}_4}+\id_{\Free V}}& K\pow\uf \mathscr{L}_4 +\Free V\ar[d]^{\delta_{\uf \mathscr{L}_4}+\id_{\Free V}}
\\
& & \pow \polyFunc[\omega]^* \uf \mathscr{L}_4+\Free V\ar[d]^{\pow \hat{\delta}\inv_{\mathscr{L}_4}+\id_{\Free V}}\\
& & \pow \uf K \mathscr{L}_4 +\Free V\ar[d]_{\pow\uf (\alpha\circ r_{\mathscr{L}_4})+\pow\uf\eta_V^L\circ \eta_{\Free V}}\\
\Free_K\Free V\ar[r]_{q_{(4)}\circ\xi_V}\ar@/_2pc/[rr] _{\lsem - \rsem_{\pow\uf \mathscr{L}_4}}& \mathscr{L}_4\ar[r]_{\eta_{\mathscr{L}_4}} & \pow \uf \mathscr{L}_4
}
\]
We know from the previous step and from the commutativity of the diagram above that all the formulas of $(4)^*$ are valid in the coalgebra 
\[
\uf \mathscr{L}_4\xto{\uf(\alpha\circ r_{\mathscr{L}_4})}\uf K\mathscr{L}_4\xto{\hat{\delta}\inv_{\mathscr{L}_4}}\polyFunc[\omega]^*\uf\mathscr{L}_4
\]
together with the canonical valuation arising from $\tilde{v}$ by adjointness: $\hat{v}: \uf \mathscr{L}_4\to \uf\Free V=\mathcal{Q}V$.
For $a\in\Phi$, since $r$ is epi there exists a pre-image $a'$ under $\xi_V$. It is clear from the diagram above that 
\begin{align*}
\lsem a'\rsem_{\pow\uf\mathscr{L}_4}&=\eta_{\mathscr{L}_4}\circ q_{(4)}\circ \xi_V(a')\\
&=\{u\in \uf \mathscr{L}_4\mid a\in u\}
\end{align*}
and thus $\langle\Phi\rangle\eup\in \lsem a'\rsem_{\pow\uf\mathscr{L}_4}$. Equivalently, for every $a\in\Phi$ we can find $a'$ such that $\xi_V(a')=a$ and such that
\[
\langle\Phi\rangle\eup, \hat{\delta}\inv_{\mathscr{L}_4}\circ \uf(\alpha\circ r_{\mathscr{L}_4}),\hat{v}\models a'
\]

\item We now need to build a $\cpowf$-model with the same property. This is the most delicate step. Notice that from the previous steps we know how to build an $\polyFunc[\omega]^*-$model, but since we want to quotient our model into a $\cpowf$ model, we would have wanted a $\polyFunc[\omega]-$model and used $q$ to turn in into a $\cpowf$-model. 

We use a different trick from that developed in \cite{2013:self} to turn our $\polyFunc[\omega]^*-$model into a $\cpowf-$model. Here we simply define a extension of the epi-transformation $q: \polyFunc[\omega]\epi \cpowf$ as follows
\[
q^*_X:\polyFunc[\omega]^*X\epi \cpowf X, x\mapsto\begin{cases} q(x) &\text{if }x\neq *\\
\emptyset &\text{else}
\end{cases}
\]
We now check that the triples defined by $(K, \polyFunc[\omega]^*,\lambda^*: K\pow\to\pow \polyFunc[\omega]^*)$ and $(L,\cpowf,\delta: L\pow\to \pow\cpowf)$ - where $\lambda$ is the obvious extension of $\lambda: K\pow\to\pow \polyFunc[\omega]$ as defined in Examples \ref{ch5:ex:treef} and \ref{ch5:ex:treeOmega} - are compatible in the sense of Definition \ref{ch4:def:compSem}, i.e. that
\[
\lambda^*_X(a)\subseteq \pow q^*_X\circ \delta_X\circ r_{\pow X}(a)
\]
But as was shown in Section \ref{ch4:subsec:nabla}, we already know that
\[
\lambda_X(a)\subseteq \pow q_X\circ \delta_X\circ r_{\pow X}(a)
\]
It is clear that $\lambda_X(a)=\lambda_X^*(a)$ for any $a\in K\pow X$, since the only difference is that $\lambda_X^*$ has an additional point $*$ in its codomain (which is never reached by $\lambda_X^*$). Moreover, by definition of $q^*_X$ it is clear that $\pow q_X(U)\subseteq \pow q_X^*(U)$, with the inequality being strict when $\emptyset\in U$. We thus have
\[\lambda_X^*(a)=\lambda_X(a)\subseteq \pow q_X\circ \delta_X\circ r_{\pow X}(a)\subseteq \pow q_X^*\circ \delta_X\circ r_{\pow X}(a)
\]
and the triples above therefore define compatible semantics. We can now use Theorem \ref{ch4:thm:compSem} to show that since for each $a\in\Phi$ there exists $a'\in \Free_K\Free V$ such that $\xi_V(a')=a$ and
\[
\langle\Phi\rangle\eup, \hat{\delta}_{\mathscr{L}_4\inv}\circ \uf(\alpha\circ q_{\mathscr{L}_4}),\hat{v}\models a',
\]
it must also be the case that
\[
\langle\Phi\rangle\eup, q^*_{\uf \mathscr{L}_4}\circ \hat{\delta}_{\mathscr{L}_4\inv}\circ \uf(\alpha\circ r_{\mathscr{L}_4}),\hat{v}\models \xi_V(a')=a
\]
We have thus found a $\cpowf$-model satisfying every formula of $\Phi$, moreover, since every formula of $(4)^*$ is valid on the frame defined by $\hat{\delta}_{\mathscr{L}_4\inv}\circ \uf(\alpha\circ q_{\mathscr{L}_4}): \uf\mathscr{L}_4\to \polyFunc[\omega]^*\uf\mathscr{L}_4$ and $\hat{v}$, it follows from Theorem \ref{ch4:thm:compSem} that $(4)$ is valid on the frame defined by $q^*_{\uf \mathscr{L}_4}\circ\hat{\delta}_{\mathscr{L}_4\inv}\circ \uf(\alpha\circ q_{\mathscr{L}_4}): \uf\mathscr{L}_4\to \cpowf\uf\mathscr{L}_4$ and $\hat{v}$.


\end{enumerate}

\end{example}

\printindex

\bibliographystyle{alpha}

\bibliography{thesis}

\end{document}